# NUCLEAR REACTIONS

## MECHANISM AND SPECTROSCOPY

## VOLUME II

Prof. Ron W. Nielsen

(aka Jan Nurzynski)

Griffith University

2016

# Nuclear Reactions

## Mechanism and Spectroscopy
## Volume II

**Prof. Ron W. Nielsen**

(aka Jan Nurzynski)

Griffith University, Gold Coast, Qld, 4222, Australia

ronwnielsen@gmail.com

**About the Author:** http://home.iprimus.com.au/nielsens/ronnielsen.html

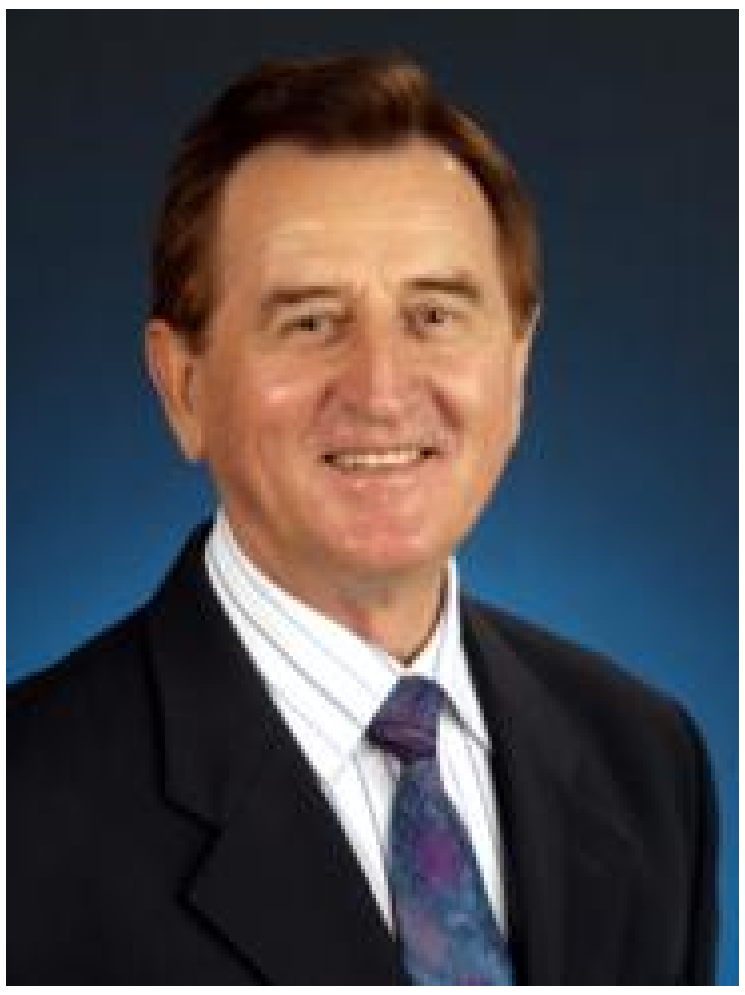

**Abstract:** Volume II of two. This document could be of interest to anyone who wants to have a comprehensive inside information about the research in nuclear physics from its early beginnings to later years. It describes highlights of my research work from the late 1950s to the late 1980s, during the best years of nuclear research, when this field of study was wide opened for its exploration. It presents a panorama of experimental and theoretical methods used in the study of nuclear reactions (their mechanism and their application to the study of nuclear structure), the panorama ranging from simple detection techniques used in the early research to more complicated in later years, from simple theoretical interpretations to more complicated descriptions. This document describes my research work in Poland, Australia, Switzerland and Germany using various particle accelerators and a wide range of experimental and theoretical techniques. It presents a typical cross section of experimental and theoretical work in the early and later stages of nuclear research in the field of nuclear reactions.

# Contents

## 23

# A Study of Spin-orbit and Tensor Interaction of Polarized Deuterons with $^{60}$Ni and $^{90}$Zr Nuclei

***Key features:***

1. We have carried out precise measurements of the differential cross sections $\sigma_0$ and all four analyzing powers $iT_{11}$, $T_{20}$, $T_{21}$ and $T_{22}$ for the elastic scattering of polarized deuterons from $^{60}$Ni and $^{90}$Zr nuclei.
2. The angular distributions were measured in the energy range of 9 - 15 MeV (lab) and for the scattering angles of $40^0$ - $160^0$ (c.m.). In addition, we have also measured excitation functions for the differential cross sections and for the $T_{20}$ tensor analyzing power.
3. We have analysed our experimental results using optical model potential that contained the central, spin-orbit, and tensor $T_R$ potentials. Each potential had both real and imaginary components.
4. We have found that five of the optical model parameters could be fixed and that three could be described using mass-dependent formulae. The remaining 10 parameters were varied to optimise the fits to the experimental data for individual energies and target nuclei.
5. Our analysis resulted in determining the central, spin-orbit and tensor potentials for the deuteron-nucleus interaction.
6. In particular, we have found that the imaginary spin-orbit component was necessary to improve the fits to the data thus confirming the earlier results of Goddard and Haeberli (1977).

***Abstract***: Angular distributions of the differential cross-sections $\sigma_0$ and analyzing powers $iT_{11}$, $T_{20}$, $T_{21}$ and $T_{22}$ have been measured for the elastic $(\vec{d},d)$ scattering from $^{60}$Ni and $^{90}$Zr over a wide range of scattering angles. The incident deuteron energies were at 9, 12 and 15 MeV for $^{60}$Ni nuclei and 10, 11, 12 and 15 MeV for $^{90}$Zr. Excitation functions for $\sigma_0$ and $T_{20}$ have been also measured at 175°(lab) in the approximate energy range of 6-13 MeV for both target isotopes. The experimental results have been analysed using the optical model with the complex central, spin-orbit and tensor $T_R$ potentials. Excellent fits to all experimental angular distributions have been obtained. The main features of the excitation functions have been also well reproduced. Out of the total of 18 parameters describing the interaction potential, five could be fixed and three could be constrained by simple mass-dependent functions. Further evidence for the presence of an imaginary component of the spin-orbit and tensor potentials is supplied by the analysis of the present data.

## Introduction

The central part of the nuclear interaction, which can be studied using angular distributions of the differential cross sections, is relatively well known. Some information on the spin-depended forces can be also obtained by analyzing differential cross sections but a more reliable way is to measure and analyse the distributions of analyzing powers. In particular distributions of tensor analyzing powers can yield information not only about spin-orbit but also about tensor forces.

Spin-orbit forces are relatively well known for deuterons but usually only a real component is used in analyses of experimental data. Goddard and Haeberli (1978) found evidence for the presence of the imaginary component. These authors measured and analysed angular distributions for deuteron scattering from medium weight nuclei in the energy range of 10-15 MeV. They have found that the fits to the data can be improved significantly if an imaginary component is included in the spin-orbit interaction. The depth of the imaginary spin-orbit potential used in their calculations was about a half of the real component. Both components had the same sign.

Much less is known about the tensor interaction because to study it one needs to have good quality data for the tensor analyzing powers. Unfortunately, such data are scarce.

In our study we have carried out precise measurements of the differential cross sections and of all four analyzing powers $iT_{11}$, $T_{20}$, $T_{21}$ and $T_{22}$ with the aim to study the details of spin-orbit and tensor interactions.

## The experimental method and results

The experimental method and procedure was described in Chapters 15 and 17. Briefly, the measurements were made using the ETH atomic beam polarized ion source (see the Appendix G) and the EN tandem electrostatic accelerator. Targets consisted of approximately 1 mg/cm$^2$ self-supporting foils of enriched isotopes with the enrichment greater than 98%. To allow for a simultaneous detection of scattered particles on both sides of the beam, the targets were mounted at 90° to the beam direction for the forward and the backward angle measurements, and rotated by 45° for the intermediate angles. Four silicon surface-barrier detectors were mounted on both sides of the beam line, 7.5° apart and 25 cm from the centre of the chamber.

The differential cross sections were derived directly from the yields obtained in the measurements of the analyzing powers. The absolute normalization was determined by measuring Rutherford scattering of 4 MeV deuterons at a number of angles below 60° (lab). An overall normalization factor was included in the optical-model calculation and was allowed to vary during the search procedure. The normalization of the 12 MeV data for $^{60}$Ni had to be changed by 10%. For all other measurements, the change was less than 3%.

The experimental angular distributions are shown in Figures 23.1-23.3 and the excitation functions in Figures 23.4-23.7. They are compared with the optical-model calculations discussed in the next section. Where no error bars are shown, the uncertainty is smaller than the size of the experimental points.

## The optical-model analysis

We have carried out optical model analysis using nuclear potential, which included not only the usual central part but also spin-orbit and tensor interactions. The parameterization of the optical model potential has been described earlier but it is convenient to list explicitly the components used in our analysis of $^{60}$Ni and $^{90}$Zr data.

The components of the optical model potential incorporate the form factor $f(r, r_i, a_i)$, which is defined as:

$$f(r, r_i, a_i) = \frac{1}{1 + e^{x_i}}$$

where

$$x_i = \frac{r - r_i A^{1/3}}{a_i}$$

Parameters $r_i$ and $a_i$, together with the depths of the potentials (see below) define completely each component of nuclear interaction.

The nuclear potential used in our calculations had the following form:

$$U(r) = U_0(r) + U_{s.o.}(r) + U_T(r)$$

where $U_0(r)$ is the central part of the optical model potential, $U_{s.o.}(r)$ the spin-orbit part, and $U_T(r)$ the tensor part.

$$U_0(r) = -Vf(r, r_0, a_0) - i4a'_0 W_D \frac{d}{dr} f(r, r'_0, a'_0)$$

$$U_{s.o.}(r) = \lambda_\pi^2 \frac{1}{r}\left[V_{s.o.} \frac{df(r, r_{s.o.}, a_{s.o.})}{dr} + iW_{s.o.} \frac{df(r, r'_{s.o.}, a'_{s.o.})}{dr}\right] \mathbf{S} \cdot \mathbf{L}$$

where $\lambda_\pi^2$ is the square of the pion Compton wavelength,

$$\lambda_\pi^2 = \left(\frac{\hbar}{m_\pi c}\right)^2 = 2\, fm^2$$

The tensor interaction can be constructed using the following three terms:

$$T_R = (\mathbf{S} \cdot \mathbf{r})^2 - \frac{2}{3}$$

$$T_L = (\mathbf{L} \cdot \mathbf{S})^2 + \frac{1}{2}(\mathbf{L} \cdot \mathbf{S}) - \frac{2}{3}\vec{L}^2$$

$$T_P = (\mathbf{S} \cdot \mathbf{r})^2 - \frac{2}{3}\mathbf{p}^2$$

where $\mathbf{p}$ is the relative momentum.

Earlier studies (Goddard 1977; Keaton and Armstrong 1973; Stamp 1970) suggested that the important tensor interaction is represented by the $T_R$ term. We have therefore used only this term in our calculations. The form of the tensor part of our optical model potential was as suggested by Keaton and Armstrong (1973):

$$U_T(r) = -\lambda_\pi^2 r \left\{ V_{TR} \frac{d}{dr}\left[\frac{1}{r} \frac{df(r, r_{TR}, a_{TR})}{dr}\right] + iW_{TR} \frac{d}{dr}\left[\frac{1}{r}\frac{d}{dr}\left(\frac{4e^{x'_{TR}}}{(1 + e^{x'_{TR}})^2}\right)\right]\right\} T_R$$

where

$$x'_{TR} = \frac{r - r'_{TR} A^{1/3}}{a'_{TR}}$$

It is convenient to summarize the full complement of the 18 parameters describing the nuclear interaction as used in our calculations. This summary is presented in Table 23.1

Table 23.1

Summary of the parameters describing nuclear interaction used in the optical model analysis of the elastic scattering of polarized deuterons from $^{60}Ni$ and $^{90}Zr$ nuclei

| | Real component | Imaginary component |
|---|---|---|
| Central potential | $V, r_0, a_0$ | $W_D, r'_0, a'_0$ |
| Spin-orbit potential | $V_{s.o.}, r_{s.o.}, a_{s.o.}$ | $W_{s.o.}, r'_{s.o.}, a'_{s.o.}$ |
| Tensor potential | $V_{TR}, r_{TR.}, a_{TR}$ | $W_{TR}, r'_{TR.}, a'_{TR}$ |

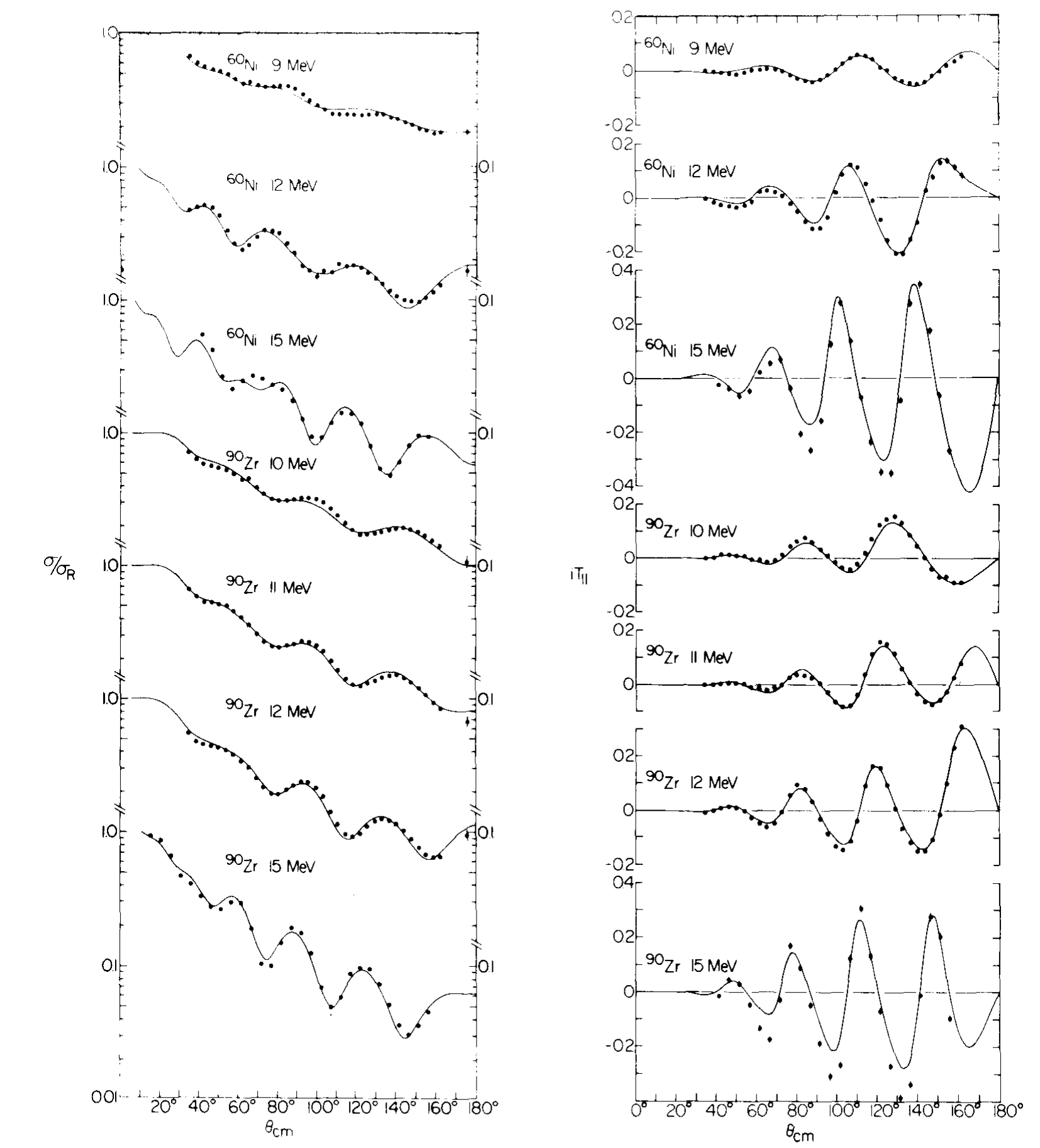

Figure 23.1. Angular distributions of the differential cross sections (left-hand side) and vector analyzing powers, $iT_{11}$, (right-hand side) for the elastic scattering of deuterons from $^{60}Ni$ and $^{90}Zr$ nuclei. Experimental results (points) are compared with the optical-model calculations (solid lines) generated by the potential parameters listed in Tables 23.2 and 23.3. If not shown, the error bars are not larger than the experimental points.

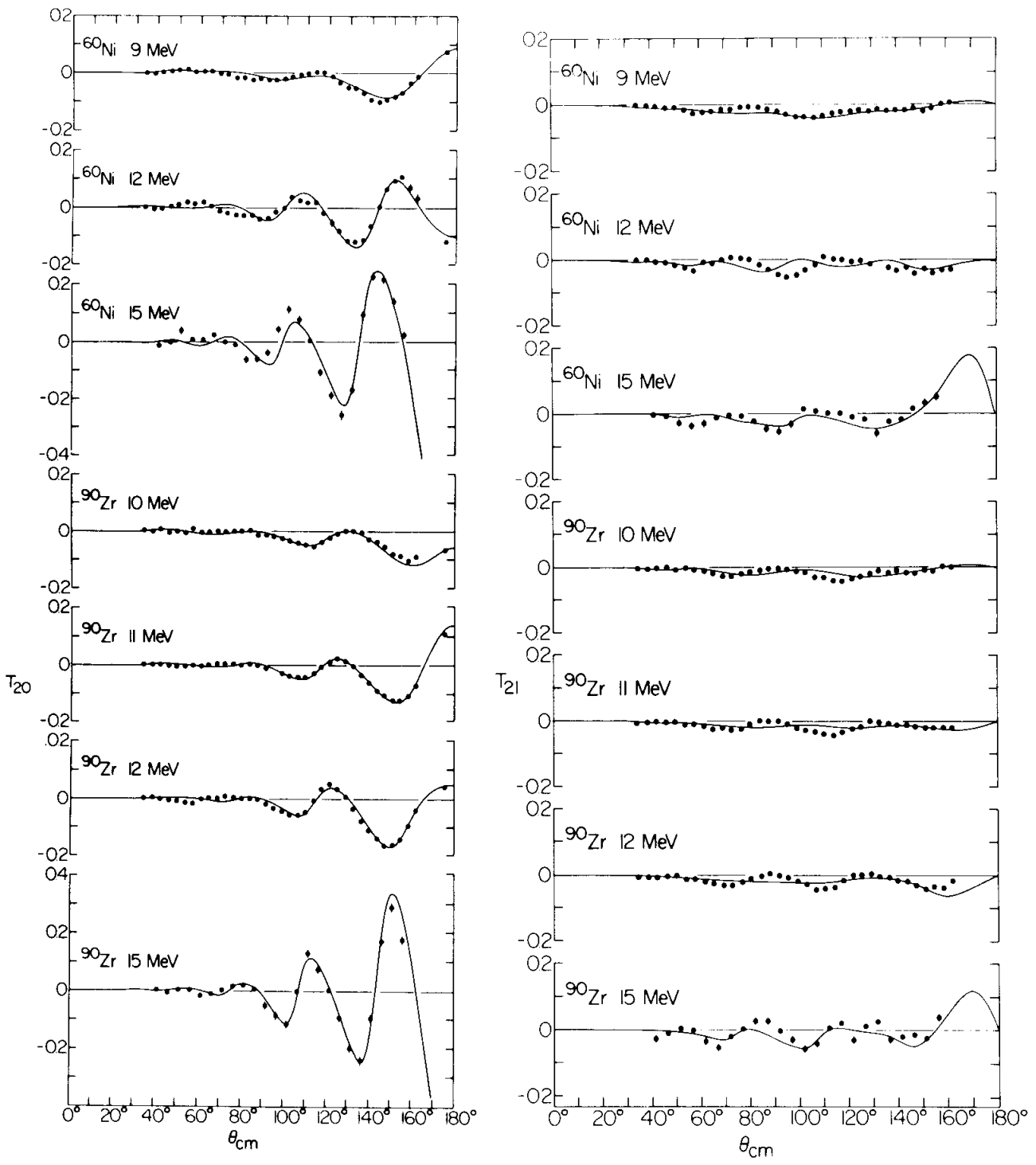

Figure 23.2. Angular distributions of the tensor analyzing powers $T_{20}$, $T_{21}$ for the elastic scattering of polarized deuterons from $^{60}$Ni and $^{90}$Zr nuclei. See the caption to Figure 23.1.

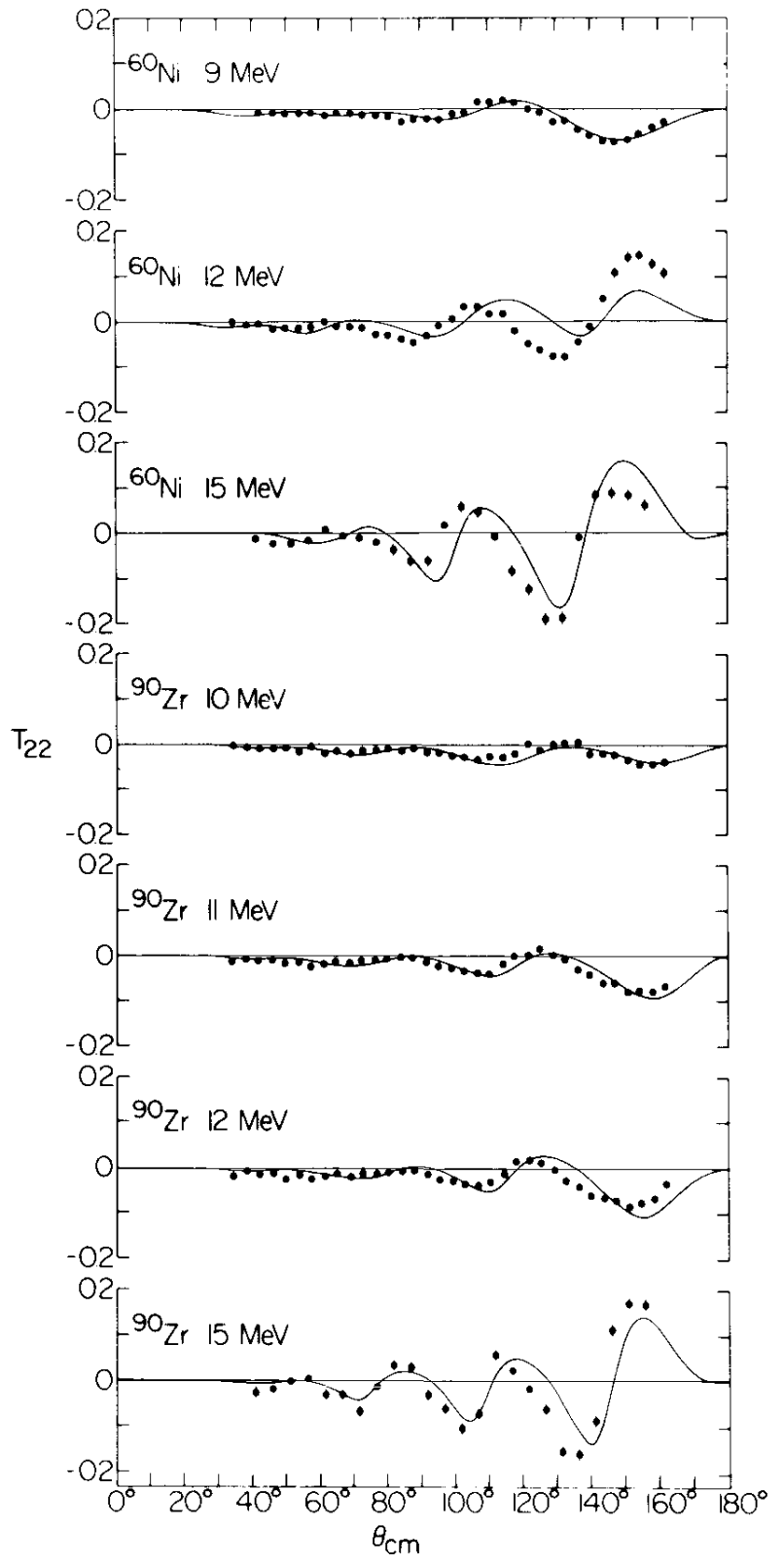

Figure 23.3. Angular distributions of the tensor analyzing powers $T_{22}$ for the elastic scattering of polarized deuterons from $^{60}$Ni and $^{90}$Zr nuclei. See the caption to Figure 23.1.

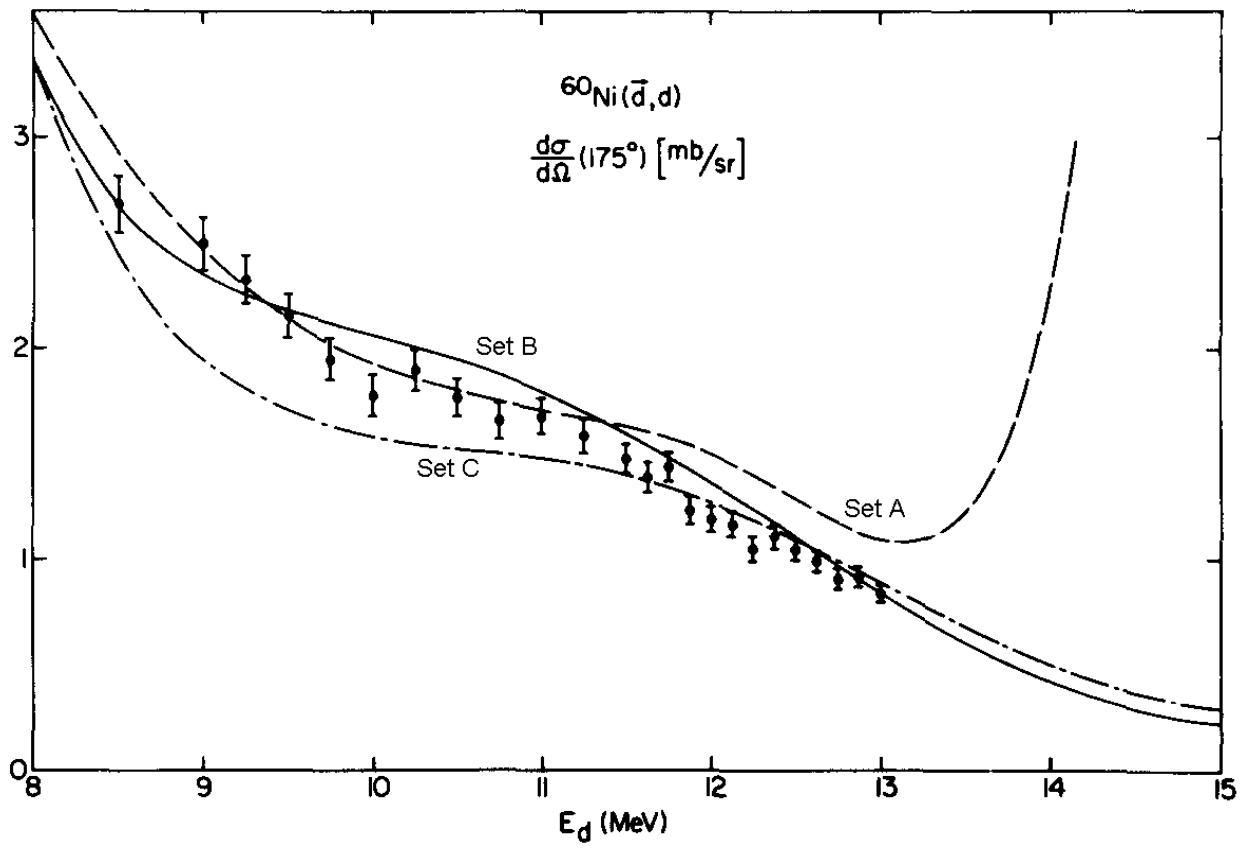

Figure 23.4. Energy dependence of the differential cross sections for the elastic $(\vec{d},d)$ scattering from $^{60}$Ni measured at 175° are compared with the optical-model calculations. Theoretical curves correspond to parameters listed in Tables 23.2 and 23.3.

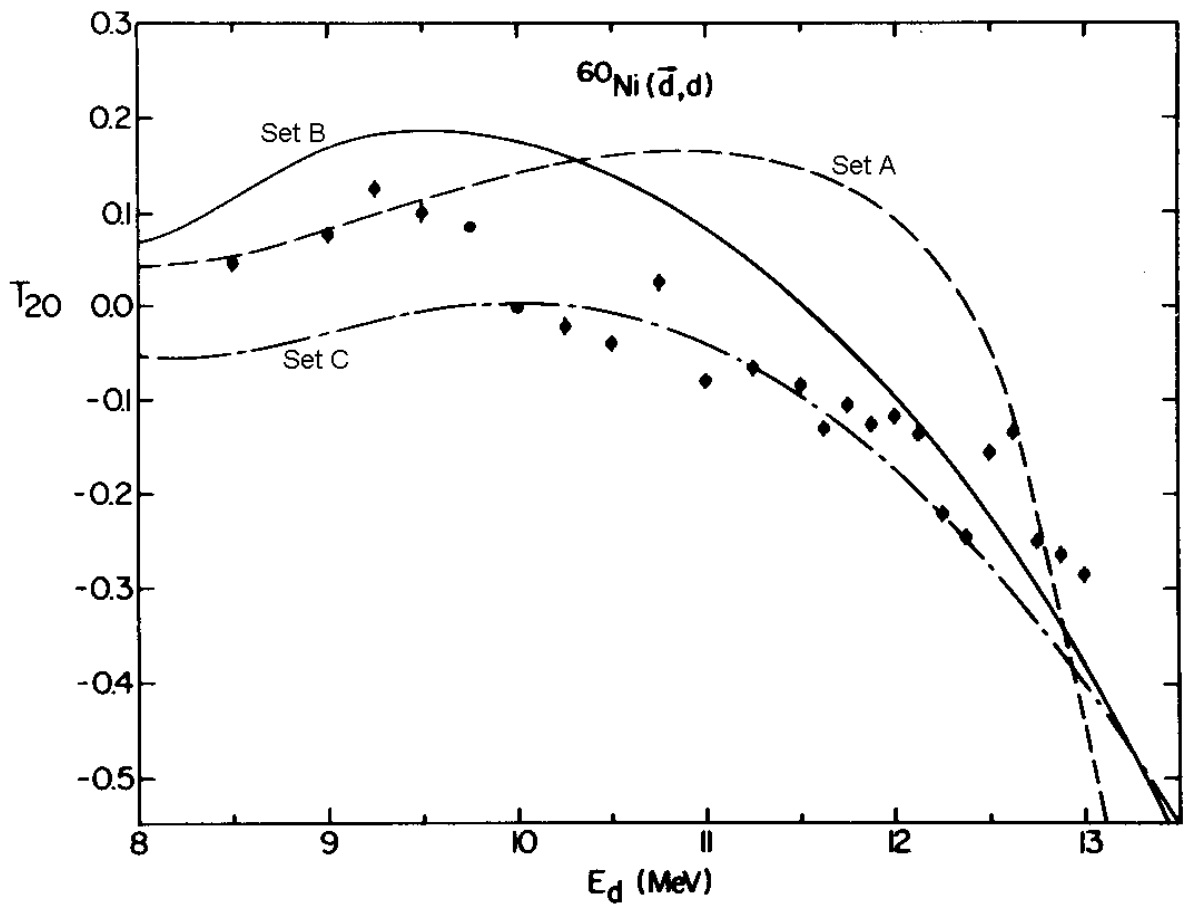

Figure 23.5. Energy dependence of the tensor analyzing power $T_{20}$ for the elastic $(\vec{d},d)$ scattering from $^{60}$Ni. See the caption to Figure 23.4.

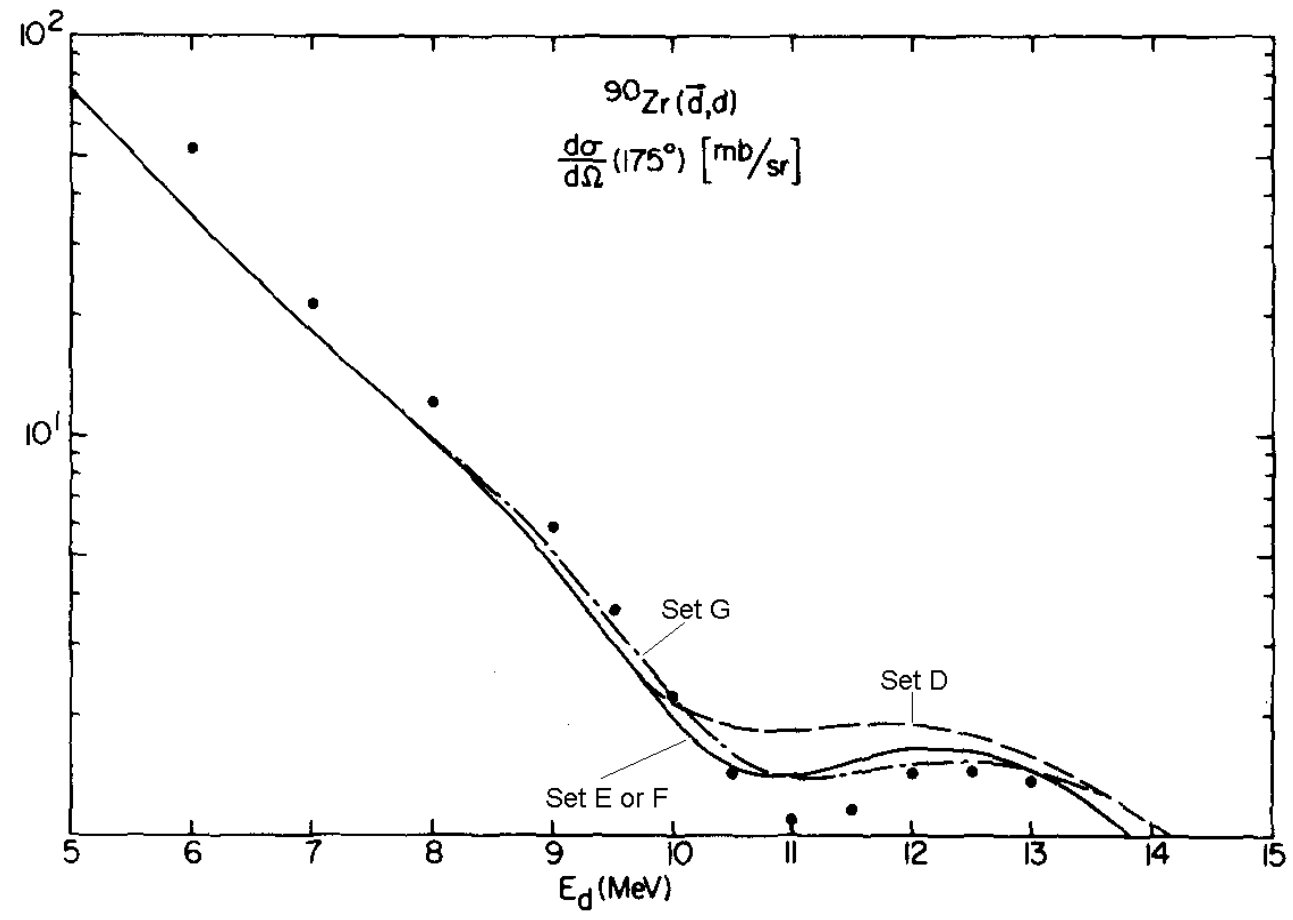

Figure 23.6. Energy dependence of the differential cross sections for the elastic $(\vec{d},d)$ scattering from $^{90}$Zr measured at 175° are compared with the optical-model calculations. Theoretical curves correspond to parameters listed in Tables 23.2 and 23.3.

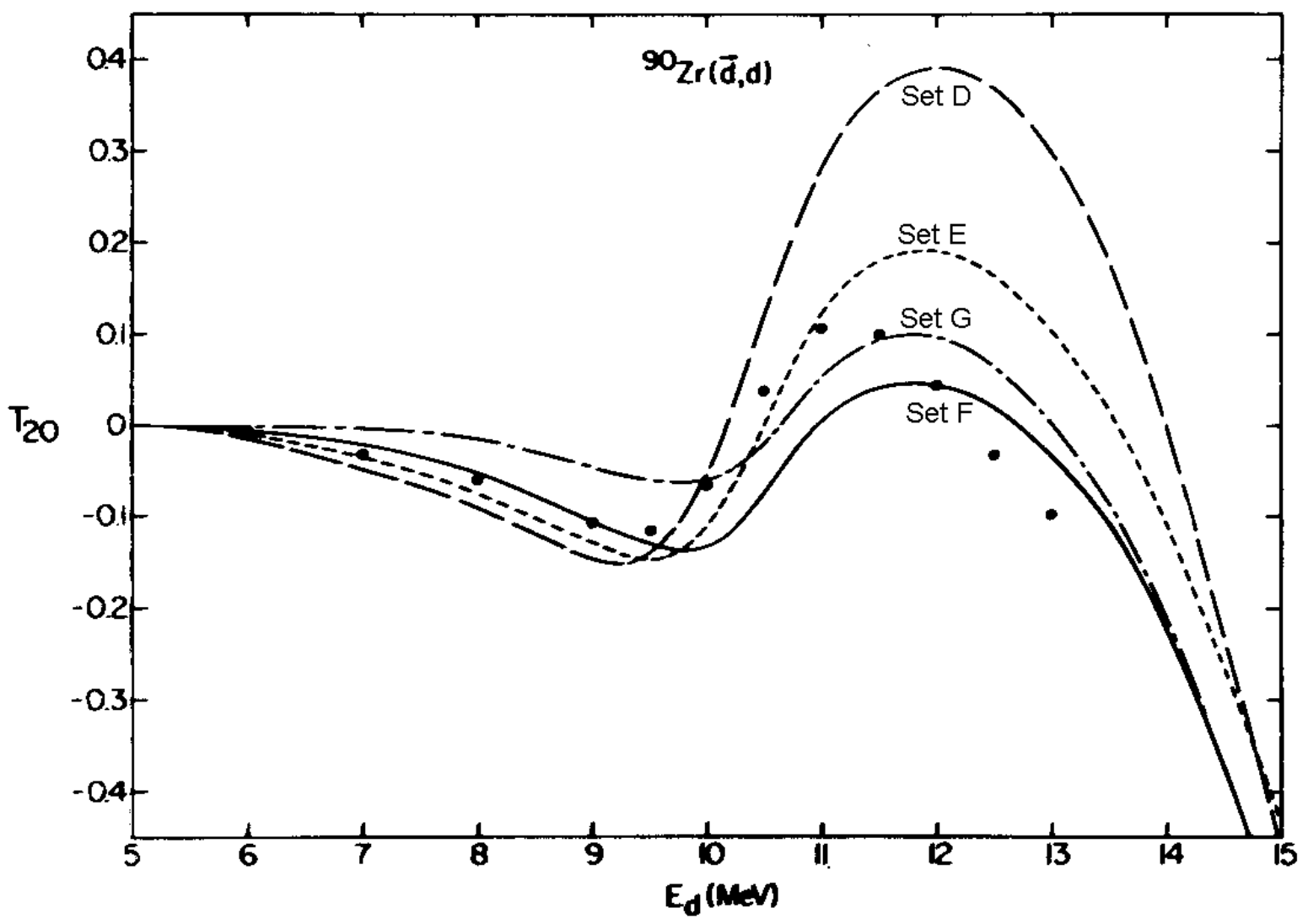

Figure 23.7. Energy dependence of the tensor analyzing power $T_{20}$ for the elastic $(\vec{d},d)$ scattering from $^{90}$Zr measured at 175° are compared with the optical-model calculations. Theoretical curves correspond to parameters listed in Tables 23.2 and 23.3.

Table 23.2

Fixed parameters and parameters described by mass-dependent formulae as used in the optical model analysis of the elastic scattering of polarized deuterons from $^{60}$Ni and $^{90}$Ze nuclei

| | | Real Component | | Imaginary Component |
|---|---|---|---|---|
| Central | $V$ | * | $W_D$ | * |
| | $r_0$ | 1.14 | $r_0'$ | $1.69-0.07A^{1/3}$ |
| | | | | $1.20+0.85A^{-1/3}$ |
| | $a_0$ | 0.8 | $a_0'$ | $0.20+0.14A^{1/3}$ |
| | | | | $1.29-2.10A^{-1/3}$ |
| Spin-orbit | $V_{s.o.}$ | * | $W_{s.o.}$ | 2.0 |
| | $r_{s.o.}$ | * | $r_{s.o.}'$ | 0.9 |
| | $a_{s.o.}$ | * | $a_{s.o.}'$ | 0.55 |
| Tensor | $V_{TR}$ | * | $W_{TR}$ | * |
| | $r_{TR}$ | $1.89-0.075A^{1/3}$ | $r_{TR}'$ | * |
| | | $1.29+1.16A^{-1/3}$ | | |
| | $a_{TR}$ | * | $a_{TR}'$ | * |

The asterisk (*) refers to parameters that could be neither fixed nor described by mass-dependent formulae. These parameters are listed in Table 23.3. The two, alternative mass-dependent formulae for $r_0'$, $a_0'$, and $r_{TR}$ give almost identical values for these parameters (see Table 23.4).

Table 23.3

Individually adjusted optical model parameters used in the optical model analysis of the elastic scattering of polarized deuterons from $^{60}$Ni and $^{90}$Ze nuclei

| Parameter | $^{60}$Ni | | | $^{90}$Zr | | | |
|---|---|---|---|---|---|---|---|
| | A<br>9 MeV | B<br>12 MeV | C<br>15 MeV | D<br>10 MeV | E<br>11 MeV | F<br>12 MeV | G<br>15 MeV |
| $V_0$ | 97.43 | 94.18 | 91.97 | 99.60 | 98.88 | 98.05 | 95.97 |
| $W_D$ | 18.03 | 14.61 | 14.45 | 11.62 | 11.12 | 10.82 | 11.28 |
| $V_{s.o.}$ | 7.60 | 4.40 | 5.05 | 5.75 | 4.10 | 4.13 | 5.52 |
| $r_{s.o.}$ | 0.62 | 0.76 | 0.77 | 0.66 | 0.63 | 0.67 | 0.95 |
| $a_{s.o.}$ | 0.28 | 0.25 | 0.39 | 0.20 | 0.22 | 0.42 | 0.37 |
| $V_{TR}$ | 2.22 | 1.73 | 0.47 | 0.66 | 0.71 | 0.72 | 0.83 |
| $a_{TR}$ | 0.41 | 0.20 | 0.20 | 0.23 | 0.38 | 0.38 | 0.21 |
| $W_{TR}$ | 9.84 | 2.98 | 0.56 | 6.08 | 4.52 | 2.43 | 0.45 |
| $r'_{TR}$ | 1.03 | 0.99 | 0.82 | 1.09 | 1.07 | 1.07 | 0.80 |
| $a'_{TR}$ | 0.31 | 0.37 | 0.20 | 0.36 | 0.38 | 0.40 | 0.22 |

The final sets of parameters, which yield the best fits to the experimentally measured differential cross section and the analyzing powers are listed in Tables 23.2 and 23.3. Parameters which could be fixed or which could be described by simple analytic formulae are listed in Table 23.2. All the remaining parameters, which had to be individually adjusted to optimise the fits at the relevant energies and for given target nuclei are shown in Table 23.3

Results of our optical model calculations are compared with the experimental angular distributions in Figures 23.1-23.3. As can be seen, there is a generally excellent agreement between the theoretical calculations and the experimental results.

Calculations for the excitations functions are shown in Figures 23.4-23.7. Here, the agreement is less satisfactory but in general the main features are reproduced.

The imaginary spin-orbit interaction has a strong influence on the calculations. By introducing this component, we were able to reduce the $\chi^2$ values by an average of about 40%, mainly by improving the fits to the differential cross sections, which was both unexpected and surprising. The only observed quantity with an equal or worse $\chi^2$ value is $T_{21}$, which is expected to be sensitive mainly to tensor forces, but less to the central or spin-orbit potentials (Hooton and Johnson 1971).

Table 23.4

Parameters calculated using two alternative mass-dependent formulae listed in Table 23.2

| | $^{60}$Ni | $^{90}$Zr |
|---|---|---|
| $r'_0 = 1.69 - 0.07A^{1/3}$ | 1.42 | 1.38 |
| $r'_0 = 1.20 + 0.85A^{-1/3}$ | 1.42 | 1.39 |
| $a'_0 = 0.20 + 0.14A^{1/3}$ | 0.75 | 0.83 |
| $a'_0 = 1.29 - 2.10A^{-1/3}$ | 0.75 | 0.82 |
| $r_{TR} = 1.89 - 0.075A^{1/3}$ | 1.60 | 1.55 |
| $r_{TR} = 1.29 + 1.16A^{-1/3}$ | 1.59 | 1.55 |

In the present analysis, it has been possible to reduce the number of the free parameters from 18 to 10. The resulting fixed or mass-dependent parameters are listed in Table 23.2. This table contains *five* energy- and mass-independent parameters and *three* mass-dependent parameters. The two alterative formulae for $r_0'$, $a_0'$, and $r_{TR}$ give almost the same values for these parameters (see Table 23.4)

Our mass-dependent formula for $r_0'$ may be compared with the formulae derived by Perrin *et al.* (1977) and Griffith *et al.* (1977):

$$r_0' = 1.20 + 0.85A^{-1/3} \qquad \text{(our formula)}$$

$$r_0' = 1.15 + 0.75A^{-1/3} \qquad \text{(Perrin et al. 1977)}$$

$$r_0' = 1.25 + 0.85A^{-1/3} - 0.004E \qquad \text{(Griffith et al. 1977)}$$

The formula of Griffith *et al.* (1977) shows negligible energy-dependence, in agreement with our results.

Both the present and previous analyses show that the diffuseness $a_0'$ of the imaginary central potential is energy independent. Initially this parameter was allowed to vary. However, when the convergence was achieved, it was found that the resulting values could be described by a simple linear relation expressed in terms of $A^{1/3}$ or $A^{-1/3}$.

It is worth mentioning that this parameter influences mainly the normalization of the differential cross section. This correlation has been discussed Dickens and Perey (1965).

The radius $r_{TR}$ of the real component of the tensor potential decreases slightly with the increasing energy. The fits, however, have been found to be rather insensitive to this parameter. The mass-dependent but energy-independent formula for this parameter gives good representation of this parameter.

The remaining *ten* parameters were allowed to vary during the final search. As the fits to the experimental data are sensitive to small changes in the depth $V$ of the central potential, this parameter was never kept fixed. However, the depth $V$ can be described by an analytic expression, which gives an excellent overall description of its mass dependence:

$$V = 95 + 1.4ZA^{-1/3} - 0.8E_{c.m.}$$

where $E_{c.m.}$ is the centre-of-mass energy in MeV. This relation is in agreement with the formulae found by Perrin *et al.* (1977) and Griffith *et al.* (1977)

The other nine parameters depend on the mass of the target nucleus and on the incident deuteron energy. Some systematic behaviour can be noticed for the depth and the radius of the imaginary tensor potential. They both decrease with the increasing energy. The depth of the imaginary central potential also increases slightly with the increasing energy but its value is around 13 MeV in agreement with the results of Perrin *et al.* (1977).

## Summary and conclusion

We have carried out precise measurements of the angular distributions for the five observables: $\sigma_0(\theta)$, $iT_{11}(\theta)$, $T_{20}(\theta)$, $T_{21}(\theta)$ and $T_{22}(\theta)$. Our energy range was 9-15 MeV and angular range $40^0$-$160^0$ (c.m.) We have also measured excitations functions for the differential cross sections, $\sigma_0$, and the tensor analyzing power, $T_{20}$.

We have carried out optical model analysis of our experimental results using a six-component optical model potential made of the central part with the surface absorption; the spin orbit part containing both real and imaginary components; and the tensor $T_R$ part also containing both real and imaginary components. The potential was described by a total of 18 components. However, we have found that *five* of them ($r_0$, $a_0$, $W_{s.o.}$, $r'_{s.o.}$ and $a'_{s.o.}$) could be fixed for the two target nuclei and for the incident deuterons energies. *Three* additional parameters $r'_0$, $a'_0$ and $r_{TR}$ could be described by mass-dependent formulae. We have found two alternative but equivalent formulae for each of these three parameters depending either on $A^{1/3}$ or $A^{-1/3}$.

Our spin-orbit parameters of the imaginary component are similar to those used by Goddard and Haeberli (1978). These parameters have a significant effect on the quality of fits to the experimental distributions. Thus, we have confirmed the earlier finding (Goddard and Haeberli 1978) that the imaginary spin-orbit component plays a significant role in optical model analyses of experimental data.

Measured tensor analyzing powers are reproduced very accurately by including a complex tensor $T_R$ potential. Its parameters, however, could not be determined very precisely.

Fits to the $T_{22}$ tensor analyzing powers get worse with the increasing energy. It has been suggested earlier that theoretical angular distributions of $T_{22}$ depend strongly on the central and the spin-orbit potentials (Hooton and Johnson 1971; Johnson 1977). Consequently, we suggest that these potentials may have some additional or an alternative structure. A possible way of improving the fits to the $T_{22}$ distributions would be to use different shapes for these potentials or to introduce an $l$ - dependent potential (Rawitscher 1977).

In general, we have obtained excellent fits to the angular distributions. Our study has resulted in important information not only about the central nuclear interaction but also about spin dependent potentials, including tensor interaction.

# 24

# Global Analysis of Deuteron-nucleus Interaction

***Key features:***

1. This study represents the first and the most extensive study of the deuteron-nucleus interaction. It includes not only the differential cross sections $\sigma_0$ but also the analyzing powers, $iT_{11}$, $T_{20}$, $T_{21}$ and $T_{22}$ measured and analysed for a wide range of target nuclei, $A$ = 40 – 90.
2. Experimental results have been analysed using the optical model containing not only the usual complex central part but also complex spin-orbit and tensor components.
3. In general, we have obtained excellent fits to the experimental data.
7. We have found global description for *all* 18 optical model parameters. All optical model parameters can be represented either by fixed values or the values calculated using simple mass-dependent formulae. Such global description is useful in analyses of transfer reaction data.
4. We have also found that both components of the central potential depend on the gamma transition probabilities.

***Abstract***: Angular distributions of the differential cross-sections $\sigma_0(\theta)$ and analyzing powers $iT_{11}(\theta)$, $T_{20}(\theta)$, $T_{21}(\theta)$ and $T_{22}(\theta)$ for the elastic scattering of 12 MeV polarized deuterons were measured in the angular range of 20°-175° (lab) using a wide range of spin-zero target nuclei with mass numbers $A$ = 40 – 90. The data were analysed using optical model potential with complex central, spin-orbit and tensor terms. With a few exceptions, excellent fits have been obtained to all measured angular distributions, yielding a set of global optical model parameters. The depths of the central part of the optical potentials have been found to depend on the structure of the target nuclei. Contrary to the results for selenium, which exhibit clear shell-closure effects, the data for $N$ = 28 nuclei do not exhibit any clear shell-closure correlation. This feature is attributed to interactions with higher configurations in this region. Irregularities in the optical-model parameters and problems in fitting the experimental results are discussed. Possible ways of improving theoretical description is to include coupling between elastic and reaction channels.

## 1. Introduction

Encouraged by our successful analysis of experimental data for the $^{60}$Ni and $^{90}$Zr nuclei (see Chapter 23) we have extended our study of spin-dependent forces to other target nuclei. To support this study we have carried out measurements of the angular distributions of the differential cross sections $\sigma_0(\theta)$ and of all four analyzing powers $iT_{11}(\theta)$, $T_{20}(\theta)$, $T_{21}(\theta)$ and $T_{22}(\theta)$ for $^{46, 48, 50}$Ti, $^{52, 54}$Cr, $^{54, 56}$Fe and $^{58}$Ni isotopes. In our global analysis of the data we have also included the previously measured distributions for $^{60}$Ni, $^{90}$Zr and $^{76,78,80,82}$Se isotopes.

## Experiment

The experimental method and procedure have been already fully described in Chapters 15 and 17. The targets in the present measurements were in the form of isotopically enriched self-supporting foils (see Table 24.1). The $^{46}$Ti and $^{50}$Ti targets contained substantial admixtures of $^{48}Ti$. Scaled contributions from this isotope, known from independent measurements, were subtracted from the measured values.

Some targets were found to have small (0.2%-0.7%) additional impurities of heavy elements such as W, Pt or Hg. These impurities were determined quantitatively by an analysis of proton induced X-rays (Bonani *et al.* 1978). Corrections due to the heavy-element impurities are only necessary at forward angles where the elastic peaks for the medium-heavy and the heavy nuclei could not be resolved. For the analyzing powers, the contributions from heavy elements are insignificant in this region.

Table 24.1
Isotopes used in our study

| Isotope | Enrichment | Thickness ($mg/cm^2$) |
|---|---|---|
| $^{46}Ti$ | 81.2 | 0.90 |
| $^{48}Ti$ | 99.4 | 0.96 |
| $^{50}Ti$ | 83.2 | 0.88 |
| $^{52}Cr$ | 99.9 | 0.80 |
| $^{54}Fe$ | 97.5 | 1.06 |
| $^{56}Fe$ | 99.9 | 1.23 |
| $^{58}Ni$ | 99.9 | 0.95 |
| $^{60}Ni$ | 99.8 | 0.83 |
| $^{76}Se$ | 86.0 | 0.48 |
| $^{78}Se$ | 98.0 | 0.64 |
| $^{80}Se$ | 94.0 | 0.58 |
| $^{82}Se$ | 97.0 | 0.86 |
| $^{90}Zr$ | 97.6 | 0.95 |

The absolute normalization of the differential cross section was determined from measurements of the Rutherford scattering of low-energy deuterons at forward angles. Corrections for the impurities in the target materials have been taken into account in the evaluation of the absolute values of the cross sections. The uncertainty in the absolute normalization varies in the range of 5-10% for various targets.

For the analyzing powers the statistical error was kept below 0.005. The uncertainties due to the inaccuracy in the scattering angles, the final geometry and the absolute values of the beam polarization have also been included. All these contributions, however, were small.

## Experimental results

The experimental results together with optical-model calculations are shown in Figures 24.1 to 24-5.

Figure 24.1 shows nuclei with the same number of protons ($Z = 22$) but different number of neutrons ($N = 24$, 26 and 28). In the simple shell model description, two protons are in the $1f_{7/2}$ orbit and neutrons are filling in the $1f_{7/2}$ sub-shell.

Figure 24.2 shows nuclei with the same number of neutrons ($N = 28$, which fill in the sub-shell $1f_{7/2}$) but with different number of protons ($Z = 22$, 24 and 26) in the $1f_{7/2}$ orbits.

Figure 24.3 shows nuclei with the same number of neutrons ($N = 30$, which fill in the$1f_{7/2}$ sub-shell and with two neutrons in the $2p_{3/2}$ orbit) and with different number of protons ($Z = 24$, 26 and 28) in the $1f_{7/2}$ orbits.

Finally, Figures 24.4 and 24.5 shows results for selenium isotopes from our previous study (see Chapter 17). These isotopes contain a fixed number of protons ($Z$ = 34, with the last six being outside the closed $1f_{7/2}$ sub-shell, i.e. in configurations $2p_{3/2}$ and $1f_{5/2}$) and with different number of neutrons ($N$ = 42, 44, 46 and 48) filling in the $1g_{9/2}$ sub-shell.

The selenium measurements are the only results that show clearly the influence of neutron shell-closure in the form of an enhancement of the oscillations of the analyzing powers for nuclei approaching neutron number $N$ = 50 (see Chapter 17). This is contrary to the usual mass dependence in which the oscillations decrease in amplitude with the increasing mass of the target nuclei. As discussed in Chapter 17, the observed amplitude enhancement effect for these isotopes has been explained as being due to contributions from two-step reaction mechanism.

The oscillations for the titanium isotopes have approximately equal amplitudes. The same is true for nuclei with neutron number $N$ = 30. For nuclei with $N$ = 28, the measured amplitudes clearly decrease with the increasing mass number. Thus shell-closure effects are much less pronounced in the $Z$ = $N$ = 28 region than for the $N$ = 50 shell.

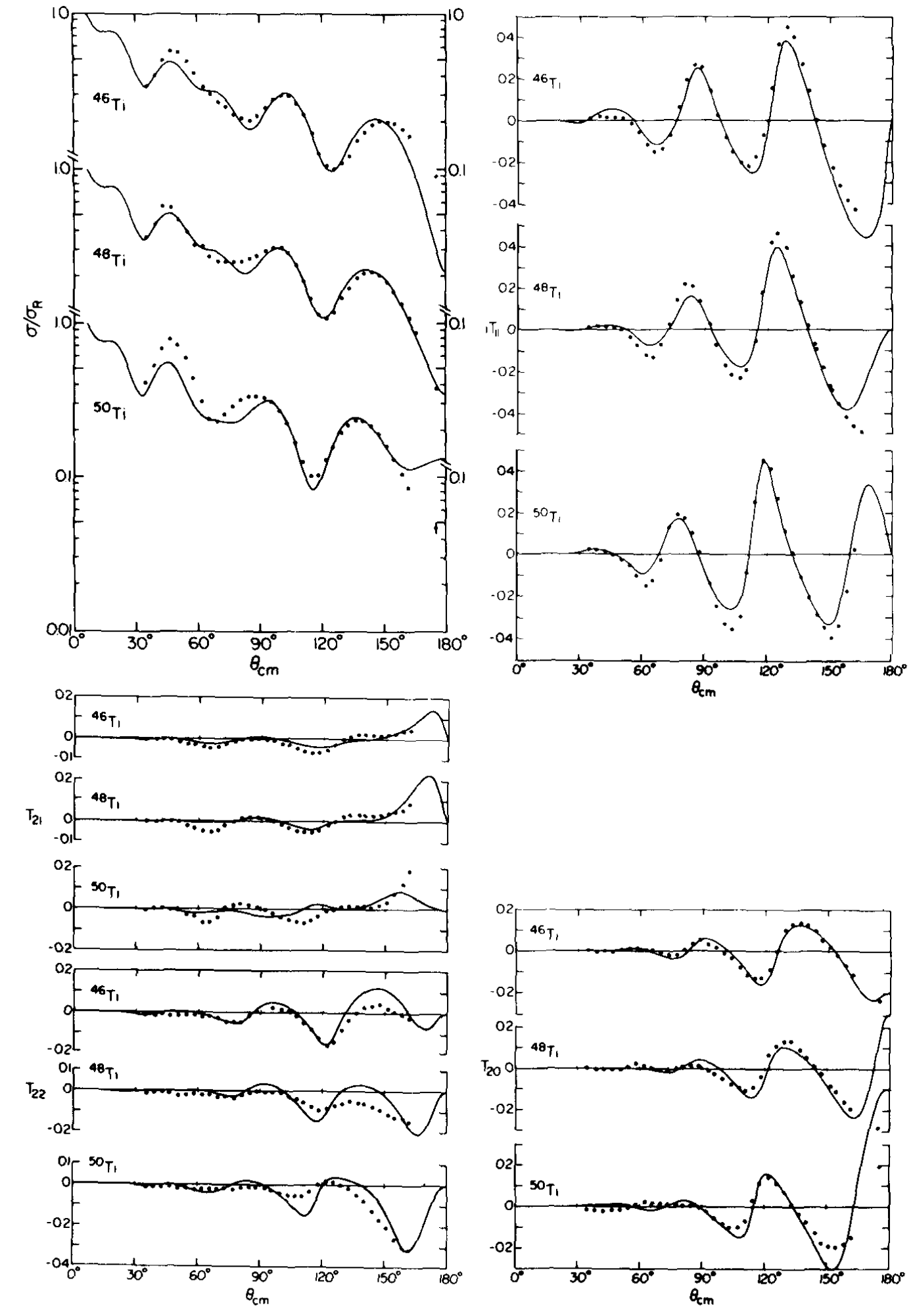

Figure 24.1. The differential cross section and analyzing powers for the elastic scattering of 12 MeV polarized deuterons from the $^{46,48,59}$Ti isotopes at 12 MeV. These nuclei contain the same number of protons ($Z$ = 22) but different numbers of neutrons ($N$ = 24, 26, and 28). In the simple shell model description, two protons are outside the sd-shell and neutrons are filling in the $1f_{7/2}$ sub-shell. The full lines are the results of our optical-model calculations.

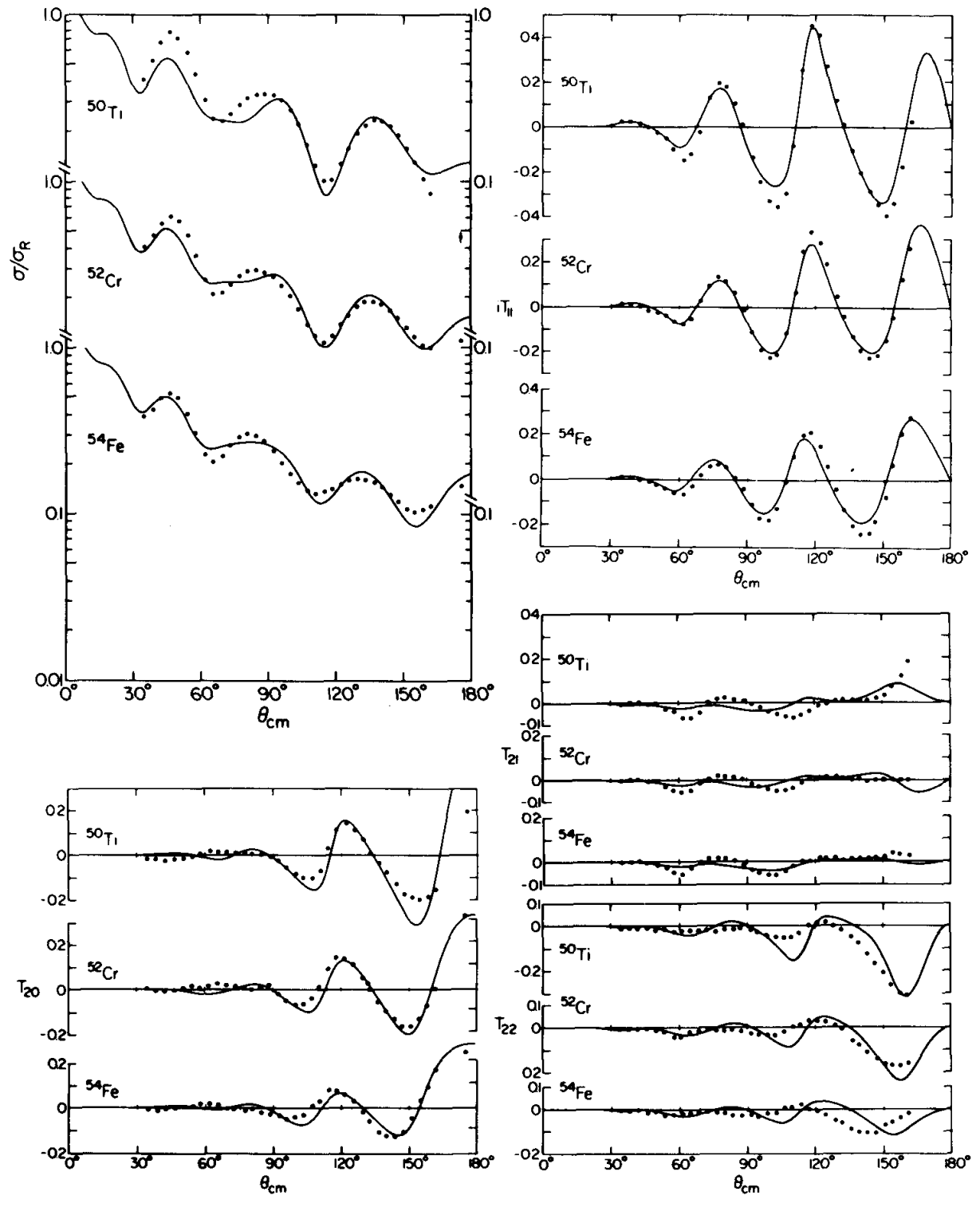

Figure 24.2. The differential cross section and analyzing powers for the elastic scattering of 12 MeV polarized deuterons from the $^{50}$Ti, $^{52}$Cr, and $^{54}$Fe isotopes containing a fixed number of neutrons $N$ = 28, which are closing the 1f$_{7/2}$ sub-shell, but different numbers of protons ($Z$ = 22, 24 and 26), which are filling in the 1f$_{7/2}$ orbits. The full lines are the results of our optical-model calculations.

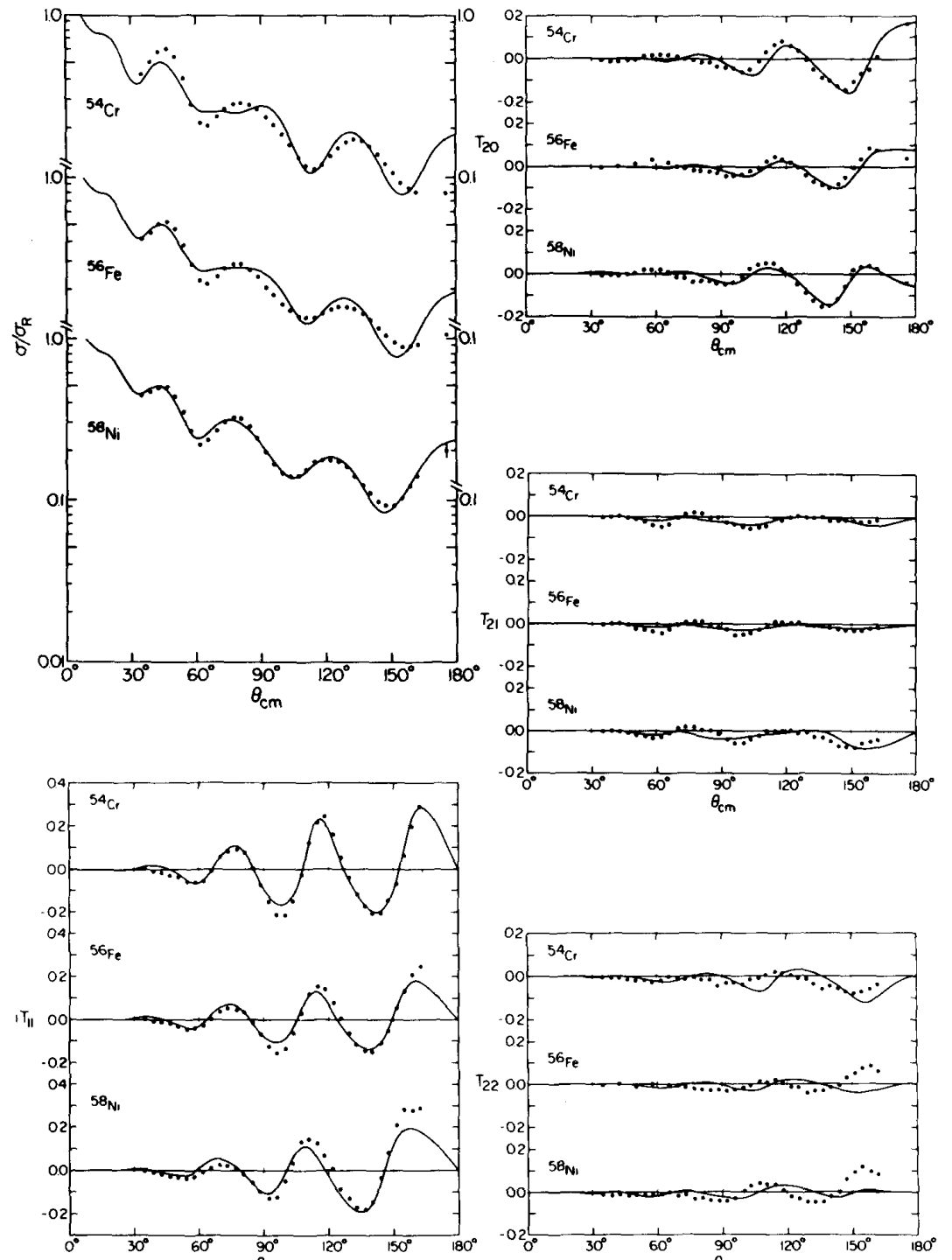

Figure 24.3. The differential cross section and analyzing powers for elastic scattering of 12 MeV polarized deuterons from the $^{54}$Cr, $^{56}$Fe, and $^{58}$Ni isotopes containing a fixed number of neutrons ($N$ = 30, with the filled-in 1f$_{7/2}$ sub-shell and with two neutrons in the 2p$_{3/2}$ orbit) and different number of protons ($Z$ = 24, 26 and 28) in the 1f$_{7/2}$ orbits. The full lines are our optical-model calculations.

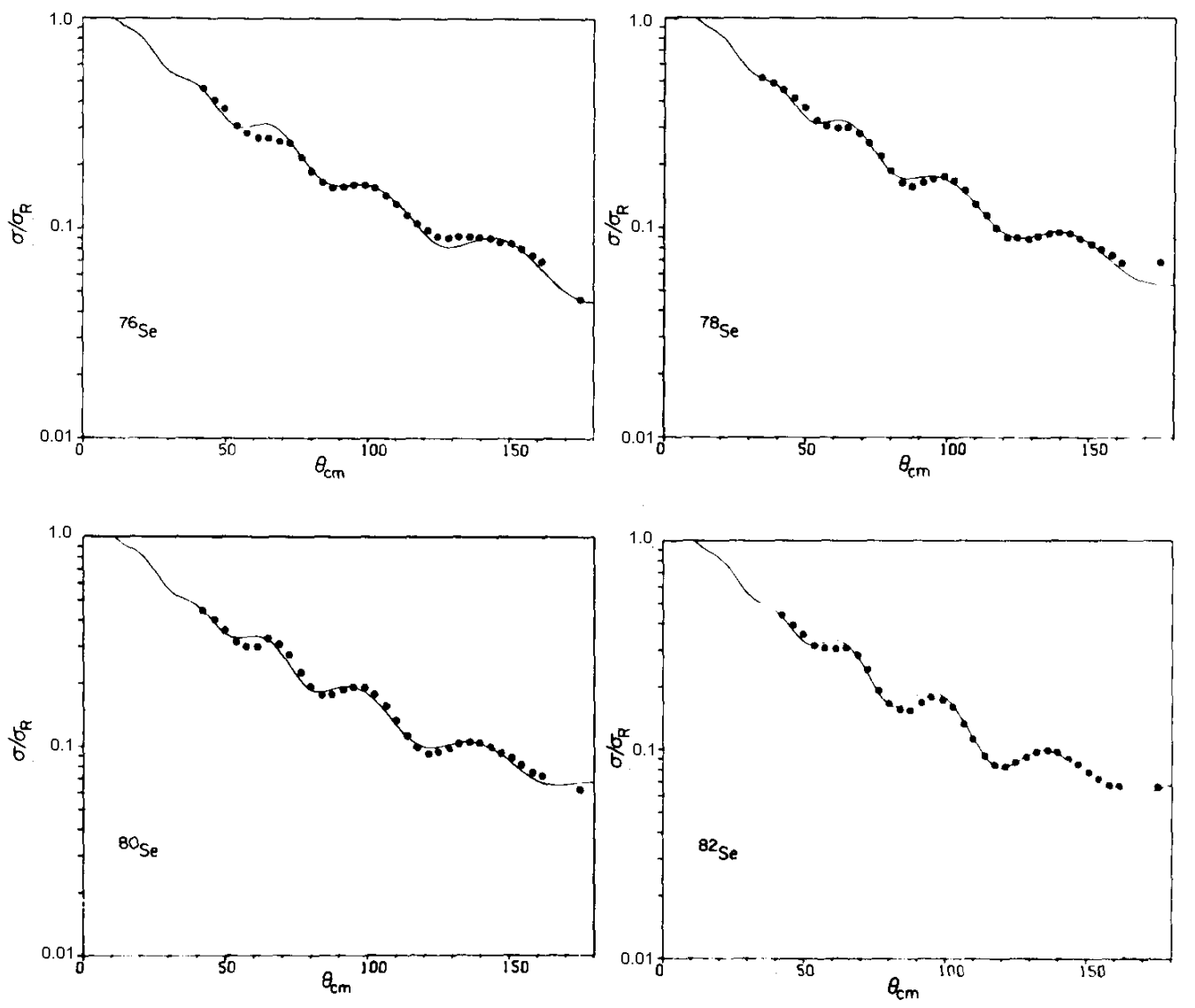

Figure 24.4. Differential cross section for the elastic scattering of 12 MeV polarized deuterons from the $^{76,78,80,82}Se$ isotopes. These isotopes contain a fixed number of protons ($Z$ = 34, with the last six being outside the closed $1f_{7/2}$ sub-shell, i.e. in configurations $2p_{3/2}$ and $1f_{5/2}$) and with different number of neutrons ($N$ = 42, 44, 46 and 48) filling in the $1g_{9/2}$ sub-shell. The full lines are our optical-model calculations.

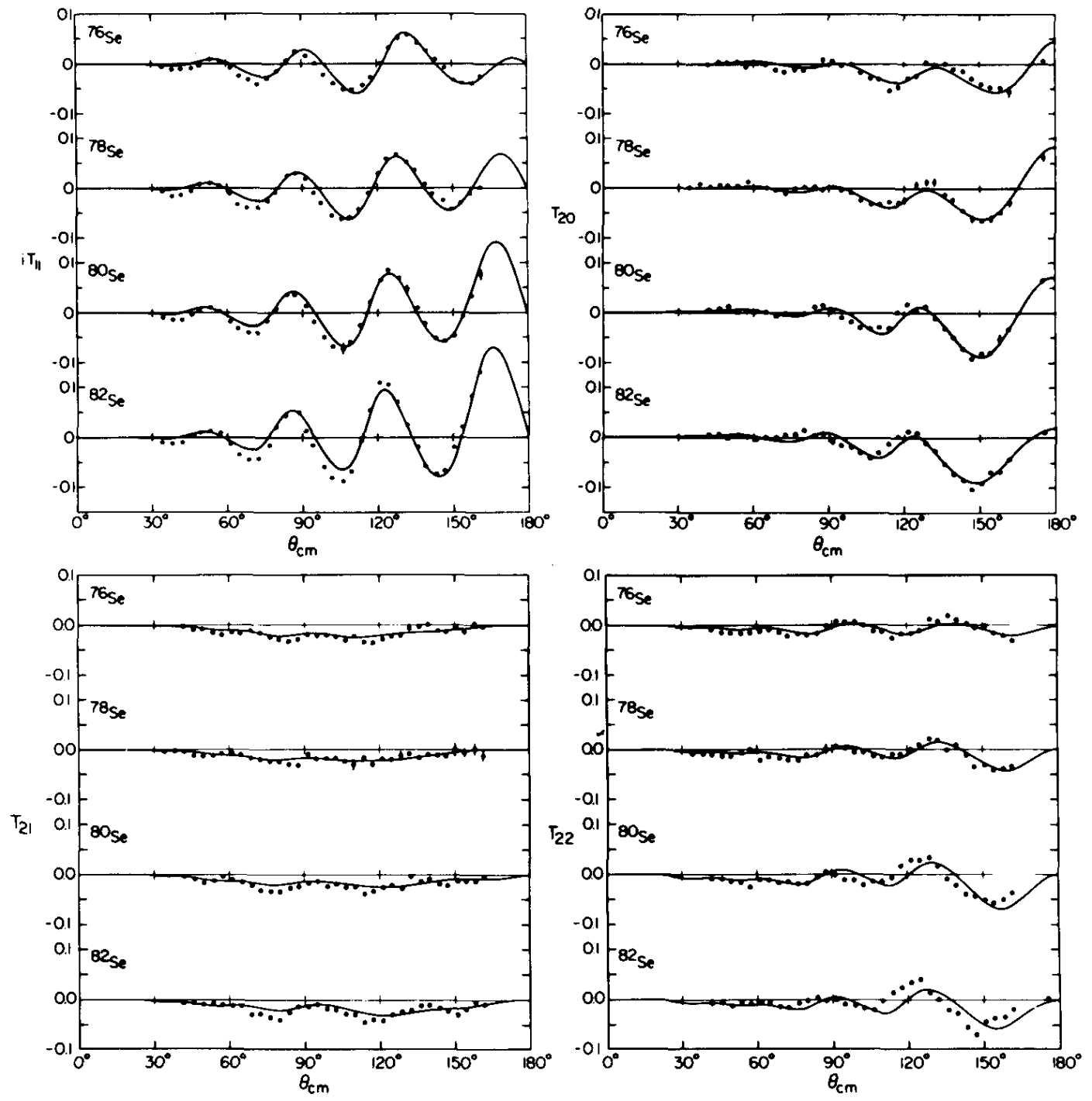

Figure 24.5. The vector and tensor analyzing powers for the elastic scattering of 12 MeV polarized deuterons from selenium isotopes. See the caption to Figure 24.4

The lack of shell-closure effects in all target nuclei but selenium isotopes could be due to an interplay of various effects such as screening by the Coulomb potential, variation in the nuclear shapes and irregularities in the filling in of the shells. A recent systematic study (England *et al.* 1982) of 25 MeV α-particle scattering from $A$ = 51-80 nuclei revealed a breakdown in the shell-closure for $^{54}Fe$. Studies of the neutron pick-up reactions, (p,d) (Suehiro, Finck and Nolen 1979) and (d,t) (England *et al.* 1980) confirmed this result indicating the presence of $p_{3/2}$ configuration in the ground-state wave function of $^{54}Fe$. (The $^{54}Fe$ nucleus has 28 neutrons and it should have a closed $1f_{7/2}$ shell.) It is likely that similar irregularities in filling in the $1f_{7/2}$ sub-shell occur for other neighbouring nuclei. This could explain why no shell closing effects are observed for nuclei other than selenium isotopes.

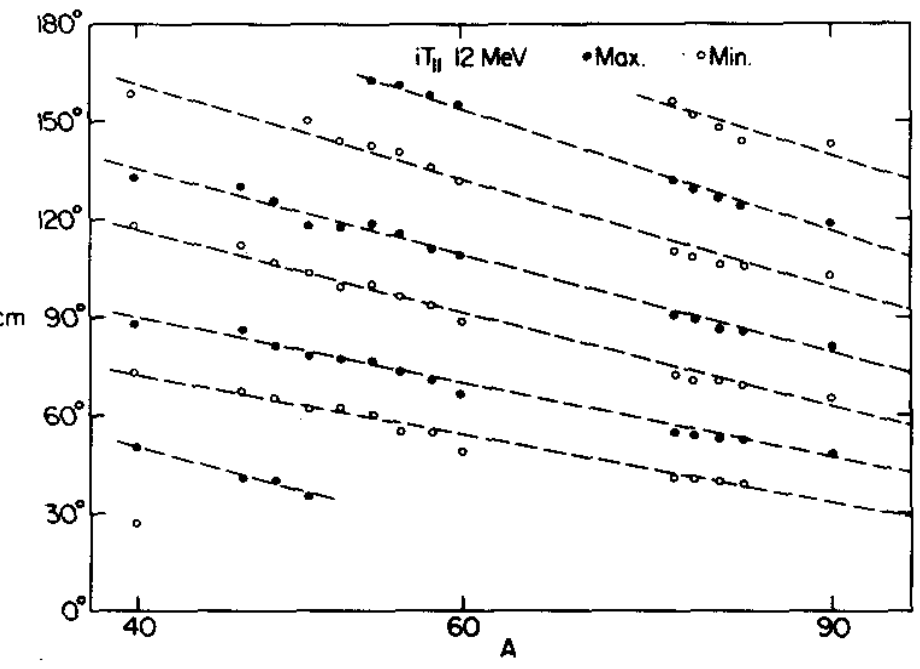

Figure 24.6. The mass dependence of the positions of the measured maxima and minima of the vector analyzing power $iT_{11}$ at 12 *MeV*. The dashed lines are to guide the eye.

In order to see whether the closure of shells can modify the relative phases of the analyzing powers, the positions of the maxima and minima for the $iT_{11}$ component have been plotted against the mass number $A$ in Figure 24.6. It can be seen that even for selenium isotopes, the positions of the diffraction patterns follow a linear dependence on $A$ indicating that the phases of the analyzing powers are not affected by the shell closure.

## Theoretical interpretation of the data

Details of the optical-model analysis have been described in Chapter 23. Briefly, the potential contained complex central, spin-orbit, and tensor parts. There are 18 parameters defining the nuclear potential (see Figures 24.7 and 24.8 and Table 24.1) but only 10 of these were varied to fit the data for each isotope. The remaining 8 parameters were either fixed or adjusted using simple mass-dependent formulae.

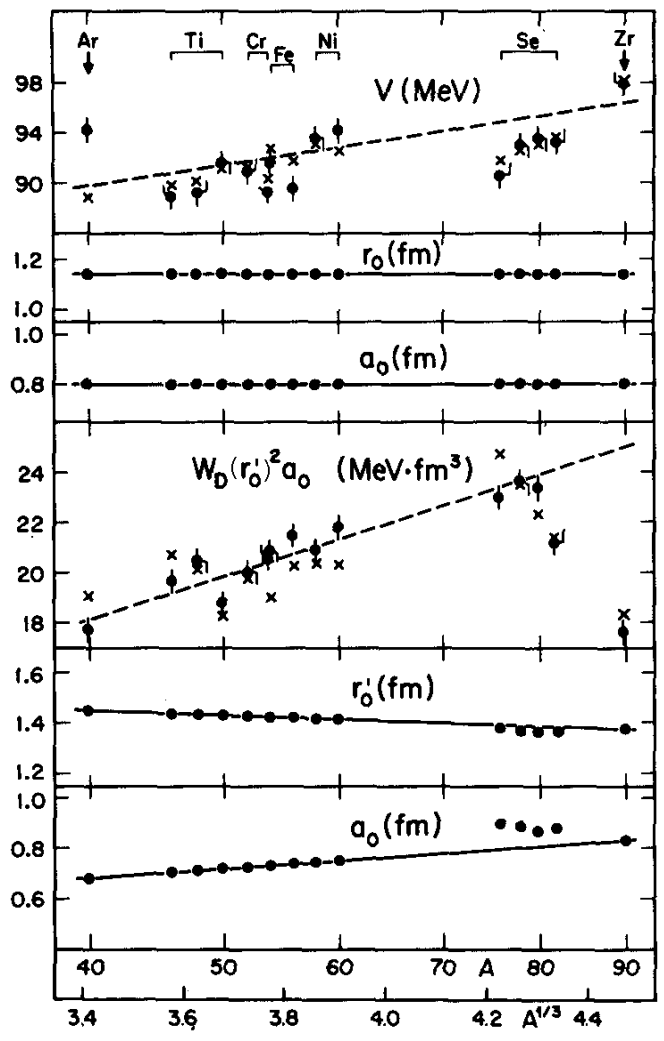

Figure 24.7. Optical-model parameters for the central potential. Where no errors bars are shown the uncertainties in the parameters are smaller than the size of the points. The lines show the mass-dependent trends. The crosses are the calculated values using the dependence on the $B(E2)$ gamma transition probabilities (see the text).

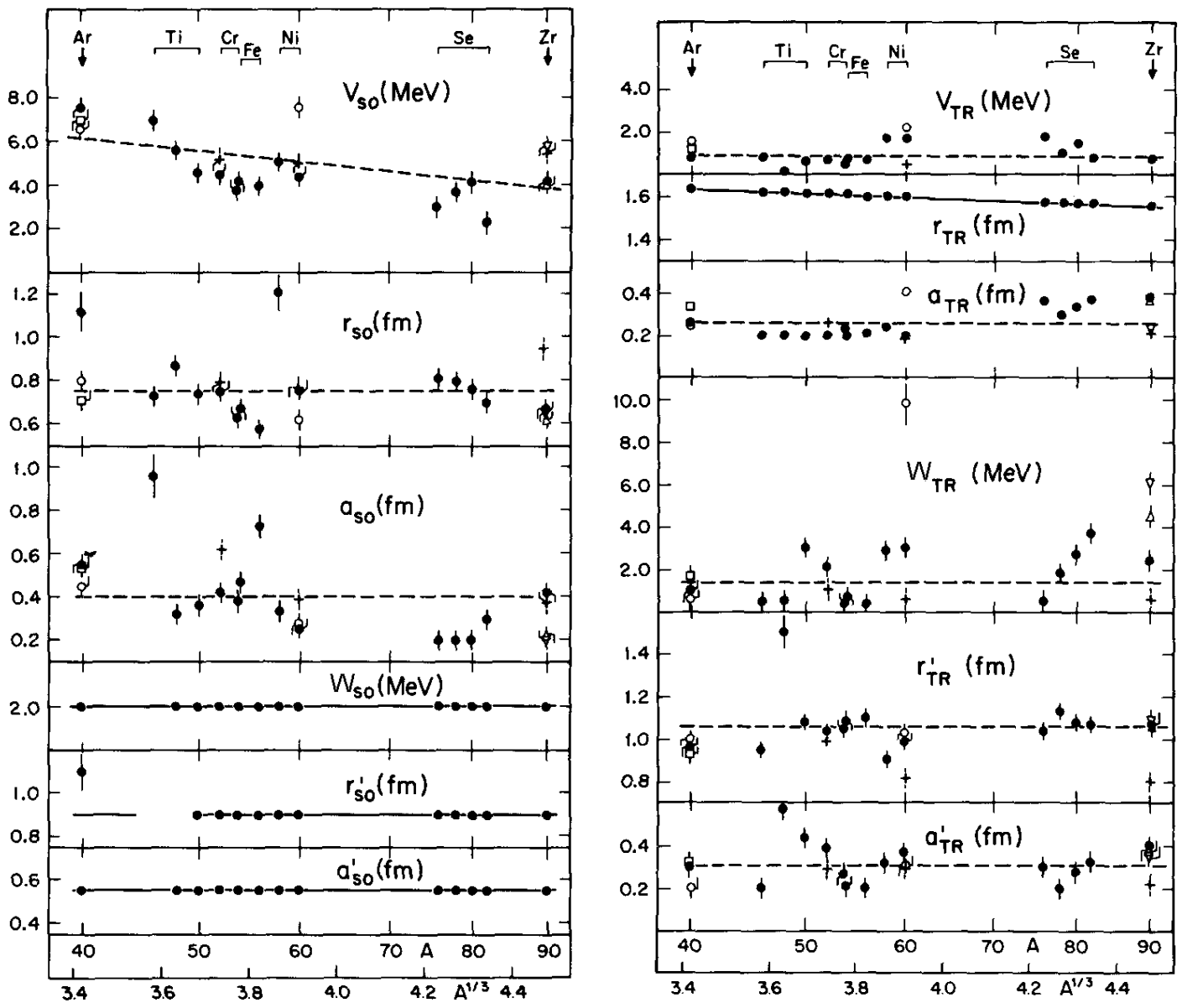

Figure 24.8. Optical-model parameters for the spin-dependent components (spin-orbit on the left-hand side and tensor on the right-hand side of the figure). The lines show the mass-dependent trends.

Table 24.1

The global parameters of the optical model potential for the elastic scattering of 12 MeV polarized deuterons in the mass range $A$ = 40 - 90

| | | Real Component | | Imaginary Component |
|---|---|---|---|---|
| Central | $V$ | $V = 84.57 + 0.13A$ | $W_D$ | $W_D = 9.91 + 0.067A$ |
| | | $V = 68.43 + 6.27A^{1/3}$ | | $W_D = 1.72 + 3.182A^{1/3}$ |
| | $r_0$ | 1.14 | $r'_0$ | $1.51 - 0.002A$ |
| | | | | $1.69 - 0.072A^{1/3}$ |
| | $a_0$ | 0.80 | $a'_0$ | $0.57 + 0.003A$ |
| | | | | $0.20 + 0.137A^{1/3}$ |
| Spin-orbit | $V_{s.o.}$ | $V_{s.o.} = 7.91 - 0.045A$ | $W_{s.o.}$ | 2.0 |
| | | $V_{s.o.} = 13.35 - 2.12A^{1/3}$ | | |
| | $r_{s.o.}$ | 0.75 | $r'_{s.o.}$ | 0.90 |
| | $a_{s.o.}$ | 0.40 | $a'_{s.o.}$ | 0.55 |
| Tensor | $V_{TR}$ | 0.89 | $W_{TR}$ | 1.45 |
| | $r_{TR.}$ | $1.85 - 0.003A$ | $r'_{TR}$ | 1.06 |
| | | $1.89 - 0.075A^{1/3}$ | | |
| | $a_{TR}$ | 0.26 | $a'_{TR}$ | 0.31 |

Additional formulae: $W(r'_0)^2 a' = 12.693 + 0.137A$ and $W(r'_0)^2 a' = 3.995 + 6.483A^{1/3}$.

The best agreement between the theory and experiment has been found for the selenium isotopes. However, it should be noted that for these nuclei the parameters do not follow the same pattern as those obtained for other target nuclei. The results for $^{50}$Ti, $^{52}$Cr and $^{54}$Fe corresponding to $N$ = 28 were hard to reproduce. Similar difficulties have been observed by Goddard and Haeberli (1978) at other deuteron energies.

In Chapter 23, it was shown that out of the eighteen parameters describing the nuclear potential *five* ($r_0$, $a_0$, $W_{s.o.}$, $r'_{s.o.}$ and $a'_{s.o.}$) could be fixed and *three* ($r'_0$, $a'_0$ and $r_{TR}$) could be expressed in terms of simple mass-dependent formulae. This feature has been confirmed in the present analysis of the results for a much wider range of target nuclei. However, we have now also found that *all* 18 parameters can be represented by either fixed values or the values calculated using simple mass-dependent formulae. This global form of parameters is summarised in Table 24.1. Such global expressions are useful in analyses of transfer reaction data.

Hjorth, Lin and Johnson (1968) and Lohr and Haeberli (1974) suggested that the imaginary part of the central optical model potential is related to the $B(E2) \equiv B(E2, 0_1^+ \to 2_1^+)$ $\gamma$ - transition probabilities. Taking the $B(E2)$ values (expressed in fm$^4$) from Stelson and Grodzins (1965) and from Endt and Van der Leun (1978) we have found that the parameters for the imaginary components of the central potential can be described closely by the following relation:

$$W_D (r'_0)^2 a'_0 = 10.9 + 0.56 Z A^{-1/3} + 10.5 [B(E2)]^{1/2} A^{-1} \qquad \text{(in MeV·fm}^3\text{)}$$

A similar dependence has been obtained by Hjorth, Lin and Johnson (1968). However, their coefficient (417) for the $[B(E2)]^{1/2} A^{-1}$ term appears to be incorrect. Examination of their results and their fig. 4 indicates that this coefficient should have a value close to 13, which would agree well with our results.

The values of $W_D (r'_0)^2 a'_0$ extracted from the above formula are represented by crosses in Figure 24.7. As can be seen, these values follow closely the corresponding values obtained from the optical-model analysis.

During our investigation, it became clear that the depth of the real part of the central potential, $V$, is correlated with the depth $W_D$ in the sense that large values of $W_D$ are associated with small values of $V$ and vice-versa. Figure 24.7 shows the dependence of $V$ on $A^{1/3}$ and as can be seen, the values obtained from the optical-model analyses (dots) show departures from the straight line. An attempt has been made to reproduce these changes using a suitable analytic form for $V$. If the dependence on $B(E2)$ is ignored, then the trend is expressed in the form of the dashed line in Figure 24.7. However, if the dependence on $B(E2)$ and $ZA^{-1/3}$ are included explicitly, then the least square analysis leads to the following relation for $V$:

$$V = 92.5 + 2.1 Z A^{-1/3} - 1.0 E_{c.m.} - 6.7 [B(E2)]^{1/2} A^{-1} \qquad \text{(in MeV)}$$

The results of this relation are shown as crosses in Figure 24.7. This formula provides a good description of the fluctuations in the values of $V$ for various targets except for $^{40}$Ar where calculated $V$ value is much too small.

The above relations demonstrate clearly a strong correlation between potential depths $V$ and $W_D$. Using these two relations it is easy to see that the correlation between $V$ and

$W_D$ in the mass range of $A$ = 40 - 90 can be expressed conveniently by the following approximate relation:

$$V + W_D = 100 + 2.5ZA^{-1/3} - E_{c.m.} \qquad \text{(in MeV)}$$

The parameter values for the spin-orbit and the tensor potentials are shown in Figure 24.8. Some parameter values are found to fluctuate with mass number of the target nucleus. The dashed lines indicate the trends for these parameters. It is possible that the observed fluctuations arise partly from an attempt to compensate for shape effects which are inadequately described by the form factors for the central potential.

## Summary and conclusions

The work described here represents the most extensive and systematic study of the elastic scattering of polarized deuterons from medium-heavy target nuclei. This study demonstrates that suitable sets of optical-model parameters can be found which describe accurately all five quantities $\sigma_0$, $iT_{11}$, $T_{20}$, $T_{21}$ and $T_{22}$ measured for nuclei in the mass range of $A$ = 40 – 90.

Of the eighteen parameters used to define nuclear interaction, *five* were kept fixed during the search and *three* were calculated using simple mass-dependent formulae. However, we have also found that all 18 parameters can be conveniently represented either using fixed values of the values calculated using simple mass-dependent formulae. Such global representation of the optical model parameters is particularly useful in analyses of transfer reaction measurements. It simplifies the process of selecting suitable parameters, which need to be used in such studies.

Both, the depths $V$ and $W_D$ of the central potential were found to be strongly dependent on the structure of the target nuclei. These parameters can be expressed conveniently in terms of the $B(E2, 0_1^+ \rightarrow 2_1^+)$ $\gamma$ - transition probabilities. We have also derived a mathematical correlation between $V$ and $W_D$.

We have found that it was difficult to describe the results for nuclei with $N$ *or* $Z$ near 28. In particular, the fits for the $N$ = 28 targets, $^{50}$Ti, $^{52}$Cr and $^{54}$Fe, are poorer than for other investigated nuclei. A similar problem with $^{52}$Cr and $^{54}$Fe targets has been reported by Goddard and Haeberli (1978). There is now sufficient evidence indicating that the $1f_{7/2}$ neutron shell is not closed for $^{54}$Fe ($N$ = 28). Thus, the poor theoretical description of the data may be associated with structure irregularities in this mass region, and a better fit might be obtained by including structure effects, e.g. two-step processes, explicitly in the calculation as suggested by my analysis of selenium data (see Chapter 17).

It is possible that the fits could be improved by using different form factors for the central potential, other than the conventional Wood-Soxon and its derivative. The inclusion of $l$ - dependent potentials could be also considered. However, additional parameters in already parameter-rich descriptions appears undesirable. It has been also shown (Kobos and Mackintosh 1979) that introducing an $l$ - dependent component is equivalent to the imitation of a coupling between elastic and reaction channels. Such a coupling could be considered explicitly in the theoretical analysis as described in the next chapter.

# 25

# Collective Excitation Effects in the Elastic Scattering

***Key features:***

1. Conventional optical model analyses consider only the single-step $(d,d)$ elastic scattering. This has resulted in significant complications when describing the interaction potential. For instance, in our analysis (see Chapter 24) we had to use nine components of the optical model potential with the total of 18 parameters.
2. Following my successful study of selenium isotopes (see Chapter 17), I have decided to extend my coupled-channels calculations to a wider range of nuclei by including explicitly the two-step $(d,d')2_1^+(d',d)$ contributions to the elastic scattering. This procedure has now simplified considerably the description of the deuteron-nucleus interaction.
   a. The interaction potential can now be described using only three components with the total of only 9 parameters.
   b. The imaginary spin-orbit component, which has been found essential in our conventional optical model analysis, is no longer required.
   c. Five of the nine parameters describing the deuteron-nucleus interaction can now be either fixed for all target nuclei or described using simple mass-dependent formulae, leaving only four parameters that need to be adjusted to optimise the fits to the experimental angular distributions.
   d. The dependence of the potential depths on the $B(E2;0_1^+ \rightarrow 2_1^2)$ $\gamma$- transition probabilities, which has been found repeatedly in various conventional optical model analyses, is now eliminated. This dependence, therefore, reflects the presence of the second-order processes, which normally are not included in analyses of elastic scattering.

**Abstract:** The differential cross sections, $\sigma_0(\theta)$ and vector analyzing powers, $iT_{11}(\theta)$ for 12 MeV vector polarized deuterons scattered elastically from $^{40}$Ar, $^{46,48.50}$Ti, $^{52,54}$Cr, $^{54,56}$Fe, $^{58,60}$Ni, $^{76,78,80,82}$Se and $^{90}$Zr were analysed using the coupled-channels formalism, which included the two-step scattering via the first $2_1^+$ states in the target nuclei. In contrast with the results of the conventional optical model analysis for the same isotopes, there was now no need to include the imaginary spin-orbit component in the description of the interaction potential. The parameterization of the optical model potential has been significantly simplified and the dependence on the $\gamma$-transition probabilities, $B(E2;0_1^+ \rightarrow 2_1^2)$, has been removed.

## Introduction

As described in Chapters 23 and 24, in our analysis of polarization data we had to use the spin-orbit potential with an imaginary component. However, this component influenced also significantly the parameter values of the central potential. We have suggested, that the source of this unexpected effect might be due to the generally adopted mathematical description of the shape of the nuclear potential using the Woods-Saxon function. Such representation might not be sufficiently accurate. However, there is also another alternative: the problem might be associated with the assumed simplified reaction mechanism for the elastic scattering. The mechanism might be more complex than the simple shape scattering used in conventional optical-model calculations.

Indeed, in Chapter 17 I have shown that the two-step scattering via the $2_1^+$ inelastic scattering channel plays an important role in the elastic scattering. Without considering this process, the depth of the imaginary component of the central potential has to be varied to fit the data, which is a crude way of accounting for contributions from indirect scattering. I have shown that without coupling to the first excited states, the depth $W_D$ of the imaginary component of the optical model potential depends linearly on the quadrupole deformation parameters $\beta_2$ derived from the $\gamma$ -transition probabilities, $B(E2;0_1^+ \rightarrow 2_1^2)$. However, if the two-step scattering via the inelastic channels is considered explicitly in the calculations the need for adjusting the potential depth $W_D$ is removed and the data can be fitted using a fixed set of the optical model parameters for all four selenium isotopes.

In an earlier work, Rawitscher (1978) considered a folding model with non-symmetric break-up processes, which suggested an $l$ - dependent potentials. Unfortunately, this kind of approach does not seem practical unless the $l$ - dependence can be predicted by theoretical considerations. The number of adjustable parameters, which have to be used to describe the elastic scattering of deuterons, is already too large and to increase their number is undesirable. Furthermore, in the case of proton scattering, it has been pointed out (Kobos and Mackintosh 1979) that the physical meaning of the $l$ - dependent potentials is simply to simulate the coupling to reaction channels. Consequently, straightforward coupled-channel calculations would appear more appropriate. My successful coupled-channels analysis of selenium data (Chapter 17) supports this approach.

The observed dependence of the optical model potentials on the electric quadrupole transition probabilities $B(E2) = B(E2;0_1^+ \rightarrow 2_1^2)$ or equivalently on the quadrupole deformation parameter $\beta_2$ also suggests that it would be better to include explicitly the coupling to inelastic channels in analyses of elastic scattering. The $B(E2)$ dependence becomes particularly clear for measurements carried out using polarized beams.

A correlation between the $B(E2)$ values and the parameters of the central imaginary potential for deuterons was first suggested by Hjorth, Lin, and Johnson (1968). They measured the differential cross sections for the elastic scattering of 14.5 MeV deuterons from isotopically enriched targets in the mass range of $A$ = 54 - 124 and found that the product $W_D(r_0')^2 a_0'$ (see the next section) depends linearly on $[B(E2)]^{1/2}/A$.

Lohr and Haeberli (1974) carried out measurements of angular distributions of the cross sections and vector analyzing powers for the elastic scattering of vector polarized deuterons. They had found that the volume integral of the imaginary central potential followed a reasonably clear linear dependence on the quadrupole deformation parameter $\beta_2$ for deuterons in the energy range of 7-13 MeV and over the target mass range of 27 - 208. A systematic study of proton elastic scattering on a wide range of nuclei also indicated a linear dependence of the depth of the surface absorption potential on $\beta_2$ (Fabrici *et al.* 1980).

Our study in the mass range of $A$ = 40 - 90 suggested that not only the imaginary but also the real central potential component depends on $B(E2)$ (see Chapter 24). Thus, all these results appear to suggest a detectable influence of the inelastic, $2_1^+$, channel on the elastic scattering.

It is reasonable to expect that the dependence of the optical model parameters on the quadrupole properties of the target nuclei could be accounted for by coupling between the elastic and inelastic channels. Consequently, by following the same procedure as outlined in Chapter 17 for selenium isotopes it should be possible to reduce or even to remove the $B(E2)$ dependence. My aim therefore was to extend my investigation of the two-step mechanism to a wider range of nuclei and hopefully to simplify the parameterization of the interaction potential.

## The coupled-channels analysis

The differential cross sections, $d\sigma(\theta)/d\Omega$ and vector analyzing powers, $iT_{11}(\theta)$ for the elastic scattering of 12 MeV vector polarized deuterons from $^{40}$Ar, $^{46,48}$.$^{50}$Ti, $^{52,54}$Cr, $^{54,56}$Fe, $^{58,60}$Ni, $^{76,78,80,82}$Se and $^{90}$Zr nuclei have been analysed by considering both the direct shape scattering $(d,d)$ and the indirect $(d,d')2_1^+(d',d)$ scattering via the first excited states $2_1^+$ in the target nuclei (see Chapter 17). The analysis was done using the computer code CHUCK and the Australian National University UNIVAC 1100/82 computer.

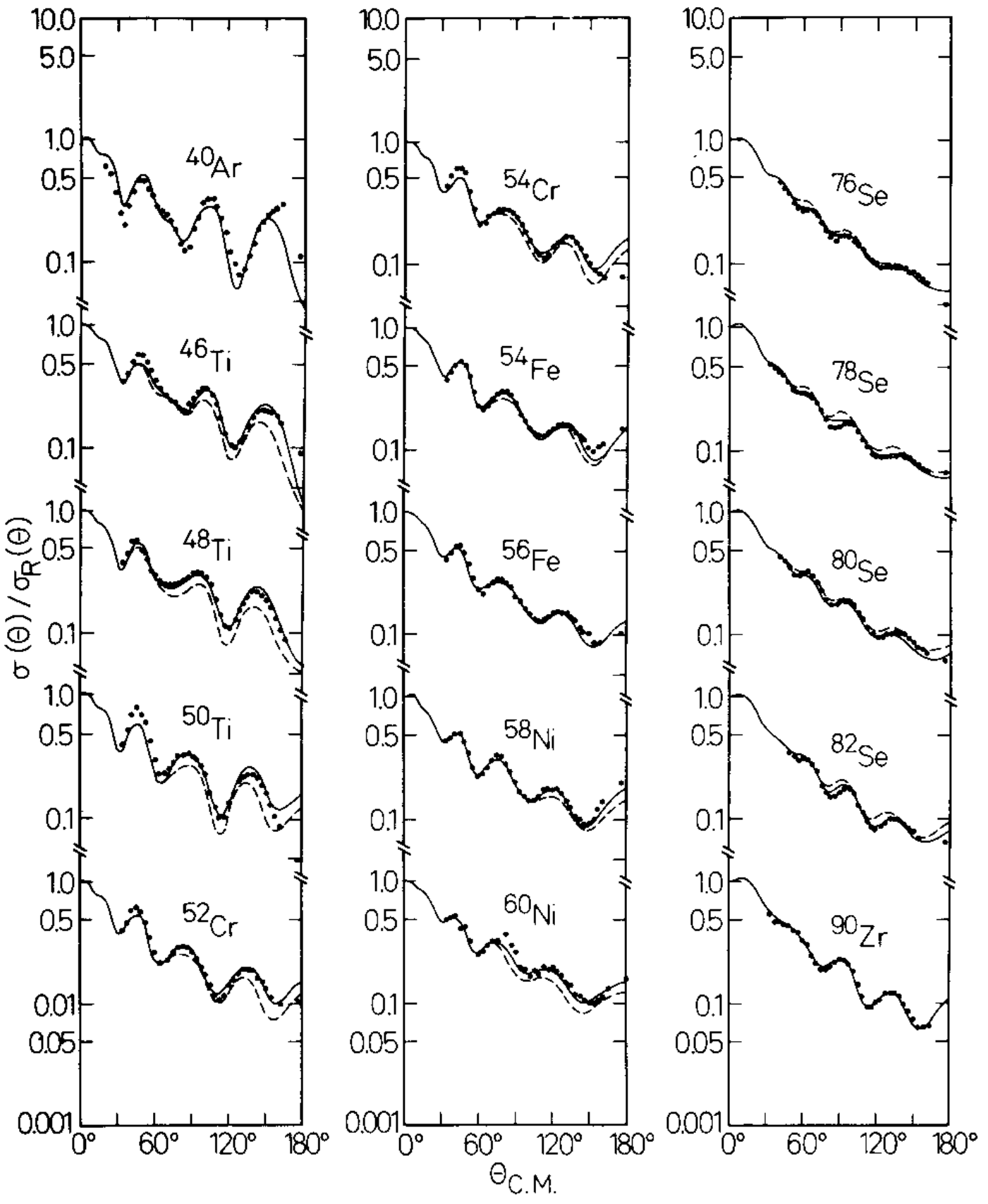

Figure 25.1. The experimental angular distributions of the differential cross sections for the elastic scattering of 12 MeV vector polarized deuterons are compared with the coupled channels calculations, which include explicitly contributions of the two-step $(d,d')2_1^+(d',d)$ mechanism. The full lines were calculated using five fixed parameters and searching for the remaining four ($V$, $W_D$, $r_0'$

and $a_0'$). The dotted lines represent the calculations in which only two parameters, *V* and $W_D$, were varied to optimise the fits to the angular distributions.

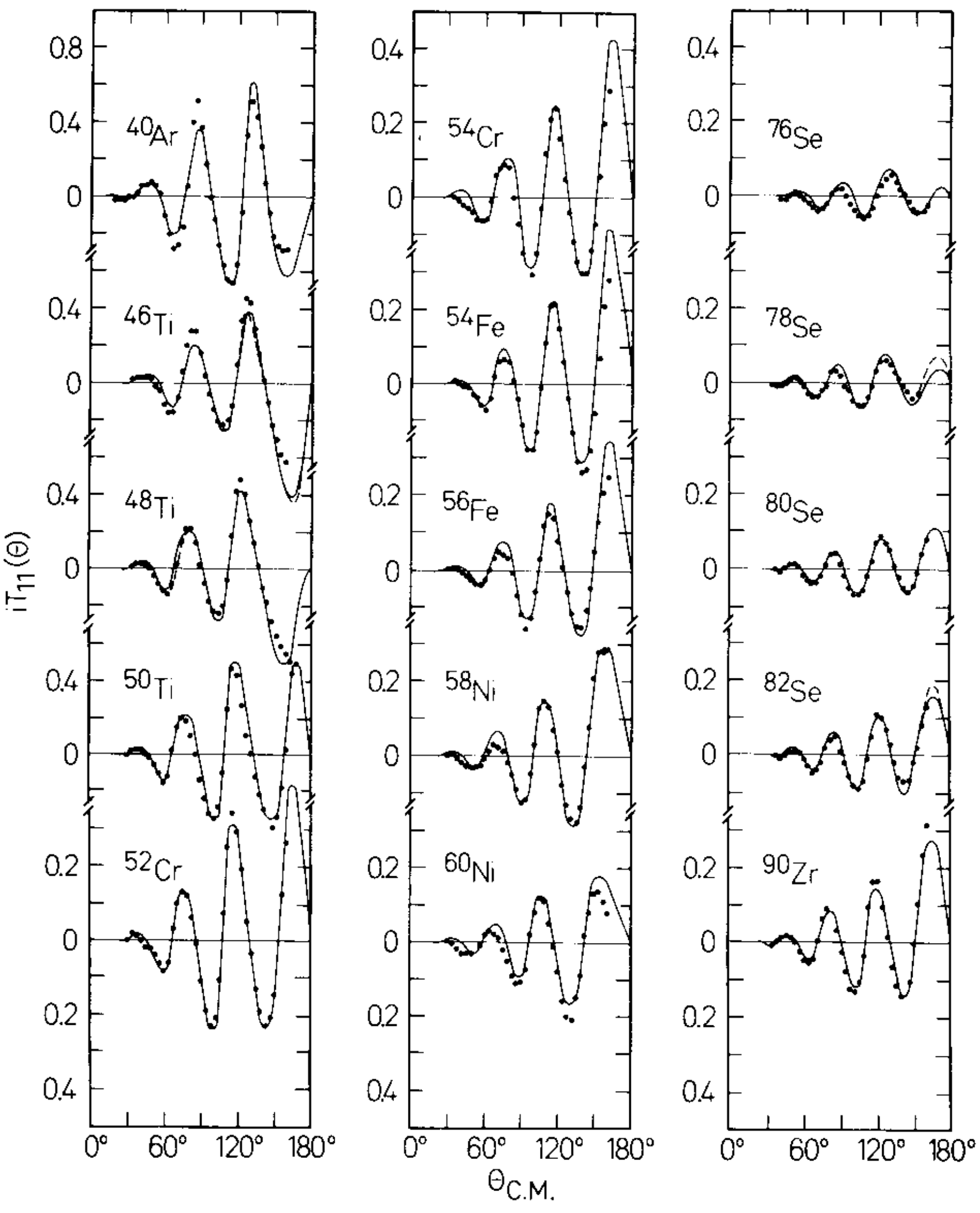

Figure 25.2. The experimental angular distribution of the vector analyzing powers $iT_{11}(\theta)$ for the elastic scattering of 12 MeV vector polarized deuterons are compared with the coupled channels calculations, which included explicitly contributions of the two-step $(d,d')2_1^+(d',d)$ mechanism. See the caption to Figure 25.1.

In my preliminary analysis, I have used both real and imaginary spin-orbit components. I have found that the imaginary component had no significant effect on improving the fits to the angular distributions. Consequently, in the remaining calculations I have used a simpler potential:

$$U(r) = U_0(r) + U_{s.o.}(r)$$

where

$$U_0(r) = -Vf(r, r_0, a_0) - i4a_0'W_D \frac{d}{dr} f(r, r_0', a_0')$$

$$U_{s.o.}(r) = \lambda_\pi^2 \frac{1}{r}\left[V_{s.o.} \frac{df(r, r_{s.o.}, a_{s.o.})}{dr}\right]\mathbf{S}\cdot\mathbf{L}$$

$$f(r, r_i, a_i) = \frac{1}{1+e^{x_i}}$$

$$x_i = \frac{r - r_i A^{1/3}}{a_i}$$

Table 25.1

Formulae for the potential depth $V$

| | $V$ (MeV) | $\chi^2$ |
|---|---|---|
| OM | $V = 92.5 + 2.1ZA^{-1/3} + 6.7[B(E2)]^{1/2} A^{-1} - E_{c.m.}$ | |
| CC4 | $V = 92.7 + 1.9ZA^{-1/3} + 1.1[B(E2)]^{1/2} A^{-1} - E_{c.m.}$ | 0.66 |
| | $V = 93.2 + 1.9ZA^{-1/3} - E_{c.m.}$ | 0.67 |
| CC2 | $V = 89.2 + 2.2ZA^{-1/3} + 2.5[B(E2)]^{1/2} A^{-1} - E_{c.m.}$ | 0.74 |
| | $V = 90.4 + 2.2ZA^{-1/3} - E_{c.m.}$ | 0.76 |

OM – The formula based on the conventional optical model analysis, which neglects contributions of the two-step scattering.
CC4 – The two optional formulae based on the coupled-channels analysis, which includes the two-step scattering. In these calculations, five optical model parameters were fixed and four were searched for. The two formulae are obtained by fitting the resulting $V$ values with and without the $B(E2)$ component. Both functions resulted in the equivalent descriptions of $V$ (*cf* the $\chi^2$ values).
CC2 – As for CC4 but now coupled-channels analysis was carried out using 7 fixed parameters and two searched for.
$\chi^2$ – The parameter, which describes the quality of the fit to the $V$ values.
The $B(E2)$ values are in $e^2$fm$^4$

Table 25.2

Formulae for $W_D(r_0')^2 a_0'$

| | $W_D(r_0')^2 a_0'$ (in MeV·fm$^3$) | $\chi^2$ |
|---|---|---|
| OM | $W_D(r_0')^2 a_0' = 10.9 + 0.56ZA^{-1/3} + 10.5[B(E2)]^{1/2} A^{-1}$ | |
| CC4 | $W_D(r_0')^2 a_0' = 11.0 + 0.82ZA^{-1/3} + 3.1[B(E2)]^{1/2} A^{-1}$ | 1.1 |
| | $W_D(r_0')^2 a_0' = 12.5 + 0.86ZA^{-1/3}$ | 1.2 |
| CC2 | $W_D(r_0')^2 a_0' = 12.7 + 0.68ZA^{-1/3} + 2.2[B(E2)]^{1/2} A^{-1}$ | 1.2 |

$$W_D(r_0')^2 a_0' = 13.7 + 0.68ZA^{-1/3} \qquad 1.3$$

Initially I have searched for all nine parameters. However, I have found that five parameters follow the values we have determined earlier in the conventional optical model analysis of the data in this mass region. Keeping these parameters fixed at their appropriate values, I have repeated the analysis of the data searching on only four parameters. Finally, I have fixed two additional parameters and search for only two.

In the four parameter search, parameters $V$, $W_D$, $r_0'$ and $a_0'$ were searched for while the remaining five parameters, $r_0$, $a_0$, $V_{s.o.}$, $r_{s.o.}$ and $a_{s.o.}$ were fixed at the following values: $r_0$=1.14 fm, $a_0$ = 0.8 fm, $V_{s.o.} = 6.3 - 0.4A^{-1/3}$ MeV, $r_{s.o.}$ = 0.75 fm and $a_{s.o.}$ = 0.4 fm. Theoretical predictions corresponding to this series of calculations are shown in the form of the full lines in Figures 25.1 and 25.2. As can be seen, this simplified potential resulted in excellent fits to the angular distributions. There was no need to complicate it by adding an imaginary spin-orbit component.

In the two-parameter search, I have kept also the $r_0'$ and $a_0'$ parameters fixed at the values given by the following mass-dependent formulae: $r_0' = 1.20 + 0.85A^{-1/3}$ and $a_0' = 1.29 - 2.10A^{-1/3}$. This set of calculations was reduced to searching only for $V$, and $W_D$. Results are shown as dotted lines in Figures 25.1 and 25.2. As expected for such severe restrictions, the fits to the data were in some cases less satisfactory than for the four-parameter search, but surprisingly in many cases they resulted in nearly equivalent representations of the experimental distributions.

Having determined the new sets of parameters based on the coupled-channels analysis it was interesting to see how they depended on the electric quadrupole transition probabilities. Using the determined parameters, I have carried out the least-squares analysis assuming the dependence on $ZA^{-1/3}$ and $[B(E2)]^{1/2}A^{-1}$. The resulting formulae are listed in Tables 25.1 and 25.2 together with the formulae derived earlier (see Chapter 24) using conventional optical model analysis, i.e. without considering contributions from the two-step scattering. As indicated by the $\chi^2$ values, the dependence on $B(E2)$ can be removed if two-step contributions are included explicitly in the analysis of experimental results.

## Discussion and conclusions

The formulae listed in Tables 25.1 and 25.2 show that the coupled channels calculations affect essentially only the $B(E2)$ part of the functions. Consequently, to see the differences between the conventional optical-model and the coupled-channels calculations it is convenient to separate the $B(E2)$ dependence from the $ZA^{-1/3}$ dependence and write $V$ and $W_D(r_0')^2 a_0'$ functions in the following form:

$$V = V_1 + V_2$$

$$W_D(r_0')^2 a_0' = w_1 + w_2$$

where

$$V_1 = a_1 + a_2[B(E2)]^{1/2} A^{-1}$$

$$V_2 = a_3 Z A^{-1/3} - E_{c.m.}$$

$$w_1 = b_1 + b_2[B(E2)]^{1/2} A^{-1}$$

$$w_2 = b_3 Z A^{-1/3}$$

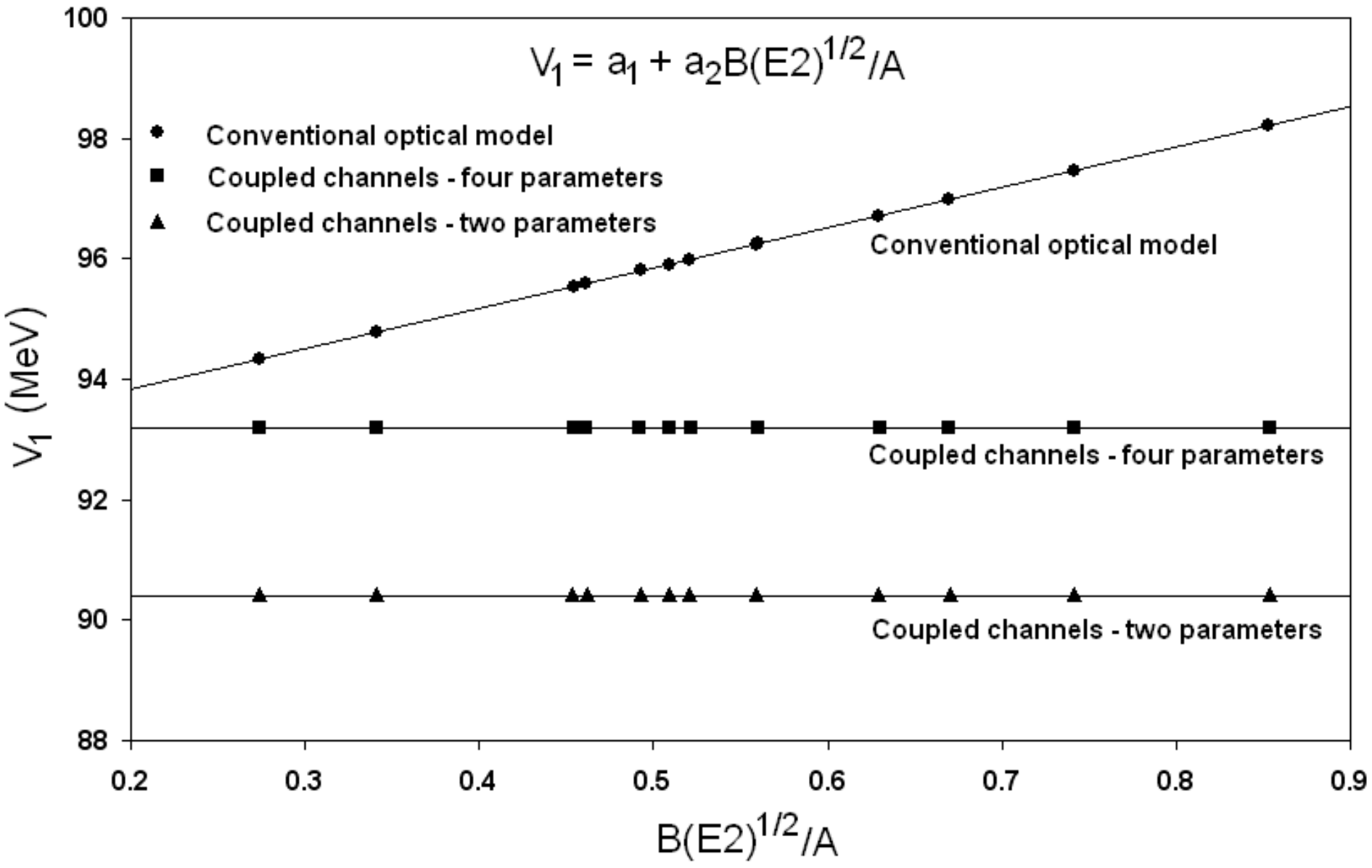

Figure 25.3. The $B(E2)$ - dependent component of the potential depth $V$. This plot shows that by including the two-step scattering mechanism $(d,d')2_1^+(d',d)$ explicitly in the calculations, the dependence on the $B(E2)$ $\gamma$- transition probabilities can be eliminated.

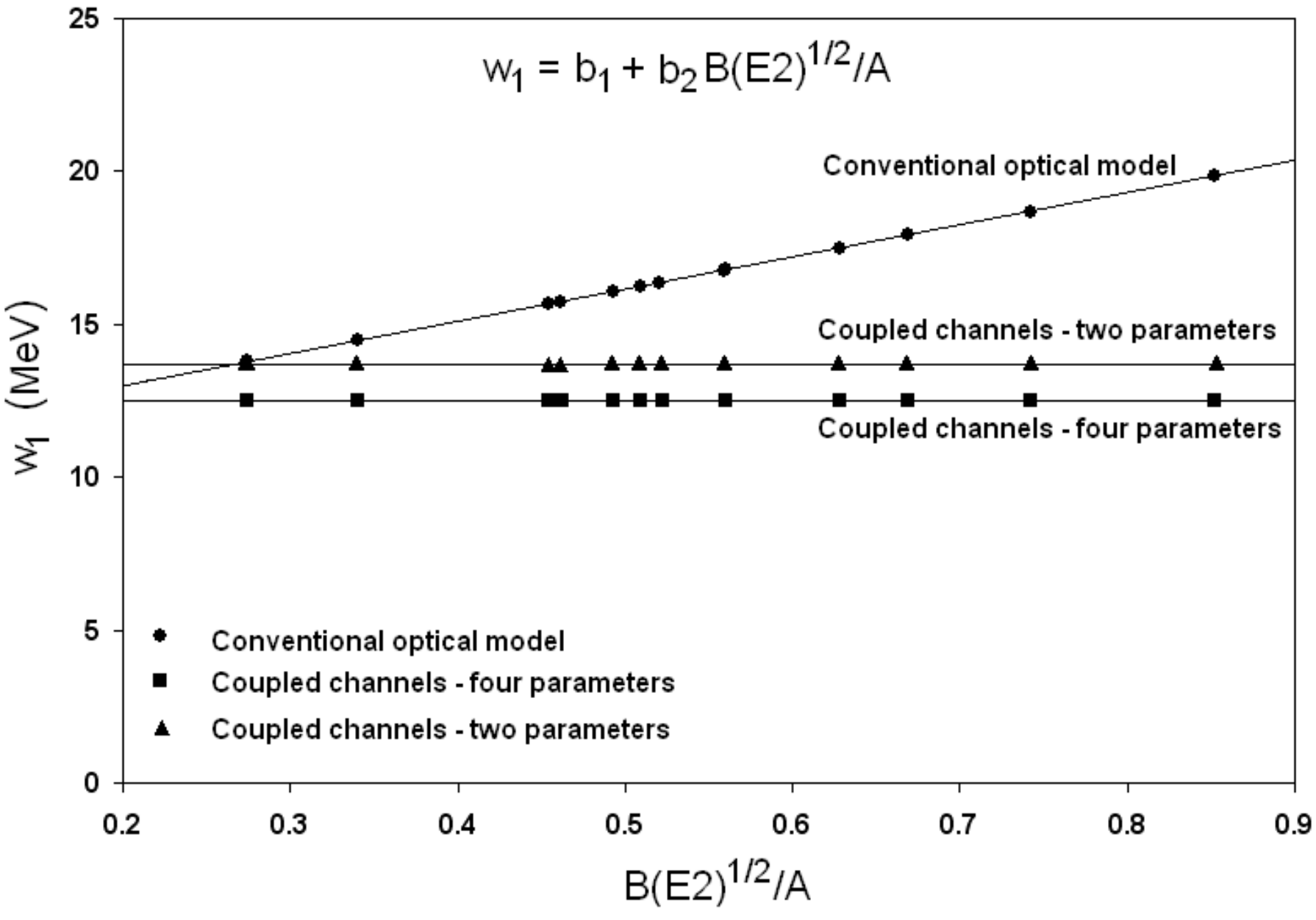

Figure 25.4. The $B(E2)$ -dependent component of $W_D(r_0')^2 a_0'$. See the caption to Figure 25.5.

The plots of the $B(E2)$-dependent components of $V$ and $W_D(r_0')^2 a_0'$, i.e. of $V_1$ and $w_1$, are shown in Figures 25.3 and 25.4. These plots show clearly that when the two-step mechanism is included explicitly in the calculations, the dependence of the optical model parameters on $B(E2)$ can be removed.

In summary, I have found that if both the direct $(d,d)$ and two-step $(d,d')2_1^+(d',d)$ scattering are considered in the analysis of experimental data, the potential describing the deuteron-nucleus interaction can be considerably simplified. The imaginary component of the spin-orbit interaction, which has been found necessary in the previous analysis using the conventional optical model procedure (see Chapter 24), is now no longer required. Even though the tensor analyzing powers were not included in my calculations, it may be recalled that the main effect of the imaginary spin orbit component was in improving the fits to the distributions of the differential cross sections. In the present coupled-channels analysis, excellent fits to these distributions were obtained without this component. Thus, by including the two-step scattering mechanism, the deuteron-nucleus interaction can be described using a simple potential containing only three components and a total of only 9 parameters.

I have found that the potential can be simplified even further by fixing 4 of the 9 parameters ($r_0$=1.14 fm, $a_0$ = 0.8 fm, MeV, $r_{s.o.}$ = 0.75 fm and $a_{s.o.}$ = 0.4 fm) and by using a simple mass-dependent formula for one ($V_{s.o.} = 6.3 - 0.4A^{-1/3}$). Thus, out of the total of 9 parameters, 5 can be constrained and only 4 need to be individually adjusted to optimise the fits to the experimental angular distributions.

I have then constrained two additional parameters ($r_0' = 1.20 + 0.85A^{-1/3}$ and $a_0' = 1.29 - 2.10A^{-1/3}$) and searched for only two, $V$ and $W_D$. In many cases, the resulting fits to the experimental angular distributions were as good as for the four-parameter search.

Comparing the present coupled channels analysis with the earlier conventional optical model calculations I have found that parameters $r_0$, $a_0$, $V_{s.o.}$, $r_{s.o.}$ and $a_{s.o.}$ have the same values in both cases. Parameters $r_0'$ and $a_0'$ also have similar values. The essential difference between the two analyses is in the values of the potential depths $V$ and $W_D$ and in particular, in their dependence on $B(E2)$.

If the two-step $(d,d')2_1^+(d',d)$ elastic scattering process is included explicitly in the analysis of experimental data, the dependence of $V$ and $W_D$ on the $B(E2)$ $\gamma$-transition probabilities can be eliminated. The interpretation of the $B(E2)$ dependence observed in earlier conventional analyses appears now to be clear: it simply reflects the previously unaccounted for effects of higher order processes in the elastic scattering, which are mainly due to the two-step scattering via the first $2_1^+$ states in the target nuclei.

# 26

# A Study of $^{97,101}$Ru Nuclei

***Key features:***

1. This study resulted in extensive spectroscopic information about the $^{97}$Ru and $^{101}$Ru isotopes.
2. A total of 38 states have been observed, with 30 of them for the first time. Excitation energies to these states have been determined with an accuracy of ±7 keV.
3. A total of 30 angular distributions have been measured and analysed using the distorted wave theory of direct nuclear reactions. Spectroscopic factors and orbital angular momenta have been determined for all of these states. In addition, spins and parities were assigned to states with $l$ = 0, 4, and 5.
4. The ground state $Q_0$ - value for the $^{96}$Ru(d,p)$^{97}$Ru reaction have been determined for the first time with high accuracy of ±0.003 MeV. The determined value is 5.886±0.003 MeV

***Abstract:*** The neutron single-particle strength distributions for the nuclei $^{97}$Ru and $^{101}$Ru have been investigated using the (d,p) reaction at deuteron energy of 11.5 MeV, with an overall experimental resolution of approximately 25 keV. Angular distributions of proton groups leading to sixteen final states in both nuclei were measured in the angular range of 15.0$^0$ to 67.5$^0$. The measured cross sections are analysed in the framework of the distorted waves Born approximation to deduce the $l$ - values and spectroscopic factors of the states in the residual nuclei. The ground-state $Q_0$ - value for the $^{96}$Ru(d,p)$^{97}$Ru reaction has been determined to a much-improved accuracy.

## Introduction

The level structure of the odd ruthenium isotopes has been investigated using $\gamma$- and $\beta$- ray spectroscopy techniques (NDS 1973, 1974a-1974c), and has been shown to be complex. Theoretical calculations using various models for these nuclei have been attempted and met with varying degrees of success (Goswami and Sherwood1967; Imanishi, Fujiwara and Nishi 1973; Kisslinger and Sorensen 1963).

The interpretation of level structure in terms of theoretical models is enhanced by information on the location and distribution of the single-particle strengths among the levels. The single-particle neutron strength distributions for the heavier isotopes $^{103}$Ru and $^{105}$Ru have been investigated via the (d,p) reaction by Fortune *et al*. (1971). The objective of our study was to obtain similar and much needed information for the $^{97}$Ru and $^{101}$Ru isotopes.

## Experimental procedure and the Q-value determination

The 11.5 MeV deuteron beam was produced by the Australian National University EN Tandem accelerator. The targets consisted of isotopically enriched $^{96}$Ru (98%) or $^{100}$Ru (97 %), evaporated using electron beam bombardment onto thin carbon backings. The targets were approximately 20 - 50 $\mu$g/cm$^2$ areal density, as determined by comparing elastically scattered 4 MeV deuterons at forward angles with the Rutherford cross section. The absolute cross sections correct to better than 15%.

Two $\Delta E$-$E$ detector telescopes were used to detect the protons emitted in the (d,p) reaction. These detectors were cooled to approximately -20°C. A target monitor detector was placed at 90°. All detectors were of the silicon surface barrier type and manufactured at the local ANU laboratory. Overall experimental energy resolution was approximately 25 keV. Examples of spectra from the two reactions are shown in Figure 26.1 for the lab angle of 25°.

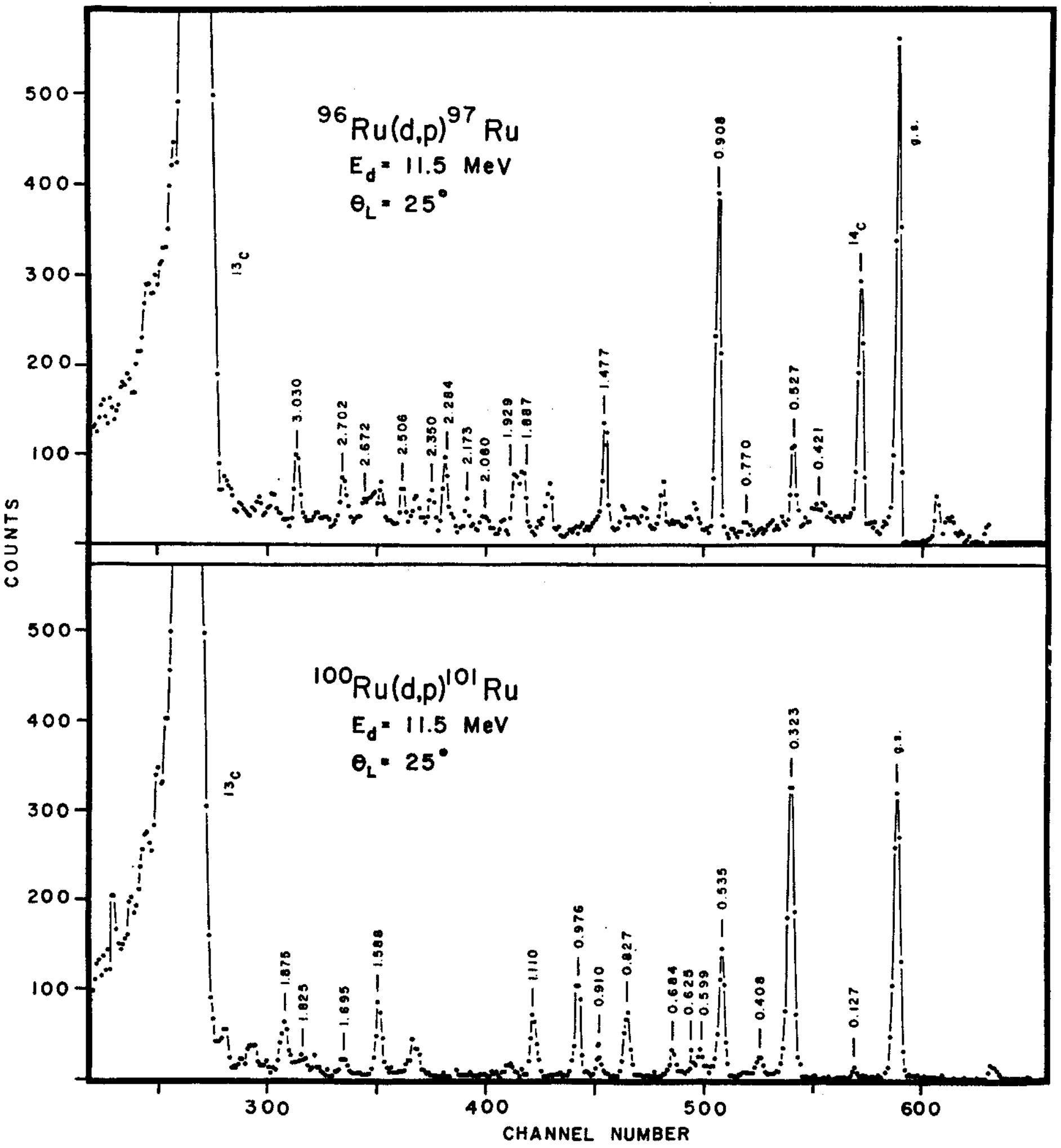

Fig. 26.1. Proton spectra for the $^{96,100}$Ru(d,p)$^{97,101}$Ru reactions at the incident deuteron energy of 11.5 MeV and lab angle of 25°.

Contaminants identified in the targets are $^{16}$0, $^{13}$C and $^{12}$C. Other light contaminants are also present in small quantities. The presence of all these contaminants prevented the extraction of yields for some states and angles.

The very thin nature of the ruthenium targets on the thin carbon backings, enabled us to make a more accurate determination of the ground-state $Q_0$ - value for the $^{96}$Ru(d,p)$^{97}$Ru reaction. At certain angles, protons from the $^{13}$C(d,p)$^{14}$C ground-state reaction were observed to have nearly the same energy as those from the $^{97}$Ru ground state. The ground-state $Q_0$ - value for the $^{96}$Ru(d,p)$^{97}$Ru reaction was determined to be 5.886±0.003 MeV, which is within the limits of the previously listed value of 5.816±0.100 MeV (Wapstra and Gove 1971) but now its accuracy has been significantly improved. The $Q_0$ - value determined for the $^{100}$Ru(d,p)$^{101}$Ru reaction is consistent with the Wapstra and Gove value of 4.581±0.004 MeV.

The excitation energies for the levels in both isotopes were determined at several angles using protons from the (d,p) reactions on $^{16}$O, $^{13}$C and $^{12}$C as calibration points. Angular distributions for fifteen levels in each of $^{97}$Ru and $^{101}$Ru were measured from 15° to 45°, and from 52.5° to 67.5° in 5° steps.

## The distorted wave analysis and results

Angular distributions for the (d,p) reaction were analysed using the DWBA formalism and the computer code DWUCK on the ANU UNIVAC 1108 computer. The calculations included a finite-range correction factor of $R$ = 0.657, as well as corrections for the non-locality of the optical potentials using non-locality lengths of 0.54 for the deuteron channel and 0.85 for the proton channel (see the Appendix E).

The distorted waves for the incident and exit channels were calculated using optical model potentials of the conventional form with a surface absorptive term and a real Thomas-type spin-orbit term as defied in Chapter 25.

The neutron bound-state wave functions were calculated using the same geometry as that of the real part of the Woods-Saxon potential for the proton channel in the distorted wave calculation. The potential also included a Thomas-type spin-orbit term. The depth of the real potential was adjusted to reproduce the experimentally determined separation energy of each level. The potential parameters for deuterons, protons and captured neutrons are listed in Table 26.1.

Table 26.1

Potential parameters used in the distorted wave analysis of the $^{96,100}$Ru(d,p)$^{97,101}$Ru angular distributions

| Particle | Target | $V$ (MeV) | $r_0$ (fm) | $a_0$ (fm) | $W_D$ (MeV) | $r_0'$ (fm) | $a_0'$ (fm) | $V_{s.o.}$ (MeV) | $r_{s.o.}$ (fm) | $a_{s.o.}$ (fm) | $r_C$ (fm) |
|---|---|---|---|---|---|---|---|---|---|---|---|
| neutron | | $^{a}$) | 1.17 | 0.75 | | | | 6.2 | 1.01 | 0.75 | |
| deuteron | $^{96}$Ru | 112.27 | 1.05 | 0.86 | 10.40 | 1.43 | 0.773 | 7.0 | 0.75 | 0.50 | 1.30 |
| | $^{100}$Ru | 111.98 | 1.05 | 0.86 | 10.12 | 1.43 | 0.780 | 7.0 | 0.75 | 0.50 | 1.30 |
| proton | $^{97}$Ru | $^{b}$) | 1.17 | 0.75 | $^{c}$) | 1.32 | 0.568 | 6.2 | 1.01 | 0.75 | 1.25 |
| | $^{101}$Ru | $^{d}$) | 1.17 | 0.75 | $^{e}$) | 1.32 | 0.594 | 6.2 | 1.01 | 0.75 | 1.25 |

$r_C$ – The Coulomb potential used in the calculations is assumed to be caused by a uniformly charged sphere with the radius of $r_C = A^{1/3}$.

$^{a}$) – The depth of the potential is adjusted to give the correct value of the separation energy for a given energy level.

$^{b}$) – Calculated using the energy-dependent formula $V = 59.84 - 0.32E_p$.

$^{c}$) – Calculated using the energy-dependent formula $W_D = 12.80 - 0.25E_p$.

$^{d}$) – Calculated using the energy-dependent formula $V = 60.87 - 0.32E_p$.

$^{e}$) – Calculated using the energy-dependent formula $W_D = 13.34 - 0.25E_p$.

The relationship between the experimental and calculated cross sections is given by

$$\left(\frac{d\sigma(\theta)}{d\Omega}\right)_{\exp} = 1.55 S(l,j) \left(\frac{d\sigma(\theta)}{d\Omega}\right)_{th}$$

where $S(l,j)$ is the spectroscopic factor, $l$ is the transferred orbital angular momentum, $j$ is the transferred total angular momentum, and 1.55 is a zero-range coefficient calculated using the Hulthén wave function for deuterons (see the Appendix E).

The $^{96}$Ru and $^{100}$Ru isotopes have 2 and 6 neutrons outside the closed $N$ = 50 shell, respectively. The shell just above $N$ = 50 is made of $1g_{7/2}$, $2d_{5/2}$, $2d_{3/2}$, $3s_{1/2}$, and $1h_{11/2}$ configuration, in that order for the undeformed potential. The stripped neutron can be deposited to any of these orbitals.

As can be seen from Figures 25.2 and 26.3, the majority of transitions for the $^{96}Ru(d,p)^{97}Ru$ and $^{100}Ru(d,p)^{101}Ru$ reactions display, as expected, the $l$ = 0 and 2 angular distributions but $l$ = 4 and 5 angular distributions are also present. In general, the fits are good for all the measured distributions.

A few weakly excited states were observed in the spectra and while their excitation energies were extracted, the corresponding complete angular distributions could not be obtained.

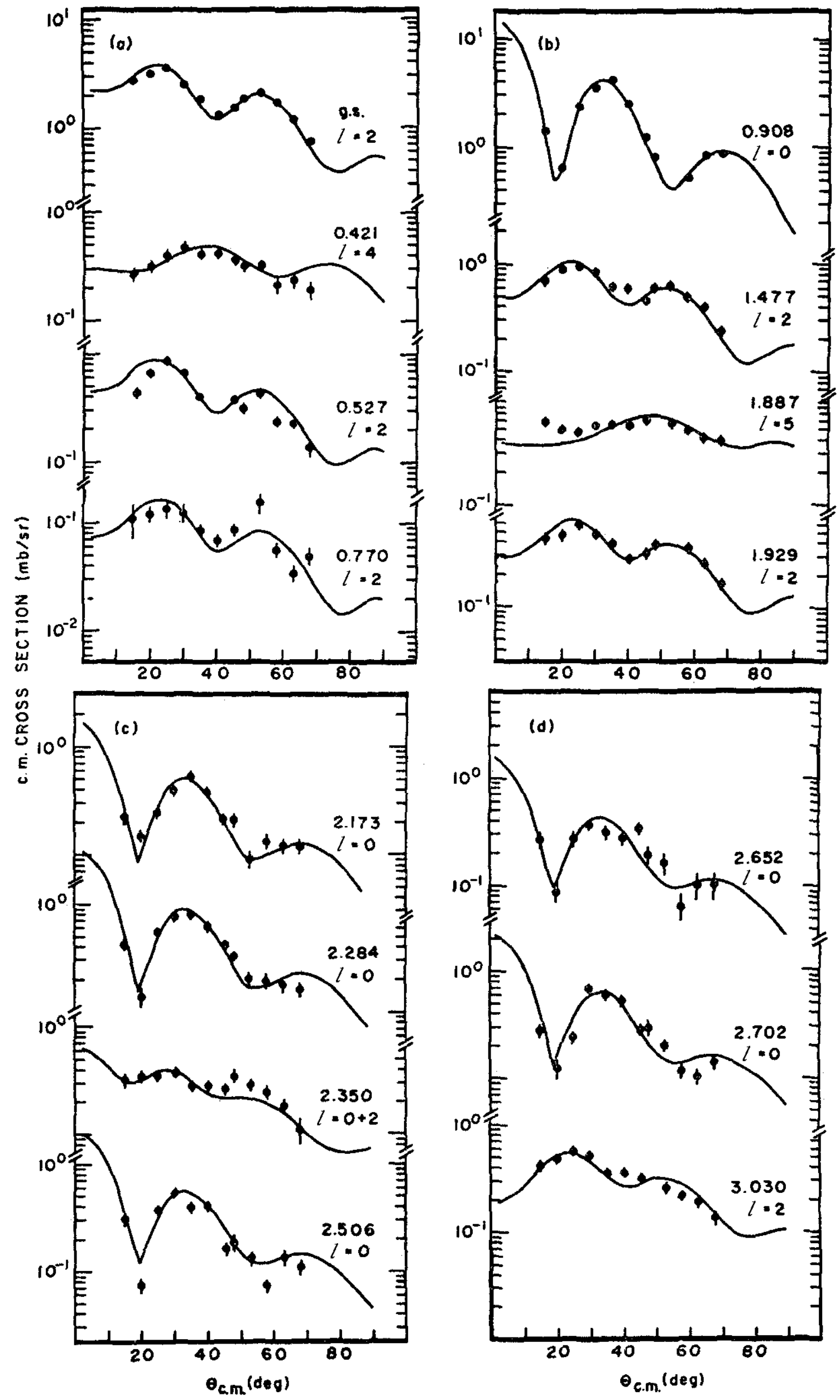

Figure 26.2. The distorted wave fits to the $^{96}Ru(d,p)^{97}Ru$ angular distribution data. The $l$ - value for transferred neutrons and excitation energies in MeV are indicated. Where error bars are not shown, the size of the data point indicates the approximate statistical error in the cross section.

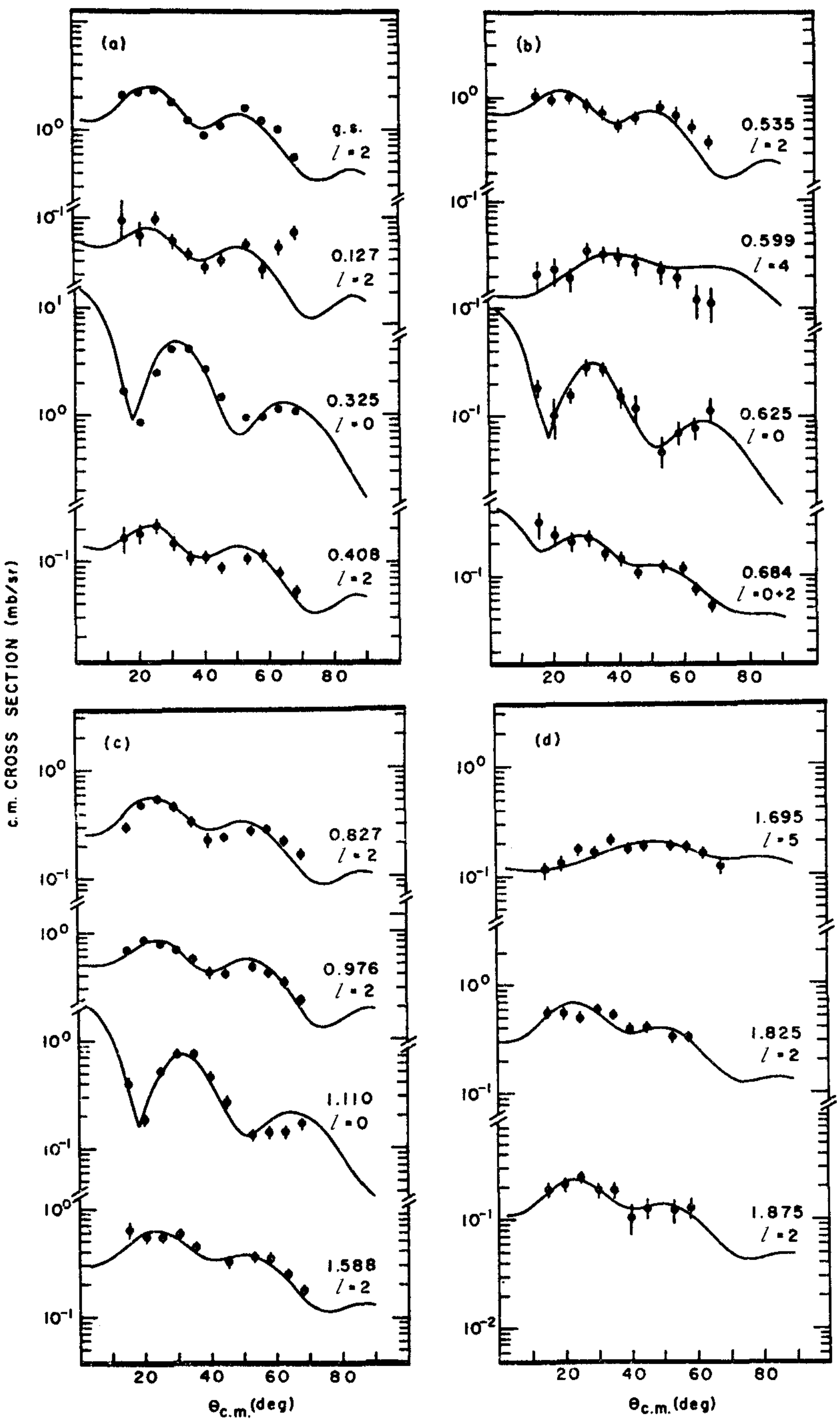

Figure 26.3. The distorted wave fits to the $^{100}Ru(d,p)^{101}Ru$ angular distribution data. See the caption to Figure 26.2

## Discussion

Spectroscopic information extracted from the $^{96,100}Ru(d,p)^{97,101}Ru$ reactions is summarized in Tables 26.2 and 26.3.

The two target nuclei, $^{96}Ru$ and $^{100}Ru$, have the ground-state spin-parity values of $0^+$. Thus, the orbital angular momenta of the states formed in the (d,p) reaction are uniquely determined by comparing their measured angular distributions with the distorted wave calculations.

For $l$ = 2, there are two configuration options available in the shall outside the closed shell $N$ = 50: $2d_{3/2}$ and $2d_{5/2}$. The distorted wave analysis does not allow to distinguish between these two configurations, so unless the spin $j$ is known from earlier studies, two values are listed in Table 26.2. For other $l$ values, only single configurations are available, so unique spin $j$ assignment can be made with a high degree of confidence to the relevant states on the basis of the $l$ values determined by the distorted wave analysis.

### $^{97}$Ru

Table 26.2 lists the excitation energies, $E_x$ (in MeV), orbital angular momenta, $l$, as determined by the distorted wave analysis, the possible neutron single-particle configurations, the total angular momenta, $j$, and the spectroscopic factors $S(l,j)$ determined using the $^{96}$Ru(d,p)$^{97}$Ru reaction.

Table 26.2

Spectroscopic information for $^{97}$Ru obtained using the $^{96}$Ru(d,p)$^{97}$Ru reaction

| $E_x$ | $l$ | *Conf.* | $j$ | $S(l,j)$ | $E_x$ | $l$ | *Conf.* | $j$ | $S(l,j)$ |
|---|---|---|---|---|---|---|---|---|---|
| 0.000 | 2 | $2d_{5/2}$ | $5/2^+$ | 0.57 | 2.080 | ? | | | |
| 0.189 | | $2d_{3/2}$ | $3/2^+$ | | 2.173 | 0 | $3s_{1/2}$ | $1/2^+$ | 0.06 |
| 0.421 | 4 | $1h_{7/2}$ | $7/2^+$ | 0.61 | 2.284 | 0 | $3s_{1/2}$ | $1/2^+$ | 0.11 |
| 0.527 | 2 | $2d_{3/2}$, $2d_{5/2}$ | $3/2^+$, $5/2^+$ | 0.13 | 2.350 | 0 | $3s_{1/2}$ | $1/2^+$ | 0.02 |
| 0.770 | 2 | $2d_{3/2}$, $2d_{5/2}$ | $3/2^+$, $5/2^+$ | 0.05 | | 2 | $2d_{3/2}$, $2d_{5/2}$ | $3/2^+$, $5/2^+$ | 0.05 |
| 0.908 | 0 | $3s_{1/2}$ | $1/2^+$ | 0.58 | 2.506 | 0 | $3s_{1/2}$ | $1/2^+$ | 0.07 |
| 1.477 | 2 | $2d_{3/2}$, $2d_{5/2}$ | $3/2^+$, $5/2^+$ | 0.19 | 2.605 | ? | | | |
| 1.887 | 5 | $1h_{11/2}$ | $11/2^-$ | 0.57 | 2.652 | 0 | $3s_{1/2}$ | $1/2^+$ | 0.05 |
| 1.929 | 2 | $2d_{3/2}$, $2d_{5/2}$ | $3/2^+$, $5/2^+$ | 0.13 | 2.702 | 0 | $3s_{1/2}$ | $1/2^+$ | 0.08 |
| 2.005 | ? | | | | 3.030 | 2 | $2d_{3/2}$, $2d_{5/2}$ | $3/2^+$, $5/2^+$ | 0.09 |

Excitation energies are determined with the accuracy of ±7 keV.

The ground state and the states with excitation energies 0.527, 0.770, 1.477, 1.929 and 3.030 MeV are populated with an $l$ = 2 transfer. The states with excitation energies 0.908, 2.173, 2.284, 2.506, 2.652 and 2.702 MeV are populated with an $l$ = 0 transfer and are assigned the spin-parity $1/2^+$. The states with 0.421 and 1.887 MeV excitation energies are populated with $l$ = 4 and 5 transfers, and are assigned the spins $7/2^+$ and $11/2^-$, respectively.

The angular distribution for the state with 2.350 MeV excitation energy could be fitted using both an $l$ = 0 and 2 transfer, and thus is presumed to be an unresolved doublet. The states with 0.189, 2.005, 2.080 and 2.605 MeV excitation energies were only weakly populated, and reliable angular distributions could not be extracted.

The ground state of $^{97}$Ru, with spin-parity $5/2^+$ consistent with the $\gamma$-spectroscopy measurements of Ohya (1974), is strongly excited in the $^{96}$Ru(d,p)$^{97}$Ru reaction and carries most of the observed $l$ = *2* strength. Its spectroscopic factor is similar to that for the ground states of the two isotones $^{93}$Zr and $^{95}$Mo. The first excited state at 0.189 MeV, with known spin-parity $3/2^+$ was observed at only a few angles and was too weakly excited to determine the spectroscopic factor, indicating that this state has little $d_{3/2}$ single-particle component in its wave function. Similar results are known for the first excited states of $^{93}$Zr and $^{95}$Mo.

The 0.421 MeV $g_{7/2}$ state is observed to have an $l$ = 4 stripping pattern and is the only $l$ = 4 transition located in our study for this isotope. Ohya (1974) suggests the presence of $^7/2^+$ or $^9/_2{}^+$ at 0.839, 0.879,1.229,1.932, 1.970, 2.186 and 2.755 MeV. In the isotones $^{93}$Zr and $^{95}$Mo, the state carrying most of the $l$ = 4 strength is found at 1.477 and 0.768 MeV respectively. The addition of protons is seen to cause a lowering of the energy of the $g_{7/2}$ neutron orbital.

The state at 0.908 MeV is strongly excited and carries most of the $l$ = 0 strength. In $^{93}$Zr and $^{95}$Mo, the $l$ = 0 strength is concentrated in states at 0.947 and 0.789 MeV, respectively.

The state at 1.887 MeV excitation energy is the only state exhibiting an $l$ = 5 stripping pattern for this reaction and is presumed to have $1h_{11/2}$ configuration. In $^{93}$Zr and $^{95}$Mo, the major $h_{11/2}$ strength is found in the states at 2.040 and 1.949 MeV, respectively. The effect of the extra protons on the $h_{11/2}$ neutron orbital appears to be not as large as on the $g_{7/2}$ configuration.

### ***$^{101}$Ru***

Spectroscopic information for the $^{100}$Ru(d,p)$^{101}$Ru reaction is summarised in Table 26.3.

Table 26.3

Spectroscopic information for $^{101}$Ru obtained using the $^{100}$Ru(d,p)$^{101}$Ru reaction

| $E_x$ | $l$ | *Conf.* | $j$ | $S(l,j)$ | $E_x$ | $l$ | *Conf.* | $j$ | $S(l,j)$ |
|---|---|---|---|---|---|---|---|---|---|
| 0.000 | 2 | $2d_{5/2}$ | $^5/_2{}^+$ | 0.35 | 0.714 | ? | | | |
| 0.127 | 2 | $2d_{3/2}$ | $^3/_2{}^+$ | 0.02 | 0.827 | 2 | $2d_{3/2}, 2d_{5/2}$ | $^3/_2{}^+, ^5/_2{}^+$ | 0.12 |
| 0.325 | 0 | $3s_{1/2}$ | $^1/_2{}^+$ | 0.65 | 0.910 | ? | | | |
| 0.408 | 2 | $2d_{3/2}, 2d_{5/2}$ | $^3/_2{}^+, ^5/_2{}^+$ | 0.05 | 0.976 | 2 | $2d_{3/2}, 2d_{5/2}$ | $^3/_2{}^+, ^5/_2{}^+$ | 0.18 |
| 0.535 | 2 | $2d_{3/2}, 2d_{5/2}$ | $^3/_2{}^+, ^5/_2{}^+$ | 0.26 | 1.110 | 0 | $3s_{1/2}$ | $^1/_2{}^+$ | 0.10 |
| 0.599 | 4 | $1g_{7/2}$ | $^7/_2{}^+$ | 0.45 | 1.588 | 2 | $2d_{3/2}, 2d_{5/2}$ | $^3/_2{}^+, ^5/_2{}^+$ | 0.12 |
| 0.625 | 0 | $3s_{1/2}$ | $^1/_2{}^+$ | 0.05 | 1.695 | 5 | $1h_{11/2}$ | $^{11}/_2{}^-$ | 0.18 |
| 0.684 | 0 | $3s_{1/2}$ | $^1/_2{}^+$ | 0.02 | 1.825 | 2 | $2d_{3/2}, 2d_{5/2}$ | $^3/_2{}^+, ^5/_2{}^+$ | 0.04 |
| | 2 | $2d_{3/2}, 2d_{5/2}$ | $^3/_2{}^+, ^5/_2{}^+$ | 0.04 | 1.875 | 2 | $2d_{3/2}, 2d_{5/2}$ | $^3/_2{}^+, ^5/_2{}^+$ | 0.12 |

Excitation energies are determined with the accuracy of ±7 keV.

The ground state and the states with excitation energies of 0.127, 0.408, 0.535, 0.827, 0.976, 1.588, 1.825 and 1.875 MeV are populated with an $l$ = 2 transfer. The states with excitation energies of 0.325, 0.625 and 1.110 MeV are populated with an $l$ = 0 transfer and are assigned a spin-parity of $^1/_2{}^+$. The states with 0.599 and 1.695 MeV excitation energies are populated with $l$ = 4 and 5 transfers, respectively.

The angular distribution for the state at 0.684 MeV could be reproduced by assuming both an $l$ = 0 and an $l$ = 2 transfer, and is presumed to be an unresolved doublet. Additional states were observed with 0.714 and 0.910 MeV excitation energies but angular distributions could not be extracted.

The ground state of $^{101}$Ru, determined previously (Fuller and Cohen 1969) to have spin-parity $^5/_2{}^+$ is strongly excited in this reaction. The first excited state of $^{101}$Ru is at 0.127 MeV excitation energy and has known spin-parity $^3/_2{}^+$. This state is more strongly excited than the 0.189 MeV state of $^{97}$Ru, indicating a larger $d_{3/2}$ single-particle component in its wave function. Its spectroscopic factor is similar to that of the first excited state of $^{103}$Pd.

The state at 0.325 MeV excitation energy is populated strongly and carries most of the $s_{1/2}$ single-particle strength. The state carrying most of the $s_{1/2}$ strength in $^{103}$Pd is found somewhat higher, at 0.500 MeV excitation energy.

The only state exhibiting an $l$ = 4 angular distribution for this isotope was located at 0.599 MeV excitation energy. The $\gamma$-ray measurements indicate many $l$ = 4 states, some with low excitation energies. In particular, one with 0.307 MeV excitation energy, which if present, would be masked by the strong $l$ = 0 state at 0.325 MeV. In $^{103}$Pd the strongest $l$ = 4 state occurs at an excitation energy of 0.245 MeV.

The state with 1.695 MeV excitation energy was the only state exhibiting an $l$ = 5 angular distribution and is assumed to have $1h_{11/2}$ configuration. In $^{103}$Pd, the major $l$ = 5 strength is found lower, in the state with 0.787 MeV excitation energy.

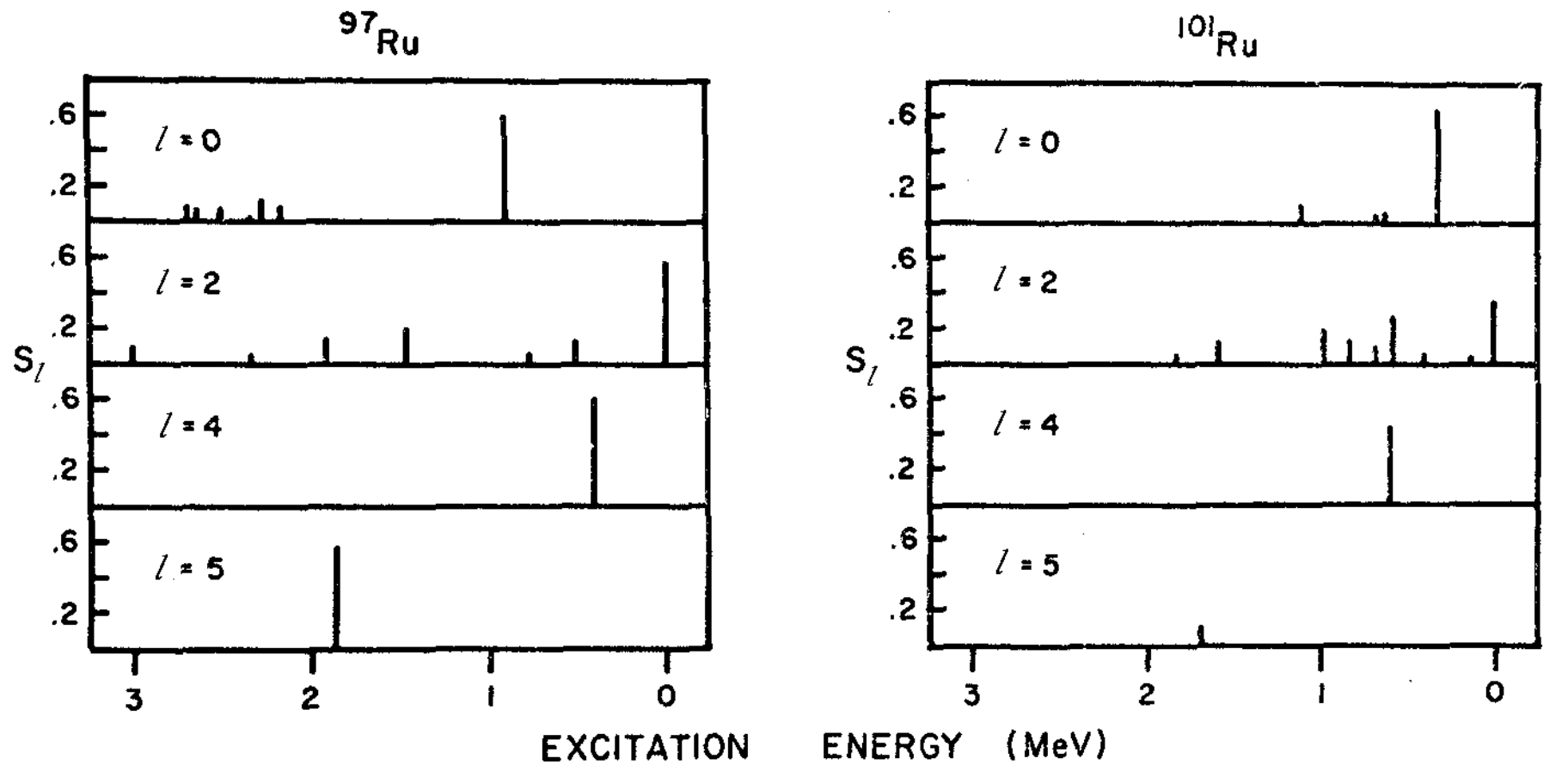

Figure 26.4. The distributions of spectroscopic strength for $l$ = 0, 2, 4 and 5 angular momentum transfers in the $^{96,100}$Ru (d, p)$^{97,101}$Ru reactions.

The spectroscopic strengths for the $^{96,100}$Ru(d,p)$^{97,101}$Ru reactions are plotted against the excitation energies in Figure 26.4 for various observed $l$ - transfers. The effect of the four additional neutrons in the $^{100}$Ru core is mainly in lowering the position of the $l$ = 0 strength and in compressing the $l$ = *2* strength distribution.

## Summary

The ground-state $Q_0$-value for the $^{96}$Ru(d,p)$^{97}$Ru reaction has been determined to much improved accuracy. Twenty states in $^{97}$Ru with excitation energies up to 3.030 MeV, sixteen not previously observed, and eighteen states in $^{101}$Ru with excitation energies up to 1.875 MeV, fourteen not previously observed, have been identified. Orbital angular momentum transfer values and spectroscopic factors have been obtained for states in both $^{97}$Ru and $^{101}$Ru. In $^{97}$Ru six $l$ = 0, six $l$ = 2, one $l$ = 4, one $l$ = 5 and one admixture of $l$ = 0 and $l$ = 2 have been assigned. In $^{101}$Ru three $l$ = 0, nine $l$ = 2, one $l$ = 4, one $l$ = 5 and one admixture of $l$ = 0 and $l$ = 2 have been assigned.

The distorted wave calculations, using optical model parameters from global analyses, provided a good description of the measured angular distributions. The neutron single-particle strength distribution for $^{97}$Ru is similar to those obtained for the isotones $^{93}$Zr and $^{95}$Mo. The neutron single-particle strength distribution for $^{101}$Ru is similar to that of the isotone $^{103}$Pd, while differing markedly from that of $^{99}$Mo.

# 27

# Spectroscopy of the $^{53,55,57}$Mn Isotopes and the Mechanism of the ($^4$He,p) Reaction

***Key features:***

1. A total of 122 excited states have been identified in the $^{53,55,57}$Mn isotopes and the corresponding excitation energies have been assigned to all of them. Many of the states have never been observed before, particularly in $^{57}$Mn where we have identified 38 new states.
2. We have found that the reaction mechanism depends strongly on the incident $^4$He energy. Most of the angular distributions measured using 18 MeV $^4$He projectiles are associated with an indirect reaction mechanism. In contrast, angular distributions measured at 26 MeV show clear direct reaction features.
3. A total of 95 distributions have been measured using 26 MeV $^4$He projectiles and were analysed using the distorted wave formalism. We have found that at this energy, the ($^4$He,p) reaction can be interpreted as a direct transfer of three-nucleon cluster.
4. The $J$ – dependence have been observed for both $L$ = 1 and $L$ = 3 angular momentum transfer.
5. A total of 46 $J^\pi$ - values have been assigned to states in $^{53,55,57}$Mn nuclei.
6. We have found that many states, which are weakly excited in single transfer reaction, are excited strongly in the ($^4$He,p) reaction. New states, which were not previously observed in the ($^3$He,d) reaction, have been also accessed by the ($^4$He,p) reaction. Thus, this reaction offers an important alternative way to study nuclear structure.

***Abstract***: The $^{50,52,54}$Cr($^4$He,p)$^{53,55,57}$Mn reactions have been studied at 18 and 26 MeV $^4$He bombarding energies. From the 26 MeV data, angular distributions for 95 levels were obtained, nearly all of which could be described by the distorted wave procedure assuming a quasi-triton transfer process. In contrast, at 18 MeV very few angular distributions could be adequately described using the direct reaction mechanism. The $J$ - dependence was observed for both $L$ = 1 and $L$ = 3 transfers and used to assign $J^\pi$ values for many states in $^{53,55}$Mn. In $^{57}$Mn, 38 new states (in a total of 57) were observed and $J^\pi$ assignments were made for many of them. The ($^4$He,p) reaction mechanism and nuclear structure are discussed.

## Introduction

The use of the ($^4$He,p) or (p,$^4$He) reactions in nuclear spectroscopy has often been limited by an inadequate knowledge of the reaction mechanism and by the need for using simplifying procedures in distorted wave analyses of experimental data. However, these factors are less restrictive in obtaining spectroscopic information if detailed single-proton transfer measurements are available to the same final states. In such cases the great attractions of these multi-particle transfer reactions can be more fully utilized. Use can be made of their selectivity associated with seniority and isospin, of their ability to access complex configurations, and of their applicability to resolve the $J^\pi$ ambiguity by using the $J$ - dependence (Bucurescu *et al.* 1972), which is particularly clear for the $L$ = 1 angular momentum transfer. Our measurements of the $^{50,52,54}$Cr($^4$He,p) reactions at 18 and 26 MeV were undertaken to study the spectroscopy of the $^{53,55,57}$Mn isotopes and to examine the ($^4$He,p) reaction mechanism in this energy-mass region.

Both $^{53}$Mn and $^{55}$Mn have been studied earlier. Katsanos and Huizenga (1967) used the (p,p') scattering to study the $^{55}$Mn isotope. Tarara *et al.* (1976) used the $^{56}$Fe(p,$^4$He) reaction at 14-16 MeV to examine the $^{53}$Mn isotope. The single-proton structure of $^{53}$Mn has been studied with the ($^3$He,d) reaction by O'Brien *et al.* (1969). The same reaction together with the $J$ - dependence for the ($^7$Li,$^6$He) reaction was used by Gunn, Fix, and Kekelis (1976) to make $J^\pi$ assignments. Rapaport *et al.* (1969) have used the ($^3$He,d) reaction in a study of $^{55}$Mn. The preponderance of measured $l_p$ values in these ($^3$He,d) studies have been for $l_p$ = 1 and 3. Hence the $J$ - dependence for the $L$ = 1 transfer in the ($^4$He,p) reactions can be used to make $J^\pi$ assignments. The availability of $L$ = 3 transitions permits also an experimental test for any similar $J$ - dependence for $L$ = 3 transfer. The (p,$^3$He) and (d,$^4$He) reactions at 27 and 16.5 MeV, respectively, have been used to study low-lying states in $^{55}$Mn by Peterson, Pittel and Rudolph (1971) and by Peterson and Rudolph (1972).

At the commencement of the present work no information was available on the excited states of $^{57}$Mn. However, in the course of our study, Mateja *et al.* (1976) published their results for the $^{54}$Cr($^4$He,p$\gamma$)$^{57}$Mn reaction at 15, 21 and 24 MeV, and Mateja *et al.* (1977) for the $^{55}$Mn(t,p)$^{57}$Mn reaction at 17.0 MeV. They have made several spins assignments on the basis of their p-$\gamma$ angular correlation studies but no proton angular distributions were reported.

The $^{50,52,54}$Cr isotopes have 24 protons and 26, 28, and 30 neutrons, respectively. In the simple shell model description, the four protons are outside the $Z$ = 20 shell and occupy the $1f_{7/2}$ orbits. The neutrons assume an interesting set of configurations spaning the $N$ = 28 shell: in $^{50}$Cr two neutrons are missing to close the $N$ = 28 shell; in $^{52}$Cr the $N$ = 28 shell is closed; and in $^{54}$Cr there are two neutrons outside the $N$ = 28 shell.

Thus, in the ($^4$He,p) reaction to low-lying states, the stripped proton may be expected to be transferred preferentially to the $N$ = 28 shell, which is made of $1f_{7/2}$ orbitals. For the stripped neutrons, the most likely transfer for the low-lying states excited in the $^{50}$Cr($^4$He,p)$^{53}$Mn reaction is to the $1f_{7/2}$ orbital. However, the participation of the configurations in the $N$ = 50 shell, i.e. $2p_{3/2}$,$1f_{5/2}$, $2p_{1/2}$ and even $1g_{9/2}$ are also possible. For the $^{52,54}$Cr($^4$He,p)$^{55,57}$Mn reactions, the two stripped neutrons are less likely to be transferred to the $1f_{7/2}$ orbital but rather to any configurations in the $N$ = 50 shell, i.e. $2p_{3/2}$,$1f_{5/2}$, $2p_{1/2}$ and $1g_{9/2}$.

## Experimental procedure

The measurements were carried out using two particle accelerators. For the 18 MeV measurements, we used 400 nA $^4$He beam delivered by the ANU EN tandem accelerator. Measurements at 26 MeV were carried out using the ANU 14UD Pelletron accelerator.

For 18 MeV measurements, chromium targets of around 50 $\mu$g/cm$^2$ on gold backings were produced from enriched material while for 26 MeV self-supporting targets of around 200 $\mu$g/cm$^2$ were used. The reaction products were detected using cooled surface-barrier detector telescopes consisting of two or three detectors depending on the beam energy and the ($^4$He,p) $Q$ - values. The overall experimental resolution was typically around 32 keV at 18 MeV and varied from 45 to 60 keV at the higher energy with the forward angle data having the better resolution. Special attention was given to minimizing, as far as possible, contributions to the resolution from kinematics and target thickness effects. The consistency of angular distribution shapes for the same known transitions in all isotopes is a good indicator that, even at the higher excitation energies, there is little contribution from unresolved components.

Examples of proton spectra are shown in Figures 27.1 and 27.2. Figure 27.1 shows the 26 MeV spectra for the $^{50,52}Cr(^4He,p)^{53,55}Mn$ reactions while Figure 27.2 shows the proton spectra for the $^{54}Cr(^4He, p)^{57}Mn$ reaction at both 18 and 26 MeV.

Energy calibrations for the $^{53,55}Mn$ isotopes were made using the well-known levels up to around 5 MeV in each isotope. The $Q$ - values were determined with accuracy generally better than 10 keV.

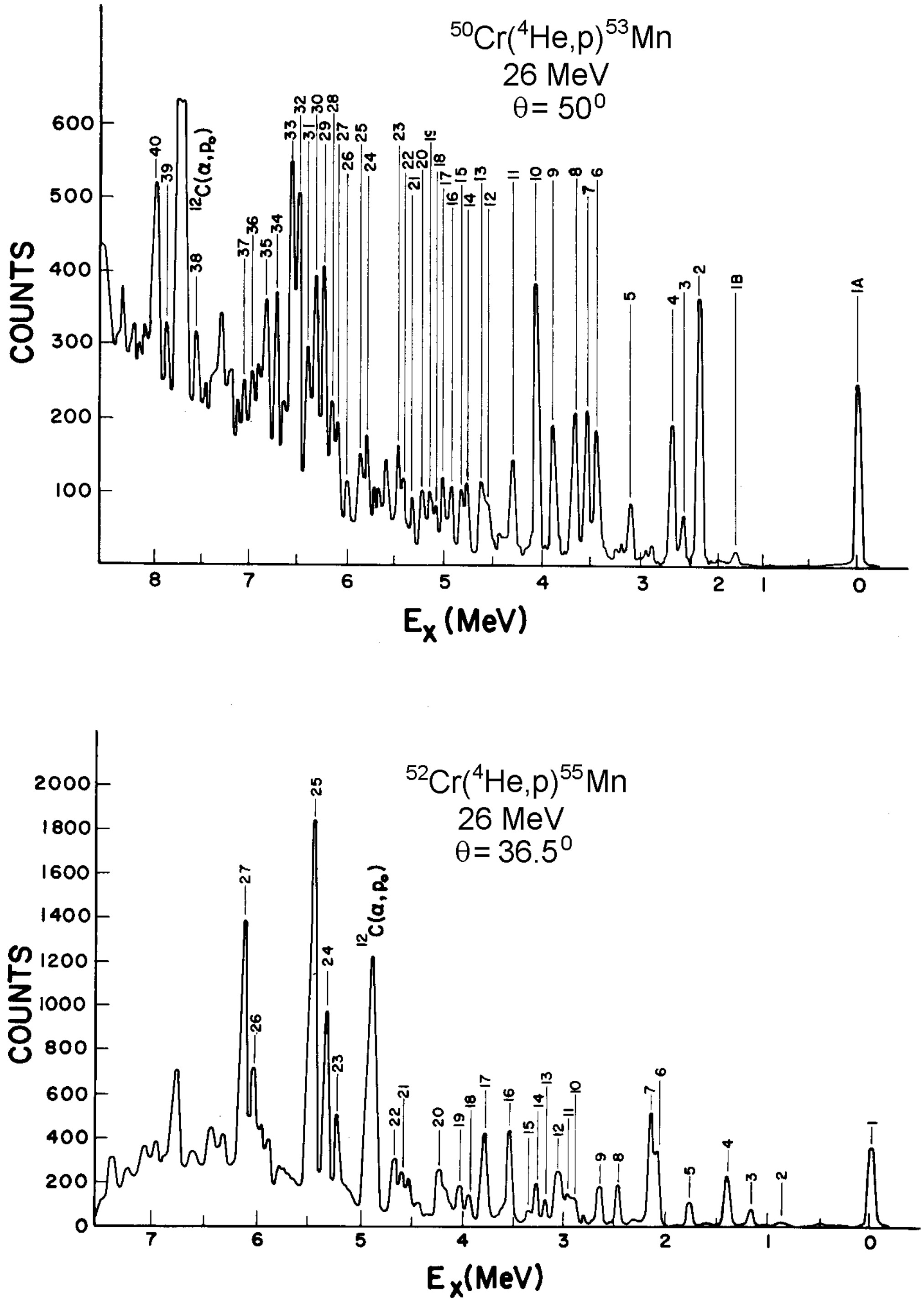

Figure 27.1. Examples of proton spectra for the reactions $^{50}Cr(^4He,p)^{53}Mn$ and $^{52}Cr(^4He,p)^{55}Mn$ at 26 MeV.

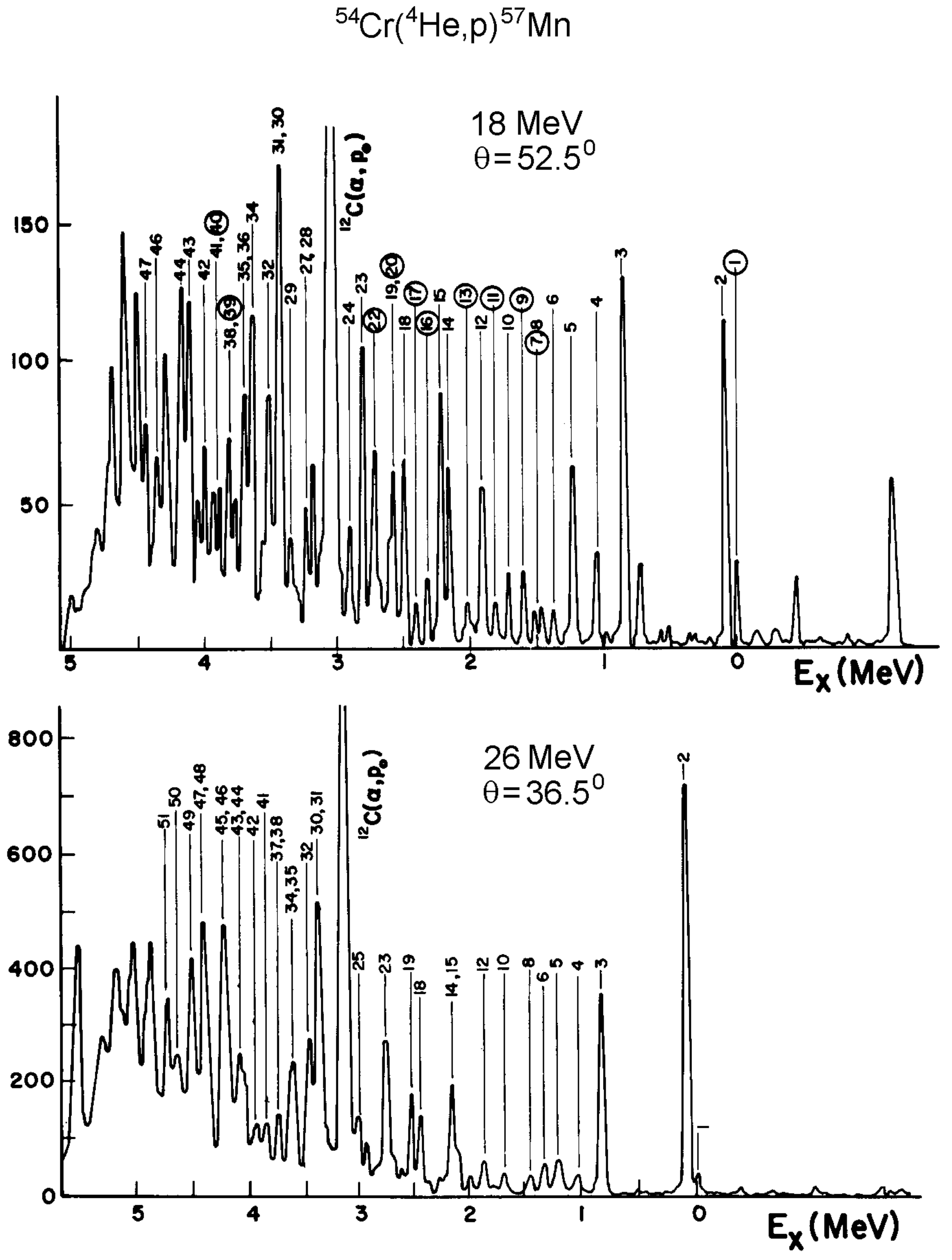

Figure 27.2. Examples of proton spectra for the $^{54}Cr(^4He,p)^{57}Mn$ reaction at 18 and 26 MeV. The circled numbers indicate the peaks that either disappear or show considerably reduced intensity at 26 MeV.

At the commencement of the present work no levels were known in $^{57}Mn$ so that special care was exercised in determining the $^{57}Mn$ $Q$ - values. Since $^{52}Cr$ was the major contaminant (7%) in the $^{54}Cr$ targets, the $^{54,52}Cr(^4He,p)$ reactions were run consecutively at each angle at both energies. In this way, the $^{57}Mn$ spectra were calibrated using $^{55}Mn$ levels and contaminants due to $^{52}Cr$ were removed by subtracting the normalized

$^{52}Cr(^{4}He,p)$ spectra. Allowance was made from elastic scattering yields for differences in target thickness. The $Q$ - values were determined from the better resolution data at 18 MeV and have errors of the same order as the levels in $^{55}Mn$.

Angular distributions were in general measured from 15° to 150° at 18 MeV and from 20° to 90° at 26 MeV. The differential cross sections at both energies were determined by measuring them relative to known elastic scattering cross sections. For normalization, a fixed monitor detector was placed at 45° and counted both the elastic events and the inelastic scattering to the first $2^+$ state of the target. At 26 MeV, the elastic scattering at each angle was also recorded in the first detector of each telescope simultaneously with the reaction. The absolute elastic scattering cross sections were determined from a comparison with Rutherford scattering. The absolute cross section is accurate to around 7%.

## Data analysis

The angular distributions for the $^{50,52,54}Cr(^{4}He,p)^{53,55,57}Mn$ reactions were analysed using the distorted wave theory and the computer code DWUCK. The calculations assumed a quasi-triton cluster transfer. The transferred triton was assumed to be bound in a Woods-Saxon potential and the customary separation energy prescription was used. A spin-orbit term was tried in the triton well but did not affect the shapes of the calculated angular distributions.

The optical model parameters used in both channels were taken from the Perey and Perey compilation (1974). The $^{4}He$-particle parameters were determined by measuring elastic scattering angular distributions and analysing them using the optical model with volume absorption. An example of fits for the elastic scattering of $^{4}He$ at 26 MeV is given in Figure 27.3. The $^{4}He$ and triton parameters used in the distorted wave analysis are listed in Table 27.1.

Table 27.1

The $^{4}He$ and triton parameters used in the distorted wave analysis of the $^{50,52,54}Cr(^{4}He,p)^{53,55,57}Mn$ angular distributions at 18 and 26 MeV

| Particle-energy | Isotope | $V$ (MeV) | $r_0$ (fm) | $a_0$ (fm) | $W$ (MeV) | $r_0'$ (fm) | $a_0'$ (fm) | $r_c$ (fm) |
|---|---|---|---|---|---|---|---|---|
| α 18 MeV | all | 189.3 | 1.36 | 0.57 | 24.9 | 1.36 | 0.57 | 1.40 |
| α 26 MeV | all | 185.6 | 1.40 | 0.57 | 25.4 | 1.40 | 0.57 | 1.40 |
| t 18 MeV | 53 | a) | 1.38 | 0.23 | | | | |
| | 55 | a) | 1.40 | 0.35 | | | | |
| | 57 | a) | 1.40 | 0.30 | | | | |
| t 26 MeV | all | a) | 1.45 | 0.35 | | | | |

a) Adjusted to match the triton separation energy.

The magnitudes and shapes of the $^{50,52,54}Cr(^{4}He,p)^{53,55,57}Mn$ angular distributions have been found sensitive to the triton well geometry. Good fits to almost all angular distributions at 26 MeV were obtained using $r_0$ = 1.45 fm and $a_0$ = 0.35 fm for tritons. However, in a few cases, to improve the fits it was necessary to deviate from this triton geometry. In particular, $a_0$ had to be reduced considerably. These exceptional cases are indicated by the dashed lines in Figures 27.4 - 27.8, which show the $^{50,52,54}Cr(^{4}He,p)^{53,55,57}Mn$ angular distributions. These cases were not used in extracting the reduced strength information. Furthermore, except for the well-known $L$ = 1 transitions characterized by a very distinct $J$ - dependence, spins associated with calculated distributions for these cases have not been considered as unambiguous assignments even for states with known $l_p$ values.

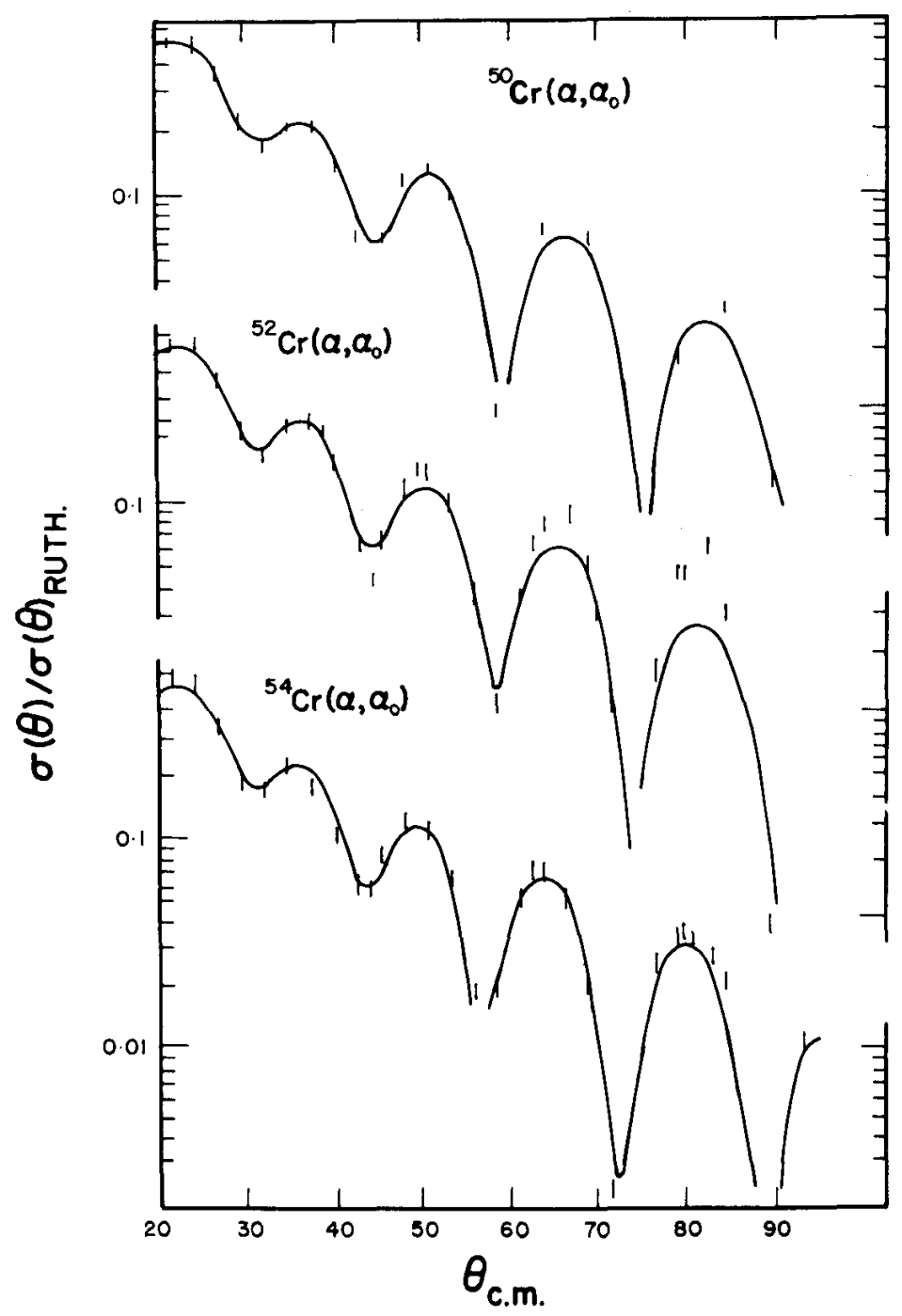

Figure 27.3. Examples of the optical model fits to the elastic scattering of $^4He$ particles at 26 MeV. The parameter sets for all these calculations are listed in Table 27.1.

The calculated distributions shown in Figures 27.4 - 27.9 have been normalized to the experimental data. The errors shown include both the statistical error and an estimate of the uncertainty introduced by the spectrum fitting and normalization procedures. The similarities in the shapes of the angular distributions made it in general difficult to assign definite spin values unless the $l_p$ transfer was already known.

## Results and discussion

In general, the ($^4He$,p) distributions measured at 18 MeV were difficult to describe using the distorted wave formalism. The distributions, which could be fitted, are displayed in Figures 27.4 and 27.5. The displayed $l_p$ values were supplied by the ($^3He$,d) measurements. States with large spectroscopic factors in ($^3He$,d) are strongly excited in the ($^4He$,p) reaction and even at 18 MeV their angular distributions exhibit the direct mechanism over a wide range of angles of at least for $\theta$ < 90°. The notable examples of this are the strong transitions to the $^7/_2^-$ and $^3/_2^-$ states at 0 and 2.413 MeV in $^{53}Mn$ and at 0.128 and 2.258 MeV in $^{55}Mn$. However, both the 18 and 26 MeV data reveal that it is not a necessary condition for a state to have a large spectroscopic factor in a single-proton transfer reaction in order to be strongly excited by the ($^4He$,p) reaction. In fact, we have found that many states, which are weakly excited in single transfer reaction, are excited strongly in the ($^4He$,p) reaction. New states, which were not previously observed in the ($^3$He,d) reaction, have been also accessed by the ($^4$He,p) reaction.

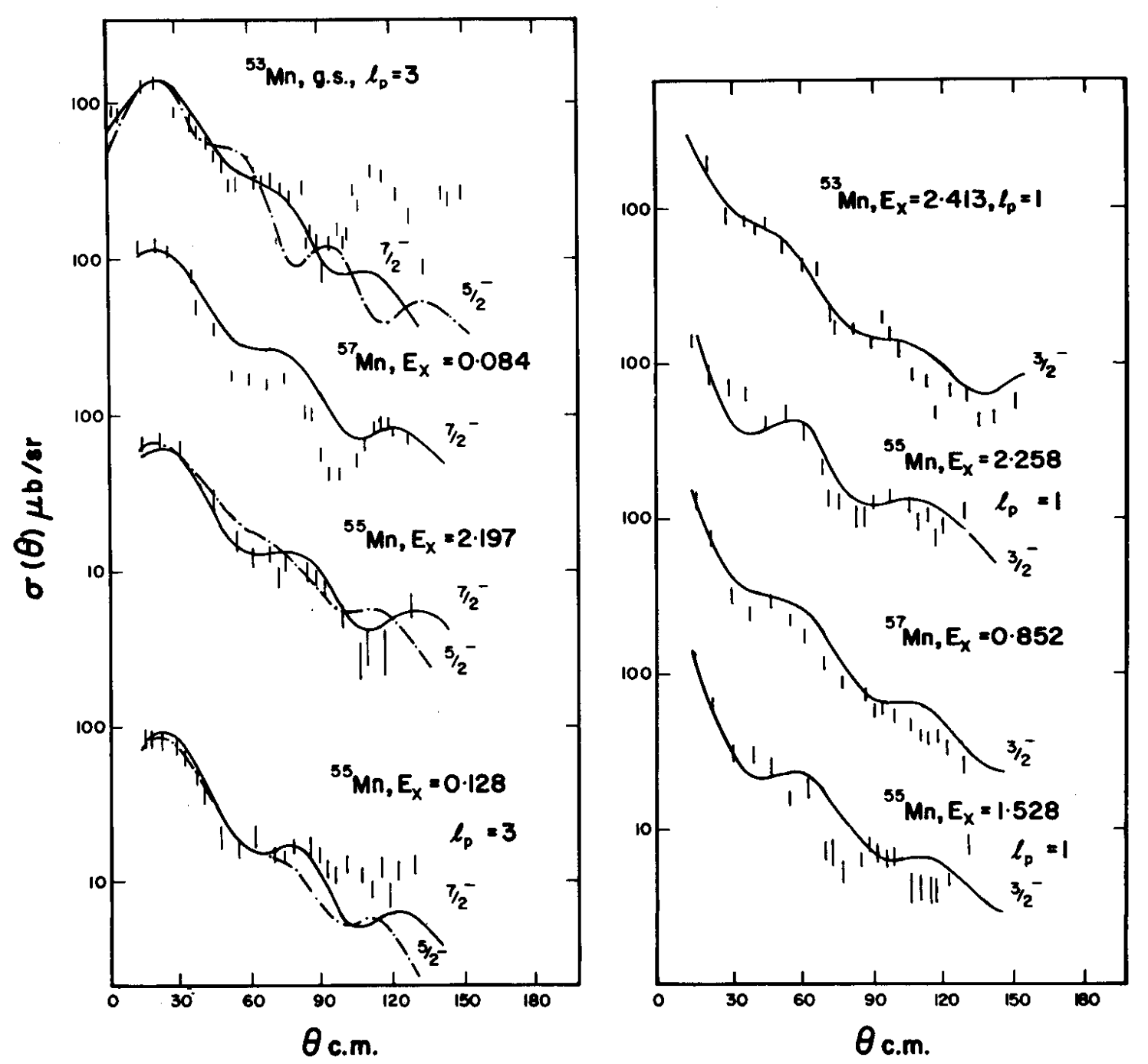

Figure 27.4. Examples of angular distributions at 18 MeV that could be interpreted using the distorted wave formalism. The displayed $l_p$ values were taken from ($^3$He,d) results of O'Brien *et al.* (1969) and Rapaport *et al.* (1969).

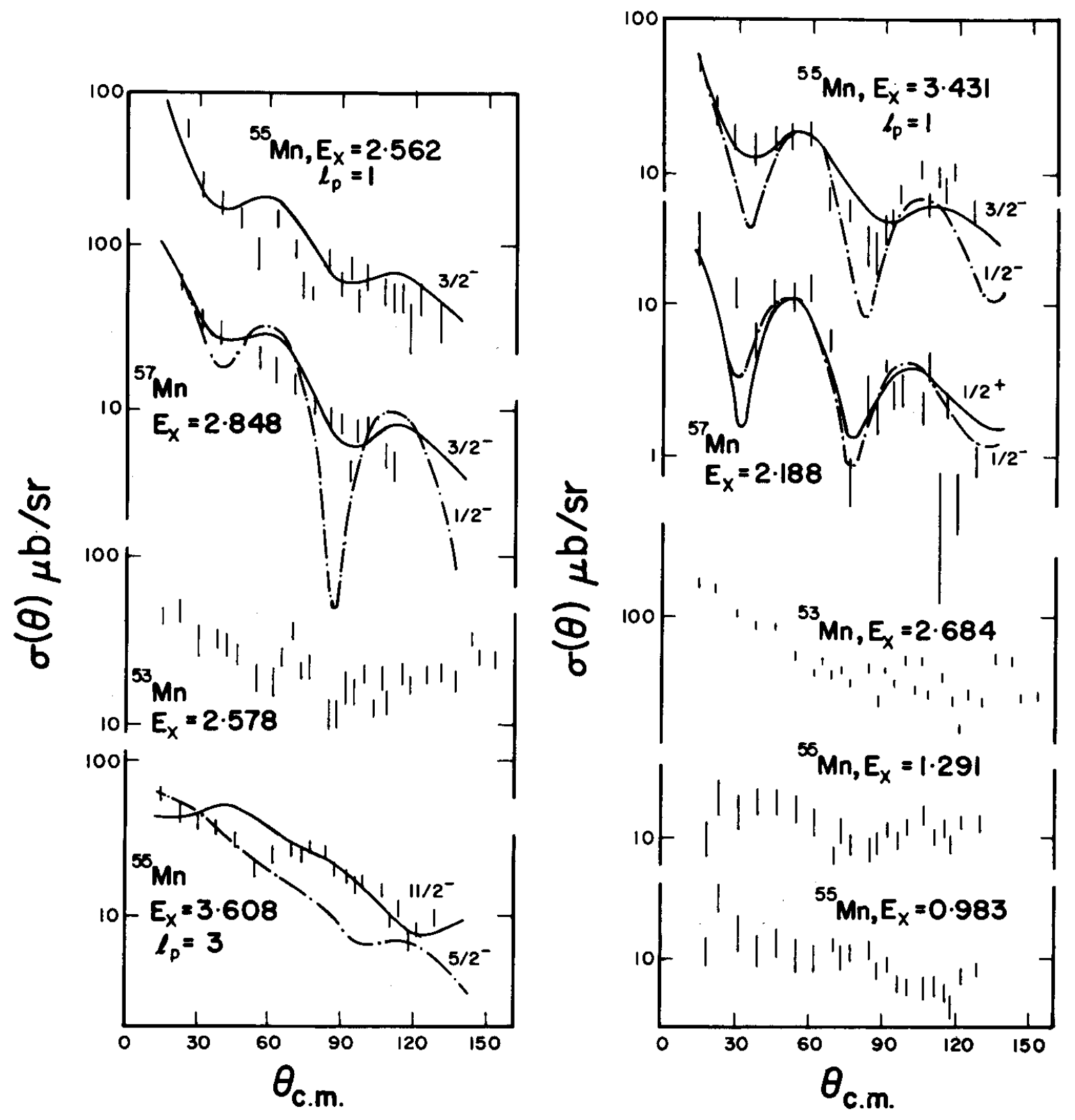

Figure 27.5. More examples of angular distributions at 18 MeV that could be interpreted using the distorted wave formalism. The figure also shows examples of angular distributions that could not be described theoretically.

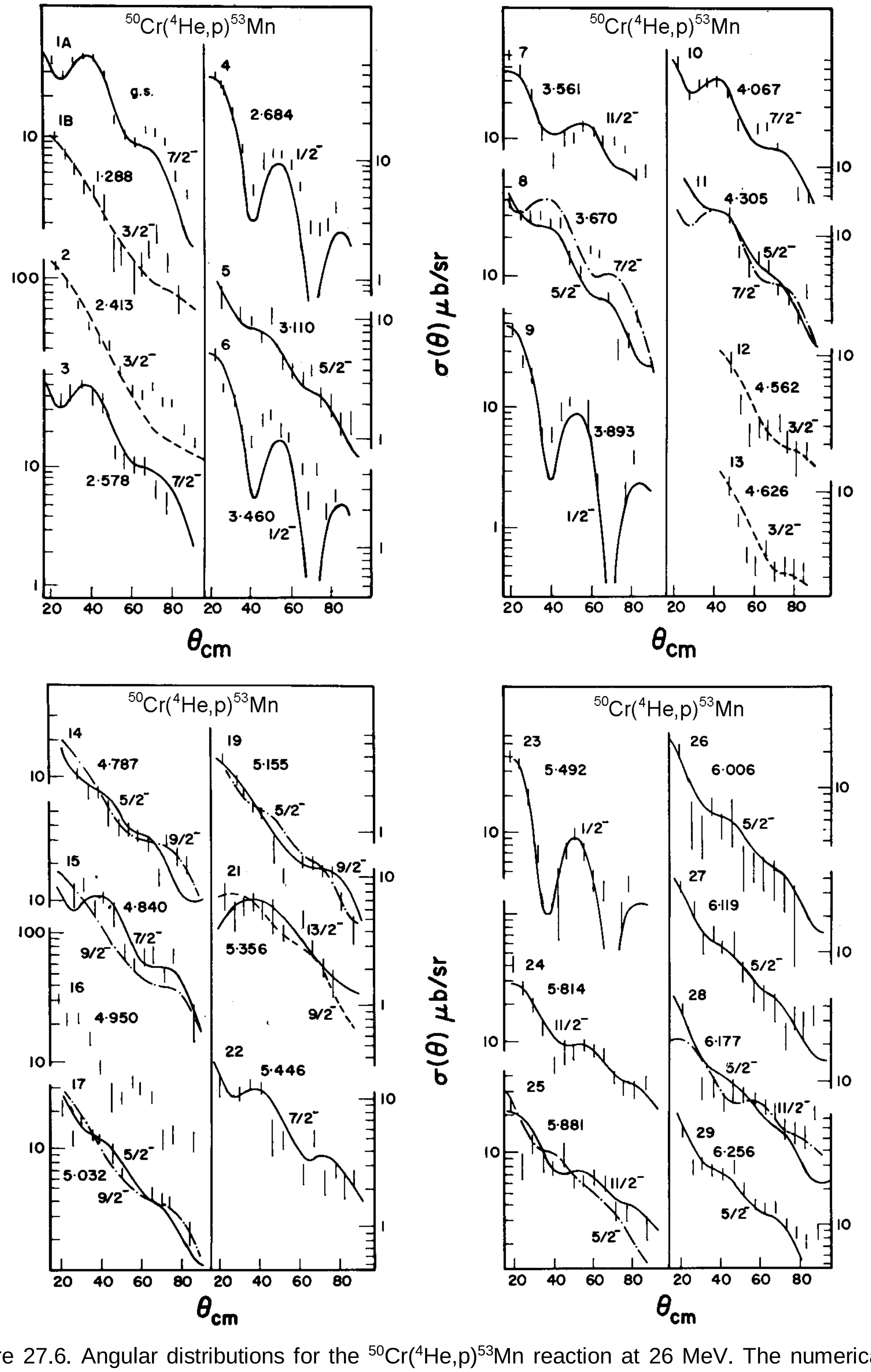

Figure 27.6. Angular distributions for the $^{50}Cr(^{4}He,p)^{53}Mn$ reaction at 26 MeV. The numerical labels correspond to the labels used in the proton spectra (see Figure 27.2). The dashed lines are calculated using different geometrical parameters (mainly $a_0$) from those listed in Table 27.1 for tritons (see the text). The dashed curves were not used to extract the reduced strengths.

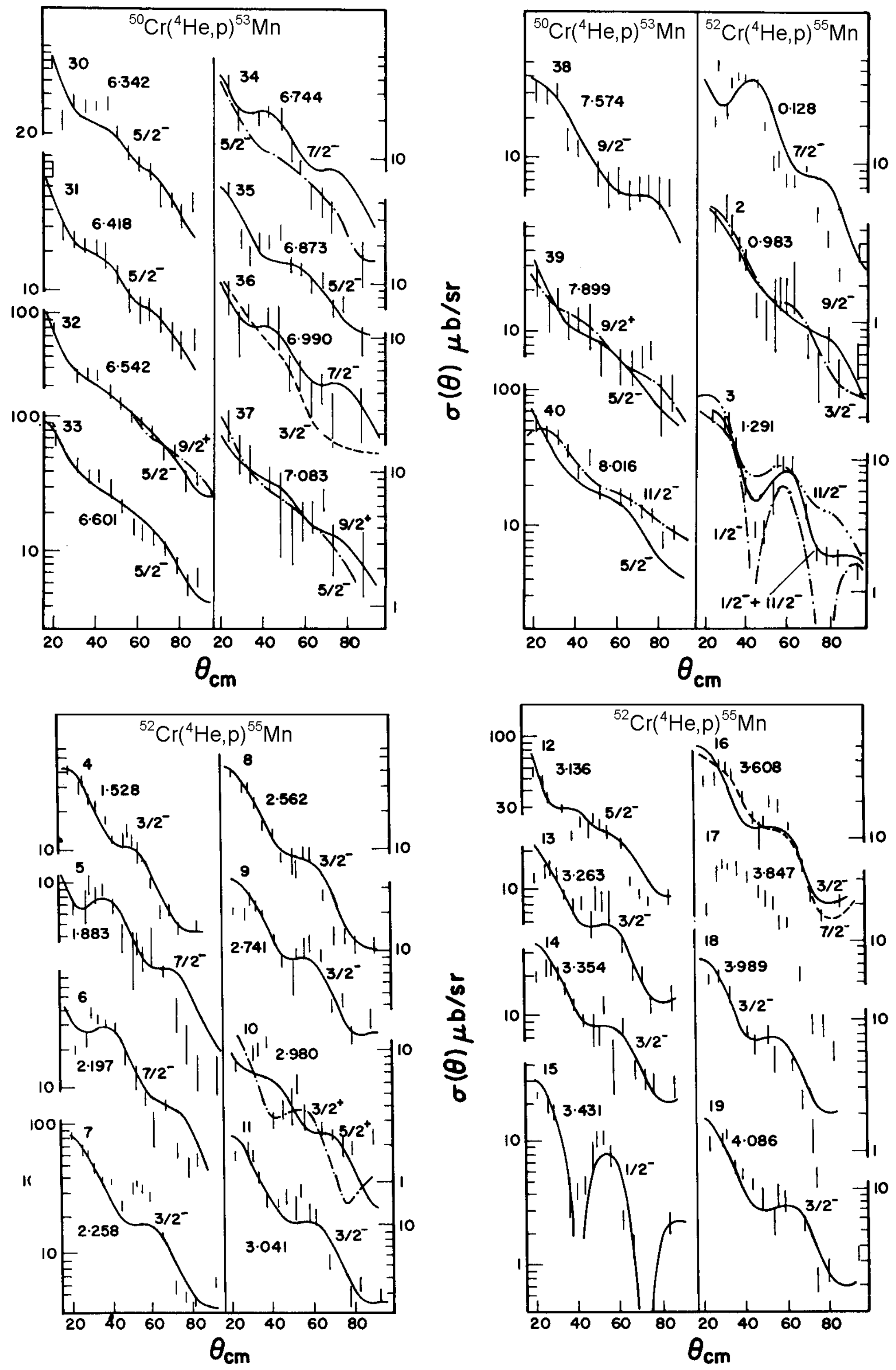

Figure 27.7. Examples of angular distributions for the $^{50,52}Cr(^4He,p)^{53,55}Mn$ reactions at 26 MeV. See the caption to Figure 27.6.

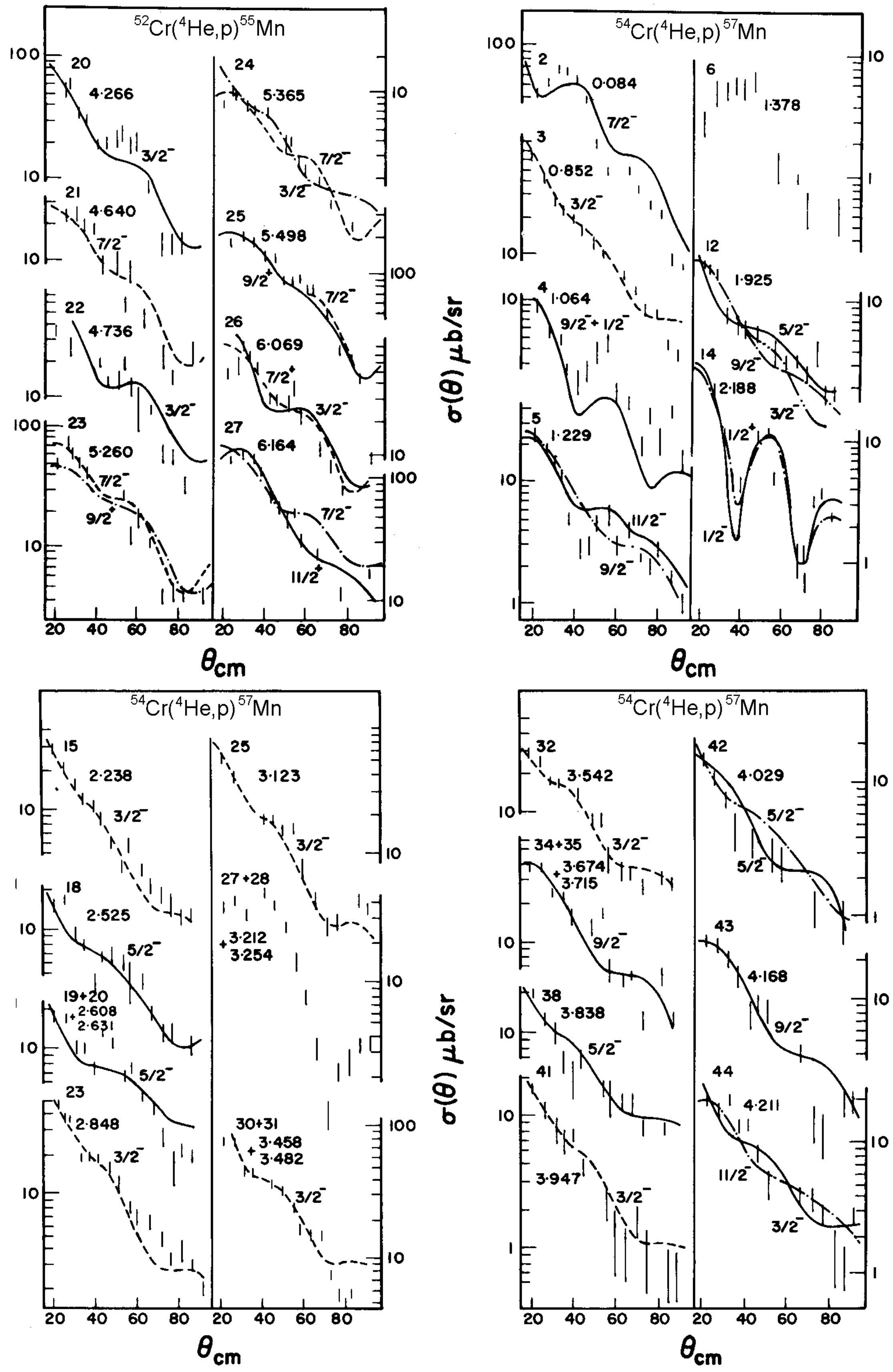

Figure 27.8. Examples of angular distributions for the $^{52,54}Cr(^4He,p)^{55,57}Mn$ reactions at 26 MeV. See the caption to Figure 27.6.

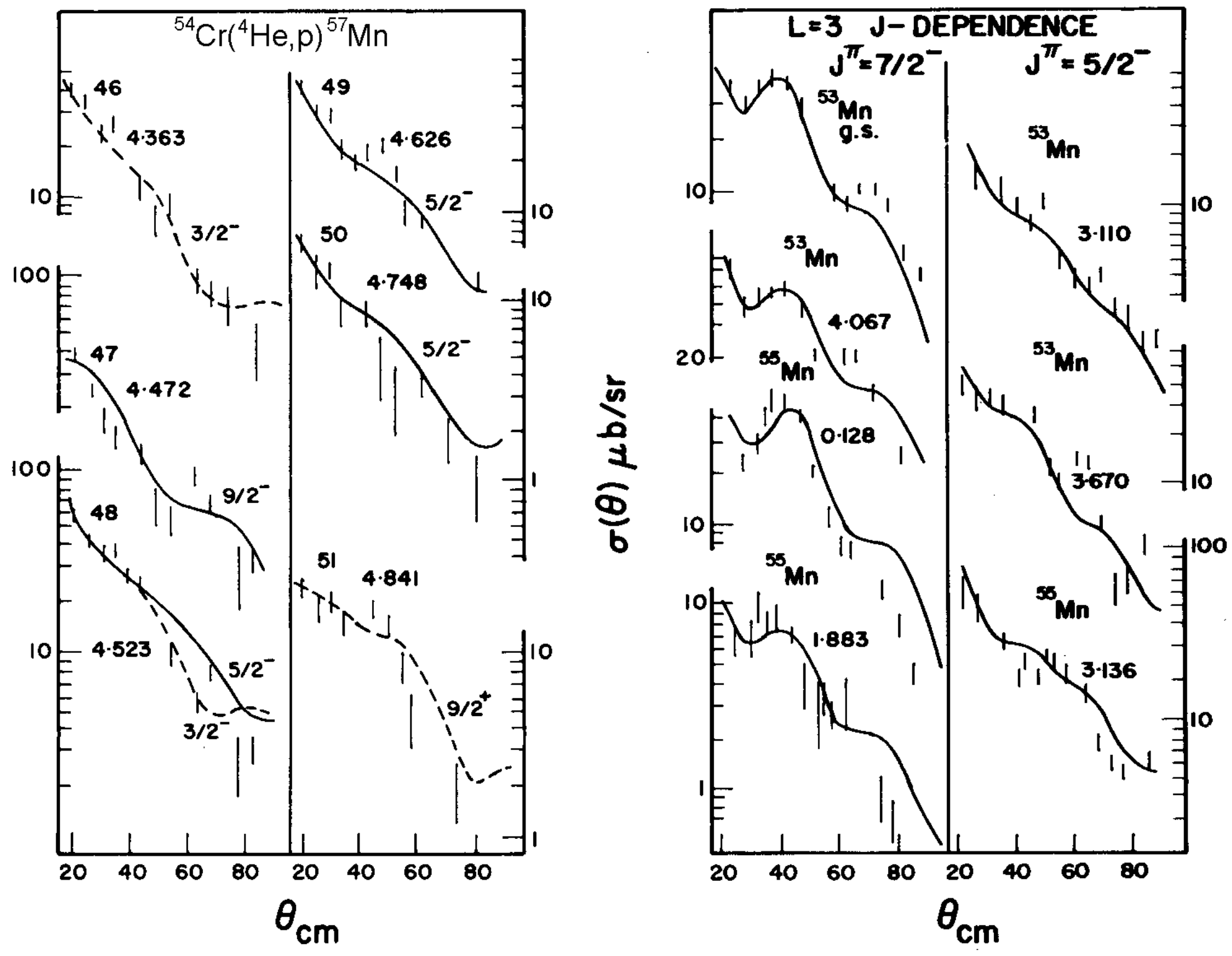

Figure 27.9. Left-hand side: Angular distributions for the $^{54}Cr(^{4}He,p)^{57}Mn$ reaction at 26 MeV. (See also the caption to Figure 27.6.) The right-hand side: Examples of the $J$ - dependence for $L$ = 3.

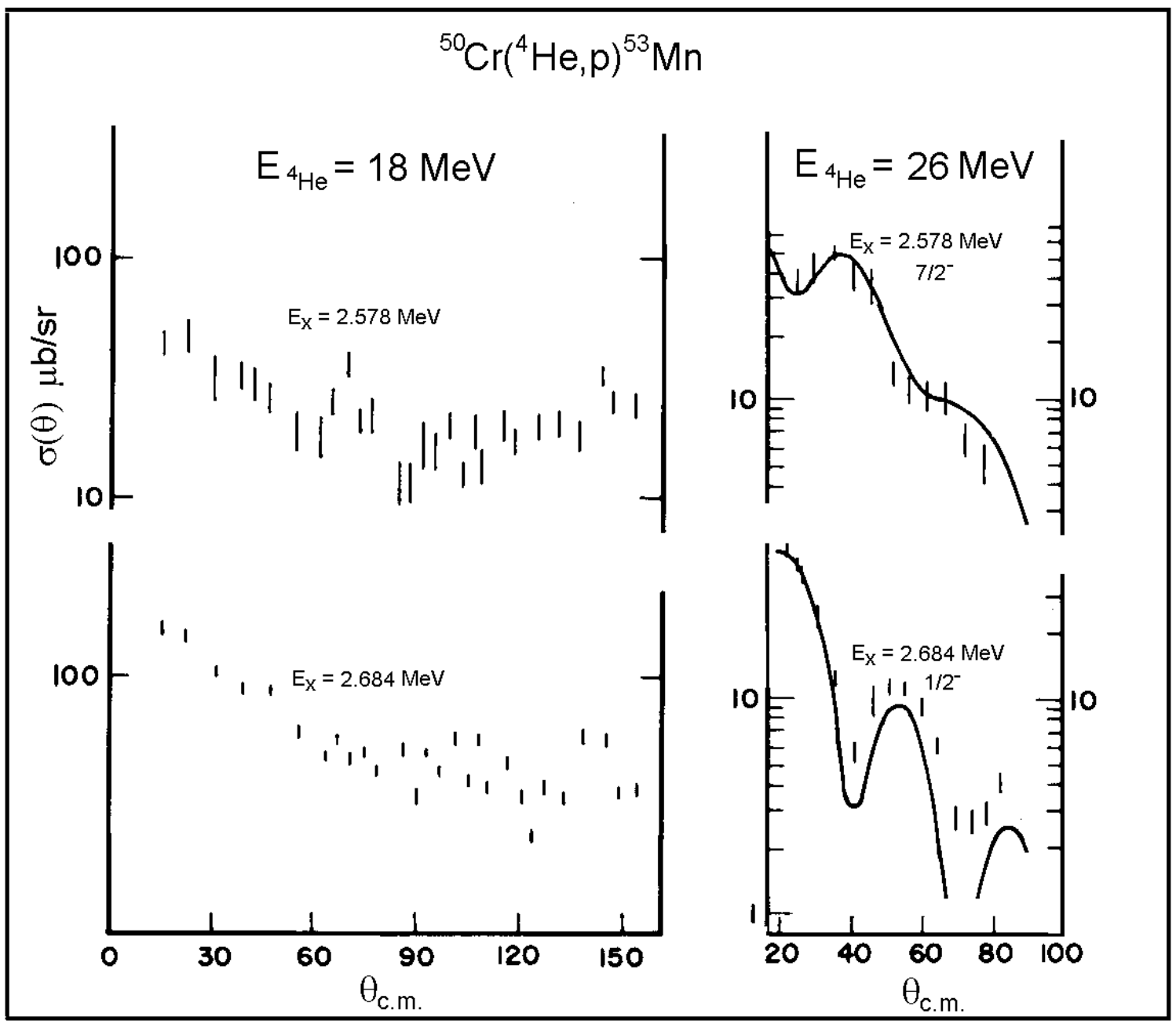

Figure 27.10. Two examples of strong energy dependence of the reaction mechanism for the ($^{4}$He,p) reaction. Featureless distributions at 18 MeV display clear direct transfer mechanism at 26 MeV.

At 26 MeV, nearly all the angular distributions could be described well by the distorted wave theory. These angular distributions are presented in Figures 27.6-27.9. Our study revealed that the reaction mechanism for ($^{4}$He,p) reactions depends strongly on the incident energy of $^{4}$He particles. Many featureless distributions measured at 18

MeV displayed clear characteristics of direct mechanism at 26 MeV. Examples are the distributions for the 2.578 and 2.684 MeV in $^{53}$Cr shown in Figures 27.10 for 18 and 26 MeV.

The 26 MeV data also show a clear $J$ – dependence of the shape of the angular distributions for $L$ = 3 (see Figure 27.9).

Referring to the results for the ($^{3}$He,d) reactions, $l_p$ = 3 transfers have been located at 0.0, 3.110, 3.670 and 4.067 MeV in $^{53}$Mn and at 0.128, 1.883, 3.136 and 3.608 MeV in $^{55}$Mn. The ($^{4}$He,p) angular distributions for these states can be categorized into two groups. One group consists of the ground and 4.067 MeV states in $^{53}$Mn shown in Figure 27.6 and the 0.128 and 1.883 MeV states in $^{55}$Mn shown in Figure 27.7. This group is characterized by a peak in the cross section at about 35°.

The second group contains the 3.110 and 3.670 MeV states in $^{53}$Mn shown in Figure 27.6 and the 3.136 MeV state in $^{55}$Mn shown in Figure 27.7. The angular distributions for these states show less structure than do those of the first group. In particular, they do not exhibit a maximum in the cross section at $\theta$ = 35°. All these distributions are grouped together in Figure 27.9. Based on the distorted wave calculations the first group consists of $J^{\pi} = {}^{7}/_{2}^{-}$ transfers and the second of $J^{\pi} = {}^{5}/_{2}^{-}$ transfers.

We have also observed a clear $J$ -dependence for $L$ = 1 transfers. This can be seen by comparing distributions for $J^{\pi} = {}^{1}/_{2}^{-}$ and ${}^{3}/_{2}^{-}$ in Figures 27.6 and 27.7. The distribution for $J^{\pi} = {}^{1}/_{2}^{-}$ display well-defined diffraction structure, which is reproduced reasonably well using the distorted wave theory. In contrast, the differential cross sections for $J^{\pi} = {}^{3}/_{2}^{-}$ decrease virtually smoothly with the increasing reaction angle.

### *The $^{50}$Cr($^{4}$He,p)$^{53}$Mn reaction*

The results of the present work are compared with those from previous measurements in Table 27.2. The high-resolution $^{56}$Fe(p,$^{4}$He)$^{53}$Mn measurements of Tarara *et al.* (1976) indicate that there are at least 81 states in $^{53}$Mn with the excitation energies of up to 5.092 MeV, whereas the ($^{4}$He,p) reaction at 26 MeV excites only nineteen resolvable groups.

By comparing theoretical and experimental cross sections one can derive the reduced strength $R$ values for the ($^{4}$He,p) transitions. The reduced strength is defined as

$$R = \frac{\sum_{\theta} \sigma_{\exp}(\theta)}{\sum_{\theta} \sigma_{th}(\theta)}$$

where $\sigma_{\exp}(\theta)$ is the experimental differential cross section at the reaction angle $\theta$ and $\sigma_{th}(\theta)$ is the cross section calculated theoretically using the distorted wave formalism. The reduced strengths for the $^{50}$Cr($^{4}$He,p)$^{53}$Mn reaction at 26 MeV are shown in Figure 27.11.

Several theoretical studies of the $^{53}$Mn nuclear structure have been made (Benson and Johnstone 1975; Lips and McEllistrem 1970; Malik and Scholz 1966; McCullen, Bayman, and Zamick 1964; Osnes 1971; Saayman and Irvine 1976; Scholz and Malik 1967). The most extensive study was by Benson and Johnstone (1975). Their study included the $1f_{7/2}^{-3}$, $1f_{7/2}^{-4}$, $2p_{3/2}$, $2p_{1/2}$, and $1f_{5/2}$ configurations. They concluded that the lowest 1p-4h states belong to predominantly neutron excitations.

Table 27.2

Spectroscopic information about $^{53}$Mn extracted from our study of the $^{50}$Cr($^{4}$He,p)$^{53}$Mn reaction at 26 MeV compared with earlier results

| Peak a) no. | $E_x$ b) (keV) | $E_x$ c) (keV) | $l_p$ e) | $J^\pi$ f) | $J^\pi$ g) | $\sigma(\theta_0)$ h) ($\mu$b/sr) |
|---|---|---|---|---|---|---|
| 1A | 0 | 0 | 3 | $\frac{7}{2}^-$ | $\frac{7}{2}^-$ | 36.7 |
| 1B | 1288 d) | 1296 | 1 | $\frac{3}{2}^-$ | $\frac{3}{2}^-$ | 10.9 |
| 2 | 2413 d) | 2413 | 1 | $\frac{3}{2}^-$ | $\frac{3}{2}^-$ | 119 |
| 3 | 2578 | | | $\frac{7}{2}^-$ | $\frac{7}{2}^-$ | 7.0 |
| 4 | 2684 | 2678 | 1 | $\frac{1}{2}^-$ | $\frac{1}{2}^-$ | 50.1 |
| 5 | 3110 | 3061 | 3 | $\frac{5}{2}^-$ | $\frac{5}{2}^-$ | 13.7 |
| | | 3104 | 1 | $\frac{3}{2}^-$ | | |
| 6 | 3460 | 3484 | 1 | $\frac{1}{2}^-$ | $\frac{1}{2}^-$ | 40.5 |
| 7 | 3561 | | | | $(\frac{11}{2}^-)$ | 44.4 |
| 8 | 3670 | 3669 | 3 | $\frac{5}{2}^-$ | $\frac{5}{2}^-$ | 37.2 |
| 9 | 3893 | 3900 | 1 | $\frac{3}{2}^-$ | $\frac{1}{2}^-$ | 43.2 |
| 10 | 4067 | 4070 | 1+3 | $(\frac{3}{2}^-)+\frac{5}{2}^-$ | $(\frac{7}{2}^-)$ | 65.5 |
| | | 4278 | | $\frac{5}{2}^-$ | | |
| 11 | 4305 | 4304 | 3 | | | 14.1 |
| 12 | 4562 d) | 4569 | 1+3 | $(\frac{1}{2}^-)+(\frac{5}{2}^-)$ | | 8.4 |
| 13 | 4626 d) | | | | | 10.4 |
| 14 | 4787 | 4788 | (0), (1) | | | 13.2 |
| 15 | 4840 | | | | | 9.9 |
| 16 | 4950 | 4936 | 3 | $(\frac{5}{2}^-)$ | | 26.6 |
| | | 4963 | 1 | $(\frac{1}{2}^-)$ | | |
| 17 | 5032 | | | | | 20.4 |
| 18 | 5112 | 5085 | 1 | $(\frac{1}{2}^-)$ | | |
| 19 | 5155 | | | | | 38.4 |
| 20 | 5240 | | | | | |
| 21 | 5356 | | | | $\geqq \frac{9}{2}$ | 7.3 |
| 22 | 5446 | | | | $(\frac{7}{2}^-)$ | 14.1 |
| 23 | 5492 | 5485 | 1 | $(\frac{1}{2}^-)$ | $\frac{1}{2}^-$ | 42.8 |
| 24 | 5814 | 5800 | | | $(\frac{11}{2}^-)$ | 36.5 |
| 25 | 5881 | 5886 | | | | 22.7 |
| 26 | 6006 | 6005 | | | | 18.0 |
| 27 | 6119 | | | | | 29.7 |
| 28 | 6177 | 6150 | | | | 33.1 |
| 29 | 6256 | 6240 | | | | 52.2 |
| 30 | 6342 | 6320 | | | | 56.4 |
| 31 | 6418 | 6410 | | | | 67.5 |
| 32 | 6542 | 6540 | (4) | | | 80.4 |
| 33 | 6601 | | | | | 75.0 |
| 34 | 6744 | 6730 | | | | 34.4 |
| 35 | 6873 | 6870 | | | | 50.3 |
| 36 | 6990 | 6970 | 1 | $\frac{3}{2}^-$ | | 19.8 |
| 37 | 7083 | 7100 | (4) | | | 22.3 |
| 38 | 7574 | 7540 | 1 | $\frac{1}{2}^-$ | | 29.8 |
| 39 | 7899 | 7910 | 4 | $\frac{9}{2}^+$ | | 23.9 |
| 40 | 8016 | 8030 | 3 | $\frac{5}{2}^-$ | | 61.0 |

a) See Figure 27.1.

b) Excitation energies determined by our measurements.

c) O'Brien *et al.* (1969) for $E_x \leq 6$ MeV; Gunn, Fox, and Kekelis (1976) for $E_x > 6$ MeV.

d) Distributions fitted using altered geometrical parameters for tritons (see the text).

e) Compilation of the orbital angular momentum values for the transferred proton (Armstrong and Blair 1965; Čujec and Szöghy 1969; Gunn, Fox, and Kekelis 1976; O'Brien *et al.* 1969).

f) Compilation of previous spin assignments (Auble and Rao 1970; Armstrong, Blair, and Thomas, 1967; Gunn, Fox, and Kekelis 1976; Schulte, King, and Taylor (1975); Wiest *et al.*1971).

g) Our $J^\pi$ assignments.

h) Differential cross sections for the $^{50}$Cr($^{4}$He,p)$^{53}$Mn at 26 MeV measured at the most forward angle $\theta_0$. For most states $\theta_0 \approx 20.8^0$ (c.m.).

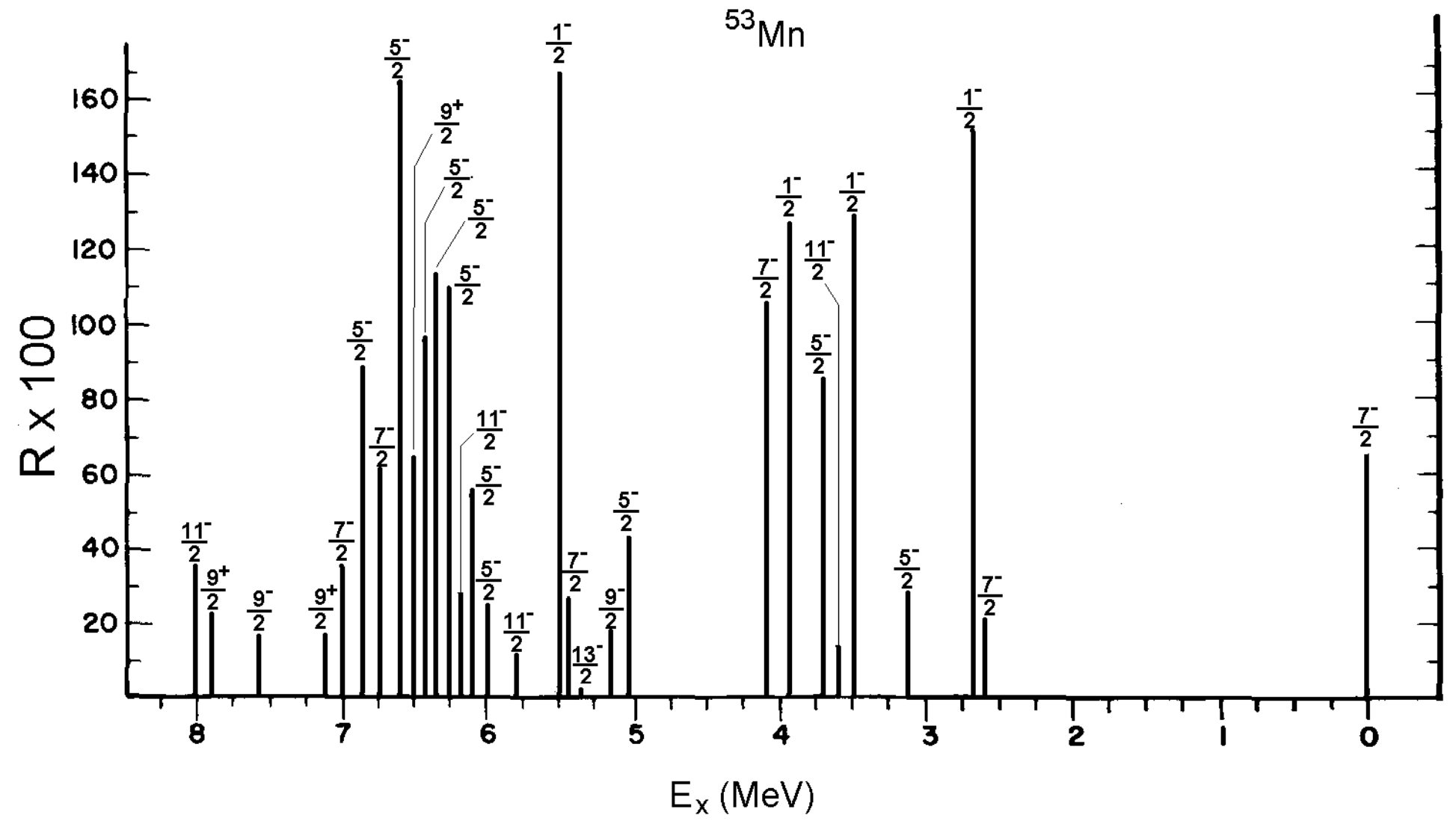

Figure 27.11. Reduced strengths $R$ for the $^{50}$Cr($^4$He,p)$^{53}$Mn reaction at 26 MeV. The figure shows also the assigned $J^\pi$ values.

The first five states of $^{53}$Mn at 0, 0.376, 1.288, 1.438 and 1.618 MeV consist mainly of the $(1f_{7/2}^{-3})_J$ configuration with well-known spins of $^5/_2{}^-$, $^3/_2{}^-$, $^{11}/_2{}^-$, and $^9/_2{}^-$. As such, the proton seniority[1] ($v_p$) of the $1f_{7/2}^{-3}$ components is necessarily $v_p$ = 3 for these states.

The excitation of the 1.288 MeV state in both the ($^3$He,d) and ($^4$He,p) reactions might be associated with a $(1f_{7/2}^{-4})_0 2p_{3/2}$ component present due to the close proximity of the $2p_{3/2}$ single-particle state. The angular distribution for this state could not be fitted satisfactorily with the distorted wave procedure and as a consequence cannot rule out a more complicated excitation process. However, the same fitting problems were encountered with all $^3/_2{}^-$ distributions in $^{53}$Mn, including the distribution for the state at 2.413 MeV, which can be identified with the 2.5 MeV state predicted by Benson and Johnstone.

According to the same calculations, the 2.578 MeV level, assigned $^7/_2{}^-$, is probably formed from two-neutron excitation out of the $1f_{7/2}$, shell. This could explain both the excitation of this state in the ($^4$He,p) reaction and its non-population in the single-proton transfer reaction. The calculation also predicts two $^1/_2{}^-$ states at 2.56 and 3.4 MeV and a 5/2$^-$ state at 3.1 MeV. These could be identified with the levels observed at 2.684, 3.460 and 3.110 MeV, respectively. The distorted wave calculations have reproduced these angular distributions satisfactorily.

### *The $^{52}$Cr($^4$He,p)$^{55}$Mn reaction*

Spectroscopic information for the $^{55}$Mn nucleus is summarized in Table 27.3.

A comparison of proton spectra in Figure 27.1 shows that many more states are excited in $^{55}$Mn than in $^{53}$Mn for levels of up to 4.1 MeV excitation energy. In fact, there are 19 groups in the $^{55}$Mn spectrum up to 4.086 MeV (peak 19) compared to 10 groups in the $^{53}$Mn spectrum. The larger density of states in $^{55}$Mn is not surprising considering that $^{53}$Mn has a closed $N$ = 28 neutron core whereas the two added neutrons in $^{55}$Mn have access to the *f-p* orbitals. The prominent gap in the $^{53}$Mn spectrum between the ground state and the 2.413 MeV level (peaks 1A and 2) is broken only by the weakly excited 1.288 MeV level (peak IB). However, in $^{55}$Mn there are five states excited between the strongly excited $^7/_2{}^-$ and $^3/_2{}^-$ levels (peaks 1 and 7). In both nuclei, the low-lying $^5/_2{}^-$ states at 0.376 MeV in $^{53}$Mn and for the ground state in $^{55}$Mn are absent in the 26 MeV spectra.

[1] The number of unpaired fermions.

Table 27.3

Spectroscopic information about $^{55}$Mn extracted from our study of the $^{52}$Cr($^4$He,p)$^{55}$Mn reaction at 26 MeV compared with earlier results

| Peak a) no. | $E_x$ b) (keV) | $E_x$ c) (keV) | $l_p$ c) | $J^\pi$ e) | $J^\pi$ f) | $\sigma(\theta_0)$ g) ($\mu$b/sr) |
|---|---|---|---|---|---|---|
| 1 | 128 | 127 | 3 | $\frac{7}{2}^-$ | $\frac{7}{2}^-$ | 18.9 |
| 2 | 983 | | | $(\frac{9}{2})^-$ | | 5.3 |
| 3 | 1291 h) | | | | $\frac{1}{2}^{(-)}+(\frac{11}{2}^-)$ | 22.3 |
| 4 | 1528 | 1527 | 1 | $\frac{3}{2}^-$ | $\frac{3}{2}^-$ | 46.0 |
| 5 | 1883 | 1881 | 3 | $(\frac{5}{2},\frac{7}{2})^-$ | $\frac{7}{2}^-$ | 6.5 |
| 6 | 2197 | 2198 | | $\frac{7}{2}^{(-)}$ | $\frac{7}{2}^-$ | 18.6 |
| 7 | 2258 | 2250 | 1 | $(\frac{1}{2},\frac{3}{2})^-$ | $\frac{3}{2}^-$ | 89.3 |
| 8 | 2562 | 2560 | 1 | $\frac{3}{2}^-$ | $\frac{3}{2}^-$ | 36.6 |
| 9 | 2741 h) | 2742 | | $\frac{7}{2}+(\frac{5}{2},\frac{7}{2})^-$ | | 19.4 |
| 10 | 2980 | 2984 | 2 | $(\frac{3}{2},\frac{5}{2})^+$ | | 7.6 |
| 11 | 3041 | 3028 | 1 | $(\frac{1}{2},\frac{3}{2})^-$ | $\frac{3}{2}^-$ | 30.0 |
| 12 | 3136 | 3147 | 3 | $(\frac{5}{2},\frac{7}{2})^-$ | $\frac{5}{2}^-$ | 56.0 |
| 13 | 3263 | 3260 | | | $(\frac{3}{2}^-)$ | 12.1 |
| 14 | 3354 | | | | $(\frac{3}{2}^-)$ | 19.1 |
| 15 | 3431 | 3429 | 1 | $(\frac{1}{2},\frac{3}{2})^-$ | $\frac{1}{2}^-$ | 22.9 |
| 16 | 3608 d) | 3608 | 3 | $(\frac{7}{2}^-)$ | $(\frac{3}{2}^-)$ | 28.2 |
| 17 | 3847 | | | | | 17.2 |
| 18 | 3989 | 3998 | 1 | $(\frac{1}{2},\frac{3}{2})^-$ | $\frac{3}{2}^-$ | 20.6 |
| 19 | 4086 | | 1 i) | $(\frac{1}{2},\frac{3}{2})^-$ | $\frac{3}{2}^-$ | 20.6 |
| 20 | 4266 | | | | $(\frac{3}{2}^-)$ | 43.0 |
| 21 | 4640 d) | 4638 | 1 | $(\frac{1}{2},\frac{3}{2})^-$ | | 15.9 |
| 22 | 4736 | 4742 | 1 | $(\frac{1}{2},\frac{3}{2})^-$ | $\frac{3}{2}^-$ | 34.7 |
| 23 | 5260 d) | | | | | 39.7 |
| 24 | 5365 d) | 5366 | 1 | $(\frac{1}{2},\frac{3}{2})^-$ | | 87.4 |
| 25 | 5498 | 5498 | 3 | $(\frac{9}{2}^+)$ | | 172.7 |
| 26 | 6069 d) | | | | $(\frac{3}{2}^-)$ | 39.9 |
| 27 | 6164 | | | | | 132.3 |

a) See Figure 27.1.
b) Our work.
c) Katsanos and Huizenga (1967).
d) Could be fitted only by readjusting triton parameters. See the text.
e) Katsanos and Huizenga (1967) and Kocher (1976).
f) $J^\pi$ assignments based on our work.
g) $\theta_0$ = 23.4$^0$ (c.m.).
h) Suspected doublet (Kocher 1976).
i) Discussed in the text.
j) Assignment based on the (d,n) reaction (Kocher 1976).

An interesting feature of the $^{55}$Mn spectrum is the presence of several strongly excited states above 4.7 MeV (peaks 23 to 27). At all angles measured in our study, the state at 5.498 MeV (peak 25) dominated the 26 MeV spectra. It is interesting to note that similar dominant peaks have been observed in $^{61,63,65,67}$Cu isotopes following the ($^4$He,p) reaction at 19.3 MeV bombarding energy (Bucurescu *et al.* 1972). Convincing $J^\pi$ = $^9/_2{}^+$ assignments were made for these states. It can be seen in Figure 27.8 that in our study, the angular distribution for the 5.498 MeV state can also be well reproduced by $J^{\pi} = 9/2^{+}$. A state at the same excitation energy has been previously assigned (Rapaport et al. 1969) an $l_p = 3$ transfer. This would suggest spin $J^{\pi} = 3/2^{-}$ or $5/2^{-}$ for this state. However, in our study no fit to the data could be obtained using such values unless the triton diffuseness was reduced to 0.1 fm.

Reduced strengths for states in $^{55}$Mn excited via $^{52}$Cr($^{4}$He,p)$^{55}$Mn reaction at 26 MeV are displayed in Figure 27.12.

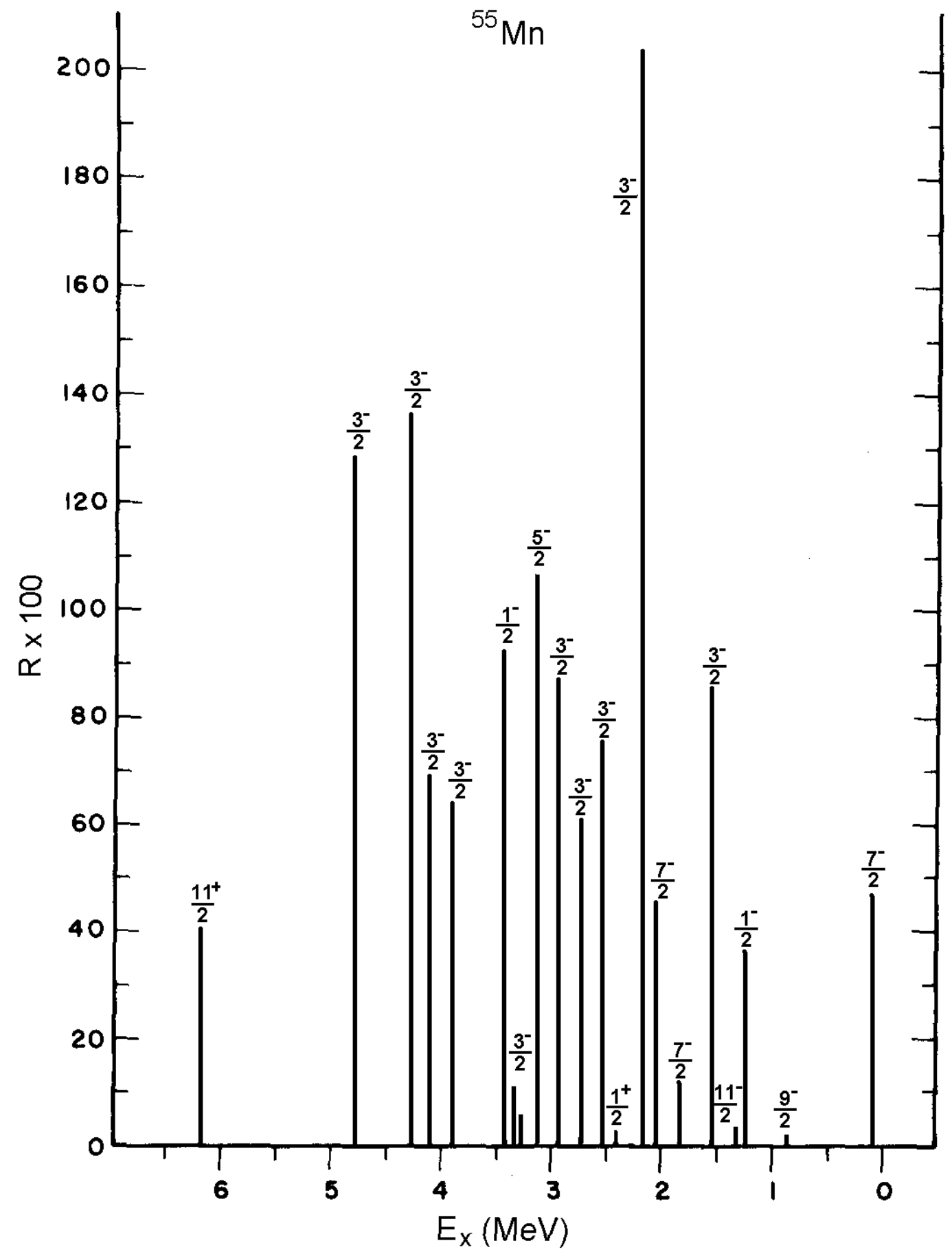

Figure 27.12. Reduced strengths $R$ for the reaction $^{52}$Cr($^{4}$He,p)$^{55}$Mn reaction at 26 MeV. The figure shows also the assigned $J^{\pi}$ values.

The energy level at 1.291 MeV excitation energy (peak 3) is of particular interest. There is a considerable uncertainty about the nature of this level (see Chapter 28). Our ($^{4}$He,p) measurements at 26 MeV indicate that there is a doublet with $J^{\pi} = 1/2^{-}$ and $11/2^{-}$ at this excitation energy. As can be seen in Figure 27.7, the angular distribution for the $^{52}$Cr($^{4}$He,p)$^{55}$Mn reaction leading to this state show a pronounced maximum and minimum at the same angles as observed for other $J^{\pi} = 1/2^{-}$ transitions. We have calculated two angular distributions for this state using the distorted wave formalism, one corresponding to $J^{\pi} = 1/2^{-}$ and one for $J^{\pi} = 11/2^{-}$. We have then applied a least squares procedure to fit the experimental angular distribution for this state using the two calculated theoretical shapes. The resulting fit is displayed in Figure 27.7 by a full line. The absolute values of the theoretical cross-sections at $\theta_{c.m.} = 23°$ are 12 and 10 μb/sr for the $J^{\pi} = 1/2^{-}$ and $11/2^{-}$ components, respectively

The 2.197 MeV state was not populated strongly in single-proton transfer reactions. Our study shows that angular distribution for the $^{52}Cr(^4He,p)^{55}Mn$ reaction leading to this state can be fitted using $J^\pi = 7/2^-$ (see Figures 27.4 and 27.7) In contrast with its weak excitation in the ($^3$He,d) reaction, its intensity (peak 6) in the ($^4$He,p) reaction is comparable to the intensity of the $7/2^-$ level at 0.128 MeV (peak 1).

The adopted level scheme (Kocher 1976) shows two states at 2.727 and 2.753 MeV with spins of $7/2^-$ and ($5/2^-$, $7/2^-$), respectively. In our study, the peak observed at 2.741 MeV could therefore be regarded as a doublet. However, the ($^4$He,p) angular distribution has been fitted with just one $J^\pi$ value ($J^\pi = 7/2^-$). No improvement to the fit was obtained by using combinations of $J^\pi = 3/2^-$ and $5/2^-$ or $3/2^-$ and $7/2^-$.

The level at 2.980 MeV is supposed to have spins $3/2^+$ or $5/2^+$ (Kocher 1976; Rapaport *et al.* 1969). We have tried both of them without any success.

Early shell-model calculations (McGrory 1967; Vervier 1966) of the structure of $^{55}$Mn confined protons to the $1f_{7/2}$ shell. This is an inadequate treatment as indicated by the number of easily reproducible $J^\pi = 3/2^-$ transitions observed in the earlier ($^3$He,d) data and in our ($^4$He,p) measurements. Moreover, the existence of a $1/2^-$ level at about 1.291 MeV, which is needed to reproduce our ($^4$He,p) angular distribution and the data of Peterson, Pittel, and Rudolph (1971) and Peterson and Rodolph (1972) can be only predicted if proton excitation into the *f-p* shell is included. However, their 2p-3h plus 3p-4h calculations fail to predict the close proximity of the strong $3/2^-$ state at 2.258 MeV and the $7/2^-$ state at 2.197 MeV. They also predict that much of the $2p_{1/2}$ strength is at 2.46 MeV while the ($^4$He,p) data indicate that the $L = 1$ transfers around this energy correspond to $J^\pi = 3/2^-$.

### *The $^{54}Cr(^4He,p)^{57}Mn$ reaction*

The $Q$ - values for states in $^{57}$Mn were determined using the 18 and 26 MeV data and the energy calibrations obtained for the $^{55}$Mn spectra, which were taken under the same experimental as for the $^{57}$Mn measurements. From the 18 MeV data, the ground-state $Q_0$ - value for the $^{54}Cr(^4He,p)^{57}Mn$ reaction is

$$Q_0 = -4.302 \pm 0.008 \text{ MeV}.$$

This value is in agreement with the value reported by Mateja *et al.* (1976). These authors measured proton spectra at a few angles for the $^{54}Cr(^4He,p)^{57}Mn$ reaction. The discrepancy between the $Q_0$-value determined from our ($^4$He,p) data and the value reported by Gove and Wapstra (1972) can be removed if the data of Ward, Pile, and Kuroda (1969) on the $\beta$ - decay of $^{57}$Mn is reinterpreted by requiring the 83% branching mode to populate the 136 keV level in $^{57}$Fe instead of the 14 keV level.

During the course of our investigation, Mateja *et al.* (1976) reported on low-lying levels in $^{57}$Mn by using the ($^4$He,p) and ($^4$He,p$\gamma$) reactions at 15, 21 and 24 MeV. They have identified levels in $^{57}$Mn for up to 2.234 MeV excitation energy. The spin assignments made on the basis of the ($^4$He,p$\gamma$) measurements are in agreement with our spin assignments. Of particular interest in the results of Mateja *et al.* is a pair of states at about 1.06 MeV separated by 15 keV and assigned spins of ($1/2^-$) and ($9/2^-$). This feature parallels the suspected $1/2^-$ and $11/2^-$ doublet at 1.292 MeV in $^{55}$Mn.

Table 27.4
Spectroscopic information for $^{57}Mn$

| Peak [a] no. | $-(Q \pm \Delta Q)$ [b] (keV) 18 MeV | | $-(Q \pm \Delta Q)$ [b] (keV) 26 MeV | | $E_x$ [b] (keV) | $E_x$ [c] (keV) | $E_x$ [d] | $J^\pi$ [e] | $J^\pi$ [b] | $\sigma(\theta_0)$ [f] ($\mu$b/sr) |
|---|---|---|---|---|---|---|---|---|---|---|
| 1 | 4302 | 8 | | | 0 | 0 | 0 | $\frac{5}{2}^-$ | | |
| 2 | 4386 | 6 | 4395 | 7 | 84 | 84 | 84 | $(\frac{7}{2}^-)$ | $\frac{7}{2}^-$ | 40.0 |
| 3 | 5154 | 6 | 5162 | 7 | 852 | 851 | 851 | $(\frac{3}{2}^-)$ | $(\frac{3}{2}^-)$ | 63.5 |
| 4 | 5366 | 6 | 5373 | 8 | 1064 | {1059, 1074} | {1057, 1071} | $(\frac{1}{2}^- + \frac{9}{2}^-)$ | $(\frac{1}{2}^- + \frac{9}{2}^-)$ | 9.5 |
| 5 | 5531 | 7 | 5540 | 9 | 1229 | 1227 | 1229 | $(\frac{11}{2}^-)$ | $(\frac{11}{2}^-)$ | 22.2 |
| 6 | 5680 | 6 | 5680 | 7 | 1378 | 1376 | 1375 | | | 2.4 |
| 7 | 5789 | 11 | | | 1487 | | {1477, 1493} | | | |
| 8 | 5835 | 10 | 5824 | 19 | 1533 | | 1536 | | | |
| 9 | 5932 | 10 | | | 1630 | | | | | |
| 10 | 6036 | 10 | | | 1734 | | 1726 | | | |
| 11 | 6132 | 11 | | | 1830 | | 1837 | $\frac{5}{2}^-$ | | |
| 12 | 6227 | 7 | 6233 | 11 | 1925 | | {1916, 1928} | | | 12.6 |
| 13 | 6318 | 14 | | | 2016 | | 2008 | | | |
| 14 | 6490 | 14 | 6499 | 8 | 2188 | 2188 | 2185 | | $\frac{1}{2}^{(-)}$ | 35.8 |
| 15 | 6540 | 10 | 6544 | 11 | 2238 | 2234 | 2232 | | | 28.1 |
| 16 | 6645 | 8 | | | 2343 | | 2340 | | | |
| 17 | 6729 | 7 | | | 2427 | | 2417 | | | |
| 18 | 6827 | 10 | 6831 | 10 | 2525 | 2520 | | | $(\frac{5}{2}^-)$ | 16.1 |
| 19 | 6910 | 9 | 6926 | 8 | 2608 | 2607 | | | | 17.9 |
| 20 | 6933 | 11 | | | 2631 | 2640 | | | | |
| 21 | 7002 | 11 | 7022 | 25 | 2700 | | | | | |
| | 7043 | 7 | | | 2741 | | | | | |
| 22 | 7074 | 12 | | | 2772 | | | | | |
| 23 | 7150 | 14 | 7143 | 7 | 2848 | | | | | 38.3 |
| 24 | 7230 | 14 | | | 2928 | | | | | |
| | 7389 | 12 | 7381 | 15 | 3087 | | | | | |
| 25 | 7425 | 10 | 7429 | 15 | 3123 | | | | | 47.0 |
| 26 | 7466 | 12 | | | 3164 | | | | | |
| 27 | 7514 | 10 | 7511 | 9 | 3212 | | | | | |
| 28 | 7556 | 13 | 7558 | 12 | 3254 | | | | | 32.3 |
| 29 | 7673 | 11 | 7682 | 15 | 3371 | | | | | |
| 30 | 7760 | 11 | 7748 | 17 | 3458 | | | | | |
| 31 | 7784 | 10 | 7777 | 13 | 3482 | | | | | 74.9 |
| 32 | 7844 | 10 | 7846 | 15 | 3542 | | | | | 29.0 |
| 33 | 7910 | 14 | | | 3608 | | | | | |
| 34 | 7976 | 13 | 7987 | 11 | 3674 | | | | | |
| 35 | 8017 | 9 | 8031 | 17 | 3715 | | | | | 35.6 |
| 36 | 8059 | 9 | | | 3757 | | | | | |
| 37 | 8098 | 7 | | | 3796 | | | | | |
| 38 | 8140 | 10 | 8139 | 16 | 3838 | | | | $(\frac{5}{2}^-)$ | 19.9 |
| 39 | 8172 | 9 | | | 3870 | | | | | |
| 40 | 8213 | 10 | | | 3911 | | | | | |
| 41 | 8249 | 9 | 8247 | 17 | 3947 | | | | | 15.7 |
| | 8303 | 11 | 8316 | 17 | 4001 | | | | | |
| 42 | 8331 | 4 | 8363 | 20 | 4029 | | | | $(\frac{5}{2}^-, \frac{9}{2}^-)$ | 14.7 |
| 43 | 8470 | 10 | 8455 | 15 | 4168 | | | | $(\frac{9}{2}^-)$ | 26.0 |
| 44 | 8513 | 10 | 8507 | 17 | 4211 | | | | $(\frac{3}{2}^-, \frac{11}{2}^-)$ | 20.8 |
| 45 | 8594 | 15 | 8609 | 13 | 4292 | | | | | |
| 46 | 8665 | 11 | 8653 | 18 | 4363 | | | | | 37.9 |
| 47 | 8774 | 9 | 8778 | 18 | 4472 | | | | $(\frac{9}{2}^-)$ | 38.5 |
| 48 | | | 8825 | 23 | 4523 | | | | | 58.3 |
| 49 | | | 8928 | 15 | 4626 | | | | $(\frac{5}{2}^-)$ | 43.5 |
| 50 | | | 9050 | 15 | 4748 | | | | $(\frac{5}{2}^-)$ | 19.9 |
| 51 | | | 9143 | 19 | 4841 | | | | | 20.5 |
| | | | 9311 | 13 | 5009 | | | | | |
| | | | 9362 | 19 | 5060 | | | | | |
| | | | 9469 | 13 | 5167 | | | | | |

[a]) See Figure 27.2. [b]) Our assignments. [c]) Mateja *et al.* (1976). [d]) Mateja *et al.* (1977). [e]) Mateja at al. (1976, 1977); Ward, Pile, and Kuroda (1969). [f]) $\theta_0 = 23.4^0$, $E_\alpha = 26$ MeV.

Results of our work for $^{57}$Mn are tabulated in Table 27.4 where they are compared with results of Mateja *et al.* (1976, 1977). A comparison of the spectra presented in Figure 27.2 with the ($^4$He,p) spectra of Mateja *et al.* (1976) reveals a discrepancy in the number of states excited between 1.378 and 2.188 MeV (peaks 6 and 14). The present work indicates the existence of seven levels in this energy interval while Mateja *et al.* (1976) found none. This may be explained partly by the differences in bombarding energies and partly by statistics. It should be noted, however, that all these additional states have been excited in their subsequent work (Mateja *et al.* 1977) using the (t,p) two-neutron transfer reaction and have been well reproduced by the distorted wave theory.

In general, it was difficult to make unambiguous spin assignments in $^{57}$Mn. This was partly due to the absence of known $l_p$ values and partly because of the close similarity between angular distributions calculated for different values of $J^\pi$.

## Summary and conclusions

The $^{50,52,54}$Cr($^4$He,p)$^{53,55,57}$Mn reactions were studied at 18 and 26 MeV bombarding energy. Analysis of the proton angular distributions revealed that contributions from compound nucleus reactions were significant at 18 MeV.

In contrast, at 26 MeV, nearly all the proton groups detected had angular distributions, which could be described using the distorted wave theory and assuming a triton cluster transfer mechanism. The angular distributions of 41 proton groups were obtained for $^{53}$Mn, 27 groups for $^{55}$Mn and 27 for $^{57}$Mn. Nearly all the calculated $J^\pi$ transfers were for $L$ = 1 or $L$ = 3. Using the $L$ = 1 $J$ - dependence for the ($^4$He,p) reaction, many spin assignments could be made for states whose $l_p$ values had already been determined from previous ($^3$He,d) studies of $^{53,55}$Mn. In addition, an $L$ = 3 $J$ - dependence was also found for ($^4$He,p) in this energy-mass region and used to assign the relevant $J^\pi$ to the observed states.

A summary of energy levels in $^{53,55,57}$Mn for which angular distributions have been measured at 26 MeV incident $^4$He energy is shown in Figure 27.13. The figure contains the excitation energies and $J^\pi$ assignments based on our study. A summary of all spectroscopic information is presented in Tables 27.2-27.4.

The $^7/_2{}^-$ strength is spread out among more states in $^{55}$Mn than in $^{53}$Mn, although not all the $^7/_2{}^-$ strength in $^{55}$Mn can be associated with the single particle $^7/_2{}^-$ configuration. The addition of two neutrons increases the number of configurations that the ($^4$He,p) reaction can excite as opposed to the configurations accessible in single proton transfer. These extra degrees of freedom also allow for an explanation of the larger number of states seen in $^{55,57}$Mn below 2.4 MeV as compared with $^{53}$Mn. The $^9/_2{}^-$ and $^{11}/_2{}^-$ states at 0.983 and 1.292 MeV in $^{55}$Mn are presumably excited through

$$\left[\left(\pi f_{7/2}^{-3}\right)_{7/2}\left(\nu j_1 j_2\right)_{J_n}\right]_J$$

components, where $j_1$, $j_2$ are $2p_{3/2}$ or $1f_{5/2}$ and $J_n$ = 2 or 4.

The importance of ($3p$-$4h$) configurations in the low-lying states of $^{55}$Mn is demonstrated by the $L$ = 1, $J^\pi$ = $^3/_2{}^-$ state at 1.528 MeV. The ($^4$He,p) reaction at 26

MeV also corroborates the assignment of a $^1/_2{}^-$ state, which is nearly degenerate with the $^{11}/_2{}^-$ state at 1.292 MeV as suggested by Peterson, Pittel, and Rudolph (1971).

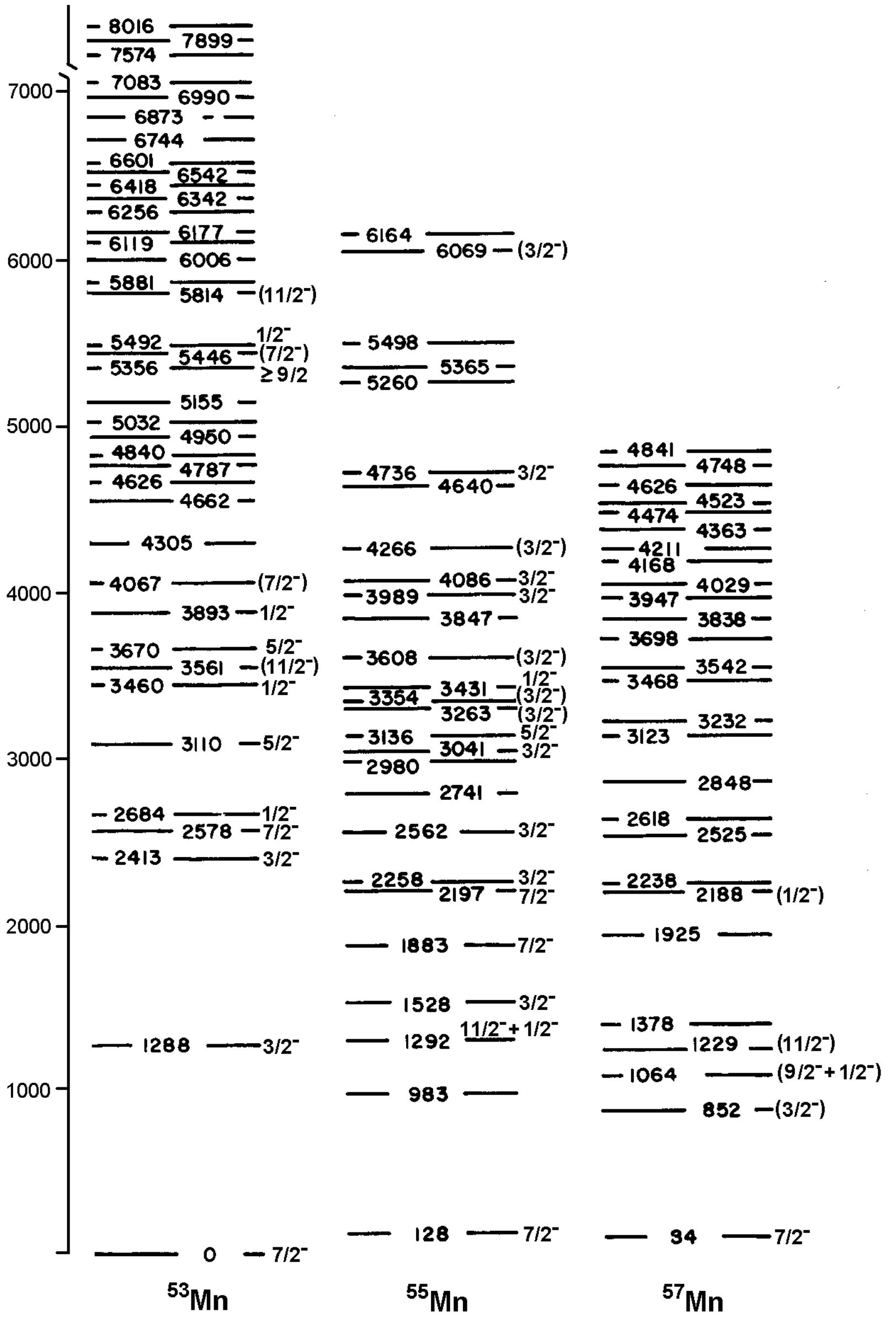

Figure 27.13. Excitation energies and $J^\pi$ assignments based on our study of the $^{50,52,54}Cr(^4He,p)^{53,55,57}Mn$ reactions. Only states for which angular distributions at 26 MeV were measured are shown.

In $^{57}Mn$, $Q$ - values for 57 states were determined using both the 18 and 26 MeV data. The ground state $Q_0$ - value was determined to be -4.302 ± 0.008 MeV, in agreement with the value reported by Mateja *et al.* (1976). The states measured in ($^4He$,p) for excitation energies greater than 2.2 MeV have not been reported previously.

The simple model of triton cluster transfer accounts very well for the shapes of the angular distributions. Our measurements can serve as a basis for more refined calculations to understand the details of reaction mechanism of three-nucleon transfer and the structure of the Mn isotopes.

We have found that many states, which were not observed in single-nucleon transfer reactions were accessible via the three-nucleon transfer in the ($^4$He,p) reaction. Likewise, many states that were weakly excited in single nucleon transfer reactions were strongly excited in the ($^4$He,p) reaction. The ($^4$He,p) is a useful tool for uncovering and studying new configurations in states of residual nuclei.

# 28

# Gamma De-excitation of $^{55}$Mn Following the $^{55}$Mn(p,p'γ)$^{55}$Mn Reaction

***Key features:***

1. The gamma de-excitation scheme of $^{55}$Mn has been determined for states of up to 3385 keV excitation energy.
2. A total of 45 γ transitions have been identified between 27 states in $^{55}$Mn.
3. The enigma of the 1292 keV level has been studied.

    e. Our results confirm the γ de-excitation scheme of Hichwa et al. (1973a) and Hichwa, Lawson, and Chagnon (1973b) around this excitation energy but do not confirm the expected γ transition (Kulkarni 1976) to the ground state from the 1292 level.

    f. The incident proton energy of 7.975 MeV was chosen to coincide with the energy used by Katsanos and Huizenga (1967) who reported a doublet of states at the 1292 keV excitation energy in (p,p') scattering. Our high-resolution measurements do not confirm their results.

    g. Combining the results of this study, with the results of our earlier study of the $^{52}$Cr($^{4}$He,p)$^{55}$Mn reaction at 18 and 26 MeV and the existing results by other authors we conclude that there is a doublet of states at the 1292 keV excitation energy in $^{55}$Mn with the separation energy of less than 10 keV. However, the excitation of its 1/2$^-$ member appears to be strongly selective.

**Abstract:** Gamma de-excitation scheme has been studied using the $^{55}$Mn(p,p'γ)$^{55}$Mn reaction induced by the 7.975 MeV protons. A total of 45 γ transitions between 27 states in $^{55}$Mn, extending up to the 3385 keV excitation energy, have been observed in both the singles and p-γ coincidence spectra. The enigmatic 1292 keV level has been studied using both the gamma and the high-resolution proton spectra.

## Introduction

The study of gamma de-excitation of $^{55}$Mg was prompted by the enigma of the 1.29 MeV level. A great deal of confusion surrounded the level structure of $^{55}$Mn at about this excitation energy. In a high resolution (p,p') study using 7.975 MeV protons Katsanos and Huizenga (1967) claimed to have observed a doublet at this excitation energy.

Peterson, Pittel and Rudolph (1971) studied the $^{57}$Fe(p,$^{3}$He)$^{55}$Mn and $^{57}$Fe(d,$^{4}$He) $^{55}$Mn reactions at $E_p$ =27 MeV and $E_d$ = 16.5 MeV. They claimed the existence of a $^1/_2{}^-$ state unresolved from a state with $^{11}/_2{}^-$ at 1.29 MeV. This study is in agreement with our study of $^{52}$Cr($^{4}$He,p)$^{55}$Mn reaction using 18 and 26 MeV $^{4}$He particles (see Chapter 27).

Using $^{4}$He-particles with the energy of 5 to 7 MeV, Kulkarni and Nainan (1974) found γ-ray yields for a 1293 keV γ-ray which could be fitted by first order Coulomb excitation theory for a $\lambda = 2$ transition. They claimed that they were exciting a state at 1.293 MeV with $^1/_2{}^- \leq J^\pi \leq {}^9/_2{}^-$. Observation of γ-ray transitions to the $^7/_2{}^-$ and $^9/_2{}^-$ states at 0.128 and 0.983 MeV limited their spin assignments to $^5/_2{}^- \leq J^\pi \leq {}^9/_2{}^-$. They also found that the yields for the 307 and 1167 keV γ-rays, which originate from the 1.293 MeV level could be accounted for by Coulomb excitation theory.

Unfortunately, a degree of uncertainty about their interpretation Coulomb excitation data as pointed out later by Kulkarni (1976) who extended Coulomb excitation measurements of $^{55}$Mn to 8 MeV. He observed the same $\gamma$-ray spectrum as in the earlier publication (Kulkarni and Nainan 1974) except for the 307 and 1167 keV $\gamma$-rays, which could be identified with the 304, and 1164 keV $\gamma$-rays observed at 8 MeV. However, he found that the yield for the 1164 keV $\gamma$-rays rose too steeply to be accounted for by a single E2 excitation and thus this $\gamma$-rays could not be identified as the 1167 keV $\gamma$-rays of Kulkarni and Nainen (1974). He concluded that there were two levels separated by 3 keV: a $J^{\pi}$ = $^{11}/_{2}{}^{-}$ state at 1.290 MeV and a $^{1}/_{2}{}^{-}$ state at 1.293 MeV. The assignment of spin $^{1}/_{2}{}^{-}$ for the state at 1.293 MeV was made on the basis of both the $\gamma$-rays yields and a limited $\gamma$-rays angular distribution measured at $\theta_{\gamma}$ = 0° and 90°. According to Kulkarni (1976) the 1.293 MeV (1/2$^-$) level de-excites 100% to the ground state.

Unfortunately, a 1293 keV $\gamma$-ray finds no confirmation in the publication of Kocher (1976). In an earlier study, Hichwa *et al.* (1973a) and Hichwa, Lawson, and Chagnon (1973b) saw no evidence for the population of a $^{1}/_{2}$ level at 1.29 MeV in their ($^{4}$He,p$\gamma$) measurements at $E$ =10.5 and 11.1 MeV. Their spin assignment agreed with the previously assigned $^{11}/_{2}$ value but they did not observe a 1292 keV $\gamma$-ray transition to the ground state, which according to Kulkarni (1976) should be the only decay pathway open from the alleged $^{1}/_{2}{}^{-}$ state.

The branching ratios for the 1.29 MeV state $\gamma$-decays claimed by Kulkarni and Nainen (1974) and Kulkarni (1976) are in serious disagreement with those of Hichwa et al. (1973a) and Hichwa, Lawson, and Chagnon (1973b). It is also not clear whether the 1293 keV $\gamma$-ray observed by Kulkarni and Nainen (1974) and Kulkarni (1976) originates from a level at 1.29 MeV because they did not report particle-gamma coincidence measurements or observe any $\gamma$-rays in coincidence with the 1293 keV $\gamma$-ray.

In our study of $^{52}$Cr($^{4}$He,p)$^{55}$Mn reactions (see Chapter 27), the data at 26 MeV showed a pronounced minimum and maximum at the same angles as observed for other $J^{\pi}$ = $^{1}/_{2}{}^{-}$ transitions. A least-squares combination of $J^{\pi}$ = $^{1}/_{2}{}^{-}$ and $^{11}/_{2}{}^{-}$ produced a good fit to the measured angular distribution thus indicating a transition to a doublet state at this excitation energy. The determined intensities of the two components at $\theta$ =23° are 12 µb/sr for the $J^{\pi}$ = $^{1}/2^{-}$ state and 10 µb/sr for $J^{\pi}$ = $^{11}/_{2}{}^{-}$.

## Experimental procedures and results

We have carried out our measurements using a 7.975 MeV proton beam from the ANU EN tandem accelerator. The energy was chosen to coincide with the energy used by Katsanos and Huizenga (1967) in their measurements of (p,p') scattering. The target contained a minimum of 99% $^{55}$Mn, with maximum limits of iron 0.002%, lead 0.001%, nickel 0.002% and zinc 0.05%. Preliminary measurements were performed using the 24" double focussing spectrometer. The best resolution attained was 12 keV. Measurements of $^{55}$Mn(p,p') spectra were made at 30°, 40°, 50° and 140° (lab). In contrast with the claim of Katsanos and Huizenga (1967) the particle spectra showed no evidence of a doublet at the 1293 keV excitation energy. This could either mean that the separation of the doublet states is significantly less than 12 keV, thus confirming the claim of Kulkarni (1976) or that one of the components of the doublet is not excited in (p,p') scattering.

We have then carried out the $^{55}$Mn(p,p'$\gamma$) measurements using the same proton bombarding energy. A diagram of the experimental arrangement is shown in Figure 28.1. The $\gamma$-ray detector was made of 62 cm$^3$ Ge(Li) crystal (4.9 cm diameter, 3.3 cm long). The detector was located at 90° with respect to the beam to minimize Doppler shift and the total acceptance angle was ±32°. To detect protons, Si surface barrier detector was used, with a diameter of 1.9 cm. The detector was set at -80°. Magnetic electron suppression was used and the detector was cooled by Cu strap connected to cold finger. We have used a strip $^{55}$Mn target, 2 mm wide, approximately 150 $\mu$g/cm$^2$, on a thin carbon backing.

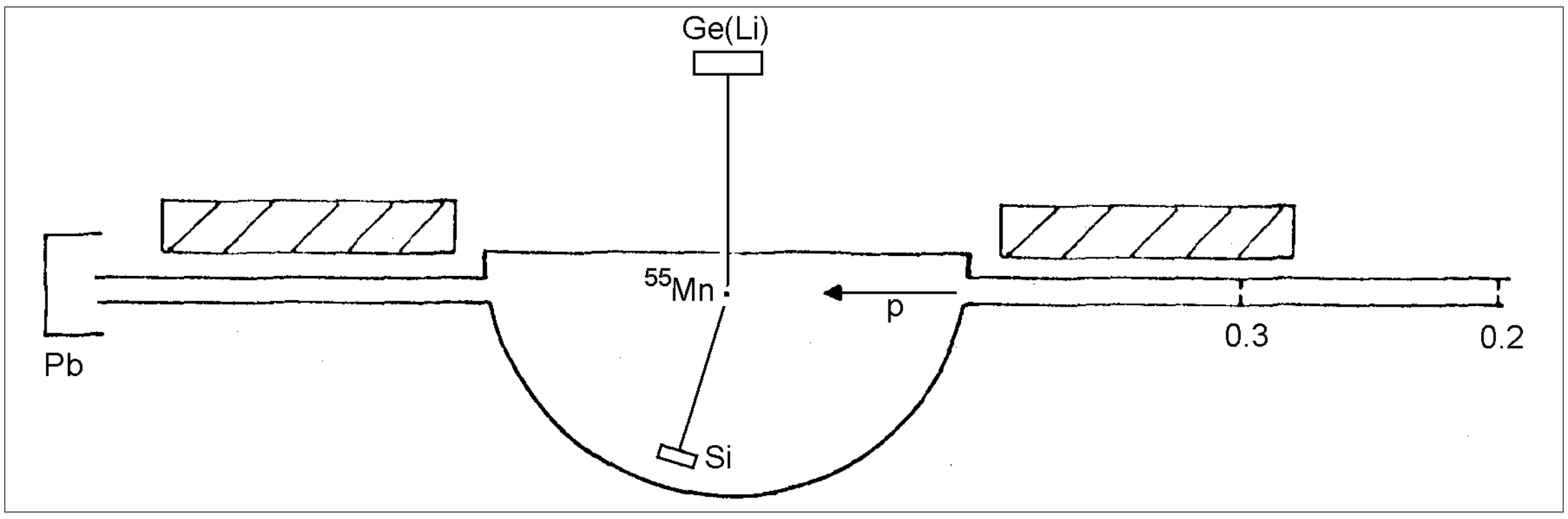

Figure 28.1. Experimental arrangement for the $^{55}$Mn(p,p'$\gamma$)$^{55}$Mn measurements

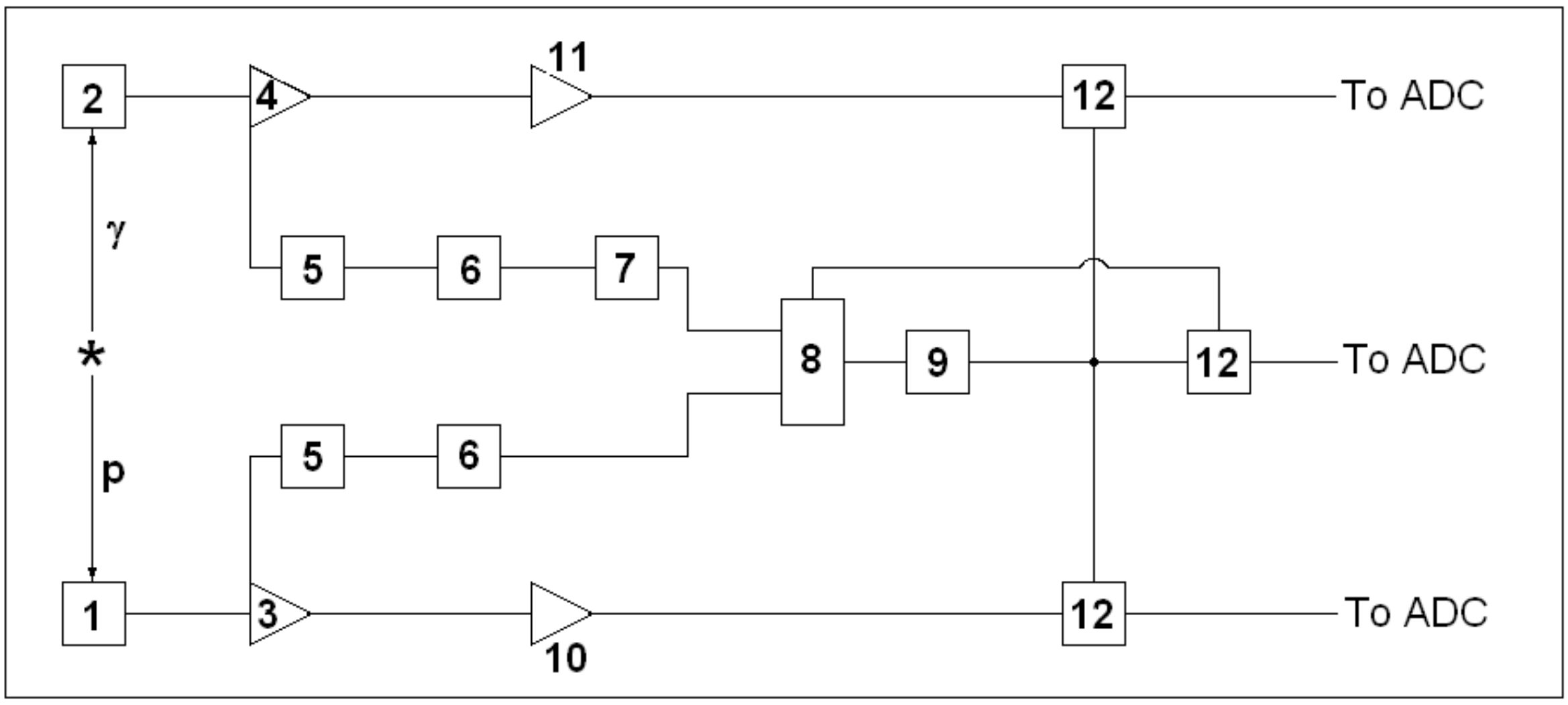

Figure 28.2. Electronics for the $^{55}$Mn(p,p'$\gamma$)$^{55}$Mn measurements. 1. Si surface barrier detector; 2. Ge(Li) 62 cm$^3$ Seforad $\gamma$-detector; 3. ORTEC 125 preamplifier; 4. Seforad Sr 100 preamplifier; 5. ORTEC 454 timing filter; 6. ORTEC 463 constant fraction discriminator; 7. Nanosecond delay; 8 Canberra 1443 time to amplitude converter; 9. Logic shaper and delay; 10. Tennelec TC 203 BLR amplifier; 11. Tennelec TC 205A amplifier; 12. Canberra 1454 linear gate and stretcher.

Singles $\gamma$-ray spectra were collected with the target in and out of the beam. The spectra were calibrated using a $^{152}$Eu source located in the same position as the target. A sample of singles spectrum is shown in Figure 28.3 for $\gamma$ energies around the alleged 1292 transition from the 1292 level to the ground state. No 1292 keV γ-ray was observed.

According to Kulkarni and Nainan (1974) the branch to the ground state of the 1292 keV state should be five times more likely than the branch to the first excited state which produces an 1165 keV γ-ray. The observed 1280 keV γ-ray has about 1/5 the intensity of the 1165 keV γ-ray in Figure 28.3. If a 1292 keV γ-ray is present in this spectrum its intensity is less than 1/5 that of the 1165 keV transition.

A list of γ-rays seen in singles, along with their relative intensities is presented in Table 28.1. Gamma rays, which were present in the singles spectrum when the target was removed from the beam, have not been included in the table.

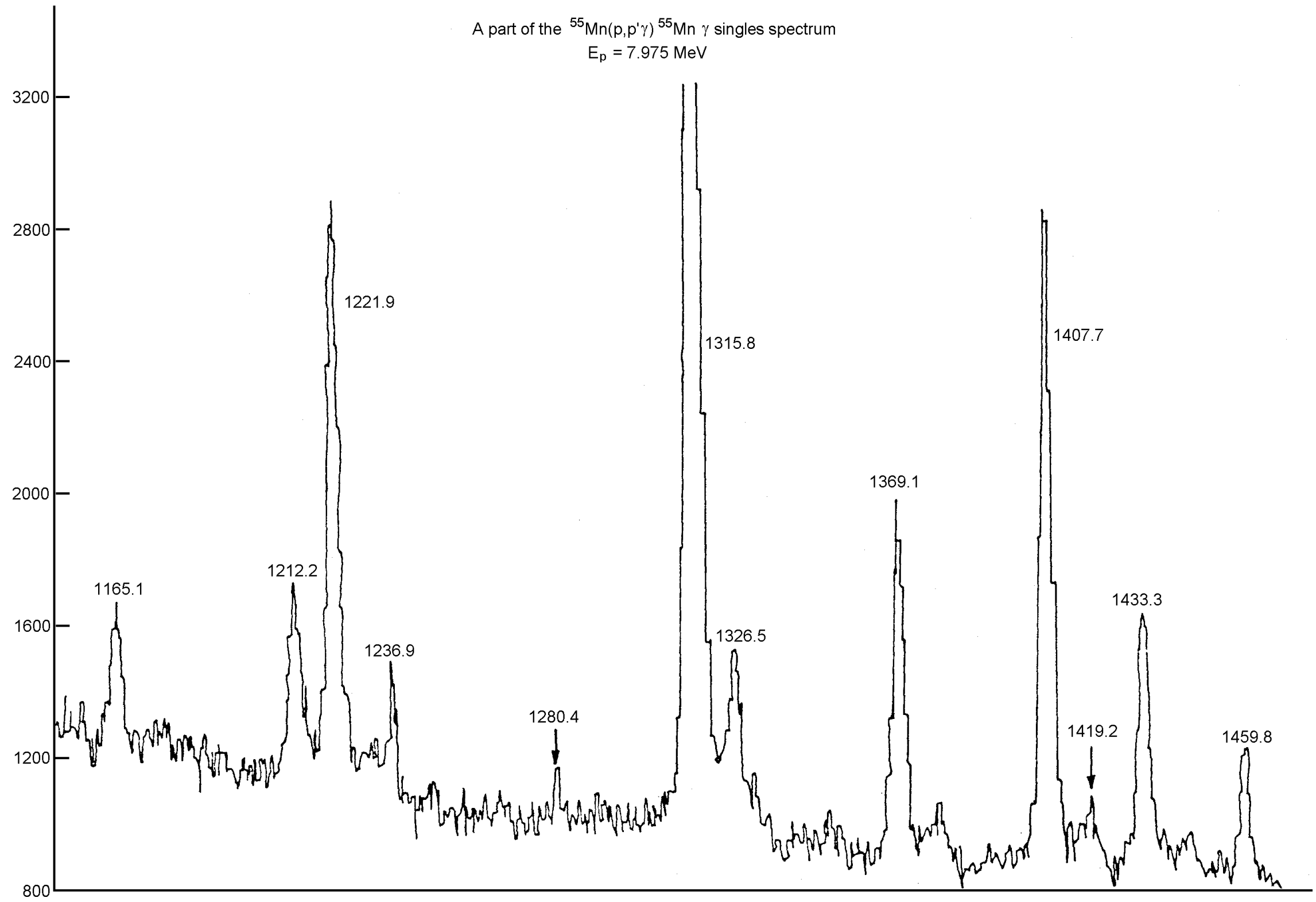

Figure 28.3. A part of the $^{55}$Mn(p,p'γ)$^{55}$Mn γ singles spectrum at $E_p$ = 7.975 MeV around the alleged 1292 keV γ de-excitation to the ground state from the 1292 MeV level in $^{55}$Mn. The γ energies are in keV.

The p-γ coincidence data were recorded event by event on magnetic tape and the measurements took approximately 45 hours of running time with 2 to 8 nA on the 150 μg/cm$^2$ 55Mn target. Figure 28.4 shows the proton projection and an example of the γ-ray spectrum in coincidence with the 1292 keV proton group. The TAC time window was set to 200 ns and the true to chance ratio was 20/1 with a 12 ns FWHM for the TAC peak.

From the γ coincidence spectrum presented in Figure 28,4 it is clear that no γ-ray was detected at 1292 keV. Our measurements disagree with the conclusion of Kulkarni (1976) but confirm the decay scheme of Hichwa et al. (1973a) and Hichwa, Lawson, and Chagnon (1973b) for the 1292 keV level. We have also carried out the coincidence γ-ray measurements for the other proton peaks but we have seen no γ transition to the ground state from the 1292 keV level.

The p-γ coincidence scheme obtained from our measurements is shown in Figure 28.5. Strong and firmly assigned transitions are indicated by solid lines. Weak transition, corresponding to low number of counts, are shown as dashed lines.

The p-γ coincidence scheme obtained from our work is in substantial agreement with the scheme determined by Hichwa et al. (1973a) and Hichwa, Lawson, and Chagnon (1973b) for levels below approximately 2800 keV. We have extended the scheme to 3385 excitation energy.

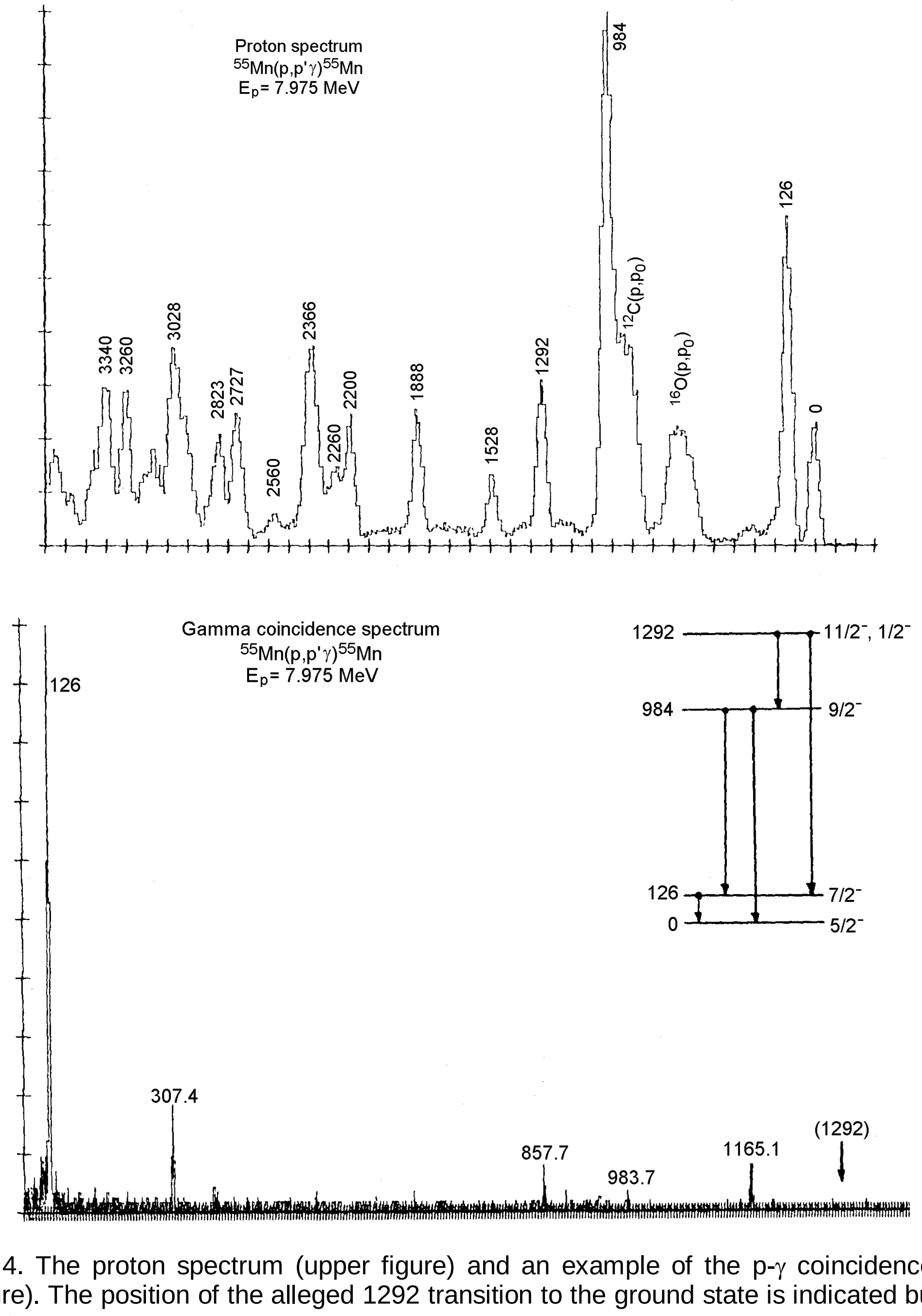

Figure 28.4. The proton spectrum (upper figure) and an example of the p-γ coincidence spectrum (lower figure). The position of the alleged 1292 transition to the ground state is indicated but is absent in the spectrum.

Table 28.1

Gamma rays observed in singles measurements for the $^{55}Mn(p,p'\gamma)^{55}Mn$ reaction at $E_p$ = 7.975 MeV [a])

| $E\gamma$ (keV) | ± $\Delta E\gamma$ (keV) | $RI$ | ± $\Delta RI$ | $E\gamma$ (keV) | ± $\Delta E\gamma$ (keV) | $RI$ | ± $\Delta RI$ |
|---|---|---|---|---|---|---|---|
| 129.7 | .2 | 301 | 13 | 983.7 | .5 | 250 | 35 |
| 158.2 | .2 | 99 | 11 | 1165.1 | .3 | 333 | 49 |
| 238.0 | .2 | 147 | 22 | 1212.2 | .3 | 567 | 75 |
| 273.3 | .2 | 688 | 40 | 1221.9 | .4 | 1622 | 162 |
| 307.4 | .2 | 109 | 15 | 1236.9 | .5 | 154 | 31 |
| 385.2 | .2 | 534 | 44 | 1280.4 | .5 | 68 | 20 |
| 411.3 | .2 | 224 | 36 | 1315.8 | .3 | 5542 | 512 |
| 416.7 | .5 | 8.7 | 2.5 | 1326.5 | .3 | 412 | 64 |
| 442.0 | .2 | 221 | 33 | 1369.1 | .2 | 1067 | 116 |
| 477.0 | .2 | 3094 | 214 | 1378.6 | 1.3 | 105 | 48 |
| 482.9 | .7 | 10 | 3 | 1407.7 | .2 | 2067 | 215 |
| 532.0 | .2 | 209 | 29 | 1419.2 | .7 | 169 | 51 |
| 743.8 | .6 | 65 | 21 | 1433.3 | .3 | 958 | 127 |
| 765.1 | .2 | 76 | 16 | 1459.8 | .4 | 375 | 62 |
| 803.2 | .2 | 2172 | 132 | 1505.2 | .8 | 191 | 52 |
| 810.6 | .2 | 352 | 38 | 1527.7 | .4 | 678 | 103 |
| 826.7 | .2 | 400 | 44 | 1554.7 | .8 | 280 | 72 |
| 846.0 | .2 | 1014 | 79 | 1572.0 | .8 | 356 | 84 |
| 857.7 | .2 | 1000 | | 1620.7 | .5 | 315 | 68 |
| 895.1 | .2 | 252 | 55 | 1638.9 | .4 | 531 | 94 |
| 910.5 | .8 | 68 | 28 | 1663.4 | .5 | 287 | 63 |
| 930.8 | .2 | 7964 | 491 | 1882.4 | .5 | 379 | 93 |
| 962.9 | .8 | 103 | 30 | | | | |

[a]) Known impurities or non-prompt $\gamma$-rays have been excluded.

$RI$ – Relative intensity normalized to the 857.7 keV $\gamma$-ray.

## Summary and conclusions

We have carried out a study of the gamma de-excitation of $^{55}Mn$ using the $^{55}Mn(p,p'\gamma)^{55}Mn$ reaction at 7.975 MeV. We have observed 47 gamma transitions between 27 states in $^{55}Mn$ extending to 3385 keV excitation energy. In addition, we have measured proton spectra using a 24" double focusing spectrometer. The incident proton energy was chosen to coincide with the energy used by Katsanos and Huizenga (1967) in their (p,p') measurements. In contrast with their claim, our high-resolution data do not show a doublet of states at 1292 keV excitation energy.

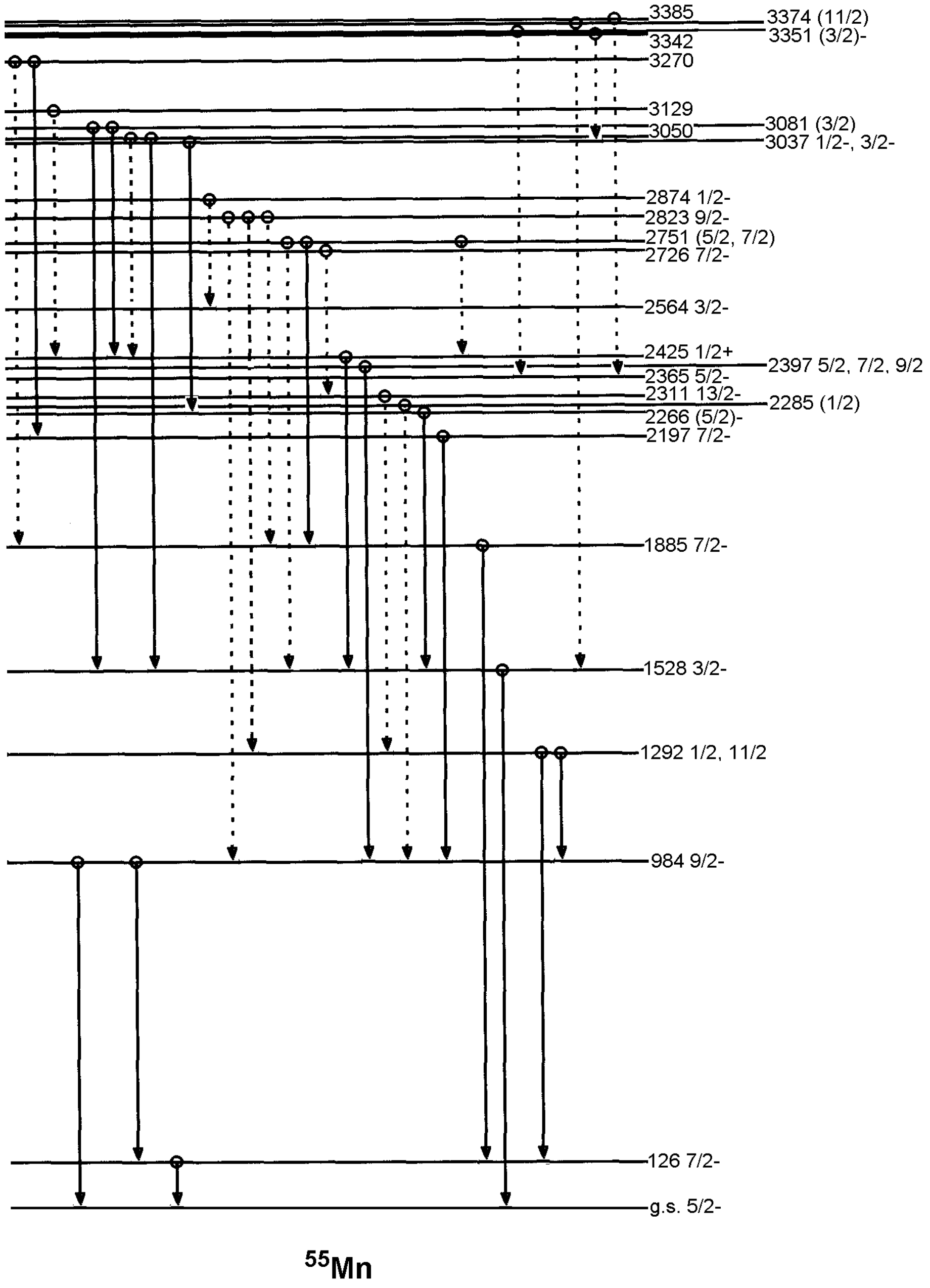

Figure 28.5. The gamma de-excitation scheme of $^{55}$Mn determined by our measurements. The relatively strong and firmly determined γ transitions are indicated by solid lines. The dotted lines show weak transitions.

Our gamma de-excitation measurements are in good agreement with the results of Hichwa et al. (1973a) and Hichwa, Lawson, and Chagnon (1973b) around this excitation energy. However, in contrast with the conclusion of Kulkarni (1976), our results clearly demonstrate that there is no gamma transition between the 1292 keV level and the ground state.

Combining our results for the gamma de-excitation, with our previous results on the $^{52}$Cr($^{4}$He,p)$^{55}$Mn reaction at 18 and 26 MeV and with available information based on research by other authors we conclude that there must be a closely spaced doublet of states at 1292 keV excitation energy. The energy spacing between the two members of the doublet appears to be less than 10 keV. The excitation of the $^{1}/_{2}^{-}$ member of the doublet appears to be strongly selective and to depend on the reaction used to access energy levels in $^{55}$Mn.

## 29

# The $^{138}$Ba($^{7}$Li,$^{6}$He)$^{139}$La and $^{140}$Ce($^{7}$Li,$^{6}$He)$^{141}$Pr Reactions at 52 MeV

***Key features:***

4. Spectroscopic applicability of the single-proton ($^{7}$Li,$^{6}$He) reaction has been studied.
5. The target nuclei of $^{138}$Ba and $^{140}$Ce and have been selected to have a clean access to a wide range of single-proton configurations: $1g_{7/2}$, $2d_{5/2}$, $2d_{3/2}$, $3s_{1/2}$, and $1h_{11/2}$.
6. A distinct $j$ – dependence has been observed for the $l$ = 2 transfer, which allowed to distinguish between the $2d_{5/2}$ and $2d_{3/2}$ configurations.
7. Spin assignments have been made and spectroscopic factors have been extracted to states in the residual nuclei $^{139}$L and $^{141}$Pr. They compare well with earlier studies and thus show that the ($^{7}$Li,$^{6}$He) reaction can serve as a useful spectroscopic tool.

***Abstract***: Angular distributions have been measured for transitions to low-lying states in $^{139}$La and $^{141}$Pr populated by the $^{138}$Ba($^{7}$Li,$^{6}$He)$^{139}$La and the $^{140}$Ce($^{7}$Li,$^{6}$He)$^{141}$Pr reactions at $E_{7Li}$ = 52 MeV. Elastic scattering of $^{7}$Li at 52 MeV on $^{138}$Ba and $^{140}$Ce, and $^{6}$Li at 48 MeV on $^{139}$La and at 47 MeV on $^{141}$Pr were measured to determine the interaction potentials in the incident and outgoing channels. Optical-model parameters extracted from fits to the scattering data were used in a finite-range distorted wave analysis of the measured angular distributions for levels below 2.40 MeV excitation energy in $^{139}$La and 1.65 MeV in $^{141}$Pr. Final-state spins have been assigned to levels in $^{139}$La and $^{141}$Pr. The reaction cross sections exhibit less structure than predicted by the distorted wave calculations, but the extracted spectroscopic factors are generally in good agreement with light-ion results.

## Introduction

Heavy-ion-induced, single-nucleon stripping reactions can be used to extract spectroscopic information complementary to that obtained from light-ion work. However, as the mass of the target-projectile system increases, bell-shaped angular distributions centred at the grazing angle are observed and they have only limited spectroscopic usefulness. Furthermore, heavy-ion studies employing $^{16}$O, $^{14}$N and $^{12}$C projectiles have been limited by energy resolution to residual nuclei with large level spacing.

To explore the spectroscopic applicability of heavy-ion-induced reactions we have selected a lighter projectile $^{7}$Li. Combined with the available good resolution, reactions induced by this projectile was be expected to provide useful spectroscopic information for closely spaced states.

The shell model indicates that systems with 50 or 82 nucleons constitute unusually tightly bound cores. Consequently, by using single-nucleon transfer reactions on such nuclei one should be able to study conveniently states corresponding to configurations outside closed cores.

For our study, we have chosen a single-proton transfer reaction ($^{7}$Li,$^{6}$He) and we have selected $^{138}$Ba and $^{140}$Ce as target nuclei. These two isotopes have an $N$ = 82 neutron core. They also contain 56 and 58 protons, respectively. Thus, $^{138}$Ba contains 6 protons outside the closed $Z$ = 50 proton core, and $^{140}$Ce contains 8. The stripped proton may be expected to be deposited to any of the following orbitals: $1g_{7/2}$, $2d_{5/2}$, $2d_{3/2}$, $3s_{1/2}$, and $1h_{11/2}$. However, the $1g_{7/2}$ orbitals are nearly full in both isotopes, particularly in $^{140}$Ce so it might be expected that most protons will be transferred to other configurations. Nevertheless, the $1g_{7/2}$ should be also accessible.

## Experimental method and results

To have a complete set of data for the intended theoretical analysis we had to measure not only the differential cross sections for the ($^7$Li,$^6$He) reactions but also the elastic scattering cross sections in the incident and outgoing channels. However, since $^6$He particles cannot be used as projectiles, we have measured the elastic scattering of $^6$Li on the relevant target nuclei.

Beams of $^6Li^-$ and $^7Li^-$ from a General Ionex sputter source were injected into the Australian National University 14UD Pelletron accelerator. Beam currents of up to 300 nA of $^6Li^{+3}$ and $^7Li^{+3}$ were obtained on the target.

Targets of enriched $^{138}$Ba (> 99%) and $^{140}$Ce (> 99%) and natural $^{139}$La and $^{141}$Pr, comprised of metal on thin carbon backings, proved to be extremely fragile and many ruptured before they could be removed from the vacuum system in which they were prepared. Others broke whilst standing in vacuum storage. Fortunately, at least one target of each material survived both preparation and beam bombardment. However, the thickness of surviving targets was small, only about 25 μg/cm$^2$. We could have had acceptable resolution with significantly thicker targets of about 100 -150 μg/cm$^2$.

Reaction data and elastic scattering data were measured with an Enge split-pole spectrograph using a resistive-wire gas proportional detector (Ophel and Johnston 1978) located in the focal plane. From the energy loss *($\Delta E$)* and the position signal ($\propto B\rho$) of the focal plane detector, a mass identification signal ($M^2 = (B\rho)^2 \Delta E$) was obtained. The difference in magnetic rigidity between $^6Li^{3+}$ and $^6He^{2+}$ was sufficient to allow for unambiguous mass identification. Additionally, the high field necessary to bend the $^6$He particles onto the detector completely removed the $^7Li^{3+}$ elastic events from the detector, allowing high beam currents to be used at forward angles. Fixed monitor detectors at 15° and 30° were used for normalization between runs.

To obtain the best possible information about the wave functions in the incident and outgoing channels the following elastic scattering measurements have been carried out: $^{138}$Ba($^7$Li,$^7$Li)$^{138}$Ba and $^{140}$Ce($^7$Li,$^7$Li)$^{140}$Ce at E($^7$Li) = 52 MeV and $^{139}$La($^6$Li,$^6$Li)$^{139}$La at E($^6$Li) = 48 MeV and $^{141}$Pr($^6$Li,$^6$Li)$^{141}$Pr at E($^6$Li) = 47 MeV. The energies for $^6$Li projectiles were chosen to correspond to the average outgoing $^6$He energy.

Absolute cross sections were obtained by normalizing the forward angle elastic scattering to the Rutherford cross sections. The error in the absolute normalization is estimated to be 5% for the elastic scattering, resulting mainly from angle setting and dead time uncertainties. Based on the reproducibility of the ($^7$Li,$^6$He) data, the absolute cross sections for the transfer reactions are accurate to ±12%. The relative errors in the cross sections are shown by the error bars on the individual data points where these are larger than the plotted points.

Figures 29.1 and 29.2 show spectra for the $^{138}$Ba($^7$Li,$^6$He)$^{139}$La reaction at $\theta_{lab}$ = 27°, and the $^{140}$Ce($^7$Li,$^6$He)$^{141}$Pr reaction at $\theta_{lab}$ = 20°, respectively. The resolution is 70 keV FWHM and little background is evident at these angles. However, at angles forward of 8° (lab), background from impurities in the target was larger, but it did not prevent extraction of the data.

The distributions for the elastic scattering and for the ($^{7}Li$,$^{6}He$) reaction are shown in Figures 29.3 – 29.7.

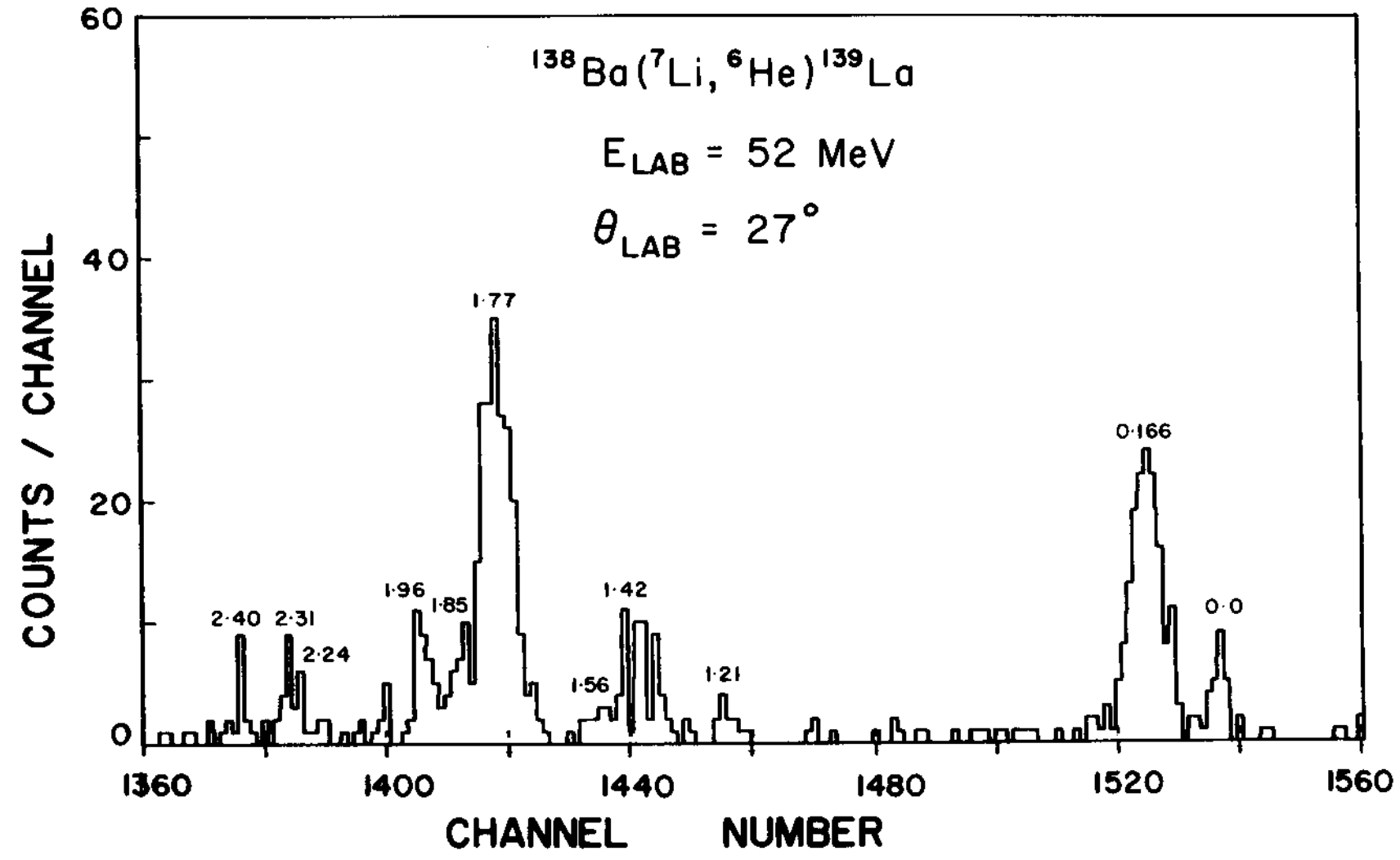

Figure 29.1. A $^{6}He$ spectrum for the $^{138}Ba(^{7}Li,^{6}He)^{139}La$ reaction at $27^{0}$. States in $^{139}La$ are labelled with the appropriate excitation energies.

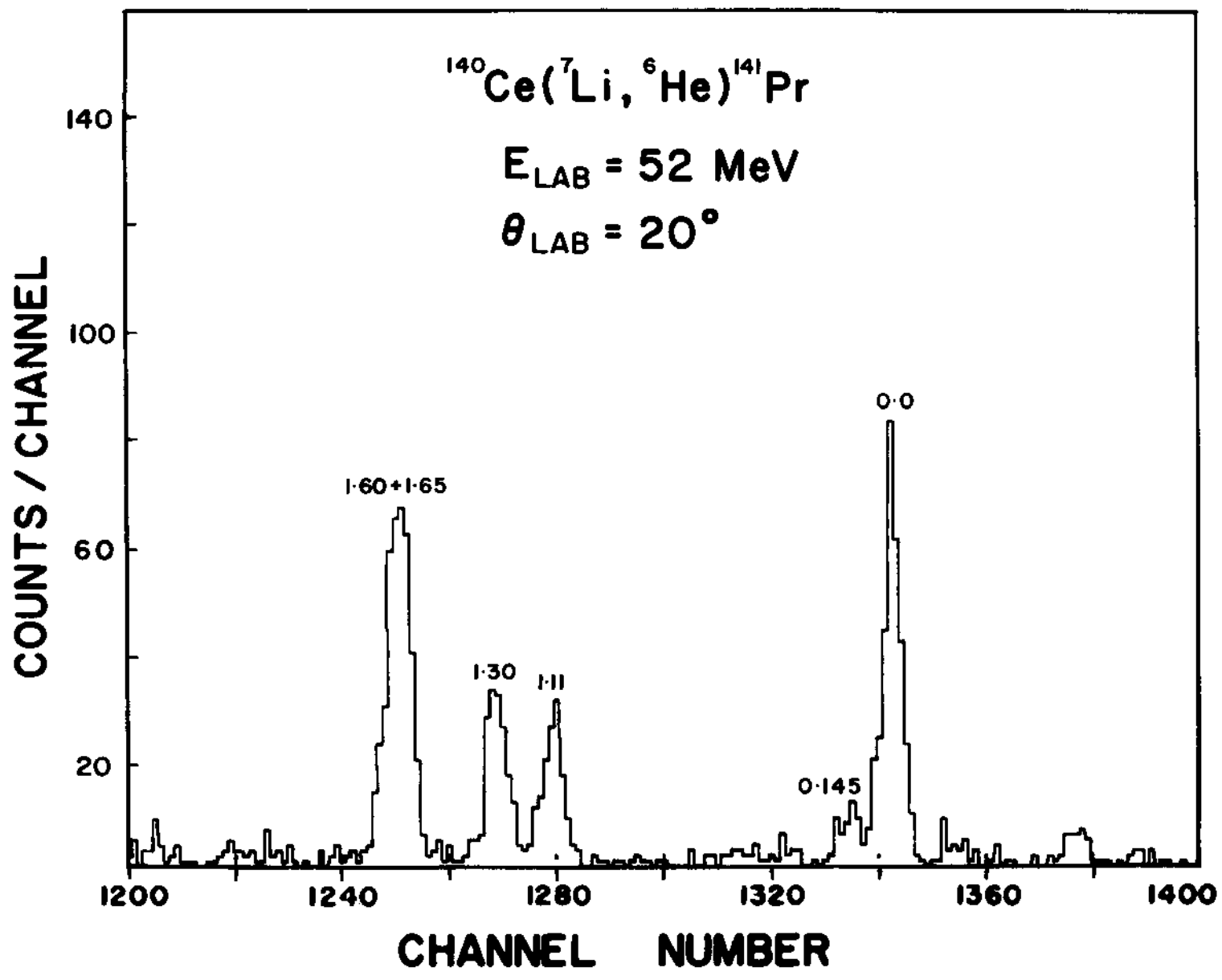

Figure 29.2. A $^{6}He$ spectrum for the $^{140}Ce(^{7}Li,^{6}He)^{141}Pr$ reaction at $20^{0}$. States in $^{141}Pr$ are labelled with the appropriate excitation energies.

## Theoretical analysis

The elastic scattering data were analysed using a simple (central) optical model potential combined with the usual description of the Coulomb interaction:

$$U(r) = V_c(r) - Vf(r, r_0, a_0) - iWg(r, r_0', a_0')$$

where,

$$f(r,r_0,a_0)=\left\{1+\left[\left(r-r_0A_t^{1/3}\right)/a_0\right]\right\}^{-1}$$

$$g(r,r_0',a_0')=\left\{1+\left[\left(r-r_0'A_t^{1/3}\right)/a_0'\right]\right\}^{-1}$$

$V_c(r)$ the Coulomb potential between a point projectile charge and the field of a uniformly charged sphere of radius $R_c=r_cA^{1/3}$.

$$V_c(r)=\frac{Z_pZ_te^2\left[3-\left(r/r_cA_t^{1/3}\right)\right]}{2r_cA_t^{1/3}} \qquad \text{for } r\le r_cA_t^{1/3}$$

$$V_c(r)=\frac{Z_pZ_te^2}{r} \qquad \text{for } r>r_cA_t^{1/3}$$

$Z_p$ and $Z_t$ are the projectile and target charge.

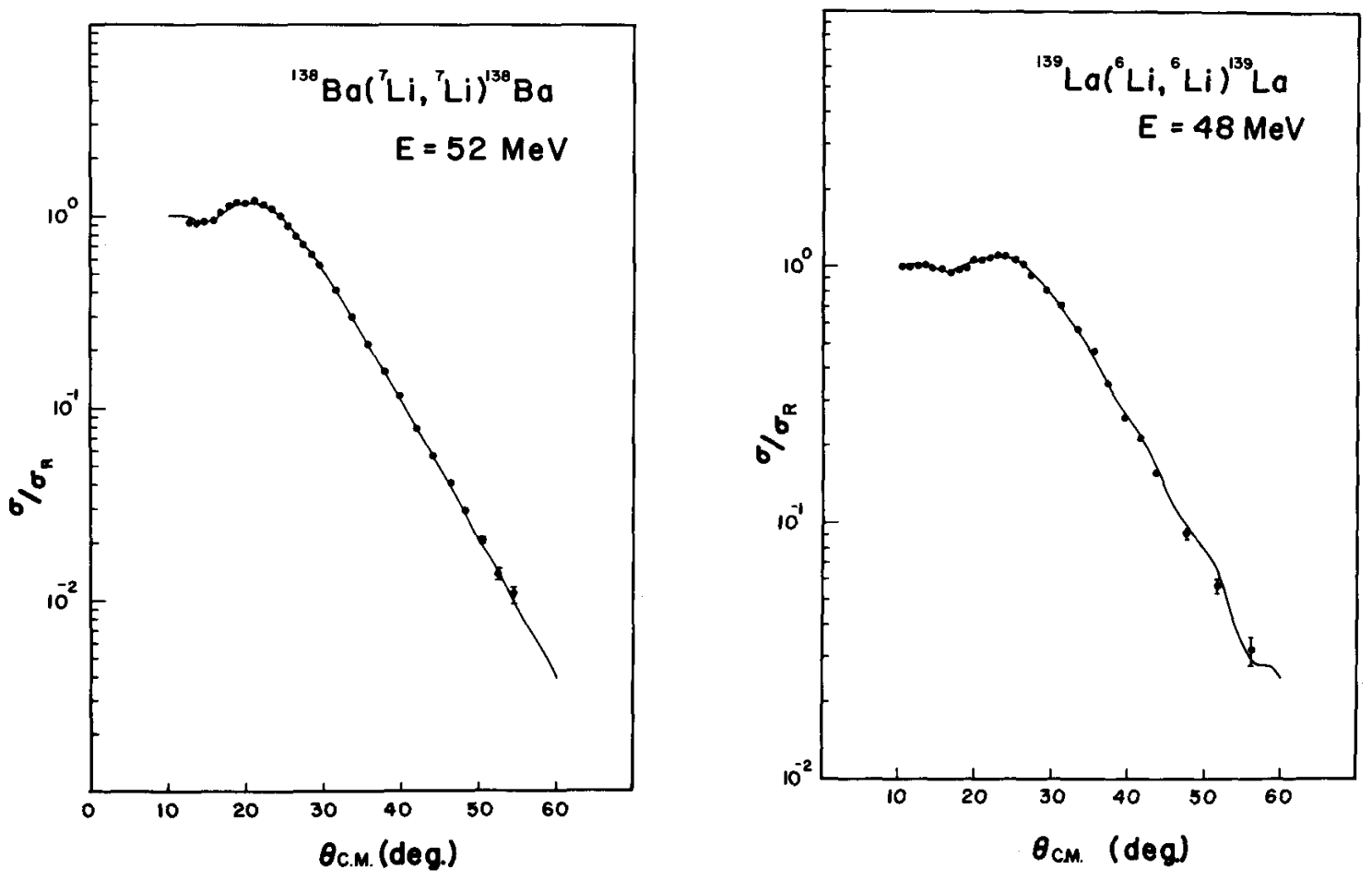

Fig. 3. Angular distributions for the $^{138}$Ba($^7$Li,$^7$Li)$^{138}$Ba elastic scattering at 52 MeV and $^{139}$La($^6$Li,$^6$Li)$^{139}$La at 48 MeV. The solid lines are the optical-model fits to the data.

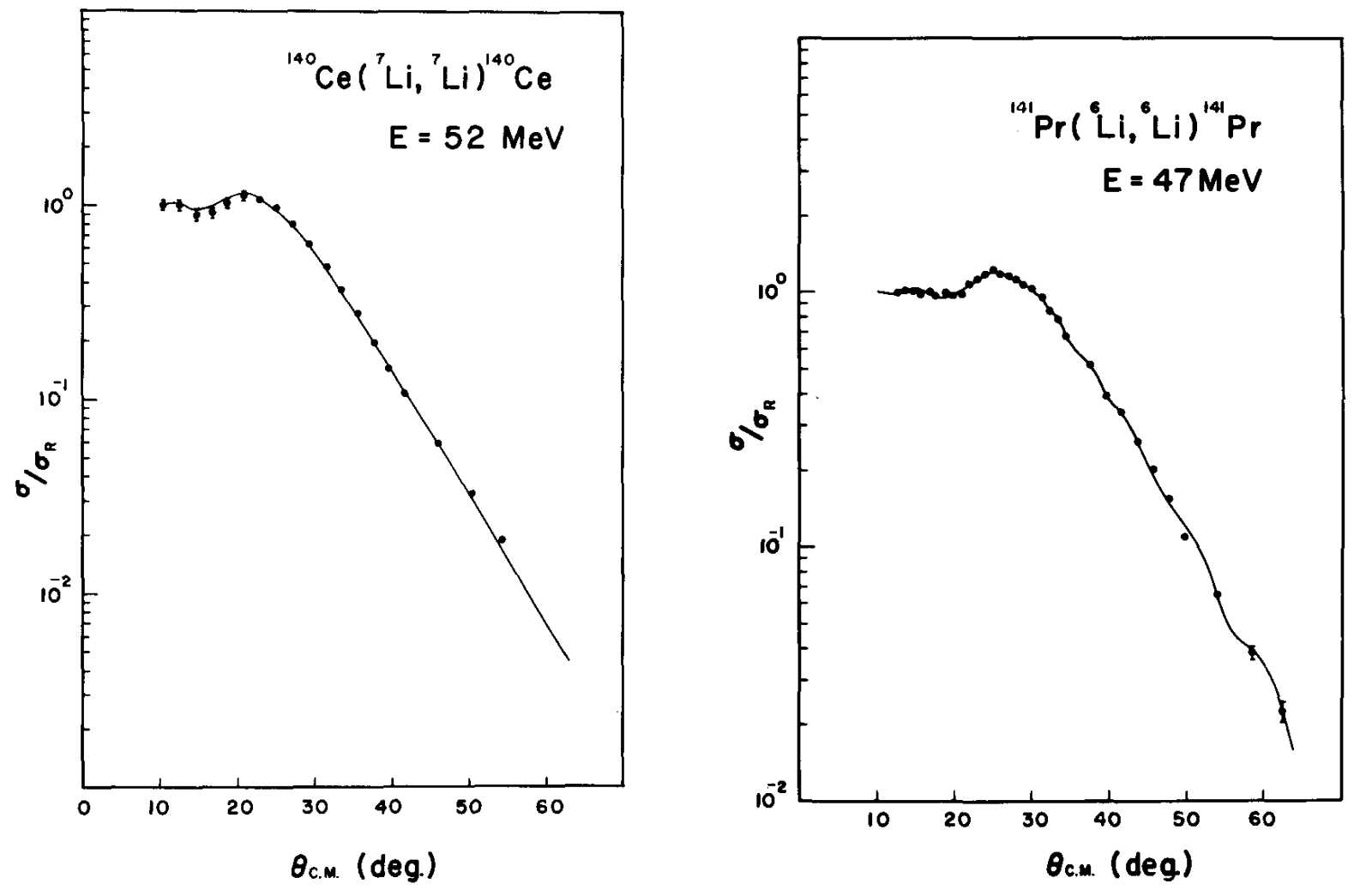

Figure 29.4. Angular distributions for the $^{140}$Ce($^7$Li,$^7$Li)$^{140}$Ce elastic scattering at 52 MeV and for $^{141}$Pr($^6$Li,$^6$Li)$^{141}$Pr at 47 MeV. The solid lines are the optical-model fits to the data.

Table 29.1
Nuclear and Coulomb potential parameters used to fit the elastic scattering and the ($^7$Li,$^6$He) reaction angular distributions

| Reaction | $E_{lab}$ (MeV) | $V$ (MeV) | $r_0$ (fm) | $a_0$ (fm) | $W$ (MeV) | $r_0'$ (fm) | $a_0'$ (fm) | $r_c$ (fm) |
|---|---|---|---|---|---|---|---|---|
| $^{138}$Ba($^7$Li, $^7$Li)$^{138}$Ba | 52 | 277.7 | 1.276 | 0.752 | 31.91 | 1.536 | 0.780 | 1.400 |
| $^{139}$La($^6$Li, $^6$Li)$^{139}$La | 48 | 246.6 | 1.351 | 0.644 | 8.38 | 1.804 | 0.796 | 1.400 |
| $^{140}$Ce($^7$Li, $^7$Li)$^{140}$Ce | 52 | 276.1 | 1.276 | 0.752 | 48.20 | 1.536 | 0.780 | 1.400 |
| $^{141}$Pr($^6$Li, $^6$Li)$^{141}$Pr | 47 | 249.1 | 1.339 | 0.673 | 8.91 | 1.714 | 0.707 | 1.400 |

The computer code JIB (Perey 1967), which I have earlier modified and adapted to run at ANU and which I have used in analyses of scattering of light projectiles, was now used to fit the elastic scattering data for heavy projectiles. The parameters were varied two at a time until a minimum $\chi^2$ was obtained. The experimental angular distributions and the optical-model fits are shown in Figures 29.3 and 29.4. The extracted parameters are listed in Table 29.1.

These parameters were then used in the exact finite-range (EFR) distorted wave calculations using the computer code LOLA (DeVries 1973) for transitions to the strongly populated states observed in the $^{138}$Ba($^7$Li,$^6$He)$^{139}$La and $^{140}$Ce($^7$Li,$^6$He)$^{141}$Pr reactions.

The wave functions of the bound proton were generated with Woods-Saxon potentials whose depths were adjusted to give the correct binding energies. The ground-state binding energies are -9.978 MeV for p + $^6$He, -6.201 MeV for p+$^{138}$Ba and - 5.227 MeV for p + $^{140}$Ce (ND 1971). The shape parameters were fixed at $r_0$ = 1.25 fm and $a_0$= 0.65 fm. The experimental angular distributions and the EFR-distorted wave fits to the data are shown in Figures 29.5 – 29.7. Generally, it is easy to distinguish angular distributions belonging to different orbital angular momenta. However, transfer of a proton to orbitals outside the closed shell $Z$ = 50 involves two $l$ = 2 configurations, $2d_{3/2}$ and $2d_{5/2}$. Fortunately, there is a clear $j$ - dependence for these configurations. Calculations for $2d_{3/2}$ and $2d_{5/2}$ are shown in Figures 29.5 and 29.6 for states at 0.166 MeV, 1.56 MeV, 1.85 MeV and 1.96 MeV in $^{139}$La, and in Figure 29.7 for the ground state in $^{141}$Pr. Clearly the data forward of 6° (c.m.) allow unambiguous distinction between $2d_{3/2}$ and $2d_{5/2}$ final states.

Spectroscopic factors are extracted using the following relationship (DeVries 1973)

$$\left(\frac{d\sigma}{d\Omega}\right)_{\exp} = \frac{2J_B+1}{l}\sum_l (2l+1)W^2(l_1 j_1 l_2 j_2;\frac{1}{2}l)C^2S(^7Li)C^2S_B\left(\frac{d\sigma}{d\Omega}\right)_{LOLA}$$

where $B$ refers to the states in the final nuclei, and subscripts 1 and 2 refer to the projectile and final nuclei. The $l$ and $W$ are the transferred orbital angular momentum and the appropriate Racah coefficient, respectively. The $C^2S(^7Li)$ is the overlap of $^7$Li with $^6$He + p in a $p_{3/2}$ state and was taken from Cohen and Kurath (1967) to be 0.59. Extracted spectroscopic factors are obtained by normalization in the region of the grazing angle (25°-29° for $^{138}$Ba and 26°-30° for $^{140}$Ce).

The absolute spectroscopic factors obtained from the analysis are listed in Tables 29.2 and 29.3, which also show the spectroscopic factors obtained from the ($^3$He,d) reactions (Wildenthal, Newman, and Auble 1971). The errors in the absolute spectroscopic factors include the uncertainty in the absolute normalization of the experimental data and statistical errors. Relative spectroscopic factors, normalized to the ground state, are also listed in Tables 29.2 and 29.3.

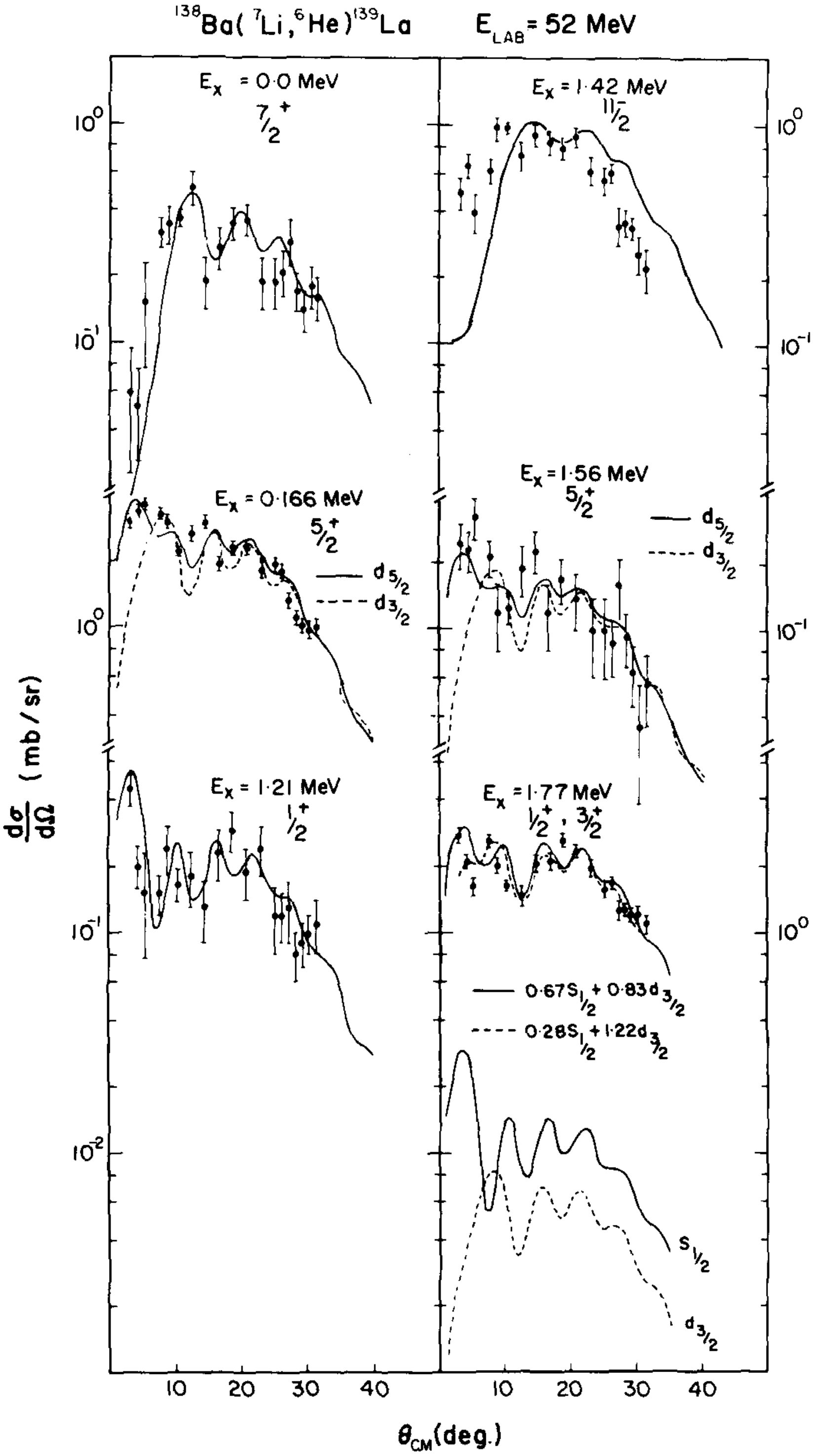

Figure 29.5. Angular distributions for states populated in the $^{138}Ba(^{7}Li,^{6}He)^{139}La$ reaction. The solid and dashed lines are the EFR-distorted wave calculations normalized to the data. The 1.77 MeV doublet is shown using spectroscopic strengths obtained from the ($^{3}He$,d) work of Wildenthal, Newman, and Auble (1971) (solid line) and spectroscopic strengths obtained from a best fit to the data (dashed line). The pure $s_{1/2}$ and $d_{3/2}$ components are shown below the data.

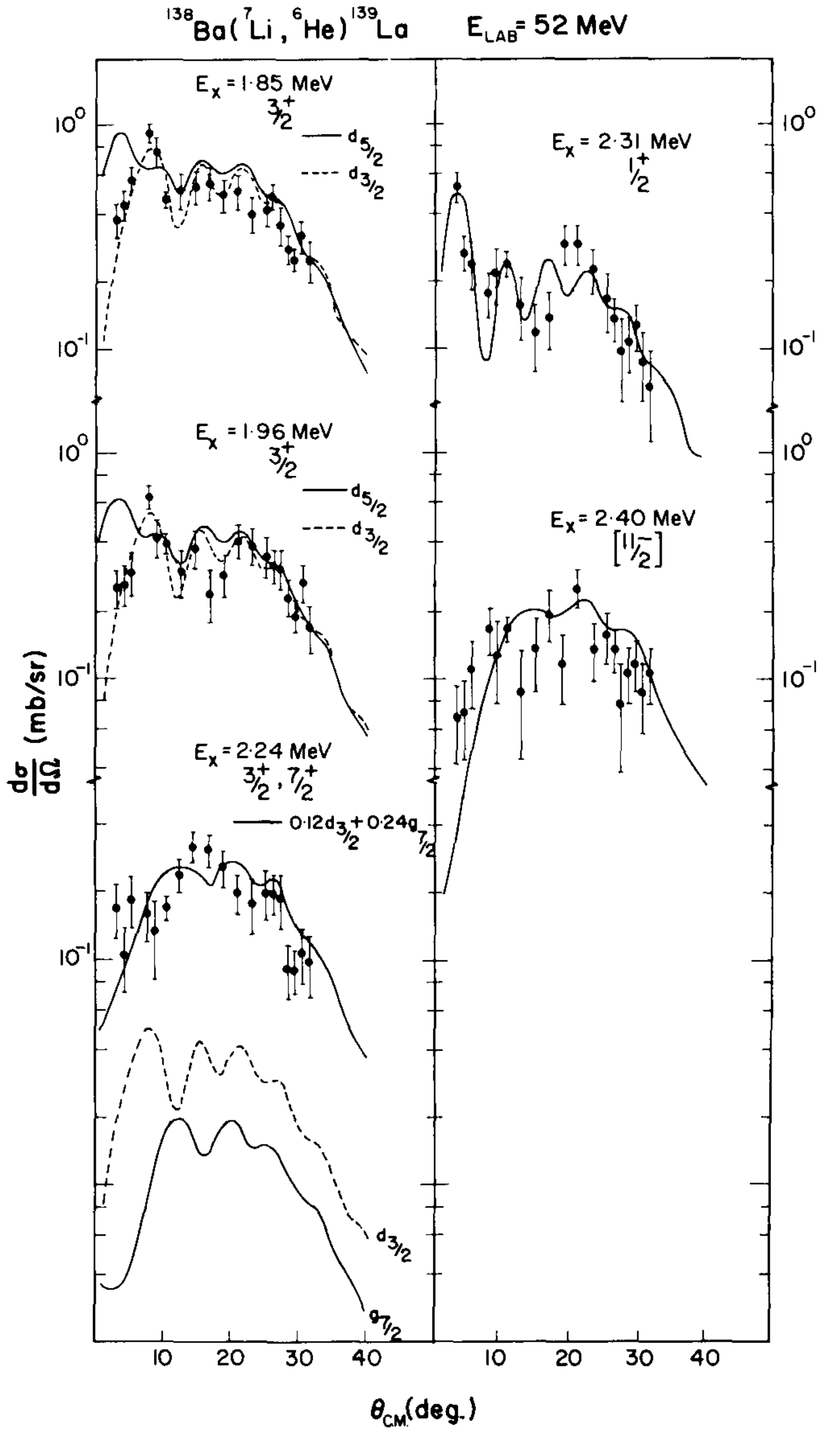

Figure 29.6. Angular distributions for states populated in the $^{138}Ba(^{7}Li,^{6}He)^{139}La$ reaction. The solid and dashed lines are the EFR-distorted wave calculations normalized to the data. The 2.24 MeV data are shown with the spectroscopic strengths obtained from a best fit to the data using a sum of $d_{3,2}$ and $g_{7/2}$ contributions. The pure $d_{3/2}$ and $g_{7/2}$ components are shown below the data.

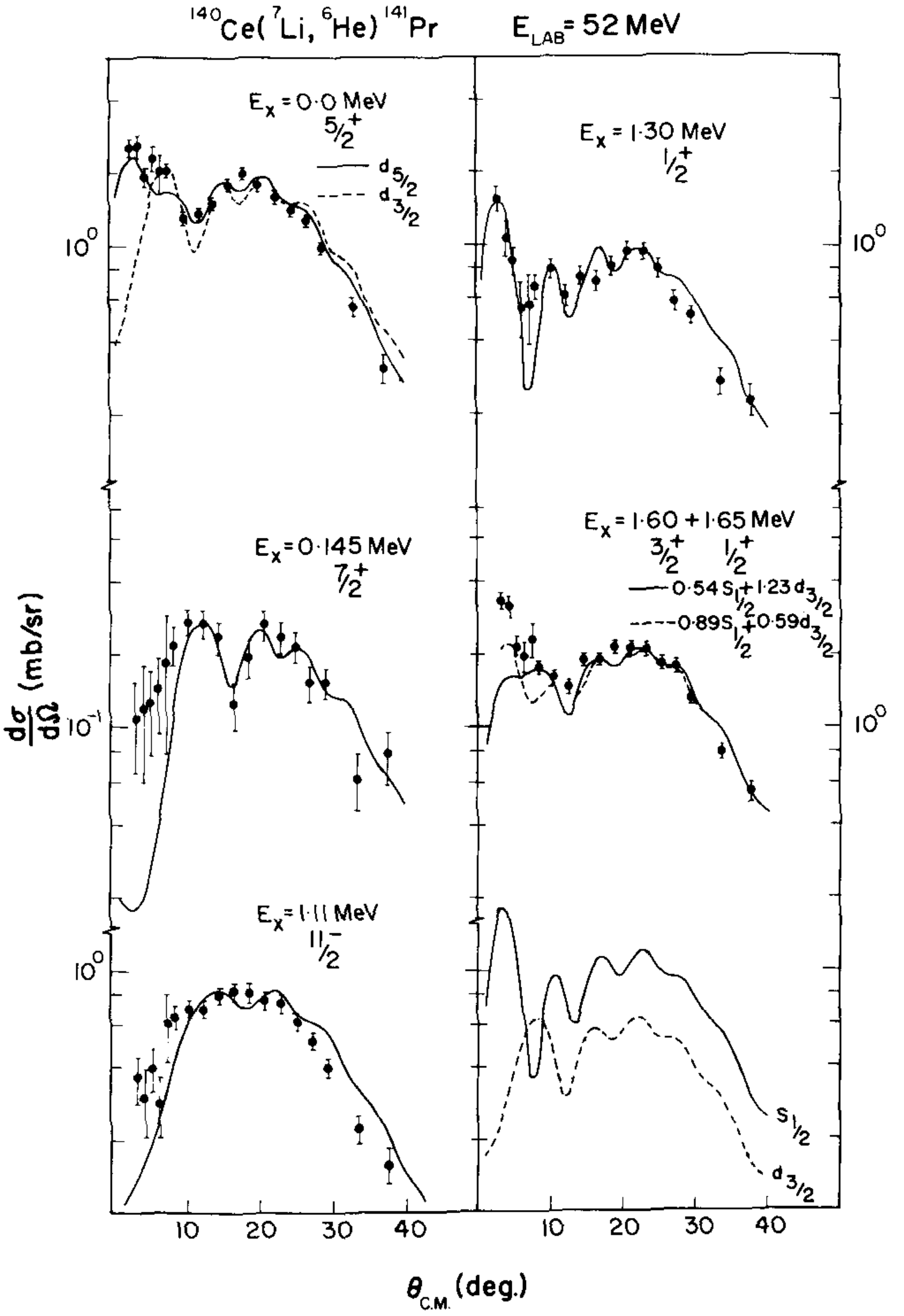

Figure 29.7. Angular distributions for states populated in the $^{140}Ce(^{7}Li,^{6}He)^{141}Pr$ reaction. The solid and dashed lines are the EFR-distorted wave calculations normalized to the data. The 1.60 MeV and 1.65 MeV states could not be resolved, and are shown using spectroscopic strength ratios obtained from the ($^{3}He$,d) work of Wildenthal, Newman, and Auble (1971) (solid line) and spectroscopic strengths obtained from a best fit to the data (dashed line). The pure $s_{1/2}$ and $d_{3/2}$ components are shown below the data.

As can be seen from Figures 29.5 – 29.7, the EFR-distorted wave calculations describe the data well. Angular distributions for transitions to $3s_{1/2}$, levels via the $l$ = 1 transfer at 1.21 MeV and 2.31 MeV in $^{139}La$, and 1.30 MeV in $^{141}Pr$ are well reproduced by the calculations but are slightly out of phase, a problem which has been observed in other ($^{7}Li$,$^{6}He$) reactions (Moore, Camper, and Charlton 1970) where the data and the calculations are seriously out of phase.

The states of the unresolved doublet at 1.77 MeV in $^{139}La$ are described as $3s_{1/2}$, and $2d_{3/2}$ in the ($^{3}He$,d) reaction (Wildenthal, Newman, and Auble 1971). The solid line in Figure 29.5 represents the fit to the data obtained by summing the calculated differential cross sections for the two configurations and using spectroscopic strengths as derived in the ($^{3}He$,d) work. The dashed line represents a similar fit but obtained by allowing the spectroscopic strengths to be adjusted by the least-squares fitting procedure.

Table 29.2
Spectroscopic information for $^{139}$La

| $E$ a) (MeV) | $E$ b) (MeV) | $nl_j$ | Absolute ($^3$He, d) a) 40.3 MeV | Absolute ($^7$Li, $^6$He) b) 52 MeV | Relative ($^3$He, d) a) 40.3 MeV | Relative ($^7$Li, $^6$He) b) 52 MeV |
|---|---|---|---|---|---|---|
| 0.0 | 0.0 | $1g_{7/2}$ | 0.54 | 0.72±0.17 | 1.00 | 1.00±0.24 |
| 0.166 | 0.166 | $2d_{5/2}$ | 0.90 | 1.03±0.15 | 1.67 | 1.43±0.21 |
| 1.21 | 1.21 | $3s_{1/2}$ | 0.10 | 0.14±0.04 | 0.19 | 0.19±0.06 |
| 1.42 | 1.42 | $1h_{11/2}$ | 0.71 | 1.19±0.22 | 1.31 | 1.65±0.31 |
| 1.56 | 1.56 | $2d_{5/2}$ | 0.07 | 0.06±0.02 | 0.13 | 0.08±0.03 |
| 1.78 | 1.77 | $3s_{1/2}$ | 0.67 | 0.28±0.04 | 1.24 | 0.39±0.06 |
| | | $2d_{3/2}$ | 0.83 | 1.22±0.18 | 1.54 | 1.69±0.25 |
| 1.85 | 1.85 | $2d_{3/2}$ | 0.30 | 0.43±0.08 | 0.56 | 0.60±0.12 |
| 1.96 | 1.96 | $2d_{3/2}$ | 0.19 | 0.28±0.06 | 0.35 | 0.39±0.08 |
| 2.24 | 2.24 | $2d_{3/2}$ | 0.10 | 0.12±0.04 | 0.19 | 0.17±0.06 |
| | | $1g_{7/2}$ | | 0.24±0.08 | | 0.33±0.11 |
| 2.31 | 2.31 | $3s_{1/2}$ | 0.13 | 0.15±0.05 | 0.24 | 0.21±0.07 |
| | 2.40 | ($1h_{11/2}$) | | 0.21±0.07 | | 0.29±0.10 |

a) Wildenthal, Newman, and Auble (1971). b) Our work.
$E$ – Excitation energy. $nl_j$ – Single particle configuration. Absolute/Relative – The absolute and relative values of the spectroscopic factors, respectively.

Table 29.3
Spectroscopic information for $^{141}$Pr

| $E$ a) (MeV) | $E$ b) (MeV) | $nl_j$ | Absolute ($^3$He, d) a) 40.3 MeV | Absolute ($^7$Li, $^6$He) b) 52 MeV | Relative ($^3$He, d) a) 40.3 MeV | Relative ($^7$Li, $^6$He) b) 52 MeV |
|---|---|---|---|---|---|---|
| 0.0 | 0.0 | $2d_{5/2}$ | 0.64 | 0.72±0.10 | 1.00 | 1.00±0.14 |
| 0.145 | 0.145 | $1g_{7/2}$ | 0.35 | 0.45±0.10 | 0.55 | 0.63±0.14 |
| 1.11 | 1.11 | $1h_{11/2}$ | 0.84 | 0.98±0.15 | 1.31 | 1.36±0.21 |
| 1.30 | 1.30 | $3s_{1/2}$ | 0.65 | 0.63±0.09 | 1.02 | 0.88±0.13 |
| 1.60 | 1.60+ | $2d_{3/2}$ | 1.23 | 0.59±0.09 | 1.92 | 0.82±0.13 |
| 1.65 | 1.65 | $3s_{1/2}$ | 0.54 | 0.89±0.14 | 0.84 | 1.24±0.19 |

See notes to Table 29.2.

States at 1.60 and 1.65 MeV in $^{141}$Pr could not be resolved in our experiment. These states are described as $2d_{3/2}$ (1.60 MeV) and $3s_{1/2}$ (1.65 MeV) in the ($^3$He,d) reaction (Wildenthal, Newman, and Auble 1971). Figure 29.7 shows the best fits obtained either by taking a sum of the calculated distributions with the spectroscopic strengths as derived in the ($^3$He,d) work for the two configurations (the solid line) or by adjusting the spectroscopic strengths in the least-squares fitting procedure (the dashed line). If the calculated distribution for the $3s_{1/2}$, transition to the 1.65 MeV state is out of phase with the experimental distribution, as it is for the $l$ = 1 transitions to the 1.21 and 2.31 MeV states (see above), this would strongly affect the fitting procedure and hence the spectroscopic factors. Thus, the spectroscopic factors for these states should be viewed with caution.

The state at 2.24 MeV in $^{139}$La is described as a $d$ - state in the ($^3$He,d) reaction (Wildenthal, Newman, and Auble 1971). However, neither calculations for a $2d_{3/2}$ nor a $2d_{5/2}$, state gave a satisfactory fit. A level at 2.232 MeV has been observed in ($\gamma$,$\gamma$') work (Moreh and Nof 1970) and assigned tentative spins of either $^7/_2{}^+$ or $^{11}/_2{}^-$. The best fit for the studied here ($^7$Li,$^6$He) reaction has been obtained using a sum of $2d_{3/2}$ and $1g_{7/2}$ (see Figure 29.6).

## Summary and conclusions

We have carried out high resolution measurements for the $^{138}$Ba($^7$Li,$^6$He)$^{139}$La and $^{140}$Ce($^7$Li,$^6$He)$^{141}$Pr reactions at 52 MeV incident $^7$Li energy. In addition, we have also measured angular distributions for the elastic scattering $^{138}$Ba($^7$Li,$^7$Li)$^{138}$Ba and $^{140}$Ce($^7$Li,$^7$Li)$^{140}$Ce at E($^7$Li) = 52 MeV and $^{139}$La($^6$Li,$^6$Li)$^{139}$La at E($^6$Li) = 48 MeV and $^{141}$Pr($^6$Li,$^6$Li)$^{141}$Pr at E($^6$Li) = 47 MeV.

We have analysed the elastic scattering data using standard optical model potential and we have then applied the derived parameters in our exact finite range distorted wave analysis of transfer reaction data. The calculations described the transfer angular distributions well and only slight phasing problems were encountered for $s_{1/2}$ states.

The absolute spectroscopic factors obtained in our study are generally larger than those obtained from the light-ion ($^3$He,d) reactions (Wildenthal, Newman, and Auble 1971), but the relative spectroscopic factors are in good agreement.

Heavy-ion forward-angle $j$ - dependence has been observed and used to assign the following spins to $d$ - states in $^{139}$La; 0.166 ($^5/_2{}^+$), 1.56 ($^5/_2{}^+$), 1.85 ($^3/_2{}^+$), 1.96 ($^3/_2{}^+$); and to the ground state of $^{141}$Pr, ($^5/_2{}^+$). The spin of the *d* - state at 1.60 MeV in $^{141}$Pr could not be assigned with certainty because levels at 1.60 MeV and 1.65 MeV were unresolved. Previous spin assignments for the $d$ - state levels (Lederer and Shirley 1978) are given as 0.166 ($^5/_2{}^+$), 1.56 ($^3/_2{}^+$), 1.85 ($^3/_2{}^+$) and 1.96 ($^3/_2{}^+$) in $^{139}$La; and 0.0 ($^5/_2{}^+$), 1.60 ($^3/_2{}^+$) in $^{141}$Pr.

The state at 2.40 MeV in $^{139}$La was not observed in the ($^3$He,d) work (Wildenthal, Newman, and Auble 1971) but has been seen in ($\alpha$,$\alpha$') scattering (Baker and Tickle 1972) and assigned a negative parity. Our distorted wave calculation with $^{11}/_2{}^-$ spin for this state and corresponding to the only allowed negative-parity configuration $1h_{11/2}$, is shown in Figure 29.6. However, the statistical errors in the data points prohibit the definite assignment of any spin to this state.

In conclusion, the ($^7$Li,$^6$He) has been shown to be a functional tool which can be used in determining final-state spins and spectroscopic factors.

# 30

# Nuclear Molecular Excitations

***Key features:***

1. We have measured angular distributions and excitation functions for the $^{24}Mg(^{16}O,^{12}C)^{28}Si$ reaction. The aim was to study nuclear molecular excitations.
2. Three new broad resonances have been discovered in the excitation functions for the ground state and the first excited $2_1^+$ state in $^{28}Si$.
3. The analysis of our experimental results combined with a compilation of earlier results shows that the observed resonances can be explained as nuclear molecular excitations in either incident or exit channels, i.e. as orbiting nuclear states of the $^{24}Mg+^{16}O$ or $^{28}Si+^{12}C$ systems.

***Abstract***: Excitation functions for the reaction $^{24}Mg(^{16}O,^{12}C)^{28}Si$ leading to the ground state and the first $2_1^+$ excited state in $^{28}Si$ were measured at 5°(lab) in the energy range 32.4 < $E_{c.m.}$ < 48.6 MeV. Although the resonant structure, previously observed at lower energies, becomes progressively weaker, three new correlated maxima have been observed at $E_{c.m.}$ = 37.5, 40.2 and 43.5 MeV. Attempts to find a consistent optical-model fit to the elastic scattering in the entrance channel and an exact finite-range distorted wave fit to the transfer reaction cross sections in this energy range were unsuccessful. Such a failure is to be expected if strong coupling between the elastic and inelastic channels in either the initial or final system is present. By comparing the angular distribution with the Legendre polynomial distributions, $P_J^2(\cos\theta)$, spin assignments $J$ = 27, 29 and 31 were made for the three observed resonances. The observed resonant behaviour can be explained as nuclear molecular excitations.

## Introduction

Considerable resonance structure has been observed in heavy-ion reactions. Typical gross structure (with width ~2 MeV) has been seen in the excitation functions for the ($^{16}O,^{16}O$) elastic scattering at energies above the Coulomb barrier (Maher *et al.* 1969). Optical-model analyses (Gobbi *et al*. 1973; Maher *et al*. 1969) of these data led to either a shallow and weakly absorbing four-parameter potential or a surface-transparent six-parameter potential in which the imaginary well has smaller radius and diffuseness parameters than the real well. Such potentials allow the two colliding ions to retain their individual structure during a grazing collision and give rise to the possibility of so-called *orbiting molecular states.* However, because such collisions are essentially a direct process, the ions soon pass through so that they stay in the molecular orbit for perhaps only about $^1/_3$ of a revolution. Nevertheless, even such brief nuclear molecular configuration appears to have significant influence on the measured excitation functions and angular distributions. Quantum mechanically, such states correspond to virtual broad shape resonances in the ion-ion potential. Consequently, they give rise to a *broad* resonance structure of the type seen in the $^{16}O$ + $^{16}O$ elastic scattering.

The surface-transparent potentials cause the lower partial waves to be strongly absorbed while trajectories corresponding to grazing collisions are only weakly absorbed. These properties lead to two interesting phenomena: (i) typical diffraction effects associated with the strong absorption and (ii) the "glory" effect arising from orbiting trajectories in the weakly absorbing surface region.

The diffraction model of Austern and Blair (1965) has been employed (Phillips *et al.* 1979) to show that diffraction effects can lead to gross structure of the kind observed in the $^{12}C$ +$^{12}C$ and $^{16}O$ + $^{16}O$ inelastic scattering excitation functions. However, as pointed out by Friedman, McVoy, and Nemes (1979), the presence of strong internal absorption does not exclude the possibility of resonance effects. Indeed, in a full quantum mechanical treatment, the gross structure should arise from both the diffraction and orbiting resonance effects of the optical potentials involved.

The glory effect arises through interference of a normal backward scattered wave with one, which has orbited through a negative deflection angle of about 180°. This leads to large back-angle elastic scattering cross sections which display large oscillatory character such as is observed in $^{16}O$ + $^{28}Si$ elastic scattering at $E_{c.m.}$ = 35 MeV (Braun-Muzinger *et al.* 1977). These large-angle oscillations often follow the square of a Legendre polynomial, $P_J^2(\cos\theta)$, suggesting a resonating partial wave of the order of $J$. The $J$ - values extracted in this way lie close to the grazing angular momenta predicted by the appropriate surface-transparent optical potential.

Since the discovery of such phenomena, several descriptions involving Regge poles (Braun-Muzinger *et al.* 1977), angular momentum dependent absorption potentials (Chatwin *et al*. 1970), resonances (Barrette *et al.* 1978; Malmin *et al.* 1972) or parity-dependent optical potentials (Dehnhard, Shlolnik and Franey 1978) have been proposed. All these approaches, some of which are closely related, are designed to enhance one or more partial waves close to the grazing angular momentum.

In some heavy-ion reactions, the cross sections are rapidly fluctuating functions of energy and the question arises whether these are true resonance structure or simply statistical Ericson fluctuations. In some cases, e.g. for the $^{12}C$ + $^{14}N$ system (Hansen *et al*. 1974; Olmer *et al.* 1974), statistical calculations using the Hauser-Feshbach method (Hauser and Feshbach 1952) give a good description of such fine structure provided one subtracts out the underlying gross structure. However, in several cases, e.g. in the $^{12}C$ +$^{16}O$ system at $E_{c.m.}$ = 19.7 MeV, there exists a strong correlation between the excitation functions of the elastic scattering at several angles, suggesting a resonance. Such resonance has a width of ~ 0.4 MeV and is an example of *intermediate* structure.

Correlated structure of this kind with the width of ~ 0.1 MeV was found in the very first precision heavy-ion measurements (Bromley, Kuehner, and Almqvist 1960) for the $^{12}C$ + $^{12}C$ scattering. In addition to the widths, the spacing of the observed resonances in this system are too small to be readily described in terms of simple shape resonances associated with quasi-bound states in a molecular-type potential between the two ions. Such intermediate structure has been interpreted (Imanishi 1968, 1969; Michaud and Voght 1969, 1972) in terms of "doorway" states in which the incident channel couples to another degree of freedom of the resonating system. In particular, Imanishi (1968, 1969) proposed that the incident elastic channel may be strongly coupled to a channel in which one of the $^{12}C$ nuclei is excited to its first $2^+$ state at 4.4 MeV. This concept was extended by Scheid, Greiner and Lemmer (1970) and Fink, Scheid and Greiner (1972) in their double resonance mechanism in which a virtual state in the entrance channel is excited by a grazing partial wave and acts as a doorway state which feeds quasi-bound states in inelastic channels corresponding to excitation of collective states of the individual nuclei. In these approaches, the intermediate structure is described in terms of coupling between the elastic and inelastic channels and arises naturally in appropriate coupled-channels calculations (Fink,

Scheid and Greiner 1972; Imanishi 1968, 1969; Scheid, Greiner and Lemmer 1970).

In order to predict the energies and spins of possible intermediate structure resonances arising from such coupling, Abe (1977) and Kondo, Matsuse and Abe (1978) have suggested a schematic band-crossing model. In this picture, the resonance structure arises whenever two quasi-rotational bands of states in the composite system, corresponding to the elastic and inelastic channels, cross each other and the molecular-type states involved are neither too narrow nor too broad.

If the above intermediate structure occurs in the grazing partial waves, then similar resonance-like behaviour is to be expected even at forward angles in direct reactions, which are strongly surface peaked. Indeed, pronounced resonance structure has been observed by Paul *et al.* (1978) for the reaction $^{24}Mg(^{16}O,^{12}C)^{28}Si$ leading to the ground state and 1.77 MeV $2^+$ state of $^{28}Si$ at two forward angles, 0° and 11° (lab), for $23 \leq E_{c.m.} \leq 38$ MeV.

The purpose of our work was to extend these results for the $^{24}Mg(^{16}O,^{12}C)^{28}Si$ reaction to higher bombarding energies to determine if the strong resonance structure persists and in this way to study further the nature of these resonances.

## Experimental method and results

Thin targets (~ 100 μg/cm$^2$) were made by evaporating enriched $^{24}Mg$ (99.92 % $^{24}Mg$, 0.06 % $^{25}Mg$ and 0.02 % $^{26}Mg$) onto a thin carbon backing (~ 10 μg/cm$^2$). The targets were bombarded with a $^{16}O$ beam from the Australian National University 14 UD Pelletron accelerator. The reaction products were momentum analysed using an Enge split-pole magnetic spectrometer and were detected in a multi-electrode focal plane detector. The particles were identified using the ratio $(B\rho)^2/E$ where $B\rho$ is the magnetic rigidity and $E$ is the energy of the detected ions.

The data were recorded event by event onto magnetic tapes using a HP-2100 data acquisition system. The incident beam intensity was recorded using a beam current integrator. In addition, two solid-state detectors at 15° and 30° were employed as monitors.

The elastically scattered $^{16}O^{7+}$ ions and the most intense group of $^{12}C$ ions ($^{12}C^{6+}$) from the transfer reaction have a similar magnetic rigidity. Thus, the transfer reaction measurements at forward angles were carried out using a 20-cm slot in front of the focal plane detector to stop all groups of elastically scattered $^{16}O$ projectiles while allowing the $^{12}C^{6+}$ ions to enter the detector. In the energy and angular regions studied in the present experiment, the $^{12}C^{6+}$ group contains 82-98% of the total intensity of the $^{12}C$ charge state distribution.

Excitation functions for the reaction $^{24}Mg(^{16}O,^{12}C)^{28}Si$ were measured at 5° (lab) in the energy range 54-81 MeV (lab) with a horizontal acceptance of 4.5°(lab). Test measurements carried out by varying the reaction angle around 5° for various incident $^{16}O$ energies indicated that the second maximum in the ground-state angular distribution was located well within the acceptance angle. Figure 30.1 shows our results together with earlier data of Paul *et al.* (1978), which were taken at 0° and 11°(lab) for the ground-state transition at lower $^{16}O$ bombarding energies. It can be seen that the resonant structure becomes progressively weaker at higher energies. Nevertheless, three additional correlated maxima are evident near $E_{c.m.}$ = 37.5, 40.2 and 43.5 MeV.

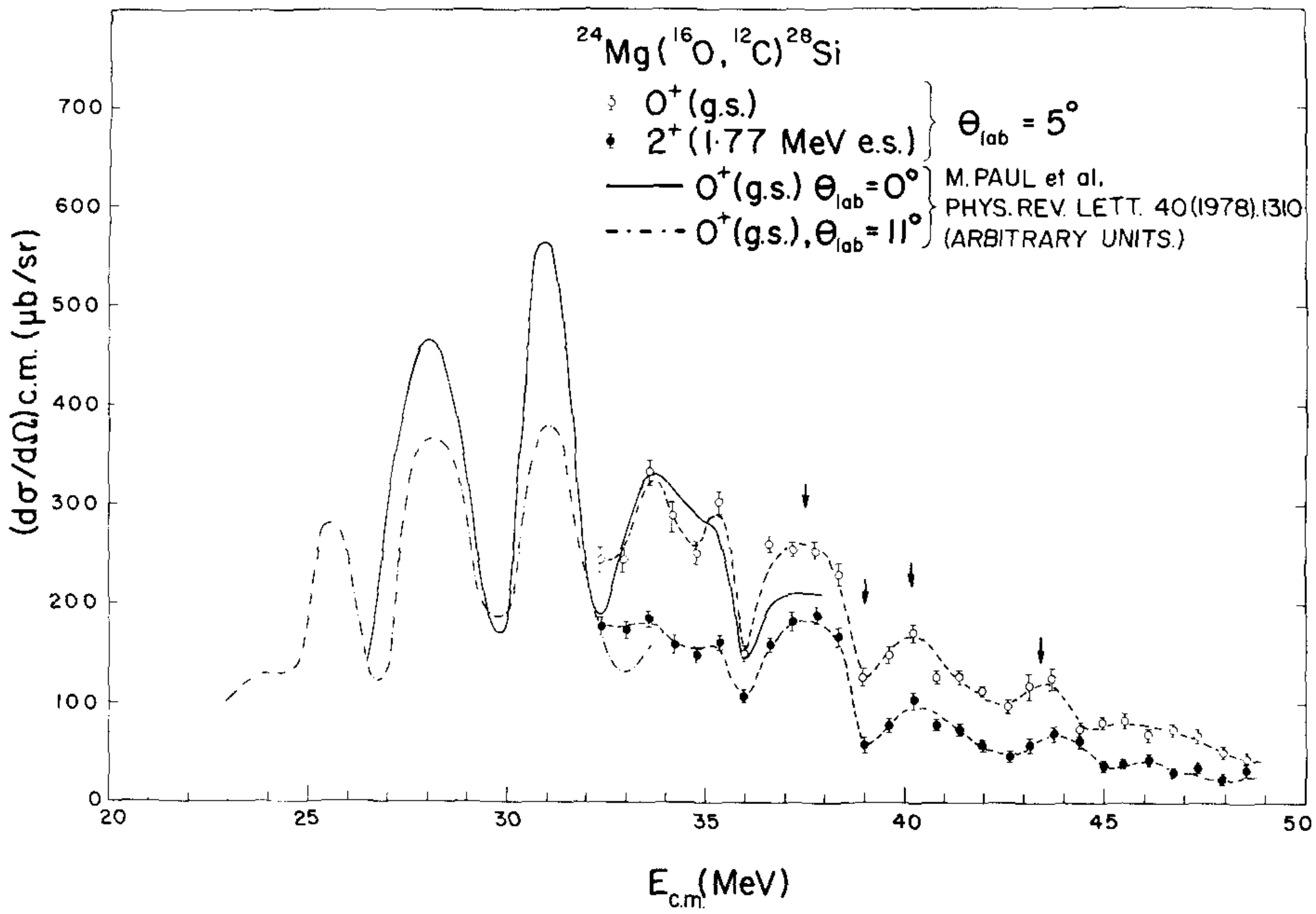

Figure 30.1. Excitation functions (dots) for the reaction $^{24}Mg(^{16}O,^{12}C)^{28}Si$ leading to the ground state and the first excited state in $^{28}Si$, measured in the range of energies 32.4-48.6 MeV (c.m.) at 5° (lab) are compared with the excitation functions measured at lower energies (Paul *et al.* 1978) at 0° (full line) and 11° (dash-dot line). Results of Paul *et al.* (1978) are expressed in arbitrary units. The dashed lines are used to guide the eye. Arrows indicate the energies at which angular distributions were measured. The vertical scale is for the elastic scattering. Data for the inelastic scattering have been displaced to show the structure. The cross sections at around 32 MeV (c.m.) are nearly the same for both excitation functions.

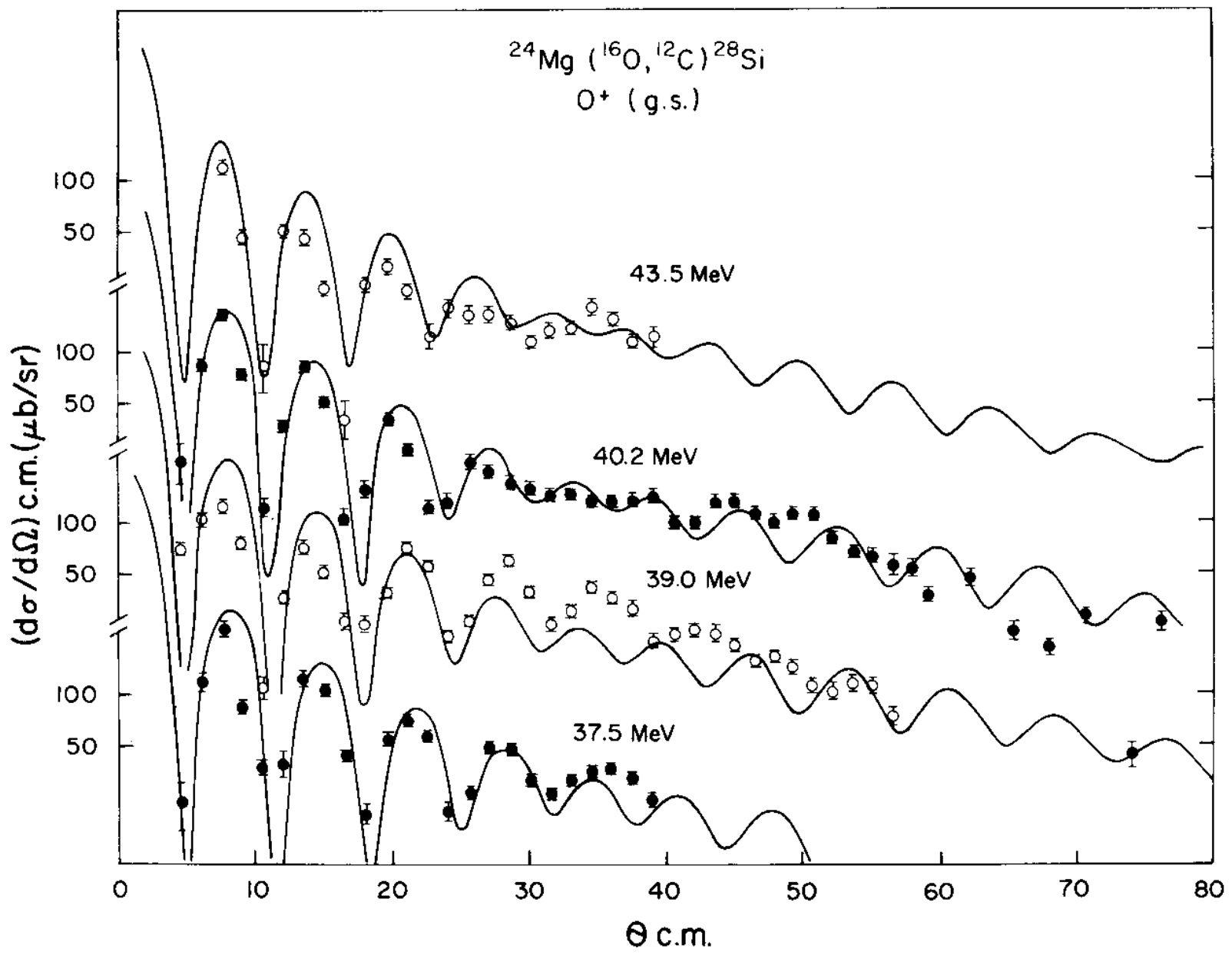

Figure 30.2. Angular distributions for the reaction $^{24}Mg(^{16}O,^{12}C)^{28}Si$ ($0^+$, g.s.) measured at the indicated c.m. energies (points) are compared with the distorted wave calculations (full lines).

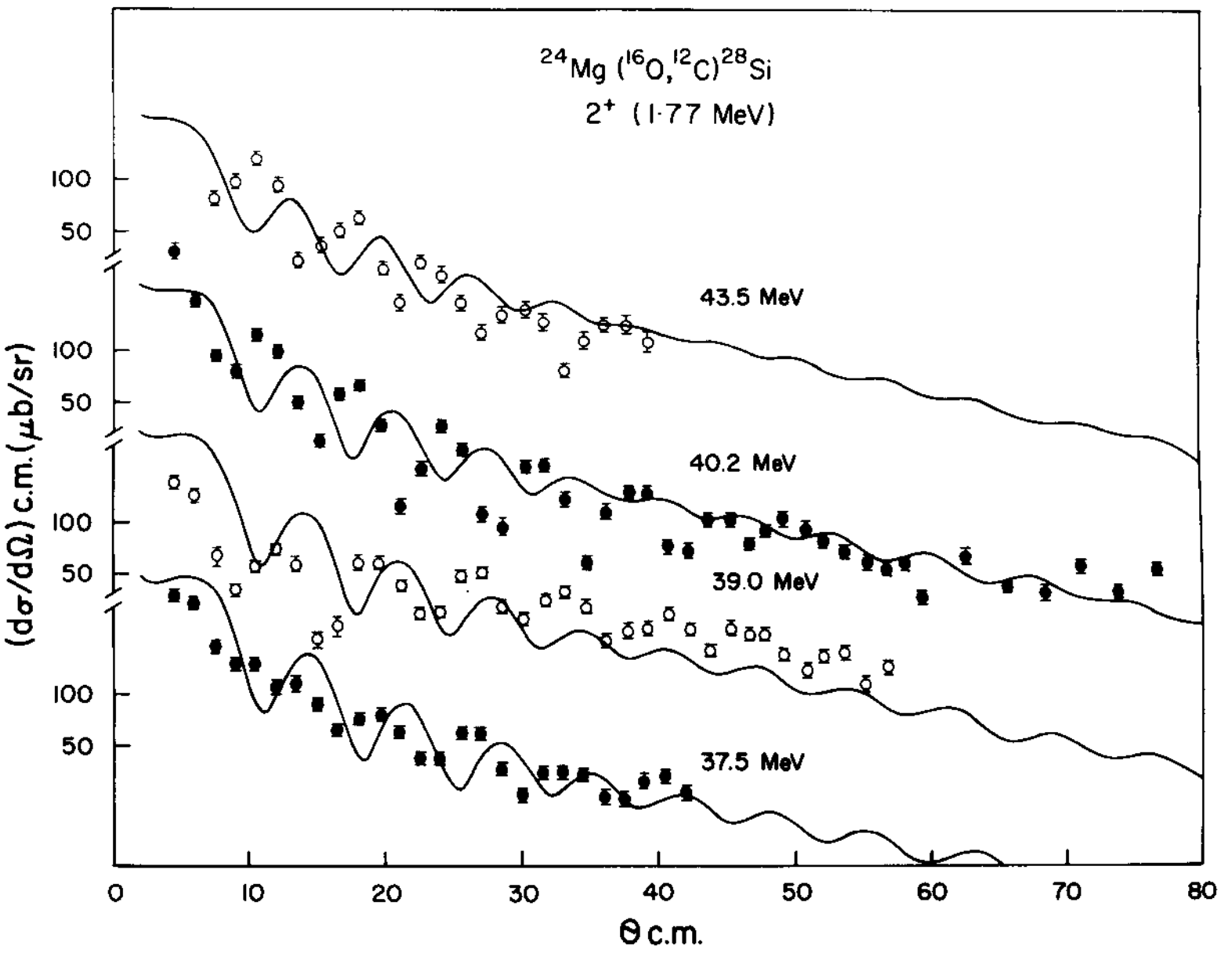

Figure 30.3. Angular distributions for the reaction $^{24}Mg(^{16}O,^{12}C)^{28}Si$ ($2^+$, 1.77 MeV) measured at the indicated c.m. energies (points) are compared with the distorted wave calculations (full lines).

Angular distributions for the reactions were measured at these three energies and at $E_{c.m.}$ = 39.0 MeV, corresponding to a trough between two maxima in the excitation functions. The energies at which angular distributions were measured are indicated by arrows in Figure 30.1. An acceptance angle of 1° (lab) was employed for these measurements. The results are shown in Figures 30.2 and 30.3 where diffraction patterns are clearly evident for both transitions. However, some irregularities are present. In particular, a prominent distortion of the simple oscillatory structure occurs for the ground state distribution at 40.2 MeV in the angular range of 25-40°(c.m.). This irregularity does not occur for the transition to the $2^+$ state.

The elastic scattering cross sections for the reaction $^{24}Mg(^{16}O,^{16}O)^{24}Mg$ were measured at $E_{c.m.}$ = 37.5 and 40.2 MeV with the whole detector exposed to the reaction products. The data are shown in Figure 30.4. It can be seen that while the 37.5 MeV measurements exhibit oscillating structure for $\theta_{c.m.}$ > 40°, the 40.2 MeV data are relatively structureless in this angular region.

## Analysis of the elastic scattering

As pointed out by Siemssen (1977), two schools of thought exist regarding the description of heavy-ion scattering in terms of the nuclear optical model. The first argues that as heavy ions are complex loosely bound particles, they must disintegrate upon impact so that only strongly absorbing potentials are acceptable. The second school of thought, however, simply attempts to determine empirically which features can or cannot be consistently described by the optical model. This second approach has typically led to a range of potentials, which are weakly absorbing for surface partial waves. Such "surface-transparent" potentials are often characterized by a smaller geometry for the absorption potential than the real potential i.e. $r_0' < r_0$ and $a_0' < a_0$.[2]

[2] Optical model parameters are as defined in Chapter 29.

Previously, elastic scattering for the $^{16}O$ + $^{24}Mg$ system has been studied for various energies ranging from the Coulomb barrier up to $E_{c.m.}$. ≈ 43 MeV. In general, analyses above the Coulomb barrier favour a moderately shallow real potential depth, which increases linearly with energy ($V$ ≈ 5 + 0.5$E_{c.m.}$ MeV), and an imaginary potential strength, which increases quadratically with energy.

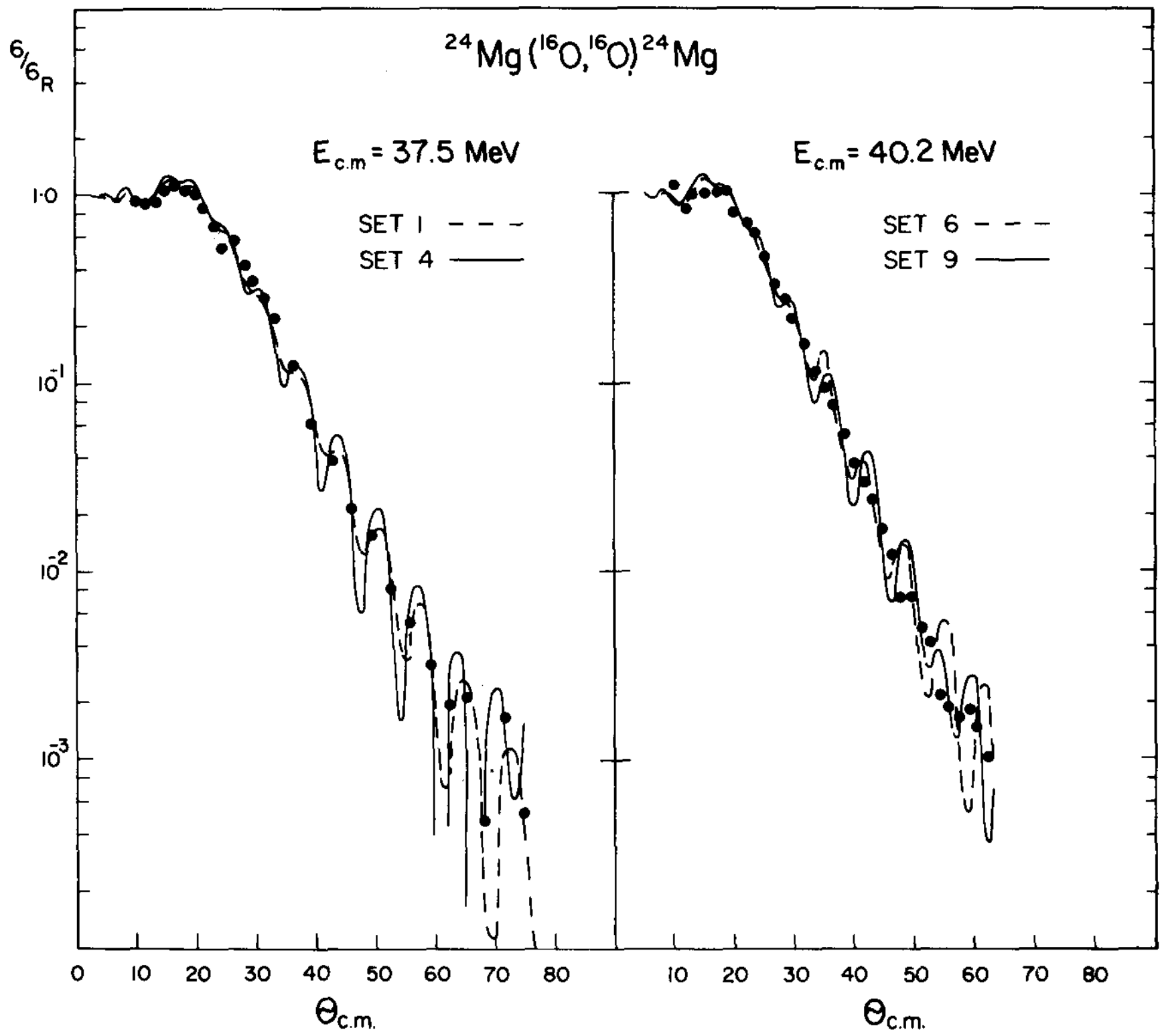

Figure 30.4. Angular distributions for the $^{24}Mg(^{16}O,^{16}O)^{24}Mg$ elastic scattering measured at two energies (dots) are compared with the optical-model calculations. Parameter sets are listed in Table 30.2.

These and similar optical-model analyses for neighbouring mass systems have led to parameters, which can be classified broadly into three groups:

(i) Potentials, which have similar real and imaginary radii and diffuseness (i.e. $r_0' \approx r_0$ and $a_0' \approx a_0$) and are strongly absorbing ($W$ is relatively large).

(ii) Potentials, which have similar real and imaginary potential geometries but have weak absorption strengths.

(iii) Potentials which have $r_0' < r_0$ and $a_0' < a_0$ , and moderate absorption strength.

Both potentials (ii) and (iii) give rise to surface transparency for the grazing partial waves. Examples of these three types of potentials is given in Table 30.1.

In our work, the sets $A$, $E$, and $F$ of Table 30.1 were taken as representative potentials for the three types of interactions and were used as starting sets in parameter searches to fit the $^{16}O$ + $^{24}Mg$ elastic scattering angular distributions at $E_{c.m.}$ = 37.5 and 40.2 MeV. These data were analysed using the optical-model parameter search code SOPHIE (Delic 1975) to obtain parameter sets for the distorted wave analysis of the $^{24}Mg(^{16}O,^{16}O)^{24}Mg$ transfer reaction. The searches were *unconstrained* and had led to potentials shown as sets 1-3 and 6 and 8 in Table 30.2.

Table 30.1

Classification of heavy-ion optical-model potentials [a])

| Set | $V$ (MeV) | $r_0$ (fm) | $a_0$ (fm) | $W$ (MeV) | $r_0'$ (fm) | $a_0'$ (fm) | $r_c$ (fm) |
|---|---|---|---|---|---|---|---|
| | | | | *Group (i)* | | | |
| *A* | 100 | 1.22 | 0.50 | 30 | 1.25 | 0.40 | 1.22 |
| *B* | 100 | 1.14 | 0.58 | 20 | 1.20 | 0.60 | 1.35 |
| | | | | *Group (ii)* | | | |
| *C* | $5+0.5E_{c.m.}$ | 1.37 | 0.53 | $0.008E^2_{c.m.}$ | 1.37 | 0.36 | 1.37 |
| *D* | $7.5+0.5E_{c.m.}$ | 1.31 | 0.49 | $0.4+0.15E_{c.m.}$ | 1.31 | 0.30 | 1.31 |
| *E* | 32.8 | 1.25 | 0.50 | 8.11 | 1.25 | 0.50 | 1.25 |
| | | | | *Group (iii)* | | | |
| *F* | 35.286 | 1.307 | 0.4926 | 19.67 | 1.242 | 0.204 | 1.307 |
| *G* | 17.0 | 1.26 | 0.42 | 12.48 | 1.20 | 0.25 | 1.26 |

[a]) Potentials are as defined in Chapter 29. Set 1: Braun-Muzinger *et al.* 1973; Set 2: Tserruya *et al.* 1975; Set 3: Siwek-Wilczynska, Wilczynski, and Christensen 1974: Set 4: Siemssen 1971; Set 5: Ball 1975; Set 6: Lemaire *et al.* 1974a; Set 7: Siemssen *et al.* 1969.

Table 30.2

Optical-model potentials for the $^{24}Mg(^{16}O,^{12}C)^{28}Si$ reaction

| *Set* | $E_{c.m.}$ (MeV) | $V$ (MeV) | $r_0$ (fm) | $a_0$ (fm) | $W$ (MeV) | $r_0'$ (fm) | $a_0'$ (fm) | $r_c$ (fm) | $\chi^2$ |
|---|---|---|---|---|---|---|---|---|---|
| 1 | 37.5 | 85.320 | 1.245 | 0.525 | 38.099 | 1.309 | 0.322 | 1.22 | 196.7 |
| 2 | 37.5 | 32.831 | 1.319 | 0.526 | 8.223 | 1.184 | 0.556 | 1.25 | 212.8 |
| 3 | 37.5 | 39.744 | 1.274 | 0.498 | 22.929 | 1.291 | 0.279 | 1.307 | 422.7 |
| 4 | 37.5 | 35.471 | 1.313 | 0.515 | 8.930 | 1.160 | 0.221 | 1.25 | 274.4 |
| 5 | 37.5 | 15.799 | 1.302 | 0.637 | 11.790 | 1.308 | 0.697 | 1.26 | 119.7 |
| 6 | 40.2 | 82.725 | 1.242 | 0.532 | 43.119 | 1.316 | 0.311 | 1.22 | 1063.8 |
| 7 | 40.2 | 33.389 | 1.349 | 0.497 | 8.085 | 1.218 | 0.600 | 1.25 | 1056.4 |
| 8 | 40.2 | 46.872 | 1.321 | 0.518 | 35.019 | 1.351 | 0.304 | 1.307 | 1033.4 |
| 9 | 40.2 | 35.534 | 1.334 | 0.493 | 8.668 | 1.156 | 0.225 | 1.25 | 1132.4 |
| 10 | 40.2 | 13.221 | 1.442 | 0.518 | 24.470 | 1.342 | 0.442 | 1.26 | 831.8 |
| 11 | 37.5–43.5 | 32.0 | 1.30 | 0.50 | 15.0 | 1.20 | 0.20 | 1.25 | |

Sets 1,2, and 3 were obtained using sets *A*, *E*, and *F* of Table 30.1 as starting values. The same applies to sets 6, 7, and 8. Sets 4 and 9 were obtained using sets 2 and 7, respectively, as the starting values. Sets 5 and 10 were obtained using set G of Table 30.1 as starting values. Set 11 gives the best description of the angular distributions for the $^{24}Mg(^{16}O,^{12}C)^{28}Si$ reaction as shown in Figures 30.2 and 30.3.

In attempting to fit the 40.2 MeV data with a shallow imaginary well, the automatic search converged on an unphysically large value for the imaginary diffuseness parameter $a_0' = 1.12$ fm. The set 7 shown in Table 30.2 was obtained by requiring that $a_0' \leq 0.6$ fm.

The 37.5 MeV data are described equally well by either a strongly absorbing potential (set 1) or a weakly absorbing surface-transparent interaction (set 2). Both sets of parameters generate the necessary oscillatory structure for $\theta_{c,m.}$ > 40° (Figure 30.4).

The fits obtained for the 40.2 MeV data are distinctly poorer in that all the potentials predict oscillatory structure in the differential cross section for $\theta_{c,m.} > 40°$, which unlike the 37.5 MeV measurements, is not observed at the higher energy.

From other data, such as few nucleon transfer reactions, there is an evidence (Lemaire *et al.* 1974b) of a preference for surface transparency of the optical potential with $a_0' < a_0$. This was taken into account in the present analysis when using sets 2 and 7 (Table 30.2) as starting values for further searches in which the value of $a_0'$, was constrained to be ≤ 0.3 fm. The resulting potentials are given in Table 30.2 as sets 4 and 9. The predictions for these sets together with the corresponding strongly absorbing potentials (sets 1 and 6) are shown in Figure 30.4. It can be seen that both types of the $^{16}O$ + $^{24}Mg$ interaction describe the 37.5 MeV data well while neither reproduces the structureless 40.2 MeV angular distribution.

In some analyses, very shallow real potentials have been found. This possibility was investigated by employing the shallow potential set $G$ of Table 30.1 as a starting point for a search in which the real strength $V$ was constrained to have a value < 20 MeV. The resulting potentials from these searches are presented as sets 5 and 10 in Table 30.2. These interactions give excellent fits to the data, particularly to the structureless angular distribution at 40.2 MeV. However, the shallow potential, which describes the 37.5 MeV data, is significantly different from the potential, which fits the 40.2 MeV measurements. In particular, the absorption strengths are 11.79 and 24.47 MeV, respectively. Such a rapid increase in the imaginary potential is much larger than one expects even if a quadratic energy dependence (Siwek-Wilczynska, Wilczynski, and Christensen 1974) is assumed.

## The distorted wave analysis

Exact finite-range (EFR) distorted wave calculations were carried out for the four-nucleon transfer reaction $^{24}Mg(^{16}O,^{12}C)^{28}Si$ leading to the ground and 1.77 MeV states of $^{28}Si$ using the code LOLA (DeVries 1972). In the calculations, an $\alpha$-cluster transfer from $^{16}O$ to $^{24}Mg$ nuclei was assumed. The corresponding bound-state wave functions in the projectile and residual nuclei were calculated in a Woods-Saxon potential with radius $1.25A^{1/3}$ fm and diffuseness 0.65 fm, the depths being adjusted to obtain the experimental $\alpha$-particle separation energies in $^{16}O$ and $^{28}Si$.

In our analysis, an attempt was made firstly to describe the reaction data for the ground-state transition at 40.2 MeV using the optical-model parameters obtained from the elastic scattering analysis. For simplicity, the same parameters were employed for both the entrance and exit channels. Such calculations gave a poor description of the measurements. However, a better fit to the angular distribution was obtained by adjusting the parameters of set 9 (Table 30.2) although similar attempts based upon parameter sets 6 and 8 were unsuccessful.

Figure 30.2 shows the best fits to the experimental angular distributions. The corresponding set of parameters is listed as set 11 in Table 30.2. The theoretical cross sections were normalized using factors listed in Table 30.3 to bring them in line with the experimental data. These numbers, which correspond to a product of spectroscopic factors (see Chapter 29), are expected to be energy independent if the distorted wave theory is valid. As can be seen from Table 30.3, the normalization factors are not constant.

The significant energy dependence of the spectroscopic factors, as implied by such normalization factors, has been previously observed for this reaction at lower energies (Peng *et al.* 1976). While it is possible that the observed energy dependence could be at least partially removed by allowing the optical-model parameters to be smoothly energy dependent it is unlikely that the over-all shapes of the angular distributions could be reproduced at the same time. Moreover, the excitation functions calculated using such a smoothly energy-dependent potential would not exhibit rapid fluctuations of the type displayed in Figure 30.1.

Figure 30.2 shows that the calculated curves describe qualitatively the strong oscillatory character at very forward angles and the smoother angular distribution for $\theta_{c.m.} > 25°$. However, parameter set 11 gives a poor description of the elastic scattering data at 40.2 MeV so there is a consistency problem.

Table 30.3
Normalization factors (DeVries 1973) for the $^{24}$Mg($^{16}$O, $^{12}$C)$^{28}$Si reaction

| $E_x(^{28}\mathrm{Si})$ (MeV) | 37.5 MeV | 39.0 MeV | 40.2 MeV | 43.5 MeV |
|---|---|---|---|---|
| 0.00 | 0.52 | 0.45 | 0.32 | 0.33 |
| 1.77 | 0.68 | 0.45 | 0.33 | 0.32 |

Figure 30.3 shows the distorted wave results for the $^{24}$Mg($^{16}$O,$^{12}$C)$^{28}$Si reaction leading to the 1.77 MeV $2_1^+$ state in $^{28}$Si. These curves were also calculated using the parameter set 11. As can be seen, in this case, theory and experiment are out of phase. In order to remove this gross discrepancy, large changes in the optical-model parameters, away from those that describe the ground-state transition, would be necessary. In view of this problem and the over-all inconsistency of the optical model and distorted wave analyses, it was not considered worthwhile to attach much significance to the normalization factors of Table 30.3 or to attempt any determination of the alpha-nucleus spectroscopic factors.

## Discussion

The angular distributions for the $^{24}$Mg($^{16}$O,$^{12}$C)$^{28}$Si reaction to the ground state of $^{28}$Si at $E_{c.m.}$ = 37.5, 40.2 and 43.5 MeV, corresponding to peaks in the excitation function for $\theta = 5°$(lab), exhibit strong oscillatory structure at forward angles ($\theta_{c.m.} < 25°$), which can be described reasonably well using the distorted wave formalism. However, it is easy to see that they can be also well described by the squares of Legendre polynomials, $P_J^2(\cos\theta)$, with $J$ = 27, 29 and 31, respectively (see Figure 30.5).

The well-pronounced resonances in the excitation functions of Figure 30.1 at $E_{c.m.}$ = 28.2, 31.2 and 34.2 MeV have been similarly described (Paul *et al.* 1978) using $J$ = 21, 23 and 25, respectively. If these peaks arise from the enhancement of a single partial wave, one can use such Legendre polynomial comparisons to assign spin $J$ to the corresponding maxima in the excitation functions. It is interesting to note, however, that the positions of forward-angle maxima in the angular distribution corresponding to the ***trough*** in the excitation function at $E_{c.m.}$ = 39.0 MeV can also be well reproduced by a $P_J^2(\cos\theta)$ distribution.

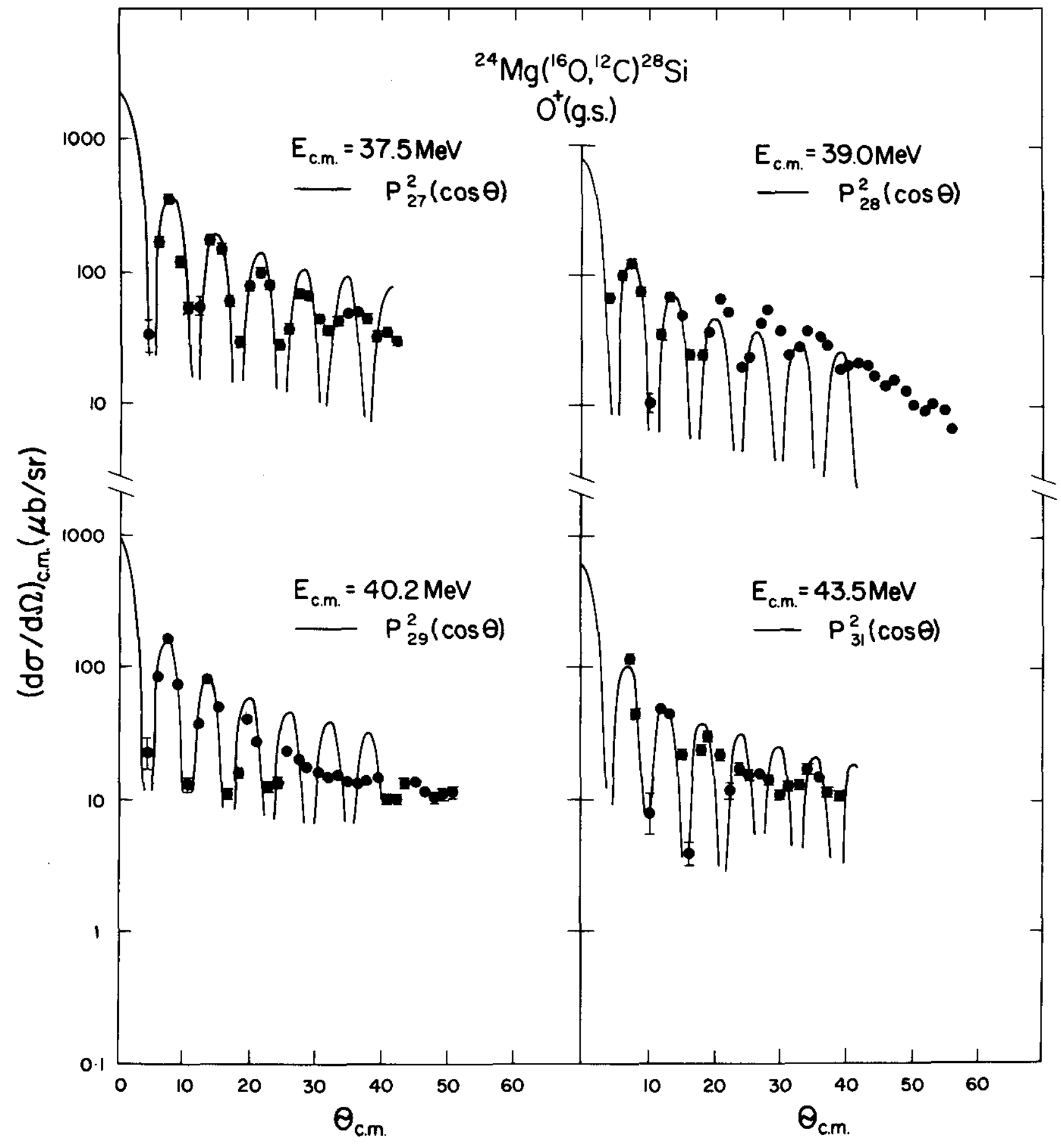

Figure 30.5. Legendre polynomial fits to the forward-angle cross sections of the experimental angular distributions measured at the indicated c.m. energies.

The above values of $J$ have been deduced assuming that only a single partial wave dominates the reaction. However, the deviations in Figure 30.5 between the $P_J^2(\cos\theta)$ fits and the data suggests that other partial waves, from either a non-resonant reaction mechanism or overlapping resonances, also contribute. Indeed, for the energy range 24 ≤ $E_{c.m.}$ ≤ 40 MeV, a study of the transfer reaction at backward angles (Paul *et al.* 1980) indicates that this is the case. It appears that the Legendre polynomial fitting procedure allows, at best, a determination of the dominant $J$ - value to $\pm 1\hbar$, and may be less precise in those cases where there is significant partial wave interference.

For the energy region 37.5 - 43.5 MeV, the values of $J$ extracted from our experiment by optimal $P_J^2(\cos\theta)$ are shown in Figure 30.5 and are about two units greater than the grazing angular momenta (defined as the $J$ - value of that partial wave which is 50% absorbed) for both the entrance and exit channels for the potential 11 of Table 30.2. The spin assignments for the resonant states at lower energies (Paul *et al.* 1978) have shown a similar correlation to the grazing angular momentum.

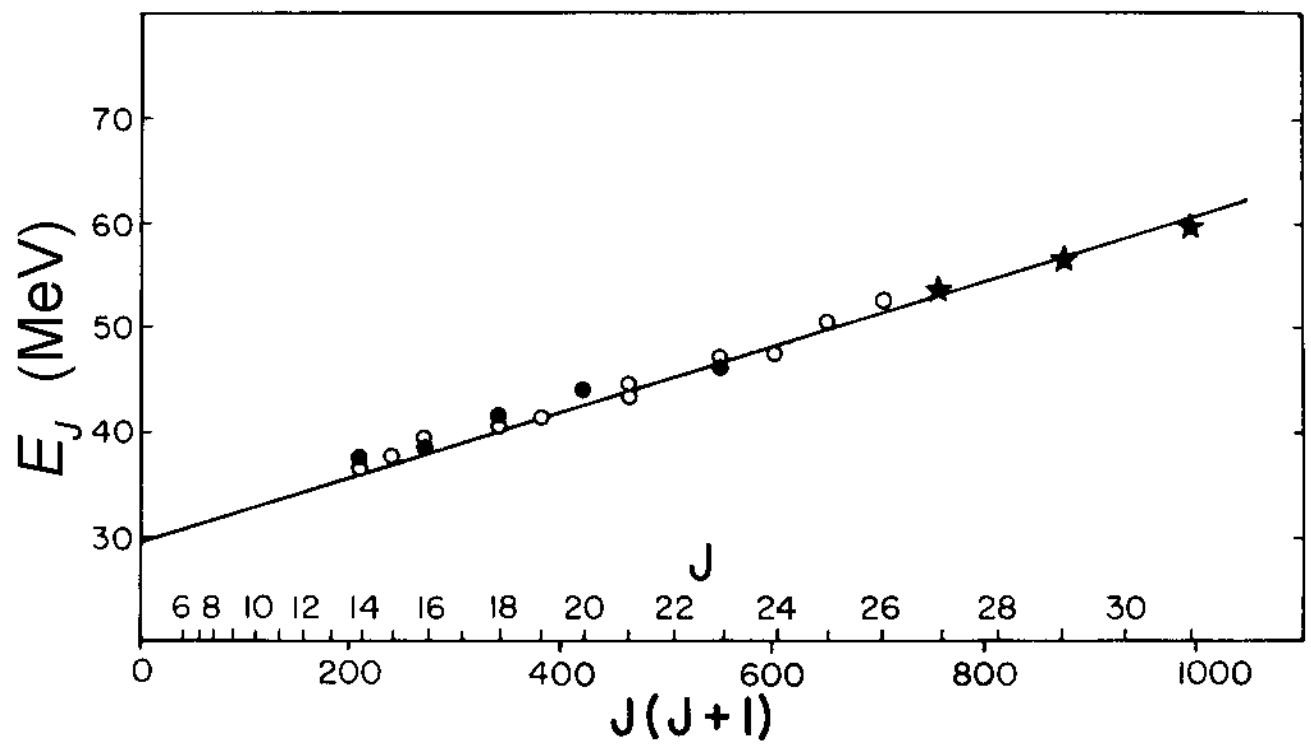

Figure 30.6. A plot of energies $E_J$ for observed resonances against $J(J+1)$. The $J$ assignments made in our work are indicated by stars, those obtained from previous studies are given as open and closed circles (see the text). The best fit (straight line) has a band head of 29,8 MeV and an effective moment of inertia of 7,08 $\times 10^{-42}$ MeV·s$^2$.

A surface-transparent interaction of the type favoured by the distorted wave analysis discussed earlier allows the two colliding ions to retain their individual structure during a grazing collision and gives rise to the possibility of orbiting molecular states in which two ions rotate about the centre of mass of the composite system. To examine this possibility, we show in Figure 30.6 a compilation of $J$ values for observed resonances. The compilation is based on the measurements by Clover *et al.* (1979) for the $^{24}$Mg($^{16}$O,$^{16}$O)$^{24}$Mg elastic scattering (closed circles), and by Paul *et al.* (1978, 1980), Peng *et al.* (1976), Clover *et al.* (1979), and Lee *et al.* (1979) for the $^{24}$Mg($^{16}$O,$^{12}$C)$^{28}$Si and $^{28}$Si($^{12}$C,$^{16}$O)$^{24}$Mg reactions (open circles). Stars denote the three new resonances we have observed.

It can be seen that all the resonances lie close to a straight line suggesting a rotational-like band, which can be described by the following equation:

$$E_J = \frac{\hbar^2}{2\Im} J(J+1) + E_0$$

where $\Im$ is the effective moment of inertia of the system and $E_0$ is the band head.

The gradient of the fitted straight line to the points corresponds to an effective moment of inertia of 7.08 x $10^{-42}$ MeV · s$^2$ and the projected $J$ = 0 resonance lies at 29.8 MeV. These values are in good agreement with the moment of inertia of 7.05 x 10~$^{42}$ MeV · s$^2$ and the energy 31.6 MeV for the lowest resonance, $J$ = 0, when $^{24}$Mg and $^{16}$O nuclei are just touching each other with no relative rotational energy as predicted by Cindro and Počanić (1980). The series of quasi-bound and virtual states (or shape resonances) of a molecular-like potential between two heavy ions are expected to form such a rotational-like band.

## Summary and Conclusions

We have measured excitation functions for the $^{24}$Mg($^{16}$O,$^{12}$C)$^{28}$Si (g.s., $2_1^+$) reaction in the energy range of $E_{c.m.}$ = 32.4 – 48.6 MeV at $\theta$ = 5$^0$ (lab). The chosen angle corresponds to a maximum in the angular distributions for the $^{24}$Mg($^{16}$O,$^{12}$C)$^{28}$Si reaction leading to the ground state in $^{28}$Si. We have observed three correlated maxima in the excitation functions corresponding to $E_{c.m.}$ = 37.5, 40.2, and 43.5 MeV. We have measured angular distributions for the $^{24}$Mg($^{16}$O,$^{12}$C)$^{28}$Si (g.s. and $2_1^+$ excited state) at these resonance energies and at $E_{c.m.}$ = 39.0 MeV which corresponds to the trough between the first and the second observed maxima in the excitation function. In addition, we have measured angular distributions for the elastic scattering, $^{16}O$ + $^{24}Mg$ at $E_{c.m.}$ = 37.5 and 40.2 MeV.

The optical model analysis of the elastic scattering data yielded the required parameters, which we used in our distorted wave analysis of the transfer reaction data. We have found that while the theory reproduces reasonably well the measured distributions for the $^{24}Mg(^{16}O,^{12}C)^{28}Si$ reaction leading to the ground state, the calculated and experimental distributions for the first excited state, $2_1^+$, are out of phase. However, the theoretical distributions follow closely the angular trend of the measured cross sections.

We have then analysed the angular distributions for the $^{24}Mg(^{16}O,^{12}C)^{28}Si$ (g.s.) using the square of the Legendre polynomials, $P_J^2(\cos\theta)$, and have found that the three observed resonances located at energies $E_{c.m..}$ = 37.5, 40.2, and 43.5 MeV can be assigned spins $J$ = 27, 29 and 31, respectively. A compilation of earlier data combined with our new measurements has shown that the resonance structure observed in the reaction $^{24}Mg(^{16}O,^{12}C)^{28}Si$ can be described as nuclear molecular excitations of binary nuclear systems made of interacting heavy ions either in the incident or outgoing channels.

# 31

# The Interaction of $^7$Li with $^{28}$Si and $^{40}$Ca Nuclei

***Key features:***

1. Interaction of projectiles with $4 < A < 12$ is difficult to describe theoretically. New data were needed to help to investigate and hopefully understand this problem.
2. We have measured angular distributions for the elastic and inelastic scattering from $^{28}$Si and $^{40}$Ca. These two target nuclei have been selected because of the differences in their collective excitations: rotational and vibrational, respectively.
3. We have carried out standard and double folding optical model analysis of the elastic scattering. Both give similar and satisfactory description of the experimental angular distributions. We have found that the absorptive potential dominates the interaction of $^7$Li with both $^{28}$Si and $^{40}$Ca nuclei.
4. Both the elastic and inelastic scattering data for the $^7$Li + $^{40}$Ca system can be reproduced well using coupled-channels formalism. However, for the $^7$Li + $^{28}$Si interaction, the theory reproduces the trends of the differential cross sections but cannot fit simultaneously the elastic or inelastic distributions unless different deformation parameters are used. The problem might be associated with the mutual excitations of projectile and target nuclei and thus involving different collective structures, which are not accounted for in the available theoretical formalism.

**Abstract:** Elastic and inelastic scattering of 45 MeV $^7$Li from $^{28}$Si and $^{40}$Ca have been measured and analysed. The inelastic scattering distributions corresponded to the excitation of the 1.78 MeV state in $^{28}$Si and to the 3.73 and 3.90 states in $^{40}$Ca. Double folding model calculations using a realistic effective nucleon-nucleon interaction similar to that used for the $^9$Be + $^{28}$Si and $^9$Be + $^{40}$Ca scattering have been carried out for the elastic angular distributions. The real potential had to be renormalized to yield agreement with the measured cross sections. Coupled channels calculations using a Woods-Saxon potential were performed in an effort to describe both the elastic and inelastic angular distributions. The extracted deformation parameters are in reasonable agreement with those obtained from light and heavier ion scattering from the same target nuclei. The effect of strong excitation of the 0.48 MeV state in $^7$Li and of mutual excitation of target and projectile is considered in a qualitative manner.

## Introduction

The region between light ($A < 4$) and heavy ($A > 12$) projectiles is challenging because of the problems encountered in theoretical interpretations of experimental data. In particular, the double folding model (Satchler 1976; Satchler and Love 1979) with a realistic nucleon-nucleon interaction, which has been successful in describing the elastic $\alpha$ and heavy ion scattering, is not as successful in describing the elastic scattering of $^6$Li, $^7$Li, and $^9$Be from the same target nuclei unless the real double folding potential is reduced by a factor of about 0.4 to 0.6.

Furthermore, the optical model using either the calculated double folding potential or one of Woods-Saxon shape, indicates that the effective interaction distance for these intermediate nuclei is larger than for $\alpha$ and heavy-ions (Balzer *at al.* 1977; Ungricht *et al.* 1979). Perhaps more important is a transition from the dominance of the real potential to the dominance of the imaginary potential as one proceeds through this intermediate region between $A = 4$ and $A = 12$ projectiles. In particular, the interaction of $^6$Li with $^{28}$Si suggests weak absorption (DeVries *et al.* 1977) but a study of elastic, inelastic, and fusion interactions of $^9$Be with $A = 20$ - $60$ targets indicates the dominance of a strong absorption (Zisman *et al.* 1980).

In order to investigate this transition in the absorption strength, we have measured the elastic and inelastic scattering cross sections for the excitation of the 1.78 MeV state in $^{28}Si$, the 3.73 and 3.90 MeV states in $^{40}Ca$, and the 0.48 MeV state in $^{7}Li$ using $^{7}Li$ projectiles at a bombarding energy of 45 MeV. In addition, we have measured the cross section for mutual excitation of the 0.48 MeV state in $^{7}Li$ and the 1.78 MeV state in $^{28}Si$. The experimental details are given below.

The measured elastic cross sections were fitted using both the standard optical model as well as the double folding procedure (Satchler and Love 1979), which was successful in describing the elastic scattering of $\alpha$ particles, and heavy ions ($A > 12$). The inelastic cross sections were fitted using the coupled channels formalism. Deformation lengths and deformation parameters are extracted from the fitted cross sections and compared with those obtained from measurements using other projectiles (Hendrie 1973; Thompson and Eck 1977). The effect of the absorptive potential on the calculated cross sections is investigated and compared to the dominance it exhibits in the case of $^{9}Be$ scattering from the same target nuclei (Zisman *et al.* 1980).

## Experimental procedure

The $^{7}Li$ beam was extracted from a sputter source in the form of $^{7}Li^{-}$ and was injected into the ANU 14D Pelletron accelerator. The targets consisted of ~200 $\mu/cm^2$ self-supporting $SiO_2$ (greater than 99.5% enriched in $^{28}Si$) or ~160 $\mu/cm^2$ Ca (greater than 99.8% enriched in $^{40}Ca$) evaporated onto 10 $\mu/cm^2$ C foils. The scattered $^{7}Li$ ions were detected using the Enge split pole magnetic spectrograph and a focal plane detector, which was operated in the light-ion mode (Ophel and Johnston 1978). The states in $^{28}Si$, $^{40}Ca$, and $^{7}Li$ were clearly resolved for up to about 7 MeV excitation energy by gating on the $^{7}Li$ mass. A typical spectrum for $^{7}Li$ + $^{28}Si$ is shown in Figure 31.1

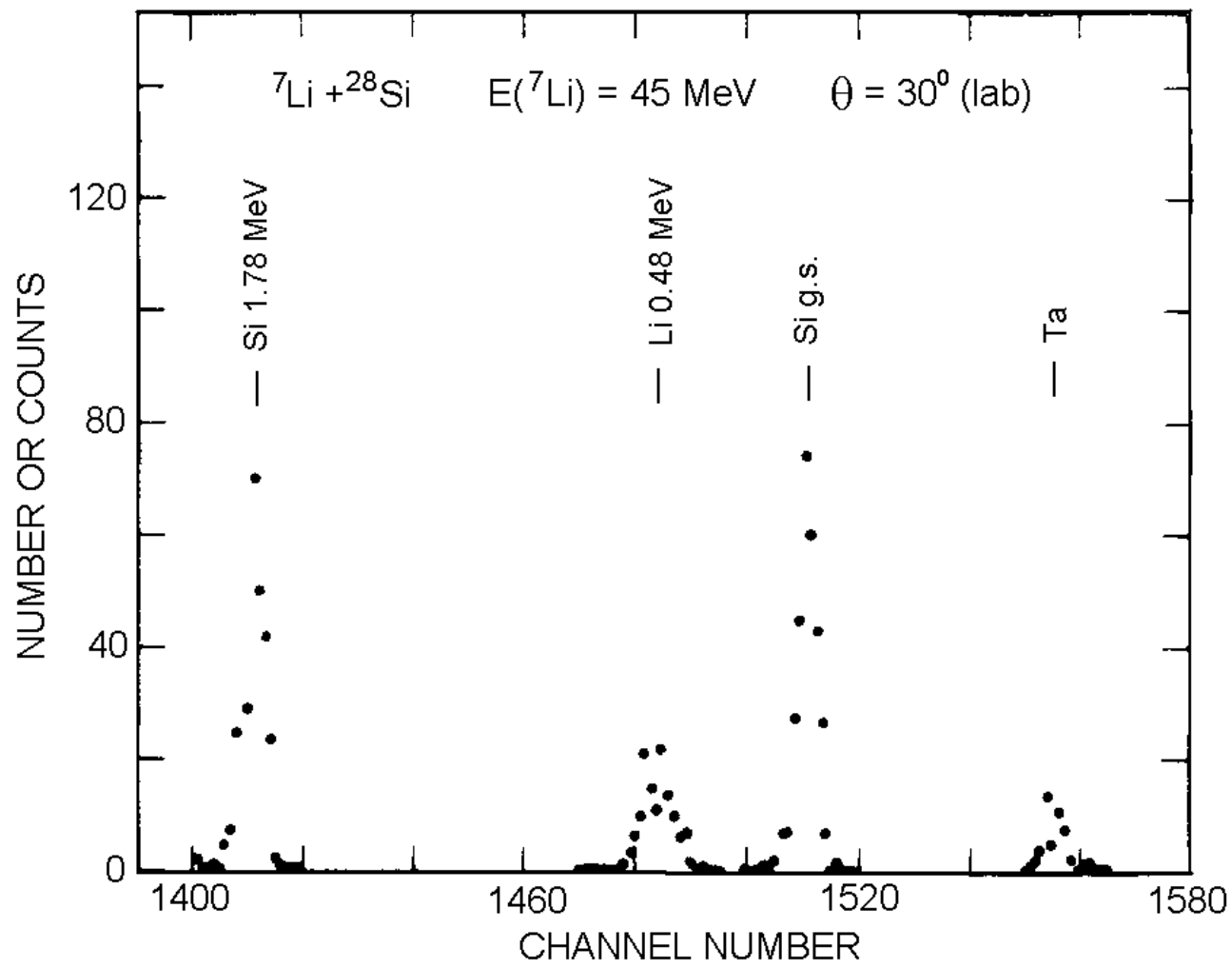

Figure 31.1. Typical position spectrum gated on the $^{7}Li$ mass for the $^{7}Li$ + $^{28}Si$ scattering at $\theta = 30°$ (lab) obtained using the Enge split pole spectrometer and the focal plane detector.

The relative normalization was obtained by using a monitor detector placed at a laboratory angle of 15°. The absolute normalization was carried out by normalizing the measured cross sections to Rutherford scattering at a bombarding energy of 25 MeV and a laboratory scattering angle of 7.5°. The absolute normalization is accurate to about 10%. The measured angular distributions are shown in Figures 31.6 – 31.9.

The elastic and inelastic angular distributions fall off quickly with increasing angle and show strong diffraction-like patterns. In the measured angular range (10° < $\theta_{c.m.}$ < 80°), there is no indication of any levelling off or rising of the cross section with increasing angle as has been observed for $^{12}C$ + $^{28}Si$ and $^{16}0$ + $^{28}Si$ (Braun-Muzinger *et al.* 1977; Clover *et al.* 1978).

## Theoretical analysis

### *The optical-model analysis*

Prior to carrying out the coupled channels calculations, the elastic scattering angular distributions were fitted using a standard optical model Woods-Saxon potential defined in Chapter 29.

The initial parameters were chosen to be those of Cramer *et al.* (1976), which were used to fit $^{l6}O$ + $^{28}Si$ scattering over a wide energy range. Only $V$ and $W$ were varied and the best-fit parameters are listed in Table 31.1. The calculated best-fit cross sections are shown in Figure 31.2 and 31.3 for $^{7}Li$ + $^{28}Si$ and $^{7}Li$ + $^{40}Ca$, respectively.

Table 31.1

Best-fit optical-model parameters

| Standard Woods-Saxon Potential | | | | | | | | | | | |
|---|---|---|---|---|---|---|---|---|---|---|---|
| Target | $V$ MeV | $r_0$ fm | $a_0$ fm | $W$ MeV | $r_0'$ fm | $a_0'$ fm | $L_{1/2}$ | $D_{1/2}$ fm | $-ReU(D_{1/2})$ MeV | $-ImU(D_{1/2})$ MeV | $Im/Re$ |
| $^{28}Si$ | 7.17 | 1.35 | 0.618 | 17.69 | 1.20 | 0.552 | 20.55 | 5.95 | 5.27 | 9.943 | 1.89 |
| $^{40}Ca$ | 7.96 | 1.35 | 0.618 | 19.26 | 1.20 | 0.552 | 22.74 | 6.84 | 4.69 | 8.29 | 1.77 |

| Double Folding Potential | | | | | | | | | | |
|---|---|---|---|---|---|---|---|---|---|---|
| Target | $N$ | $W$ MeV | $r_0'$ fm | $a_0'$ fm | $\rho_R$ mb | $L_{1/2}$ | $D_{1/2}$ fm | $-ReU(D_{1/2})$ MeV | $-ImU(D_{1/2})$ MeV | $Im/Re$ |
| $^{28}Si$ | 0.571 | 14.4 | 1.188 | 0.786 | 1688 | 21.5 | 8.0 | 0.68 | 0.91 | 1.33 |
| $^{40}Ca$ | 0.586 | 13.6 | 1.241 | 0.795 | 1877 | 24.3 | 8.7 | 0.58 | 0.93 | 1.60 |

The ratio of the real to the imaginary potential was calculated at $r = D_{l/2}$ where $D_{l/2}$ is the distance of closest approach for the Rutherford orbit for which $l = L_{1/2}$. $L_{1/2}$ is the value of $l$ at which the transmission coefficient $T$ = 1/2. The $D_{l/2}$, $L_{1/2}$, and the ratio of $W/V$ evaluated at $D_{l/2}$ are listed in Table 31.1.

The next step was to refine the calculations by using a double folding model. The double folding procedure used in our study is the same as that used by Satchler (1976, 1979) and Satchler and Love (1979) to describe $^{6}Li$, $^{12,13}C$, $^{14,15}N$, and $^{16,17,18}O$ scattering from $^{28}Si$, $^{40}Ca$, and other nuclei in the same mass region.

The real component of the interaction potential may be written as:

$$U_F(R) = \int \rho(r_1)\rho(r_2)V_{NN}(\vec{R} - \vec{r}_1 + \vec{r}_2)d\vec{r}_1 d\vec{r}_2$$

where $R$ is the distance between the centres of the nuclei, $\rho_1$ and $\rho_2$ are the nucleon distributions in the interacting nuclei, and $V_{NN}(\vec{r})$ is the effective nucleon-nucleon interaction.

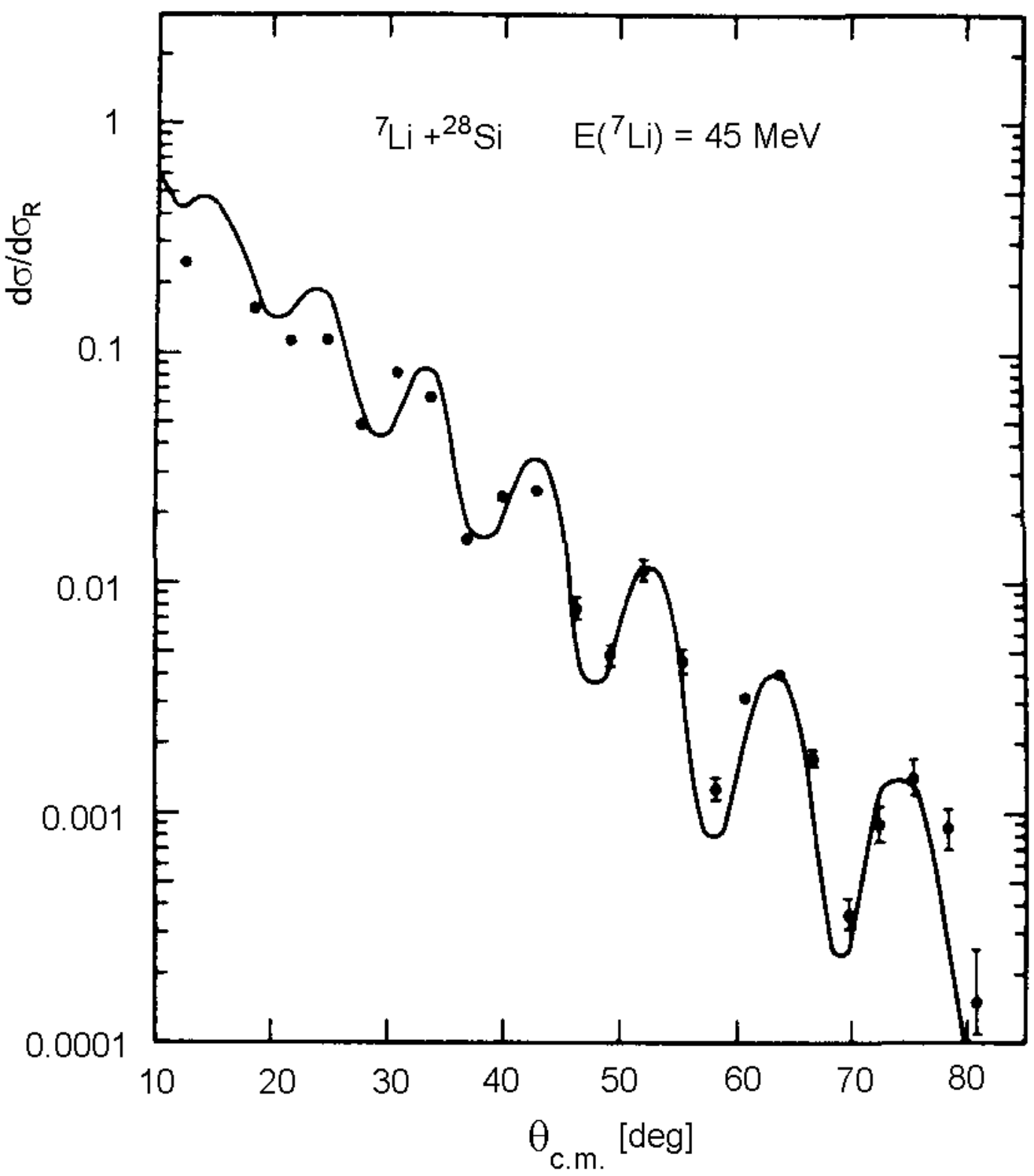

Figure 31.2. The calculated angular distribution for the $^7$Li + $^{28}$Si elastic scattering at $E$($^7$Li) = 45 MeV using the Woods-Saxon optical model potential and parameters of Table 31.1 are compared with our experimental data.

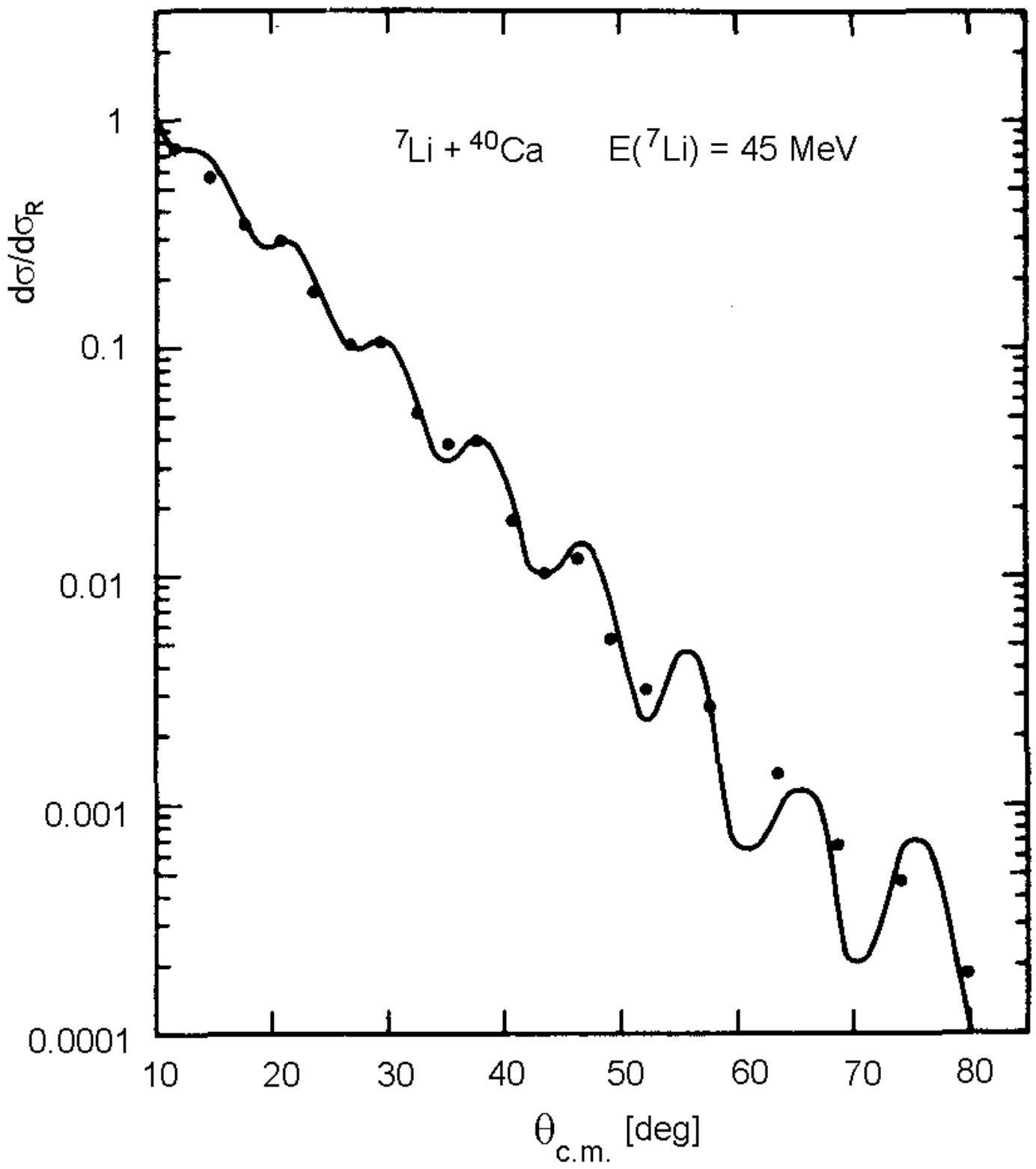

Figure 31.3. The calculated angular distribution for $^7$Li + $^{40}$Ca elastic scattering at $E$($^7$Li) = 45 MeV using the Woods-Saxon optical model potential and parameters of Table 31.1 are compared with our experimental data.

The most widely used forms for the $V_{NN}(\vec{r})$ interaction are Reid or Paris (Brandan and Satchler 1997; Lacombe *et al.* 1981; Reid 1968; Satchler and Love 1979). In our calculations, we have used the soft-core Reid potential. The method for constructing the density distributions is described by Satchler and Love (1979). The real folded potential $U_F(R)$ thus constructed is multiplied by a normalizing factor $N$, where $N$ is varied to obtain the best fit.

A Woods-Saxon imaginary term with $R' = r_0'(A_p^{1/3} + A_t^{1/3})$, where $p$ stands for projectile and $t$ for target, is added to give the total potential. In our work, $r_0'$ was fixed at the value of 1.3 fm. The normalization parameter $N$ for the real potential, the imaginary potential strength $W$, and the diffuseness of the imaginary well $a_0'$, were then varied to obtain the best fit. The best fit parameters are given in Table 31.1 and the best folding model fit calculations for the measured elastic scattering angular distributions are shown in Figures 31.4 and 31.5. As can be seen from Table 31.1, the $L_{l/2}$ values are slightly different for each potential for both the $^7$Li + $^{28}$Si and $^7$Li + $^{40}$Ca scattering. The ratios of $W/V$ indicate the dominance of strong absorption at $r$ = $D_{l/2}$ in both cases.

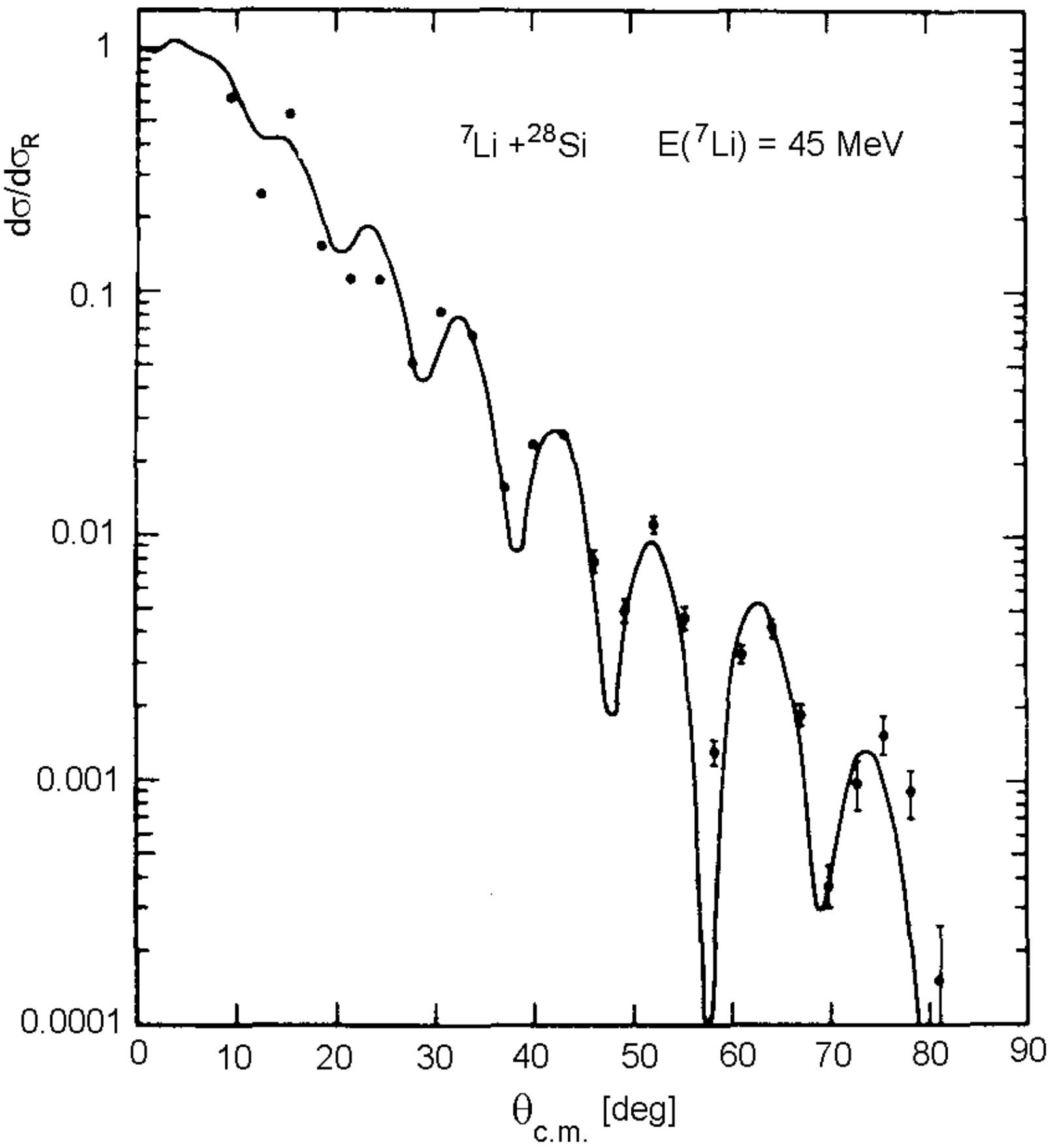

Figure 31.4. Folding model calculation (solid curve) for the $^7$Li + $^{28}$Si elastic scattering at $E$($^7$Li) = 45 MeV using the realistic nucleon-nucleon potential. See the text for details. Experimental cross sections are shown as closed circles.

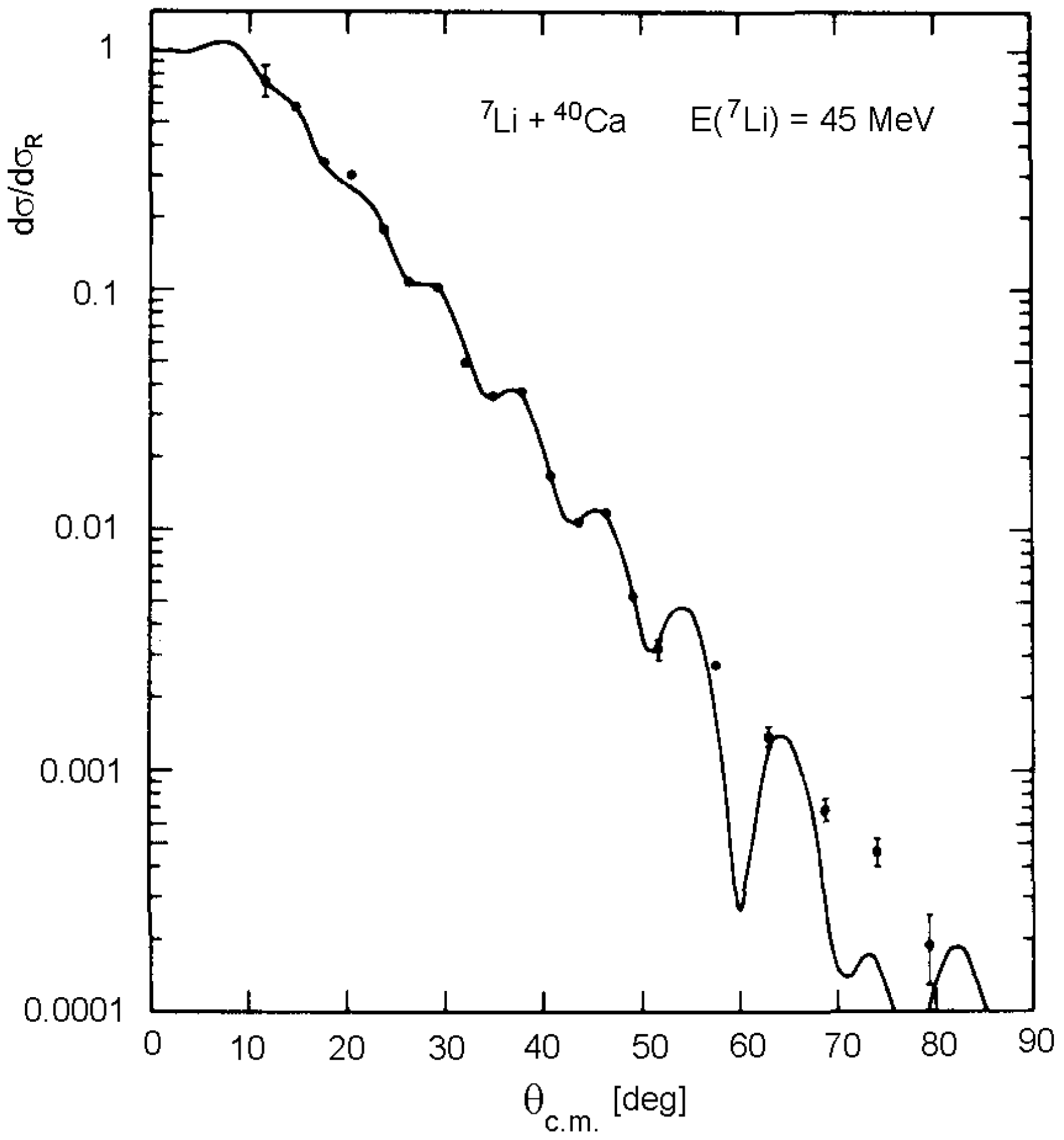

Figure 31.5. Folding model calculation (solid curve) for the $^7$Li + $^{40}$Ca elastic scattering at $E$($^7$Li) = 45 MeV using the realistic nucleon-nucleon potential. See the text for details. Experimental cross sections are shown as closed circles.

In order to fit the cross sections, it was necessary to renormalise the potential by a factor of $N \approx 0.58$. This result is similar to that obtained for $^6$Li and $^9$Be scattering from the same target nuclei. It has been shown recently that the necessity for this renormalization is eliminated for the case of $^7$Li and $^9$Be scattering if quadrupole effects (and therefore reorientation) are included in the data analysis (Hinzdo, Kemper, and Szymakowski 1981). A difficulty with this approach is that it does not eliminate the necessity for renormalization of the potential for $^6$Li scattering (Satchler and Love 1979).

### *Coupled-channels calculations*

The coupled channels calculations were performed using the computer code ECIS79 (Raynal 1972, 1981), which I have adapted earlier to run on an ANU computer. The calculations were carried out using 80 partial waves and assuming that the low excited states of $^{28}$Si can be described using a rotational model while the lowest states of $^{40}$Ca can be described by a vibrational model. Radial integrations were carried out to 40 fm to account properly for Coulomb excitation. The optical model parameters obtained from fitting the elastic cross section were utilized except that $W$ was reduced by about 10% and the quadrupole deformation parameter $\beta_2$ was varied to obtain the optimal fit.

The fits produced by this technique were not totally satisfactory. The fit for $\beta_2$ = -0.l5 and $W$ = 11.69 is shown in Figure 31.6. For these parameters, the fit to the elastic cross section is good but for the inelastic cross section is too low in magnitude and exhibits too much structure. Increasing the absolute magnitude of the quadrupole deformation parameter to $\beta_2$ = -0.25 improves the fit to the inelastic cross section but causes a deterioration in the elastic fit due to a reduction in the calculated diffraction structure, especially at back angles. This can be remedied slightly by decreasing $W$ but this procedure exaggerates the diffraction oscillations. The fit obtained for $\beta_2$ = -0.25 and $W$ = 15.00 MeV is shown in Figure 31.7.

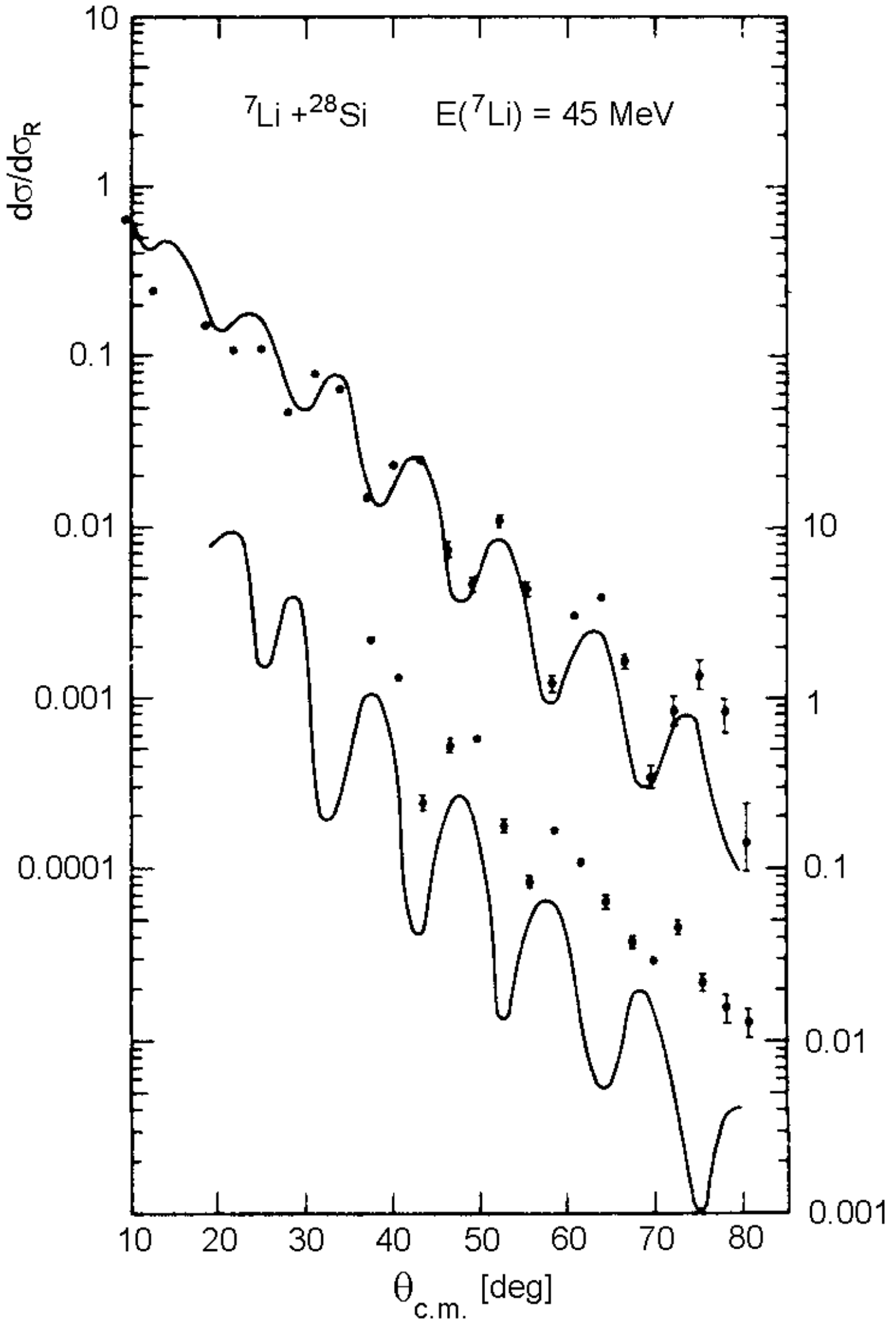

Figure 31.6. Angular distributions for the elastic and inelastic scattering (to the 1.78 MeV state in $^{28}$Si) of 45 MeV $^7$Li projectiles. The elastic scattering cross section is shown as a ratio to the Rutherford by the upper set of closed circles. The inelastic cross section is given in absolute units and is shown by the lower set of closed circles. The solid curves are the cross sections calculated using the coupled channels theory assuming the deformation parameter $\beta_2$ = -0.15 and $W$ = 11.69 MeV. See the text for details.

Inelastic scattering to the 0.48 MeV state in $^7$Li and the mutual excitation of the 0.48 and 1.78 MeV states have not been included in the calculation. These cross sections are shown in Figure 31.8 and comparison with Figure 31.6 or 31.7 shows that these neglected cross sections are similar in magnitude to the cross section for exciting the 1.78 MeV state and need to be included explicitly in the coupled channels calculation if better results are to be expected. Unfortunately, this kind of calculations has been inaccessible within the available theoretical framework.

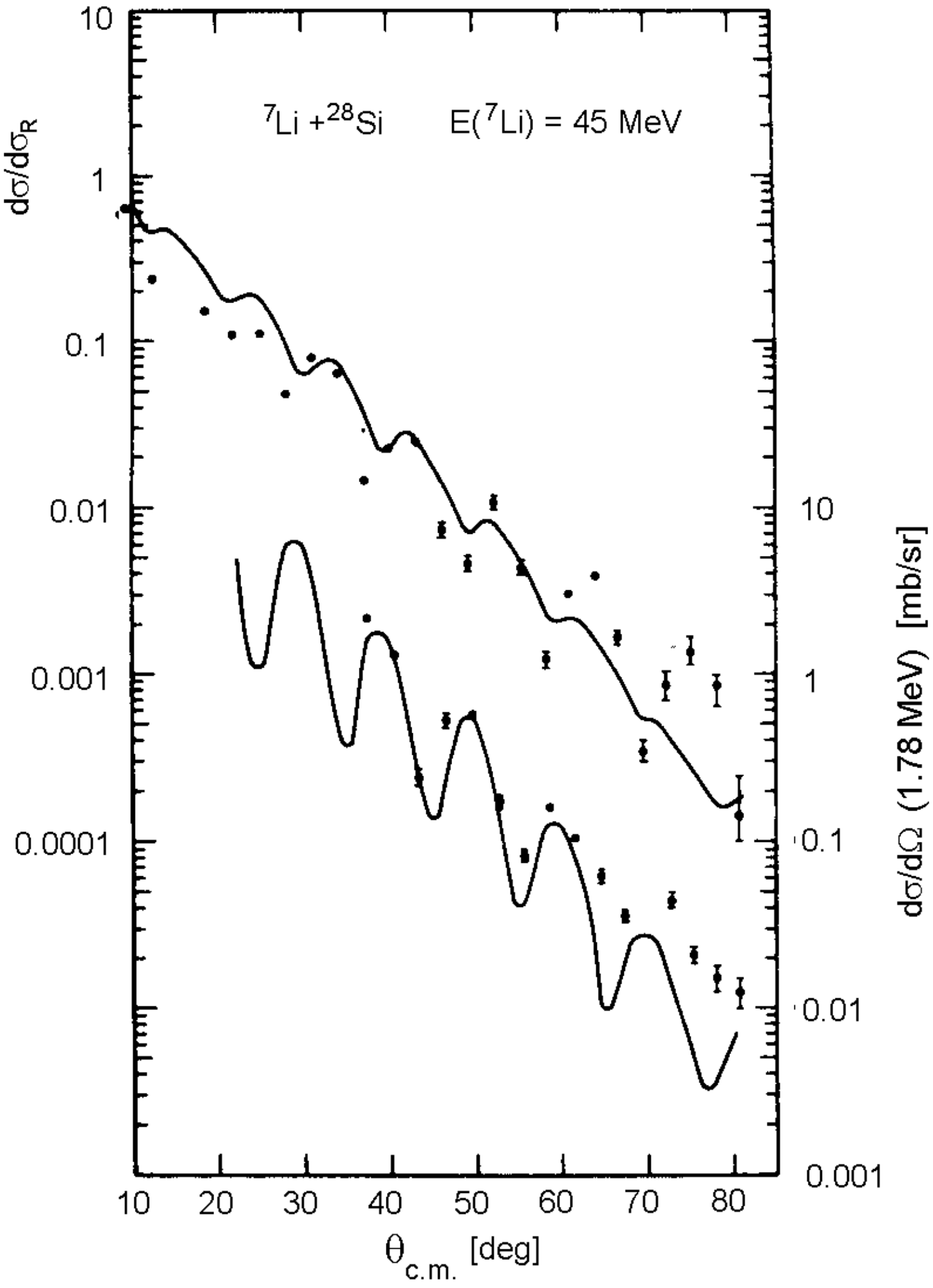

Figure 31. 7. Experimental angular distributions are as displayed in Figure 31.6. The calculated curves are for the deformation parameter $\beta_2$ = -0.25 and $W$ = 15 MeV. See the Caption to Figure 31.6 and the text for details.

The $^7$Li + $^{40}$Ca elastic and inelastic cross sections were fitted using a second order vibrational model to describe the low-lying $2^+$ and $3^-$ states of $^{40}$Ca. The 3.73 ($3^-$) state is assumed to be the excitation of one octupole phonon and the 3.90 ($2^+$) state is assumed to be the excitation of a single quadrupole phonon.

By reducing $W$ and adjusting $\beta_2$ and $\beta_3$ the obtained fits are shown in Figure 31.9. For this case $W$ = 18.00 MeV, $\beta_2$ = 0.06, and $\beta_3$ = 0.15. As can be seen, the fits to both elastic and inelastic scattering angular distributions are quite satisfactory. For $^7$Li + $^{40}$Ca the excitation of the 0.48 MeV state in $^7$Li is also relatively strong but its exclusion from the calculation does not appear to affect strongly the final fits. The values of $\beta_2$ and $\beta_3$ can be compared with those obtained from $\alpha$ particle scattering from $^{40}$Ca at $E_\alpha$ = 29 MeV by calculating the deformation distance $\delta_l$ defined as $\delta_l = \beta_l R$, where the potential radii from α + $^{40}$Ca and $^7$Li + $^{40}$Ca scattering are 4.76 and 7.20 fm, respectively. The deformation distances are compared in Table 31.2. The agreement for $\beta_2 R$ is fair but there is a significant difference for $\beta_3 R$.

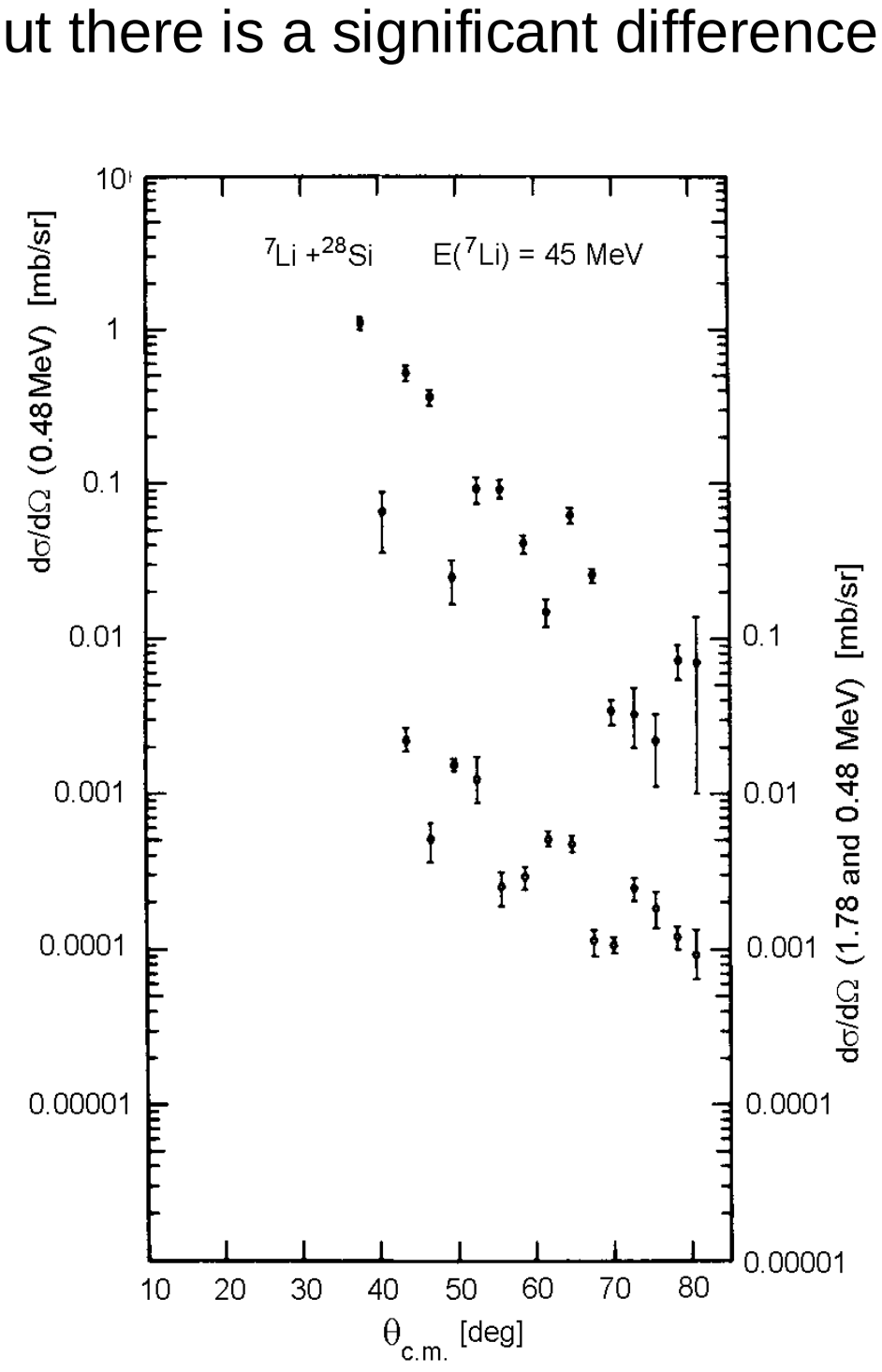

Figure 31.8. Measured cross sections for excitation of the 0.48 MeV state of $^7$Li (upper set of points) and for mutual excitation of the 0.48 MeV state of $^7$Li and the 1.78 MeV state of $^{28}$Si (lower set of points) by $^7$Li + $^{28}$Si scattering at $E$($^7$Li) = 45 MeV.

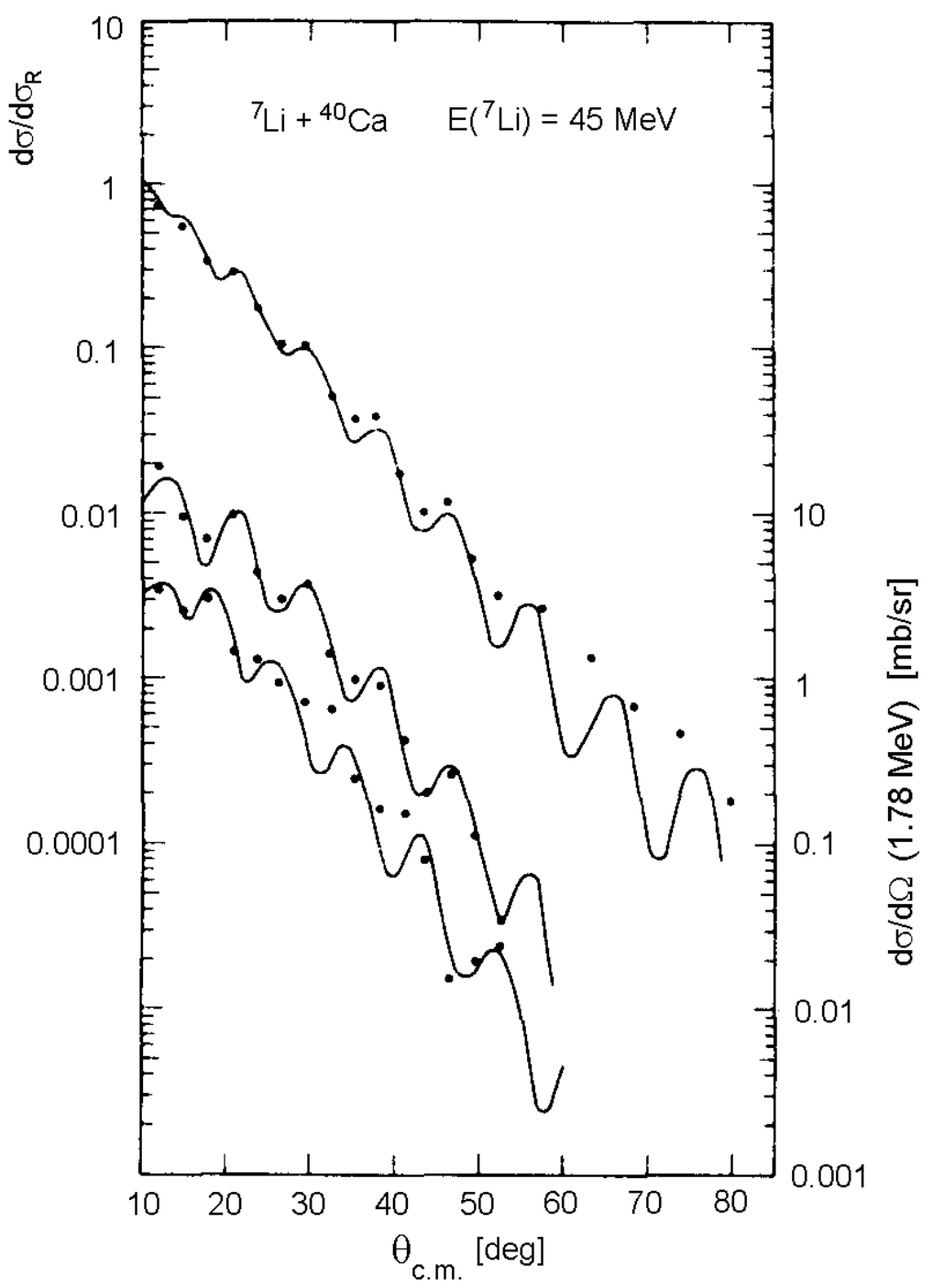

Figure 31.9. Angular distributions for the elastic scattering and inelastic scattering (to the 3.73 and 3.90 MeV states in $^{40}$Ca) of 45 MeV $^7$Li projectiles. Elastic scattering shown as a ratio to the Rutherford by the upper set of points. Inelastic cross sections for the excitation of the 3.73 and 3.90 MeV states are given in absolute units and are indicated by the middle and bottom sets of dots, respectively. The solid curves are cross sections calculated using the coupled channels formalism and the deformation parameters $\beta_2$ = 0.06, and $\beta_3$ = 0.15.

Table 31.2
Deformation lengths for $^{40}$Ca

| Experiment | $\beta_2 R$ | $\beta_3 R$ |
|---|---|---|
| $\alpha + ^{40}$Ca | 0.57 | 1.19 |
| $^7$Li + $^{40}$Ca | 0.43 | 1.80 |

## Summary and conclusions

Angular distributions for the elastic and inelastic scattering cross sections for $^7$Li scattering from $^{28}$Si and $^{40}$Ca targets have been measured and analysed. The low-lying excited states of both target and projectile are strongly excited and in the case of $^7$Li + $^{28}$Si scattering mutual excitations of both target and projectile are significant.

Double-folding model calculations of the elastic scattering cross sections yield potentials, which must be renormalized by a factor of $\approx 0.6$ in order to fit adequately the measured cross sections.

Coupled channels calculations for $^7$Li + $^{28}$Si reproduce the trends of the experimental data. However, the theory requires different deformation parameters to fit either the elastic or inelastic distributions. In the case of $^7$Li + $^{40}$Ca scattering, the cross sections for the ground state, the first excited state ($3^-$) at 3.73 MeV, and the second excited state ($2^+$) at 3.90 MeV are well described using the coupled channels formalism. Comparison of the deformation lengths obtained here and those obtained from $\alpha$ +$^{40}$Ca scattering are in fair agreement for $\beta_2$ but not for $\beta_3$.

Considering the fairly good agreement between the theory and experiment for the $^7$Li + $^{40}$Ca interaction the problems encountered with the theoretical interpretation of the $^7$Li + $^{28}$Si scattering is hard to explain. In both cases, the same projectile is used, so the argument based on the weakly bound nature of $^7$Li appear unconvincing. The problem might be associated with the mutual excitation of $^7$Li and the target nucleus, which are not accounted for by the available coupled channels formalism. Even though the projectile is the same for the $^7$Li + $^{28}$Si and $^7$Li + $^{40}$Ca, the target nucleus is different. Excited states in $^{28}$Si can be described using rotational model, whereas $^{40}$Ca is a spherical nucleus, which exhibits vibrational excitations. The $2^+$ state in $^{28}$Si has a sizable quadrupole moment of around +($16 \pm 3$) $e$fm$^2$ (Stone 2001). The mutual excitations of different collective structures might have a substantial effect on the coupling to the observed transitions and thus influence the character of the measured angular distributions.

# 32

# Triaxial Structures in $^{24}$Mg

***Key features:***

1. High-resolution measurements were carried out for the elastic and inelastic scattering of $^{16}$O from $^{24}$Mg nuclei using 72.5 MeV (lab) $^{16}$O beam. The doublet of states, 4.12/4.24 MeV, essential in a study of triaxial deformation of $^{24}$Mg has been resolved for the first time for heavy ion projectiles.
2. Energy level schemes for $^{24}$Mg have been calculated assuming a rigid asymmetric ($\gamma \neq 0$) rotor. We have used both the standard and extended Davydov-Filoppov model.
3. Inclusion of hexadecapole deformation resulted in a better description of reduced transition rates.
4. Coupled channels analysis of experimental angular distributions were carried out using three types of the interaction potential containing the quadrupole and hexadecapole deformations for both the real and imaginary components. Nearly prefect fits were obtained for a surface-transparent potential with a moderate real depth.
5. Deformation distances for the quadrupole and hexadecapole deformations compare well with the lengths obtained for light projectiles.
6. The effect of hexacontatetrapole component in the interaction potential has been investigated and found to be negligible.

**Abstract:** Angular distributions for the elastic and inelastic scattering of $^{16}$O from $^{24}$Mg, exciting the $2_1^+$, 1.37 MeV, $4_1^+$, 4.12 MeV and $2_2^+$, 4.24 MeV states, have been measured at $E_{c.m.}$ = 43.5 MeV, the energy at which resonance-like structure has been observed previously in the related $^{24}$Mg($^{16}$O,$^{12}$C)$^{28}$Si reaction. The $4_1^+$, $2_2^+$ doublet has been resolved for the first time for heavy-ion projectiles. The data have been well described by coupled-channels calculations within the framework of the Davydov-Filippov asymmetric rotor model for the low-lying states of $^{24}$Mg, which has been extended to include a symmetric hexadecapole shape component. The optical model potential for the $^{16}$O + $^{24}$Mg interaction was found to have a moderate real well depth and surface transparency. The shape parameters for the nuclear potential were determined to be $\beta_2^{(N)}$ = 0.25, $\gamma$ = 22° and $\beta_4^{(N)}$ = -0.065 and the corresponding deformation distances are in good agreement with earlier light-ion results. The inclusion of a negative symmetric hexadecapole component leads to an improved description of the reduced transition rates. The triaxial structure of $^{24}$Mg is discussed.

## Introduction

The possible existence of triaxial structures in some 2s-1d shell nuclei has been discussed since the early 1960's. Of specific interest in this work was the nucleus $^{24}$Mg, which was considered during that early period (Batchelor *et al.* 1960; Cohen and Cookson 1962) within the framework of the Davydov-Filippov (1958) model.

In this model the nucleus is considered to be a rigid ellipsoid with quadrupole deformation rotating about its centre of mass. The conclusions of that early work were summarized by Robinson and Bent (1968) who pointed out that the Davydov-Filippov model, when evaluated for $\gamma$ = 22°, which gives the best description of the energy separations for the low-lying states of $^{24}$Mg, was unable to predict many of the branching ratios and magnitudes of the reduced transition rates.

Measurements of lifetimes and branching ratios for $^{24}$Mg (Branford, McGough, and Wright 1975) are in closer agreement with the expectations of a Davydov-Filippov model with $\gamma$ = 21.5° for the intra-band transitions although some of the cross-band transitions are still poorly described. Branford, McGough, and Wright (1975) have pointed out that both sets of transitions are better described with $\gamma$ = 14°, a value which unfortunately places the $2_2^+$ state at 10.8 MeV excitation energy. It is also not clear whether one should associate the $4_2^+$ model state with the level observed at 6.01 MeV or with that at 8.44 MeV, although energy considerations favour the latter.

The calculations of Aspelund (1982) using the rotation-vibration model of Faessler and Greiner (1962) indicate non-negligible band-mixing for the 4$^+$ states but give the level at 8.44 MeV as the third member of the $K$ = 2 band. The existence of two 4$^+$ levels between 6 and 9 MeV excitation energy (in addition to several other states) contrary to the rigid asymmetric rotor model indicates that this model may not provide a good description of these higher 4$^+$ states in $^{24}$Mg.

A revived interest in the use of an asymmetric rotor model for $^{24}$Mg is motivated by two considerations. First, the increased sophistication of coupled-channels analyses of elastic and inelastic scattering experiments has prompted extensive use of such approaches to determine nuclear shape parameters of low-lying levels of light nuclei. For $^{24}$Mg, Hartree-Fock and other calculations (Abgrall *et al.* 1969; Grammaticos 1975; Kurath 1972) indicate triaxial deformations for the ground (*K* = 0) and *K* = 2 rotational bands, and analyses of proton (Eenmaa *et al.* 1974; Lombard, Escudié and Soyeur 1978; Lovas *et al.* 1977) and $\alpha$ - particle (Kokame et al. 1964; Tamura 1965; van den Borg, Harakeh, and Nilsson 1979) inelastic scattering from $^{24}$Mg have been carried out within an asymmetric rotor model framework.

The second reason arises from the observation of considerable resonance-like structure in excitation functions for heavy-ion reactions, particularly those involving the nuclei $^{12}$C, $^{16}$O, $^{24}$Mg and $^{28}$Si. It has been proposed, in some cases at least, that the observed resonances, which are too narrow to be readily described in terms of simple shape resonances associated with quasi-bound states in a molecular-like potential between the two ions, should be interpreted as intermediate structure which arises in a "doorway-state" model in which either the initial or final channel couples to another degree of freedom of the system (Abe, Kondo, and Matsuse 1980; Fink, Scheid, and Greiner 1972; Imanishi 1968, 1969; Kondo, Matsuse, and Abe 1978; Michaud and Vogt 1969, 1972; Scheid, Greiner, and Lemmer 1970). In particular, it has been suggested (Nurzynski *et al.* 1981) that resonant structure observed in the $^{24}$Mg($^{16}$O,$^{12}$C)$^{28}$Si reaction may be a consequence of strong couplings between the elastic and inelastic channels of either the initial or final systems.

The first step in the testing of the initial-system coupling hypothesis is a detailed coupled-channels analysis of the $^{24}$Mg + $^{16}$O system. The light-ion work mentioned above suggests that an asymmetric rotor model is an appropriate starting point for such analyses. An earlier attempt (Eck *et al.* 1981) at such an analysis was hampered by an experimental inability to resolve the $4_1^+$ and $2_2^+$ states in $^{24}$Mg and the lack of forward-angle data.

The discussion presented in this chapter is about the measurements of elastic and inelastic scattering at 43.5 MeV (c.m.) of $^{16}$O from $^{24}$Mg, with particular emphasis on resolving the closely-spaced $4_1^+$, $2_2^+$ doublet near 4.2 MeV. The bombarding energy was chosen to correspond to a peak in the related $^{24}Mg(^{16}O,^{12}C)^{28}Si$ excitation function (see Chapter 30) in order to study the nature of the $^{24}Mg$ + $^{16}O$ interaction at an energy close to one of its possible shape resonances. Since the inelastic scattering at forward angles is expected to be much less sensitive than the transfer reaction to a single resonating partial wave, the data should also permit a reasonable extraction of $^{24}Mg$ shape parameters.

## Experimental procedure and results

In this experiment, special effort was made to resolve the doublet near 4.2 MeV excitation energy in $^{24}Mg$. For this purpose, very thin targets were used and a high-resolution delay-line focal-plane detector (Leigh and Ophel 1982) was employed to detect the reaction products at the focal plane of a split-pole Enge spectrometer. The targets were prepared by vacuum evaporation of a thin (~ 5 μg/cm$^2$) layer of enriched (99.92%) $^{24}Mg$ on to a comparable thickness of carbon backing. A beam of 72.5 MeV $^{16}O$ projectiles was provided by the Australian National University 14 UD Pelletron accelerator. The incident $^{16}O$ energy corresponded to one of the resonances observed earlier (Nurzynski *at al.* 1981) in a study of the reaction $^{24}Mg(^{16}O,^{12}C)^{28}Si$.

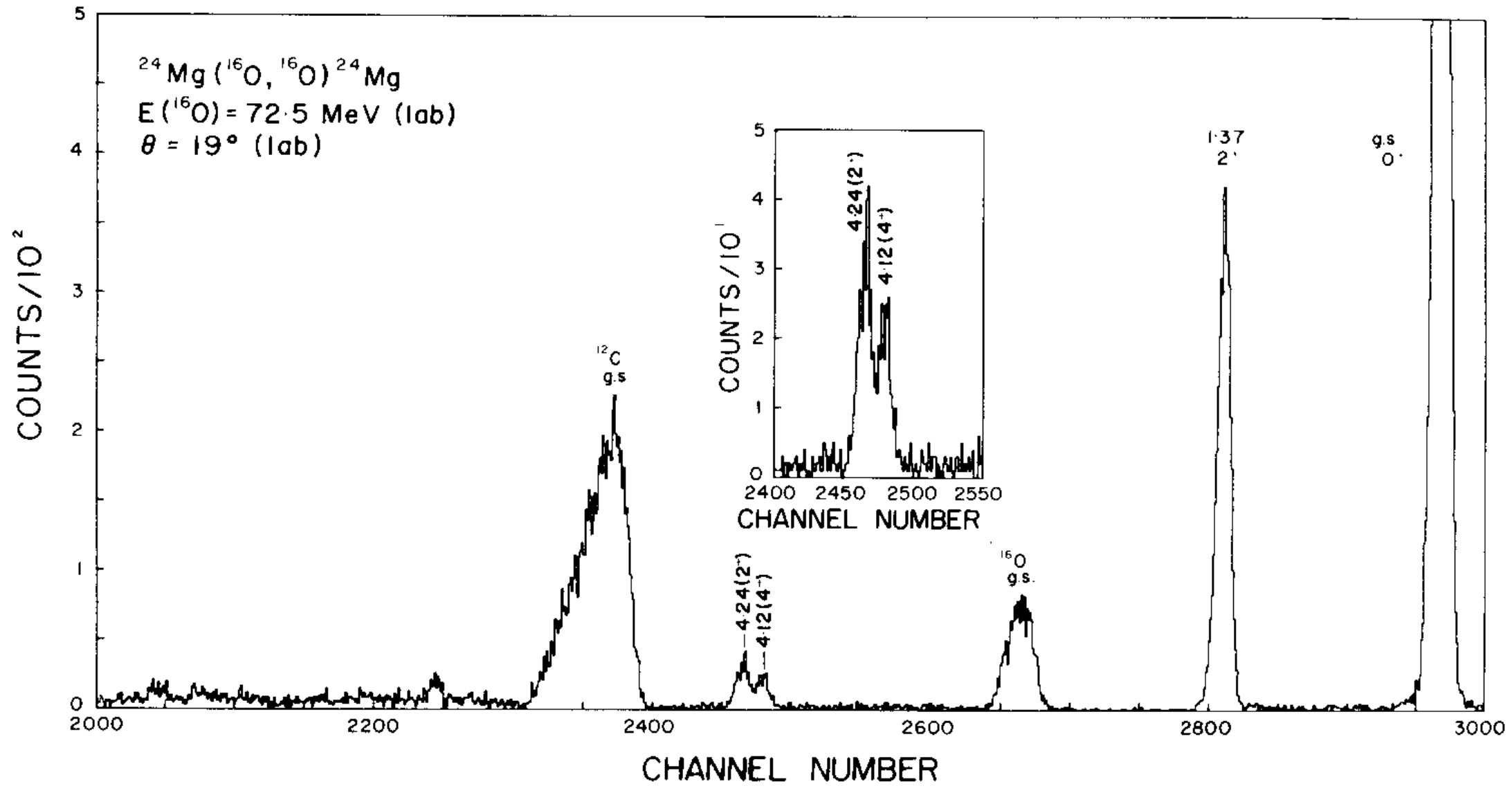

Figure 32.1. Energy spectrum taken at 19° (lab) for the scattering of $^{16}O$ from $^{24}Mg$ and contaminating elements at 72.5 MeV (lab) bombarding energy.

Data were recorded in event-by-event mode on magnetic tapes using a HP-2100 data acquisition system. Measurements of the angular distributions for both elastic scattering and inelastic scattering to the $2_1^+$ (1.37 MeV), $4_1^+$ (4.12 MeV) and $2_2^+$ (4.24 MeV) states in $^{24}Mg$ were carried out in steps of 1° for the angles 4° - 29° (lab). The horizontal acceptance angle of the magnetic spectrometer was 1°. At the most forward angles, the elastic and inelastic scattering cross sections were measured separately. For the inelastic groups, good statistics were obtained by blocking the elastic group and maintaining high beam intensities. Between about 300 and 2000 counts were obtained for each member of the doublet and the overall resolution of the detector system was maintained at 80 - 100 keV (FWHM) over the data collection periods, which were as long as 4 hours for some angles.

Events corresponding to $^{16}O$ + $^{24}Mg$ scattering were selected from the magnetic tapes using two-dimensional windows and a typical $^{16}O$ particle spectrum as shown in Figure 32.1. The doublet near 4.2 MeV was analysed by fitting skewed Gaussian distributions obtained from the elastic and the $2_1^+$ inelastic peaks. Absolute cross sections were determined by normalization of the data to Rutherford scattering at forward angles and the resultant angular distributions are displayed in Figures 32.3, 32.4 and 32.6. For certain angles, the inelastic scattering data were obscured by $^{12}C$ and $^{16}O$ elastic scattering. The main sources of error in the individual points of the angular distributions are statistical errors and uncertainties arising from inaccuracy of the peak fitting.

## Extended rigid asymmetric rotor model of $^{24}Mg$

The rigid asymmetric rotor model of Davydov and Filippov (1958), which describes only quadrupole deformation, has been extended by Baker (1979) and Barker *at al.* (1979) to include hexadecapole deformation. This extended model is employed in the present work for the analysis of $^{16}O$ scattering from $^{24}Mg$.

It is assumed that the nuclear charge density is uniform inside a radius given in the intrinsic coordinate system by

$$R(\theta',\phi') =$$
$$R_0\left\{1+\beta_2\cos\gamma Y'_{20}+\sqrt{1/2}\beta_2\sin\gamma(Y'_{22}+Y'_{2-2})\beta_4 Y'_{40}+a_{42}\left(Y'_{42}+Y'_{4-2}\right)+a_{44}\left(Y'_{44}+Y'_{4-4}\right)\right\}$$

where $Y'_{\lambda\mu}=Y_{\lambda\mu}(\theta',\phi')$. The rigid rotor Hamiltonian for such a distribution has been derived by Baker (1979) assuming that the inertial parameters ($B_2$ and $B_4$) satisfy the irrotational relation (Strutt 1926) (i.e. $B$ = $B_2$ = $2B_4$). In our study, for simplicity, we have considered only symmetric hexadecapole shapes (i.e. $a_{42}$ = $a_{44}$ = 0). The Schrödinger equation for the triaxial system has the form:

$$\left\{\left(\frac{1}{4}\hbar^2/B\beta_2^2\right)\sum_{k=1}^{3}\frac{1}{2}J_k^2\left[\sin^2\left(\gamma-\frac{2}{3}\pi k\right)+\frac{5}{4}\left(\frac{\beta_4}{\beta_2}\right)^2\left(1-\delta_{k,3}\right)\right]^{-1}-\varepsilon\right\}\psi=0$$

where $J_k$ are the operators of the projection of the nuclear angular momentum on the axes of the body-fixed coordinate system. The wave function $\psi$ for the $n$th state of spin $I$ and projection $M$ can be conveniently expanded in terms of basis states $|IMK\rangle$, i.e.

$$\psi_{IM}^{(n)}=\sum_{\substack{K=0\\ even}}^{I}A_{IK}^{(n)}|IMK\rangle$$

where

$$|IMK\rangle=\left[\frac{2I+1}{16\pi^2(1+\delta_{K0})}\right]^{1/2}\left\{D_{MK}^{I}+\left(-1\right)^{I}D_{M-K}^{I}\right\}$$

$$\sum_{\substack{K=0 \\ even}}^{I} A_{IK}^{(n)} A_{IK}^{(n')} = \delta_{nn'}$$

and the eigenvalues and eigenvectors can be obtained by standard techniques (Eisenberg and Greiner 1975).

To evaluate transition probabilities and moments of the nucleus, the reduced matrix elements of the electric multipole operators $\mathrm{M}(E\lambda,\mu)$ are required in space-fixed coordinates. We have

$$\mathrm{M}(E\lambda,\mu) = \sum_{\nu} \mathrm{M}'(E\lambda,\nu) D_{\mu\nu}^{\lambda}$$

where $\mathrm{M}'(E\lambda,\mu)$ are the corresponding $E\lambda$ spherical tensor operators in the body-fixed coordinate frame given in terms of a volume integral over the nuclear charge density $\rho$ by the relation

$$\mathrm{M}'(E\lambda,\mu) = \int Y'_{\lambda\nu} r^{\lambda} \rho(r,\theta',\phi') d\tau'$$

For the standard Davydov-Filippov model (i.e. $\beta_4 = 0$), explicit expressions for the energies of some of the low-lying energy levels and first-order terms for the reduced $E2$ transition probabilities have been given (Davydov and Filippov 1958; Davydov and Rostovskii 1959). In particular, the ratio of the energies of the $2_2^+$ and $2_1^+$ states is a function of the angle $\gamma$ only. One has

$$\frac{\varepsilon(2_2^+)}{\varepsilon(2_1^+)} = \frac{\left\{3 + \left[9 - 8\sin^2(3\gamma)\right]^{1/2}\right\}}{\left\{3 - \left[9 - 8\sin^2(3\gamma)\right]^{1/2}\right\}}$$

which gives $\gamma \approx 22°$ for $^{24}$Mg.

Normalization of the energy scaling $\hbar^2/4B\beta_2^2$ to the experimental value of the $2_1^+$ state yields spectrum *A* of Figure 32.2. It is seen that good agreement with experimental values (Endt and van der Leun 1978) is obtained for the low-lying levels and that the $4_2^+$ predicted level lies much closer to the $4^+$ state at 8.44 MeV rather than that at 6.01 MeV.

It should be noted that at 8.44 MeV excitation energy, a close ($4^+$,$1^-$) doublet has been identified (Ollerhead *et al.* 1968). The corresponding $B(E2)$ and quadrupole moment of the $2_1^+$ state ($Q_{2_1^+}$) predictions using the Davydov-Filippov expressions (Davydov and Filippov 1958; Davydov and Rostovskii 1959) are presented in the fourth column of Table 32.1. Comparison with the experimental values (Branford, McGough, and Wright 1975; Fewell *et al.* 1979) shows good agreement for the intra-band transitions and $Q_{2_1^+}$. However, some of the cross-band transition rates [$2_2^+ \rightarrow 2_1^+$, $3^+ \rightarrow 4_1^+$ and $4^+$(6.01 MeV)$\rightarrow 4_1^+$] are predicted to be too high. Similar discrepancies have been obtained in both shell-model (Lombard, Escudié and Soyeur 1978) and Hartree-Fock (Branford, McGough, and Wright 1975) studies. In Table 32.1, the decay properties of the $4_1^+$ model state are compared with the available data (Branford, McGough, and Wright 1975; Meyer, Reinecke, and Reitmann 1972) for both the 6.01 and 8.44 MeV levels.

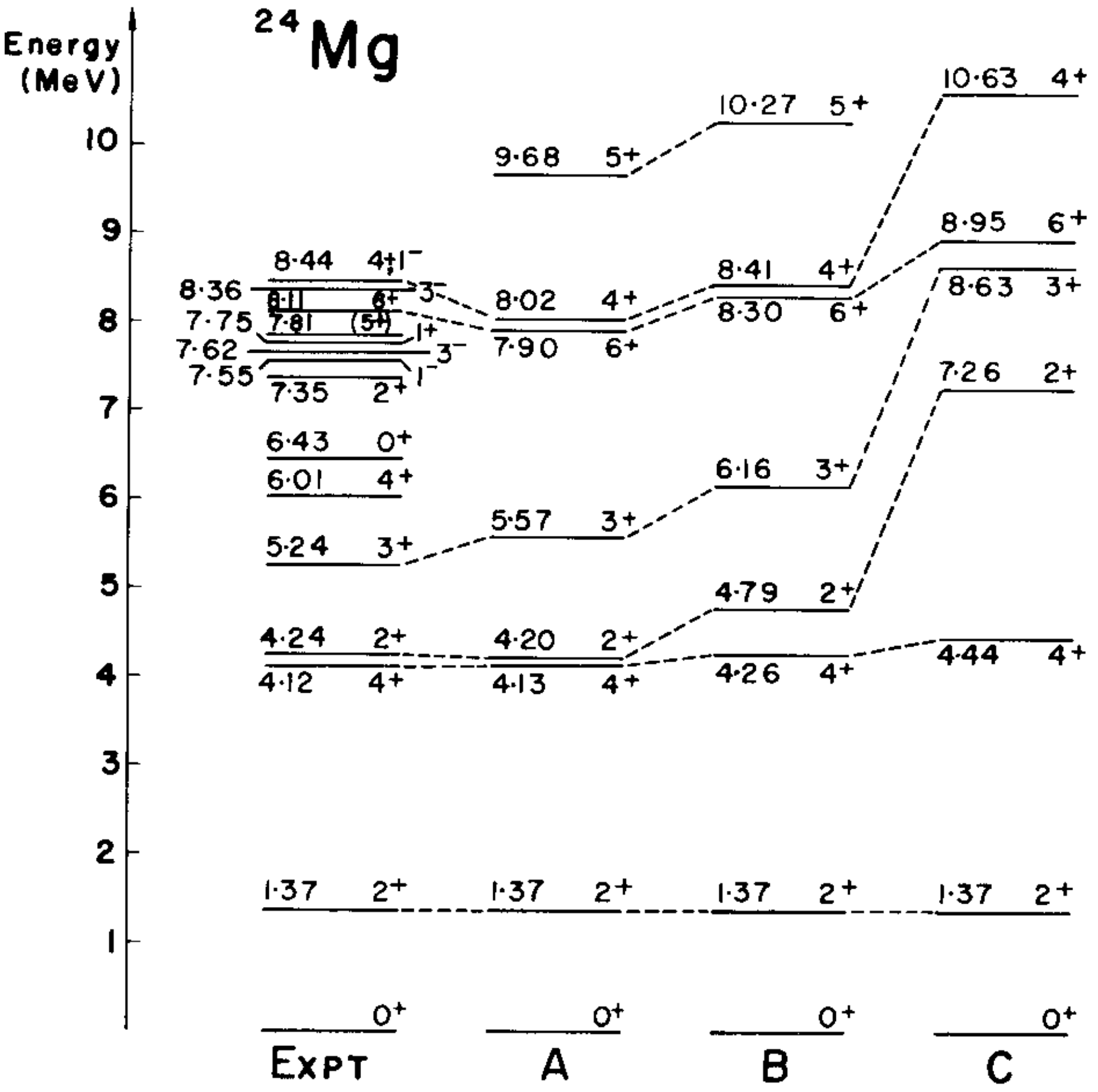

Figure 32.2. Energy levels for $^{24}$Mg predicted by rigid triaxial rotor models are compared with experimental values (Endt and van der Leun 1978). Result *A* is for the standard Davydov-Filippov (1958) model with $\gamma$ = 22°. Results B and C include a symmetric hexadecapole deformation $\beta_4$ = - 0.26$\beta_2$, with $\gamma$ = 22° and 18°, respectively. In each case, the energy scale is normalized by matching the energy of the $2_1^+$ (1.37 MeV) state.

Figure 32.2 and Table 32.1 also show the results when a symmetric hexadecapole deformation ($\beta_4/\beta_2$ = - 0.26) is included. The values of the parameters $\beta_2$ and $\beta_4$ for this ratio were determined by a coupled-channels analysis of the $^{24}$Mg($^{16}$O,$^{16}$O') data discussed in the next section. The resultant level spectrum, assuming $\gamma$ = 22° (result *B*), gives a somewhat poorer separation of the $4_1^+$ and $2_2^+$ states. However, the $B(E2)$ predictions for the $2_2^+ \to 2_1^+, 3^+ \to 4_1^+$ and $4^+$(6.01 MeV)$\to 4_1^+$ cross-band transitions are significantly improved. The predictions for these transitions can be brought into close agreement with experimental values by retaining the non-zero $\beta_4$ deformation but reducing the value of $\gamma$ to 18° (last column of Table 32.1). On the other hand, Figure 32.2 shows (result *C*) that the corresponding energies for the $K$ = 2 band states now lie far too high relative to those of the ground-state $K$ = 0 band so the smaller value of $\gamma$ seems unsatisfactory.

Table 32.1

The values of the $\gamma$ transition probabilities $B(E2)$ and quadrupole moments $Q_{2_1^+}$ for $^{24}Mg$

| Initial state | Final state | Experiment ($e^2 \cdot fm^4$) | Davydov-Filippov $\gamma = 22°$ | $\beta_4/\beta_2 = -0.26$ $\gamma = 22°$ | $\beta_4/\beta_2 = -0.26$ $\gamma = 18°$ |
|---|---|---|---|---|---|
| $K = 0$ band | | | | | |
| $2_1^+$ 1.37 MeV | $0^+$ g.s. | 84.3 ± 2.5 | 81.92 | 83.99 | 88.39 |
| $4_1^+$ 4.12 MeV | $2_1^+$ 1.37 MeV | 95 ± 16 | 119.57 | 120.37 | 127.66 |
| $6^+$ 8.11 MeV | $4_1^+$ 4.12 MeV | 140 + 148 − 42 | 145.32 | 139.75 | 143.92 |
| $K = 2$ band | | | | | |
| $3^+$ 5.24 MeV | $2_2^+$ 4.24 MeV | 140 ± 25 | 146.30 | 149.99 | 157.85 |
| $4^+$ 6.01 MeV | $2_2^+$ 4.24 MeV | 66 ± 12 | 39.69 | 44.65 | 50.30 |
| Cross band | | | | | |
| $2_2^+$ 4.24 MeV | $0^+$ g.s. | 5.8 ± 1.2 | 5.55 | 2.35 | 2.34 |
| $2_2^+$ 4.24 MeV | $2_1^+$ 1.37MeV | 11.1 ± 1.6 | 44.88 | 22.95 | 10.80 |
| $3^+$ 5.24 MeV | $2_1^+$ 1.37 MeV | 8.6 ± 1.2 | 9.91 | 4.19 | 4.18 |
| $3^+$ 5.24 MeV | $4_1^+$ 4.12 MeV | < 18 | 50.42 | 30.86 | 13.02 |
| $4^+$ 6.01 MeV | $2_1^+$ 1.37 MeV | 4.1 ± 0.8 | 0.79 | 0.98 | 0.001 |
| $4^+$ 6.01 MeV | $4_1^+$ 4.12 MeV | 4.1 ± 4.1 | 29.65 | 19.33 | 12.02 |
| $4^+$ 8.44 MeV | $2_1^+$ 1.37 MeV | 2.5 ± 0.8 | 0.79 | 0.98 | 0.001 |
| $Q_{2_1^+}$ ($e \cdot fm^2$) | | −17.8 ± 1.3 | −15.17 | −16.98 | −18.47 |

All calculated values assume $R_0 = 1.25 A_t^{1/3}$ fm and $\beta_2 R_0^2 = 7.3$ fm$^2$.

Davydov-Filippov – Only quadrupole deformation is considered. $\beta_4/\beta_2$ = -0.26 – The effect of including both quadrupole and hexadecapole deformations.

## Coupled-channels analysis

Coupled-channels calculations for the elastic and inelastic scattering of $^{16}O$ from $^{24}Mg$ at a bombarding energy of 72.5 MeV ($E_{c.m.}$ = 43.5 MeV) were performed using the computer code ECIS79 (Raynal 1972, 1981)[3]. The $0^+$ (g.s.), $2_1^+$ (1.37 MeV), $4_1^+$ (4.12 MeV) and $2_2^+$ (4.24 MeV) states in $^{24}Mg$ were included in all the calculations and select computations to obtain the final results were carried out including also the $3^+$ (5.24 MeV), $6^+$ (8.11 MeV) and $4^+$ (8.44 MeV) states. The calculations involving the full set of six excited states took an order of magnitude longer to perform than those in which only the lowest three excited states were taken into account. Thus, most of the preliminary calculations were carried out employing the smaller set of states. All the calculations employed the Coulomb correction technique developed by Raynal (1980,1981) with matching at a radius of approximately 14 fm. Comparison with conventional calculations with matching near 40 fm indicated agreement to better than 2% in the predicted angular distributions for all states included in the coupling scheme.

The deformed optical potential was of the form

$$U(r,\theta,\phi) = U_N(r,\theta,\phi) + V_C(r,\theta,\phi)$$

where

$$U_N(r,\theta,\phi) = -Vf_V(r,\theta,\phi) - iWf_W(r,\theta,\phi)$$

with

$$f_V(r,\theta,\phi) = [1 + \exp\{[r - R_V(\theta,\phi)]/a_0\}]^{-1}$$

$$f_W(r,\theta,\phi) = [1 + \exp\{[r - R_W(\theta,\phi)]/a_0'\}]^{-1}$$

[3] See Chapter 16.

and the Coulomb potential is generated from the deformed charge density according to the relation

$$V_C(r,\theta,\phi) = Z_p Z_t e^2 \int^{R_C(\theta,\phi)} \frac{d\vec{r}'}{|\vec{r}-\vec{r}'|} / \int^{R_C(\theta,\phi)} d\vec{r}'$$

Here $Z_pe$ and $Z_te$ denote the charges of the incident and target nuclei, respectively. The deformed radius for the triaxial rotor under consideration has the form

$$R_V(r,\theta,\phi) = r_0\left(A_p^{1/3} + A_t^{1/3}\right)\left[1 + \beta_2^{(V)}\cos\gamma Y_{20} + \sqrt{\frac{1}{2}}\beta_2^{(V)}\sin\gamma\left(Y_{22} + Y_{2-2}\right) + \beta_4^{(V)} Y_{40}\right]$$

$$R_W(r,\theta,\phi) = r_0'\left(A_p^{1/3} + A_t^{1/3}\right)\left[1 + \beta_2^{(W)}\cos\gamma Y_{20} + \sqrt{\frac{1}{2}}\beta_2^{(W)}\sin\gamma\left(Y_{22} + Y_{2-2}\right) + \beta_4^{(W)} Y_{40}\right]$$

$$R_C(r,\theta,\phi) = r_C\left(A_p^{1/3} + A_t^{1/3}\right)\left[1 + \beta_2^{(C)}\cos\gamma Y_{20} + \sqrt{\frac{1}{2}}\beta_2^{(C)}\sin\gamma\left(Y_{22} + Y_{2-2}\right) + \beta_4^{(C)} Y_{40}\right]$$

The quantities $A_p$, and $A_t$ are the atomic mass numbers of the projectile and the target nucleus, respectively.

A number of optical potentials have been considered in the present work. These are presented in Table 32.2 and are discussed in the following section. The angle $\gamma$ was set at 22° in accord with the requirement to describe the energy level spectrum for $^{24}$Mg as discussed earlier. For simplicity, the real and imaginary nuclear potential deformations were constrained to be identical (i.e. $\beta_2^{(V)} = \beta_2^{(W)} = \beta_2^{(N)}$ and $\beta_4^{(V)} = \beta_4^{(W)} = \beta_4^{(N)}$) but otherwise were treated as free parameters. The quadrupole Coulomb deformation $\beta_2^{(C)}$, was constrained to satisfy the relation $\beta_2^{(N)} R_C^2 = 7.3$ fm$^2$, a value determined by matching the relevant $B(E2)$ and $Q_{2_1^+}$ experimental values for $^{24}$Mg, as discussed in the previous section.

In each case, the $\beta_4^{(C)}$ value was determined by requiring that the ratio $\beta_4^{(C)}/\beta_2^{(C)}$ be identical to the corresponding ratio, $\beta_4^{(N)}/\beta_2^{(N)}$, for the nuclear deformation. It should be noted that the predicted angular distributions were found to be relatively insensitive to the inclusion of the $\beta_4^{(C)}$ deformation.

Finally, in the coupled-channels calculations, it is necessary to specify the wave functions for the relevant $^{24}$Mg states expressed as an expansion in the standard $|IMK\rangle$ basis states (see the previous section). In each case, these wave functions were determined using $\gamma$ and $\beta_4/\beta_2$ values consistent with those used in the deformed nuclear and Coulomb potentials.

**Optical model potential and shape deformation parameters**

Previously reported (Nurzynski *et al.* 1981) optical model analysis of $^{16}$O scattering from $^{24}$Mg and studies for neighbouring mass systems have led to three broad classifications of possible potentials (see Chapter 30):

(i) Deep real potential wells (~ 80 MeV) with strong absorption strength and nearly the same geometrical parameters ($r_0' \approx r_0$ and $a_0' \approx a_0$).

(ii) Moderate real potential wells (~ 30 MeV) with weak absorption strength and $r_0' \leq r_0$, $a_0' \leq a_0$.

(iii) Shallow real potential wells with moderate absorption strength and $r_0' < r_0$ and $a_0' < a_0$.

Potentials of types (ii) and (iii) give rise to surface transparency. In the work described in this chapter, potentials in each of these classes have been examined.

Table 32.2

Optical model potentials

| Potential | Type | $V$ (MeV) | $r_0$ (fm) | $a_0$ (fm) | $W$ (MeV) | $r_0'$ (fm) | $a_0'$ (fm) | $r_C$ (fm) |
|---|---|---|---|---|---|---|---|---|
| 1 | (i) | 83.0 | 1.24 | 0.530 | 45.0 | 1.32 | 0.420 | 1.25 |
| 2 | (ii) | 32.0 | 1.30 | 0.500 | 12.0 | 1.30 | 0.400 | 1.25 |
| 3 | (iii) | 10.0 | 1.35 | 0.618 | 23.4 | 1.23 | 0.552 | 1.00 |

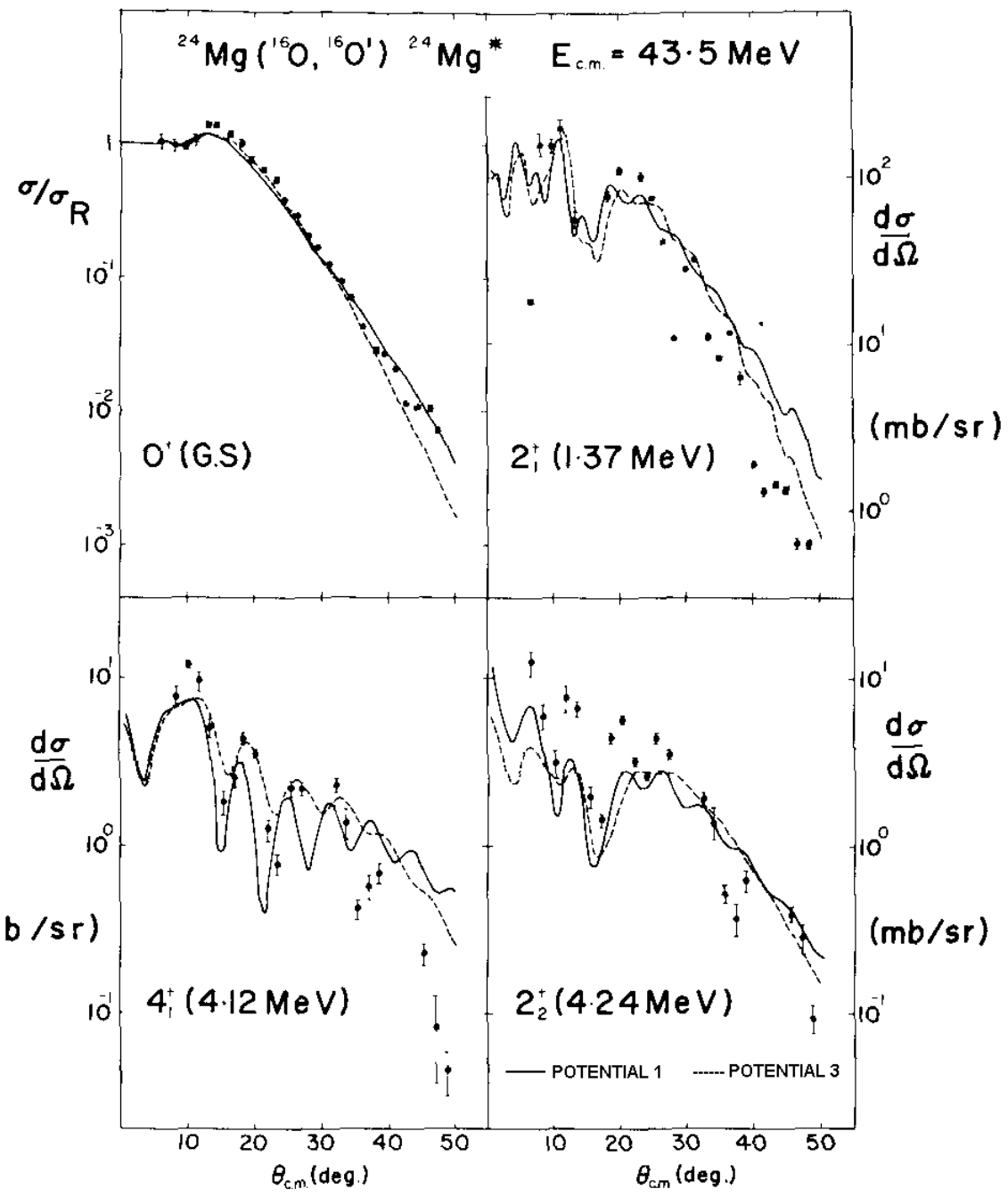

Figure 32.3. Angular distributions (dots) for the reactions $^{24}$Mg($^{16}$0,$^{16}$O')$^{24}$Mg* (0$^+$,g.s.; $2_1^+$, 1.37 MeV; $4_1^+$, 4.12 MeV; $2_2^+$, 4.24 MeV) measured at $E_{c.m.}$ = 43.5 MeV are compared with coupled-channels calculations using potential 1 (solid curves) of Table 32.2 and potential 3 (dashed curves). The deformation parameters are $\beta_2^{(N)}$ = 0.30 and $\beta_4^{(N)}$ = - 0.065 (for the solid curves) and $\beta_2^{(N)}$ = 0.35 and $\beta_4^{(N)}$ = - 0.10 (for the dashed curves).

Figure 32.3 shows typical results for a deep real potential well (potential 1) of Table 32.2, which is based upon potential 6 of Nurzynski *et al.* (1981) with $\beta_2^{(N)}$ = 0.30 and $\beta_4^{(N)}$ = - 0.065 (solid curves). The figure contains also results for the shallow real potential (potential 3 of Table 32.2). In these calculations, represented by dashed lines in Figure 32.3, the best fits were obtained for the deformation parameters $\beta_2^{(N)}$= 0.35 and $\beta_4^{(N)}$= - 0.10. Potential 3 is similar to that used by Cramer *et al.* (1976) in their analysis of $^{16}$O scattering from $^{28}$Si and by Eck *et al.* (1979) in analysis of the $^{24}$Mg($^{16}$0,$^{16}$O')$^{24}$Mg* scattering at 67 MeV bombarding energy. It can be seen that both sets of calculations produce unsatisfactory fits to the experimental angular distributions.

Figure 32.4 shows the results obtained for potential 2 of Table 32.2 with $\beta_2^{(N)}$= 0.25 and $\beta_4^{(N)}$= - 0.065. For this moderate real potential well, nearly perfect agreement with the data is obtained for the magnitudes and shapes of the four angular distributions, although some of the details are not well described. While the whole of parameter space could not be searched, it seems unlikely that these fits could be improved significantly.

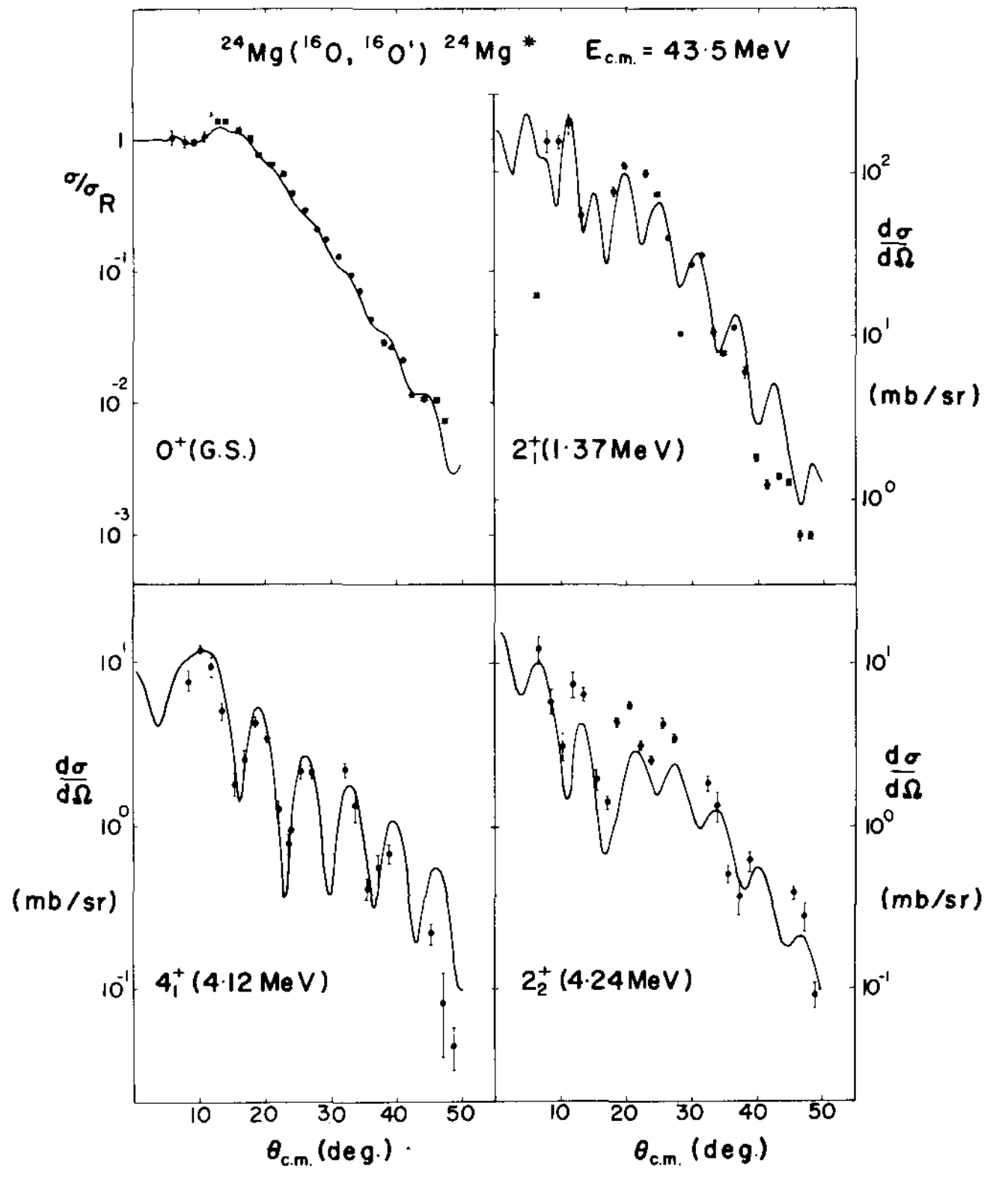

Figure 32.4. Angular distributions (dots) for the reactions $^{24}$Mg($^{16}$O,$^{16}$O')$^{24}$Mg* (0$^+$,g.s.; $2_1^+$, 1.37 MeV; $4_1^+$, 4.12 MeV; $2_2^+$, 4.24 MeV) measured at $E_{c.m.}$ = 43.5 MeV are compared with coupled-channels calculations (solid curves) using potential 2 of Table 32.2, $\beta_2^{(N)}$= 0.25 and $\beta_4^{(N)}$= - 0.065.

Some of the discrepancies arise probably from insufficiencies of the extended asymmetric rotor model employed for the calculations. For example, for simplicity and consistency, the deformation parameters for the states of the $K$ = 2 band were taken to be identical with those for the ground-state band although the analysis of van der Borg *et al.* (1979) suggests otherwise. Furthermore, the optical model potentials were constrained to have Woods-Saxon form factors and no parity-dependent term (Dehnhard, Shkolnik, and Franey 1978).

Potential 2 is similar to potential 11 of Nurzynski *et al.* (1981, see Chapter 30), which was found to give a satisfactory description of the ground-state transition for the $^{24}Mg(^{16}O,^{12}C)^{28}Si$ transfer reaction. However, it is less surface transparent since the geometries of the real and imaginary potentials are similar. Attempts to fit the present data with more surface transparent interactions led to too much structure at larger angles. It is possible that the use of an explicit $J$ - dependent absorption potential (Chatwin *et al.* 1970) rather than a potential with $r_0' \leq r_0$, $a_0' \leq a_0$ which is supposed to simulate such an effect may lead to an improved description of the data. These two kinds of surface transparent potentials are not equivalent (Kondo and Tamura 1982).

Figure 32.5 shows the corresponding results for the $3^+$ (5,24 MeV), $6^+$ (8.11 MeV) and $4^+$ (8.44 MeV) states. These cross sections are considerably smaller than those for the lower states. We have found that the inclusion of these three states in the coupling scheme affected only the cross section for the $2_2^+$ state. This effect arises from the intra-band coupling of the $3^+$ and $4^+$ levels with the $2_2^+$ state and is only slightly dependent upon whether the $4_1^+$ model state is associated with the level at 8.44 MeV or with that at 6.01 MeV. Figure 32.5 also shows the result, which include a hexacontatetrapole deformation (a $\beta_6^{(N)}Y_{60}$ term) in the deformed optical potential. For $\beta_6^{(N)}$ = 0.0267, a value which gives a potential deformation distance $\delta_6 \equiv \beta_6^{(N)} r_0 (A_p^{1/3} + A_t^{1/3})$ = 0.188 fm, approximately equivalent to that employed by Lombard, Escudié and Soyeur (1978) for proton scattering by $^{24}Mg$ (i.e. $\delta_6 = \beta_6^{(N)} r_0 A_t^{1/3}$=0.179 fm), it is seen that the peak cross section for the $6^+$ state is only increased by about 25%. The effect of the finite $\beta_6^{(N)}$ deformation on the other states was much smaller.

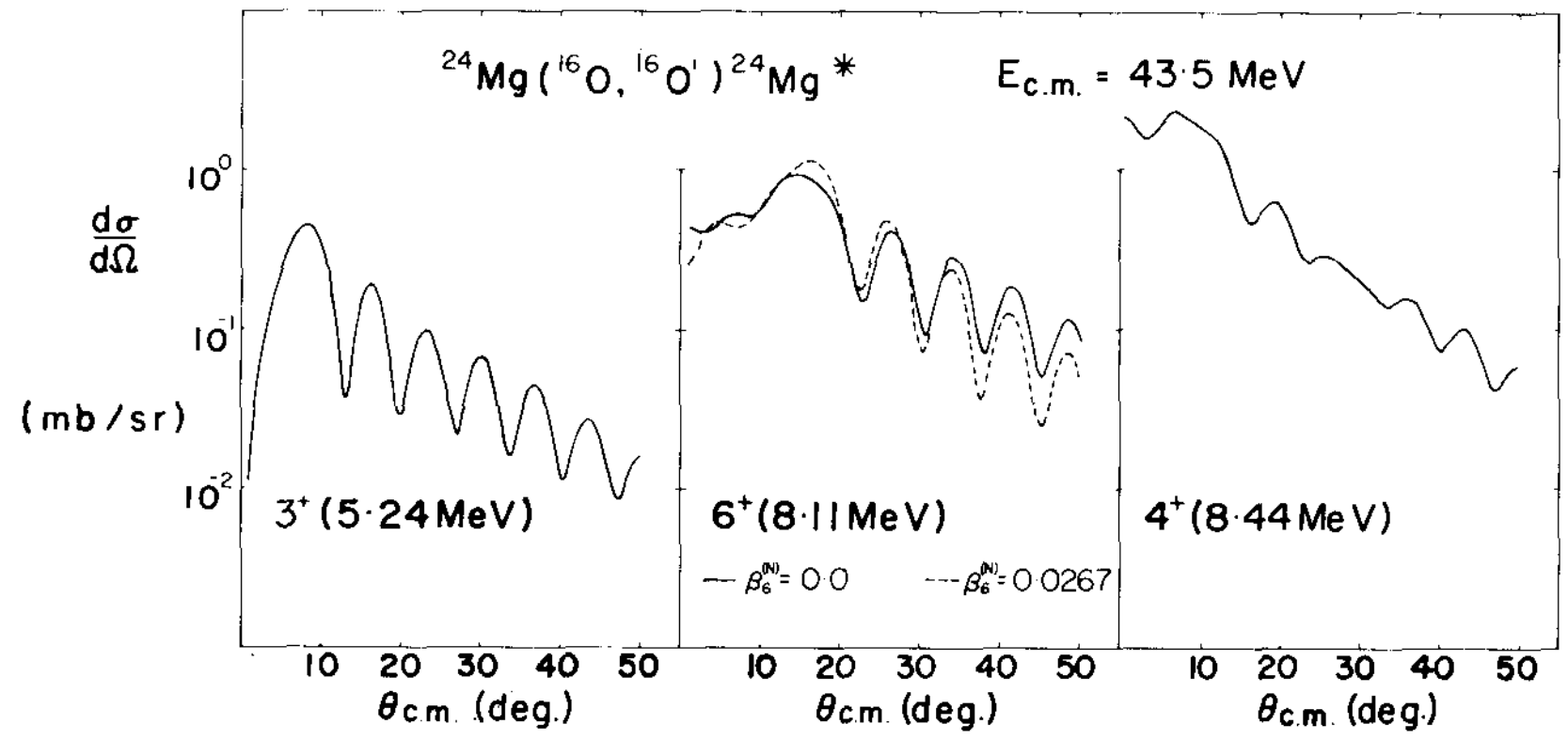

Figure 32.5. Coupled-channels calculations (solid curves) of angular distributions for the reactions $^{24}Mg(^{16}O,^{16}O')^{24}Mg^*$ ($3^+$, 5.24 MeV; $6^+$, 8.11 MeV; $4^+$, 8.44 MeV)'at $E_{c.m.}$ = 43.5 MeV using potential 2 of Table 32.2, $\beta_2^{(N)}$ = 0.25 and $\beta_4^{(N)}$ = - 0.065. The dashed curve shows the effect of including a $\beta_6^{(N)}$ = 0.0267 term in the optical potential.

We have also studied the effect of either removing the hexadecapole deformation or changing its sign. Figure 32.6 exhibits the result for $\beta_4^{(N)}$ = 0 (solid curve) and $\beta_4^{(N)}$ = +0.065 (dashed curve), respectively. For $\beta_4^{(N)}$ = 0, the fit to the angular distribution for the $4_1^+$ state is less satisfactory both in magnitude and shape than the fit for $\beta_4^{(N)}$= -0.065 of Figure 32.4, while for $\beta_4^{(N)}$= +0.065 an additional oscillation is introduced into the calculated cross sections. Thus, the negative value of $\beta_4^{(N)}$ is strongly favoured by the present analysis.

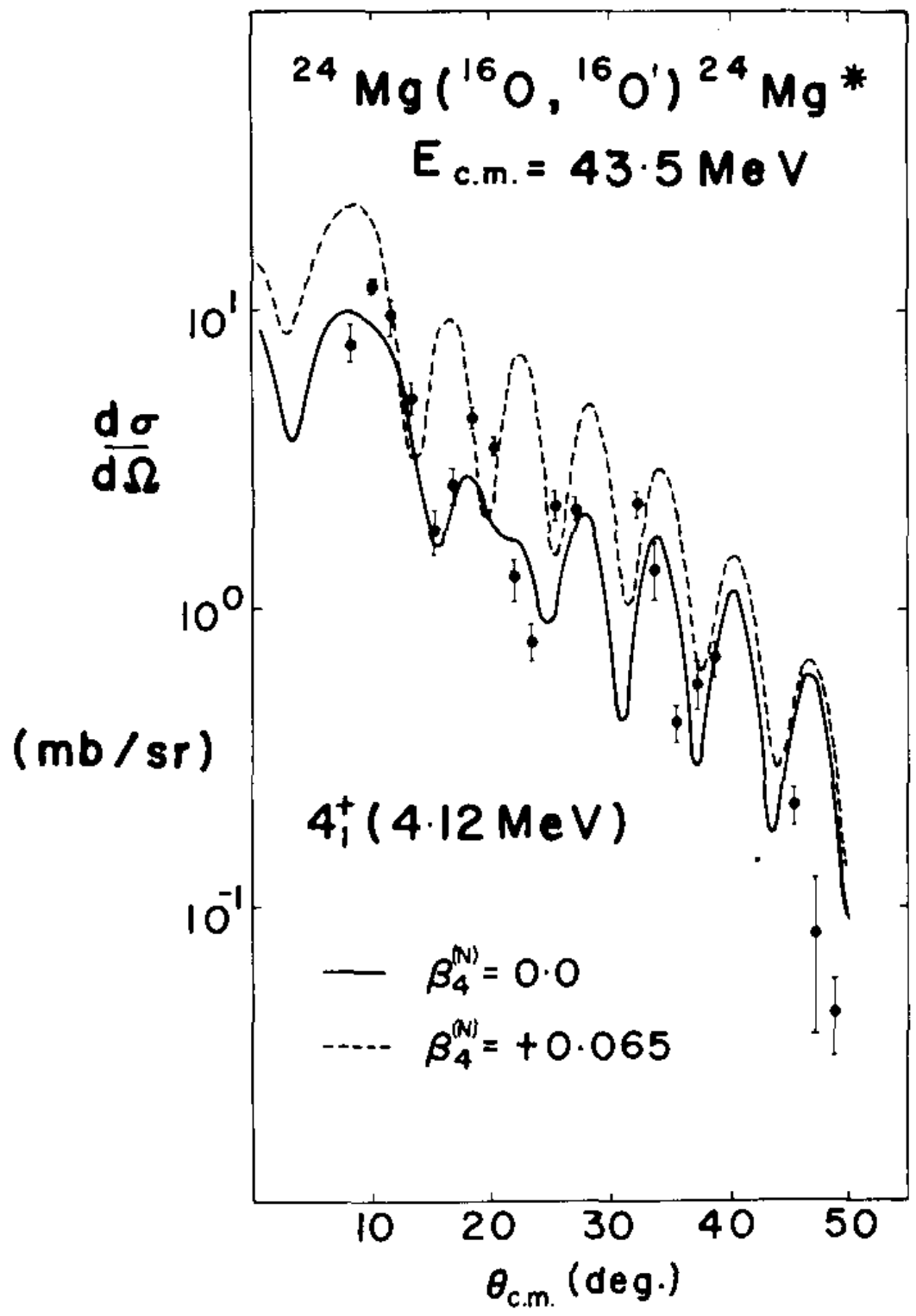

Figure 32.6. Coupled-channels calculations of the angular distribution for the reaction $^{24}Mg(^{16}O,^{16}O')^{24}Mg*$ ($4_1^+$, 4.12 MeV) at $E_{c.m.}$ = 43.5 MeV using potential 2 of Table 32.2 with $\beta_2^{(N)}$= 0.25. The hexadecapole deformation was assumed to be either $\beta_4^{(N)}$= 0 (solid curve) or positive, $\beta_4^{(N)}$ = +0.065, (dashed curve).

Table 32.3
Potential deformation distances for the asymmetric rotor model for $^{24}Mg$

| Ref. | Reaction | $\gamma$ (deg) | $\delta_2$ (fm) | $\delta_4$ (fm) |
|---|---|---|---|---|
| a) | (p, p′) | 23.0 | 1.72 | 0.00 |
| b) | (p, p′) | 22.4 h) | 1.76 h) | 0.00 |
| c) | (p, p′) | 21.5 | 1.56 | −0.20 |
| d) | (α, α′) | 35.0 | 1.73 | 0.00 |
| e) | (α, α′) | 21.8 | { 1.31<br>{ 1.59 i) | { −0.11<br>{ −0.48 i) |
| f) | ($^{16}$O, $^{16}$O′) | 22.0 | 2.63 | −0.73 |
| g) | ($^{16}$O, $^{16}$O′) | 22.0 | 1.76 | −0.46 |

a) Eenmaa *et al.* (1974); b) Lovas *et al.* (1977); c) Lombard *et al.* (1978);
d) Tamura (1965); e) van der Borg *et al.* (1979); f) Eck *et al.* (1981);
g) Our present work; h) Average values;
i) $\delta_2$ = 1.59 fm employed for coupling of the $K$ = 2 band states; $\delta_4$ = -0.48 employed for coupling of the g.s. with $4_2^+$ model state; $\delta_6$ = 0.179 fm also included.

Table 32.3 displays the various potential deformation distances $\delta_2$ and $\delta_4$ obtained by analyses of both light- and heavy-ion inelastic scattering data for $^{24}$Mg. This comparison takes into account (Hendri 1973) the different parameterization of the potential radii in terms of $A_t^{1/3}$ and ($A_p^{1/3} + A_t^{1/3}$) for light and heavy ion scattering, respectively. It is seen that the present results are in good agreement with the light-ion work. The earlier ($^{16}$O,$^{16}$O') analysis (Eck *et al.* 1981) employed a shallow real potential well and the larger deformation distances obtained in this case are a further argument against such an interaction.

## Discussion and conclusion

Angular distributions for the elastic and inelastic scattering of $^{16}$O from $^{24}$Mg, exciting the $2_1^+$, 1.37 MeV, $4_1^+$, 4.12 MeV and $2_2^+$, 4.24 MeV states have been measured at $E_{lab}$= 72.5 MeV ($E_{c.m.}$ = 43.5 MeV) bombarding energy. The data have been described well by the coupled-channels calculations within the framework of the Davydov-Filippov (1958) asymmetric rotor model for the low-lying states of $^{24}$Mg, which has been extended to include a symmetric hexadecapole shape component. The analysis included coupling to the higher $3^+$ (5.24 MeV), $6^+$ (8.11 MeV) and $4^+$ (8.44 MeV) states. We have found that for these higher states only the intra-band couplings between the $2_2^+$, $3^+$ and $4^+$ (8.44 MeV) states are significant.

The optical model potential for the $^{16}$O + $^{24}$Mg interaction was determined to have a moderate real well depth confirming the earlier study by Nurzynski *et al.* (1981). Furthermore, the potential appears to be sufficiently transparent for it to be consistent with the possibility of a formation of quasi-molecular states for the $^{16}$O + $^{24}$Mg system. The overall success of the extended asymmetric rotor model for $^{24}$Mg indicates the possibility of excited states, additional to those described by a simple symmetric rotor model as discussed in our earlier work (Nurzynski *et al.* 1981), being strongly coupled to the elastic channel.

The shape parameters of the nuclear potential for the $^{16}$O + $^{24}$Mg system were found to be $\beta_2^{(N)}$ = 0.25, $\gamma$ = 22° and $\beta_4^{(N)}$ = -0.065. The corresponding deformation distances are in good agreement with earlier light-ion results (see references in Table 32.3). The negative $\beta_4^{(N)}$ deformation parameter is well established by the analysis of the $4_1^+$ angular distribution. The scattering analysis shows little sensitivity to the parameter $\gamma$, although the $^{24}$Mg level spectrum constrains the value to be close to $\gamma$ = 22°.

The question, which arises as a consequence of the good description of $^{16}$O scattering, obtained here and of previous works using light ions, is whether this agreement means that $^{24}$Mg is a rigid rotor. As Yamakazi (1963) has pointed out, it is difficult to distinguish whether the nuclear equilibrium shape is axially symmetric ($\gamma = 0$) or asymmetric ($\gamma \neq 0$) by consideration of predictions for only the ground-state $K$ = 0 band and a $K$ = 2 band based upon either $\gamma$ - vibrations or a fixed $\gamma$ - deformation, respectively. The Hartree-Fock calculations of Grammaticos (1975) show that $^{24}$Mg may be soft to vibrations in the $\gamma$ - direction, depending upon the effective interaction employed. If the $K$ = 2 band arises from $\gamma$ - vibrations, the results of such a model are similar to those obtained for a rigid triaxial nucleus with an effective intermediate value of $\gamma$, which should be considered as a "freezing approximation" of the $\gamma$ -vibrations (Lombard, Escudié and Soyeur 1978). In this picture, the description of the $^{24}$Mg states in terms of a rigid asymmetric rotor, parameterised by $\beta_2$, $\gamma$ and $\beta_4$, is a convenient and relatively realistic way of modelling the important couplings involved in the analysis of light- or heavy-ion scattering to the low-lying levels of $^{24}$Mg.

# 33

# Spin Assignments for the $^{143}$Pm and $^{145}$Eu Isotopes

***Key features:***

1. Using $j$ – dependence, spin assignments have been made for states belonging to the $2d_{5/2}$ and $2d_{3/2}$ configurations in $^{143}$Pm and $^{145}$Eu nuclei. Other spin assignments have been also made using the unique $1g_{7/2}$, $3s_{1/2}$, and $1h_{11/2}$ proton configurations accessible via the selected target nuclei.
2. Spectroscopic factors have been extracted and compared with the factors determined using the ($^3$He,d) reaction.

**Abstract**: Angular distributions have been measured for transitions to low-lying states in $^{143}$Pm and $^{145}$Eu populated by the $^{142}$Nd($^7$Li,$^6$He)$^{143}$Pm and the $^{144}$Sm($^7$Li,$^6$He)$^{145}$Eu reactions at $E$($^7$Li) = 52 MeV. Elastic scattering of $^7$Li at 52 MeV on $^{142}$Nd and $^{144}$Sm, and $^6$Li at 46 MeV on $^{142}$Nd and at 45 MeV on $^{144}$Sm, were measured. Optical-model parameters extracted from fits to the scattering data were used in the finite-range distorted waves analysis of the angular distributions for levels below 1.40 MeV excitation energy in $^{143}$Pm and 1.84 MeV in $^{145}$Eu. The reaction cross sections forward of 6° (c.m.) allow unambiguous distinction between $2d_{5/2}$ and $2d_{3/2}$ states. Final-state spins have been assigned to $d$ - states in $^{143}$Pm and in $^{145}$Eu. Existing assignments to other levels in both residual nuclei have been confirmed.

## Introduction

As discussed in Chapter 29, lithium-induced, single-nucleon, stripping reactions present a useful tool for extracting spectroscopic information complementary to that obtained from light-ion work. For most cases the data can be well described by the exact finite-range (EFR) distorted-waves Born approximation formalism. The ($^7$Li,$^6$He) reaction involves the transfer of a proton from a $p$ - wave orbit in the projectile, and not from a predominantly $s$ - state as is the case of light-ion reactions such as (d,n), ($^3$He,d) and ($\alpha$,t). The observed $j$ - dependence for such reactions can be used to distinguish between spins belonging to $2d_{5/2}$ and $2d_{3/2}$ configurations.

In the work described in this chapter, the $^{142}$Nd($^7$Li,$^6$He)$^{143}$Pm and $^{144}$Sm($^7$Li,$^6$He)$^{145}$Eu reactions have been studied. In addition, the elastic scattering of $^7$Li and $^6$Li on $^{142}$Nd and $^{144}$Sm has been measured to obtain optical-model parameters for the distorted waves calculations.

The $^{142}$Nd nucleus has 82 neutrons and 60 protons. Its 1g/2d/3s/1h neutron shell is closed. Its $N$ = 50 proton shell is also closed and the remaining 10 protons are in the next, 1g/2d/3s/1h, shell. In the simple shell-model description, 8 of these protons would occupy the $1g_{7/2}$ orbitals, and 2 would have a $2d_{5/2}$ configuration. However, considering the residual interaction, there will be a mixture of other configurations. Nevertheless, the $1g_{7/2}$ will be almost full but other configurations ($2d_{5/2}$, $2d_{3/2}$, $3s_{1/2}$, and $1h_{11/2}$) will be available for the stripped proton. Thus, the strongest transitions could be expected to belong to these almost empty configurations in the $^{143}$Pm nucleus.

The $^{144}$Sm nucleus also has 82 neutrons and thus its 1g/2d/3s/1h neutron shell is also closed. However, this isotope has 62 protons. The presence of the extra two protons should be expected to reduce the probability for transfers to the $1g_{7/2}$ orbitals but should not affect significantly the transfers to other configurations.

## Experimental method and results

Beams of $^{6}Li^{-}$ and $^{7}Li^{-}$ from a General Ionex sputter source were injected into the Australian National University 14UD Pelletron accelerator. Beam currents of up to 300 nA of $^{6}Li^{3+}$ and $^{7}Li^{3+}$ were obtained on target. Targets of enriched $^{142}$Nd (> 96%) and $^{144}$Sm (> 96%), comprised of metal on thin carbon backings, were used. Target thicknesses were ~ 25 $\mu$/cm$^{2}$, although thicker targets (~ 100-150 $\mu$/cm$^{2}$) could be used for the required resolution of the $^{6}$He groups corresponding to observed states in the residual nuclei.

Reaction data and elastic scattering data were measured with an Enge split-pole spectrograph using a resistive-wire gas proportional detector (Ophel and Johnston 1978) located at the focal plane. From the energy loss ($\Delta E$) and the position signal ($\propto B\rho$) of the focal plane detector, a mass identification signal ($M^2 = (B\rho)^2 \Delta E$) was obtained as shown in Figure 33.1. The difference in magnetic rigidity between $^{6}Li^{3+}$ and $^{6}He^{2+}$ is sufficient to allow unambiguous mass identification. In our measurements for transfer reactions, $^{6}Li^{3+}$ ions did not enter the detector and therefore were not interfering with the collection of data. Additionally, the high field necessary to place the $^{6}$He particles onto the detector removed completely the $^{7}Li^{3+}$ elastic events from the detector, allowing high beam currents to be used at forward angles. The angular acceptance of the spectrograph was 1°. Fixed monitor detectors at 15° and 30° were used for normalization between runs and to check on the target deterioration.

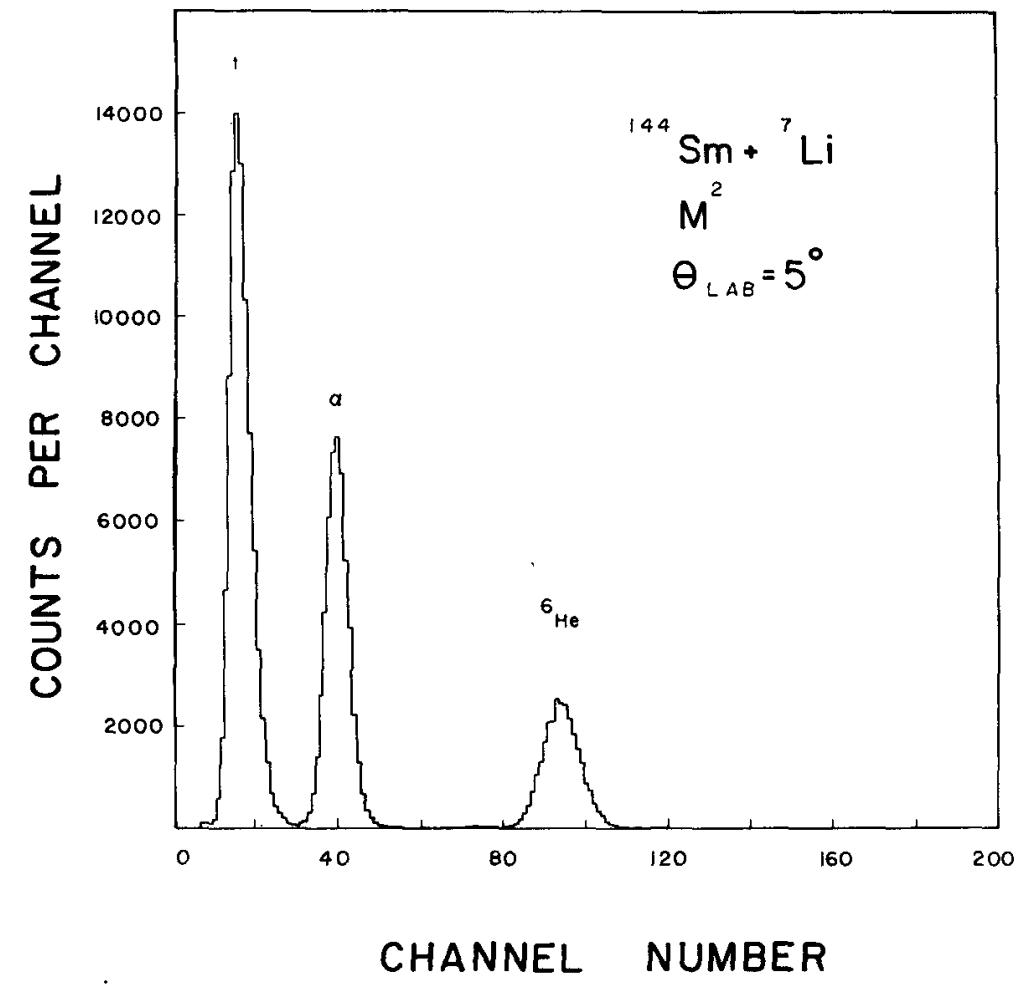

Figure 33.1. A mass identification signal (mass squared) for the $^{7}$Li + $^{144}$Sm reaction at 5°. The $^{6}$Li particles are excluded because their magnetic rigidity is such that they do not enter the detector.

To obtain information on the elastic scattering wave functions, needed in the distorted waves analysis, we have also measured the $^{142}$Nd($^{7}$Li,$^{7}$Li)$^{142}$Nd and $^{144}$Sm($^{7}$Li,$^{7}$Li)$^{144}$Sm at $E$($^{7}$Li) = 52 MeV and $^{142}$Nd($^{6}$Li,$^{6}$Li)$^{142}$Nd at $E$($^{6}$Li) = 46 MeV and $^{144}$Sm($^{6}$Li,$^{6}$Li)$^{144}$Sm at $E$($^{6}$Li) = 45 MeV. The use of $^{6}$Li optical parameters to describe $^{6}$He distorted waves has been shown to work well in the analysis of other ($^{7}$Li,$^{6}$He) reactions (see Chapter 29).

Absolute cross sections were obtained by normalizing to the forward-angle elastic scattering, where the cross section is purely Rutherford. The error in the absolute normalization is estimated to be 5% for the elastic scattering, resulting mainly from possible angle setting errors and uncertainties in the dead-time corrections. Based on the reproducibility of the ($^7$Li, $^6$He) data, the absolute cross sections of the transfer reactions are accurate to ± 12%. The relative errors in the cross sections are shown by the error bars on the individual data points where these are larger than the plotted points.

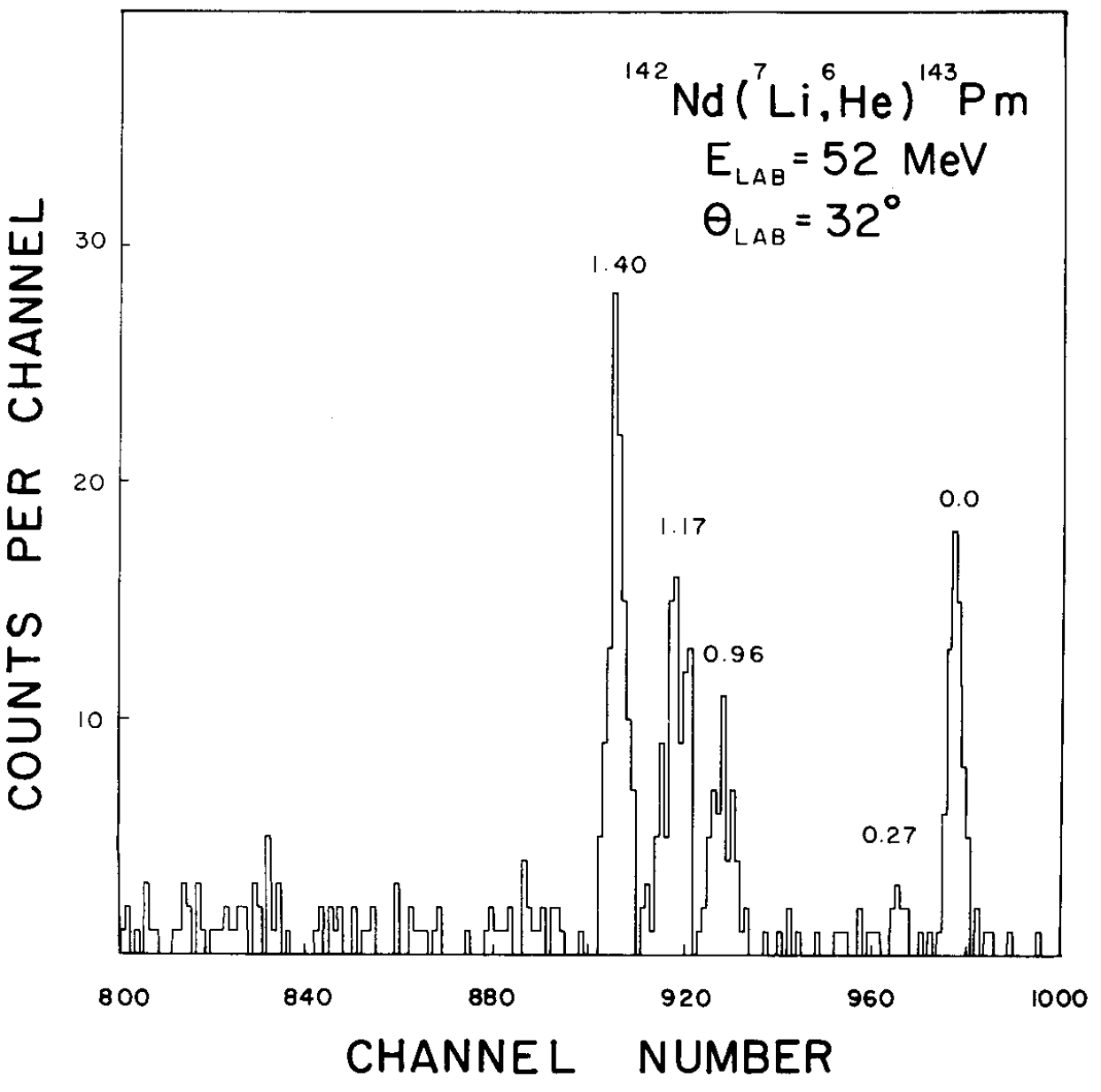

Figure 33.2. A $^6$He spectrum for the $^{142}$Nd($^7$Li,$^6$He)$^{143}$Pm reaction at 32°. States in $^{143}$Pm are labelled with the appropriate excitation energies.

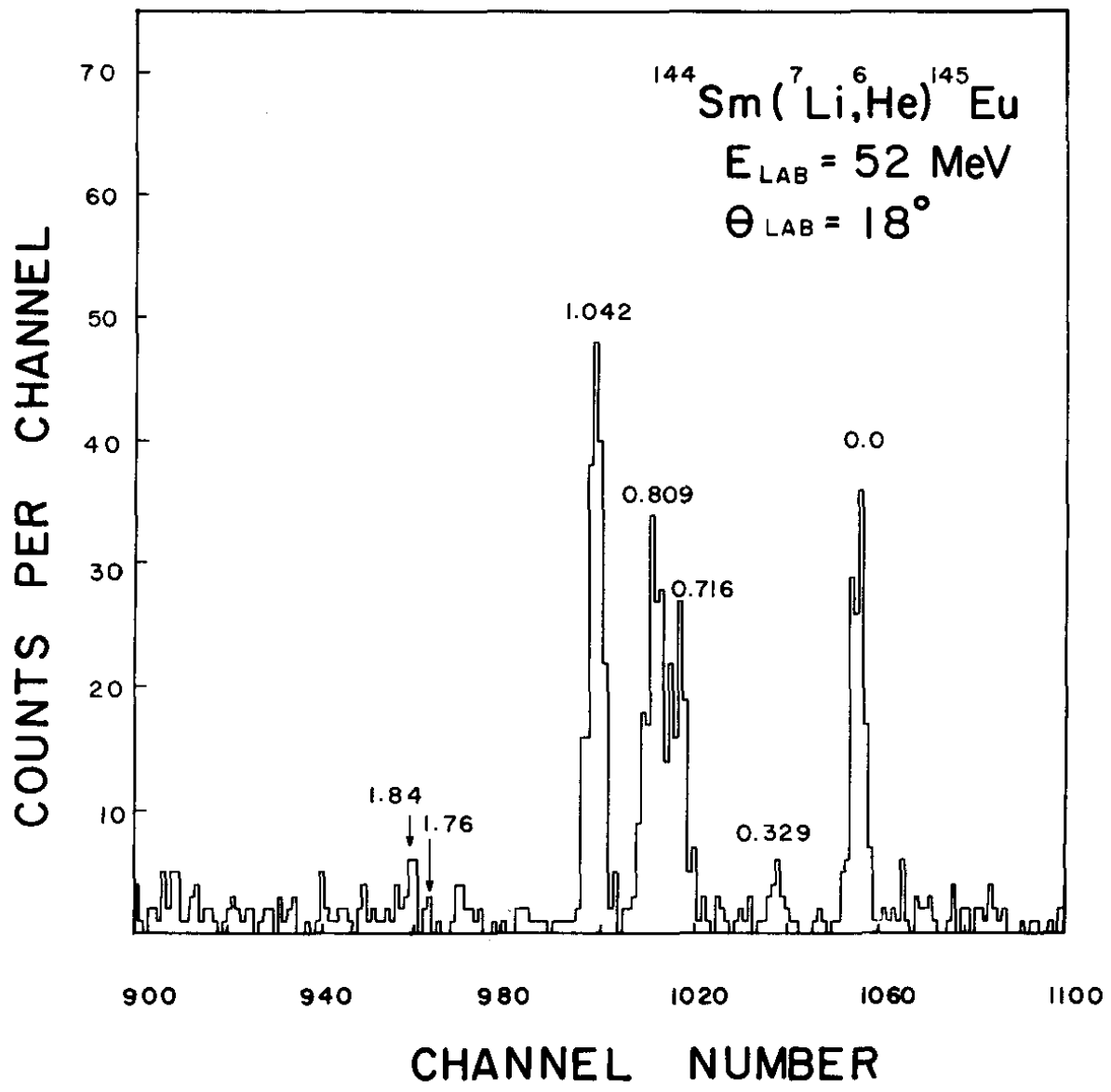

Figure 33.3. A $^6$He spectrum for $^{144}$Sm($^7$Li,$^6$He)$^{145}$Eu reaction at 18°. States in $^{145}$Eu are labelled with the appropriate excitation energies.

Figures 33.2 and 33.3 show spectra of the $^{142}$Nd($^7$Li,$^6$He)$^{143}$Pm reaction at $\theta_{lab}$ = 32°, and the $^{144}$Sm($^7$Li,$^6$He)$^{145}$Eu reaction at $\theta_{lab}$ = 18°. The resolution is 70 keV FWHM and little background is evident at these angles. Unfortunately, the $Q$ - value for the $^{12}$C($^7$Li,$^6$He)$^{13}$N reaction is such that the ground-state group obscures the states at 1.76 and 1.84 MeV in $^{145}$Eu at angles forward of 10° (lab). The $^{13}$C($^7$Li,$^6$He)$^{14}$N reaction was a less serious contaminant. A group, corresponding to the excitation of the 3.95 MeV level in $^{14}$N, prevented the extraction of the cross section for the 1.17 MeV state of $^{143}$Pm at 3° lab.

## Theoretical analysis

The elastic scattering data were analysed using the standard optical-model potential as described in Chapter 29. The computer code JIB (Perey 1967)[4] was used to fit the data, starting with the parameters used to fit the $^{140}$Ce($^{7}$Li,$^{7}$Li) and $^{141}$Pr($^{6}$Li,$^{6}$Li) elastic scattering data at 52 and 47 MeV, respectively. The parameters were varied two at a time until a minimum $\chi^2$ was obtained. The experimental angular distributions and the optical-model fits are shown in Figures 33.4 and 33.5. The extracted parameters are listed in Table 33.1.

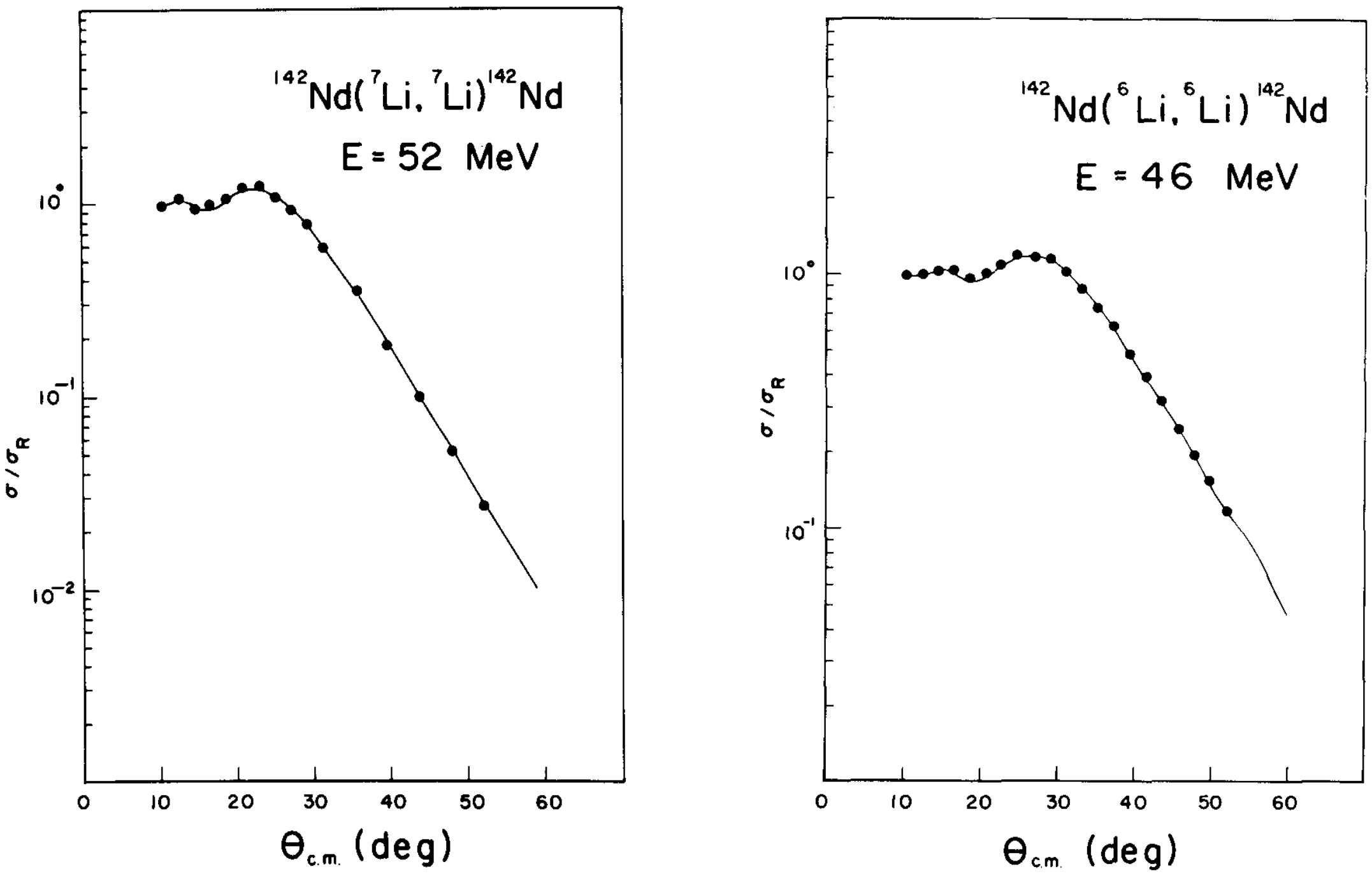

Figure 33.4. Angular distribution for the $^{142}$Nd($^{7}$Li,$^{7}$Li)$^{142}$Nd at 52 MeV and $^{142}$Nd($^{6}$Li,$^{6}$Li)$^{142}$Nd at 46 MeV elastic scattering. The solid lines are the optical-model fits to the data.

Table 33.1
Optical-model parameters

| Channel | $E$ (MeV) | $V$ (MeV) | $r_0$ (fm) | $a_0$ (fm) | $W$ (MeV) | $r_0'$ (fm) | $a_0'$ (fm) | $r_c$ (fm) |
|---|---|---|---|---|---|---|---|---|
| $^{142}$Nd($^{7}$Li, $^{7}$Li)$^{142}$Nd | 52 | 263.5 | 1.281 | 0.765 | 33.46 | 1.589 | 0.687 | 1.400 |
| $^{142}$Nd($^{6}$Li, $^{6}$Li)$^{142}$Nd | 46 | 256.6 | 1.359 | 0.686 | 10.44 | 1.786 | 0.534 | 1.400 |
| $^{144}$Sm($^{7}$Li, $^{7}$Li)$^{144}$Sm | 52 | 277.3 | 1.395 | 0.616 | 20.64 | 1.415 | 0.917 | 1.400 |
| $^{144}$Sm($^{6}$Li, $^{6}$Li)$^{144}$Sm | 45 | 247.5 | 1.332 | 0.677 | 8.10 | 1.752 | 0.762 | 1.400 |

The radii are defined using $A_t^{1/3}$ i.e. $R_0 = r_0 A_t^{1/3}$, $R_0' = r_0' A_t^{1/3}$ and $R_c = r_c A_t^{1/3}$.

[4] The same program, which I have modified and adapted to run at ANU, and which I have used to support my study of nuclear reactions induced by light projectiles.

Exact finite-range (EFR) distorted waves calculations using the optical-model parameters listed in Table 33. 1 were performed with the computer code LOLA (DeVries 1973) for transitions to the strongly populated states observed in the $^{142}$Nd($^{7}$Li,$^{6}$He)$^{143}$Pm and $^{144}$Sm($^{7}$Li,$^{6}$He)$^{145}$Eu reactions. Fifty-six partial waves were used in the calculations and the radial integrations were carried out to a radius of 30 fm in steps of 0.13 fm. The single-particle bound-state wave functions for $^{7}$Li, $^{143}$Pm and $^{145}$Eu were generated by the code assuming a volume Woods-Saxon form for the interaction potential with $r_0$ = 1.25 fm and $a_0$ = 0.65 fm, and the spin-orbit factor $\lambda$ = 25. No spin-orbit potential was used in the distorted waves. The depths of the potentials were adjusted so that the binding energy for the transferred proton in $^{7}$Li, $^{143}$Pm and $^{145}$Eu were equivalent to the correct separation energies. The experimental angular distributions and the EFR-distorted waves fits to the data are shown in Figures 33.6 and 33.7.

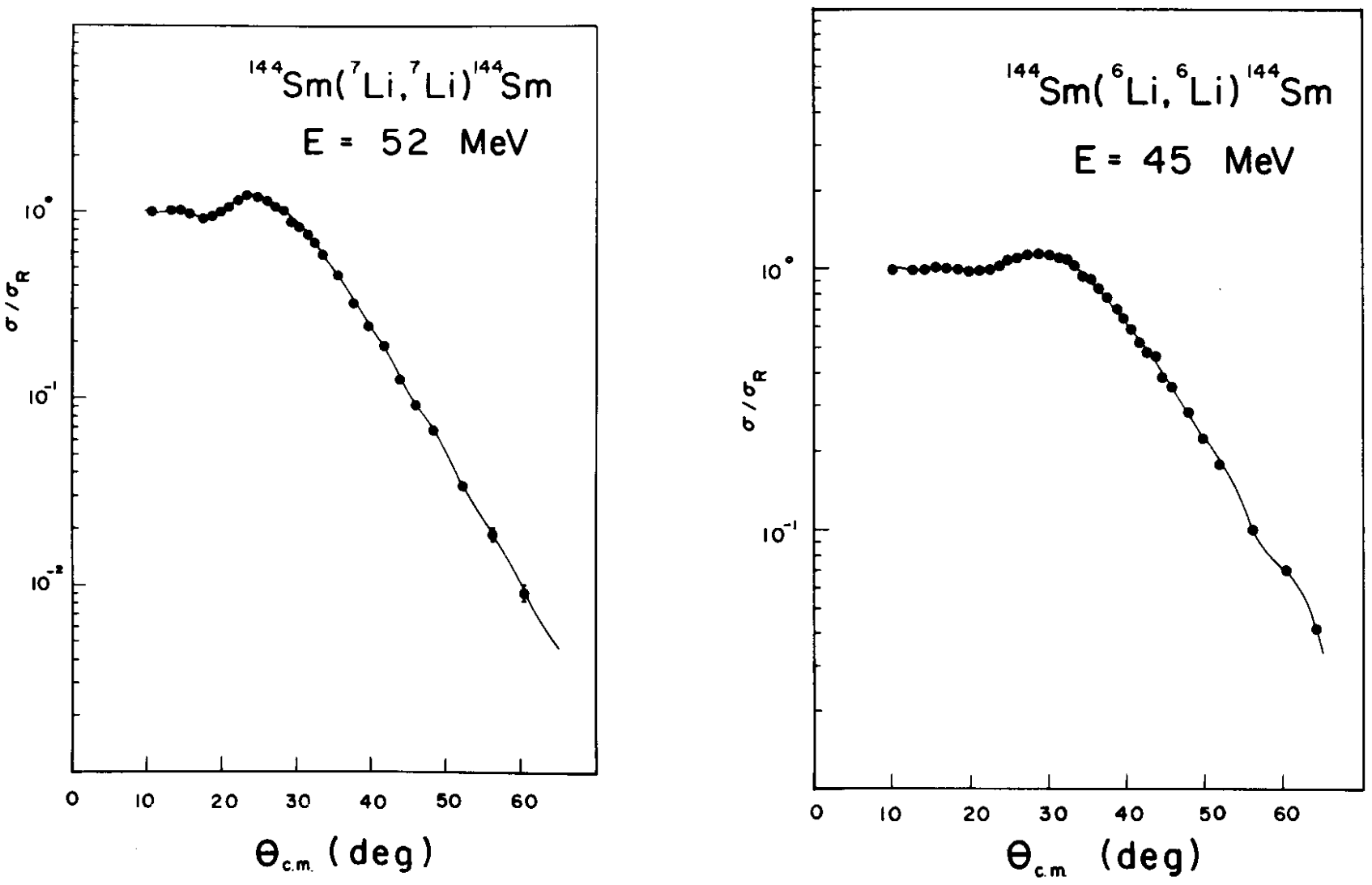

Figure 33.5. Angular distribution for the $^{144}$Sm($^{7}$Li,$^{7}$Li)$^{144}$Sm at 52 MeV and $^{144}$Sm($^{6}$Li,$^{6}$Li)$^{144}$Sm at 45 MeV elastic scattering. The solid lines are the optical-model fit to the data.

It has been shown in Chapter 29 that the angular distributions for ($^{7}$Li,$^{6}$He) leading to $2d_{5/2}$ and $2d_{3/2}$ final states can be distinguished at forward angles using *j* -dependence. This distinction can be made even though *l* = 1, 2 and 3 are allowed for both final states assuming a $p_{3/2}$ transferred proton, because the Racah coefficient multiplying the distorted waves cross section weights the transfers differently for the two states. Thus, the *l* = 1 component is 8 times stronger for a $2d_{5/2}$ state than for a $2d_{3/2}$ state so that the forward-angle cross section for a $2d_{5/2}$ state is larger than for a $2d_{3/2}$ state. Calculations for pure $2d_{5/2}$ and $2d_{3/2}$ are shown in Figure 33.6 for states at 0.0 MeV and 1.40 MeV in $^{143}$Pm, and in Figure 33.7 for states at 0.0 MeV, 1.042 MeV, 1.76 MeV, and 1.84 MeV in $^{145}$Eu. Clearly the data forward of 6° c.m. allow unambiguous distinction to be made between $2d_{5/2}$ and $2d_{3/2}$ final states.

The absolute spectroscopic factor, extracted as described in Chapter 29, are listed in Table 33.2 and 33.3, which also show spectroscopic factors obtained using ($^{3}$He,d) reactions (Wildenthal, Newman, and Auble 1971; Ishimatsu *et al.* 1970; Newman *et al.* 1970). The errors in the absolute values of the spectroscopic factors include the uncertainty in the absolute normalization of the experimental data and statistical errors. Uncertainties resulting from the choice of optical-model parameters and from the use of standard values for the bound-state potential parameters are not included. Relative spectroscopic factors, normalized to the ground state, are also listed in Tables 33.2 and 33.3.

Angular distributions for transitions to $3s_{1/2}$ levels via the $l$ = 1 transfer at 1.17 MeV in $^{143}$Pm and 0.809 MeV in $^{145}$Eu are well reproduced by the calculations but are slightly out of phase by 1° to 2°, a problem which has been observed in the analysis described in Chapter 29 and by Morre, Kemper, and Chalton (1975).

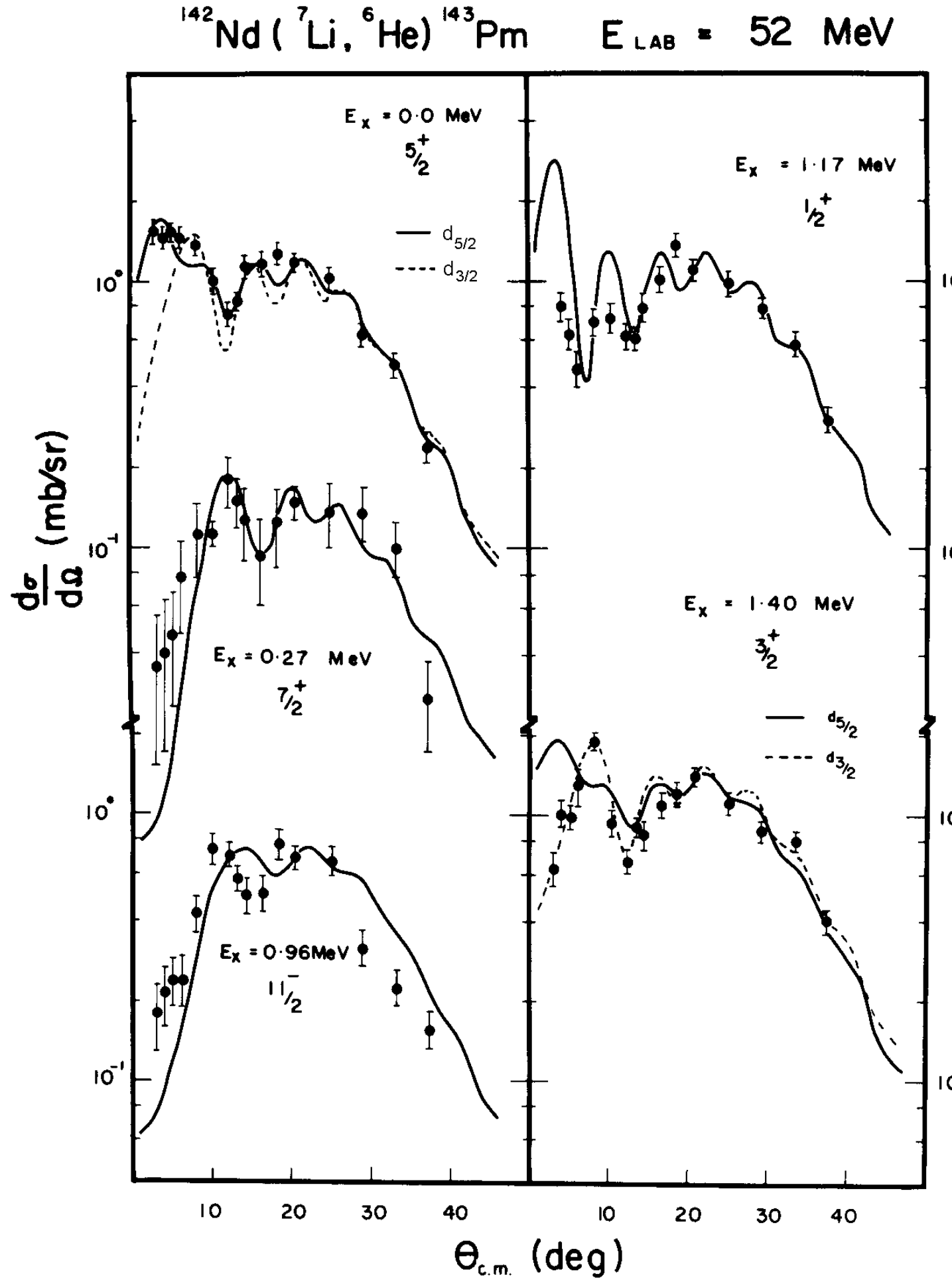

Figure 33.6. Angular distributions populated in the $^{142}$Nd($^{7}$Li,$^{6}$He)$^{143}$Pm reaction. The solid and dashed lines are the EFR-distorted waves calculations normalized to the data.

Table 33.2
Spectroscopic factors for states in $^{143}$Pm

| *E* (MeV) | *Conf.* | Absolute ($^3$He, d) [a] 40.3 MeV | Absolute ($^3$He, d) [b] 27.3 MeV | Absolute ($^7$Li, $^6$He) [c] 52 MeV | Relative ($^3$He, d) [a] 40.3 MeV | Relative ($^3$He, d) [b] 27.3 MeV | Relative ($^7$Li, $^6$He) [c] 52 MeV |
|---|---|---|---|---|---|---|---|
| 0.00 | $2d_{5/2}$ | 0.52 | 0.48 | 0.41±0.07 | 1.00 | 1.00 | 1.00±0.17 |
| 0.27 | $1g_{7/2}$ | 0.32 | 0.36 | 0.25±0.07 | 0.62 | 0.75 | 0.61±0.17 |
| 0.96 | $1h_{11/2}$ | 0.71 | 0.85 | 0.77±0.15 | 1.37 | 1.77 | 1.88±0.37 |
| 1.17 | $3s_{1/2}$ | 1.12 | 0.96 | 0.81±0.13 | 2.15 | 2.00 | 1.98±0.32 |
| 1.40 | $2d_{3/2}$ | 1.31 | 1.02 | 0.93±0.14 | 2.52 | 2.13 | 2.27±0.34 |

[a]) Wildenthal *et al.* (1971); [b]) Ishimatsu *et al.* (1970); [c]) Our work.

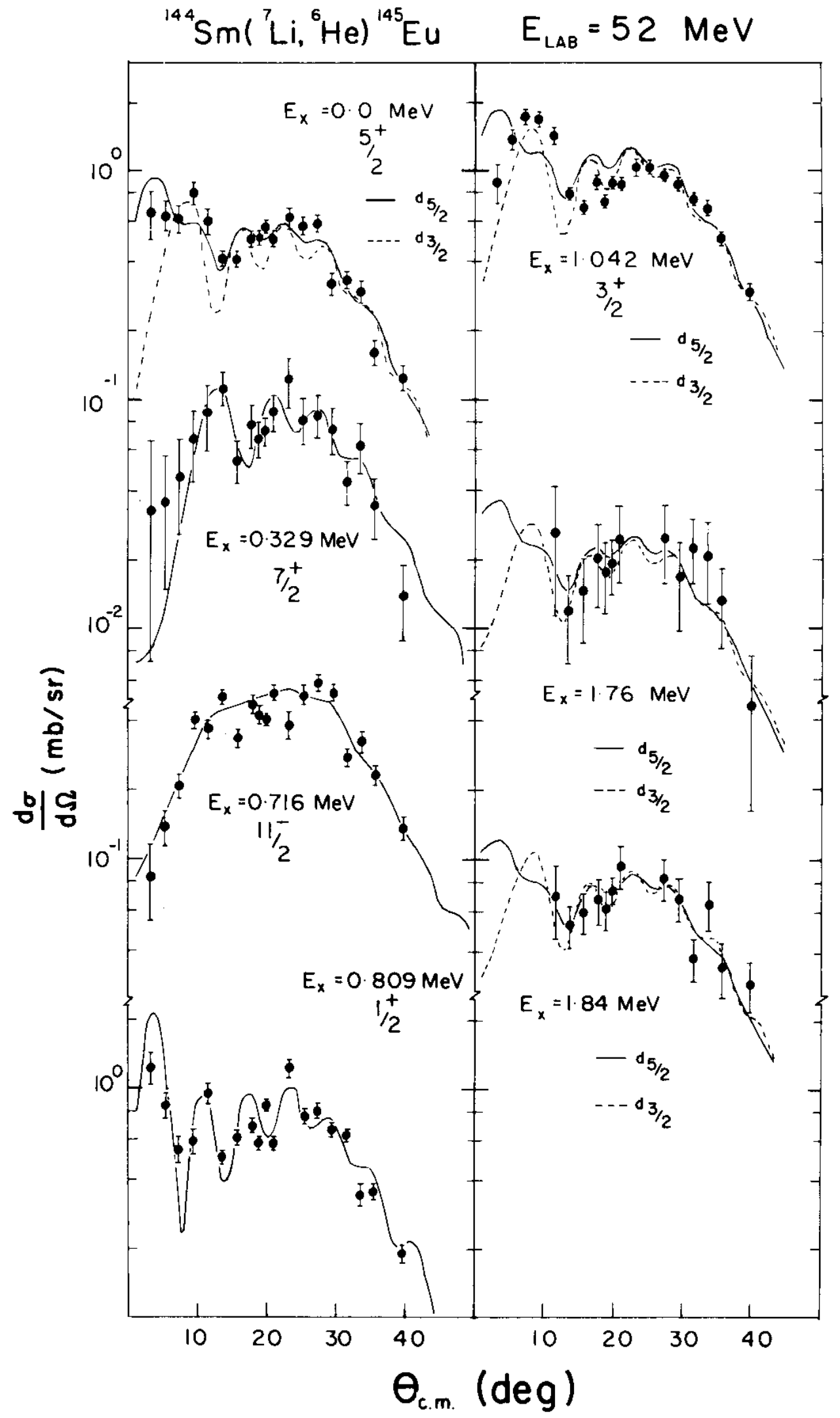

Figure 33.7. Angular distributions populated in the $^{144}Sm(^{7}Li,^{6}He)^{145}Eu$ reaction. The solid and dashed lines are the EFR-distorted waves calculations normalized to the data.

Table 33.3

Spectroscopic factors for states in $^{145}Eu$

| E (MeV) | Conf. | Absolute | | | Relative | | |
|---|---|---|---|---|---|---|---|
| | | ($^3$He, d) [a] 40.3 MeV | ($^3$He, d) [b] 40.33 MeV | ($^7$Li, $^6$He) [c] 52 MeV | ($^3$He, d) [a] 40.3 MeV | ($^3$He, d) [b] 40.33 MeV | ($^7$Li, $^6$He) [c] 52 MeV |
| 0.00 | $2d_{5/2}$ | 0.33 | 0.33 | 0.19 ±0.03 | 1.00 | 1.00 | 1.00±0.16 |
| 0.329 | $1g_{7/2}$ | 0.22 | 0.17 | 0.12 ±0.03 | 0.67 | 0.52 | 0.63±0.17 |
| 0.716 | $1h_{11/2}$ | 0.72 | 0.82 | 0.45 ±0.07 | 2.18 | 2.48 | 2.37±0.37 |
| 0.808 | $3s_{1/2}$ | 1.05 | 0.98 | 0.60 ±0.09 | 3.18 | 2.97 | 3.16±0.47 |
| 1.042 | $2d_{3/2}$ | 1.21 | 1.01 | 0.66 ±0.09 | 3.67 | 3.06 | 3.47±0.47 |
| 1.76 | $(2d_{3/2})$ | 0.05 | 0.02 | 0.014±0.006 | 0.15 | 0.06 | 0.07±0.03 |
| 1.84 | $(2d_{3/2})$ | 0.12 | 0.10 | 0.05 ±0.013 | 0.36 | 0.30 | 0.26±0.07 |

[a]) Wildenthal *et al.* (1971); [b]) Newman *et al.* (1970); [c]) Our work.

Distorted waves calculations for transitions to $1g_{7/2}$, levels at 0.27 MeV in $^{143}Pm$ and 0.329 MeV in $^{145}Eu$ fit the experimental angular distributions extremely well except at very forward angles where the calculated cross sections are smaller than the experimental cross sections. However, this discrepancy could be due to the large statistical errors present in the forward-angle data for $g_{7/2}$ states, since, as expected (see the Introduction) these states are not strongly populated.

Angular distributions are also shown for transitions to $1h_{11/2}$ levels at 0.96 MeV in $^{143}$Pm and at 0.716 MeV in $^{145}$Eu. The distorted waves calculation agrees well with the experimental points for the level at 0.716 MeV in $^{145}$Eu, but is out of phase by 3°-4° for the level at 0.96 MeV in $^{143}$Pm.

## Summary and conclusions

We have measured angular distributions for the differential cross sections of the proton transfer reactions $^{142}$Nd($^7$Li,$^6$He)$^{143}$Pm and $^{144}$Sm($^7$Li,$^6$He)$^{145}$Eu at 52 MeV incident $^7$Li energy. We have also measured the elastic scattering $^{142}$Nd($^7$Li,$^7$Li)$^{142}$Nd and $^{144}$Sm($^7$Li,$^7$Li)$^{144}$Sm at $E$($^7$Li) = 52 MeV and $^{142}$Nd($^6$Li,$^6$Li)$^{142}$Nd at $E$($^6$Li) = 46 MeV and $^{144}$Sm($^6$Li,$^6$Li)$^{144}$Sm at $E$($^6$Li) = 45 MeV.

We have carried out theoretical analysis of the elastic scattering using the conventional central, spherical optical model with volume absorption. We then used the determined interaction parameters in the theoretical analysis of proton transfer reactions using the exact finite-range distorted waves formalism.

The elastic scattering data were well described by the optical model. The distorted waves calculations generally described the corresponding transfer angular distributions well, although a slight phasing problem was encountered with $s_{1/2}$ states and more significantly with the $1h_{11/2}$ state at 0.96 MeV in $^{143}$Pm.

The absolute spectroscopic factors obtained here are slightly lower than those obtained from the light-ion ($^3$He,d) reactions, but the relative spectroscopic factors are in good agreement.

Heavy-ion forward-angle $j$ - dependence has been used to assign the following spins to $d$ - states in $^{143}$Pm: 0.0 MeV (5/2$^+$), 1.40 MeV (3/2$^+$); and in $^{145}$Eu: 0.0 MeV (5/2$^+$), 1.042 MeV (3/2$^+$). Spins could not be assigned to the $d$ - states at 1.76 MeV and 1.84 MeV in $^{145}$Eu due to the lack of forward-angle data. Previous spin assignments for these levels are given in the compilation by Lederer and Shirley (1978) as 0.0 (5/2$^+$) and 1.40 (3/2$^+$) in $^{143}$Pm; and 0.0 (5/2$^+$), 1.042 (3/2$^+$), 1.76 (3/2$^+$) and 1.84 (3/2$^+$) in $^{145}$Eu. The ($^7$Li,$^6$He) single-proton stripping reactions have been confirmed as a useful spectroscopic tool.

# 34

# Search for Structures in the $^{16}O$ + $^{24}Mg$ Interaction

***Key features:***

1. Our measurements of excitation functions at forward angles for the $^{24}Mg(^{16}O,^{12}C)^{28}Si$ $\alpha$ - transfer reaction has led to a discovery of three broad resonances. We have argued that these resonant structures, together with similar features at lower incident energies, are associated with nuclear molecular excitations.
2. In order to study further nuclear molecular excitations, we have now carried out measurements of excitation functions for the $^{16}O$ + $^{24}Mg$ elastic and inelastic scattering over a wide range of the incident $^{16}O$ energies at a forward angle.
3. We have found no correlation between excitation functions for the $^{24}Mg(^{16}O,^{12}C)^{28}Si$ reaction and $^{16}O$ + $^{24}Mg$ scattering. Thus, the forward angle excitation functions for the $^{16}O$ + $^{24}Mg$ system display no evidence of nuclear molecular excitations.
4. We have found that the excitation functions for the elastic and inelastic $^{16}O$ + $^{24}Mg$ scattering can be well described using coupled channels formalism.
5. Curious irregularities, which cannot be reproduced by coupled channels calculations, have been observed in the excitation functions corresponding to the 8.11 MeV excited state in $^{24}Mg$ and at 8.4 MeV excitation energy, which represents a group of excitations belonging to both $^{16}O$ and $^{24}Mg$. We argue that these irregularities cannot be associated with nuclear molecular excitations.
6. Our work shows that the resonances observed for the $^{24}Mg(^{16}O,^{12}C)^{28}Si$ reaction are most probably associated with processes in the exit channel.

**Abstract**. Excitation functions for the scattering of $^{16}O$ from $^{24}Mg$ with excitation energies ($E_x$) up to 8.4 MeV have been measured for $\theta_{lab}$ = 19.5° and 33.6 MeV < $E_{c.m.}$ < 49.2 MeV. Strong energy dependence is observed for states above the 8 MeV excitation energy. Coupled-channels calculations, which predict smooth energy variations, give good agreement with the data for $E_x$ < 8 MeV.

## Introduction

The $^{24}Mg(^{16}O,^{12}C)^{28}Si$ reaction, discussed in Chapter 30, has led to uncovering interesting and challenging features, described by a number of authors (Nurzynski *et al.* 1981; Paul *et al.* 1978; Sanders *et al.* 1985). In particular, gross structures (with widths $\Gamma_{c.m.}$~1-3 MeV) have been observed in excitation functions at forward angles in the case of the transition to the $^{28}Si$ ground state for the energy range 23 MeV ≤ $E_{c.m.}$ ≤ 53 MeV. These structures have been assigned parities, and in some cases spins, by analysis of both angular distributions (Nurzynski *et al.* 1981; Paul *et al.* 1978) and excitation functions at $\theta_{c.m.}$ = 0°, 90° and 180° (Paul *et al.* 1980; Sanders *et al.* 1980a, 1985). Furthermore, it has been found that the forward-angle structures are correlated for a number of $^{28}Si$ states up to 10 MeV excitation (Sanders *et al.* 1980b).

In view of these previous investigations, it seemed natural to look for correlations between the resonant structures at forward angles for the $^{24}Mg(^{16}O,^{12}C)^{28}Si$ reaction and the $^{16}O$ + $^{24}Mg$ interaction. Unfortunately, most of available data at that time were for backward angles (Clover *et al.* 1979; Lee *at al.* 1979; Paul *et al.* 1980), which were difficult to interpret. Structures at these angles are highly fractionated and show little apparent correlation with the forward-angle $\alpha$ - transfer structures (Sanders *et al.* 1985).

An alternative way is to look for correlations in elastic and inelastic scattering data at forward angles. One set of measurements (Mitting *et al.* 1974) has been reported for $^{16}O$ + $^{24}Mg$ elastic and inelastic scattering at $\theta_{lab}$ = 35° and 50° for 15 MeV ≤ $E_{c.m.}$ ≤ 39 MeV which may contain some resonant structure. Fulton *et al.* (1983) have measured excitation function for the $^{24}Mg(^{16}0,^{16}O')^{24}Mg^*$ ($2^+$ 1.37 MeV) scattering at 20° < $\theta_{lab}$ < 40° and 25 MeV < $E_{c.m.}$ < 39 MeV, which showed no resonant structure. However, it should be noted that their work covered only a part of the energy range for which structure has been observed for the $\alpha$ - transfer reaction. Furthermore, the simple band-crossing model predicts (Nurzynski *et al.* 1981) that the $2^+$ 1.37 MeV channel monitored by Fulton *et al.* should be active at $E_{c.m.}$ ~18MeV (i.e. ~7 MeV below the region investigated by Fulton *et al.*) and hence their measurement may be insensitive to the type of effect proposed by Nurzynski *et al.* (1981).

In the study described in this chapter, we have carried out measurements of excitation functions at $\theta_{lab}$ = 19.5° for the $^{16}O$ + $^{24}Mg$ elastic and inelastic scattering, which complement the work of Fulton *et al.* (1983) by extending the data up to $E_{c.m.}$ ~ 50 MeV for excitation energies up to 8.4 MeV. We have also carried out coupled channels analysis of the data.

## Experimental methods and results

In the measurements described here, the experimental arrangement was similar to that described by Nurzynski *et al.* (1981, 1982 see also Chapters 30 and 32). A beam of $^{16}O$ ions was provided by the Australian National University 14D Pelletron accelerator. Particles scattered from a thin (~5 μg/cm$^2$) $^{24}Mg$ (enriched to 99.92%) target were detected using a multi-element detector in the focal plane of a split-pole Enge spectrometer. The forward angle of 19.5° (lab) was chosen to be as close as possible to 5° (lab), the angle selected for the earlier ($^{16}O$,$^{12}C$) measurements (Nurzynski *et al.* 1981), whilst at the same time minimising the interference caused by scattering from carbon and oxygen contaminants in the target. The horizontal acceptance angle of the spectrometer was 1°, which ensured good resolution of contaminant peaks.

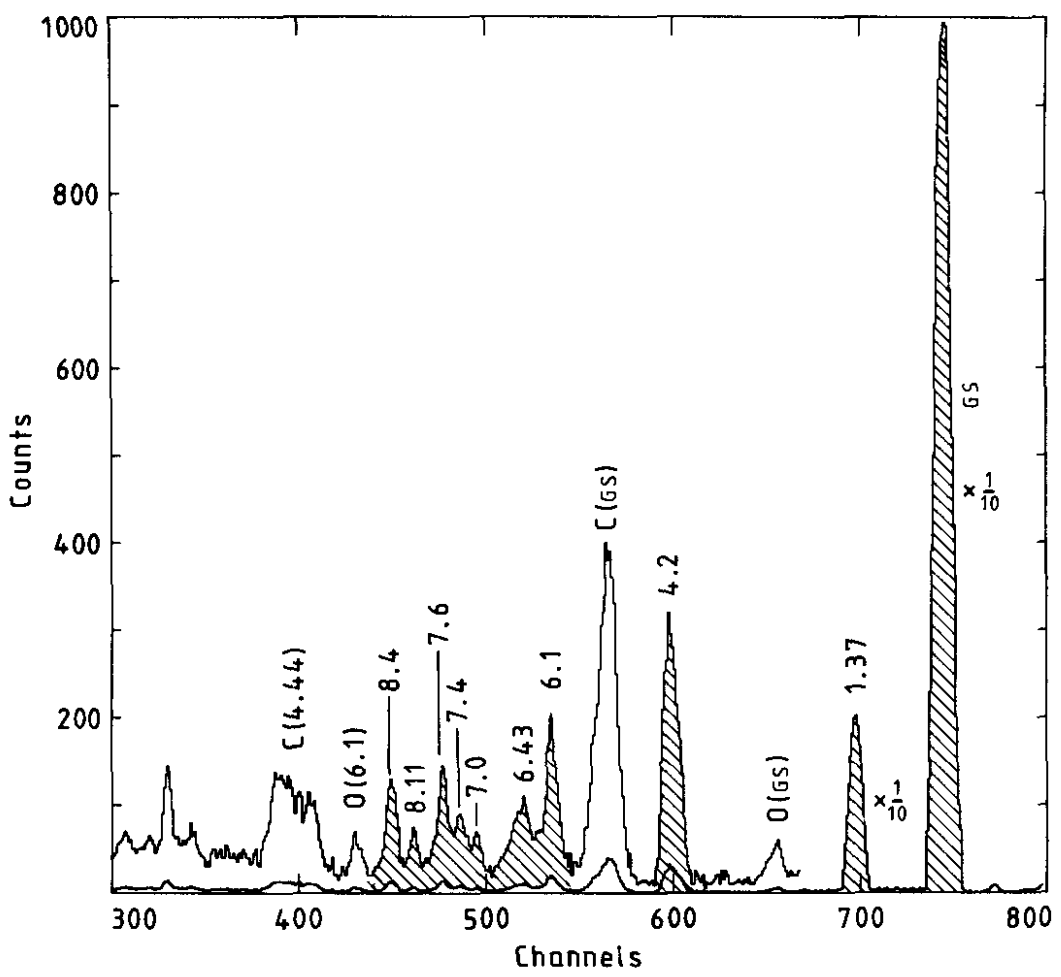

Figure 34.1. The energy spectrum of $^{16}O$ particles at $\theta_{lab}$ = 19.5° and $E_{c.m}$ = 41.4 MeV. Shaded peaks correspond to the $^{16}O$ + $^{24}Mg$ scattering and indicate groups of particles (labelled according to their excitation energies) for which excitation functions were measured. Scattering from carbon and oxygen contaminants in the target is also indicated in the spectrum.

The energy resolution of peaks corresponding to scattering from $^{24}Mg$ was ≤140 keV. A typical $^{16}O$ spectrum is shown in Figure 34.1. Oxygen contamination in the target was small. However, some interference form $^{16}O$ + $^{16}O$ ($E_x$ = 6.05/6.13 MeV) scattering was present in the 8.4 MeV group at $E_{c.m.}$<37.2 MeV.

The incident beam intensity was measured using a beam current integrator. A solid-state detector at 15° was also used as a target monitor. The data were recorded event by event on to magnetic tapes and care was taken to obtain acceptable statistics for weakly excited states. The peak intensities for close-lying groups were extracted by fitting Gaussian distributions.

Table 34.1

The observed particle groups identified by their $|Q|$ - values for the $^{16}O$ + $^{24}Mg$ scattering and their contributing components.

| Group $\|Q\|$-value (MeV) | Possible states $^{16}O$ $J^\pi$ energy (MeV) | $^{24}Mg$ $J^\pi$ energy (MeV) | Total excitation energy (MeV) |
|---|---|---|---|
| 0.00 | $0^+$ 0.00 | $0^+$ 0.00 | 0.00 |
| 1.37 | $0^+$ 0.00 | $2^+$ 1.37 | 1.37 |
| 4.2 | $0^+$ 0.00 | $4^+$ 4.12 | 4.12 |
| | $0^+$ 0.00 | $2^+$ 4.24 | 4.24 |
| 6.1 | $0^+$ 0.00 | $4^+$ 6.01 | 6.01 |
| | $0^+$ 6.05 | $0^+$ 0.00 | 6.05 |
| | $3^-$ 6.13 | $0^+$ 0.00 | 6.13 |
| 6.43 | $0^+$ 0.00 | $0^+$ 6.43 | 6.43 |
| 7.0 | $2^+$ 6.92 | $0^+$ 0.00 | 6.92 |
| | $1^-$ 7.12 | $0^+$ 0.00 | 7.12 |
| 7.4 | $0^+$ 0.00 | $2^+$ 7.35 | 7.35 |
| | $0^+$ 6.05 | $2^+$ 1.37 | 7.42 |
| 7.6 | $3^-$ 6.13 | $2^+$ 1.37 | 7.50 |
| | $0^+$ 0.00 | $1^-$ 7.55 | 7.55 |
| | $0^+$ 0.00 | $3^-$ 7.62 | 7.62 |
| 8.11 | $0^+$ 0.00 | $6^+$ 8.11 | 8.11 |
| 8.4 | $2^+$ 6.92 | $2^+$ 1.37 | 8.29 |
| | $0^+$ 0.00 | $3^-$ 8.36 | 8.36 |
| | $0^+$ 0.00 | $4^+$ 8.44 | 8.44 |
| | $0^+$ 0.00 | $1^-$ 8.44 | 8.44 |
| | $1^-$ 7.12 | $2^+$ 1.37 | 8.49 |

Figure 34.2 shows the measured excitation functions for the scattering of $^{16}O$ from $^{24}Mg$. The displayed excitation energies ($E_x$) are for the groups of particles identified by their $|Q|$ - values in Table 34.1. The error bars include both statistical and background subtraction errors. The known excited states contributing to these groups of $^{16}O$ particles are listed in Table 34.1.

In general, the data display smooth energy dependences in agreement with earlier lower energy measurements at forward angles for the 1.37 MeV state (Fulton *et al.* 1983). However, the excitation functions corresponding to $E_x$ = 8.11 and 8.4 MeV display strong energy-dependent structures. A data point at 42.6 MeV (c.m.) for $E_x$ = 8.11 MeV cannot be shown. At this energy, no peak corresponding to $E_x$ = 8.11 MeV in the spectrum could be distinguished from the background. The observed intensities for this group were also small at energies < 31 MeV. The vertical lines in Figure 34.2 indicate the energies at which peaks occur in the forward-angle $^{24}Mg(^{16}O,^{12}C)^{28}Si$(g.s.) yield (Nurzynski *et al.* 1981, Sanders *et al.* 1985). As can be seen there is no obvious correlation between the present measurements for the $^{16}O$ + $^{24}Mg$ scattering and the results for the $^{24}Mg(^{16}O,^{12}C)^{28}S$ reaction.

An attempt was made to measure excitation functions for $^{16}O$ + $^{24}Mg$ scattering at other forward angles. However, the yield for the $E_x$ = 8.11 MeV group was small at these anglers and the contaminant interference was prohibitively high.

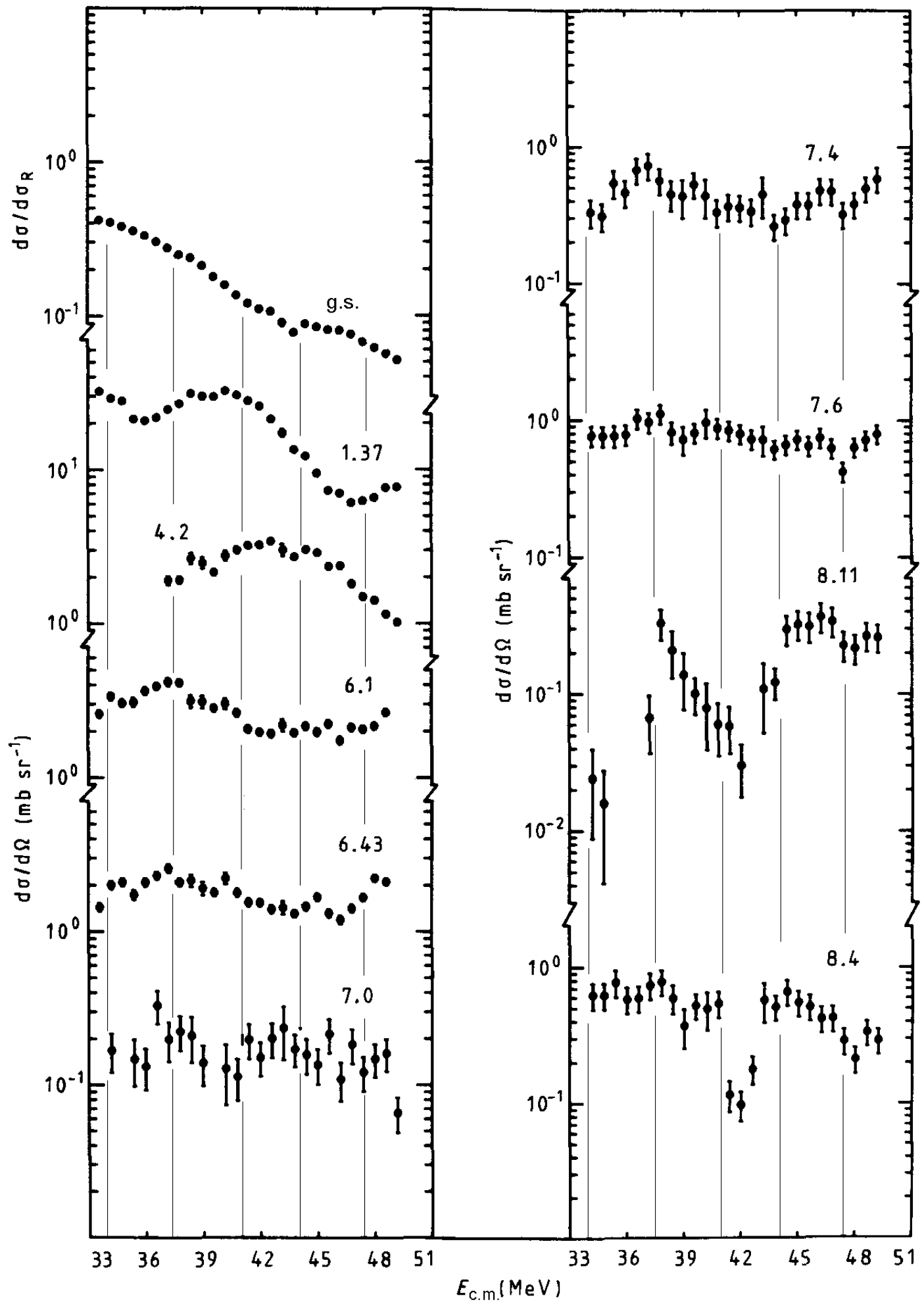

Figure 34.2. Measured excitation functions for the scattering of $^{16}O$ from $^{24}Mg$ with 0 ≤ $E_x$ ≤ 8.4 MeV (see Table 34.1) at $\theta_{lab}$ = 19.5°. The vertical lines indicate the energies at which peaks occur in the forward-angle excitation function for the $^{24}Mg(^{16}O,^{12}C)^{28}Si$ (g.s.) reaction (Nurzynski *et al.* 1981, Sanders *et al.* 1985).

## Theoretical analysis

Coupled-channels calculations for the elastic and inelastic scattering of $^{16}O$ from $^{24}Mg$ were performed using the computer code ECIS (Raynal 1972, 1981)[5]. The $0^+$ (g.s.), $2^+$ (1.37 MeV), $4^+$ (4.12 MeV), $2^+$ (4.24 MeV), $3^+$ (5.24 MeV), $6^+$ (8.11 MeV) and $4^+$ (8.44 MeV) states in $^{24}Mg$ were included in the calculations. The deformed optical potential employed for each channel was taken to be potential 2 of Nurzynski *at al.* (1983, see also Chapter 32), which was determined by fitting angular distributions at $E_{c.m.}$ = 43.5 MeV for the lowest four levels of $^{24}$Mg. The shape parameters for the nuclear potential were taken to be the same as determined by Nurzynski *et al.* (1983) i.e. $\beta_2$ = 0.25, $\gamma$ = 22° and $\beta_4$ = -0.065 (see Chapter 32). The wave functions for the $^{24}$Mg states were obtained using the rigid asymmetric rotor model of Davydov and Filippov (1958) as extended by Baker (1979) and Barker *et al.* (1979) to include a symmetric hexadecapole deformation, with shape parameters consistent with those employed in the deformed optical potential. All the calculations employed the Coulomb correction technique developed by Raynal (1980, 1981) with matching at a radius of 15.2 fm.

[5] See Chapter 16.

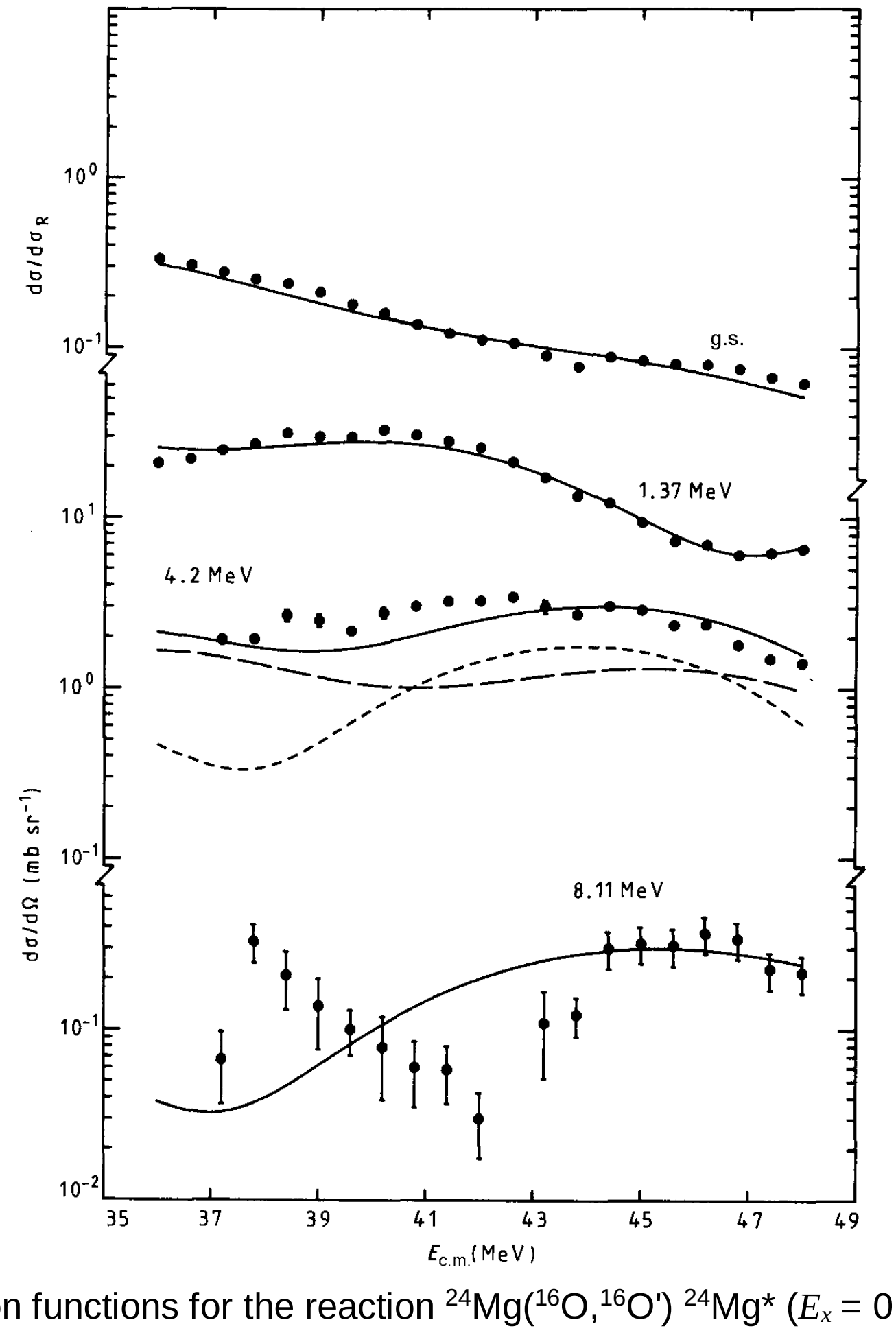

Figure 34.3. Excitation functions for the reaction $^{24}$Mg($^{16}$O,$^{16}$O') $^{24}$Mg* ($E_x$ = 0, 1.37, 4.2 and 8.11 MeV) at $\theta_{lab}$ = 19.5°. Coupled-channels calculations (full lines) are compared with the data. For the unresolved 4.2 MeV doublet, the full curve is the sum of the calculated contributions from the 4.12 MeV (dotted curve) and the 4.24 MeV (broken curve) states.

Figure 34.3 shows the coupled-channels predictions (full curves) for several excitation functions compared with the data at $\theta_{lab}$ = 19.5° and for 36 MeV ≤ $E_{c.m.}$ ≤ 48 MeV. In the case of the unresolved 4.2 MeV doublet, the full curve is the sum of the calculated contributions from the 4.12 MeV (dotted curve) and the 4.24 MeV (broken curve) states of $^{24}$Mg. It should be noted that these calculations employed the same potential parameters as given by Nurzynski *et al.* (1983) with no modification for changes in the bombarding energy.

As can be seen, coupled channels calculations produce smooth functions with no indication of any resonant structure. The broad maxima and minima arise purely from angular effects associated with the oscillatory nature of the angular distributions. In these calculations, nuclear potential effects are dominant. This can be inferred from the angular distribution data (Nurzynski *et al.* 1983) at $E_{c.m.}$ = 43.5 MeV, which show that at $\theta_{lab}$ = 19.5° ($\theta_{c.m.}$ = 32.36°) the Rutherford ratio has fallen to approximately 0.1.

Figure 34.3 shows that coupled-channels calculations give a good description of both the magnitudes and the general features of the data for the lowest three excitation energies. For the 8.11 MeV state, the calculations predict only the general magnitude of the cross section. However, the observed structure of the data is not reproduced and, within the present model, the structure does not arise from an angular effect.

The calculated excitation functions for the 5.24 MeV and 8.44 MeV states of $^{24}$Mg (not shown in Figure 34.3) were also found to have smooth behaviours. In the case of the 5.24 MeV state, the calculated cross section was found to be <0.06 mb/sr, which is consistent with the fact that this state was generally not observed. For the 8.44 MeV state, the predicted excitation function lies below the data. This is not surprising, since the data for the 8.4 MeV excitation function are believed to include contributions from five unresolved states (see Table 34.1).

## Discussion and summary

The anomalous structure observed for the $6^+$ 8.11 MeV (and possibly for the 8.4 MeV) excitation function is not readily explained. If the structure is interpreted as arising from shape resonances in the ion-ion potentials, it is necessary for the potential associated with the $6^+$ 8.11 MeV channel to be significantly more surface transparent than that employed in the present coupled-channels calculations in order to enhance the grazing partial waves in this channel. Such a large difference between the absorptive strength for the $6^+$ channel and for channels with lower excitation energies seems unlikely and artificial. Moreover, in this picture one should expect to observe strong correlations with the resonant structures found in the $\alpha$ -transfer reaction at forward angles (Nurzynski *et al.* 1981; Paul *et al*. 1978; Sanders *et al.* 1985). However, as shown in Figure 34.2, no such correlations are present in the $^{16}$O + $^{24}$Mg excitation functions.

An alternative interpretation (Nurzynski *et al.* 1981) ascribes the forward-angle $\alpha$ -transfer structures to a fragmentation of the potential shape resonances, arising from coupling between the elastic and inelastic channels of both the initial and final systems. Since there is no evidence of a $\beta_6$ deformation parameter which could alter the $6^+$ inelastic scattering cross section (Nurzynski *et al.* 1983, see also Chapter 32), this interpretation requires the additional coupling of intermediate $2^+$ or $4^+$ states, via $\beta_2$ and $\beta_4$ terms. However, as the data indicate no significant correlated structures in the excitation functions for these intermediate states, this explanation (which again requires a more surface-transparent potential for the $6^+$ inelastic channel) appears unlikely.

Concerning the gross resonant structures observed in the $^{24}$Mg($^{16}$O,$^{12}$C)$^{28}$Si reaction at forward angles, it is interesting to note that the fusion cross sections for the $^{12}$C + $^{28}$Si system exhibit some resonance-like structures (Racca *et al.* 1983) while those for the $^{16}$O + $^{24}$Mg system show only weak and uncorrelated structures (Racca et al. 1982).

These facts combined with the present work, which shows smooth excitation functions described well by the coupled-channels calculations, suggest that the structures seen in the $\alpha$ - transfer reaction may be associated with resonating processes in the exit $^{12}C$ + $^{28}Si$ channel rather than with those in the entrance channel.

# 35

## Parity-dependent Interaction

***Key features:***

1. We have measured excitation functions for the $^{16}O$ + $^{20}Ne$ (g.s.) and $^{16}O$ + $^{20}Ne^*$ (1.634 MeV) in the energy range of 21.5 MeV < $E_{c.m.}$ < 31.2 MeV.
2. We have analysed our $^{16}O$ + $^{20}Ne$ (g.s.) experimental results and the results of Schimizu *et al.* (1982) using an extended optical model potential containing a parity-dependent interaction.
3. We have found that the interaction parameters determined in previous studies (Kondo, Robson, and Smith 1985) resulted in an inadequate description of experimental data.
4. We have found new interaction parameters (potential 3, Table 35.1) which gives excellent description of our $^{16}O$ + $^{20}Ne$ data. The potential contains a negative parity-dependent component, which means that it is deeper for the even partial waves. The potential assigns even spin shape resonances $J$ = $16^+$, $18^+$, and $20^+$ at $E_{c.m.}$ = 22.5, 25.4, and 28.5 MeV, respectively.
5. We have also carried out parity independent calculations but included explicitly contributions from $\alpha$ – transfer. These calculations reproduce the shape of the excitation functions reasonably well and thus support the claim (see Chapter 24 and references in the text) that the parity-dependent interaction simulates coupling to non-elastic channels.

**Abstract**: The excitation functions for the $^{16}O$ + $^{20}Ne$ elastic scattering at $\theta_{c.m.}$ = 90° and for the $^{16}O$ + $^{20}Ne^*$ ($E_x$ = 1.634 MeV) inelastic scattering corresponding to $\theta_{c.m.}$ = 90.95° - 91.45° have been measured over the energy range 21.5 MeV < $E_{c.m.}$ < 31.2 MeV. The $^{16}O$ + $^{20}Ne$ elastic scattering was analysed within the framework of an extended optical model, in order to place constraints on spin assignments to resonant states. Excellent description is obtained with a potential, which is deeper for the even partial waves.

## Introduction

Resonant phenomena in heavy-ion interactions have been observed for systems involving *sd* - shell nuclei for elastic, inelastic and a variety of transfer channels (see Chapter 30). One interesting case is the $^{16}O$ + $^{20}Ne$ system, for which prominent structures have been observed in the excitation functions at forward and backward angles for the elastic scattering, inelastic scattering to the $^{20}Ne(2_1^+)$ and $^{20}Ne(4_1^+)$ states, and for the $^{20}Ne(^{16}O,^{12}C)^{24}Mg$ reaction (Schimizu *et al.* 1982).

These data have been described using optical model and coupled channel techniques, the essential feature of which is that the ion-ion potentials employed include both the $J$ - dependent absorptive term and a real parity-dependent interaction. It has been argued that the $J$ - dependent absorption arises from the requirement that angular momentum and energy must be simultaneously conserved in the open reaction channels (Chatwin *et al.* 1970), whilst the parity dependence is to be expected as a consequence of the Pauli principle (Baye, Deenen, and Salomon 1977).

The parity dependent term has the effect of staggering the shape resonances, which arise for surface partial waves, to form "doublets" of resonances. The $J$ - dependent absorptive term is suitably transparent to surface partial waves so that the "doublets" appear as unresolved gross structures in the predicted $^{16}O + ^{20}Ne$ elastic scattering excitation function at backward angles, in accord with experiment.

It should be noted, however, that in determining the values of parameters used in the potentials for such analyses, it is necessary to make $J^{\pi}$ assignments to peaks observed in $^{16}O + ^{20}Ne$ scattering and $^{20}Ne(^{16}O,^{12}C)^{24}Mg$ reaction excitation-function data (Schimizu *et al.* 1982). The assignments made in optical model analyses were based upon assignments of $18^+$, $19^-$ and $24^+$ to observed peaks at $E_{c.m.}$ = 24.5, 25.4 and 35.5 MeV respectively, these assignments being consistent with values extracted by Schimizu *et al.* (1982), using $P_J^2(\cos\theta)$ comparisons with measured angular distributions. Furthermore, to avoid the possibility of having too many resonances within the energy range of interest, it was assumed that each resonance of a given spin and parity was that with the lowest energy.

An ambiguity arises, however, when it is acknowledged that a $P_J^2(\cos\theta)$ fitting technique allows an assignment of spin values to $\pm 1\hbar$ at very best. For this reason, optical potentials were also determined using alternative assignments of spins to observed peaks allowed within such a $\pm 1\hbar$ uncertainty. The main consequence of such a one-unit change in spin assignment is a reversal of the sign of the parity dependent interaction in the ion-ion potential. Investigations of the available $^{16}O + ^{20}Ne$ elastic scattering and $^{20}Ne(^{16}0,^{12}C)^{24}Mg$ (g.s.) data were unable to resolve conclusively this ambiguity in sign.

Studies by Kondo *et al*. (1985) re-examined the $^{20}Ne(^{16}0,^{16}O)^{20}Ne$ excitation function data within the framework of a model, which attributed the parity dependent term in the optical model calculations to a requirement to simulate the effects of $\alpha$ - transfer amplitudes. The results of that work supported strongly a choice of sign for the parity dependent term, which would be consistent with spin assignments of $17^-$, $18^+$ and $23^-$ to the observed peaks at $E_{c.m.}$ = 24.5, 25.4 and 35.5 MeV respectively, rather than the $18^+$, $19^-$ and $24^+$ assignments made previously.

In order to resolve the above ambiguity, we have undertaken measurements and analysis of the $\theta_{c.m.}$ = 90° excitation function for $^{16}O + ^{20}Ne$ elastic scattering over the energy range 21.5 MeV < $E_{c.m.}$ < 31.2 MeV. Such measurements and analyses place additional constraints on the resonance spins and hence constraints on the parity dependent part of the $^{16}O + ^{20}Ne$ potential. Data have been taken also for the excitation of the 1.634 MeV ($2^+$) state of $^{20}Ne$.

## Experimental procedure and results

A beam of $^{16}O$ ions from the Australian National University 14UD Pelletron accelerator was used to bombard natural neon target retained by thin windows in a small gas cell shown in Figure 35.1.

The design requirements for the gas cell and detector system were an angular aperture in the horizontal plane of <1° (lab) and an angular positioning accuracy of 0.1° or better at the required laboratory angle of 51.34°. The reaction energy at the centre of the gas target should be calculable to better than 0.1 MeV (lab) and with a range of reaction energies limited to a spread of the order of 0.2 MeV (lab). The detector should be able to resolve elastic scattering by $^{20}Ne$ from inelastic scattering leaving $^{20}Ne$ in its first excited state at 1.634 MeV and from elastic scattering by the $^{22}Ne$ component (9.2%) of natural neon. These design requirements were achieved.

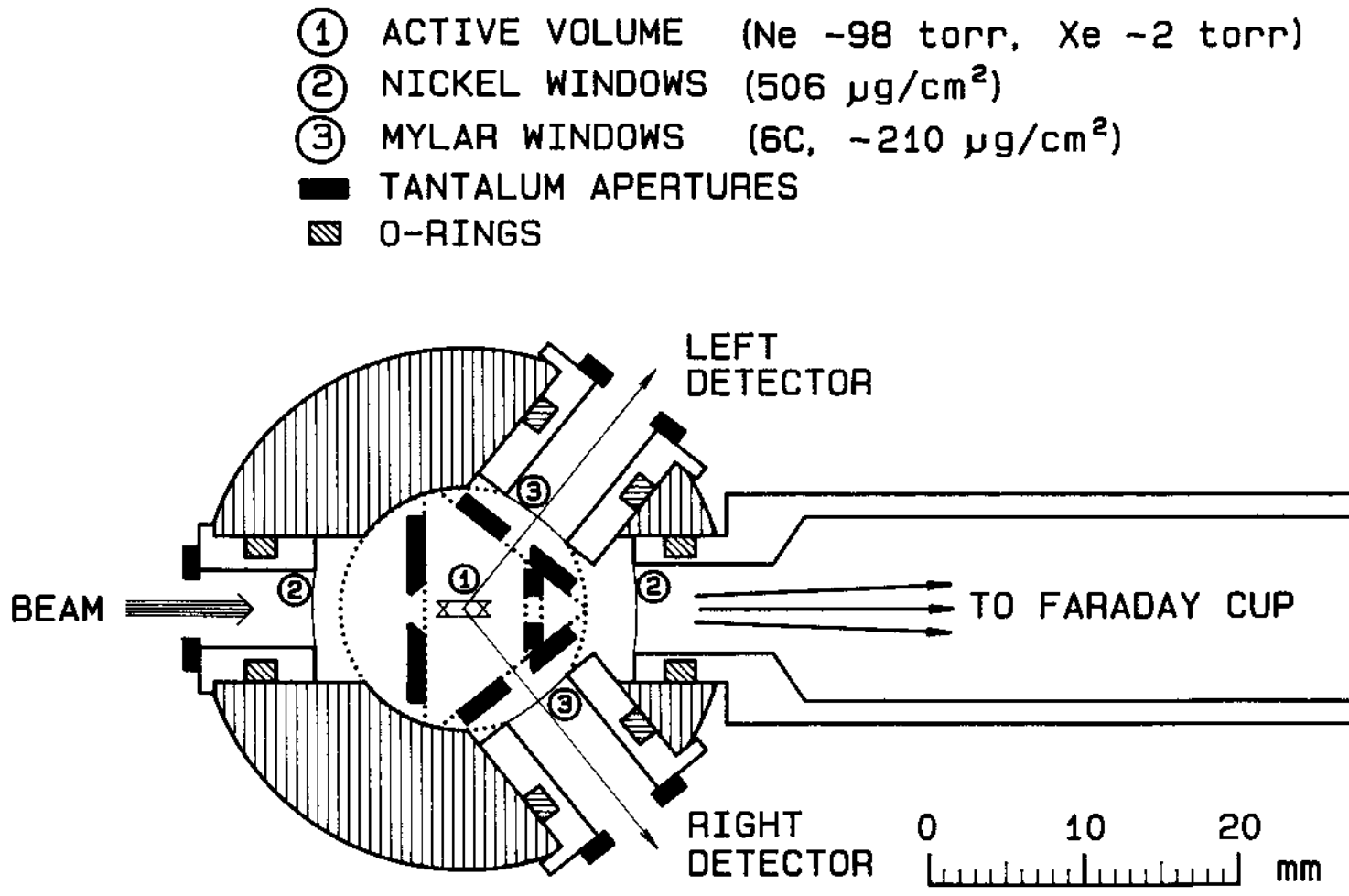

Figure 35.1. A horizontal section through the gas target at the beam height, showing the windows, the slits mounted on a cylindrical plug, anti-scatter apertures, windows and the defined reaction volume.

The incoming $^{16}O$ beam was collimated to 3 mm vertically by 1 mm horizontally by two collimators, one 700 mm and the other 106 mm upstream from the centre of the main target chamber. Optical alignment to a precision of <0.1 mm ensured the co-linearity of the two collimators and the centre of the main target chamber horizontally in the incoming beam direction. The alignment precision in the vertical direction was better than 0.2 mm.

The gas cell was then aligned, using the following procedure: The gas cell body was mounted centrally in the main target chamber at the correct height. Its beam inlet and outlet ports were mechanically aligned, by rotation, to the direction of the nearer collimator and the cell body was clamped. A cylindrical plug carrying the target length-defining slits and the anti-scatter slits was inserted along the vertical axis of the gas cell and was mechanically aligned, using the slits themselves, by rotation to the beam axis now defined by the beam inlet port. These slits, mounted on plane faces milled parallel to the cylindrical plug axis, had been previously aligned during assembly. All mechanical alignments were performed with accuracy better than 0.1 mm. The beam collimators ensured that sideways excursions of the beam were limited to less than ± 0.2 mm with an expected half-intensity at ±0.1 mm. Vertical beam excursions were limited to ±1 mm by an aperture mounted on the plug carrying the beam entrance window.

The gas target assembly was completed by the insertion into the target cell body of plugs carrying the thin windows. These were then covered by anti-scatter apertures. The $^{16}O$ beam enters the gas target through a nickel window and leaves through another nickel window. The ejectiles leave through Mylar windows (210 (μg/cm$^2$). The energy losses of these windows were determined in a separate experiment, measuring to an accuracy of 5 keV the loss after transmission of 48.83 MeV $^{16}O$ ions through the windows.

Two detectors were provided, for increased count rates and for redundancy checking. They each consisted of a tantalum entrance aperture, 5x2 mm, a 210 μg/cm$^2$ Mylar entrance window, 28mm length of isobutane gas at 50 Torr (constant flow) in a gridded ionisation chamber and a silicon surface barrier detector for residual energy measurement. The performance of these detectors was such that ejectiles of different $Z$ could be readily distinguished and the required overall energy resolution was attained. They were mounted at ±51.35 ± 0.05° (lab) with the entrance apertures of 133.1 mm from the centre of the gas target, subtending a solid angle of 0.565 msr each.

Slits inside the gas target were mounted 5 mm from the target centre with an aperture of 2 mm. Combined with the detector aperture, these defined a length along the beam axis of 2.56 mm for which the full detector aperture was visible and a further ~0.1 mm at each end with a reduced detector aperture. The available angular range distribution (taking the solid angle into account) was approximately triangular with 75% of the distribution within ±0.43° (lab) and the remainder within ±0.86°. The count rates in the individual detectors were influenced by sideways misalignment of the beam, by differences in the target defining geometries, in the detector solid angles and in the detector angular settings.

An early run disclosed unequal count rates in the two detectors for Rutherford scattering from xenon. An optical-mechanical check showed a zero error of 0.25° in the angular scale used inside the main target chamber, compared with the established beam axis. Compensation for this zero error then gave detector count rate comparisons agreeing to better than 1 %.

The evacuated gas target and ballast volume (6.5 litre) were first filled in a reproducible way to ~2 Torr by expansion of a limited volume of xenon gas at bottle pressure. The amount of xenon transferred decreased by 0.5% for each fill but the absolute pressure of xenon was only known to ±10%. The target and ballast volume were then filled to ~100 Torr total pressure with natural neon. This pressure was known to 3%. Because of a slow leak, the total pressure was recorded before and after each run to provide an average pressure from which the reaction energy was calculated. Intermittently it was necessary to top up with neon gas and at such times the partial pressure ratio of xenon and neon was recalculated.

It was important that the reaction energy be reliably calculated. A computer program, which reproduced the known beam energy loss through the nickel entry window, was used to calculate energy loss and straggling to the reaction volume and followed the ejectile through to the residual energy detector. The nickel window gave energy losses from 1.55 to 2.24 MeV over the range of beam energies used. Calculations for the individual gas fills gave a further loss of 0.5 to 0.6 MeV to the target centre over the same beam energy range. Calibrations of the residual energy detectors based on observed pulse height responses and calculated energies were linear from 5.4 to 45.5 MeV, passing through zero as expected. An unsatisfactory calibration would lead to non-linearity and a non-zero extrapolation. Therefore, we are confident that the first ~0.5 MeV of energy loss to the reaction volume has been calculated satisfactorily. The same computer program was also useful in identification of the elastic or inelastic scattering in the cases where only one of these spectrum peaks was clearly visible. The predicted position was always found to be within the FWHM range when identification was no problem. Figure 35.2 shows a spectrum from one of the detectors taken at the lowest beam energy measured, where the resolution between the particle groups was poorest but the intensity of both groups was high. This spectrum has had the identification requirement for $^{16}O$ particles imposed upon it.

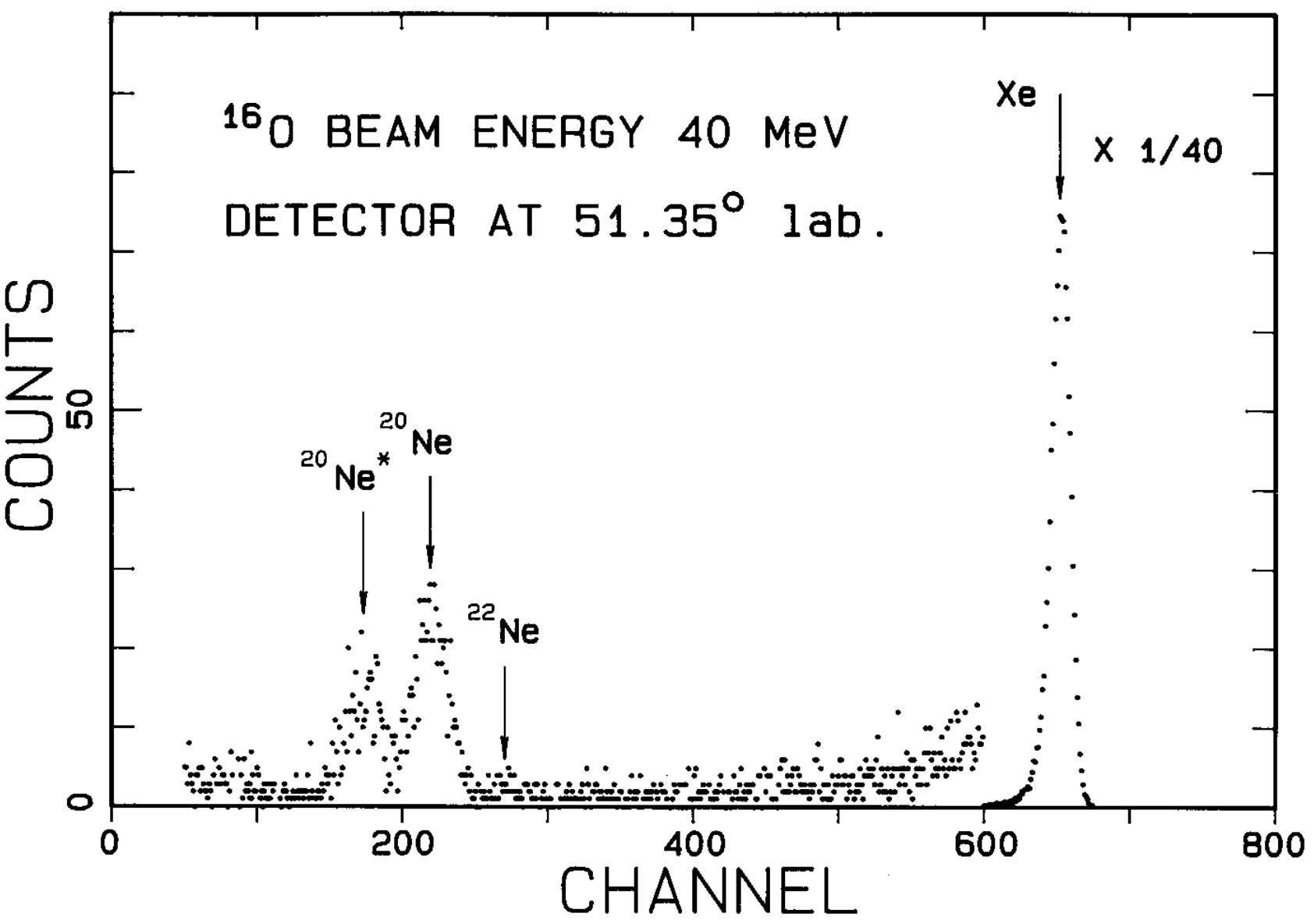

Figure 35.2. A spectrum of $^{16}O$ ions scattered from the reaction volume and entering the residual energy detector. Other ion species have been rejected by the $\Delta E - E_{residual}$ requirement. The inelastically scattered group leaving $^{20}Ne$ in its first excited state enters the residual energy detector with only 5.4 MeV for $^{16}O$ beam energy of 40.0 MeV.

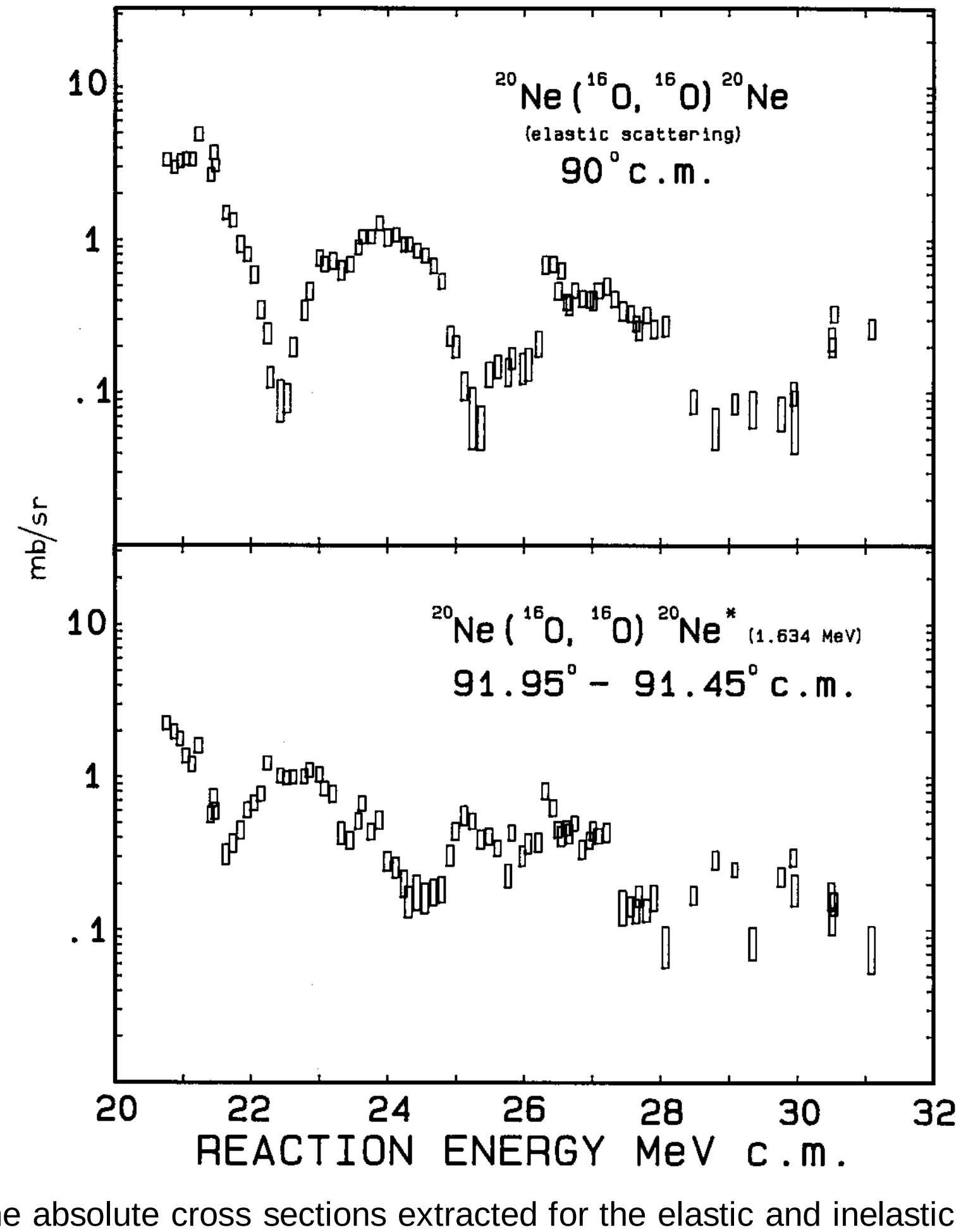

Figure 35.3. The absolute cross sections extracted for the elastic and inelastic scattering of $^{16}O$ from $^{20}Ne$. The rectangle widths show the range of energy averaging while the rectangle heights are ±1 standard deviation total error estimates. Relative errors, excluding the contribution from the absolute xenon pressure, are somewhat smaller.

From spectra such as these, the yields of scattered $^{16}O$ from $^{20}Ne$ and xenon were extracted. Data were only accepted if there were no significant differences between the estimated backgrounds or between any of the intensities of the three peaks mentioned above when spectra from the two independent detectors were compared. The absolute cross sections for elastic and inelastic scattering from $^{20}Ne$ are dependent only on the intensities of the $^{20}Ne$ peaks extracted, the intensity of Rutherford scattering from the xenon isotopes and the ratio of partial pressures for xenon and for $^{20}Ne$. The calculated mean reaction energy enters through the calculation of the Rutherford scattering cross section for the xenon. The only significant contributions to uncertainties in these absolute cross sections come from the $^{20}Ne$ peak statistics and from the uncertainty (±10%) in the absolute pressure of xenon. All other contributions are small compared with these.

Figure 35.3 shows these cross sections as a function of reaction energy as rectangles. The height of a rectangle shows ±1 standard deviation for the absolute cross section determinations while the width of a rectangle shows the calculated range of reaction energies over which the data are averaged.

## Theoretical analysis

The measured $\theta_{c.m.}$ = 90° (Figure 35.3) and $\theta_{c.m.}$ = 154°± 2° (Schimizu *et al.* 1982) excitation functions for elastic scattering of $^{16}O$ ions from $^{20}Ne$ have been analysed in terms of an extended optical model. The $^{16}O$ + $^{20}Ne$ potential was assumed to have the following form:

$$U(r) = V_c(r) + \left[V(E_{c.m.}, L) + iW(J)\right] f(r)$$

where $V_c(r)$ is the Coulomb potential for a uniform charge distribution of radius $R$ and *f(r)* is the usual Woods-Saxon form factor with diffuseness $a_0$ and radius

$$R = r_0[(16)^{1/3} + (20)^{1/3}]$$

In the above, $L$ is the relative orbital angular momentum and $J$ is the total angular momentum. For the elastic scattering of spinless projectiles $J = L$.

The depth of the real nuclear potential is given by

$$V(E_{c.m.}, L) = V_0 + V_E E_{c.m.} + (-1)^L V_\pi$$

and consists of three terms: a constant $V_0$; an energy dependent term; and an explicit parity dependence.

The depth of the imaginary nuclear potential is given by

$$W(J) = W_0\left\{1 + \exp\left[\left(J - J_c\right)/\Delta\right]\right\}^{-1}$$

where $J_c$ is a cut-off angular momentum and $\Delta$ is a diffuseness parameter. For each energy, $J_c$ is parameterised by the expression (Chatwin *et al.* 1970):

$$J_c = \overline{R}[(2\mu/\hbar^2)(E_{c.m.} - \overline{Q})]^{1/2}$$

where $\overline{R}$ and $\overline{Q}$ represent average values of the radius and the threshold energy for the predominant non-elastic reactions, respectively, and $\mu$, is the reduced mass of the system.

Parameter sets used in our theoretical analysis are listed in Table 35.1.

Table 35.1
Optical model parameters

| Potential | $V_0$ (MeV) | $V_E$ (MeV) | $V_\pi$ (MeV) | $W_0$ (MeV) | $\bar{R}$ (fm) | $\bar{Q}$ (MeV) |
|---|---|---|---|---|---|---|
| 1 | −12.99 | −0.62 | 0.99 | −7.60 | 8.48 | 15.04 |
| 2 | −12.49 | −0.47 | −0.96 | −10.50 | 8.27 | 15.70 |
| 3 | −5.51 | −0.75 | −0.82 | −8.00 | 8.21 | 14.74 |

For all potentials $r_0$ = 1.25 fm, $a_0$ = 0.60 fm and $\Delta$ = 0.80.

### *Initial optical model analysis*

As a starting point, we have used two potential sets suggested by previous optical model studies (Kondo, Robson, and Smith 1983) of $^{16}O$ + $^{20}Ne$ elastic scattering. They are listed as potentials 1 and 2 in Table 35.1. Potential 1 predicts $18^+$, $19^-$ and $24^+$ resonances at $E_{c.m.}$ = 24.5, 25.4 and 35.5 MeV, respectively, while potential 2 predicts $17^-$, $18^+$ and $23^-$ resonances at these three energies. As can be seen, potential 1 has a positive value of $V_\pi$, whereas potential 2 has a negative value of $V_\pi$.

Based upon potentials 1 and 2, two types of calculation were performed and compared with data. Firstly, optical model calculations were carried out using the full parity dependent potentials 1 and 2. These results were complemented by calculations of the type reported Kondo *et al.* (1985) where the parity-independent parts of potentials 1 and 2 of Table 35.1 (i.e. $V_\pi$ set to zero) were used, and the calculated $\alpha$ - transfer amplitude at angle ($\pi$ - $\theta$) was added coherently to the calculated elastic scattering amplitude at angle $\theta$. Such calculations allow investigation of whether the parity dependent term in optical model calculations might arise from a requirement to simulate the effects of $\alpha$-transfer amplitudes.

Using both approaches to the calculation, predictions were obtained for the $^{20}Ne(^{16}0,^{16}O)^{20}Ne$ excitation function at $\theta_{c.m.}$ = 90° and $\theta_{c.m.}$ = 154°±2°, and comparison made with present measurements and with the data reported Schimizu *et al.* (1982).

Optical model calculations of the elastic scattering excitation function for potentials 1 (dash curves) and 2 (solid curves) are compared with the measured excitation functions at $\theta_{c.m}$ = 154°±2° (panel (a)) and $\theta_{c.m}$ = 90° (panel (b)) in Figure 34.4. The short vertical arrows in the figure represent the energies of the relevant shape resonances in each case. (In panels (a) and (b), the numbers above and below the arrows are the spins of the resonating partial waves for potentials 1 and 2, respectively.)

As can be seen, both potentials provide similar predictions of the excitation function at $\theta_{c.m}$ = 154° ± 2°, with unresolved gross structures being predicted at the energies of each "doublet" of shape resonances. The excitation function predictions at $\theta_{c.m}$ =

90° however, show significant sensitivity to the optical potential used, with a preference for potential 2.

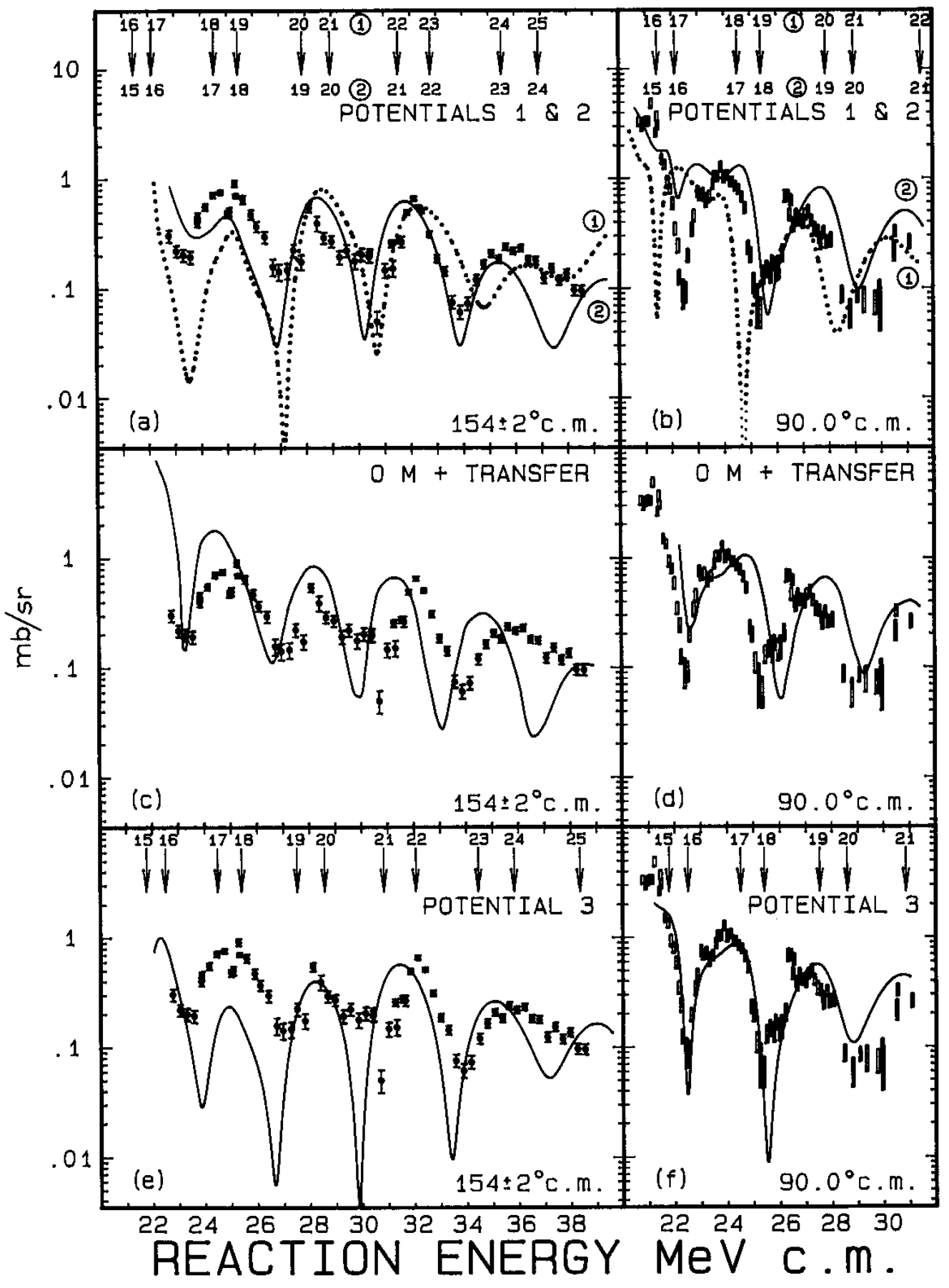

Figure 35.4. Excitation functions at $\theta_{c.m}$ = 154°± 2° (left-had side) and 90° (right-hand side) for $^{16}O$ + $^{20}Ne$ elastic scattering. Panels (a) and (b) show the theoretical results for potentials 1 and 2 compared (Table 35.1) with the data. Panels (c) and (d) show the results of a coherent sum of elastic scattering and ground state $\alpha$ - transfer amplitudes calculated using the parity-independent parts of potential 2. Panels (e) and (f) show the results for potential 3. The short vertical arrows indicate the positions of shape resonances and the respective $J = L$ values for the resonating partial waves.

### *Transfer reaction contributions*

Calculations, which explicitly take into account the ground-state $\alpha$ - transfer amplitudes in the prediction of the elastic scattering excitation function were also performed, and are compared with the measured excitation functions at $\theta_{c.m}$ = 154°±2° (panel (c)) and $\theta_{c.m}$ = 90° (panel (d)) in Figure 35.4. In these calculations, we used the parity-independent part of potential 2 of Table 35.1 (i.e. $V_\pi$ set to zero). It should be noted that equivalent calculations using potential 1 were reported (Kondo *et al.* 1985) and were rejected as they were unable to describe the $\theta_{c.m}$ = 154°± 2° excitation function satisfactorily.

Figure 35.4 shows that the transfer contribution model, which predicts the magnitude and general structure of the data at $\theta_{c.m}$ =154°±2° (panel (c)), also provides a good description of the data at $\theta_{c.m}$ = 90° (panel (d)). This strengthens the conclusions of the earlier study (Kondo *et al.* 1985), which attributed the parity dependence used in the optical model calculations to a requirement to simulate the effects of $\alpha$ - transfer amplitudes.

### *Re-examination of the optical model potential*

As a final step, we have decided to depart from the restrictions of using the potentials 1 and 2 of Kondo, Robson, and Smith (1983) and search for a potential that could give the best fit to our experimental data. This has resulted in potential 3 in Table 35.1. The real component $V(E_{c.m.}, L)$ of this potential is shown in Figure 35.5.

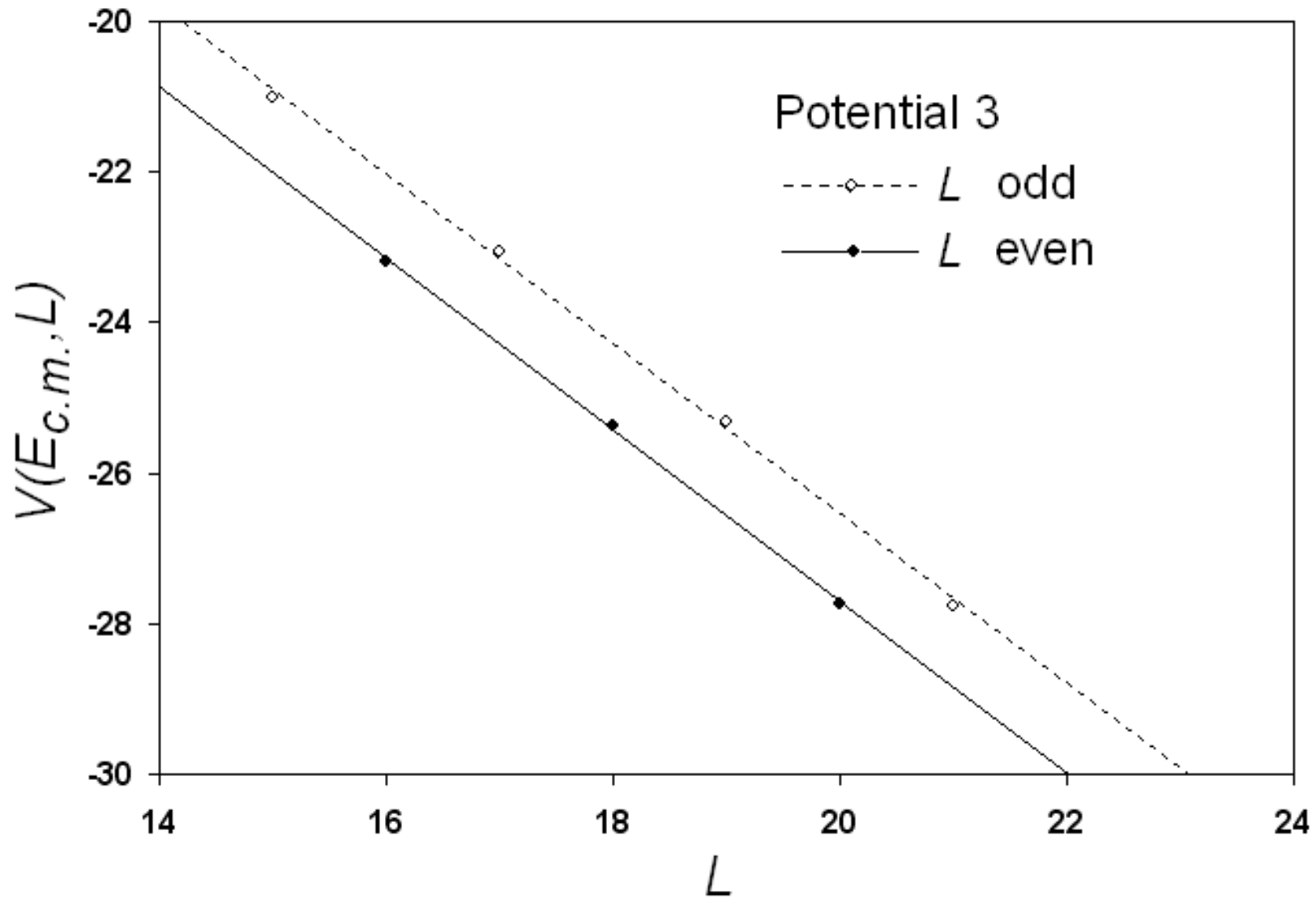

Figure 35.5. The depth of the real component of the potential 3, which gives the best fit to our excitation function (see panel (f) in Figure 35.4). The graph shows that the potential is deeper for even $L$ – values and shallower for odd $L$ – values.

The corresponding fit to the $\theta_{c.m}$ =90° data is shown in panel (f) of Figure 35.4. As can be seen, potential 3 produces excellent fit to our data.

The calculations for $\theta_{c.m}$ =154°± 2° are displayed in panel (e). The fit is less satisfactory but the calculations reproduce the positions of maxima and minima in the excitation function sufficiently well. This potential gives spins resonances, $16^+$, $18^+$, and $20^+$ at $E_{c.m.}$ = 22.5, 25.4, and 28.9 MeV.

### Summary and conclusions

We have measured excitation function for $^{16}O$ + $^{20}Ne$ elastic scattering at $\theta_{c.m}$ = 90° over the energy range 21.5 MeV< $E_{c.m.}$ < 31.2 MeV. We have then carried out an extensive theoretical analysis of our data and the data of Schimizu *et al.* (1982) for $\theta_{c.m}$ = 154° ± 2° using an extended optical model with the aim of placing additional constraints on spin assignments to resonant states, and hence constraints on the parity dependent part of the $^{16}O$ + $^{20}Ne$ potential.

We have found that one of the previously obtained potential (potential 2, Table 35.1) is favoured by the available new data. However, we have also found a new potential (potential 3, Table 35.1) which gives excellent fits to our $\theta_{c.m}$ = 90° excitation function while maintaining a satisfactory description of the excitation function at $\theta_{c.m.}$=154°±2°. Our study suggests even spin shape resonances $J$ = $16^+$, $18^+$, and $20^+$ at energies $E_{c.m.}$ ≈ 22.5, 25.4 and 28.5 MeV, respectively.

The potential 3, which gives the best fit to our excitation function, and the favoured potential 2, both have a negative parity-dependent component, which means that the real potential depth is deeper for even than for odd partial waves (see Figure 35.5).

We have also carried out an analysis without the parity dependent component but including explicitly the amplitudes for $\alpha$ – transfer reaction. These calculations, though less satisfactory than the calculations using potential 3, give sufficiently good representation of the shape and magnitude of the excitation function. Thus, as suggested by Kondo *et al.* (1985), the parity dependent interaction appears to simulate the effects of $\alpha$ -transfer amplitudes.

More generally, Baye, Deenen, and Salomon (1977), interpret parity dependences of ion-ion potentials as a consequence of the Pauli principle associated with an exchange of nucleons between two nuclei. They also suggest that the strongest parity-dependence should be for nuclei with similar masses, which applies well to the $^{16}O+^{20}Ne$ system. Thus, as also mentioned earlier in Chapter 24 for light projectiles, parity dependent interaction appears to simulate coupling to non-elastic channels. To include such coupling explicitly in the calculations would mean to study a relatively large number of channels available in heavy-ion interactions.

# Appendix A

## Semi-classical descriptions of polarization in stripping reactions

This summary presents some early descriptions of nucleon polarization in stripping reactions. In this discussion, I use the product $\vec{k}_{in} \times \vec{k}_{out}$ to define the positive direction of the vector polarization. In the early publications, the positive direction was defined using either $\vec{k}_{in} \times \vec{k}_{out}$ or $\vec{k}_{out} \times \vec{k}_{in}$. Thus, for instance, Newns (1953) used $\vec{k}_{out} \times \vec{k}_{in}$ but in the next related paper, Newns and Rafai (1958) used $\vec{k}_{in} \times \vec{k}_{out}$. In the same year, Hansel and Parkinson (1958) still used $\vec{k}_{out} \times \vec{k}_{in}$. Even later, Sitenko (1959) also used $\vec{k}_{out} \times \vec{k}_{in}$. Thus, care must be exercised when reading the early publications.

The left and right directions may be also ambiguous. For instance, Wolfenstein (1956) defines them by looking upstream.

To adhere to the uniform description, I have corrected the signs in the published formulae and in the quoted polarization values whenever necessary. The discussion is for the A(d,p)B reactions but the same formulae apply also to the A(d,n)B reactions.

A simple, classical interpretation of the nucleon polarization produced in the deuteron stripping reaction was proposed by Newns (1953). The mechanism is illustrated in Figure A.1.

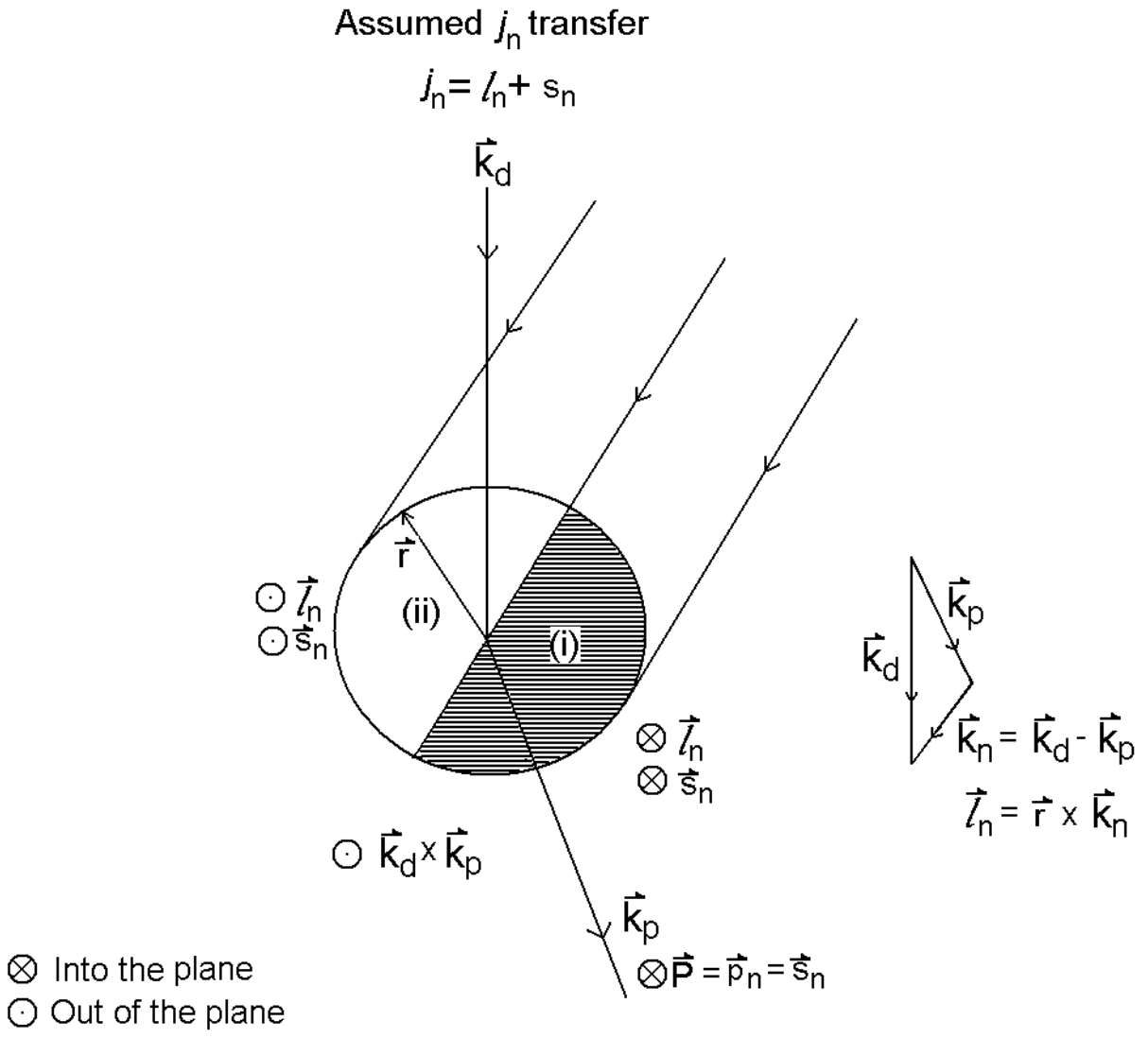

Figure A.1. Classical model (Newns 1953) of nucleon polarization in stripping reactions. This early model assumes that the nucleus is transparent to deuterons but not to protons.

If we assume that the selection rules allow for only $j = j_n = l_n + 1/2$ transfer, then the orbital momentum $\vec{l}_n$ and the spin $\vec{s}_n$ of the captured neutron must be parallel. For neutrons captured in the area (i) the direction of the spin is into the plane and in the area (ii) out of the plane. As the spins of protons and neutrons in deuterons are in the same direction, the direction of the spin $\vec{s}_p$ of protons emitted in the area (i) are in the into-the-plane direction and those emitted in the area (ii) are in the out-of-the-plane direction.

If the nucleus is opaque to protons, but transparent to deuterons, then protons emerging from the area (ii) will be absorbed more readily than protons emerging from the area (i). The net spin direction of the outgoing protons will be in the into-the-plane direction, which is in the opposite direction to $\vec{k}_d \times \vec{k}_p$. Thus, if we assume that the stripping reaction leads to a state in the final nucleus, which corresponds to the capture of the neutron with $j = l + 1/2$ the resulting polarization of the outgoing protons will be *negative*. Likewise, if $j = l - 1/2$ then the polarization of protons will be *positive*.

Using this classical model for stripping reactions, Newns (1953) gives the following formula for the polarization:

$$P = \pm \frac{\sum_{m \ge 0} m \frac{(l-m)!}{(l+m)!} \left( \frac{2^m \sqrt{\pi}}{\Gamma\left[\frac{1}{2}(l-m)+1\right] \Gamma\left[\frac{1}{2}(1-l-m)\right]} \right)^2}{3(j+1/2) \sum_{m \ge 0} \frac{(l-m)!}{(l+m)!} \left[ \frac{2^m \sqrt{\pi}}{\Gamma\left[\frac{1}{2}(l-m)+1\right] \Gamma\left[\frac{1}{2}(1-l-m)\right]} \right]^2}$$

The positive sing is for $j = l - 1/2$ and the negative sign for $j = l + 1/2$.

It should be noted that in this classical description, the polarization of the outgoing nucleons does not depend on the target nucleus but only on the transferred value of the total angular momentum.

Using this formula, the following values of the polarization can be calculated (Newns 1953):

Table A.1

The polarization of outgoing protons from the reaction A(d,p)B calculated the semi-classical description proposed by Newns (1953)

| $l$ | 0 | 1 | 2 | 3 |
|---|---|---|---|---|
| $j = l + 1/2$ | 0 | -16.67 | -13.33 | -18.75 |
| $j = l - 1/2$ | 0 | +33.33 | +20.00 | +25.00 |

The classical model of Newns (1953) is useful because it gives a simple explanation of the mechanism of the nucleon polarization in stripping reactions. However, the weakness of this model is that it neglects the interaction of deuterons with the target nucleus.

It has been first pointed out by Tobocman (1956) that if one reverses the roles of the outgoing proton and incoming deuteron, i.e. if one assumes that the nucleus is opaque to deuterons but transparent to the outgoing protons then the sign of the polarization will be reversed: it will be *positive* for $j = l + 1/2$ and *negative* for $j = l - 1/2$.

Indeed, for the strongly interacting deuterons we should reverse the shaded areas in Figure A.1. In this reversed representation, the region (ii) is darker and region (i) lighter. For protons, the whole nucleus will be transparent. In this new classical description, more deuterons will be absorbed in the region (ii) than in (i). Consequently, more protons will be coming from the region (ii) than from (i) and if we again assume that $j = j_n = l_n + 1/2$ we shall find that the net spin of the outgoing protons will be in the out-of-the-plane direction, i.e. in the direction $\vec{k}_d \times \vec{k}_p$. The polarization of the outgoing protons will be positive for $j = l + 1/2$ and negative for $j = l - 1/2$. The summary of the two models is presented in Table A.2.

Table A.2

Summary of the classical models of the nucleon polarization produced in stripping reaction of deuterons

| | Newns (1953) | Tobocman (1956) |
|---|---|---|
| Crystal ball | Transparent to deuterons<br>Cloudy for protons | Transparent to protons<br>Cloudy for deuterons |
| Deuterons | Pass easily through the ball | Start being absorbed in the area (II) |
| Stripped protons | Emitted uniformly throughout the volume of the ball | Emitted preferentially in the area (II) |
| Observed protons | Mainly emitted from the area (I) | Mainly emitted from the area (II) |
| Proton spins for $j = l + 1/2$ | Mainly in the opposite direction to $\vec{k}_d \times \vec{k}_p$ | Mainly in the direction of $\vec{k}_d \times \vec{k}_p$ |
| Proton polarization at forward angles | $\pm$ for $j = l \mp 1/2$ | $\pm$ for $j = l \pm 1/2$ |

In real life, there will be a contribution from the interactions of both, protons and deuteron, with the target nucleus. However, Satchler (1960) pointed out that for a given depth of the nuclear potential the interaction of deuterons is about twice as effective in producing the resulting polarization as the interaction of the outgoing protons. He has also shown that only if $V_p \approx 2V_d$ and $k_p = k_d$ that the two competing effects cancel and the polarization is zero. This is hardly ever the case. We know now that the depths of the potentials are in the opposite direction, i.e. that $V_d \approx 2V_p$ so if we want to apply the classical interpretation of the polarization mechanism we should expect that the observed signs of the polarization will be opposite to the signs predicted by the original model of Newns (1953).

It should be noted that the polarization depends on the reaction angle so that even for the same value of the transferred total angular momentum it can change sign depending on the reaction angle. The classical picture applies to reactions at small angles. In the more exact quantum mechanical description, the association of the sign of the polarization with the transferred total angular momentum applies to the region of the first stripping maximum, which depending on the transferred orbital angular momentum $l$ will have a peak at different angles but always in the forward direction. The polarization will change the sign when the stripping amplitude changes its sign.

Sitenko (1959) refers to the model of Newns but relates the polarization to the wave functions of deuterons and protons. He gives the following formula:

$$P = \pm \frac{2}{3(2j_n+1)} \left( \frac{\sum_m m \left| I_l^m \right|^2}{\sum_m \left| I_l^m \right|^2} \right) \quad \text{for } j_n = l \mp 1/2$$

where $j_n$ is the total spin of the transferred proton. The quantity $I_l^m$ is given by

$$I_l^m = \int \varphi_{k_p}^*(\vec{r}) \frac{t_l(k_n r)}{t_l(k_n R)} Y_{lm}^*(\theta,\phi) \varphi_{k_d}(\vec{r}) d\vec{r}$$

where

$$t_l(x) = \sqrt{\frac{\pi}{2x}} K_{l+1/2}(x)$$

is the modified Bessel function of the second kind (Abramowitz and Stegun 1964).[6]

If the target nucleus has spin zero, the polarization of the outgoing protons is directly related to the spin of the relevant energy level in the residual nucleus $J = j$. As can be seen, there are some obvious similarities between the formulae of Newns and Sitenko.

The predicted maximum of the absolute value of the nucleon polarization produced in the deuteron stripping reactions is 1/3 (Horowitz and Messiah 1953; Newns 1953). This can be easily seen from the following formulae (Huby *et al.* 1958; Satchler 1958; see also Glendenning 1963):

$$P = \frac{1}{3} \frac{\langle m_l \rangle}{l+1} \quad \text{for } j = l + 1/2$$

$$P = -\frac{1}{3} \frac{\langle m_l \rangle}{l} \quad \text{for } j = l - 1/2$$

If $\langle m_l \rangle$ assumes its maximum value of $l$ then

$$P = \frac{1}{3} \frac{l}{l+1} \quad \text{for } j = l + 1/2$$

$$P = -\frac{1}{3} \quad \text{for } j = l - 1/2$$

[6] Known also as the spherical MacDonald function, which is the name used by Sitenko (1959).

Even in the early studies of nucleon polarization it became clear that the observed values were higher than those predicted by the first theoretical calculations (Allas and Schull 1959. 1962; Bokhari *et al.* 1958; Hillman 1956; Isoya, Marrone, Michaletti, and Reber 1962).

If $l=0$ then $\langle m_l \rangle = 0$ and the polarization $P=0$. For transitions involving $l=0$, $P \neq 0$ if spin-orbit interaction is introduced.

Quantum-mechanically, the $\langle m_l \rangle$ values can be related to the amplitudes $B_l^{m_l}$ for stripping reactions[7] (Horowitz and Messiah 1953):

$$\langle m_l \rangle = \frac{\sum_{m_l} m_l \left| B_l^{m_l} \right|^2}{\sum_{m_l} \left| B_l^{m_l} \right|^2}$$

where

$$B_l^{m_l} \equiv B_l^{m_l}(\vec{k}_d, \vec{k}_p) = \frac{1}{i(2l+1)^{1/2}} \int \psi_p^*(\vec{k}_p, \vec{r}_p) \phi_l^{m*}(\vec{r}_n) V_{np}(r) \phi_d(\vec{r}) \psi_d(\vec{k}_d, \vec{R}) d\vec{r}_n d\vec{r}_p$$

$$\vec{r} = \vec{r}_n - \vec{r}_p, \quad 2\vec{R} = \vec{r}_n + \vec{r}_p.$$

In the absence of the spin-orbit interaction, $\left| B_l^{m_l} \right| = \left| B_l^{-m_l} \right|$ and $P=0$ if $l=0$. The same applies to the formula of Sitenko (1959) because $\left| I_l^m \right| = \left| I_l^{-m} \right|$.

If the spin-orbit interaction is included, then the expression for $B_l^{m_l}$ becomes more complex and there is no simple formula for the predicted polarization.

[7] $B_l^{m_l}$ is the amplitude for the absorption of a neutron with quantum numbers $(l, m_l)$.

# Appendix B

## The diffraction theory

The diffraction theory of nuclear scattering was developed by Blair (Blair 1959, 1961) assuming a sharp cutoff radius, and extended by Blair, Sharp, and Wilets (1962), by introducing a diffuse surface for target nuclei.

### The elastic scattering

The amplitude for the elastic scattering of spinless particles by spherical nuclei can be expressed in terms of partial waves:[8]

$$f(\theta) = \frac{1}{2ik_i}\sum_l (2l+1)(1-\eta_l)P_l(\cos\theta)$$

where $k_i$ is the wave number for the incoming particle, $P_l(\cos\theta)$ is the Legendre polynomial, and $\eta_l$ is the amplitude of the scattered $l$ th partial wave.

If the nucleus is represented as a black disk with a sharp cutoff radius, then

$$\eta_l = 0 \quad \text{for } l < L,$$

$$\eta_l = 1 \quad \text{for } l > L,$$

$$\eta_l = 1/2 \quad \text{for } l = L,$$

where $L$ is the orbital angular momentum or the partial wave determined by the nuclear radius $R_0$.

In the diffraction theory, the amplitude $f(\theta)$ is calculated in a similar way as in the optics using the Fraunhofer approximation:

$$f(\theta) \cong \frac{ik_i}{2\pi}\iint dA \exp(i\vec{k}_i\vec{r})$$

The integration is over the area of the projection of the sphere.

This leads to the following expression for the elastic scattering differential cross section:

$$\left(\frac{d\sigma}{d\Omega}\right)_{el} = \left(k_i R_0^2\right)^2\left[\frac{J_1(x)}{x}\right]^2$$

where $k_i \equiv 1/\lambda_i$ is the wave number for the incident particles, $R_0$ the nuclear radius, and $J_1(x)$ is the cylindrical Bessel function of the first order.

The variable $x \equiv qR_0$, where $q \equiv \left|\vec{k}_i - \vec{k}_f\right|$ with $\vec{k}_i$ and $\vec{k}_f$ being the wave vectors for the incoming and outgoing particles, respectively (see Figure B.1).

[8] The Coulomb scattering is not included.

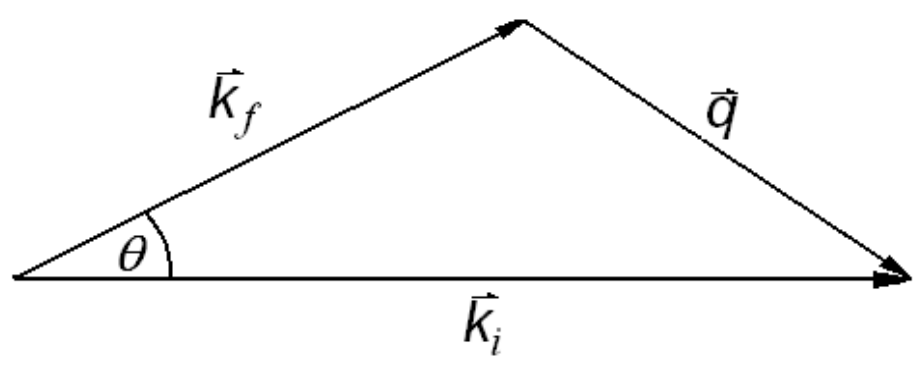

Figure B.1. The diagram showing the relation between $\vec{k}_i$ , $\vec{k}_f$ , $\vec{q}$ and the reaction angle $\theta$.

From the well-known relation for the triangle shown in Figure B.1 we find that

$$q^2 = k_i^2 + k_f^2 - 2k_i k_f \cos\theta$$

For the elastic scattering $k_i = k_f$ and since

$$sin(\theta/2) = \sqrt{\frac{1}{2}(1 - cos\,\theta)}$$

we find that $x = 2k_i R_0 \sin(\theta/2)$.

The universal function $[J_1(x)/x]^2$ is presented in Figure B.2 for $x \geq 4$, i.e. starting from the second maximum. To show the structure of this function, the first large maximum at $x$ = 0 is not displayed.

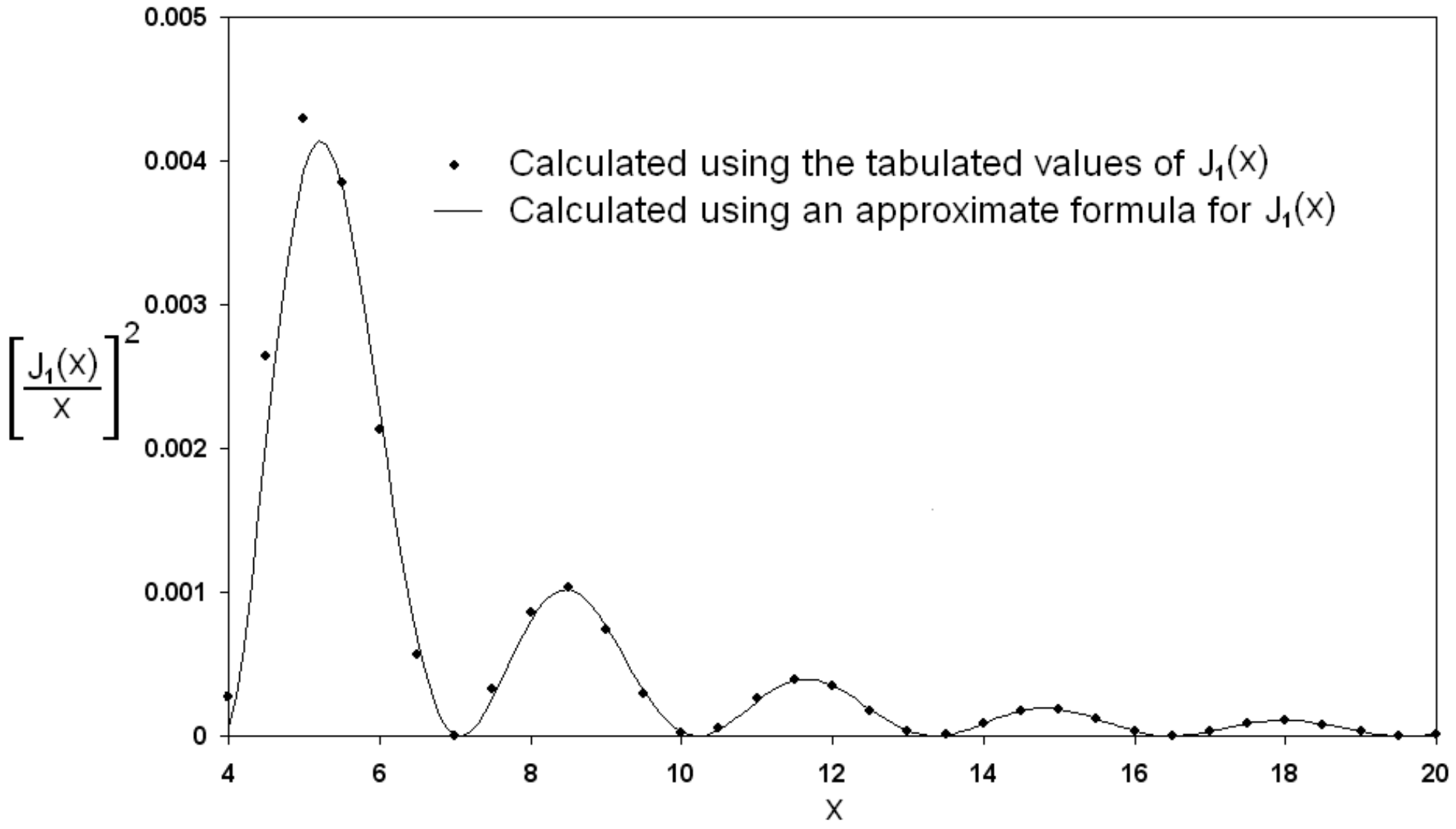

Figure B.2. The universal function $[J_1(x)/x]^2$ calculated using the tabulated values (West 1970) and the approximate formula (Bronstein and Semedleiev 1957) for the Bessel function $J_1(x)$. To show the structure of the universal function, the first large maximum at $x$ = 0 is not displayed.

For sufficiently large values of $x$, the cylindrical Bessel function of the order of $n$ can be calculated using the following approximate formula (Bronstein and Semedleiev 1957):

$$J_n(x) \approx \sqrt{\frac{2}{\pi x}}\, cos\left(x - \frac{n\pi}{2} - \frac{\pi}{4}\right)$$

It can be seen from Figure B.2 that for $x > 6$ the approximate formula for $J_1(x)$ gives a good representation of the universal function $\left[J_1(x)/x\right]^2$.

## The inelastic scattering

For the inelastic scattering from deformed nuclei, integration is also carried out over the projection area of the scattering nucleus. The integration can be reduced to integration over the projection of a sphere and integration over a projection of the deformed part. The whole contribution to the inelastic scattering comes from the integration from the later area.

The black disc formula for the inelastic scattering corresponding to the $l$ = 2 (quadrupole) excitation has the following form (Blair 1959,1961):

$$\frac{d\sigma}{d\Omega}(l = 2, I_0 \to I) = (k_i R_0)^2 \left[\frac{1}{2I_0 + 1}\frac{2l+1}{4\pi}\sum_{M_0,M,m}\left|\left\langle IM\left|\delta_{l,m}\right|I_0 M_0\right\rangle\right|^2\right]\left[\frac{1}{4}J_0^2(qR_0) + \frac{3}{4}J_2^2(qR_0)\right]$$

In this formula, $I_0$ is the spin of the ground state, $I$ spin of the excited state, $\delta_{l,m}$ the deformation distance, and $J_0$ and $J_2$ are the cylindrical Bessel functions.

The deformation distance is defined by the description of nuclear radius $R$ in the following way:

$$R(\theta', \phi') = R_0 + \sum_{lm}\delta_{lm}Y_{lm}^*(\theta', \phi')$$

where $Y_{lm}$ are the spherical harmonic functions, and $\theta'$ and $\phi'$ are the body fixed coordinates.

The cross-section formula can be rewritten as:

$$\frac{d\sigma}{d\Omega}(l = 2, I_0 \to I) = (k_i R_0)^2 \left[\frac{1}{2I_0 + 1}\frac{5}{4\pi}\sum_{M_0,M}\left|\left\langle IM\left|\delta_2\right|I_0 M_0\right\rangle\right|^2\right]\left[\frac{1}{4}J_0^2(qR_0) + \frac{3}{4}J_2^2(qR_0)\right]$$

or in even a simpler form as

$$\frac{d\sigma}{d\Omega}(l = 2, I_0 \to I) = (k_i R_0)^2 \mathrm{M} F(x)$$

where

$$\mathrm{M} \equiv \frac{1}{2I_0 + 1}\frac{5}{4\pi}\sum_{M_0,M}\left|\left\langle IM\left|\delta_2\right|I_0 M_0\right\rangle\right|^2 \equiv \frac{1}{2I_0 + 1}\mathrm{M}'$$

and

$$F(x) \equiv \left[\frac{1}{4}J_0^2(x) + \frac{3}{4}J_2^2(x)\right]$$

As in the case of the elastic scattering, which was described by the $\left[J_1(x)/x\right]^2$ function, $F(x)$ is also a model-independent universal function. The function is presented in Figure B.3 for $0 \le x \le 12$. I have calculated it using the $J_0(x)$ and $J_2(x)$ values tabulated by Abramowitz and Stegun (1964).

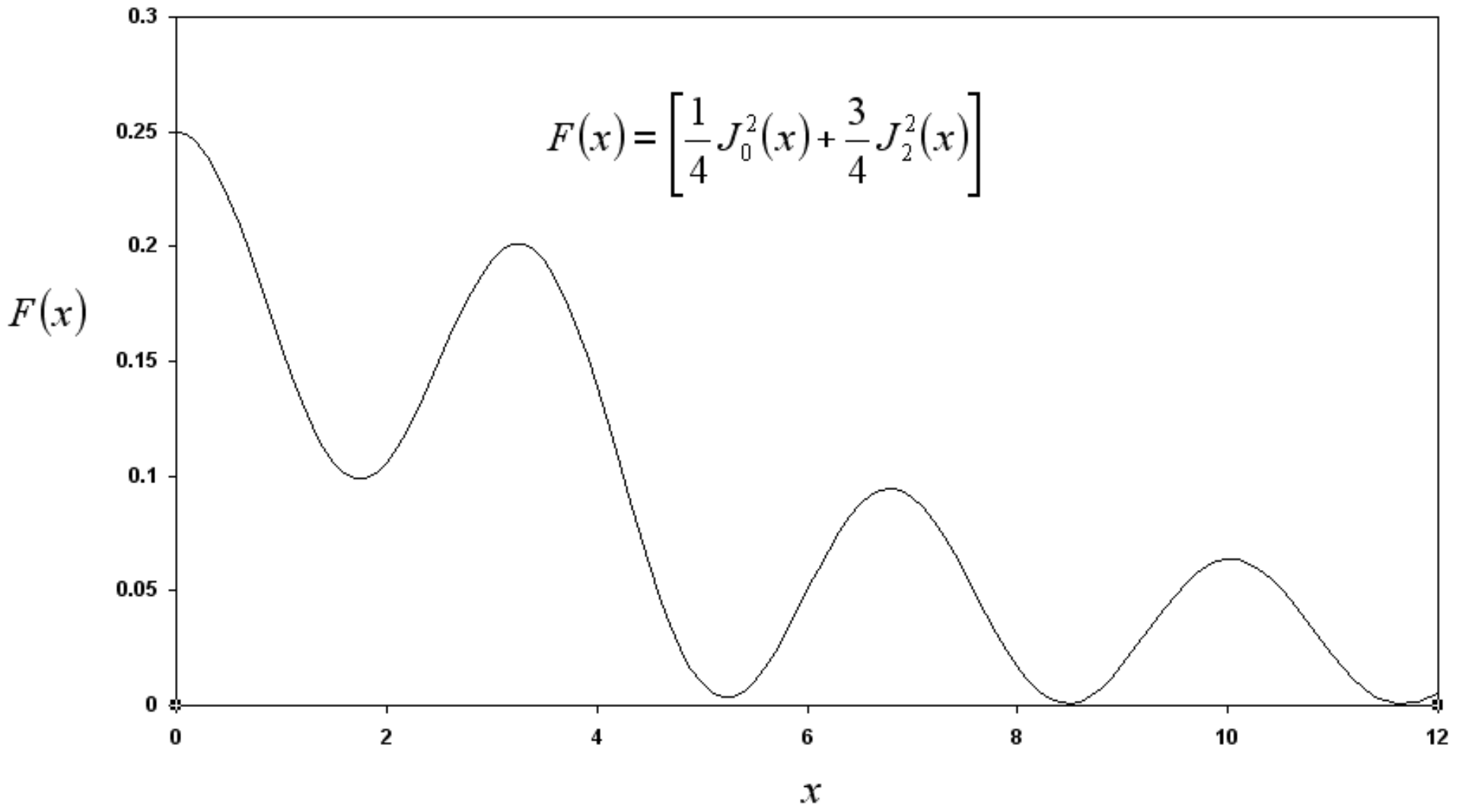

Figure B.3. The universal shape function $F(x)$ used in describing inelastic scattering.

## Procedure in theoretical analysis

Theoretical analysis of experimental data for either elastic or inelastic scattering is simple. The first step consists in determining the radius $R_0$. This is done by using the formula:

$$R_0 = \frac{x_{max}}{\tilde{q}}$$

where $x_{max}$ is the position of the second maximum of the universal function $[J_1(x)/x]^2$ or $F(x)$, $\tilde{q}$ is the $q$ - value corresponding to the position $\theta_{max}$ of the relevant maximum of the experimental angular distributions:

$$\tilde{q} = \left(k_i^2 + k_f^2 - 2k_i k_f \cos\theta_{\max}\right)^{1/2}$$

For instance, for the inelastic scattering, the position of the second maximum of the universal function $F(x)$ is at $x_{max}$ =3.25.

Using thus determined $R_0$ one can then express the function $[J_1(x)/x]^2$ or $F(x)$ as a function of $\theta$ and compare it with the experimental angular distribution.

## Model-dependent formulae

The matrix element $\mathrm{M}'$ is related to nuclear structure. For extreme collective models, $\mathrm{M}'$ can be used to calculate model-dependent parameters, $\delta_2$ or $(\hbar\omega_2 / 2C_2)$ using the following relations (Blair 1961):

***Even-even-mass nucleus***

(a) Rotational model

$$\mathrm{M}' = \frac{\delta_2^2}{4\pi}$$

where $\delta_2$ is the deformation distance parameter.

(b) Vibrational model (i.e. surface vibrations with phonon energy given by $\hbar\omega_2$)

$$\mathrm{M}' = \frac{5}{4\pi}\left(\frac{\hbar\omega_2}{2C_2}\right)R_c^2$$

where $C_2$ is the surface tension parameter and $R_c$ is a nuclear matter radius, which is close to the charge radius. It is reasonable to assume that $R_c = 1.2A^{1/3}$, where $A$ is the atomic mass of the target nucleus.

***Odd-mass nucleus***

(a) Rotational model

$$\mathrm{M}' = (I_0, l, K, 0 \mid I, K)\frac{\delta_2^2}{4\pi}$$

where $(I_0, l, K, 0 \mid I, K)$ is the Clebsch-Gordan coefficient given by (Condon and Shortley 1935):

$$(I_0, l, K, 0 \mid I, K) = K\left[\frac{3(I_0 - K + 1)(I_0 + K + 1)}{I_0(2I_0 + 1)(I_0 + 1)(I_0 + 2)}\right]^{1/2}$$

For instance, if we assume the strong coupling model for $^{27}$Al, then for the ground state band $K = {}^5/_2$ and $(I_0, l, K, 0 \mid I, K) = 0.69$.

(b) Vibrational model

$$\mathrm{M}' = \frac{2I + 1}{2I_0 + 1}\frac{1}{4\pi}\left(\frac{\hbar\omega_2}{2C_2}\right)R_c^2$$

We can use the above formulae if we really know that these models apply or if we wish to have a means of parameterising the magnitude of the nuclear matrix elements, even though the model does not apply. However, when we are just starting to analyse the data and do not know whether any of these extreme models can be applied (as in the case discussed in Chapter 3) then it is best to list just the factors $\mathrm{M}$ or $\mathrm{M}'$.

**Extended diffraction theory**

One of the biggest defects of the Fraunhofer version (i.e. the black disk with sharp cut-off radius) of the diffraction model is that the nuclear surface is taken to be sharp. An extremely simple way of making this region 'grey' is to multiply the sharp-edge cross sections by some factor $f(k_i\theta)$. A typical form factor is the Gaussian, $f(k_i\theta) = \exp[-(\alpha k_i\theta)]^2$, where $\alpha$ is some distance characteristic of the surface.

The effect of introducing a grey or diffuse region at the nuclear surface is that the amplitudes of the calculated angular distributions are progressively pushed lower with the increasing reaction angle $\theta$. This effect has been also simulated by introducing an obliquity factor $\cos^2(\theta/2)$. However, a far better way of accounting for the diffuse region of nuclear surface is to parameterise the amplitude $\eta_l$ of the scattered partial wave $l$. Such a smooth transition from a complete to no absorption has been introduced by Blair, Sharp, and Wilets (1962) by assuming that $\eta_l$ has the following form:

$$\eta_l = \{1 + exp[(L - l)]/\Delta\}$$

They introduced two parameters: the cut-off angular momentum $L = k_i R_0 - 1/2$ and the ratio $\Delta/L$, which describes the width of the transition region. The location of the maxima and minima depend on $L$ and relative heights only on $\Delta/L$. The difference between the sharp cut-off radius and smooth cut-off radius calculations is illustrated in Chapter 3 for scattering of deuterons from $^{27}$Al and $^{28}$Si nuclei.

# Appendix C

## The plane-wave theory of inelastic scattering

In its general form, the transition amplitude for the inelastic scattering from state $i$ to state $f$ is given by (Rost and Austern 1960):

$$T_{fi} = \left\langle \varphi_f \chi_f^{(-)} \middle| V \middle| \varphi_i \chi_f^{(+)} \right\rangle$$

where, $\varphi_i$ and $\varphi_f$ are the intrinsic wave functions for state $i$ and $f$, $\chi_i$ and $\chi_f$ the wave function describing relative motion of the projectile in the incident and outgoing channel, and $V$ is the interaction potential that causes the transition from state $i$ to $f$ without influencing the wave functions $\chi_i$ and $\chi_f$.

The differential cross section is then given by:

$$\frac{d\sigma(i \to f)}{d\Omega} = \left( \frac{\mu}{2\pi\hbar^2} \right)^2 \frac{k_f}{k_i} \sum_{av} \left| T_{fi} \right|^2$$

where $\mu$ is the reduced mass of the projectile and $k_i$ and $k_f$ are the wave numbers in the incident and outgoing channels.

The wave functions $\chi_i$ and $\chi_f$ are solutions of the Schrödinger equation:

$$\left[ -\left( \frac{\hbar^2}{2\mu} \right) \Delta^2 + U \right] \chi_{i,f}^{(\pm)} = E_{i,f} \chi_{i,f}^{(\pm)}$$

where $U$ is the optical model potential.

Butler (1950, 1951, 1952, 1957; Butler and Austern 1954; Butler, Austern, and Pearson 1958) was the first to introduce the concept of direct nuclear interactions. In his simple model, he ignored the nuclear interaction $U$ in both the incident and outgoing channels and thus represented the wave functions $\chi_i$ and $\chi_f$ by the undisturbed plane waves. Under this assumption, the transition matrix can be written as:

$$T_{fi} = \left\langle \varphi_f \exp(-i\vec{k}_f \vec{r}_f) \middle| V \middle| \varphi_i \exp(i\vec{k}_i \vec{r}_i) \right\rangle$$

where, $\vec{k}_i$ and $\vec{k}_f$ are the initial and final momenta.

The differential cross section for the inelastic scattering takes then a simple form:

$$\frac{d\sigma}{d\Omega} = S\left[q^2 + K^2\right]^{-1} W^2 \left[ j_l(qr), h_l(iKr) \right]_{r=R_0}$$

where $S$ is an arbitrary normalisation factor, $q = \left| \vec{k}_i - \vec{k}_f \right|$, $K = \kappa_i + \kappa_f$, $W\left[ j_l(qr), h_l(iKr) \right]$ is the Wronskian, $j_l$ and $h_l$ are the spherical Bessel and Hankel functions, respectively, and $R_0$ is the nuclear radius.

The momentum transfer $q$ is given by:

$$q = \left(k_i^{\,2} + k_f^{\,2} - 2k_i k_f \cos\theta\right)^{1/2}$$

where $\theta$ is the scattering angle.

The quantity $K = \kappa_i + \kappa_f$ is related to the projectile binding energies in the target nucleus before and after the scattering, i.e. when the target nucleus is in the ground state and excited state, respectively.

$$\kappa_i = \frac{\sqrt{2\mu B_i}}{\hbar}$$

$$\kappa_f = \frac{\sqrt{2\mu B_f}}{\hbar}$$

where $B_i$ and $B_f$ the binding energies before and after the scattering.

The Wronskian is given by the following expression:

$$W(j_l, h_l) \equiv \begin{vmatrix} j_l & h_l \\ \dfrac{dj_l}{dr} & \dfrac{dh_l}{dr} \end{vmatrix}_{r=R_0} = \left[ j_l \frac{dh_l}{dr} - h_l \frac{dj_l}{dr} \right]_{r=R_0}$$

It can be shown that

$$W[j_l(x), h_i(iy)] = G_l(y)\, j_l(x) + x j_{l-1}(x)$$

where $x = qR_0$, $y = (\kappa_i + \kappa_f)R_0$ and

$$G_l(y) = -iy \frac{h_{l-1}(iy)}{h_l(iy)}$$

For $l$ = 2 we have

$$W[j_2(x), h_2(iy)] = G_2(y)\, j_2(x) + x j_1(x)$$

and

$$G_2(y) = \frac{y^2(1+y)}{3 + 3y + y^2}$$

As with the diffraction theory (see the Appendix B), the fitting procedure consists in aligning the first maximum of the angle-dependant function [in this case the function $W^2[j_2(x), h_2(iy)] \equiv W_2^2(x,y)$] with the first maximum of the experimental distribution. However, the problem here is that unlike the universal $F(x)$ function in the diffraction theory, the function $W_2^2(x,y)$ depends on two variables, each of which depends on the radius $R_0$. Thus, the maximum of $W_2^2(x,y)$ will depend not only on the value of $x = qR_0$ but also on $y = (\kappa_i + \kappa_f)R_0$.

Fortunately, the dependence on $y$ is weak. I have calculated the $W_2^2(x,y)$ function for various values of $y$ using the tabulated values for the spherical Bessel functions (Abramowitz and Stegun 1964). Figure C.1 shows the $W_2^2(x,y)$ functions for $y$ = 14, 15, and 16, and for $x$ = 0 - 10. It can be seen that there is virtually no change in the location of the first maximum.

I have also calculated the positions of the first maximum for a wide range of $y$ = 4 - 30. Results are shown in Figure C.2. It can be seen that even for such a wide range of $y$ values the position of the first maximum changes only by a few per cent.

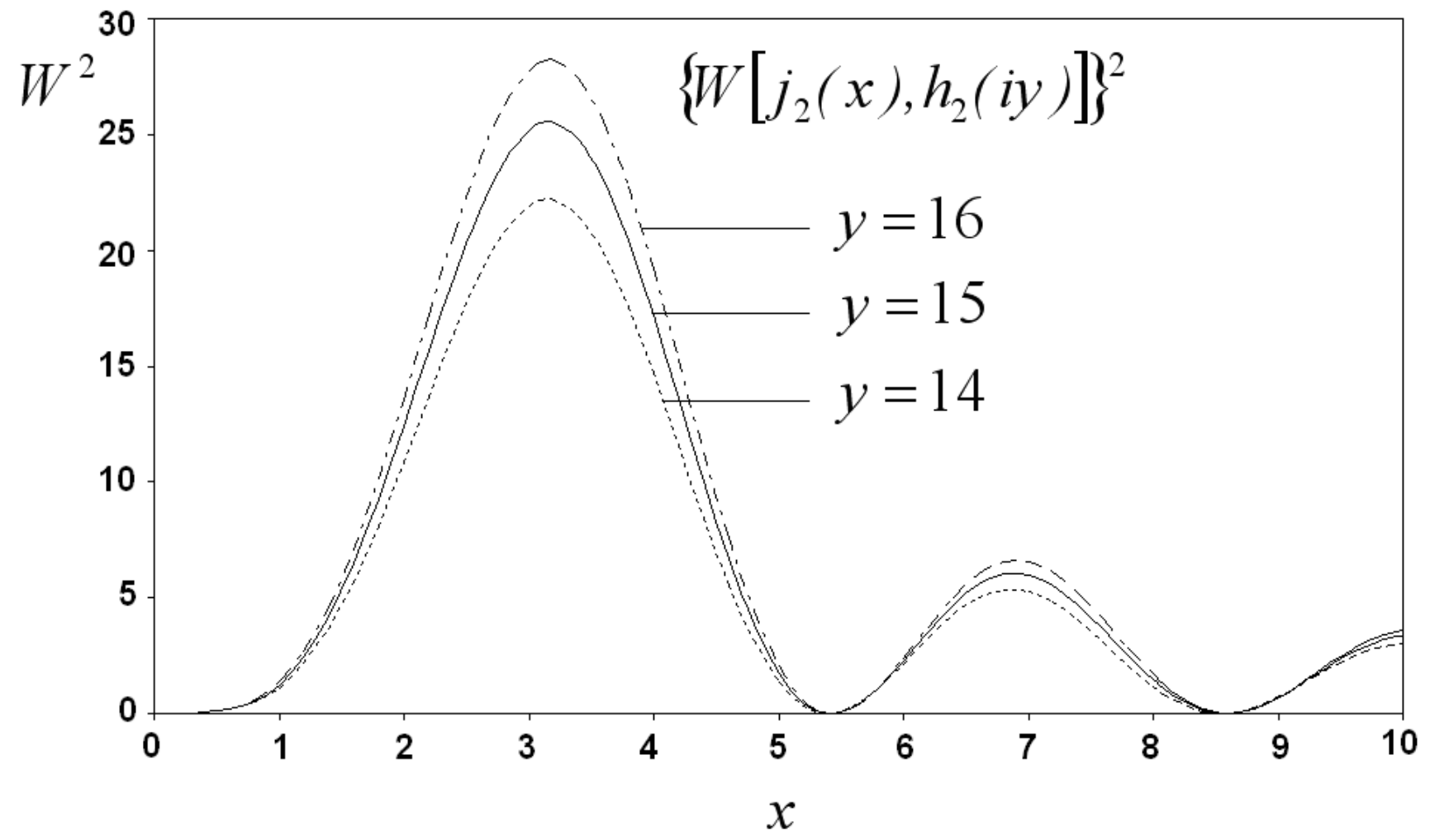

Figure C.1. Squared Wronskian of the second order plotted for three values of $y$. The functions were constructed using the tabulated spherical Bessel functions (Abramowitz and Stegun 1965).

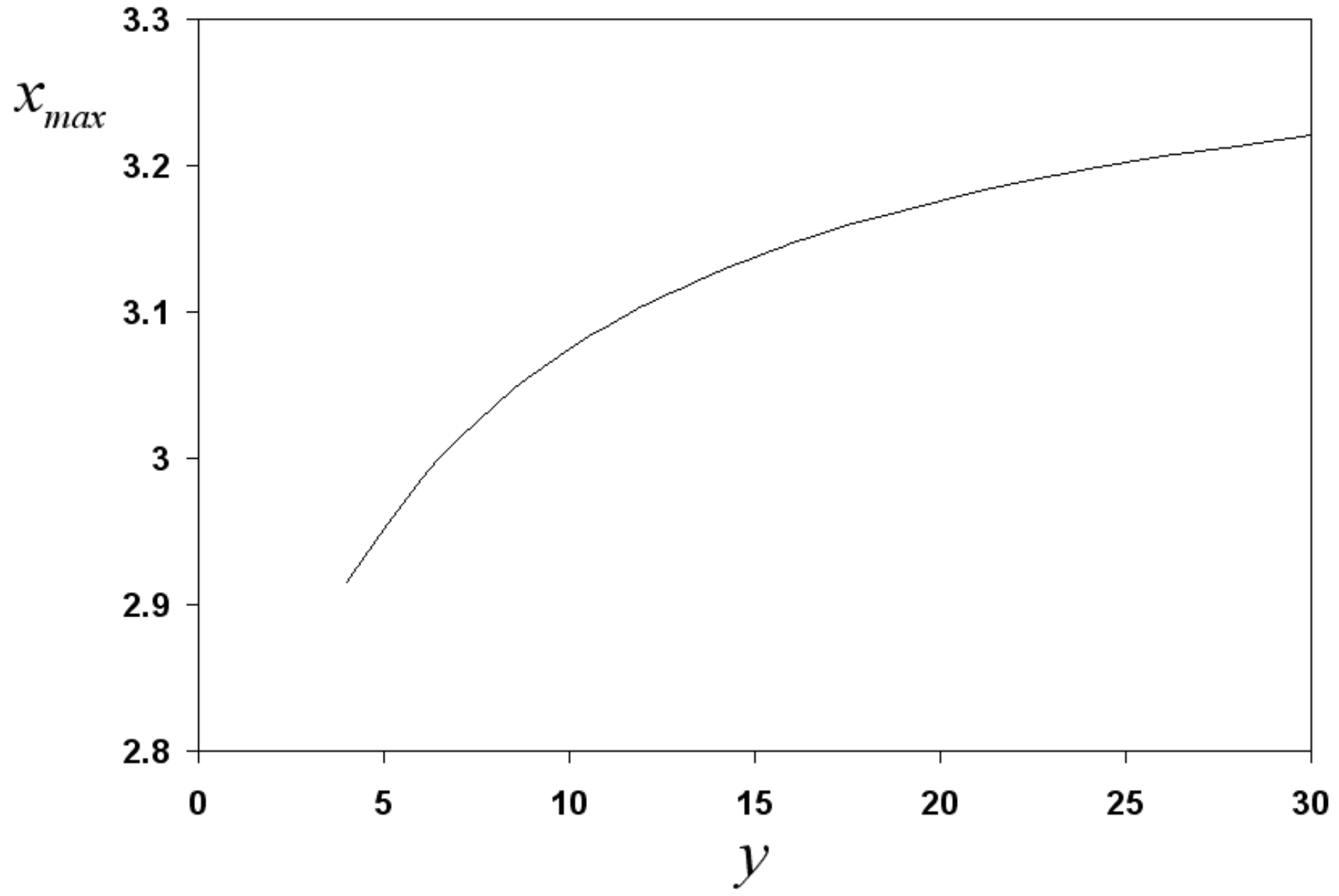

Figure C.2. The position $x_{max}$ of the first maximum of the function $W^2[j_2(x), h_2(iy)] \equiv W_2^2(x,y)$ as the function of $y$.

The traditional way of fitting experimental data with the plane wave theory is to use the tabulated cross sections (Lubitz 1957). First one has to calculate the *y/x* ratio, which does not depend on $R_0$:

$$\frac{y}{x} = \frac{\kappa_i + \kappa_f}{\tilde{q}}$$

where, as for the diffraction model, $\tilde{q}$ is the $q$ - value corresponding to the position $\theta_{max}$ of the first maximum of the experimental angular distributions.

Next, from a plot of the dependence of *y/x* on *x* (Lubitz 1957) for a given *l* one determines the value of *y*, which then one can use to find the corresponding tabulated $\sigma^{l}_{TAB}(x,y)$ cross sections. Using the tabulated $\sigma^{l}_{TAB}(x,y)$ cross sections one finds the position $x_{max}$ of the first maximum of function $W_2^2(x,y)$, and use it to calculate $R_0$:

$$R_0 = \frac{x_{max}}{\tilde{q}}$$

Finally, using the determined $R_0$ one can express $W_2^2(x,y)$ as a function of $\theta$ and compare it with the experimental angular distribution.

Alternatively, one can assume that the radius $R_0$ is given approximately by

$$R_0 \approx 1.2A^{1/3} + R_p$$

where $A$ is the mass number of the target nucleus and $R_p$ the radius of the projectile.

Using thus calculated $R_0$ one can calculate $y$ and $W_2^2(x,y)$. As the position of the first maximum of $W_2^2(x,y)$ depends weakly on $y$, one can locate the $x_{max}$ value and use it to recalculate the value of $R_0$.

In both of these options, the procedure can be repeated for different values of $y$ until satisfactory results are obtained.

# Appendix D

## The strong coupling theory

The concept of the strong-coupling theory (as used in the analysis discussed in Chapter 3) has been described in the following publications: Buck (1963), Buck, Stamp, and Hodgson (1963), and Chase, Wilets, and Edmonds (1958).

We start with the Schrödinger equation:

$$[H_n(\xi) + T + V(\vec{r},\xi)]\Psi(\vec{r},\xi) = E\Psi(\vec{r},\xi)$$

where $\vec{r}$ are the coordinates of the incident particles, $\xi$ – internal coordinates of the target nucleus, $H_n(\xi)$ – the target nuclear Hamiltonian, $T$ – the kinetic energy operator for the relative motion, $V(\vec{r},\xi)$ – the interaction energy between particle and nucleus, $\Psi(\vec{r},\xi)$ – the complete wave function for the system, and $E$ – the total energy. The system of coordinates is in the centre-of-mass frame of reference, the mass is represented by reduced mass, and the relative energies and momenta are used.

To solve this equation, we expand $\Psi(\vec{r},\xi)$ in terms of eigenstates of the total angular momentum $J$:

$$\Psi(\vec{r},\xi) = \sum_{JM} a_{JM}\psi_J^M(\vec{r},\xi)$$

Here we have to introduce notations for the angular momenta:

$I$ – target spin

$l$ – the relative orbital angular momentum of the partial wave of the incident particles

$J$ – the total angular momentum ( $\hat{J} = \hat{l} + \hat{I}$ )

All these quantities are for the entrance channel. For the exit channels, we have $I'$ and $l'$ adding also to $J$.

The function $\psi_J^M(\vec{r},\xi)$ can be expressed as a superposition of elastic and inelastic scattering components:

$$\psi_J^M(\vec{r},\xi) = \frac{1}{r} f_{Il}^J(r)\phi_{Il}^{JM}(\vec{r},\xi) + \frac{1}{r}\sum_{I'l'} f_{I'l'}^J(r)\phi_{I'l'}(\vec{r},\xi)$$

where $f_{Il}^J(r)$ are the radial wave functions for the system with the target in spin state $I$ and relative target-particle angular momentum $l$. The sum is over all $I'$ values but excluding the ground state $I$.

By inserting this function in the Schrödinger equation, we then get the following set of differential equations for the radial wave functions:

$$\left\{-\frac{\hbar^2}{2\mu}\left(\frac{d^2}{dr^2} - \frac{l(l+1)}{r^2}\right) + \varepsilon_I - E\right\} f_{Il}^J(r) + \sum_{I'l'} V_{Il:I'l'}^J(r)\, f_{I'l'}^J(r) = 0$$

$$\left\{-\frac{\hbar^2}{2\mu}\left(\frac{d^2}{dr^2}-\frac{l'(l'+1)}{r^2}\right)+\varepsilon_{I'}-E\right\}f^J_{I'l'}(r)+V^J_{I'l':Il}(r)\,f^J_{Il}(r)+\sum_{I''l''}V^J_{I'l':I''l''}(r)\,f^J_{I''l''}(r)=0$$

where $\mu$ is the reduced mass, $\varepsilon_I$ is the excitation energy of the target state with spin $I$, and $V^J_{Il:I'l'}(r)$ are the coupling matrix elements. The sum in the second set of equations is for $I''\neq I$ and $l''\neq l$.

The coupling matrix elements are defined as

$$V^J_{Il:I'l'}(r)=\left\langle\phi^{JM}_{Il}(\vec{r},\xi)\middle|V(\vec{r},\xi)\middle|\phi^{JM}_{I'l'}(\vec{r},\xi)\right\rangle$$

As mentioned in Chapter 3, the computer code used in the analysis of our experimental results was restricted to scattering from even-even nuclei and for only $0^+\rightarrow 2^+$ transitions. This assumption simplifies significantly the calculations. In this case $J = l$ for the ground state and $I' = I'' = 2$. The differential equations for the radial functions take then a simpler form:

$$\left\{\frac{\hbar^2}{2\mu}\left(\frac{d^2}{dr^2}-\frac{J(J+1)}{r^2}\right)+E-V^J_{0J:0J}(r)\right\}f^J_{0J}(r)-\sum_{l'}V^J_{0J:2l'}(r)\,f^J_{2l'}(r)=0$$

$$\left\{\frac{\hbar^2}{2\mu}\left(\frac{d^2}{dr^2}-\frac{l'(l'+1)}{r^2}\right)+E'-V^J_{2l':2l'}(r)\right\}f^J_{2l'}(r)-V^J_{2l':0J}(r)\,f^J_{0J}(r)-\sum_{l''\neq l'}V^J_{2l':2l''}(r)\,f^J_{2l''}(r)=0$$

where $E=E-\varepsilon_2$.

In principle, the equations are for an infinite number of partial waves. However, in practice, only a limited number of partial waves contributes to the reaction and consequently $J$ values are terminated at a certain $J_{\max}$ value, which is determined by examining the solutions of the differential equations (see below).

The coupling matrix elements involve the use of Clebsch-Gordan coefficients, which limit the values for $l'$ and $l''$ to three values for each $J$, namely $J$ - 2, $J$, and $J$ + 2. With these restrictions, the lower set of equations is limited to three, making (with the upper equation) the total of only four coupled equations for each value of $J$.

The solutions $f^J_{0J}(r)$, $f^J_{2J-2}(r)$, $f^J_{2J}(r)$, and $f^J_{2J+2}(r)$ are then expressed in terms of four scattering matrix elements $\alpha^J_{0J}$, $\beta^J_{2J-2}$, $\beta^J_{2J}$, and $\beta^J_{2J+2}$ determined by the boundary conditions. The scattering matrix elements are directly related to the better-known $S$-matrix elements, which enter into the expressions of the elastic and inelastic scattering amplitudes:

$$A_0(\theta)=f_c(\theta)+\frac{1}{2ik}\sum_J\exp(2i\omega_J)(2J+1)(S^J_{0J}-1)P_J(\cos\theta)$$

$$A_2^M(\theta)=\frac{1}{ik}\left(\frac{k'}{k}\right)^{1/2}\sum_{JJ'}\exp[i(\omega'_{J'}+\omega_J)](2J+1)S^J_{2J'}(-1)^M\left[\frac{(J'-M)!}{(J-M)!}\right]C^{J'J2}_{M0M}P^M_{J'}(\cos\theta)$$

where

$\theta$ – the scattering angle

$f_c(\theta)$ – the Coulomb scattering amplitude

$$f_c(\theta) = -\frac{\gamma}{2k\sin^2(\theta/2)}\exp[-2i\gamma\log\sin(\theta/2)]$$

$$\gamma = \frac{\mu z Z e^2}{k\hbar^2}$$

with $z$ and $Z$ being the atomic numbers for the projectile and target nucleus respectively

$\omega_J = \sigma_J - \sigma_0$,

with $\sigma_J$ being the Coulomb phase shifts (Buck, Maddison, and Hodgson 1960)

$J' = J-2$, $J$, $J-2$,

$S^J_{0J}$ and $S^J_{2J'}$ are the $S$ – matrix elements

As mentioned earlier, the $S$ – matrix elements are related to the four scattering matrix elements $\alpha^J_{0J}$, $\beta^J_{2J-2}$, $\beta^J_{2J}$, and $\beta^J_{2J+2}$ [which in turn are related to the four solutions $f^J_{0J}(r)$, $f^J_{2J-2}(r)$, $f^J_{2J}(r)$, and $f^J_{2J+2}(r)$] of the coupled-channel equations in the following way:

$$S^J_{0J} = 2\alpha^J_{0J} + 1$$

$$S^J_{2J-2} = \beta^J_{2J-2}$$

$$S^J_{2J} = \beta^J_{2J}$$

$$S^J_{2J+2} = \beta^J_{2J+2}$$

$C^{J'J2}_{M0M}$ – the Clebsch-Gordan coefficients in Biedenharn (1952) notation.

$P_J(\cos\theta)$ and $P^M_J(\cos\theta)$ – Legendre and associated Legendre polynomials, respectively.

The four differential equations for the radial wave functions are solved numerically for various values of $J$ until $\alpha^J_{0J}$, $\beta^J_{2J-2}$, $\beta^J_{2J}$, and $\beta^J_{2J+2}$ become negligible. This determines the maximum value $J_{max}$, which corresponds to the maximum number of partial waves contributing to the scattering.

Using the theoretically calculated elastic and inelastic scattering amplitudes and the $S$ – matrix elements one can calculate the quantities, which can be compared with the experimental results:

The differential cross section for elastic scattering:

$$\left(\frac{d\sigma}{d\Omega}\right)_{el} = |A_0(\theta)|^2$$

The differential cross section for inelastic scattering:

$$\left(\frac{d\sigma}{d\Omega}\right)(0^+ \to 2^+) = \sum_M \left|A_2^M(\theta)\right|^2$$

The total absorption cross section:

$$\sigma_A = \frac{\pi}{k^2}\sum_J (2J+1)\left(1 - \left|S_{0J}^J\right|^2\right)$$

The total elastic cross section for incident particles:

$$\sigma_{EL} = \frac{\pi}{k^2}\sum_J (2J+1)\left|1 - S_{0J}^J\right|^2$$

The total inelastic cross section:

$$\sigma_{IN}(0^+ \to 2^+) = \frac{\pi}{k^2}\frac{k'}{k}\sum_{JJ'} (2J+1)\left|S_{2J'}^J\right|^2$$

The total cross section for particles:

$$\sigma_T = \frac{2\pi}{k^2}\sum_J (2J+1)\left(1 - \mathrm{Re}\, S_{0J}^J\right)$$

In the above expressions, $J = 0, 1, 2, \ldots J_{\max}$ and $J' = J, J \pm 2$.

# Appendix E

## Theories of direct nuclear reactions

### Introduction

Theories of direct nuclear reactions have been discussed in numerous publications. Convenient sources of reference are two books: Glendening (2004) and Satchler (1983). Presented here are just general concepts of these theories.

Consider a general reaction A($a$,$b$)B, which is illustrated in Figure E.1 for the stripping reaction.

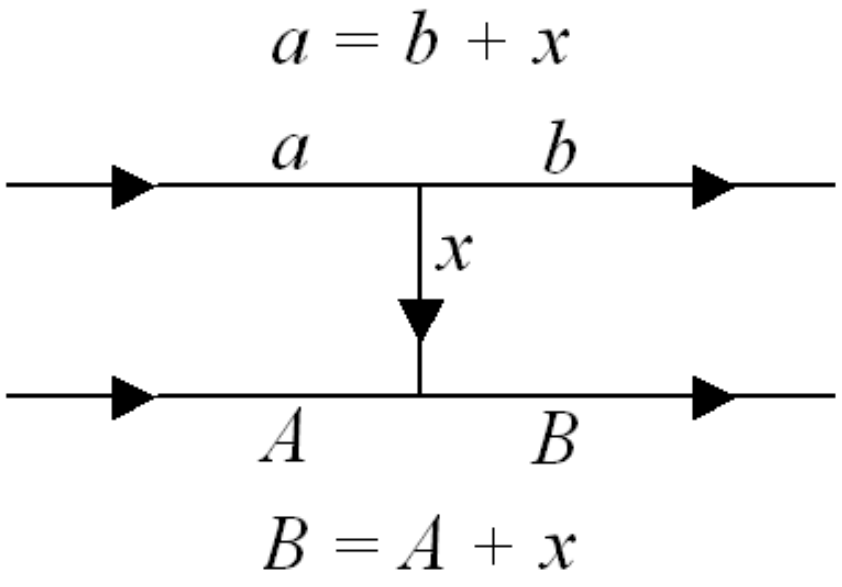

Figure E.1. The stripping reaction A($a$,$b$)B. The particle $x$ is stripped from $a$ and absorbed by $A$ to form the particle $B$.

Information about the reaction mechanism and nuclear structure is contained in the transition matrix $T_{ba}$, which in its general form can be written as

$$T_{ba} = \langle \Psi_\beta | V | \Psi_\alpha \rangle$$

where $\Psi_\alpha$ and $\Psi_\beta$ are the complete wave functions describing the incident and outgoing channels, $\alpha(a,A)$ and $\beta(b,B)$, respectively, and $V$ is the interaction potential responsible for the transition $a \rightarrow b + x$.

The wave functions $\Psi_\alpha$ and $\Psi_\beta$, describing the entire systems $a + A$ and $b + B$ respectively are the solutions of the Schrödinger equations:

$$(H_\alpha + T_\alpha + U_\alpha - E)\psi_\alpha = 0$$

$$(H_\beta + T_\beta + U_\beta - E)\psi_\beta = 0$$

where

$$H_\alpha = H_a + H_A$$

$$H_\beta = H_b + H_B$$

$H_a$ and $H_b$ are the internal Hamiltonians for the incident and outgoing particles; $H_A$ and $H_B$ are the Hamiltonians for the target and residual nuclei; $T_\alpha$ and $T_\beta$ are the kinetic energy operators in channels $\alpha$ and $\beta$; $U_\alpha$ and $U_\beta$ are the operators describing the interaction of the incident particle $a$ with the target nucleus $A$ and of the outgoing particle $b$ with the residual nucleus $B$, respectively.

It is impossible to solve the Schrödinger equations for the entire systems $a + A$ and $b + B$ so we have to introduce simplifications, which lead to various versions of theories of direct reactions.

Functions $\Psi_\alpha$ and $\Psi_\beta$ can be written as products of three functions

$$\Psi_\alpha = \psi_a \psi_A \psi_{aA}$$

$$\Psi_\beta = \psi_b \psi_B \psi_{bB}$$

where

$\psi_a$, $\psi_b$, $\psi_A$, and $\psi_B$ are the internal wave functions for the particles $a$, $b$, $A$, and $B$. These wave functions are solutions of the following Schrödinger equations:

$$(H_a - \varepsilon_a)\psi_a$$

$$(H_b - \varepsilon_b)\psi_b$$

$$(H_A - \varepsilon_A)\psi_A$$

$$(H_B - \varepsilon_B)\psi_B$$

It is usually assumed that $H_k$ ($k = a, b, A,$ or $B$) are shell model Hamiltonians of the form

$$H_A = \sum_{i=1}^{A}(T_i + U_i) + \sum_{i \neq j} V_{ij}$$

where

$$T_i = -\frac{\hbar^2}{2m_i}\nabla_i^2$$

is the kinetic energy operator of the $i$th particle, $U_i \equiv U(r_i)$ a central potential acting on a single particle, and $V_{ij} \equiv V(\vec{r}_i - \vec{r}_j)$ an interaction potential between two particles $i$ and $j$, known as the residual interaction.

The wave functions $\psi_{aA}$ and $\psi_{bB}$ describe events taking place in the systems $a + A$ and $b + B$ but do not describe the intrinsic structure of particles $a$, $b$, $A$, and $B$.

It is convenient to write the transition matrix element in the form:

$$T_{\alpha\beta} = \int\int \psi_{aA}^{(-)}(\vec{r}_a)\langle \psi_b \psi_B | V | \psi_a \psi_A \rangle \psi_{bB}^{(+)}(\vec{r}_b) d\vec{r}_a d\vec{r}_b$$

where $\vec{r}_a$ and $\vec{r}_b$ are the relative coordinates of ($a$,$A$) and ($b$,$B$).

The nuclear matrix element $\langle \psi_b \psi_B | V | \psi_a \psi_A \rangle$, also known as the form factor, represents integration over all intrinsic coordinates. This form factor contains information about nuclear structure.

**The Plane Wave Born Approximation (PWBA)**

The simplest treatment of direct nuclear reactions is to assume that there is no interaction between $a + A$ and $b + B$ systems (except for a momentary interaction changing $a$ into $b$), i.e. that $U_\alpha = U_\beta = 0$ (see the Appendix C). The wave functions $\psi_{aA}$ and $\psi_{bB}$ can then be written as plane waves

$$\psi_{aA} = \exp(i\vec{k}_a \cdot \vec{r}_a)$$

$$\psi_{bB} = \exp(i\vec{k}_b \cdot \vec{r}_b)$$

where $\vec{k}_a$ and $\vec{k}_b$ are the relative momenta of particles $a$ and $b$.

The transition matrix has then the form:

$$T_{PWBA} = \int e^{-i\vec{k}_b \cdot \vec{r}_a} \left\langle \psi_b \psi_B \middle| V \middle| \psi_a \psi_A \right\rangle e^{i\vec{k}_b \cdot \vec{r}_b} d\vec{r}_a d\vec{r}_b$$

If we assume that particle $b$ is emitted from the same point where particle $a$ is absorbed, which makes $\vec{r}_a \approx \vec{r}_b = \vec{r}$ , we have for the following simpler expression for the transition matrix

$$T_{PWBA} = \int e^{-i\vec{k}_b \cdot \vec{r}} \left\langle \psi_b \psi_B \middle| V \middle| \psi_a \psi_A \right\rangle e^{i\vec{k}_a \cdot \vec{r}} d\vec{r} = \int e^{i\vec{q} \cdot \vec{r}} \left\langle \psi_b \psi_B \middle| V \middle| \psi_a \psi_A \right\rangle d\vec{r}$$

where $\vec{q} = \vec{k}_a - \vec{k}_b$ is the momentum transfer, which depends on the reaction angle $\theta$:

$$q = \left[k_a^2 + k_b^2 - 2k_a k_b \cos\theta\right]^{1/2}$$

The form factor $\left\langle \psi_b \psi_B \middle| V \middle| \psi_a \psi_A \right\rangle$ may be written as

$$\left\langle \psi_b \psi_B \middle| V \middle| \psi_a \psi_A \right\rangle = \sum_{l,m} f_l(r) Y_l^m(\theta', \phi')$$

where $l$ is the angular momentum transfer involved in the reaction, which is usually represented by just one value. In this case,

$$T_{PWBA} \propto \int j_l(qr) f_l(r) r^2 dr$$

where $j_l(qR)$ is the spherical Bessel function of the order of $l$.

If we assume that the nucleus is a black object with the radius $R$, then the integration can be carried out for $r \geq R$. If we further assume that the reaction is confined to nuclear surface, i.e. that $r \approx R$ and thus neglect integration over $r > R$ then in its simplest form

$$T_{PWBA} \propto j_l(qR)$$

The differential cross sections, which are proportional to the square of the transition matrix for the reaction A($a$,$b$)B are in its simplest form proportional to the square of the spherical Bessel function

$$\frac{d\sigma}{d\Omega} \propto j_l^2(qR)$$

This is a convenient formula, which could be used easily in the interpretation of experimental data. By comparing the $j_l^2(qR)$ function with experimental angular distributions one can easily determine the $l$ transfer and thus the orbital angular momentum of nuclear state excited in a single particle stripping reaction or angular momentum involved in inelastic scattering.

The radius $R$ is a parameter that is obtained by matching the position of a dominant maximum in the experimental angular distribution with the relevant maximum of the $j_l^2(qR)$ function (see the Appendices B and C). This parameter may be interpreted as a nuclear radius.

## The Distorted Wave Theory

### *Formalism*

The next step in improving theoretical description of direct reactions was to introduce the concept of distorted waves. In this description, it is assumed that there is nuclear interaction between $a$ and $A$ and $b$ and $B$, that is, that the interaction potentials $U_\alpha \neq 0$ and $U_\beta \neq 0$. However, the key assumption is that the interactions in both the entrance and exit systems are dominated by the elastic scattering. Coupling between all other channels (inelastic scattering and rearrangement) which result from a collision of $a$ with $A$ are treated as perturbations and are absorbed in the imaginary components of the potentials describing the systems $a + A$ and $b + B$. The transition matrix can then be written as

$$T_{DW} = \int \psi_{aA}^{(-)*}(\vec{k}_a \vec{r}_a) \langle \psi_b \psi_B | V | \psi_a \psi_A \rangle \psi_{bB}^{(+)}(\vec{k}_b \vec{r}_b) d\vec{r}_a d\vec{r}_b$$

where $\psi_{aA}^{(-)}$ and $\psi_{bB}^{(+)}$ are the "distorted waves". They are just elastic scattering waves describing the relative motion of $a + A$ and $b + B$. These functions are generated by solving Schrödinger equations containing optical model potentials.

In principle, parameters of optical model potentials in the entrance and exit channels should be determined by fitting angular distributions for the elastic scattering of $a$ from $A$ and $b$ from $B$. However, in general, only the distributions for the elastic scattering of $a$ from $A$ can be measured and fitted using optical model procedure. To determine optical model parameters for the exit channel, $b + B$, various alternative methods are used. If $b$ can be used as a projectile and $B$ as a target and if the energy of $b$ can be made to match the energy of $b$ in the exit channel of the studied reaction, then optical model parameters for the exit $b + B$ channel can be determined by measuring the $b + B$ elastic scattering angular distribution and fitting it using optical model procedure. If the elastic scattering angular distributions for the exit channels cannot be measured, then one can use optical model parameters determined for the neighbouring masses and energies of $b$ and $B$. In fact, it may be argued that it is better to use the "average" optical model parameters both in the entrance and exit channels when fitting the A($a,b$)B angular distributions. The "average" parameters are obtained by studying elastic scattering angular distributions over a sufficiently wide range of targets $A$ and $B$ for incident particles $a$ and $b$ and for energies, which are in the vicinity of the energies for the entrance and exit channels of the reaction A($a,b$)B.

In its simplest form, the optical model potential contains only two central components, the real and the imaginary. The interaction potential $U(r)$ describing the relative motion of particles can be written as:

$$U(r) = -Vf(x) - iWg(x') + V_C(r)$$

where $V$ and $W$ are potential depths, $f(x)$ and $g(x')$ are the shape functions and $V_C(r)$ is the Coulomb potential. The function $f(x)$ has the Woods-Saxon form:

$$f(x) = (1+e^x)^{-1} \quad \text{where} \quad x = \frac{r - r_0 A^{1/3}}{a}$$

The function $g(x')$ has either the Woods-Saxon form (for the volume absorption):

$$g(x') = (1+e^{x'})^{-1} \qquad x' = \frac{r - r_0' A^{1/3}}{a'}$$

or the derivative of the Woods-Saxon shape (for the surface absorption):

$$g(x') = 4a' \frac{d}{dr}\left[(1+e^{x'})^{-1}\right]$$

The Coulomb potential $V_C(r)$ is taken as the potential due to a uniformly charged sphere of radius $R_C$:

$$V_C(r) = \frac{Ze^2}{r} \quad \text{for} \quad r > R_C$$

$$V_C(r) = \frac{Ze^2}{2R_C}\left(3 - \frac{r^2}{R_C^2}\right) \quad \text{for} \quad r \le R_C$$

where $Z$ is the charge of the target nucleus.

Other terms may be added to the optical model potential. The most common additional term is the spin-orbit potential:

$$V(r) = V_{so}\left(\frac{\hbar}{m_\pi c}\right)^2 \frac{1}{r}\frac{dh(r, r_{so}, a_{so})}{dr} \mathbf{L}\cdot\mathbf{S}$$

where $V_{so}$ is the spin-orbit potential depth and

$$h(r, r_{so}, a_{so}) = \frac{1}{1 + \exp[(r - r_{so}A^{1/3})/a_{so}]}$$

Finally, more complex tensor components are also included when necessary (see for instance Chapter 24).

The distorted waves $\psi^{(\pm)}(\vec{k},\vec{r})$ generated by the optical model potentials in the input and output channels take an asymptotic form of a plane wave with momentum $\vec{k}$ plus an outgoing (or incoming) spherical scattered wave, which in the absence of Coulomb interaction has the form

$$\psi^{(\pm)}(\vec{k},\vec{r}) \to e^{i\vec{k}\cdot\vec{r}} + f(\theta)\frac{e^{\pm ikr}}{r}$$

If we assume that the distorting potentials for the incoming and outgoing particles contain spin orbit interaction, then the distorted waves become matrices in the spin space

$$\psi^{(\pm)}(\vec{k},\vec{r})\eta_{s,m} = \sum_{m'} \psi^{(\pm)}_{m,m'}(\vec{k},\vec{r})\eta_{s,m'}$$

where $\eta_{s,m}$ are spin functions.

To solve the relevant Schrödinger equations, the usual partial wave expansion is carried out for functions $\psi^{(\pm)}_{m,m'}(\vec{k},\vec{r})$

$$\psi^{(\pm)}_{m,m'}(\vec{k},\vec{r}) = \frac{\sqrt{4\pi}}{kr}\sum_{J,L} i^L \sqrt{2L+1}\chi_{JLs}(k,r)(LsMm \mid JM')(LsM'-m'm' \mid JM')Y_L^{M'-m'}(\vec{r})d^L_{0,M'-m'}(\vec{k})$$

where $d^L_{0,m}$ are the rotations functions for integer spin (Edmonds 1957).

The radial parts of distorted waves satisfy the following equations:

$$\left\{\frac{d^2}{dr^2} + k^2 - \frac{L(L+1)}{r^2} - \frac{2\mu}{\hbar^2}\left[U(r)\right]\right\}\chi_{JLs}(k,r) = 0$$

where $U(r)$ is the potential describing elastic scattering of $a$ from $A$ or $b$ from $B$. It contains the Coulomb interaction potential and the usual optical model potential (with the spin-orbit interaction).

From here, the description of the theory becomes a little more complicated.

The form factor $\langle \psi_b\psi_B | V | \psi_a\psi_A \rangle$, which contains information about nuclear structure can be expressed as a sum containing elements $B_{lsj}$ and $f_{lsj}$. Elements $B_{lsj}$ measure the strength of the interaction and $f_{lsj}$ contain the details of reaction model. To simplify the calculations, delta function is also introduced. It describes zero-range interaction. Later simple procedure may be used to correct for the final-range interaction. The expression for the form factor $\langle \psi_b\psi_B | V | \psi_a\psi_A \rangle$ has the following form:

$$\begin{aligned}
&\langle \psi_b\psi_B | V | \psi_a\psi_A \rangle \equiv \langle J_B M_B s_b m_b | V | J_A M_A s_a m_a \rangle \\
&= \sum_{lsj} B_{lsj}(J_A j M_A M_B - M_A \mid J_B M_B)(s_a s m_b m_a - m_b \mid s_a m_a)(lsmm_a - m_b \mid jM_A - M_B) \\
&\times f_{lsj}(r_c)\delta\left(r_b - \frac{A}{B}r_a\right) i^{-l} Y_l^m(\vec{r}_a)^*
\end{aligned}$$

The angular momenta satisfy the following relations:

$$\hat{j} = \hat{J}_B - \hat{J}_A \qquad \hat{s} = \hat{s}_b - \hat{s}_a \qquad \hat{l} = \hat{j} - \hat{s}$$

In practice, usually only one set of $jls$ transfer is involved so the calculations are significantly simplified.

Using the information about the partial waves and the nuclear form factor, we can now express the transition matrix element in the following form:

$$T_{M_A M_B;m_a m_b} = \frac{\sqrt{4\pi}}{k_a k_b}\sum_{lsj}\sqrt{2l+1}B_{lsj}(J_A j M_A M_B - M_A \mid J_B M_B) S_{lsj}^{m m_a m_b}$$

This expression contains the already familiar $B_{lsj}$ element and a $S_{lsj}^{m m_a m_b}$, known as the reaction amplitude. Hidden in the reaction amplitude are the following elements: (a) the angular dependence expressed as a sum of Legendre polynomials, (b) the model-dependent radial form factors $f_{lsj}$, and (c) the distorted partial waves $\chi_{JL}$ for the entrance and exit channels.

$$S_{lsj}^{m m_a m_b} = \sum_{L_b}\beta_{lsj;L_b}^{m m_a m_b} P_{L_b}^{m_a - m - m_b}$$

where the amplitude $\beta_{lsj;L_b}^{m m_a m_b}$ has the form:

$$\begin{aligned}
&\beta_{lsj;L_b}^{m m_a m_b} \\
&= \sum_{J_a L_a J_b m}(L_a s_a 0 m_a \mid J_a m_a)(L_b s_b M m_a - m - m_b m_b \mid J_b m_a - m) \\
&\times (J_b j m_a - m m \mid J_a m_a)(2L_b + 1)(L_b l 00 \mid L_a 0) \\
&\times \sqrt{(2s_a+1)(2j+1)(2J_b+1)(2L_a+1)}\begin{Bmatrix} L_b & s_b & J_b \\ l & s & j \\ L_a & s_a & J_a \end{Bmatrix} \\
&\times I_{J_a L_a J_b L_b}^{lsj}\, i^{L_a - L_b - l}
\end{aligned}$$

This formula contains the usual Clebsch-Gordan coefficients (….|..) and 9-*j* symbol {}. It also contains the radial integrals $I$ which are related to the partial waves $\chi$ in the following way:

$$I_{J_a L_a J_b L_b}^{lsj} = \frac{CB}{A^2}\int_0^\infty \chi_{J_b L_b}^{(-)}\left(k_b, \frac{A}{B} r_a\right) f_{lsj}(r_c)\chi_{J_a L_a}^{(-)}(k_a, r_a)\,dr_c$$

where $C$ is the mass of the core of the form factor.

The differential cross section for the reaction A(*a,b*)B, which can be either inelastic scattering or transfer reaction, is expressed in terms of the transition amplitude $T$ as follows:

$$\begin{aligned}
&\frac{d\sigma(\theta)}{d\Omega} \\
&= \left(\frac{\mu_b}{2\pi\hbar^2}\right)^2 \frac{\upsilon_b}{\upsilon_a}\frac{1}{(2J_A+1)(2s_a+1)}\sum_{M_A M_B m_a m_b}\left|T_{M_A M_B;m_a m_b}\right|^2 \\
&= \frac{1}{4\pi}\frac{2J_B+1}{2J_A+1}\frac{1}{E_a E_b}\frac{k_b}{k_a}\frac{1}{2s_a+1}\sum_{m_a m_b m}\left|\sum_{lsj}\sqrt{2l+1}B_{lsj}S_{lsj}^{m m_a m_b}\right|^2
\end{aligned}$$

where $E_a$ and $E_b$ are the centre of mass energies in the entrance and exit channels, respectively.

For the inelastic scattering ($a = b$) the differential cross section takes the form:

$$\frac{d\sigma(\theta)}{d\Omega}=\frac{2J_B+1}{2J_B+1}\frac{2l+1}{2j+1}\left|B_{lsj}\right|^2\sigma_{aa'}^{lsj}(\theta)$$

where

$$\sigma_{aa'}^{lsj}(\theta)=\frac{1}{4\pi}\frac{1}{E_aE_b}\frac{k_b}{k_a}\frac{1}{2s_a+1}\sum_{m_am_bm}\left|\sum_{lsj}S_{lsj}^{mm_am_b}\right|^2$$

For transfer reactions ($a\neq b$)

$$\frac{d\sigma(\theta)}{d\Omega}=\frac{2J_B+1}{2J_B+1}\frac{1}{2j+1}\frac{\left|B_{lsj}\right|^2}{10^4}\sigma_{ab}^{lsj}(\theta)$$

$$\sigma_{ab}^{lsj}(\theta)=\frac{1}{4\pi}\frac{1}{E_aE_b}\frac{k_b}{k_a}\frac{10^4}{2s_a+1}\sum_{m_am_bm}\left|\sum_{lsj}\sqrt{2l+1}S_{lsj}^{mm_am_b}\right|^2$$

In particular, if we consider a stripping reaction A($a,b$)B with $a = b + x$ and $B = A + x$ then the nuclear form factor can be written in terms of spectroscopic factors and a $D_0$ coefficient.

$$\left\langle J_BM_Bs_bm_b\middle|V\middle|J_AM_As_am_a\right\rangle=\sum_{jl}S_{jl}^{1/2}R_{jl}(r_{xA})(lsm\mu-m\mid j\mu)(s_bsm_bm_a-m_b\mid s_am_a)$$
$$\times(J_AjM_AM_B-M_A\mid J_BM_B)D(\vec{r}_{xb})Y_l^m(\vec{r}_{xb})^*$$

where $S_{jl}$ is the spectroscopic factor, $R_{jl}= f_{lsj}$ is the radial form factor for the transfer of $x$ into the target nucleus $A$, and $D(\vec{r}_{xb})$ is the product of the projectile internal function and the interaction potential between $x$ and $b$.

$$D(\vec{r}_{bx})\equiv V_{bx}(r_{bx})\psi_a(r_{bx})$$

Assuming zero-range interaction we have

$$D(\vec{r}_{bx})=D_0\delta(\vec{r}_x-\vec{r}_b)$$

which gives

$$D_0=\int V_{bx}(r_{bx})\psi_a(r_{bx})dr_{bx}$$

The reaction strength factor $B_{lsj}$ is then given by

$$B_{lsj}=S_{lj}^{1/2}D_0$$

Consequently, the expression for the differential cross section for the stripping reaction A($a,b$)B takes the form:

$$\left[\frac{d\sigma(\theta)}{d\Omega}\right]_{st}=\frac{2J_B+1}{2J_B+1}\frac{S_{lsj}}{2j+1}\frac{D_0^2}{10^4}\sigma_{ab}^{lsj}(\theta)$$

The cross section for the pickup reaction is

$$\left[\frac{d\sigma(\theta)}{d\Omega}\right]_{pk} = \left(\frac{k_a}{k_b}\right)^2 \frac{2s_a+1}{2s_b+1}\frac{2J_A+1}{2J_B+1}\left[\frac{d\sigma(\theta)}{d\Omega}\right]_{st}$$

The zero-range coefficient $D_0$ can be either estimated experimentally, if spectroscopic factors are known, or calculated. For instance, for deuterons, if we use the Hulthén function for $\psi_a(r_{bx})$, i.e.

$$\psi_d(r_{np}) \equiv \psi_d(r) = \left[\frac{\alpha\beta(\alpha+\beta)}{2\pi(\alpha-\beta)^2}\right]\left[\frac{e^{-\alpha r} - e^{-\beta r}}{r}\right]$$

we get

$$D_0^2 = \frac{8\pi\varepsilon^2}{\alpha^3}\left(\frac{\alpha+\beta}{\beta}\right)^3$$

where

$$\varepsilon = \frac{\alpha^2\hbar^2}{2\mu}$$

is the deuteron binding energy ($\varepsilon$ = 2.23 MeV) and $\mu$ is the reduced mass ($\mu$ = 1/2).

Assuming that $\beta = 7\alpha$ we can calculate that

$$D_0^2 = 1.55\times10^4 \text{ MeV}^2\text{fm}^3$$

The difference between plane wave and distorted wave calculations is illustrated in Figure E.2.

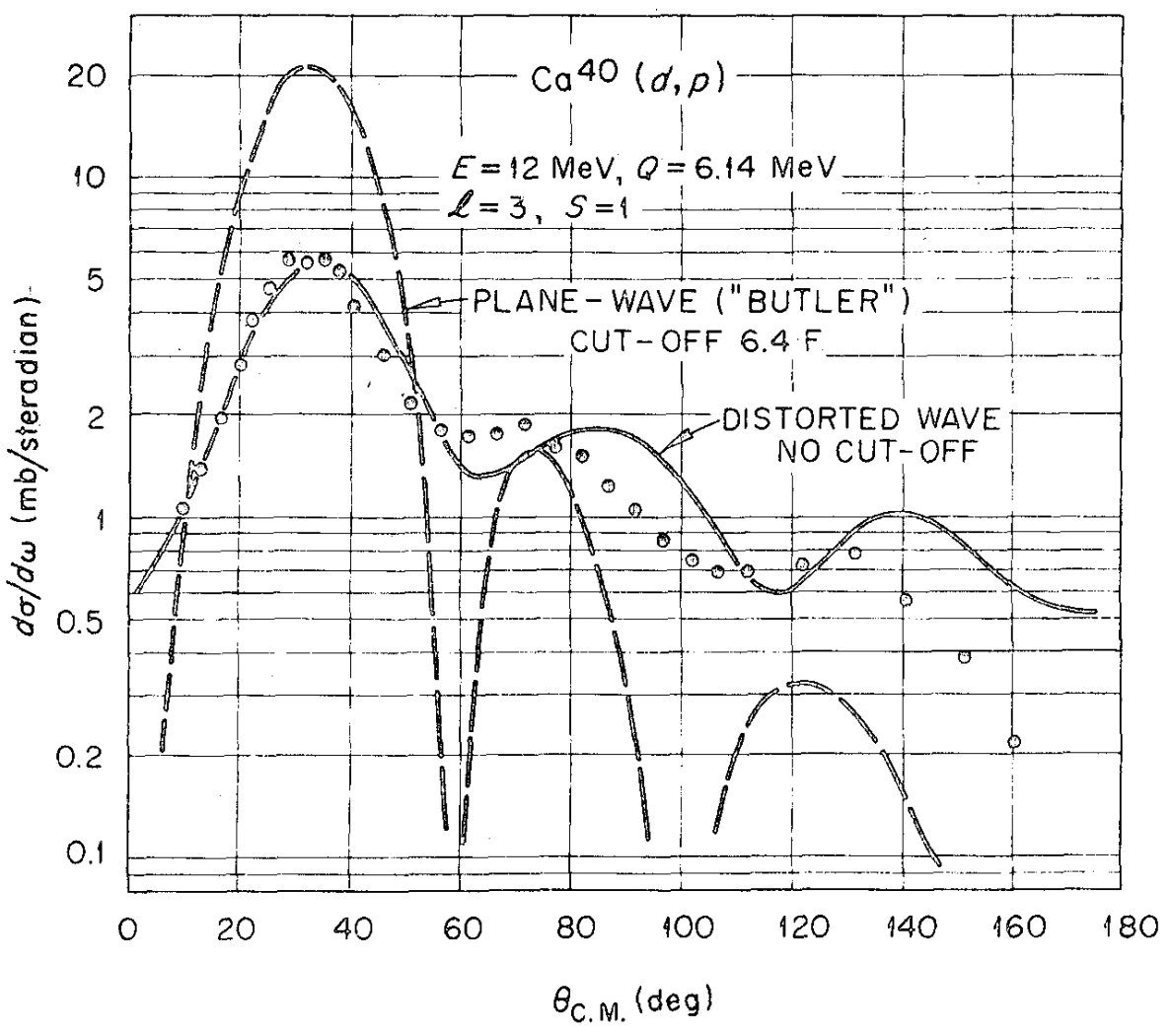

Figure E.2. Plane-wave and distorted-wave calculations are compared with experimental results for $^{40}$Ca(d,p)$^{41}$Ca reaction (Lee *at al.* 1964).

### *Finite-range and nonlocality corrections*

A correction for the finite-range interaction for single particle transfer reactions can be made by multiplying the form factor $f_{lsj}$ by one of the following functions

$$W_{FR} = \frac{1}{1 + A(r)} \qquad \text{Hulthén form}$$

$$W_{FR} = e^{-A(r)} \qquad \text{Gaussian from}$$

where

$$A(r) = \frac{2}{\hbar^2} \frac{m_b m_x}{m_a} R_{FR}^2 \left[ E_b - V_b(r_b) + E_x - V_x(r_x) - E_a + V_a(r_a) \right]$$

with $R_{FR}$ being the finite range parameter (see Table E.1), and $E_a$, $E_b$, $E_x$, and $V_a$ , $V_b$, $V_x$ the energies and potentials of particles $a$, $b$, and $x$ with respect to the target.

Table E.1

Examples of zero-range coefficients $D_0^2$ and finite range correction factors $R_{FR}$

| Reaction | $D_0^2$ | $R_{FR}$ |
|---|---|---|
| (d,p) | 1.55 | 0.621 |
| ($^3$He,d) | 4.42 | 0.770 |
| ($^3$H,d) | 5.06 | 0.845 |
| ($^4$He,$^3$He) | 24-46 | 0.700 |
| ($^4$He,$^3$H) | 24-46 | 0.700 |

The assumption that optical model potential depends only on the distance of two interacting particles does not always give a satisfactory description of experimental data. The correction for nonlocality of the interaction potentials in the entrance and exit channels, and for the transferred particle, may be made by multiplying the form factor for each particle by the following function:

$$W_{NL} = \left[ \frac{\beta_i^2}{8} \frac{2m_i}{\hbar^2} U_i(r_i) \right]$$

where $\beta_i$ is the non-locality parameter for particle $i$. For instance, typical values $\beta$ are. 0.85 fm, 0.54 fm, and 0.25 fm for nucleons, deuterons and tritons, respectively (Bassel 1966).

## Coupled channels theory

The coupled channels theory is an extension of the distorted waves formalism. In this theory, the wave function for the relative motion of interacting particles includes not only the elastic scattering but also coupling to other processes, which are present in any nuclear reaction but which are neglected in the distorted wave theory. For instance, coupling to strong inelastic channels might influence a measured angular distribution for a transfer reaction.

Coupled channels theory allows also to carry out calculations for the two-step or multi-step reactions. An example of such processes is discussed in Chapter 17 for the vector polarization of elastically scattered deuterons from Se isotopes. In that study, I have considered not only the single step (d,d) scattering but also two step processes (d,d')(d',d), (d,p)(p,d), and (d,t)(t,d). I have shown that the experimentally observed isotopic effects of deuteron polarization can by explained as an interplay between the direct (d,d) and two-step (d,d')(d',d) scattering. Figure E.3 shows a few examples of one- and two-step excitations.

In this illustration, it is assumed that states $j$ and $j'$ in the A+1 nucleus have configurations based on the ground state $I$ of the A nucleus. They therefore can be excited directly from the ground state $I$ by a single-step process. Such transitions can be treated using distorted wave theory.

However, the state $j''$ is assumed to have no parentage based on the ground state $I$ and consequently cannot be excited via a single-step transition. This state can be accessed via a two-step reaction, which involves the excitation of either a target or residual nucleus, or both. Transitions of this type have to be treated using coupled cannels theory. The coupled channels theory allows also for considering two-step contributions to transitions, which can proceed via a single-step excitation, such as for the state $j'$ in Figure E.3. This state can be excited directly from *I* and indirectly via the state *j*.

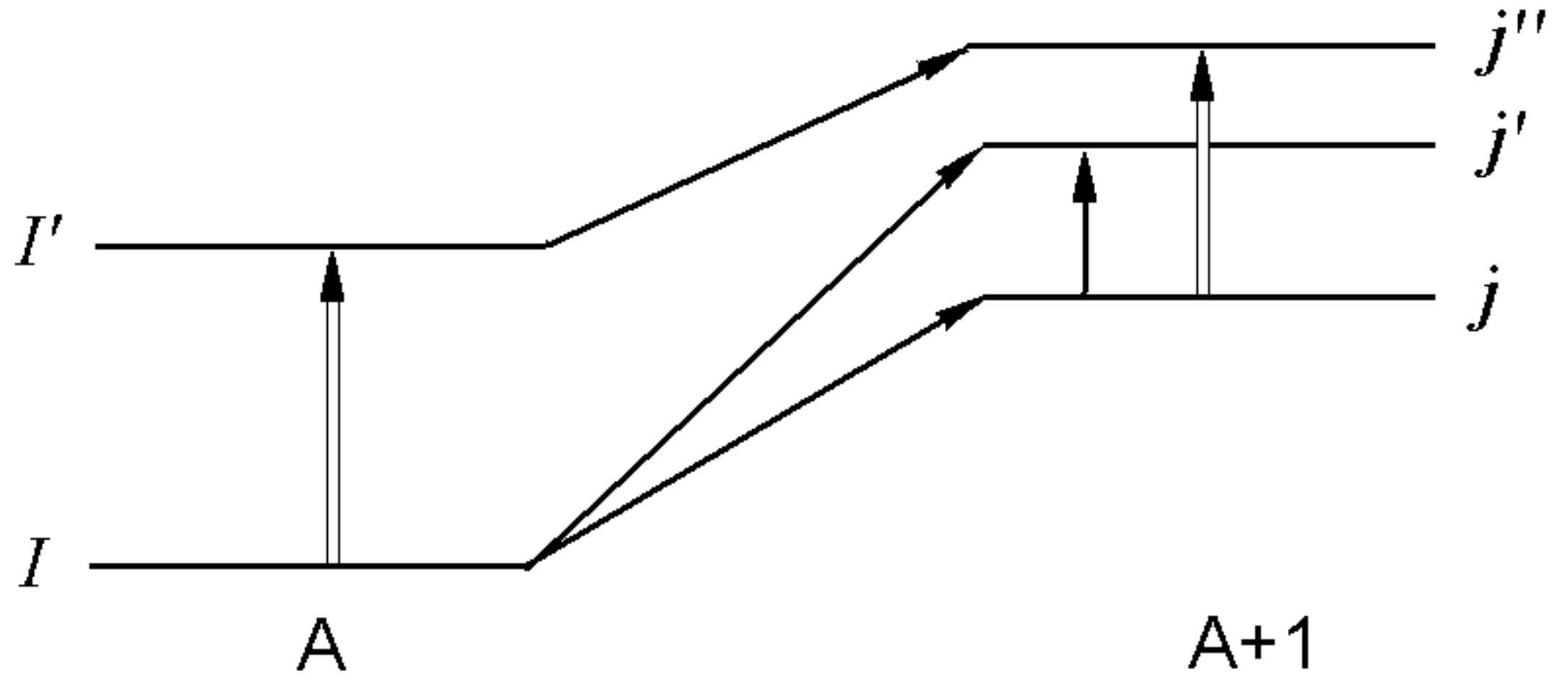

Figure E.3. Examples of single-step and two-step transitions from the target nucleus $A$ to the residual nucleus $A+1$. The double vertical lines indicate strong excitations by inelastic scattering. The state $j$ can be excited via a single-step transition. The state $j''$ can be excited only via two-step processes, which involve the excitation of the target and/or residual nucleus. The excitation of the state $j'$ can have contributions from a single-step and two-step transitions.

The concept of the coupled channels theory has been discussed in the Appendix D for the inelastic scattering. However, to see the basic difference between the coupled channels and distorted waves treatments of direct nuclear reactions it is useful to compare here the relevant Schrödinger equations for the partial wave functions.

Let us consider a wave function for a channel $c$ in a representation, which couples the orbital angular momentum of the relative motion $l_c$ to the intrinsic spin of the projectile $s_c$ to give $j_c$: $\hat{j}_c = \hat{l}_c + \hat{s}_c$.

The momentum $j_c$ is then coupled to the target spin $I_c$ to give angular momentum $J$ with the projection $M$: $\hat{J} = \hat{j}_c + \hat{I}_c$.

The wave function for all channels can then be written as

$$\Psi_{JM} = \sum_{cj_c l_c} \chi^{cJ}_{l_c j_c}(k_c, r_c)\left(\psi^c_{j_c l_c} \psi^c_{I_c}\right)_{JM}$$

The $\psi^c_{j_c l_c}$ is the wave function for coupling the relative angular momentum and intrinsic spin, and $\psi^c_{I_c}$ is the intrinsic wave function for the target. The Schrödinger equations for the radial wave function $\chi^{cJ}_{l_c j_c}$ are:

$$\left\{\frac{d^2}{dr_c^2} + k_c^2 - \frac{l_c(l_c+1)}{r_c^2} - \frac{2\mu_c}{\hbar^2}\left[U_{cc}(r_c)\right]\right\}\chi^{cJ}_{j_c l_c}(k_c, r_c) = \sum_{c'} \frac{2\mu_{c'}}{\hbar^2} U_{cc'}(r_{c'})\chi^{cJ}_{j_c l_c}(k_{c'}, r_{c'})$$

We can now compare them with the equivalent equations in the distorted waves theory:

$$\left\{\frac{d^2}{dr^2} + k^2 - \frac{L(L+1)}{r^2} - \frac{2\mu}{\hbar^2}\left[U(r)\right]\right\}\chi_{JLs}(k, r) = 0$$

We can see that the essential difference is that the zero on the right-hand side of the equations for the distorted waves is now replaces by a sum of terms describing coupling between various channels.

The diagonal term $U_{cc}$ is the potential, which describes the relative motion of particles in channel $c$. It contains both the Coulomb and optical model potentials. The off-diagonal term $U_{cc'}$ is a coupling potential, which is given by the relation:

$$U_{cc'} = \sum_{m_c m_{c'} \nu_c \nu_{c'} M_c M_{c'}} (l_c s_c m_c \nu_c \mid j_c \gamma_c)(l_{c'} s_{c'} m_{c'} \nu_{c'} \mid j_{c'} \gamma_{c'})$$
$$\times (j_c I_c \gamma_c M_c \mid JM)(j_{c'} I_{c'} \gamma_{c'} M_{c'} \mid JM)\left\langle i^{l_c} Y^{m_c}_{l_c} \phi^c_{s_c} \phi^c_{J_c M_c} \middle| V \middle| i^{l_{c'}} Y^{m_{c'}}_{l_{c'}} \phi^{c'}_{s_{c'}} \phi^{c'}_{J_{c'} M_{c'}} \right\rangle$$

The formalism contains the familiar elements, such as $f_{lsj}$ and $B_{lsj}$ but now they also have channels indices (e.g. $f^{cc'}_{lsj}$ and $B^{cc'}_{lsj}$). These and other quantities enter into the elements that are used in the calculations of relevant observables, such as the differential cross sections.

The differential cross section for channel $c$ is given by:

$$\frac{d\sigma_c(\theta)}{d\Omega} = \frac{1}{2I_{c_1}+1}\frac{1}{2s_{c_1}+1}\sum_{M_{c_1} M_c \nu_{c_1} \nu_c}\left| f_{coul}(\theta) + \sum_{l_c m_c} D^{l_c m_c}_{M_{c_1} M_c \nu_{c_1} \nu_c} P^{m_c}_{l_c}(\theta)\right|^2$$

where the subscript $c_1$ refers to the initial elastic channel, $f_{coul}(\theta)$ is the Coulomb scattering amplitude (see Appendix D), $D^{l_c m_c}_{M_{c_1} M_c \nu_{c_1} \nu_c}$ are the reaction amplitudes, which also contain the Coulombs phase shifts, and $P^{m_c}_{l_c}(\theta)$ are the Legendre polynomials.

# Appendix F

## Nuclear spin formalism

This Appendix contains a brief outline of the nuclear spin formalism for spin-1/2 and spin-1 particles. For more details see such publications as Worfenstein (1956), Gammel, Keaton, and Ohlsen (1970), and Ohlsen (1972).

### Spin-$^1/_2$ particles

A single spin-$^1/_2$ particle can be represented by a Pauli spinor

$$\chi = \begin{bmatrix} a_1 \\ a_2 \end{bmatrix}$$

The expectation value of an observable corresponding to a Hermitian operator[9] $O$ is given by[10]

$$\langle O \rangle = \chi^+ O \chi = \begin{bmatrix} a_1^* & a_2^* \end{bmatrix} \begin{bmatrix} O_{11} & O_{12} \\ O_{21} & O_{22} \end{bmatrix} \begin{bmatrix} a_1 \\ a_2 \end{bmatrix}$$

$$= |a_1|^2 O_{11} + a_1^* a_2 O_{12} + a_2^* a_1 O_{21} + |a_2|^2 O_{22} = Tr\rho O$$

where $\rho$ is the density matrix defined as

$$\rho = \begin{bmatrix} a_1 \\ a_2 \end{bmatrix} \begin{bmatrix} a_1^* & a_2^* \end{bmatrix} = \begin{bmatrix} a_1 a_1^* & a_1 a_2^* \\ a_2 a_1^* & a_2 a_2^* \end{bmatrix}$$

For an ensemble of $N$ particles, each element of the density matrix is replaced by the average value

$$\rho_{jk} = \frac{1}{N} \sum_{n=1}^{N} a_j^{(n)} a_k^{(n)*} \qquad j,k = 1,2$$

The state of polarization of an ensemble of spin-$^1/_2$ particles is specified by the expectation values of the Pauli spin operators:

$$p_X = \langle \sigma_x \rangle = Tr\rho\sigma_x$$

$$p_Y = \langle \sigma_y \rangle = Tr\rho\sigma_y$$

$$p_Z = \langle \sigma_z \rangle = Tr\rho\sigma_z$$

---

[9] An operator $O$ is Hermitian if $O^+ = O$. By definition, if $a_{ij}$ are the elements of matrix $O$ and $b_{ij}$ elements of $O^+$ then $b_{ij} = a_{ji}^*$. Consequently, for a Hermitian operator $b_{ij} = a_{ij}$. The average value of a Hermitian operator is real.

[10] The general formula is $\langle O \rangle = Tr\rho O / Tr\rho = Tr\rho O$ if $Tr\rho = 1$.

where

$$\sigma_x = \begin{bmatrix} 0 & 1 \\ 1 & 0 \end{bmatrix};\ \sigma_y = \begin{bmatrix} 0 & -i \\ i & 0 \end{bmatrix}; \sigma_z = \begin{bmatrix} 1 & 0 \\ 0 & 1 \end{bmatrix}$$

are Pauli spin operators.

If we choose the polarization axis along the quantization axis we shall have $N_+$ particles aligned with the quantization axis and $N_-$ particles aligned in the opposite direction. The density matrix is then given by

$$\rho = \frac{1}{N_+ + N_-}\begin{bmatrix} N_+ & 0 \\ 0 & N_- \end{bmatrix}$$

and the beam polarization by

$$p_Z = Tr\rho\sigma_Z = \frac{N_+ - N_-}{N_+ + N_-}$$

with $p_X = p_Y = 0$.[11]

It can be easily seen that Pauli matrices are orthogonal and normalized, i.e. that

$$Tr\sigma_i\sigma_j = 2\delta_{ij}$$

The three Pauli operators and the 2x2 unit matrix form a complete orthogonal set and consequently, any 2x2 matrix, $M$, can be expanded in terms of this basic set.

## Spin-1 particles

A spin-1 particle is represented by a three-component spinor

$$\chi = \begin{bmatrix} a_1 \\ a_2 \\ a_3 \end{bmatrix}$$

The basis spin-1 angular momentum operators are defined as

$$S_x = \frac{1}{\sqrt{2}}\begin{bmatrix} 0 & 1 & 0 \\ 1 & 0 & 1 \\ 0 & 1 & 0 \end{bmatrix};\ S_y = \frac{1}{\sqrt{2}}\begin{bmatrix} 0 & -i & 0 \\ i & 0 & -i \\ 0 & i & 0 \end{bmatrix};\ S_x = \begin{bmatrix} 1 & 0 & 0 \\ 0 & 0 & 0 \\ 0 & 0 & -1 \end{bmatrix}$$

These three vector polarization operators and a unit matrix form a set of four Hermitian operators but nine are needed to span the 3x3 space. It is therefore necessary to construct additional operators. These additional operators are defined by the following relations (Goldfarb 1958):

---

[11] I am using the capital letters for the subscripts to distinguish the polarization components in the symmetry axis frame of reference and the components in the laboratory frame of reference, which will be introduced later (see the Appendix H).

$$S_{ij} = \frac{3}{2}(S_i S_j + S_j S_i) - 2\delta_{ij} I \quad j,k = x,y,z$$

where

$$I = \begin{bmatrix} 1 & 0 & 0 \\ 0 & 1 & 0 \\ 0 & 0 & 1 \end{bmatrix}$$

Explicitly, the additional operators are

$$S_{xy} = \frac{3}{2}\begin{bmatrix} 0 & 0 & -i \\ 0 & 0 & 0 \\ i & 0 & 0 \end{bmatrix} \quad S_{xz} = \frac{3}{\sqrt{8}}\begin{bmatrix} 0 & 1 & 0 \\ 1 & 0 & -1 \\ 0 & -1 & 0 \end{bmatrix} \quad S_{yz} = \frac{3}{\sqrt{8}}\begin{bmatrix} 0 & -i & 0 \\ i & 0 & i \\ 0 & -i & 0 \end{bmatrix}$$

$$S_{xx} = \frac{1}{2}\begin{bmatrix} -1 & 0 & 3 \\ 0 & 2 & 0 \\ 3 & 0 & -1 \end{bmatrix} \quad S_{yy} = \frac{1}{2}\begin{bmatrix} -1 & 0 & -3 \\ 0 & 2 & 0 \\ -3 & 0 & -1 \end{bmatrix} \quad S_{zz} = \begin{bmatrix} 1 & 0 & 0 \\ 0 & -2 & 0 \\ 0 & 0 & 1 \end{bmatrix}$$

This definition adds six extra operators while only five are required to make a complete set of nine. We therefore have an over-complete set. However, it can be seen that

$$S_{xx} + S_{yy} + S_{zz} = 0$$

which means that there are only nine independent operators. It can be also checked that they form an orthogonal set but they are not normalized.

$$TrS_i S_j = 0 \qquad \text{for } i \neq j$$

$$TrS_i S_j = 2 \qquad \text{for } i = j$$

$$TrS_{ij} S_{kl} = 0 \qquad \text{for } ij \neq kl$$

$$TrS_{ij} S_{ij} = 9/2$$

$$TrS_{ii} S_{ii} = 6$$

We can define a new set of operators $\Omega_i$ related to the operators $S_i$ and $S_{ij}$ that will also form a complete set and for which

$$Tr\Omega_i \Omega_j = 3$$

Explicitly, this new set of nine operators is as follows:

$$\Omega_0 = I$$

$$\Omega_1 = \sqrt{\frac{3}{2}} S_x \quad \Omega_2 = \sqrt{\frac{3}{2}} S_y \quad \Omega_3 = \sqrt{\frac{3}{2}} S_z \,;$$

$$\Omega_4 = \sqrt{\frac{2}{3}} S_{xy} \quad \Omega_5 = \sqrt{\frac{2}{3}} S_{xx} \quad \Omega_6 = \sqrt{\frac{2}{3}} S_{xx}$$

$$\Omega_7 = \sqrt{\frac{1}{6}} (S_{xx} - S_{yy}) \quad \Omega_8 = \sqrt{\frac{1}{2}} S_{zz}$$

If, as for spin-$^1/_2$ particles, we chose the $z$ axis to be along the quantization axis, then the density matrix is given by

$$\rho = \frac{1}{N_+ + N_0 + N_-} \begin{bmatrix} N_+ & 0 & 0 \\ 0 & N_0 & 0 \\ 0 & 0 & N_- \end{bmatrix}$$

Using this matrix, we can calculate that

$$\rho S_z = \frac{1}{N_+ + N_0 + N_-} \begin{bmatrix} N_+ & 0 & 0 \\ 0 & N_0 & 0 \\ 0 & 0 & N_- \end{bmatrix} \begin{bmatrix} 1 & 0 & 0 \\ 0 & 0 & 0 \\ 0 & 0 & -1 \end{bmatrix} = \frac{1}{N_+ + N_0 + N_-} \begin{bmatrix} N_+ & 0 & 0 \\ 0 & 0 & 0 \\ 0 & 0 & N_- \end{bmatrix}$$

Thus

$$p_Z \equiv \langle S_z \rangle = Tr \rho S_z = \frac{N_+ - N_-}{N_+ + N_0 + N_-}$$

Likewise, we can see that

$$\rho S_{zz} = \frac{1}{N_+ + N_0 + N_-} \begin{bmatrix} N_+ & 0 & 0 \\ 0 & N_0 & 0 \\ 0 & 0 & N_- \end{bmatrix} \begin{bmatrix} 1 & 0 & 0 \\ 0 & -2 & 0 \\ 0 & 0 & -1 \end{bmatrix} = \frac{1}{N_+ + N_0 + N_-} \begin{bmatrix} N_+ & 0 & 0 \\ 0 & -2N_0 & 0 \\ 0 & 0 & N_- \end{bmatrix}$$

and thus

$$p_{ZZ} \equiv \langle S_{zz} \rangle = Tr \rho S_{zz} = \frac{N_+ + N_- - 2N_0}{N_+ + N_0 + N_-}$$

We can also calculate that

$$p_{XX} \equiv \langle S_{xx} \rangle = p_{YY} \equiv \langle S_{yy} \rangle = -\frac{1}{2} p_{ZZ}$$

and therefore

$$p_{XX} + p_{YY} + p_{ZZ} = 0$$

In general, the description of the state of polarization of spin-1 beam, which does not have an axis of symmetry, may require all three vector components and five independent tensor components. It should be also noted that expectation values of the various operators are bound by the limits ±1 for the vector polarization, ±3/2 for $p_{ij}$ components with $i \neq j$, and +1 to -2 for $p_{ii}$ components.

## Some basic definitions

*Conjugate matrix.* If $a_{ij}$ are the elements of matrix $A$ ($A = [a_{ij}]$) then $A^* = [b_{ij}]$ is called a conjugate matrix of $A$ if $b_{ij} = a_{ij}^*$, where $a_{ij}^*$ is the complex conjugate of $a_{ij}$.

*Transpose matrix.* If $a_{ij}$ are the elements of matrix $A$ ($A = [a_{ij}]$) then $A^T = [b_{ij}]$ is called a transpose matrix of $A$ if $b_{ij} = a_{ji}$.

*The conjugate transpose matrix.* If $A = [a_{ij}]$ then $A^+ = [b_{ij}]$ is called its conjugate transpose if $b_{ij} = a_{ji}^*$.

*The inverse matrix.* A matrix $A^{-1}$ is an inverse of a square matrix $A = [a_{ij}]$ if $AA^{-1} = I$, where $I$ is the *unit* (the *identity*) matrix.

*The unitary matrix.* A square matrix $A$ is called *unitary* if $AA^+ = A^+A = I$, where $I$ is the *unit* (the *identity*) matrix. For a unitary matrix $A^{-1} = A^+$.

*The orthogonal matrix.* Matrix $A$ is *orthogonal* if $AA^T = I$.

A unitary orthogonal matrix is made of a set of orthogonal unit vectors. For instance, in the case we have considered in this Appendix, the matrix $U$ is made of two two-dimensional vectors:

$$\vec{a} = (\cos\varepsilon, \sin\varepsilon)$$

$$\vec{b} = (-\sin\varepsilon, \cos\varepsilon)$$

It is easy to see that these vectors are not only orthogonal but also normalised to unity, i.e. that $\vec{a}\cdot\vec{a} = \vec{b}\cdot\vec{b} = 1$ and $\vec{a}\cdot\vec{b} = 0$. Thus, matrix $U$ is both orthogonal and unitary.

# Appendix G

## Polarized ion sources

The two main types of polarized ion sources are the atomic-beam and the Lamb-shift varieties. Excellent introduction to these experimental devices may be found in an article by Haeberli (1967) but their components and performance were also discussed in numerous other publications. Below, I am presenting their simplified description.

### The atomic beam polarized ion sources

#### *Protons*

The conventional polarized ion source is based on the Stern-Gerlach separation of atoms belonging to different hyperfine substates. This can be understood with the help of Figure G.1 showing the Zeeman effect.

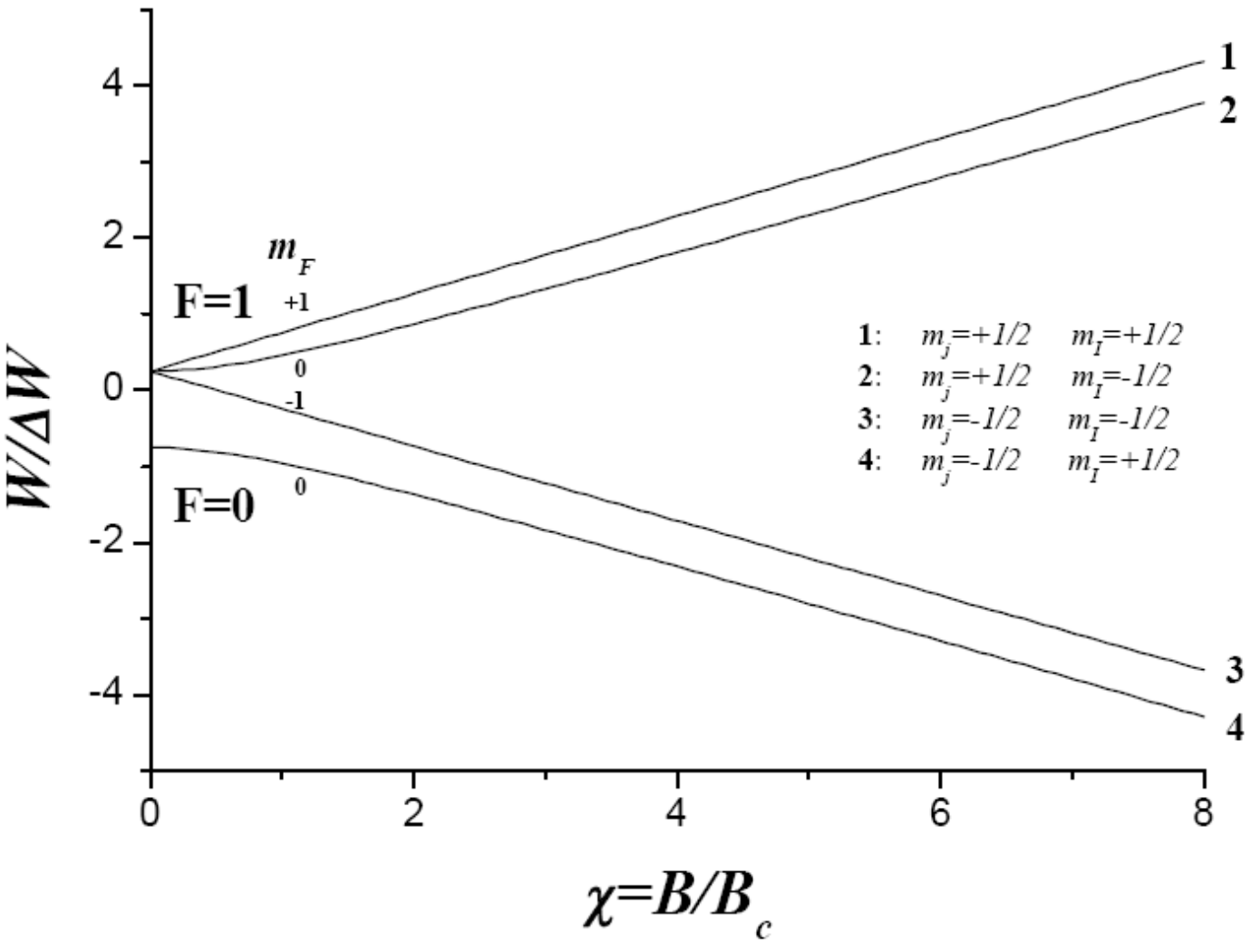

Figure G.1. Breit-Rabi energy level diagram of the $1S_{1/2}$ hyperfine states of the hydrogen atom in the magnetic field. The energy $W$ is in units $\Delta W$ and the magnetic field in $B_c$ (the critical magnetic field associated with the decoupling of the electron and nuclear moments). For the ground state of the hydrogen atom $\Delta W = h \times 1420.2$ MHz ($5.8 \times 10^{-6}$ eV) and $B_c = 507$ G. It should be noted that for weak fields ($\chi << 1$), the proton and electron spins cannot be treated as independent quantities and consequently, the hyperfine states have to be labelled using $m_F$ projection.

In the hydrogen atom, the electron with spin $j$ = 1/2 (projections $m_j$ = ±1/2) couples with the proton spin $I$ = 1/2 (projections $m_I$ = ±1/2) to form spin $F$, $\hat{F} = \hat{I} + \hat{j}$, which gives two values of $F$ ($F$ = 0, 1) with projections $m_F$ = 0, ±1.

In the absence of the magnetic field, the separation between the two substates $F$ is $\Delta W = h \times 1420.2$ MHz ($5.8 \times 10^{-6}$ eV) (Nagle, Julian, and Zacharias 1947). When the external magnetic field is applied, the substate $F$ = 1 splits into three components, which is known as the Zeeman effect.[12] The energy split is described by the Breit-Rabi formula (Ramsey 1956):

$$W = \frac{\Delta W}{2(2I+1)} + g_I \mu_B m_F B_c \chi + (-1)^{F+1}\sqrt{1 + \frac{4m_F}{2I+1}\chi + \chi^2}$$

where $g_I =$ -3.04×10$^{-3}$ and $g_j =$2.002, which are, respectively, the *g*-factors of the proton and electron in the units of the Bohr magneton $\mu_B =$-0.927×10$^{-20}$ erg/G.

The external magnetic field is measured in the units of the so-called critical field $B_c$ and expressed in terms of the ratio $\chi = B/B_c$, where

$$B_c = \frac{\Delta W}{(g_I - g_j)\mu_B}$$

which for the hydrogen atom in its ground state is 507 G.

The basic idea, used in most of the atomic beam polarized ion sources is first to reject states 3 and 4 (see Figure G.1) but accept states 1, and 2. This is done by following the original idea of the well-known Stern-Gerlach experiment but by using improved inhomogeneous magnets. In the original Stern-Gerlach experiment, a dipole magnet was used. In polarized ion sources the performance is improved by using a sextupole magnet or to a lesser degree by a quadrupole.

In an inhomogeneous magnetic field, the atom will move to the magnetic field that causes a decrease in the energy of its hyperfine substate. Thus, if we look again at the Figure G.1 we shall see that the atoms in substates 1 and 2 will move to an area where the magnetic field is weak. In contrast, the atoms with substates 3 and 4 will move in the direction of strong field.

In a sextupole (or quadrupole) magnet the gradient of the magnetic field is in the radial direction, i.e. it decreases radially towards the centre. This means that the atoms with substates 1 and 2 will move towards the centre, whereas the atoms with substates 3 and 4 will move away from the centre. Thus, they will be removed from the beam and pumped away.

This beam manipulation results in an atomic beam polarized in the electron spin ($m_j$ = +1/2) but not in the proton spin ($m_I$ = ±1/2). To polarize the atomic beam in the proton spin we have to change the population of the hyperfine states. It is possible, for instance to move the atoms in state 2 to state 4 and thus change the combination of states 1 and 2 to 1 and 4. For these two states $m_I$ = +1/2. Such transitions between states are achieved by applying an RF field.

Having achieved a desired combination of hyperfine substates, the next step is then to ionise the polarized atoms to make them ready for the acceleration. This is done in suitably designed ionisers.

There is, however, a slight complication in this relatively simple process of forming proton-polarized hydrogen beams because the degree of proton polarization for substates with antiparallel electron and proton spins depends on the external magnetic field in the ioniser. This dependence is associated with the precession of the electron and proton spins in the weak magnetic field and is shown if Figure G.2.

[12] An analogous splitting in an electric field is known as the Stark effect, or rarely as Stark-Lo Sturdo effect. It was discovered independently in 1913 by German and Italian physicists, Johannes Stark and Antonio Lo Sturdo. This effect is used in Lamb-shift sources in combination with the Zeeman effect.

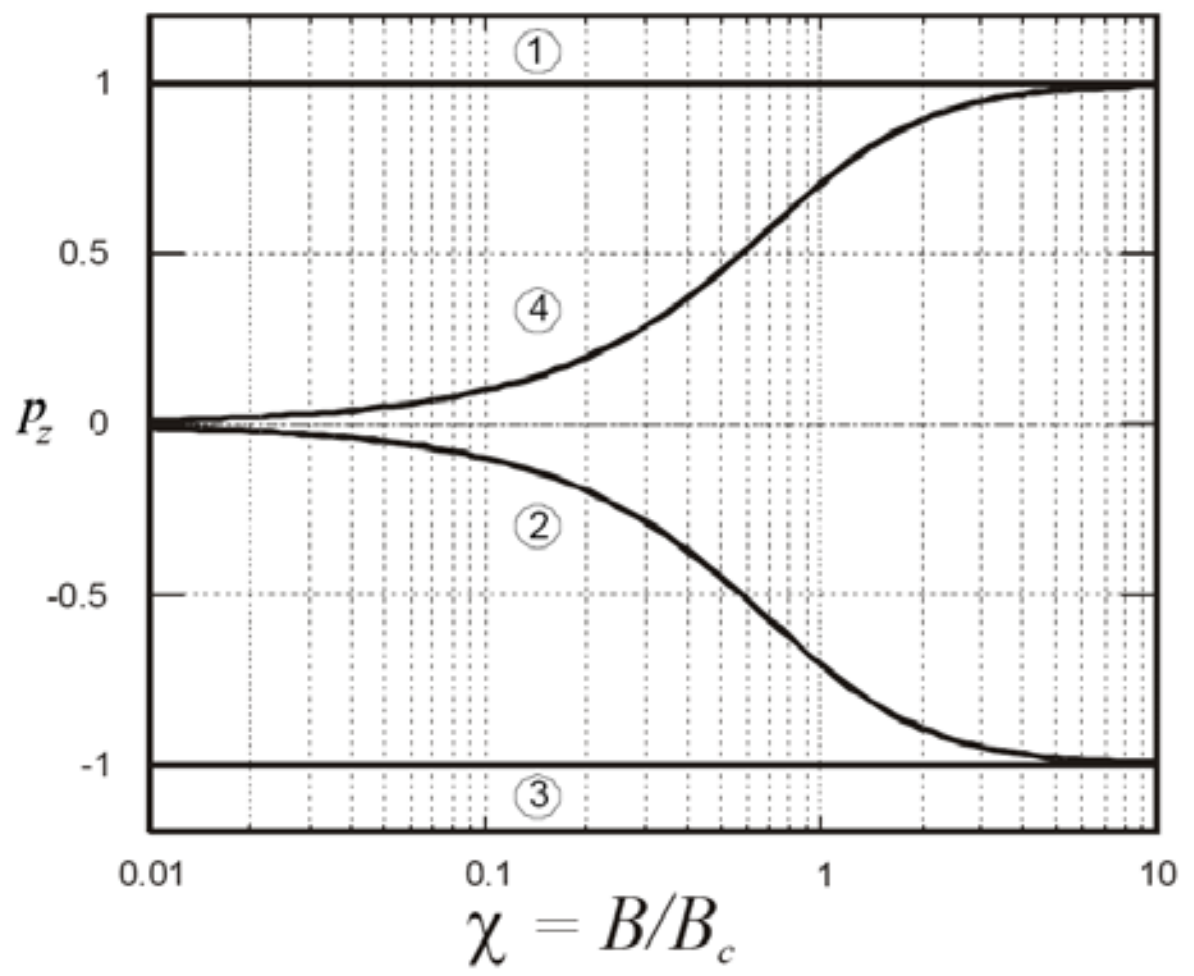

Figure G.2. The dependence of the vector polarization of protons in the hydrogen atom as the function of the external magnetic field.

We shall recall (see the Appendix F) that vector polarization $p_z$ is given by:

$$p_Z = \frac{N_+ - N_-}{N_+ + N_-}$$

where $N_\pm$ is the number of atoms with proton spin projections $m_I = \pm 1/2$.

The proton polarization of the atomic beam is an arithmetic average of the polarization of its various components. So, for instance, if we apply an RF field and move the atoms from the state 2 to 4, and if we follow this process by ionisation in a weak magnetic field then we shall have atoms in state 1 with the net polarization $p_z$ = 1 and in state 4 with polarization $p_z \approx 0$. If we assume that we have the same number of hydrogen atoms in states 1 and 4, then the resulting average vector polarization for such a beam will be

$$p_z \approx 0.5 \text{ (states 1 and 4 in the weak field)}$$

However, if we ionise the hydrogen atoms in a strong magnetic field then the average polarization will be

$$p_z \approx 1.0 \text{ (states 1 and 4 in the strong field)}$$

It should be noted that beam polarization can be created in a weak field without the use of the RF transition. If we look again at the Figure G.2 we shall see that if we use the hydrogen atoms in substates 1 and 2 (as they emerge from the sextupole magnet) and if we assume that the population of these two substates is the same, then the net polarization created in the weak magnetic field will be

$$p_z \approx 0.5 \text{ (states 1 and 2 in the weak field)}$$

Such weak-field devices were used in the late 1950s and early 1960s but they were not successful because of the low beam intensities they produced.

A simplified schematic diagram of a conventional (atomic-beam) polarized ion source is shown in Figure G.3

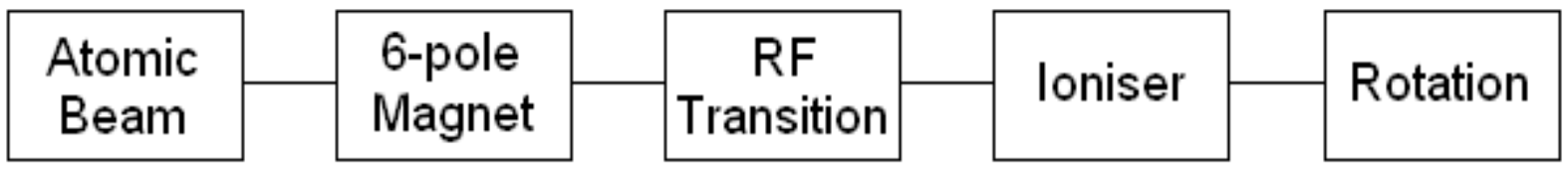

Figure G.3. A simplified schematic diagram showing the major components of the atomic beam polarized ion source. Atomic Beam – The atomic beam formation; 6-pole Magnet – An inhomogeneous magnet, sextupole or quadrupole. RF Transition – An RF unit to change the population of the hyperfine substates; Ioniser – A device used to convert the neutral atomic beam with polarized protons to negatively or positively charged ions; Rotation – A device used to change the orientation of the polarization axis.

***Deuterons***

For the deuterium atoms, the principles are the same but now $I$ = 1, $m_I$ = 0, ±1, $F$ = 1/2, 3/2, and $m_F$ = ±1/2, ±3/2. Consequently, we have three, rather than two, hyperfine substates in each branch. We also have a possibility of creating a beam with vector and tensor polarization.

The Breit-Rabi diagram for the deuterium atoms is presented in Figure G.4 and the corresponding vector and tensor polarizations for various hyperfine components in Figure G.5. In the absence of the external magnetic field $\Delta W = h \times 327.4$ MHz ($1.4 \times 10^{-6}$ eV). The critical magnetic field $B_c = 117$ G.

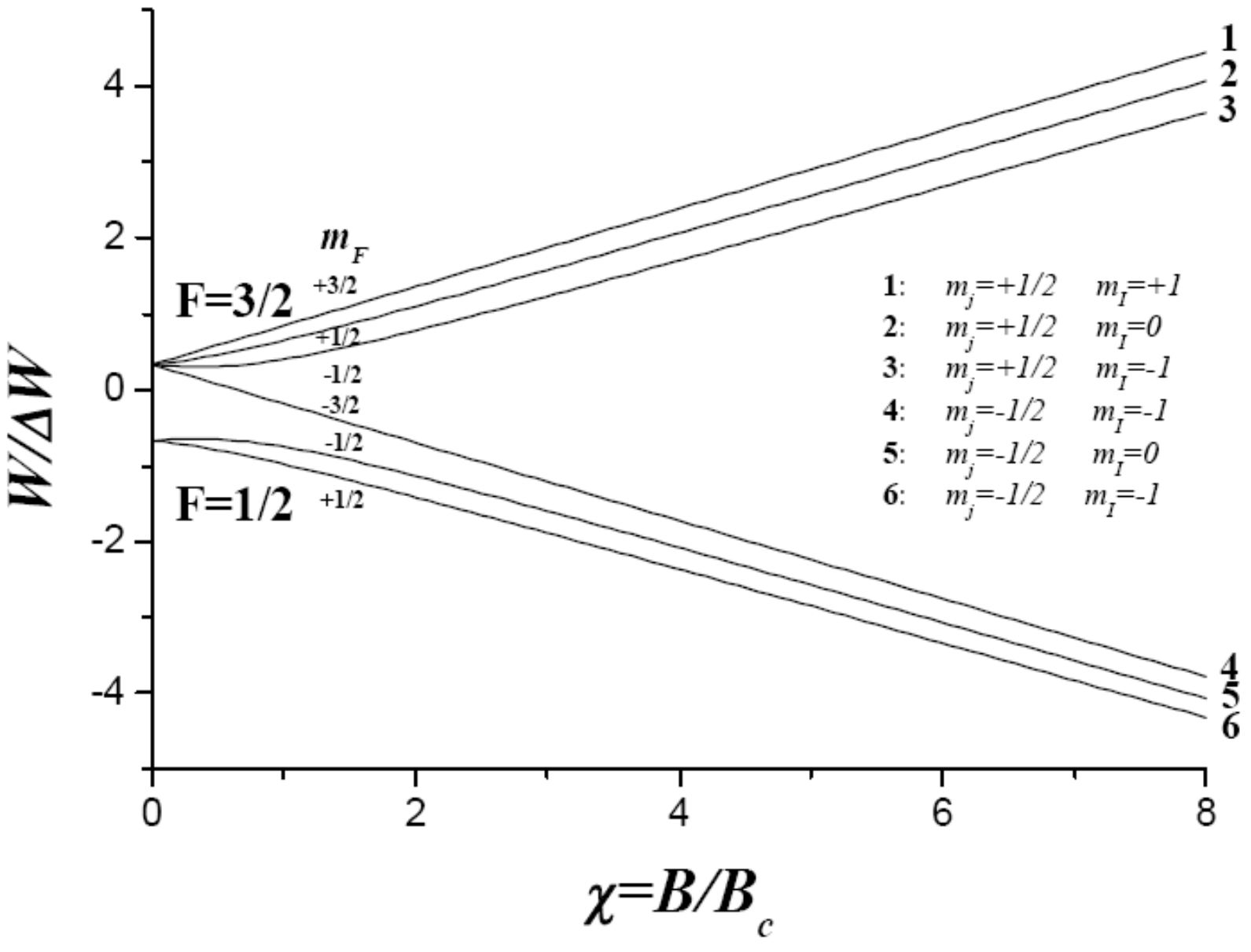

Figure G.4. The Breit-Rabi energy level diagram of the deuterium atom in the magnetic field. For the ground state of the deuterium atom $\Delta W = h \times 327.4$ MHz ($1.4 \times 10^{-6}$ eV) and $B_c = 507$ G.

After leaving the inhomogeneous magnetic field, the beam of the deuterium atoms will be composed of 1, 2, and 3 substates. Again, the polarization of their nuclei (deuterons) can be achieved by a selective transfer between the desired substates with the use of RF fields.

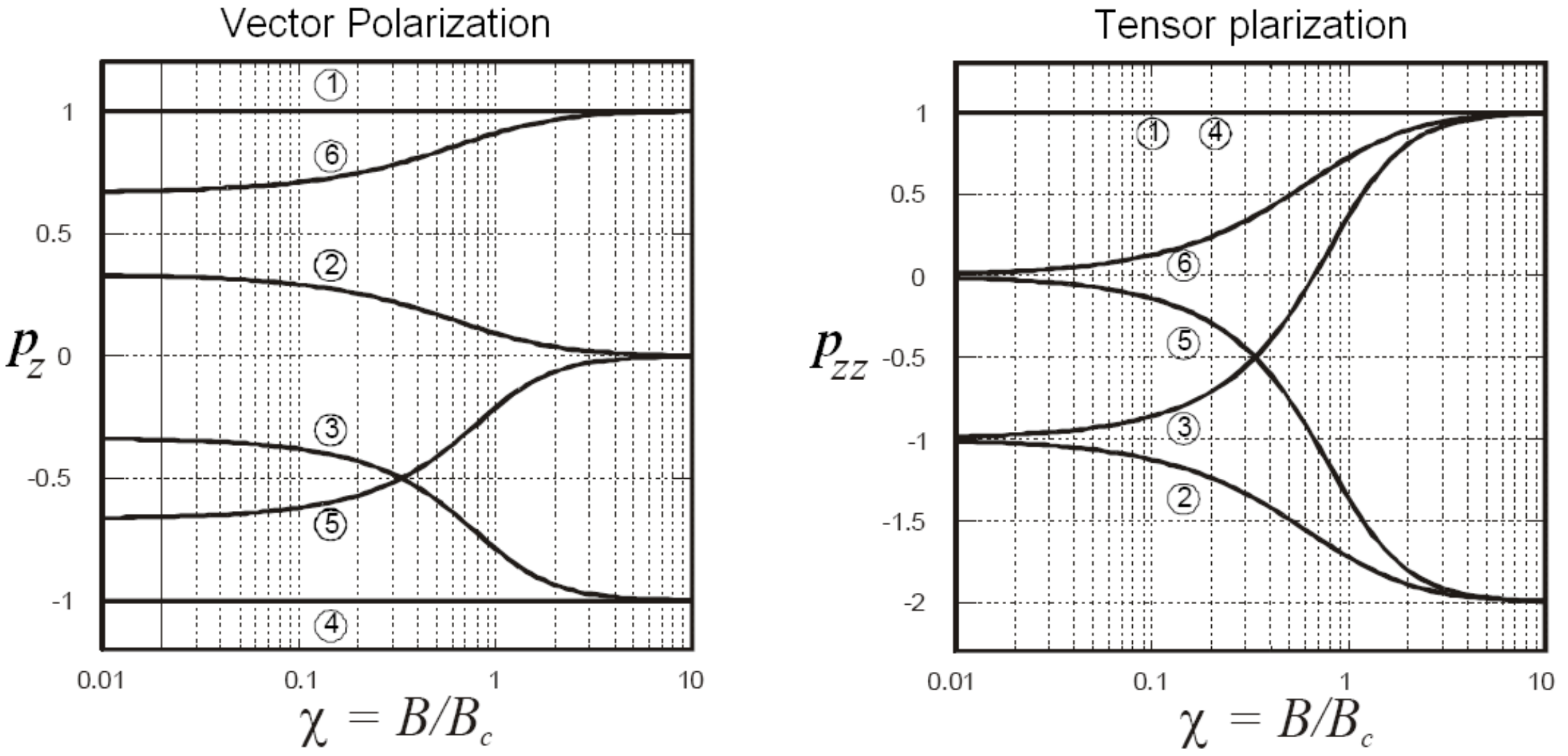

Figure G.5. The dependence of the vector and tensor polarizations of protons in the deuterium atom as the function of the external magnetic field.

Table G.1
Examples of the RF transitions the resulting beam polarizations

| | a | b | c | d | e | f | g |
|---|---|---|---|---|---|---|---|
| Strong Field | | | $2\leftrightarrow6$<br>$3\leftrightarrow5$ | $3\leftrightarrow5$ | $3\leftrightarrow5$ | $2\leftrightarrow6$ | $2\leftrightarrow6$ |
| Weak Field | | $1\leftrightarrow4$ | | | $1\leftrightarrow4$<br>$2\leftrightarrow3$<br>$5\leftrightarrow6$ | | $1\leftrightarrow4$<br>$3\leftrightarrow2$<br>$6\leftrightarrow5$ |
| Substates | 1+2+3 | 2+3+4 | 1+5+6 | 1+2+5 | 3+4+6 | 1+3+6 | 2+4+5 |
| $m_I$ | +1,0,-1 | 0,-1,-1 | +1,0,+1 | +1,0,0 | -1,-1,+1 | +1,-1,+1 | 0,-1,0 |
| $p_z$ | 0 | $-2/3$ | $+2/3$ | $+1/3$ | $-1/3$ | $+1/3$ | $-1/3$ |
| $p_{zz}$ | 0 | 0 | 0 | $-1$ | $+1$ | $+1$ | $-1$ |
| $t_{10}$ | 0 | $-\sqrt{2/3}$ | $+\sqrt{2/3}$ | $+1/\sqrt{6}$ | $-1/\sqrt{6}$ | $+1/\sqrt{6}$ | $-1/\sqrt{6}$ |
| $t_{20}$ | 0 | 0 | 0 | $-1/\sqrt{2}$ | $+1/\sqrt{2}$ | $+1/\sqrt{2}$ | $-1/\sqrt{2}$ |

Weak and strong field refer to the magnetic fields used for the RF transitions. The ionisation is performed in a strong magnetic field. The relation between Cartesian and spherical polarizations are (see Chapter 15): $t_{10} = (3/2)^{1/2} p_z$ and $t_{20} = 2^{-1/2} p_{zz}$.

As in the case of hydrogen beams, polarization can be also created without the use of RF fields but just by using the states 1,2, and 3 as they emerge from the inhomogeneous magnet.

We can recall (see the Appendix F) that for deuterons, vector $p_z$ and tensor $p_{zz}$ polarization are given by:

$$p_Z = \frac{N_+ - N_-}{N_+ + N_0 + N_-}$$

$$p_{ZZ} = \frac{N_+ + N_- - 2N_0}{N_+ + N_0 + N_-}$$

where $N_+$, $N_-$, and $N_0$ are the number of deuterium atoms with their nuclear (deuteron) spin projections $m_I$ = 1, -1, and 0, respectively.

If we now examine the Figure G.5, we shall see that in the weak field:

$p_z \approx$ 1/3 (states 1, 2, and 3 in the weak field)

$p_{zz} \approx$ -1/3 (states 1, 2, and 3 in the weak field)

In practice combinations of RF transitions are used to create desired deuteron polarizations. A few examples are listed in Table G.1.

**The Lamb-shift polarized ion sources**

As described earlier, the key procedure in creating a polarized beam is to select suitable hyperfine states of the hydrogen or deuterium atoms. The same applies to the Lamb-shift sources except that the atoms are now in their excited states and the selection of suitable hyperfine states is done in a different way.

Figure G.6 shows the energy diagram of the hydrogen atom for the principal quantum number $n$ = 2 for which there are three states $2P_{1/2}$, $2P_{3/2}$, and $2S_{1/2}$. Figure G.6 shows only the two states, $2P_{1/2}$ and $2S_{1/2}$. The $2P_{3/2}$ state is 10,968 MHz above the $2P_{1/2}$ and normally is not included in the mechanism of creating polarized beams.

The $2P_{1/2}$ and $2S_{1/2}$ states are separated by 1058 MHz ($4.4\times10^{-6}$ eV), known as the Lamb-shift (Lamb and Retherford 1950). The state $2P_{1/2}$ decays rapidly by emitting intense radiation in a discrete Lyman $\alpha$ line (1216 Å). Its lifetime is only $1.6\times10^{-9}$ s. In contrast, the $2S_{1/2}$ state decays slowly by emitting radiation over a broad continuum and its lifetime is around 0.14 s.

The basic idea of creating polarized ion beams using a Lamb-shift source is explained in the caption to Figure G.6. A schematic diagram of a Lamb-shift source is shown in Figure G.7.

The metastate hydrogen atoms can be produced by capturing electrons to a metastate orbit. This can be achieved by first producing $H^+$ ions and then passing them through a charge donor, such as cesium gas. The H(2S) source will produce not only the metastate hydrogen atoms but also the ground state H(1S) hydrogen as well as $H^+$ and $H^-$ particles. In fact, the cross section for producing the metastate hydrogen is small so most of the output will be made of the H(1S) atoms. The charged particles $H^+$ and $H^-$ can be removed by magnetic fields before passing the beam to the spin filter but the background made of the H(1S) hydrogen will remain.

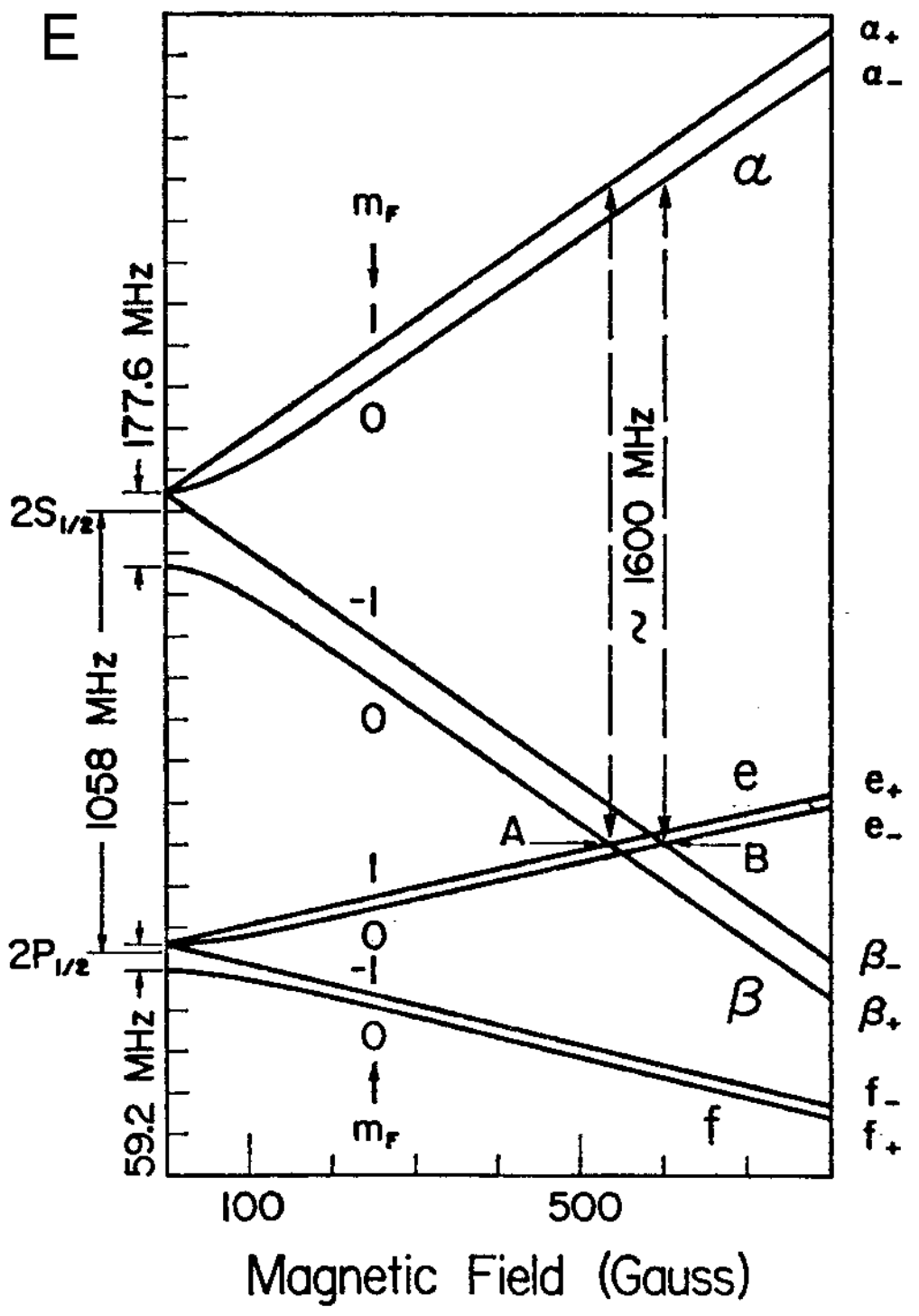

Figure G.6. The energy-level diagram for the $n$ = 2 excited states of the hydrogen atom. The $2P_{3/2}$ state, which is 10,968 MHz above the $2P_{1/2}$ state and which is not used in the Lamb-shift polarization mechanism is not shown. The short-lived $2P_{1/2}$ and metastable $2S_{1/2}$ states are separated by 1058.0 MHz (= $4.4 \times 10^{-6}$ eV), known as the Lamb-shift. The idea behind the Lamb-shift polarized ion sources is to remove the components $\beta_-$ and $\beta_+$ by mixing them with the short-lived components $e_-$ and $e_+$ components and then to remove one of the $\alpha$ components by forcing it to decay to one of the $e$ components. This operation leaves $\alpha_+$ or $\alpha_-$ each containing aligned proton spins. The basic idea for the deuteron polarization is the same but the number of hyperfine levels in each branch is then increased to three. In this diagram, labels $\alpha$, $\beta$, $e$, and $f$ refer to different orientations of the electron spin ($m_j$ = +1/2, -1/2, +1/2, and -1/2, respectively) and the subscripts + and - to different orientations of the proton spin ($m_I$ = +1/2 and -1/2, respectively).

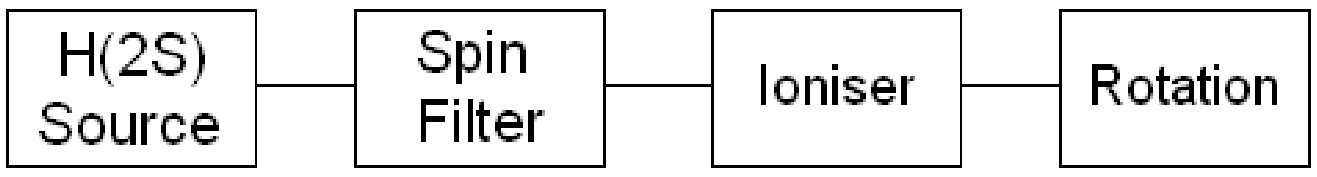

Figure G.7. Schematic diagram showing the major components of the Lamb-shift polarized ion source. H(2S) Source – a source of the metastable hydrogen (of deuterium) atoms; Spin Filter – a system of fine-tuned static magnetic, electric, and RF fields producing hydrogen (or deuterium) atoms with aligned nuclear spins; Ioniser – a selective ioniser producing positive or negative hydrogen (or deuterium) projectiles; Rotation – spin rotation unit.

The spin filter contains all the necessary units for creating a hydrogen beam with aligned nuclei. As outlined in the caption to Figure G.6, the first step is to remove the metastable components $\beta$ from the atomic beam. If the magnetic field in the direction of the beam is chosen to have a value of 535 or 605 G (see Figure G.8), and if in addition, a perpendicular electric field is applied then states $\beta$ are mixed with states $e$ and are forced to decay to the ground state. This process is described as quenching.

Quenching leaves the substates $\alpha$, which have only one projection of the electron spin, $m_j$ =+1/2, but two components of the proton spin, $m_I$ = ±1/2. As can be seen in Figure G.8, the substates $\alpha_+$ and $\alpha_-$ are separated by about 1600 MHz from the states $e_+$ and $e_-$ if the external magnetic field is 535 or 605 G. Consequently, if in addition to the external magnetic and electric fields we apply an RF field with the frequency of around 1600 MHz we shall be able to force the state $\alpha_+$ or $\alpha_-$ to decay leaving one component $\alpha_+$ or $\alpha_-$ containing positively or negatively polarized protons.

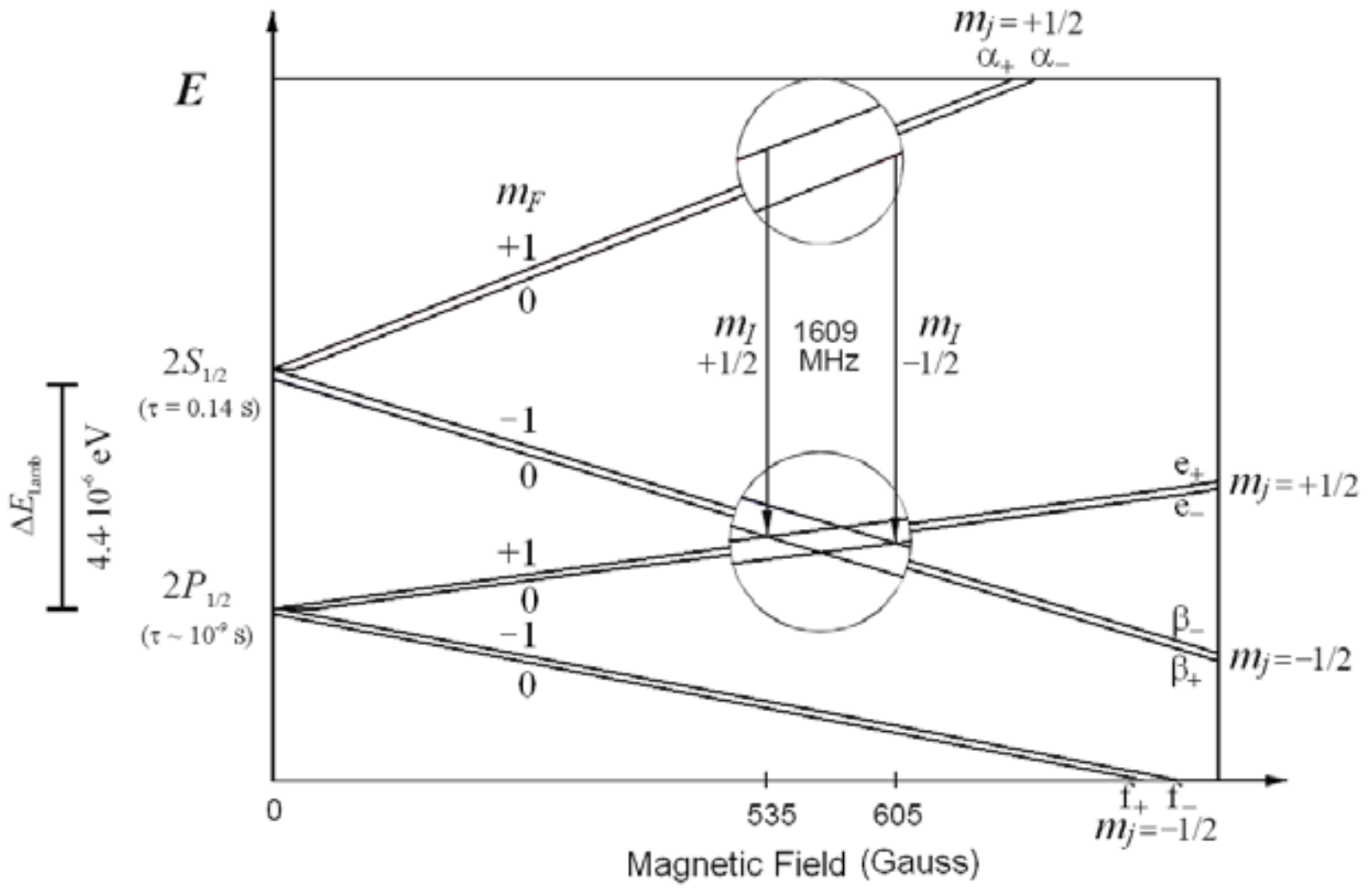

Figure G.8. Level diagram for the hydrogen atoms in the magnetic field illustrating the mechanism of producing polarized beams. The diagram for deuterium is similar but with each branch containing three substates corresponding to $m_I$ = 0, ±1. As in Figure G.6, groups labelled $\alpha$, $\beta$, $e$, and $f$ are for the electron spin orientations $m_j$ = +1/2, -1/2, +1/2, and -1/2, respectively, and the subscripts + and – for the orientations of the proton spin ($m_I$ = +1/2 and -1/2, respectively).

The atomic beam with aligned protons still contains the unpolarized H(1S) atoms. To remove them, the output of the spin filter is then passed to a selective ioniser. The charge-exchange medium for the ioniser can be made of argon or other suitable gas. The ioniser removes the unwanted H(1S) atoms and produces $H^+$ or $H^-$ ions (which are now polarized) to suit the type of the particle accelerator used in the experimental work.

The previously discusses dependence of the beam polarization on the magnetic field in the ioniser (Figures G.2 and G.5) also applies to the Lamb-shift source. However, decoupling of the electron and nuclear spins occurs at lower magnetic fields. The critical magnetic field for the metastate hydrogen or deuterium atoms is low ($B_c$ = 63.4 G for the hydrogen and 14.6 G for deuterium, as compared to 507 G and 117 G respectively for the ground state). So, if a strong field ioniser is used, the produced polarization will be close to +1 or -1 for the $\alpha_+$ or $\alpha_-$ substates, respectively.

After passing a selective ioniser, the polarized beam is suitable for the acceleration. A spin rotation device may be also added to manipulate the direction of the polarization.

# Appendix H

## Selected nuclear spin structures and the extreme values of the analyzing powers

### The $\vec{1}+\frac{1}{2}\rightarrow\frac{\vec{1}}{2}+0$ spin structure[13]

An example of this spin structure is the reaction the $^3$He$(\vec{d},p)^4$He reaction (Chapter 19).

The initial state is made of projectiles with spin-1 and the target nucleus with spin-$^1/_2$. It is therefore a direct product of spin-1 and spin-1/2 space and thus, the spin function of any initial state can be described in terms of a six-component vector. The initial spin state can be expressed as

$$\chi_i = \sum_{j=1}^{6} a_j \phi_j$$

where $\phi_j$ is a product of various combinations of spin functions of spin-1 and spin-$^1/_2$ of interacting nuclei. Explicitly

$$\phi_1 = \chi_{11}\chi_{\frac{1}{2}\frac{1}{2}} \quad \phi_2 = \chi_{10}\chi_{\frac{1}{2}\frac{1}{2}} \quad \phi_3 = \chi_{1-1}\chi_{\frac{1}{2}\frac{1}{2}} \quad \phi_4 = \chi_{11}\chi_{\frac{1}{2}-\frac{1}{2}} \quad \phi_5 = \chi_{10}\chi_{\frac{1}{2}-\frac{1}{2}} \quad \phi_6 = \chi_{1-1}\chi_{\frac{1}{2}-\frac{1}{2}}$$

where

$\chi_{1M}$ are the eigenfunctions of the spin-1 operator $S_z$ and $\chi_{\frac{1}{2}M}$ are the eigenfunctions of the spin-1/2 operator $\sigma_z$.

The final sate consists of spin-$^1/_2$ and spin-0 particles, so it can be expressed as

$$\chi_{fi} = \sum_{j=1}^{2} a_j \phi'_j$$

$\phi'_1 = \chi_{\frac{1}{2}\frac{1}{2}}$ and $\phi'_2 = \chi_{\frac{1}{2}-\frac{1}{2}}$.

The scattering amplitude represented by matrix $M$ (Wolfenstein 1956) relates the initial and final states,

$$b_j = \sum_k M_{jk} a_k \quad j = 1,2 \quad k = 1,\ldots,6$$

or

[13] See Gammel, Keaton, and Ohlsen (1970), and Ohlsen (1972)

$$\begin{bmatrix} b_1 \\ b_2 \end{bmatrix} = \begin{bmatrix} M_{11} & M_{12} & M_{13} & M_{14} & M_{15} & M_{16} \\ M_{21} & M_{22} & M_{23} & M_{24} & M_{25} & M_{26} \end{bmatrix} \begin{bmatrix} a_1 \\ a_2 \\ a_3 \\ a_4 \\ a_5 \\ a_6 \end{bmatrix}$$

The 6x6 and 2x2 density matrices that describe the initial and final states are defined, respectively, for an ensemble of $N$ particles as

$$(\rho_i)_{jk} = \frac{1}{N}\sum_{n-1}^{N} a_j^{(n)} a_k^{(n)*} \quad j,k = 1,...,6$$

$$(\rho_f)_{jk} = \frac{1}{N}\sum_{n-1}^{N} b_j^{(n)} b_k^{(n)*} \quad j,k = 1,2$$

As mentioned earlier, the initial and finial channels are related via the scattering matrix $M$

$$\chi_f = M\chi_i$$

$$\rho_f = M\rho_i M^+$$

The density matrix for the initial state can be expanded in terms of products of spin-1 and spin-$^1/_2$ operators. If we use the notations $\sigma_0 = I$ (the unit matrix), $\sigma_1 = \sigma_x$, $\sigma_2 = \sigma_y$, and $\sigma_3 = \sigma_z$ , for the Pauli spin operators, and the normalized spin-1 operators $\Omega_i$ (see the Appendix F) we can write

$$\rho_i = \frac{1}{6}\sum_{jk} \omega_j p_k \Omega_j \sigma_k \quad j = 0,1,...,8 \quad k = 0,1,2,3$$

In this expression, $\omega_j$ and $p_k$ are the *expectation values* of the *operators* $\Omega_j$ (for spin-1 particles) and $\sigma_k$ (for spin-$^1/_2$ particles), respectively, i.e. $\omega_j = \langle \Omega_j \rangle$ and $p_k = \langle \sigma_k \rangle$.

For the spin structure considered here, only spin-1 particles are polarized in the entrance channel; spin-$^1/_2$ particles are not. Consequently, $p_k = 0$ for $k$=1, 2, 3 and only $p_0 = 1$. In this case the expression for the density matrix in the entrance channel can be simplified:

$$\rho_i = \frac{1}{6}\sum_j \omega_j \Omega_j I = \frac{1}{6}\sum_j \omega_j \Omega_j$$

The $I$ is the unity matrix and therefore can be suppressed.

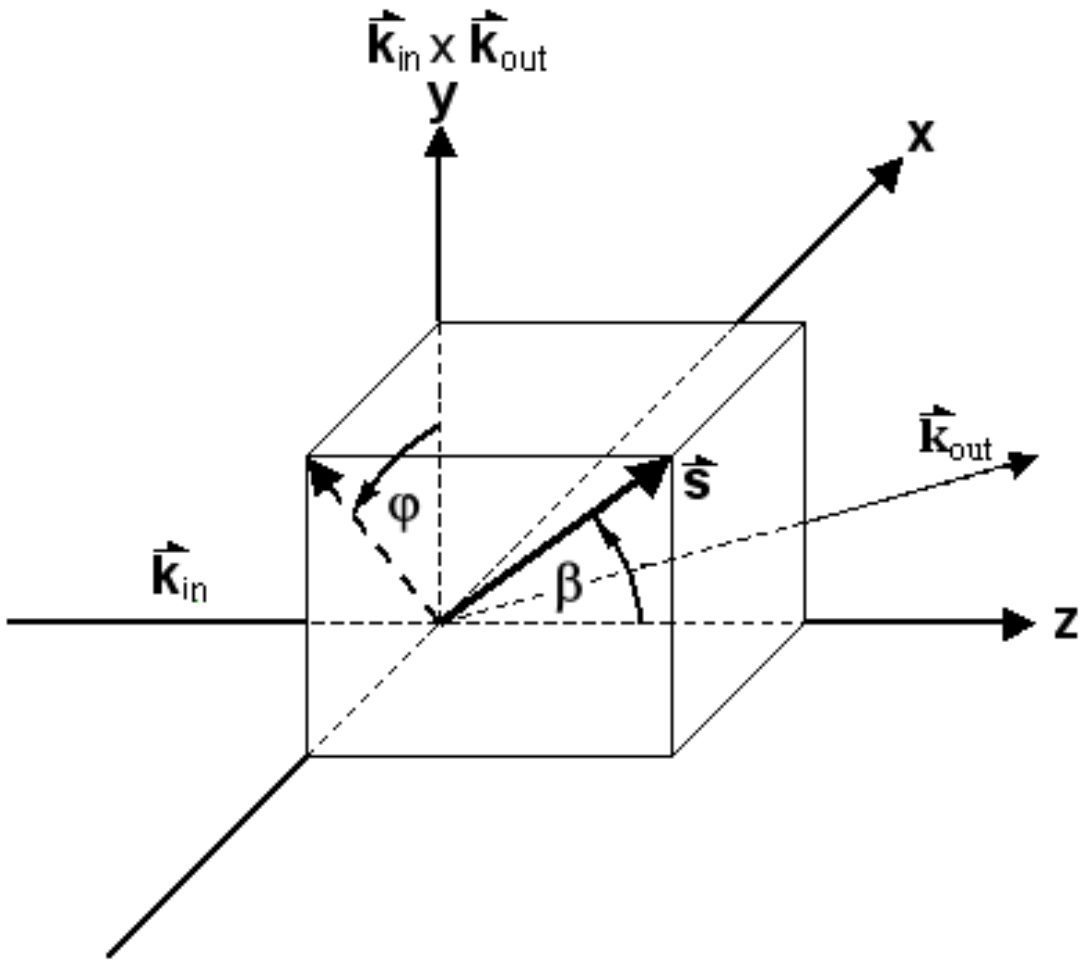

Figure H.1. The definition of angles $\beta$ and $\varphi$.

If $\rho_i$ is normalized to unity, the differential cross section for spin-1/2 particles in the exit channel is given by:

$$\sigma(\theta,\varphi) = Tr\rho_f = \frac{1}{6}\sum_j \omega_j Tr(M\Omega_j IM^+) = \frac{1}{6}\sum_j \omega_j Tr(M\Omega_j M^+)$$

where $\theta$ is the scattering angle (i.e. the angle between $\vec{k}_{in}$ and $\vec{k}_{out}$) and $\varphi$ is the angle of projection of the quantization axis on the *x-y* plane (see Figure H.1).

For an unpolarized beam

$$\rho_i = \frac{1}{6} I$$

and

$$\sigma(\theta,\varphi) \equiv \sigma_0(\theta) = \frac{1}{6} Tr(MIM^+) = \frac{1}{6} Tr(MM^+)$$

We can therefore write

$$\sigma(\theta,\varphi) = \sigma_0(\theta) \frac{\sum_j \omega_j Tr(M\Omega_j M^+)}{TrMM^+}$$

The polarization of spin-$^1/_2$ particles in the outgoing channel is given by

$$p_{k'} \equiv \langle \sigma_{k'} \rangle = \frac{Tr(\rho_f \sigma_{k'})}{Tr\rho_f} = \frac{\sum_j \omega_j Tr(M\Omega_j M^+ \sigma_{k'})}{\sum_j \omega_j Tr(M\Omega_j M^+)}$$

which can be rewritten as

$$p_{k'}\sigma(\theta,\varphi) = \sigma_0(\theta)\left[\frac{\sum_j \omega_j Tr(M\Omega_j M^+ \sigma_{k'})}{TrMM^+}\right]$$

Rotation from the spin-symmetry system of reference, where the polarization of spin-1 particles is described by just two quantities $p_Z$ and $p_{ZZ}$, to the projectile helicity system of reference as shown in Figure H.1, introduces new polarization components, which are related to angles $\beta$ and $\varphi$.

$$p_x = -p_Z \sin\beta \sin\varphi$$

$$p_y = p_Z \sin\beta \cos\varphi$$

$$p_z = p_Z \cos\beta$$

$$p_{xx} = p_{ZZ}\frac{1}{2}(3\sin^2\beta \sin^2\varphi - 1)$$

$$p_{yy} = p_{ZZ}\frac{1}{2}(3\sin^2\beta \cos^2\varphi - 1)$$

$$p_{zz} = p_{ZZ}\frac{1}{2}(3\cos^2\beta - 1)$$

$$p_{xy} = -p_{ZZ}\frac{3}{2}\sin^2\beta \cos\varphi \sin\varphi$$

$$p_{yz} = p_{ZZ}\frac{3}{2}\sin\beta \cos\beta \cos\varphi$$

$$p_{xz} = -p_{ZZ}\frac{3}{2}\sin\beta \cos\beta \sin\varphi$$

$p_{zx} = p_{xz}$ and $p_{zy} = p_{yz}$

It is easy to see that

$$p_{xx} - p_{yy} = -p_{ZZ}\frac{3}{2}\sin^2\beta \cos 2\varphi$$

In terms of these components

$$\sigma(\theta,\varphi) = \sigma_0(\theta)\left[1 + \frac{3}{2}\sum_j p_j A_j(\theta) + \frac{1}{3}\sum_{jk} p_{jk} A_{jk}(\theta)\right]$$

$$p_{l'}\sigma(\theta,\varphi) = \sigma_0(\theta)\left[P_{l'}(\theta) + \frac{3}{2}\sum_j p_j K_j^{l'} + \frac{1}{3}\sum_{jk} p_{jk} K_{jk}^{l'}\right]$$

where $j = x, y, z$ and $jk$ are all combinations of $x, y, z$.

The observables are the already mentioned polarization $p_{l'}$ of the outgoing spin-1/2 particle (induced by polarized incident beam) as well as $A_j(\theta)$, $A_{jk}(\theta)$, $P_{l'}(\theta)$, $K_j^{l'}(\theta)$, $K_{jk}^{l'}(\theta)$. They are defined by the following expressions.

Analyzing powers:

$$A_j(\theta) \equiv \frac{Tr(MS_jM^+)}{Tr(MM^+)}$$

$$A_{jk}(\theta) \equiv \frac{Tr(MS_{jk}M^+)}{Tr(MM^+)}$$

Outgoing polarization for unpolarized incident beam:

$$P_l(\theta) \equiv \frac{Tr(MM^+\sigma_{l'})}{Tr(MM^+)}$$

Polarization transfer coefficients:

$$K_j^{l'}(\theta) \equiv \frac{Tr(MS_jM^+\sigma_{l'})}{Tr(MM^+)}$$

$$K_{jk}^{l'}(\theta) \equiv \frac{Tr(MS_{jk}M^+\sigma_{l'})}{Tr(MM^+)}$$

The $M$-matrix elements are often written in terms of their expansion coefficients. For instance, the 6x2 $M$-matrix can be written as

$$M = A\chi_y^+ I + B\chi_y^+\sigma_y + C\chi_x^+\sigma_x + D\chi_x^+\sigma_z + E\chi_z^+\sigma_x + F\chi_z^+\sigma_z$$

where $A$, $B$, …, $F$ are coefficients, which depend on the nature of the investigated reaction, $I$ is the unit matrix, $\chi_i$ are a complete set of spinors in 3x1 space (see below) and $\sigma_i$ are the 2x2 Pauli spin operators.

$$\chi_x = \frac{1}{\sqrt{2}}\begin{bmatrix} -1 \\ 0 \\ 1 \end{bmatrix} \quad \chi_y = \frac{i}{\sqrt{2}}\begin{bmatrix} 1 \\ 0 \\ 1 \end{bmatrix} \quad \chi_z = \begin{bmatrix} 0 \\ 1 \\ 0 \end{bmatrix}$$

$$\chi_x^+ = \frac{1}{\sqrt{2}}\begin{bmatrix} -1 & 0 & 1 \end{bmatrix} \quad \chi_y^+ = \frac{-i}{\sqrt{2}}\begin{bmatrix} 1 & 0 & 1 \end{bmatrix} \quad \chi_z^+ = \frac{1}{\sqrt{2}}\begin{bmatrix} 0 & 1 & 0 \end{bmatrix}$$

It is straightforward to show that in terms of the amplitudes $A, B, \ldots, F$, the $M$-matrix has the following form:

$$M = \frac{1}{\sqrt{2}}\begin{bmatrix} -iA-D & \sqrt{2}F & iA+D & -B-C & \sqrt{2}E & -B+C \\ B-C & \sqrt{2}E & B+C & iA+D & -\sqrt{2}F & -iA-D \end{bmatrix}$$

In turn, we can also express the analyzing powers $A_y$ and $A_{yy}$ in terms of these amplitudes and find conditions when $A_y$ and/or $A_{yy}$ reach their extreme values. Thus, for instance, if we use the expression

$$A_{yy} = \frac{Tr(MS_{yy}M^{+})}{Tr(MM^{+})}$$

we shall find that $A_{yy} = 1$ if $A$=$B$=*0* (*cf* Seiler *at al.* 1976a).

In this case, the $M$-matrix has the form

$$M = \frac{1}{\sqrt{2}}\begin{bmatrix} -D & \sqrt{2}F & D & -C & \sqrt{2}E & C \\ -C & \sqrt{2}E & C & D & -\sqrt{2}F & -D \end{bmatrix}$$

Likewise, we can calculate

$$A_y = \frac{Tr(MS_yM^{+})}{Tr(MM^{+})}$$

and find that $A_y = \pm 1$ if not only $A$=$B$=0 but also $C = \mp iE$ and $D = \mp iF$ .

It is therefore clear that if $A_y = 1$ then $A_{yy}$ should also have its extreme value of 1 at the same energy and angle because both have to satisfy the same condition of $A$=$B$=0. However, if $A_{yy} = 1$, $A_y$ may or may not reach its extreme value of 1. If it does, the maximum should occur at the same energy and angle as $A_{yy} = 1$. However, if the $M$-matrix does not satisfy the two additional requirements of $C = \pm iE$ and $D = \pm iF$ then $A_y$ will not reach its extreme value of 1 even if $A_{yy}$ does. The full list of the observables, which can be used to check experimentally whether the conditions $A$=$B$=0, $C = \pm iE$ , and $D = \pm iF$ are satisfied is given by Seiler *at al.* (1976a).

**The $\vec{1} + 1 = 0 + 0$ spin structure**

An example of this spin structure is the $^{6}$Li$(\vec{d},\alpha)$ $^{4}$He reaction (Chapter 20).

To find the conditions for the $M$-matrix elements, which lead to the extreme values of the $A_y$ and $A_{yy}$ components of the analyzing powers we follow a similar procedure as outlined above. In terms of the expansion coefficients, the $M$-matrix for this system can be written as (Seiler *at al.* 1976b):

$$M = -\frac{1}{2}\begin{bmatrix} A-E & \sqrt{2}F & A+E \\ \sqrt{2}H & -2K & -\sqrt{2}H \\ A+E & -\sqrt{2}F & A-E \end{bmatrix}$$

If we use this expression and if we carry out the calculations using the general expressions for $A_y$ and $A_{yy}$, i.e.

$$A_y = \frac{Tr(MS_y M^+)}{Tr(MM^+)} \text{ and } A_{yy} = \frac{Tr(MS_{yy} M^+)}{Tr(MM^+)}$$

we shall find that $A_y = 1$ if $A = 0$, $H = \pm iE$, and $K = \pm iF$. In contrast, $A_{yy} = 1$ if only one condition is satisfied, i.e. if $A = 0$.

# Appendix I

## Analytic determination of the extreme $A_{yy}$ points

As an example of the analytic determination of the locations for the extreme values of the analyzing powers I use a study of the d-α scattering (Grüebler *et al.* 1975b).

The $M$-matrix for the d-α scattering can be written as

$$M(E_d,\theta)=\begin{bmatrix} M_{11} & M_{10} & M_{1-1} \\ M_{01} & M_{00} & -M_{01} \\ M_{1-1} & -M_{10} & M_{11} \end{bmatrix} \tag{1}$$

where the subscripts $ij$ are the deuteron spin projections in the incident and outgoing channels.

The elements of this matrix depend on the energy of the incident deuterons $E_d$ and on the reaction angle $\theta$. They can be calculated using experimentally determined phase shifts.

As shown in the Appendix H, tensor analyzing powers $A_{ij}$ can be expressed in terms of the $M$-matrix:

$$A_{ij}=\frac{Tr(MS_{ij}M^{+})}{Tr(MM^{+})}$$

(2)

The idea of the analytic determination of the extreme values of the analyzing powers is as follows: (a) use the experimentally determined phase-shifts to calculate the relevant matrix elements; (b) use the theoretically established relation between the elements of the $M$-matrix, which have to be satisfied for the extreme maxima to occur (see the Appendix H); (c) study the behaviour of the calculated elements to see at what points (in energy and angle) the theoretically postulated relationship between the matrix elements is satisfied.

Using the relation (2) and the matrix (1) we can find that for the $\vec{1}+0\rightarrow\vec{1}+0$ spin structure, $A_{yy}=1$ if

$$f(E_d,\theta)=M_{11}(E_d,\theta)+M_{1-1}(E_d,\theta)=0$$

Step-by-step numerical calculations can be carried out to locate the $f(E_d,\theta)=0$ points.

The function $f(E_d,\theta)=0$ if both its real and imaginary components are equal zero. Figure I1 shows the plots of $\mathrm{Re}(M_{11}+M_{1-1})$ *versus* $\mathrm{Im}(M_{11}+M_{1-1})$ with the matrix elements $M_{11}$ and $M_{1-1}$ calculated using the phase shift parameters determined by Grüebler *at al.*(1975a).

An alternative way is to calculate the $\mathrm{Re}(M_{11}+M_{1-1})$ and $\mathrm{Im}(M_{11}+M_{1-1})$ components for various energies and angles but now to identify the $(E_d,\theta)$ points where *either* $\mathrm{Re}(M_{11}+M_{1-1})=0$ *or* $\mathrm{Im}(M_{11}+M_{1-1})=0$ and plot them. Such a plot is shown in Figure I2. The points where the lines corresponding to $\mathrm{Re}(M_{11}+M_{1-1})=0$ and $\mathrm{Im}(M_{11}+M_{1-1})=0$ cross over are the points where $A_{yy}=1$

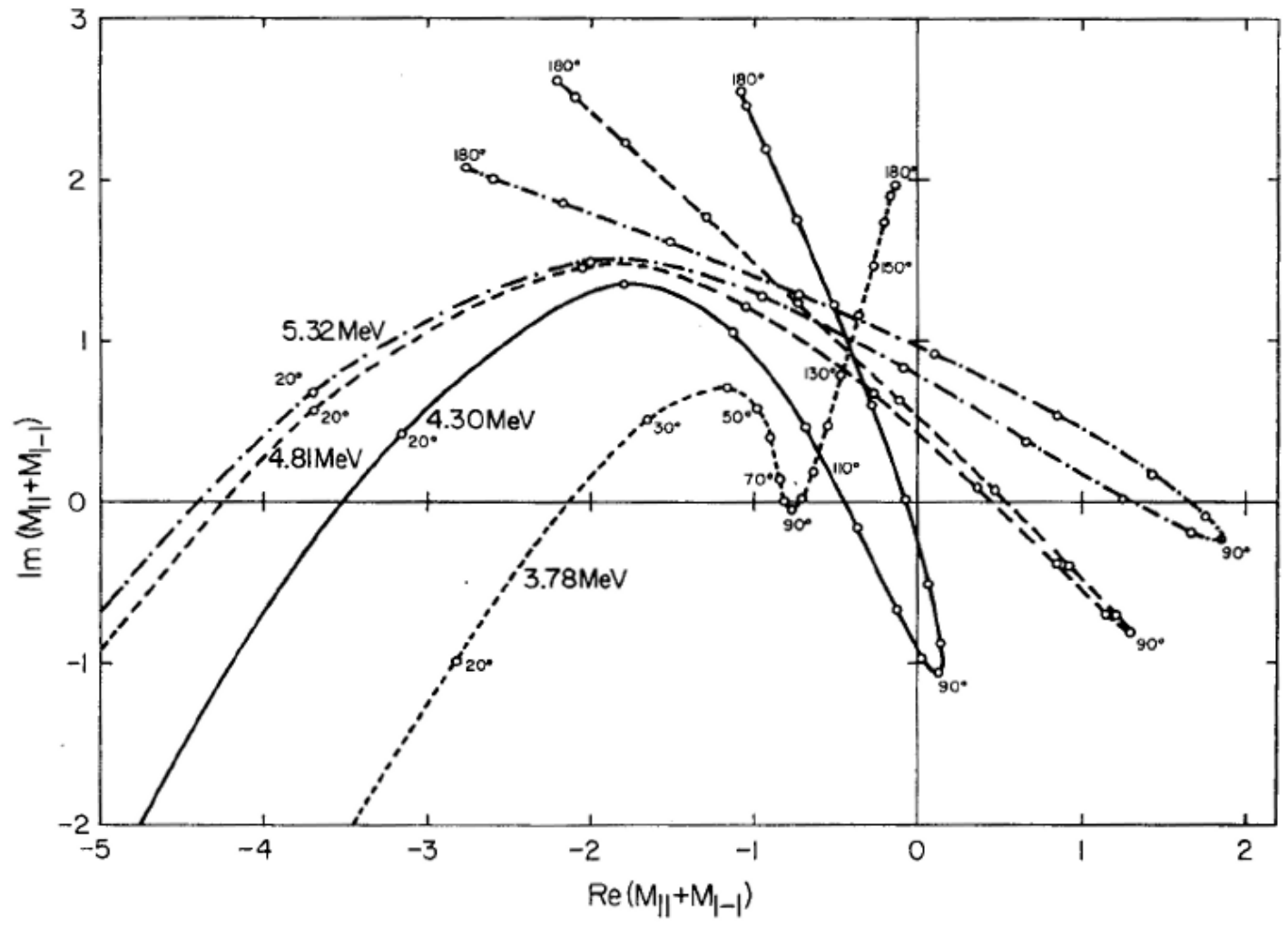

Figure I1. The trajectories of the sum of the $M_{11}$ and $M_{1-1}$ components of the $M$-matrix calculated using the experimentally determined phase shifts for the d-α reaction. This figure shows that for the incident energy of between 4.30 and 5.30 MeV, the trajectories move twice over the point of origin of the coordinate axis. This passes indicates that in this region of energies there should be two points, at around $60^0$ and $120^0$, where $A_{yy}=1$.

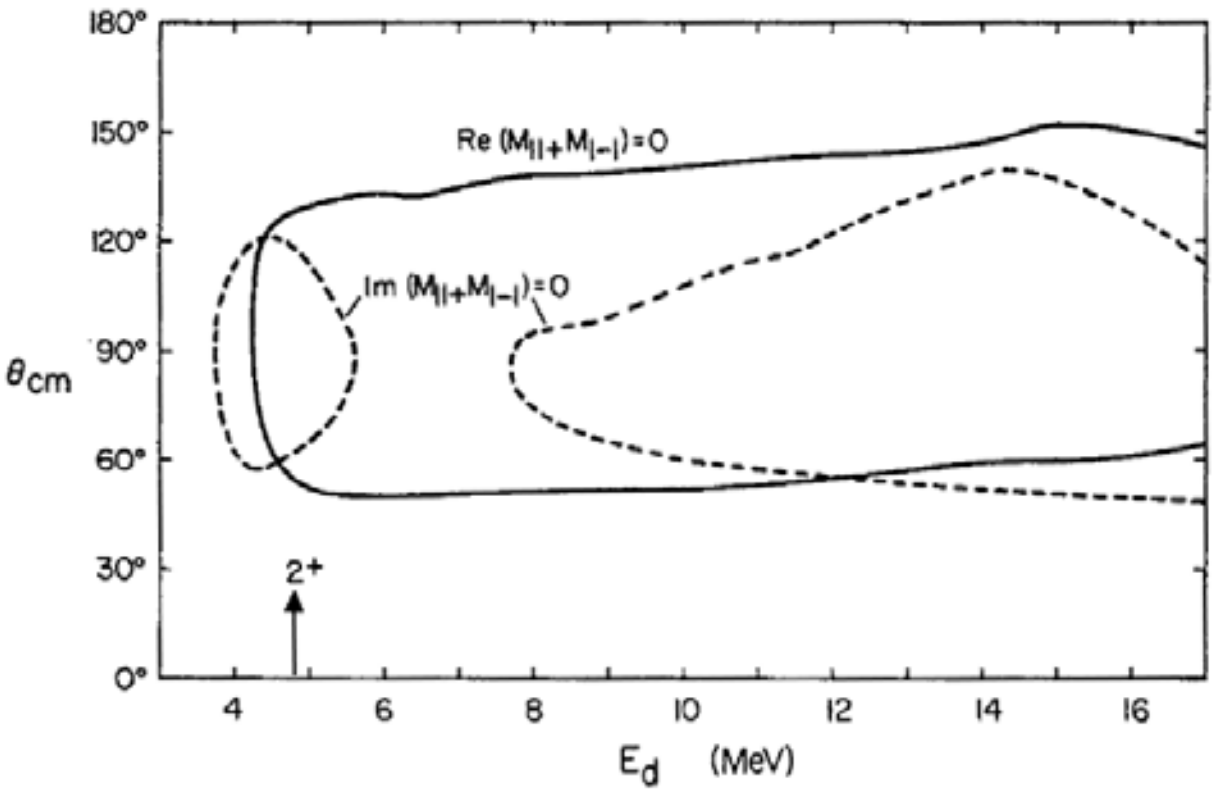

Figure I2. Contour plot of $\mathrm{Re}(M_{11}+M_{1-1})=0$ and $\mathrm{Im}(M_{11}+M_{1-1})=0$. The $A_{yy}=1$ points should be located where the contour of $\mathrm{Re}(M_{11}+M_{1-1})=0$ crosses over the contours of $\mathrm{Im}(M_{11}+M_{1-1})=0$. The figure shows that there should be two $A_{yy}=1$ points close to 5 MeV and one at around 12.5 MeV.

# Appendix J

## Phase shift analysis

The concept of phase shift analysis of experimental data can be easily understood by considering the elastic scattering. The differential cross section for the elastic scattering can be written as (Condon and Shortley 1935):

$$\sigma(\theta) = \left| f(\theta) \right|^2$$

where $f(\theta)$ is a complex function, which can be written as

$$f(\theta) = i\sqrt{\pi}ƛ^2 \sum_{l=0}^{\infty} \sqrt{(2l+1)} \left(1 - e^{2i\delta_l}\right) Y_{l,0}(\theta) \qquad (1)$$

In this expression, ƛ is the deBroglie wavelength divided by 2π,

$$ƛ = \frac{\lambda}{2\pi} = \frac{1}{k}$$

$Y_{l,0}(\theta)$ are the well-known normalized spherical harmonics as defined by Condon and Shortley (1935), and $\delta_l$ are phase shift parameters.

It should be noticed that in the expression (1) the only physical quantities, which are associated with nuclear interaction, are the phase shift parameters $\delta_l$. The remaining quantities are purely mathematical, with the exception of ƛ. However, this quantity is not related to the details of nuclear interaction but only to the energy of the incoming particles.

Phase shift analysis consists in fitting experimental angular distributions with the aim of extracting the experimentally determined parameters $\delta_l$.

In a more general case of nuclear reactions, which may or may not include elastic scattering, the aim is the same. The general idea is to express the scattering matrix in terms of the phase shift parameters and analyse experimental observables to determine the best values for these parameters. Phase shift parameters, which now define the scattering matrix, can then be used to calculate other observables in the way described in the Appendices H and I. In this brief summary, I follow the original notation of Blatt and Biedenharn (1952) who use letter $S$ for the scattering matrix.

Let us consider a general process where two particles collide and two emerge:

$$a + X = Y + b$$

We can use labels $\alpha$, $s$, and $l$ to identify the channel before the collision. They are the channel index, channel spin, and channel angular momentum, respectively. The state after the collision is labelled using $\alpha'$, $s'$, and $l'$.

The channel spin $s$ is made of the intrinsic spin $i$ of the incoming particle $a$ and the spin $I$ of the target nucleus $X$ ($\hat{s} = \hat{i} + \hat{I}$). For instance, for the n-p scattering, $i = I = 1/2$ and $s$ has two values, 1 (triplet state scattering) and 0 (singlet state scattering).

We also define the total angular momentum of the system as $\hat{J} = \hat{s} + \hat{l}$ . $J$ is preserved during the collision, i.e. $\hat{J} = \hat{s}' + \hat{l}'$. For resonance reactions, $J$ is the angular momentum of the compound nucleus.

Now we introduce the *scattering matrix* (scattering amplitude or collision matrix), which connects channels $\alpha ls$ and $\alpha' l' s'$. We write it as

$$S^{J}_{\alpha ls;\alpha' l' s'}$$

The formal definition of the scattering matrix is related to the wave function $\Psi_{\alpha s}(JM)$ in the channel $\alpha, s$ with total angular momentum quantum numbers $JM$ . At sufficiently large distance $r$, the function $\Psi_{\alpha s}(JM)$ can be expressed as a superposition of an incoming spherical wave with the amplitude $A^{JM}_{\alpha sl}$ and an outgoing spherical wave with the amplitude $B^{JM}_{\alpha sl}$ .

$$\Psi_{\alpha s}(JM) = \frac{1}{r_\alpha \sqrt{\upsilon_\alpha}} \Im^{M}_{Jls} \Phi_{\alpha s} u_\alpha(r)$$

where $\upsilon_\alpha$ is the relative velocity,

$$\Im^{M}_{Jls} = \sum_{m_l=-l}^{l} \sum_{m_s=-s}^{s} (lsm_l m_s \mid lsJM) Y_{l,m_l}(\theta,\phi) \chi_{s,m_s}$$

$(lsm_l m_s \mid lsJM)$ are Clebsch-Gordan coefficients, $Y_{l,m_l}$ are the spherical harmonics, $\chi_{s,m_s}$ spin function for spin $s$ and $z$ components $m_s$,

$$\Phi_{\alpha s} = \phi_a \phi_X$$

with $\phi_a$ and $\phi_X$ being the wave function of $a$ and $X$, $u_\alpha(r)$ is the radial wave function:

$$u_\alpha(r) = A^{JM}_{\alpha sl} \exp\left[-i\left(k_\alpha r_\alpha - \frac{1}{2} l\pi\right)\right] - B^{JM}_{\alpha sl} \exp\left[+i\left(k_\alpha r_\alpha - \frac{1}{2} l\pi\right)\right]$$

and $k_\alpha$ is the channel wave number

The scattering matrix is defined by the following equation:

$$B^{JM}_{\alpha' s' l'} = \sum_{\alpha,s,l} S^{J}_{\alpha' s' l';\alpha sl} A^{JM}_{\alpha sl}$$

The physical interpretation of the scattering matrix is that its elements determine the flux in each of the exit channels. For a reaction with $N$ channels, $S$ is an $N \times N$ matrix.

For instance, for the n-p scattering, assuming that $J$ = 1 with even parity, we can have only two combinations of $s$ and $l$, $s$ = 1 and $l$ = 0, and $s$ = 1 and $l$ = 2, which give the required $J$ value. Using the notation $^{2s+1}l_J$ we have two possibilities $^3S_1$ for \ = 0 and $^3D_1$ for \ = 2. The same applies to the exit channel. Consequently, we can have the following channels: $^3S_1 \rightarrow {}^3S_1$, $^3S_1 \rightarrow {}^3D_1$, $^3D_1 \rightarrow {}^3S_1$, $^3D_1 \rightarrow {}^3D_1$. We therefore have a 2 x 2 scattering matrix. Its elements will give the probability of flux in each of the four respective channels, i.e. the matrix will describe how probable is each of the four possible transitions.

The scattering matrix $S$ can be written as

$$S = U^{-1}\exp(2i\Delta)U \tag{2}$$

where $U$ is an orthogonal (unitary and real) $N \times N$ matrix and $\Delta$ is a diagonal matrix, whose elements are the eigen phase shifts.

The matrix $U$ is specified by $\frac{1}{2}N(N-1)$ real parameters and $\Delta$ by $N$ real eigen-phase shifts. Together the collision matrix $S$ is specified by $\frac{1}{2}N(N+1)$ real parameters.

For a two-channel reaction, $N$ = 2 and the total number of independent real parameters is 3. In this case, the matrices $U$ and $\Delta$ have the following forms

$$U = \begin{bmatrix} \cos\varepsilon & \sin\varepsilon \\ -\sin\varepsilon & \cos\varepsilon \end{bmatrix} \text{ and } \Delta = \begin{bmatrix} \delta_1 & 0 \\ 0 & \delta_2 \end{bmatrix}$$

Therefore

$$S = U^{-1}\exp(2i\Delta)U = \begin{bmatrix} \cos\varepsilon & -\sin\varepsilon \\ \sin\varepsilon & \cos\varepsilon \end{bmatrix} \begin{bmatrix} e^{2i\delta_1} & 0 \\ 0 & e^{2i\delta_2} \end{bmatrix} \begin{bmatrix} \cos\varepsilon & \sin\varepsilon \\ -\sin\varepsilon & \cos\varepsilon \end{bmatrix}$$

which gives

$$S = \begin{bmatrix} (\cos^2\varepsilon)e^{2i\delta_1} + (\sin^2\varepsilon)e^{2i\delta_2} & \frac{1}{2}(\sin 2\varepsilon)(e^{2i\delta_1} - e^{2i\delta_2}) \\ \frac{1}{2}\sin(2\varepsilon)(e^{2i\delta_1} - e^{2i\delta_2}) & (\sin^2\varepsilon)e^{2i\delta_1} + (\cos^2\varepsilon)e^{2i\delta_2} \end{bmatrix}$$

The three real parameters are the two eigen phase shifts $\delta_1$ and $\delta_2$, and the mixing (coupling) parameter $\varepsilon$. The parameter $\varepsilon$ describes mixing between allowed configurations. If the coupling between the two configurations is zero ($\varepsilon$ = 0), the off diagonal elements of the matrix S vanish and the matrix is then given by

$$S = \begin{bmatrix} e^{2i\delta_1} & 0 \\ 0 & e^{2i\delta_2} \end{bmatrix}$$

Generally, for $s$ =1 and any given $J$, we have two radial wave functions corresponding to $l = J-1$ and $l = J+1$:

$$u_1(r) = A_1\exp\left\{-i\left[kr - \frac{1}{2}(J-1)\pi\right]\right\} - B_1\exp\left\{+i\left[kr - \frac{1}{2}(J-1)\pi\right]\right\}$$

$$u_2(r) = A_2 \exp\left\{-i\left[kr - \frac{1}{2}(J+1)\pi\right]\right\} - B_2 \exp\left\{+i\left[kr - \frac{1}{2}(J+1)\pi\right]\right\}$$

The relation between the amplitudes $A_1$ and $A_2$ for the incoming waves, and the amplitudes $B_1$ and $B_2$ for the outgoing waves is given by:

$$b = Sa$$

where *b* is the column vector with components $B_1$ and $B_2$, $a$ is the column vector with components $A_1$ and $A_2$ and $S$ is the 2 x 2 scattering matrix.

Explicitly,

$$\begin{bmatrix} B_1 \\ B_2 \end{bmatrix} = \begin{bmatrix} (\cos^2\varepsilon)e^{2i\delta_1} + (\sin^2\varepsilon)e^{2i\delta_2} & \frac{1}{2}(\sin 2\varepsilon)(e^{2i\delta_1} - e^{2i\delta_2}) \\ \frac{1}{2}\sin(2\varepsilon)(e^{2i\delta_1} - e^{2i\delta_2}) & (\sin^2\varepsilon)e^{2i\delta_1} + (\cos^2\varepsilon)e^{2i\delta_2} \end{bmatrix} \begin{bmatrix} A_1 \\ A_2 \end{bmatrix}$$

In particular, for the example considered earlier, i.e. for $J$ = s = 1, $B_1 \equiv B(l=0)$, $B_2 \equiv B(l=2)$, $A_1 \equiv A(l=0)$ and $A_2 \equiv A(l=2)$.

Once the phase shifts are determined from an analysis of a certain set of experimental data, the scattering matrix $S$ can be used to calculate (predict) other observables.

# Appendix K

## Reorientation effect in Coulomb excitation

Coulomb excitation is discussed extensively by Alder *et al.* (1956) and Alder and Winther (1975). Reorientation effects were first discussed by De Boer and Eichler (1968).

### First-order perturbation theory

The process of Coulomb excitation is illustrated schematically in Figure K.1.

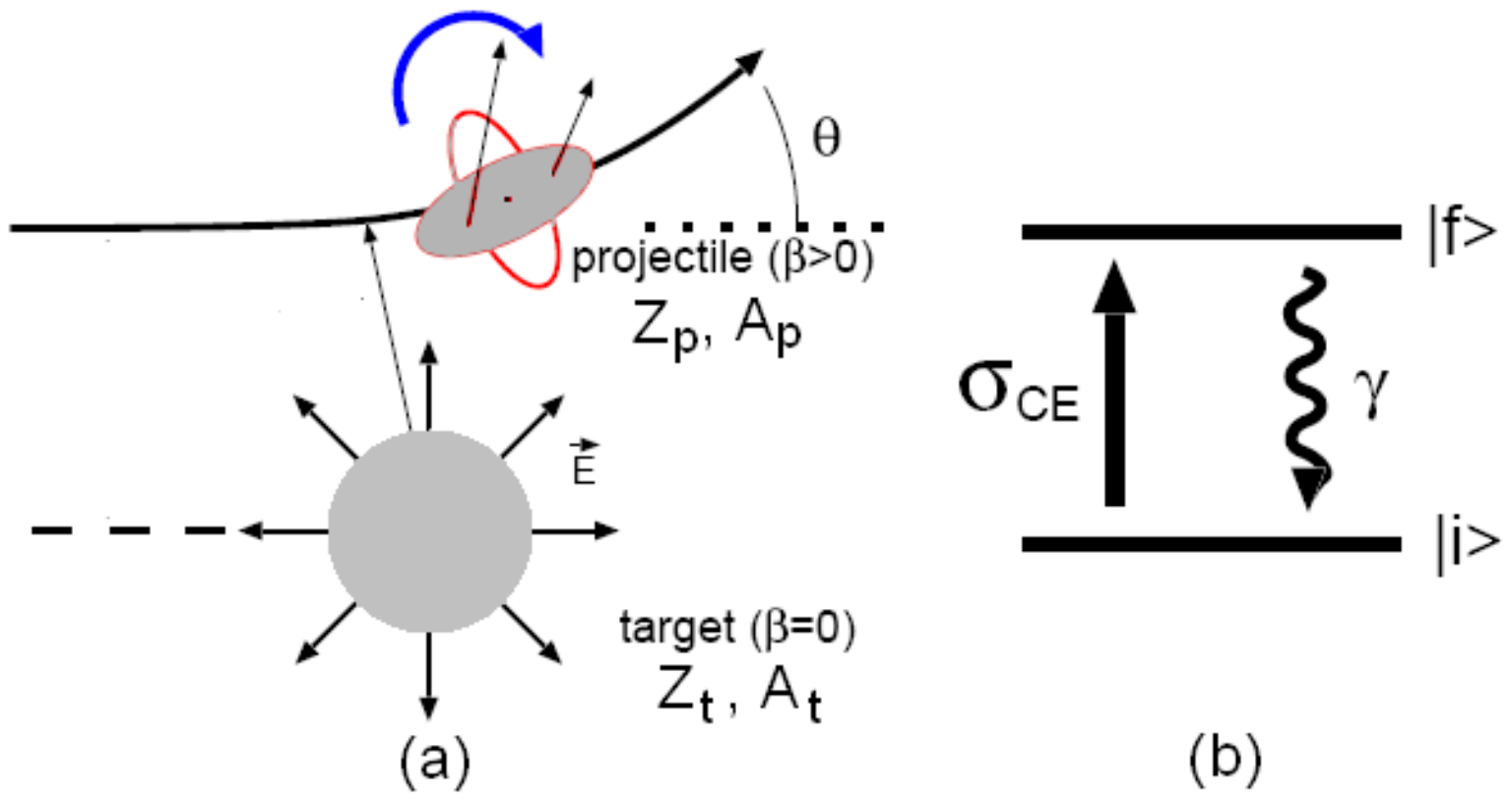

Figure K.1. Schematic diagram of Coulomb excitation. On the left-hand side, the deformed projectile is excited in the electric field of the spherical target nucleus. The arrows indicate the acting force. They cause rotation as shown in the figure. On the right-hand side, one can see the corresponding first order (direct) excitation of the projectile from state $|i\rangle$ to $|f\rangle$ and the subsequent deexcitation by emission of $\gamma$-ray.

The differential cross section for Coulomb excitation is given by:

$$\frac{d\sigma}{d\Omega} = \left(\frac{d\sigma}{d\Omega}\right)_R P_f$$

where $(d\sigma/d\Omega)_R$ is the Rutherford scattering cross section and $P_f$ is the probability to excite the final state $|f\rangle$ from the initial state $|i\rangle$.

In the first-order perturbation theory, the electromagnetic interaction between the projectile and the target nucleus is described by the time-dependent interaction potential $V(\vec{r}(t))$. The excitation amplitude is given by:

$$a_{i\to f} = \frac{1}{i\hbar}\int_{-\infty}^{\infty} e^{i\omega_{fi}t}\langle f|V(\vec{r}(t))|i\rangle dt$$

where

$$\omega_{fi} = \frac{E_f - E_i}{\hbar}$$

$E_i$ and $E_f$ are the energies of the initial and final states.

The Coulomb excitation probability is then given by:

$$P_f = \left|a_{i\to f}\right|^2$$

To evaluate the matrix element $\langle f|V(\vec{r}(t))|i\rangle$, $V(\vec{r}(t))$ is expanded into a multipole series. With the multipole expansion of the electrostatic part of $V(\vec{r}(t))$, the total Coulomb excitation cross section for the electric excitation of the order $E\lambda$ is given by

$$\sigma_{E\lambda} = \left(\frac{Z_t e}{\hbar \upsilon}\right)^2 a_0^{-2\lambda+2} B(E\lambda)\, f_{E\lambda}(\xi)$$

where $\upsilon$ is the relative velocity at large distances, $a_0$ is half the distance of closest approach in a head-on collision, $B(E\lambda)$ is the reduced transition probability associated with a radiative transition of multipole order $E\lambda$, $f_{E\lambda}(\xi)$ the total cross section function (Alder et al. 1956; Alder and Winther 1956), and $\xi$ the adiabaticity parameter.

$$f_{E\lambda}(\xi) = \frac{16\pi^3}{(2\lambda+1)^3} \sum_{\mu} \left|Y_{\lambda\mu}(\pi/2,0)\right|^2 \int_0^{\pi} \left|I_{\lambda\mu}(\theta,\xi)\right| \frac{\cos(\theta/2)}{\sin^3(\theta/2)}\, d\Theta$$

where $Y_{\lambda\mu}(\theta,\Phi)$ are the normalized spherical harmonics and $I_{\lambda\mu}(\theta,\xi)$ are given by

$$I_{\lambda\mu}(\theta,\xi) = \int_{-\infty}^{\infty} e^{i\xi(\varepsilon \sinh(w)+w)} \times \frac{[\cosh(w)+\varepsilon + i(\varepsilon^2-1)^{1/2}\sinh(w)]^{\mu}}{[\varepsilon\cosh(w)+1]^{\lambda+\mu}}\, dw$$

The cross section for the excitation by the magnetic field is given by

$$\sigma_{M\lambda} = \left(\frac{Z_t e}{\hbar c}\right)^2 a_0^{-2\lambda+2} B(M\lambda)\, f_{M\lambda}(\xi)$$

with

$$f_{M\lambda}(\xi) = \frac{16\pi^3}{(2\lambda+1)^2} \sum_{\mu} \left|Y_{\lambda+1,\mu}(\pi/2,0)\right|^2 \frac{(\lambda+1)^2-\mu^2}{\lambda^2(2\lambda+3)} \int_0^{\pi} \left|I_{\lambda+1,\mu}(\theta,\xi)\right| \cot^2(\Theta/2) \frac{\cos(\theta/2)}{\sin^3(\theta/2)}\, d\theta$$

It might be interesting to point out that the ratio $\sigma_{M\lambda}/\sigma_{E\lambda}$ is proportional to $(c/\upsilon)^2$, which means that magnetic excitations are suppressed by this factor when compared to the electric excitations.

The probabilities for electromagnetic transitions are defined by (Bohr and Mottelson 1969):

$$B(\pi\lambda; I_i \to I_f) = \sum_{\mu M_f} \left|\langle I_f M_f | \mathrm{M}(\pi\lambda\mu) | I_i M_i \rangle\right|^2 = \frac{1}{2I_i+1} \left|\langle I_f \| \mathrm{M}(\pi\lambda) \| I_i \rangle\right|^2$$

Symbol $\pi$ stands for $E$ or $M$ i.e. for the electric or magnetic transitions, $I$ and $M$ in the wave functions are the total angular momentum and the magnetic quantum numbers of the initial and final states, $\mathrm{M}(\pi\lambda\mu)$ the multipole operator, and $\langle I_f \| \mathrm{M}(\pi\lambda) | I_i | \rangle$ the reduced matrix element. The quantity $\mu$ gives the angular momentum transfer along the beam direction ($\mu = M_i - M_f$).

In the first order perturbation theory, the cross section is directly proportional to the reduced transition probability:

$$\sigma_{\pi\lambda} \propto B(\pi\lambda; I_i \to I_f)$$

Three important parameters in the Coulomb excitation are: (1) the Sommerfeld parameter $\eta$, (2) the adiabaticity parameter $\xi$, and (3) the excitation strength parameter $\chi$.

*The Sommerfeld* parameter is the ratio of the half the distance of closest approach, $a_0$, to the de Broglie wave length $\lambda\!\!\!^{-}$. As long as the ratio $\eta$ >1, there is no direct interaction of the two nuclei and a semiclassical approach for the trajectory is valid. Usually $\eta$ >>1.

*The adiabaticity* parameter $\xi$ describes how swift is the interaction. In order to excite a final state $|f\rangle$ from the initial state $|i\rangle$, the collision time must be shorter or of the same order of magnitude as the time of the internal motion of the nucleons. For a sufficiently short time of collision, the adiabaticity parameter is small and excitations are possible. The excitation probabilities given by the function $f_{E\lambda}(\xi)$ decrease with the increasing $\xi$ (see Figure K.2). As the adiabaticity parameter approaches the value of 1, the probability of the system to interact becomes smaller. For $\xi$ >1 $f_{E\lambda}(\xi)$ decreases exponentially with $\xi$.

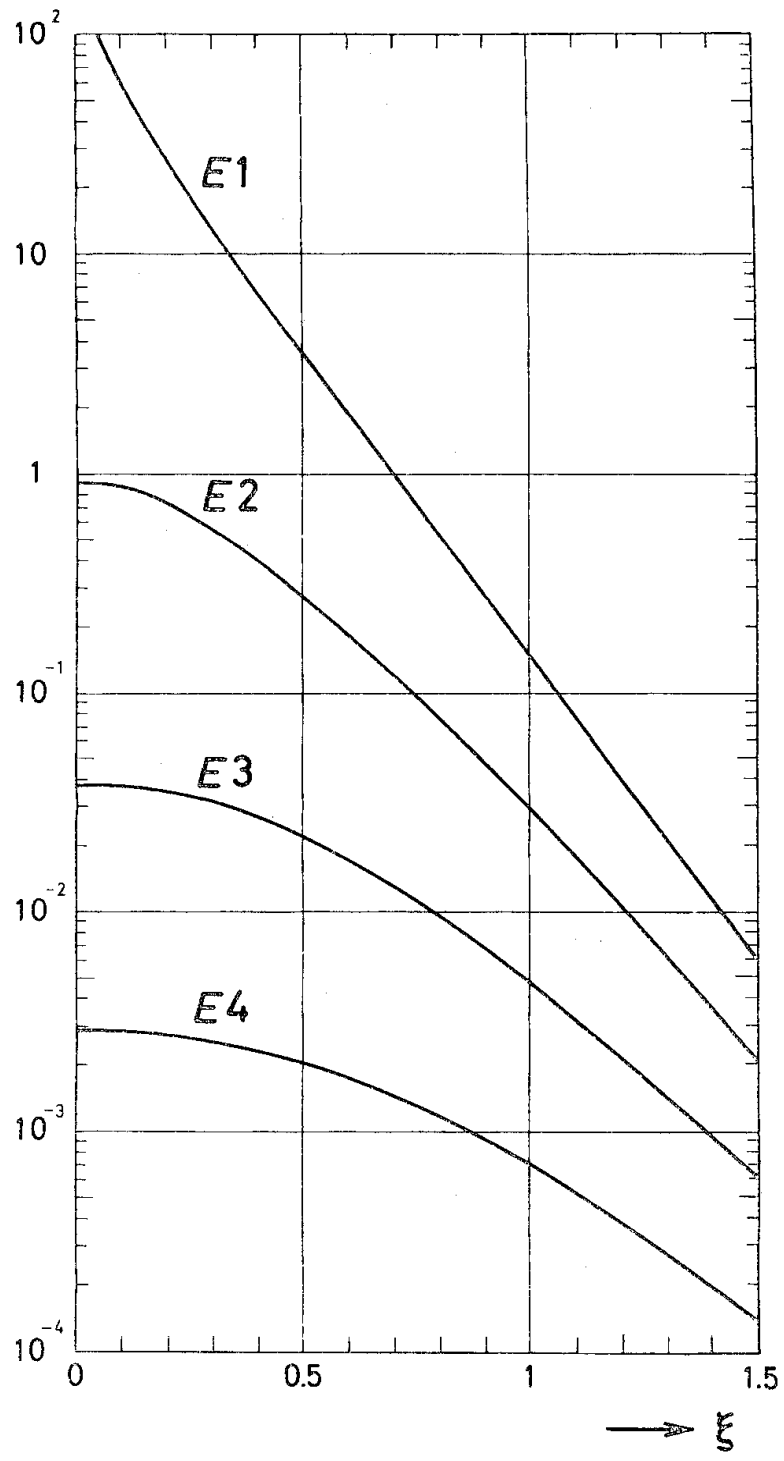

Figure K.2. Examples of the $f_{E\lambda}(\xi)$ function for $\lambda$ = 1, 2, 3, and 4 (Adler and Wither 1975).

*The excitation strength parameter* $\chi$ describes the strength of the interaction for different multipoles. As the strength increases, higher order perturbation must be taken into account.

The excitation strength parameter $\chi$ for a transition of the order $\pi\lambda$ from state $|i\rangle$ to $|f\rangle$ is defined by

$$\chi_{\pi\lambda} = \frac{\sqrt{16\pi}Z_t e}{\hbar s}\frac{(\lambda-1)!}{(2\lambda+1)!!}\frac{\langle I_f \| M(\pi\lambda) \| I_i\rangle}{a_0^\lambda\sqrt{2I_i+1}}$$

where $s = \upsilon$ for $E\lambda$ excitations and $c$ for $M\lambda$.

## Second-order perturbation theory

The first-order perturbation theory is used to describe direct excitations. Multiple excitations, which include also reorientation effects, are described by the second order perturbation theory. In this theory, the probability $P_f$ of the excitation of state $|f\rangle$ is made of three components:

$$P_f = P_f^{(1)} + P_f^{(1,2)} + P_f^{(2)}$$

where $P_f^{(1)}$ is the first order excitation probability written earlier as $P_f$, $P_f^{(2)}$ is the second-order term, and $P_f^{(1,2)}$ is the interference term between the first- and second-order amplitudes. A schematic diagram comparing the first- and second-order excitation processes is presented in Figure K.3.

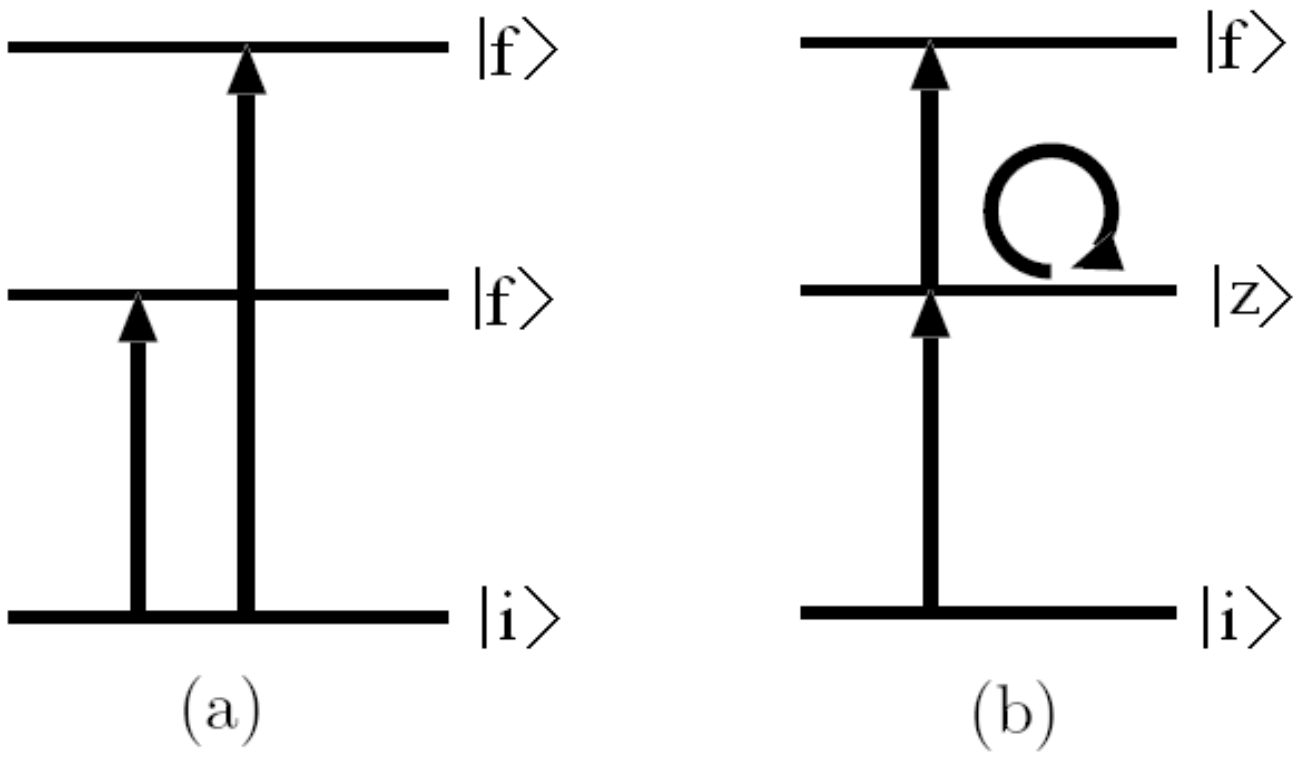

Figure K.3. A schematic diagram comparing the first- and second-order Coulomb excitation processes. The left-hand side (a) of the figure shows two excitations of states $|f\rangle$ directly from the state $|i\rangle$. The right-hand side (b) of the figure contains two examples of the second-order processes. It is assumed that a direct excitation of state $|f\rangle$ from state $|i\rangle$ is not allowed but state $|f\rangle$ can excited in a two-step process involving the intermediate state $|z\rangle$. This type of excitation is described by the second-order excitation probability $P_f^{(2)}$. The excitation of the first excited state $|z\rangle$, i.e. the transition $|i\rangle \rightarrow |z\rangle$ , can be also followed by the reorientation transition $|z\rangle \rightarrow |z\rangle$. This process is described by the second-order excitation probability $P_f^{(1,2)}$.

The $P_f^{(2)}$ term describes multiple excitations, which is analogous to multiple excitations in particle transfer reaction or scattering as discussed in Chapter 17 and in the Appendix E. This term becomes significant for transitions, which are either weak or cannot take place in single-step excitations. They involve not only the initial and final states $|i\rangle$ and $|f\rangle$ but also an intermediate state $|z\rangle$.

The $P_f^{(1,2)}$ term describes reorientation processes. In this case the intermediate state is identical with the excited state and transitions are between the magnetic substates.

Explicit expressions for the excitation probabilities from states $|i\rangle = 0$ are (Alder and Winther 1975):

$$P_f^{(1)} = \sum_{\lambda} \left|\chi_{0\to f}^{(\lambda)}\right|^2 R_\lambda^2(\theta, \xi_{f0})$$

$$P_f^{(1,2)} = \sum_{\lambda\lambda'\lambda'' I_z} \sqrt{(2I_z+1)(2\lambda+1)(2\lambda'+1)(2\lambda''+1)}(-1)^{I_0+I_f}$$
$$\times \begin{Bmatrix} \lambda & \lambda' & \lambda'' \\ I_z & I_f & I_0 \end{Bmatrix} \chi_{0\to f}^{(\lambda)} \chi_{0\to z}^{(\lambda')} \chi_{z\to f}^{(\lambda'')} \sum_{\mu} R_{\lambda\mu}^*(\theta, \xi_{f0}) G_{(\lambda'\lambda'')\lambda\mu}(\theta, \xi_{z0}, \xi_{fz})$$

$$P_f^{(2)} = \frac{1}{4} \sum_{\lambda_1\lambda_1'\lambda_2\lambda_2' I_z I_z' k} \sqrt{(2I_z+1)(2I_{z'}+1)(2\lambda_1+1)(2\lambda_1'+1)(2\lambda_2+1)(2\lambda_2'+1)}$$
$$\times (2k+1) \begin{Bmatrix} \lambda_1 & \lambda_2 & k \\ I_f & I_0 & I_z \end{Bmatrix} \begin{Bmatrix} \lambda_1' & \lambda_2' & k \\ I_f & I_0 & I_{z'} \end{Bmatrix} \chi_{0\to z}^{(\lambda_1)} \chi_{z\to f}^{(\lambda_2)} \chi_{0\to z'}^{(\lambda_1')} \chi_{z'\to f}^{(\lambda_2')}$$
$$\times \sum_{\kappa} [R_{(\lambda_1\lambda_2)k\kappa}^* R_{(\lambda_1'\lambda_2')k\kappa} + G_{(\lambda_1\lambda_2)k\kappa}^* G_{(\lambda_1'\lambda_2')k\kappa}]$$

where $R_\lambda^2(\theta,\xi)$, $R_{\lambda\mu}(\theta,\xi)$, $R_{(\lambda\lambda')k\kappa}$ and $G_{(\lambda\lambda')k\kappa}$ are the relative probabilities for electric excitations, normalized orbital integrals, real part of second-order orbital integral tensor, and imaginary part of second-order orbital integral tensor, respectively, as defined by Alder and Winther (1975).

It is seen that the last two formulae include summation over contributing $\lambda$ values and that their contributions depend on the product $\chi_{0\to z}\chi_{z\to f}$ of the excitation strength parameters and on the adiabaticity parameters $\xi_{z0}$ and $\xi_{fz}$, all of which involve intermediate states $|z\rangle$.

In a particular case of the $0^+ \to 2^+$ excitation (see Figure K.4), which is related to the study of reorientation effects in deuteron polarization as discussed in Chapter 16, Alder and Winther (1975) show that

$$P_f \equiv P_2 = \left|\chi_{0\to 2}^{(2)}\right|^2 R_2^2(\theta,\xi)[1 + \chi_{2\to 2}^2 c(\theta, s=1, \xi)]$$

where $c(\theta,s,\xi)$ is the coefficient of interference between *E2* and *E2-E2* excitation.

$$\chi^{(2)}_{2\to 2} = \frac{4}{15}\sqrt{\frac{\pi}{5}}\frac{Z_p e}{\hbar\upsilon}\frac{1}{a_0^2}\langle 2\|\mathrm{M}(E2)\|2\rangle = \frac{1}{a_0^2}\sqrt{\frac{7}{90}}\frac{Z_p e}{\hbar\upsilon}Q_2 = 8.474\frac{A_p^{1/2}E^{3/2}Q_2}{Z_p Z_t^2(1+A_p/A_t)^{1/2}}$$

where $Q_2$ is the quadrupole moment in $e\cdot 10^{-24}$ cm$^2$ and $E$ is the bombarding energy in MeV.

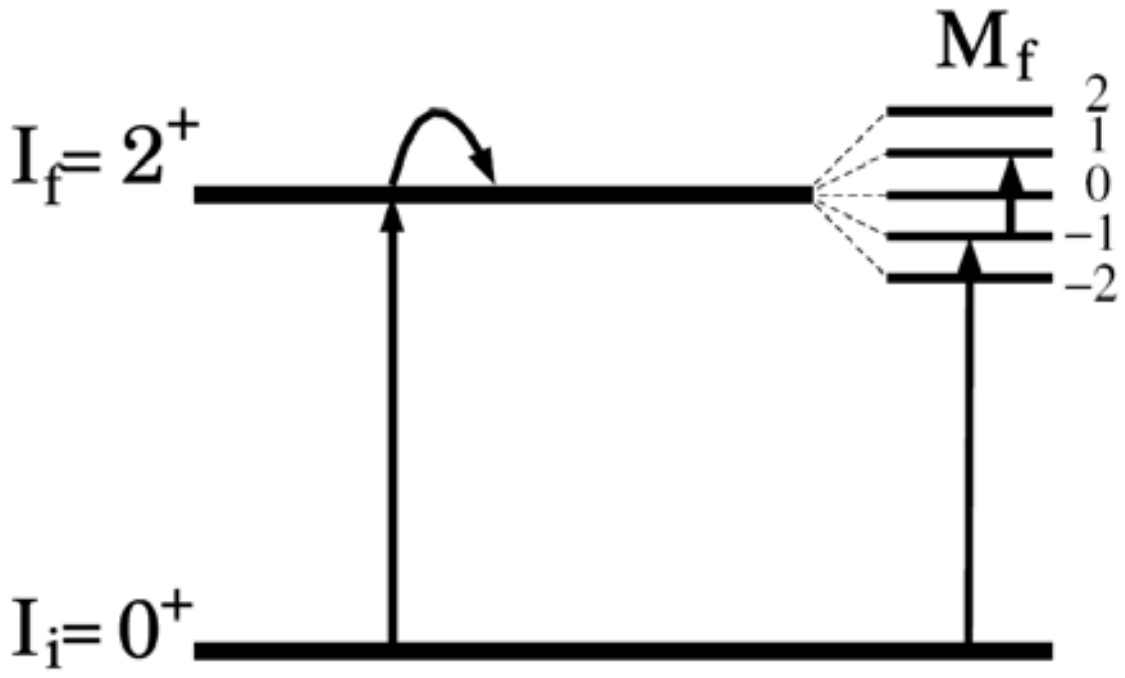

Figure K.4. The reorientation excitations involving the 2$^+$ state.

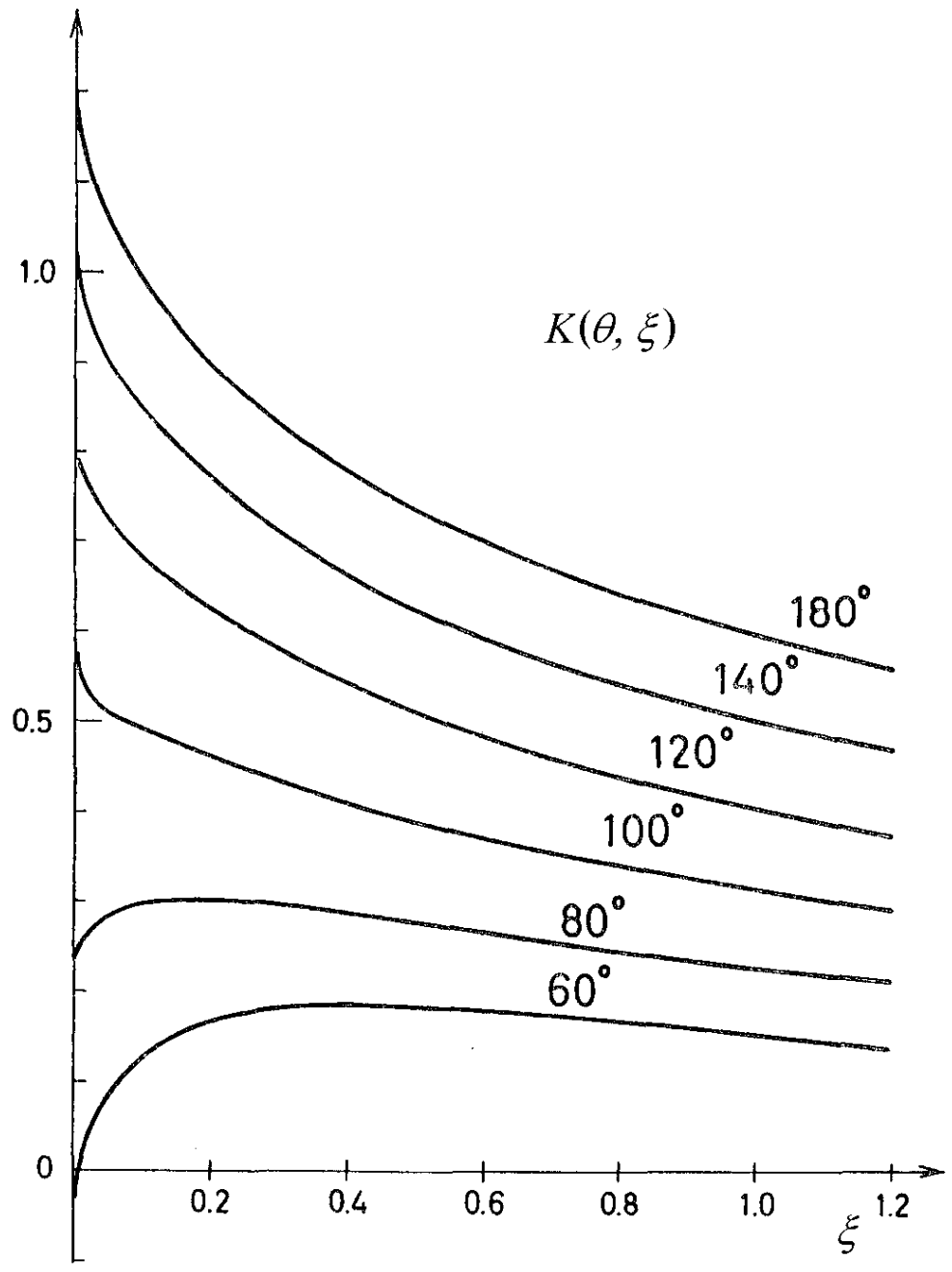

Figure K.5. The $K(\theta,\xi)$ function.

It is seen that $\chi^{(2)}_{2\to 2}$ is proportional to $Q_2$. Thus, measurements involving reorientation process can be used to determine the sign of the quadrupole moment.

Using the properties of the $c(\theta,s,\xi)$ function, the excitation probability $P_2$ can be written as (Alder and Winther 1975; De Boer and Eichler 1968):

$$P_2 = P_2^{(1)}[1 + qK(\theta,\xi)]$$

where $P_2^{(1)}$ is the first-order excitation probability,

$$q = \frac{A_p \Delta E \langle 2 \| \mathrm{M}(E2) \| 2 \rangle}{Z_t (1 + A_p / A_t)}$$

with

$$\langle 2 \| \mathrm{M}(E2) \| 2 \rangle = \sqrt{\frac{7}{2\pi}} \frac{5}{4} Q_2 = \frac{1}{0.7579} Q_2$$

and

$$K(\theta, \xi) = \frac{0.5056}{\xi} c(\theta, s = 1, \xi)$$

The function $K(\theta,\xi)$ is shown in Figure K.5. The dependence of the cross section for the excitation of $2^+$ state on the value of the quadrupole moment is shown in Figure K.6.

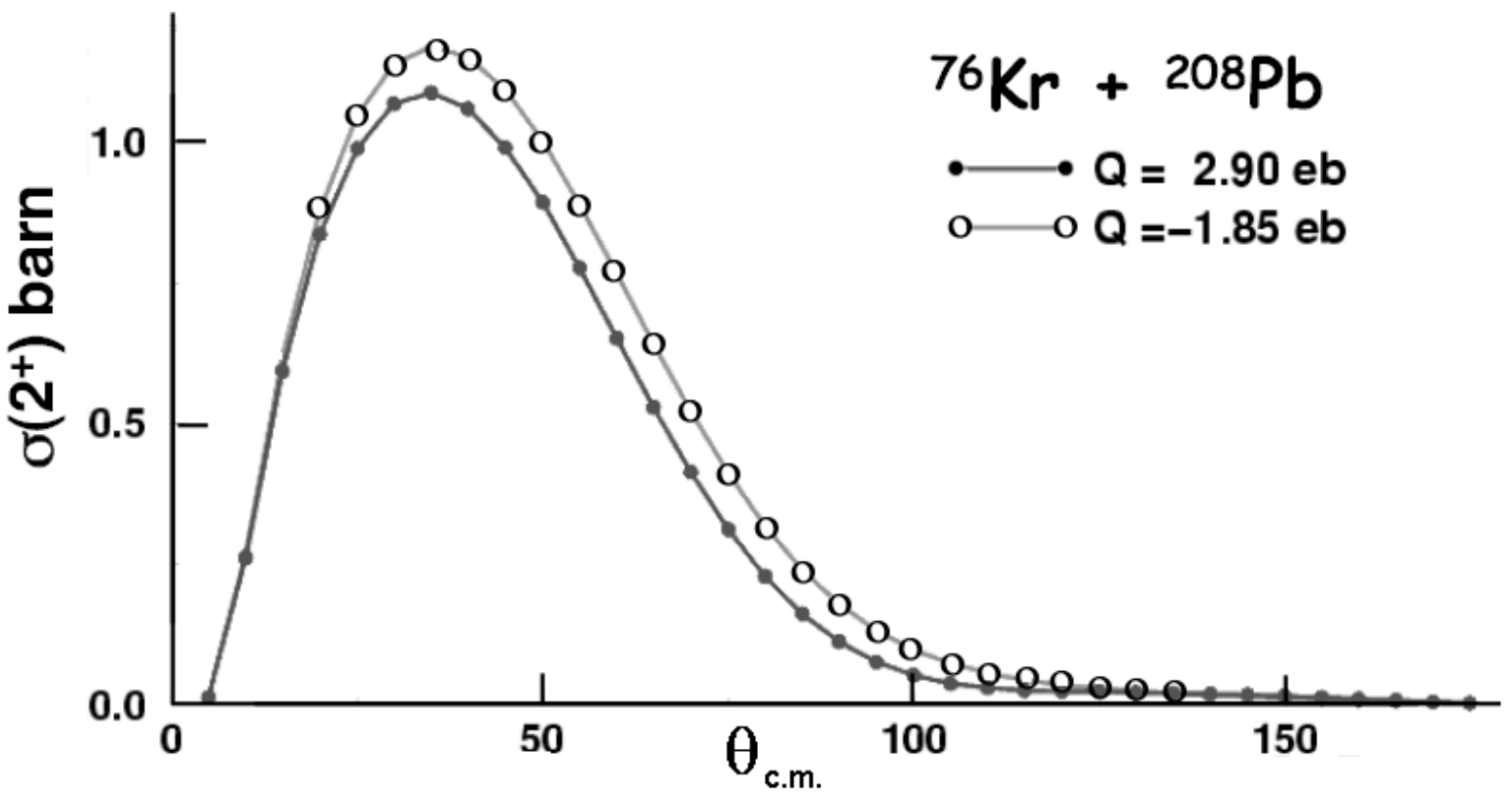

Figure K.6. Reorientation effect in Coulomb excitation. Differential cross sections calculated using the second order perturbation theory.

As in the case discussed in Chapter 16, the effect is weak. However, careful measurements can be used to distinguish between prolate and oblate deformation. Our study described in Chapter 16 shows that similar distinction between the two types of deformations can be made by studying vector polarization of inelastically scattered deuterons.

# Appendix L

## The Faddeev formalism

The Faddeev formalism describes interaction of three-body systems (Faddeev 1961a, 1961b, 1962, 1965; Schmid and Ziegelmann 1974). To understand it, I shall start with a more familiar two-body formalism.

### Two-nucleon scattering and Lippmann-Schwinger equation

The Schrödinger equation for the scattering of a particle with reduced mass $\mu$ and energy $E$ by a potential $V$ can be written as

$$(E-K)|\psi\rangle = V|\psi\rangle \tag{1}$$

where

$$K = -\frac{\hbar\nabla^2}{2\mu} \tag{1a}$$

is the kinetic energy operator.

This equation can be rewritten in a different form as (Lippmann and Schwinger 1950; Satchler 1983):

$$|\psi\rangle = |\phi\rangle + G_0 V|\psi\rangle \tag{2}$$

This new equation is known as the Lippmann-Schwinger equation or an integral form of the Schrödinger equation.

In the eq. (2), $|\varphi\rangle$ is the solution of the homogenous equation

$$(E-K)|\varphi\rangle = 0 \tag{3}$$

and

$$G_0 = \frac{1}{E-K+i\varepsilon} \tag{4}$$

is the free-particle Green operator, i.e. the operator for an undisturbed system. The parameter $\varepsilon$ is introduced to handle the problem of singularity. This is done by calculating the limit of the relevant quantities when $\varepsilon \to 0$.

As emphasised by Satchler (1983), even though the equation (2) looks like a solution of the Schrödinger equation it is in fact only an alternative but more convenient form of the equation (1). This can be seen easily by operating $(E-K)$ on the equation (2). Assuming that $i\varepsilon$ is infinitesimally small and thus can be neglected we have

$$(E-K)|\psi\rangle = (E-K)\{|\phi\rangle + G_0 V|\psi\rangle\}$$

$$(E-K)|\psi\rangle = (E-K)|\phi\rangle + (E-K)\frac{1}{E-K}V|\psi\rangle$$

but

$$(E-K)|\varphi\rangle = 0$$

and therefore

$$(E-K)|\psi\rangle = V|\psi\rangle$$

which is the original Schrödinger equation (1).

A formal solution of the Schrödinger equation can be written as (Satchler 1983):

$$|\psi\rangle = |\phi\rangle + GV|\phi\rangle \tag{5}$$

where

$$G = \frac{1}{E-H+i\varepsilon} \tag{6}$$

is the Green function for the Hamiltonian $H$.

Using operator algebra, the Green function $G$ for two interacting particles with potential $V$ can be decomposed as

$$G = G_0 + G_0 V G_0 \tag{7}$$

We can also introduce a transition operator $t$ for the potential scattering. We define it as

$$V|\psi\rangle \equiv t|\phi\rangle \tag{8}$$

It is an operator that causes a transition of the initial free state to the scattering state by means of a potential.

By multiplying the Lippmann-Schwinger equation by $V$ from left we can easily find that

$$t = V + VG_0 t \tag{9}$$

It is also easy to see that the $t$ operator can be evaluated iteratively as:

$$t = V + VG_0V + VG_0VG_0V + VG_0VG_0VG_0V + ... \tag{10}$$

This can be represented graphically as

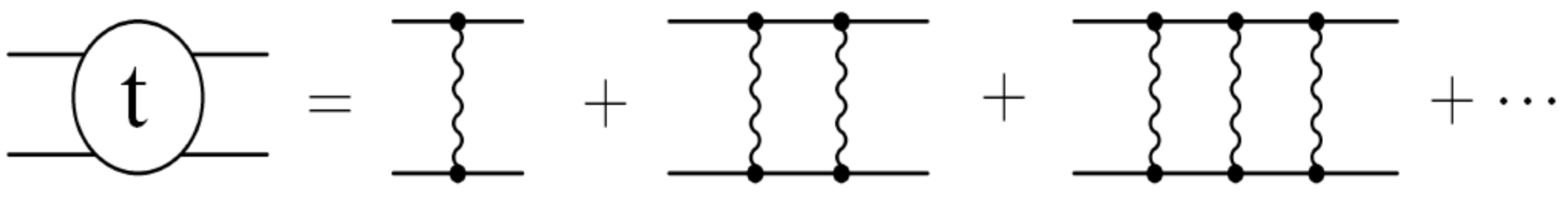

In this diagram, each wavy line represents an interaction caused by potential $V$. In between the lines, the particle moves freely as indicated by the free propagator $G_0$.

The cross section is directly related to the $t$ matrix, which is the matrix element of the transition operator in momentum space

$$\frac{d\sigma}{d\Omega} \propto \left|\langle p'|t(E+i\varepsilon)|p\rangle\right|^2 \tag{11}$$

where

$$E = \frac{p'^2}{2\mu} = \frac{p^2}{2\mu}$$

**Three-nucleon scattering and the Faddeev equations**

The formal description of three-nucleon scattering resembles the description of two-nucleon scattering even though the system is significantly more complex. Interaction of a three-body system involves not only two-body but also three-body potentials. Many forms of three-body potentials have been proposed but their study is difficult because from experimental and theoretical investigation we already know that their effects are weak when compared with the effects caused by two-body interaction. For most observables, theoretical calculations introduce insignificant differences when three-body forces are included. In this description of the Faddeev formalism I will therefore assume that the interaction in the three-body system involves only two-body potentials.

Some of the quantities used in the description of a three-body system are shown in Figure L.1. In order to handle the Schrödinger equation for the three-body interaction, Faddeev used the standard Jacobi system of coordinates.

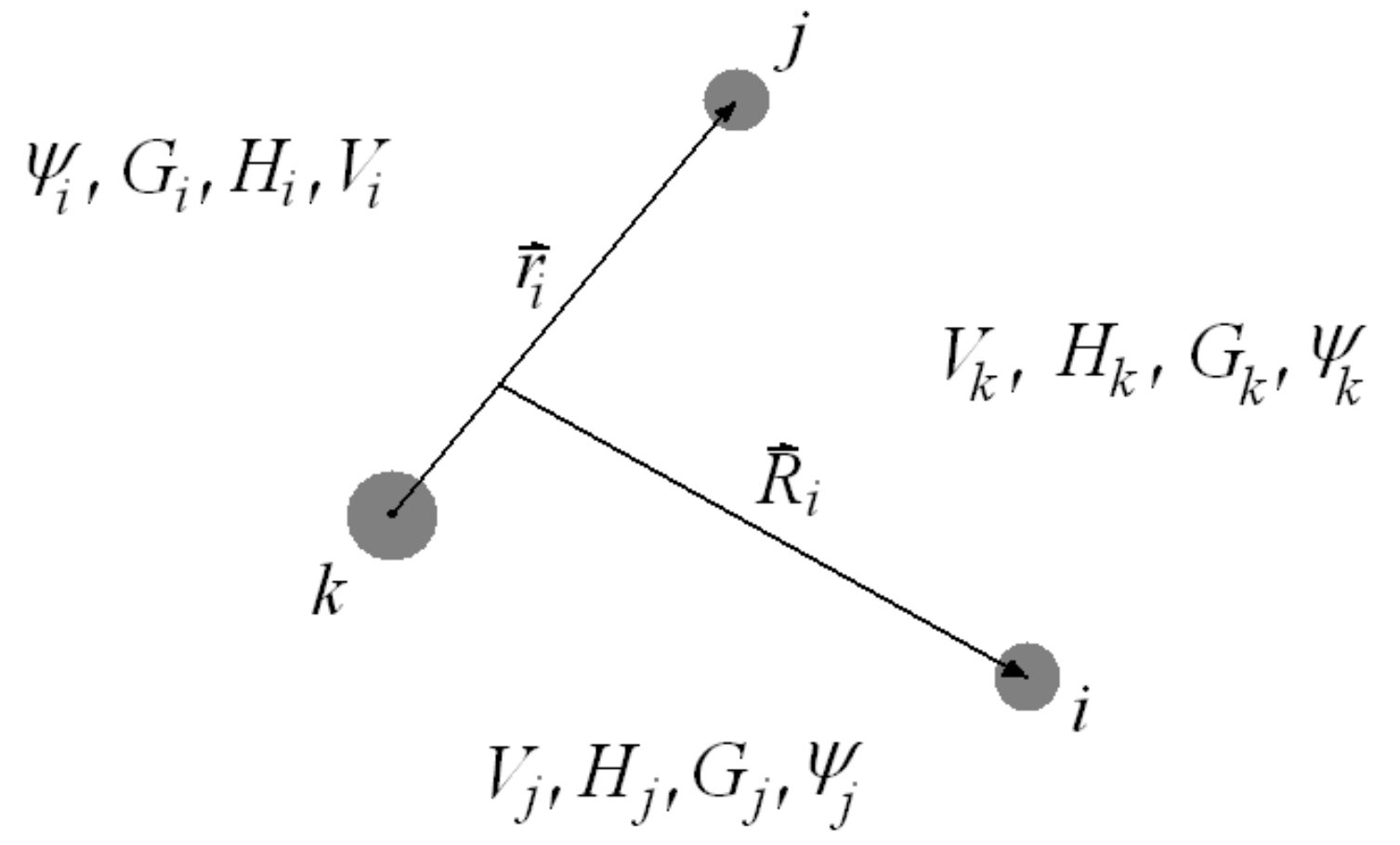

Figure L.1. A three-body system showing the labelling of the three particles, the Jacobi coordinates and some of the quantities used in respective channels to describe the system.

We have to introduce the following definitions:

$$\vec{r}_i = \vec{x}_j - \vec{x}_k ; \quad \vec{R}_i = \vec{x}_i - \frac{m_j\vec{x}_j + m_k\vec{x}_k}{m_j + m_k} ; \quad \vec{\eta} = \frac{m_i\vec{x}_i + m_j\vec{x}_j + m_k\vec{x}_k}{m_i + m_j + m_k} \tag{12}$$

$$\vec{p}_i = \frac{m_j\vec{k}_k - m_k\vec{k}_j}{m_j + m_k} ; \quad \vec{q}_i = \frac{(m_j + m_k)\vec{k}_i - m_i(\vec{k}_j + \vec{k}_k)}{m_i + m_j + m_k} ; \quad \vec{K} = \vec{k}_i + \vec{k}_j + \vec{k}_k \tag{13}$$

$$V_i = V_{jk}(|\vec{x}_j - \vec{x}_k|) ; \quad \frac{1}{\mu_i} = \frac{1}{m_j} + \frac{1}{m_k} ; \quad \frac{1}{M_i} = \frac{1}{m_i} + \frac{1}{m_j + m_k} \tag{14}$$

$$H_i = \frac{p_i^2}{2\mu_i} + \frac{q_i^2}{2M_i} + V_i\,; \quad G_i = \frac{1}{E - H_i + i\varepsilon} \tag{15}$$

$\vec{r}_i$ and $\vec{R}_i$ are as shown if Figure L.1

$\vec{K}$ – the total momentum of three particles

$\vec{p}_i$ – the relative momentum of particles $j$ and $k$

$\vec{q}_i$ – the relative momentum of particle $i$ with respect to the centre-of-mass of particles $j$ and $k$

$V_i$ – the interaction potential between particles $j$ and $k$

$H_i$ – the channel Hamiltonian

$G_i$ – the channel Green function

The configuration of a bound pair $j$-$k$ and a free particle $i$ is described by the function

$$\Phi_i = \phi_i(\vec{r}_i)e^{i\vec{q}_i \cdot \vec{R}_i} \tag{16}$$

which is the solution of

$$H_i|\Phi_i\rangle = E_i|\Phi_i\rangle \tag{17}$$

where

$$E_i = \frac{q_i^2}{2M_i} + e$$

is the total energy in the centre-of-mass frame and $e$ is the binding energy of $\phi_i$.

We can also define the channel wave function $|\psi_i\rangle$ as a solution of the Schrödinger equation

$$(H_i + V_j + V_k)|\psi_i\rangle = E|\psi_i\rangle \tag{18}$$

which can be rewritten as the Lippmann-Schwinger equation for the three-body system

$$|\psi_i\rangle = |\Phi_i\rangle + G_i(V_j + V_k)|\psi_i\rangle \tag{20}$$

and which is similar to the equation (2) for the two-body system. It can be shown (Foldy and Tobocman 1957) that this equation does not have a unique solution. Glöckle (1970) has shown that $|\psi_i\rangle$ satisfies also the following two homogenous equations:

$$|\psi_i\rangle = G_j(V_i + V_k)|\psi_i\rangle \tag{21a}$$

$$|\psi_i\rangle = G_k(V_j + V_i)|\psi_i\rangle \tag{21b}$$

The equations (20), (21a) and (21b) are called Lippmann-Schwinger triad and can be written as

$$\left|\psi_i\right\rangle = \delta_{il}\left|\Phi_i\right\rangle + G_l(V_j + V_k)\left|\psi_i\right\rangle \tag{22}$$

As in the case of two-body interaction, we can also introduce a transition operator. In analogy to the $t$ operator for the two-body interaction (see eq. (9)), the transition operator $T$ for the three-body system satisfies the following relation

$$T = V + VG_0T \tag{23}$$

where $V = V_1 + V_2 + V_3$

Faddeev splits eq. (23) into three equations:

$$T = (V_1 + V_2 + V_3) + (V_1 + V_2 + V_3)G_0T = \sum_{i=1}^{3}(V_i + V_iG_0T) = \sum_{i=1}^{3}T_i \tag{24}$$

where

$$T_i = V_i + G_0T \tag{25}$$

The eq. (25) can be expressed as

$$T_i = t_i + t_iG_0(T_j + T_k) \tag{27}$$

where by definition

$$t_i \equiv \frac{V_i}{1 - V_iG_0} \tag{28}$$

It is easy to see that eq. (28) is the same as eq. (9) and thus it is a two-body transition operator in the three-body space.

Equation (27) can be written in a matrix form as

$$\begin{bmatrix} T_1 \\ T_2 \\ T_3 \end{bmatrix} = \begin{bmatrix} t_1 \\ t_2 \\ t_3 \end{bmatrix} + \begin{bmatrix} 0 & t_1 & t_1 \\ t_2 & 0 & t_2 \\ t_3 & t_3 & 0 \end{bmatrix} G_0 \begin{bmatrix} T_1 \\ T_2 \\ T_3 \end{bmatrix} \tag{29}$$

As for the two-body system, the $T$ operator can be evaluated by iteration. If we define

$$\mathbf{T} \equiv \begin{bmatrix} T_1 \\ T_2 \\ T_3 \end{bmatrix}, \ \mathbf{t} \equiv \begin{bmatrix} t_1 \\ t_2 \\ t_3 \end{bmatrix}, \text{ and } \boldsymbol{\tau} \equiv \begin{bmatrix} 0 & t_1 & t_1 \\ t_2 & 0 & t_2 \\ t_3 & t_3 & 0 \end{bmatrix}$$

then the matrix equation for $T$ can be written as

$$\mathbf{T} = \mathbf{t} + \boldsymbol{\tau}G_0\mathbf{T} \tag{30}$$

If we now use the same iteration procedure as for the eq. (9) then we shall find that

$$\mathbf{T} = \mathbf{t} + \boldsymbol{\tau}G_0\mathbf{t} + \boldsymbol{\tau}G_0\boldsymbol{\tau}G_0\mathbf{t} + \boldsymbol{\tau}G_0\boldsymbol{\tau}G_0\boldsymbol{\tau}G_0\mathbf{t} + \ldots \tag{31}$$

# Appendix M

## Resonating group theory

Resonating group theory (also known as *method* or *model* and abbreviated as RGM) applies to an interaction of two groups of particles. RGM was first introduced by Wheeler (1937) to describe resonant transfer of a group of electrons from one group of atoms to another. Later, the method was applied to an interaction of clusters of nucleons and was improved by the introduction of three useful methods: the Peierls-Yoccoz projection method, the generating coordinate method, and the Brink alpha-cluster model (Brink 1965 and Saito 1977).

The basic assumption of the RGM is that at small distances, the interaction between individual nucleons belonging to two clusters of interacting particles is mainly defined by the Pauli principle and that the dynamic effects are described by the wave function of the relative motions of the reacting clusters.

The Schrödinger equation for the interacting groups is

$$H\psi = E\psi$$

where

$$H = \sum_{i=1}^{A} T_i + \frac{1}{2}\sum_{i \neq j}^{A} V_{ij}$$

with $T_i$ being the kinetic energy operator of the $i$th particle and $V_{ij}$ the interaction potential between particle $i$ and $j$.

The wave function $\psi \equiv \psi(A)$ for the whole system of $A$ nucleons can be written as

$$\psi(A) = \sum_{A_1 + A_2 = A} \sum_{n_1, n_2} \mathcal{A} \left[ \phi_{n_1}(A_1)\phi_{n_2}(A_2) f(n_1, n_2) \right]$$

where $\mathcal{A}$ is the antisymmetrisation operator responsible for the Pauli principle, $\phi_{n_1}(A_1)$ and $\phi_{n_2}(A_2)$ are the wave functions for the interacting nuclei, and $f(n_1, n_2)$ is the function for the relative motion.

In the algebraic version of RGM (Filippov 1989), the wave function $\psi(A)$ is expressed as

$$\psi(A) = \sum_{A_1 + A_2 = A} \sum_{n_1 n_2 n} C_{n_1 n_2 n}(A_1 A_2) \left| n_1 n_2 n \right\rangle$$

where $C_{n_1 n_2 n}(A_1 A_2)$ are the Fourier coefficients of the expansion of $\psi(A)$ in the many-particle oscillator basis functions $\left| n_1 n_2 n \right\rangle$ antisymmetric under a permutation of nucleon coordinates.

The coefficients $C_{n_1 n_2 n}(A_1 A_2)$ are determined by solving the following algebraic equations:

$$\sum_{A_1' A_2'} \sum_{n_1' n_2' n} \left\langle A_1 A_2; n_1 n_2 n \right| H - E \left| A_1' A_2'; n_1' n_2' n' \right\rangle C_{n_1 n_2 n}(A_1 A_2) = 0$$

Theoretical treatment may be expanded by including many-body forces and an interaction of three clusters of nucleons. For more information about mathematical description of the RGM and its application see for instance Aoki and Horiuchi 1982; Hofmann (2002), Lemere, Tang, and Thompson (1976), and Thompson and Tang (1973)

# Appendix N

## The R-matrix theory

The R-matrix theory has been applied successfully to nuclear reactions proceeding via formation of a compound nucleus. In this theory, the space is divided between internal and external regions. The internal region refers to the space inside the nucleus. This division is possible because of the short-range of nuclear forces and because of the Pauli principle.

Schrödinger equations are solved in both regions. To study resonances in the compound nucleus, the internal wave function is expressed in terms of a complete set of orthogonal resonant states, which are also solutions of appropriate Schrödinger equations and which obey a specific boundary condition.

The matching of the internal and external regions is done by calculating and equating logarithmic derivatives of the radial wave functions at the boundary between the two regions. The logarithmic derivative of the internal wave function leads to an expression containing $R$-function or $R$-matrix, depending on the assumptions used in the calculations. The differential cross section is expressed in terms of the collision function (or matrix). The matching of the logarithmic derivatives for the external and internal region allows for expressing the collision function (or matrix) in terms of the $R$-function (or $R$-matrix).

Discussion of the R-matrix theory may be found in the publication of Lane and Thomas (1958). Excellent outline of this theory may be also found in a set of lectures by Vogt (2004). To outline the basic ideas of R-matrix treatment, I start with the discussion of the single-channel case.

### Single-channel theory

The differential cross section is given by

$$\frac{d\sigma}{d\Omega} = \frac{1}{4k^2}\left|\sum_l (2l+1)(1-U_l)P_l(\cos\theta)\right|^2 \tag{1}$$

and integrated cross section by

$$\sigma \equiv \int \frac{d\sigma}{d\Omega} d\Omega = \frac{\pi}{k^2}\sum_l (2l+1)\left|1-U_l\right|^2 \tag{2}$$

where

$$U_l = e^{2i\delta_l} \tag{3}$$

is the collision functions with $\delta_l$ being phase shift parameters.

#### *Internal region*

The radial wave function $\varphi_l(r)$ in the internal region is a solution of the Schrödinger equation

$$-\frac{\hbar^2}{2m}\frac{d^2\varphi_l(r)}{dr^2} + V(r)\varphi_l(r) = E\varphi_l(r) \tag{4}$$

where $V(r)$ is the nuclear potential.

This function can be expressed as a sum of mutually orthogonal resonant state wave functions $X_{\lambda l}$, which are the solutions of the Schrödinger equation

$$-\frac{\hbar^2}{2m}\frac{d^2 X_{\lambda l}(r)}{dr^2} + V(r)X_{\lambda l}(r) = E_{\lambda l}X_{\lambda l}(r) \quad (5)$$

If we use radius $r = a_l$ to denote the division between the internal and external region, then the boundary condition for $X_{\lambda l}(r)$ is given by

$$a_l\left(\frac{dX_{\lambda l}(r)}{dr}\right)_{r=a_l} = b_l X_{\lambda l}(r = a_l) \quad (6)$$

where $b_l$ is an arbitrary (real) boundary condition number.

The external and internal regions are joined by matching the logarithmic derivatives of the respective radial wave functions at the nuclear surface ($r = a_l$). For the internal wave function we have[14]

$$\left(r\frac{\varphi'(r)}{\varphi(r)}\right)_{r=a_l} = \frac{1 + b_l R_l}{R_l} \quad (7)$$

where $\varphi'(r) \equiv d\varphi(r)/dr$ and $R_l$ is the so-called $R$-function, which is given by

$$R_l = \sum_\lambda \frac{\gamma_{\lambda l}^2}{E_{\lambda l} - E} \quad (8)$$

with

$$\gamma_{\lambda l}^2 = \frac{\hbar^2}{2ma_l}X_{\lambda l}^2(r = a_l) \quad (9)$$

being the reduced width. In the multi-channel theory, this function is replaced by $R$-matrix.

***External region***

The radial wave function $\varphi_l(r)$ in the external region is a solution of the Schrödinger equation containing Coulomb interaction

$$\frac{d^2\varphi(r)}{dr^2} - \left[\frac{l(l+1)}{r^2} + \left(\frac{2m}{\hbar^2}\right)\left(\frac{Z_1 Z_2 e^2}{r} - E\right)\right]\varphi(r) = 0 \quad (10)$$

Its solution can be expressed in terms of the regular $F_l$ and irregular $G_l$ solutions. At large $r$ values, they are given asymptotically by

$$F_l \rightarrow \sin[kr - \eta\log(2kr) - (1/2)l\pi + \sigma_l] \quad (11a)$$

[14] Lane and Thomas (1958) assume $b_l = 0$ (see for instance their analogous expression (IV,1.11) for this quantity).

$$G_l \to \cos\left[kr - \eta \log(2kr) - (1/2)l\pi + \sigma_l\right] \quad (11b)$$

where

$$\sigma_l = \arg[1 + l + i\eta] \quad (12)$$

is the Coulomb phase shift, and

$$\eta = \frac{Z_1 Z_2 e^2}{h\upsilon} \quad (13)$$

is the Coulomb parameter.

In terms of the collision matrix $U_l$

$$\varphi_l(r) \propto I_l - U_l O_l \quad (14)$$

where $I_l$ and $O_l$ are the incoming and outgoing wave functions, respectively,

$$I_l = (G_l - iF_l)e^{i\omega_l} \quad (15a)$$

$$O_l = (G_l + iF_l)e^{-i\omega_l} \quad (15b)$$

with

$$\omega_l = \sum_{n=1}^{l} \tan(\eta / n) \quad (16)$$

By calculating the logarithmic derivatives of the external radial wave function and by matching them with the logarithmic derivatives of the internal functions we can express the collision function $U_l$ in terms of the $R$-function:

$$U_l = \frac{I_l}{O_l}\frac{(1 - R_l L_l^*)}{(1 - R_l L_l)} \quad (17)$$

where $L_l$ contains the logarithmic derivative of $O_l$

$$L_l \equiv \frac{O_l'}{O_l} - b_l \quad (18)$$

The quantity $L_l$ can be also expressed in terms of the penetration factor $P_l$ and shift functions $S_l$:

$$L_l = S_l + iP_l - b_l \quad (19)$$

where

$$P_l = \frac{kr}{F_l^2 + G_l^2} \quad (20)$$

and

$$S_l = \frac{F_l'F + G_l'G}{F_l^2 + G_l^2} \quad (21)$$

We can also introduce the scattering phase shifts $\Omega_l$,

$$\Omega_l = \omega_l - \tan(F_l / G_l) \qquad (22)$$

which enter into the relation

$$\frac{I_l}{O_l} = e^{2i\Omega_l} \qquad (22)$$

## The multi-channel R-matrix theory

The simple case described above gives an introduction into the basic ideas of the R-matrix theory. We can now consider a more complex case where various channels are involved in the reaction. Following Vogt (2004) this is illustrated in Figure O.1 for the $^8$Be as a compound nucleus.

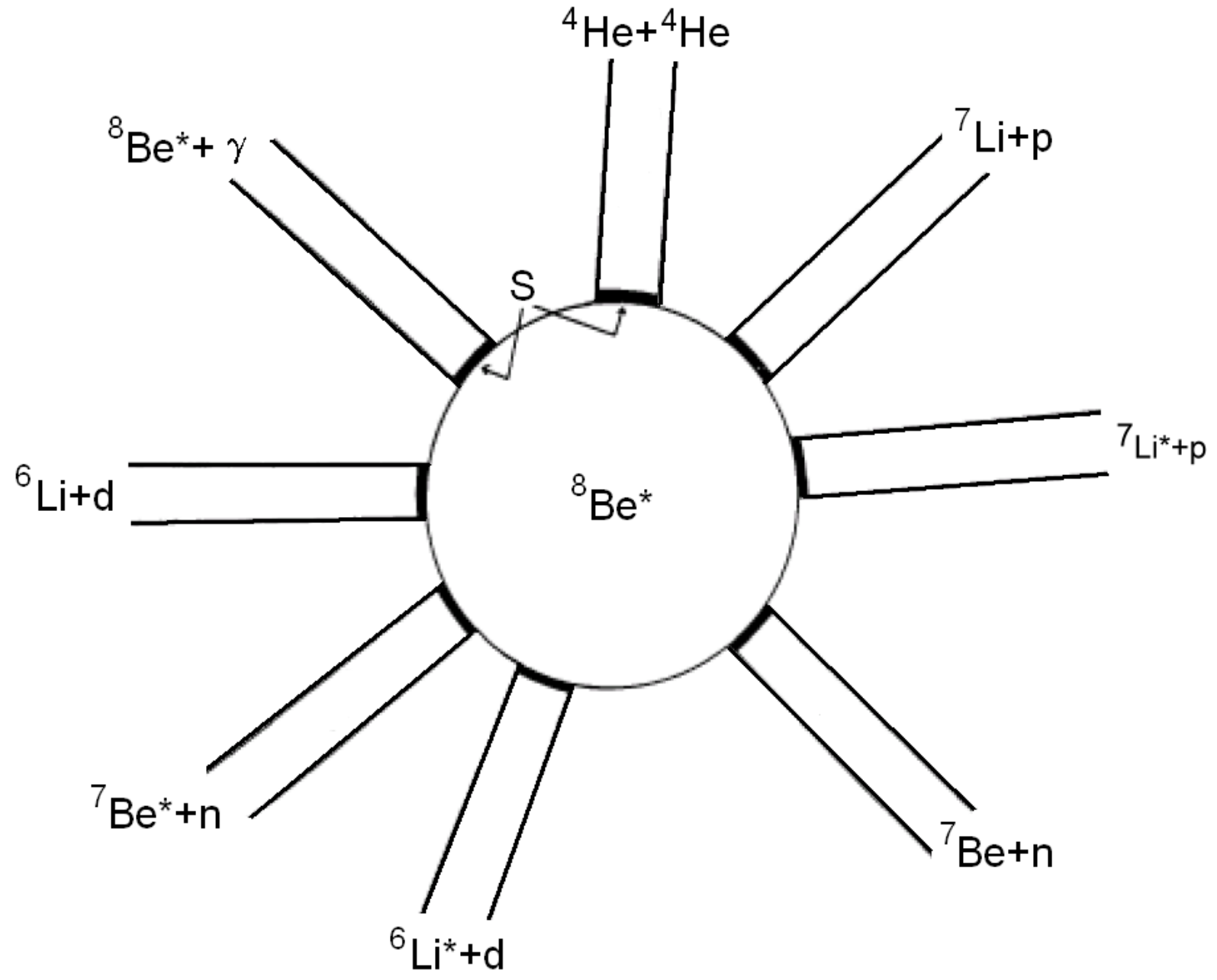

Figure O.1. A schematic diagram of a few contributing channels leading to or resulting from the formation of the compound nucleus $^8$Be*. The internal region is limited by the potential $V(r)$. The thick lines at the surface mark the surface $S$ for each channel. (Vogt 2004.)

A channel is defined by quantum numbers

$$c \equiv (\alpha, l, s, J, M_J)$$

where

$$\hat{s} = \hat{I} + \hat{i}$$

$$\hat{J} = \hat{l} + \hat{s}$$

with $\hat{I}$ and $\hat{i}$ being the intrinsic spins of the two particles in each channel.

Internal region

As before, the internal wave function $\psi$ corresponding to energy $E$ can be expressed as a sum of the resonant states $X_\lambda$ corresponding to energy $E_\lambda$,

$$\psi = \sum_\lambda C_\lambda X_\lambda \tag{24}$$

satisfying the boundary condition

$$\left[ r_c \frac{dX_\lambda(r_c)}{dr_c} \right]_{r_c = a_c} = b_c X_\lambda (r_c = a_c) \tag{25}$$

which may be compared with eqn (6).

It can be shown that coefficients $C_\lambda$ are given by

$$C_\lambda = \frac{\hbar^2}{2m_c a_c} \frac{\sum_c \gamma_{\lambda c}(r_c \varphi_c' - b_c \varphi_c)}{E_\lambda - E} \tag{26}$$

where $\gamma_{\lambda c}$ is the reduced width amplitude.

This leads to an expression containing R-matrix

$$\left( \frac{\hbar^2}{2m_c a_c} \right) \varphi_c = \sum_{c'} \left( \frac{\hbar^2}{2m_{c'} a_{c'}} \right) R_{cc'} [r_{c'} \varphi_{c'}' - b_c \varphi_{c'}] \tag{27}$$

with

$$R_{cc'} = \sum_\lambda \frac{\gamma_{\lambda c} \gamma_{\lambda c'}}{E_\lambda - E}$$

which resembles eqn (8). (28)

***External region***

The external wave function may be expressed as

$$\psi = \sum_c \psi_c \varphi_c \tag{29}$$

where $\psi_c$ are the wave functions containing spin-dependent components and functions describing internal excitations of the interacting pair in a given channel, and

$$\varphi_c = \left( \frac{1}{\upsilon_c} \right)^{1/2} (A_c I_c - B_c O_c) \tag{30}$$

with $A_c$ and $B_c$ being arbitrary coefficients and $I_c$ and $O_c$ the incoming and outgoing functions in a given channel $c$.

If we define the collision matrix as

$$B_c \equiv \sum_{c'} U_{cc'} A_{c'} \tag{31}$$

then

$$\varphi_c = \left(\frac{1}{\upsilon_c}\right)^{1/2}\left[A_c I_c - \left(\sum_{c'} U_{cc'} A_{c'}\right) O_c\right] \tag{32}$$

If we follow the same prescription as in the single-channel theory, i.e. if we match the logarithmic derivatives for the internal and external functions, we shall be able to express the collision matrix in terms of the $R$-matrix:

$$U_{c\grave{c}} = (k_c a_c)^{1/2} O_c^{-1} \sum_{c''} [1 - RL]^{-1}_{cc''} [\delta_{c''c'} - R_{c''c'} L^*_{c'}] I_{c'} (k_{c'} a_{c'})^{-1/2} \tag{33}$$

This formula is analogous to the eqn (17).

Likewise, we shall find that the differential cross section and the angle-integrated cross sections can be expressed by equations resembling the formulae (1) and (2), respectively.

The differential cross section is given by:

$$\frac{d\sigma_{\alpha's';\alpha s}}{d\Omega} = \frac{1}{k_\alpha^2 (2s+1)} \sum_{L=0}^{\infty} B_L(\alpha's' : \alpha s) P_L(\cos\theta) \tag{34}$$

where

$$\begin{aligned} B_L(\alpha's' : \alpha s) = \frac{(-)^{s'-s}}{4} & \sum_{J_1 J_2 l_1 l_2 l_1' l_2'} i^{l_1 - l_2 - L} Z(l_1 J_1 l_2 J_2, sL) i^{l'_1 - l'_2 - L'} Z(l_1' J_1 l_2' J_2, s'L) \\ \times \mathrm{Re}\Big[ & \left(\delta_{\alpha\alpha'} \delta_{l_1 l_1'} \delta_{ss'} - U_{\alpha's'l_1';\alpha s l_1^{J_1}}\right)^* \left(\delta_{\alpha'\alpha} \delta_{l_2 l_2'} \delta_{s's} - U_{\alpha's'l_2';\alpha s l_2^{J_2}}\right)\Big] \end{aligned} \tag{35}$$

with $Z$ coefficients being the products of Clebsch-Gordan coefficients.

The angle-integrated cross section can be expressed as:

$$\begin{aligned} \sigma_{\alpha's';\alpha s} &= \frac{4\pi}{(2s+1)k_\alpha^2} B_0(\alpha's';\alpha s) \\ &= \frac{\pi}{(2s+1)k_\alpha^2} \sum_{J=0}^{\infty} \sum_{l=|J-s|}^{J+s} \sum_{l'=|J-s'|}^{J+s'} (2J+1) \left|\delta_{\alpha'\alpha} \delta_{l'l} \delta_{s's} - U_{\alpha's'l';\alpha s l^J}\right|^2 \end{aligned} \tag{36}$$

The total cross section has the form:

$$\sigma_T(\alpha l) = \frac{\pi}{(2I+1)(2i+1)k_\alpha^2} \sum_{J=0}^{\infty} (2J+1) 2\,\mathrm{Re}\left[1 - U_{\alpha's'l';\alpha s l^J}\right] \tag{37}$$

## Appendix O

### List of my publications as of early December 2016

---

Note: Updates in http://home.iprimus.com.au/nielsens/RonNielsenPublications.pdf